\documentclass[10pt, reqno]{amsart}
\usepackage{amssymb}
\usepackage{graphicx}
\usepackage{xcolor} 
\usepackage{tensor}

\usepackage[margin=1in, marginpar=.6in]{geometry} 

\usepackage{enumitem}

\allowdisplaybreaks				

\usepackage[linktocpage=true]{hyperref}

\usepackage{slashed}			

\setlist[enumerate]{leftmargin=*}
\setlist[itemize]{leftmargin=*}

\definecolor{green}{rgb}{0,0.8,0} 

\newtheorem{maintheorem}{Main Theorem}
\newtheorem{theorem}{Theorem}[section]
\newtheorem{corollary}[theorem]{Corollary}

\newtheorem{lemma}[theorem]{Lemma}
\newtheorem{proposition}[theorem]{Proposition}
\theoremstyle{definition}
\newtheorem{definition}[theorem]{Definition}
\newtheorem{example}[theorem]{Example}
\theoremstyle{remark}
\newtheorem{remark}[theorem]{Remark}
\numberwithin{equation}{section}
\newcommand{\relphantom}[1]{\mathrel{\phantom{#1}}}

\makeatletter
\newcommand{\nrm}{\@ifstar{\nrmb}{\nrmi}}
\newcommand{\nrmi}[1]{\Vert{#1}\Vert}
\newcommand{\nrmb}[1]{\left\Vert{#1}\right\Vert}
\newcommand{\abs}{\@ifstar{\absb}{\absi}}
\newcommand{\absi}[1]{\vert{#1}\vert}
\newcommand{\absb}[1]{\left\vert{#1}\right\vert}
\newcommand{\brk}{\@ifstar{\brkb}{\brki}}
\newcommand{\brki}[1]{\langle{#1}\rangle}
\newcommand{\brkb}[1]{\left\langle{#1}\right\rangle}
\newcommand{\set}{\@ifstar{\setb}{\seti}}
\newcommand{\seti}[1]{\{#1\}}
\newcommand{\setb}[1]{\left\{ #1\right\}}
\makeatother

\newcommand{\td}[1]{\widetilde{#1}}
\newcommand{\br}[1]{\overline{#1}}
\newcommand{\ul}[1]{\underline{#1}}

\newcommand{\VERT}[1]{{\left\vert\kern-0.25ex\left\vert\kern-0.25ex\left\vert #1
    \right\vert\kern-0.25ex\right\vert\kern-0.25ex\right\vert}}

\DeclareMathOperator{\supp}{supp}

\newcommand{\aeq}{\simeq}
\newcommand{\aleq}{\lesssim}
\newcommand{\ageq}{\gtrsim}

\newcommand{\lap}{\Delta}

\newcommand{\ud}{\mathrm{d}}
\newcommand{\rd}{\partial}
\newcommand{\nb}{\nabla}

\newcommand{\imp}{\Rightarrow}

\newcommand{\impmi}{\Leftrightarrow}

\newcommand{\0}{\emptyset}

\newcommand{\peq}{\relphantom{=}}			

\newcommand{\alp}{\alpha}
\newcommand{\bt}{\beta}
\newcommand{\gmm}{\gamma}
\newcommand{\Gmm}{\Gamma}
\newcommand{\dlt}{\delta}

\newcommand{\eps}{\epsilon}
\newcommand{\veps}{\varepsilon}

\newcommand{\lmb}{\lambda}

\newcommand{\sgm}{\sigma}
\newcommand{\Sgm}{\Sigma}
\newcommand{\tht}{\theta}
\newcommand{\Tht}{\Theta}

\newcommand{\omg}{\omega}
\newcommand{\Omg}{\Omega}

\newcommand{\bfa}{{\bf a}}

\newcommand{\bfe}{{\bf e}}

\newcommand{\bfg}{{\bf g}}
\newcommand{\bfh}{{\bf h}}

\newcommand{\bfm}{{\bf m}}
\newcommand{\bfn}{{\bf n}}

\newcommand{\bfA}{{\bf A}}
\newcommand{\bfB}{{\bf B}}
\newcommand{\bfC}{{\bf C}}
\newcommand{\bfD}{{\bf D}}

\newcommand{\bfK}{{\bf K}}
\newcommand{\bfL}{{\bf L}}

\newcommand{\bfS}{{\bf S}}
\newcommand{\bfT}{{\bf T}}

\newcommand{\bfX}{{\bf X}}
\newcommand{\bfY}{{\bf Y}}

\newcommand{\bfalp}{\boldsymbol{\alpha}}

\newcommand{\bfgmm}{\boldsymbol{\gamma}}
\newcommand{\bfGmm}{\boldsymbol{\Gamma}}
\newcommand{\bfdlt}{\boldsymbol{\delta}}

\newcommand{\bfxi}{\boldsymbol{\xi}}

\newcommand{\bfomg}{\boldsymbol{\omega}}
\newcommand{\bfOmg}{\boldsymbol{\Omega}}

\newcommand{\bbN}{\mathbb N}

\newcommand{\bbR}{\mathbb R}
\newcommand{\bbS}{\mathbb S}

\newcommand{\bbZ}{\mathbb Z}

\newcommand{\calA}{\mathcal A}
\newcommand{\calB}{\mathcal B}
\newcommand{\calC}{\mathcal C}
\newcommand{\calD}{\mathcal D}

\newcommand{\calF}{\mathcal F}
\newcommand{\calG}{\mathcal G}
\newcommand{\calH}{\mathcal H}
\newcommand{\calI}{\mathcal I}

\newcommand{\calK}{\mathcal K}
\newcommand{\calL}{\mathcal L}
\newcommand{\calM}{\mathcal M}
\newcommand{\calN}{\mathcal N}

\newcommand{\calP}{\mathcal P}
\newcommand{\calQ}{\mathcal Q}
\newcommand{\calR}{\mathcal R}
\newcommand{\calS}{\mathcal S}
\newcommand{\calT}{\mathcal T}
\newcommand{\calU}{\mathcal U}

\newcommand{\calZ}{\mathcal Z}

\newcommand{\frkE}{\mathfrak E}

\newcommand{\frkH}{\mathfrak H}
\newcommand{\frkI}{\mathfrak I}
\newcommand{\frkJ}{\mathfrak J}

\newcommand{\frkL}{\mathfrak L}
\newcommand{\frkM}{\mathfrak M}

\newcommand{\frkR}{\mathfrak R}

\newcommand{\frkc}{\mathfrak c}

\newcommand{\frkf}{\mathfrak f}

\newcommand{\frkh}{\mathfrak h}

\newcommand{\frkm}{\mathfrak m}

\newcommand{\rst}[1]{\left. #1 \right\vert}

\newcommand{\pfstep}[1]{\smallskip \noindent {\it #1.}}

\newcommand{\chf}{\boldsymbol{1}}

\newcommand{\rem}{\varrho}
\newcommand{\tdGmm}{\td{\bfGmm}{}}

\newcommand{\rmet}{\mathring{\bfg}{}}
\newcommand{\rh}{\mathring{\bfh}{}}

\newcommand{\rbfB}{\mathring{\bfB}{}}
\newcommand{\rbfC}{\mathring{\bfC}{}}
\newcommand{\rf}{\mathring{f}{}}
\newcommand{\rF}{\mathring{F}{}}
\newcommand{\rL}{\mathring{L}{}}
\newcommand{\rV}{\mathring{V}{}}
\newcommand{\rW}{\mathring{W}{}}
\newcommand{\rPhi}{\mathring{\Phi}{}}
\newcommand{\rPsi}{\mathring{\Psi}{}}

\usepackage{accents}

\newcommand{\overbullet}[1]{\accentset{\bullet}{#1}}

\newcommand{\cPhi}{\Phi}
\newcommand{\rcPhi}{\overbullet{\cPhi}{}}

\newcommand{\rch}{\overbullet{\bfh}{}}

\newcommand{\rcF}{\overbullet{F}{}}

\newcommand{\rcH}{\overbullet{H}{}}

\newcommand{\rcu}{\overbullet{u}}

\newcommand{\crho}{\check{\rho}}
\newcommand{\hrho}[1]{{}^{[#1]} \rho}

\newcommand{\rsgmm}{\mathring{\slashed{\bfgmm}}{}}
\newcommand{\rssgm}{\mathring{\slashed{\sgm}}{}}
\newcommand{\rsdiv}{\mathring{\slashed{\mathrm{div}}}{}}
\newcommand{\rsh}{\mathring{\slashed{\bfh}}{}}

\newcommand{\rsbfB}{\mathring{\slashed{\bfB}}{}}
\newcommand{\rsbfC}{\mathring{\slashed{\bfC}}{}}
\newcommand{\rslap}{\mathring{\slashed{\lap}}{}}
\newcommand{\rsnb}{\mathring{\slashed{\nb}}{}}

\newcommand{\quasi}{\calH}
\newcommand{\semi}{\calF}

\newcommand{\csemi}{\calG}

\newcommand{\rN}{\mathring{N}{}}
\newcommand{\rquasi}{\mathring{\quasi}{}}
\newcommand{\rsemi}{\mathring{\semi}{}}
\newcommand{\rcsemi}{\mathring{\csemi}{}}

\newcommand{\far}{\mathrm{far}}
\newcommand{\med}{\mathrm{med}}
\newcommand{\near}{\mathrm{near}}
\newcommand{\ext}{\mathrm{ext}}
\newcommand{\early}{\mathrm{early}}
\newcommand{\late}{\mathrm{late}}
\newcommand{\wave}{\mathrm{wave}}

\newcommand{\nonlinear}{\mathrm{nonlinear}}
\newcommand{\linear}{\mathrm{linear}}

\newcommand{\scheq}{\hbox{``}\hskip-2px=\hskip-2px\hbox{''}}		

\newcommand{\f}{\frac}
\def \i {\infty}
\newcommand{\nB}{{\nu_\Box}}
\newcommand{\ls}{\lesssim}
\newcommand{\ep}{\epsilon}
\newcommand{\de}{\delta}
\def \th {\theta}

\newcommand{\urd}{\underline{\partial}}

\newcommand{\tprod}{\mathord{\textstyle \prod}}

\makeatletter
\newcommand{\mbrk}{\@ifstar{\mbrkb}{\mbrki}}
\newcommand{\mbrki}[1]{[ {#1} ]}
\newcommand{\mbrkb}[1]{\left[ {#1} \right]}
\makeatother

\def\XXint#1#2#3{{\setbox0=\hbox{$#1{#2#3}{\int}$}
     \vcenter{\hbox{$#2#3$}}\kern-.5\wd0}}

\begin{document}

\title[Late time tails]{Late time tail of waves \\on dynamic asymptotically flat spacetimes \\of odd space dimensions}
\author{Jonathan Luk}%
\address{Department of Mathematics, Stanford University, CA 94304, USA}%
\email{jluk@stanford.edu}%

\author{Sung-Jin Oh}%
\address{Department of Mathematics, UC Berkeley, Berkeley, CA 94720, USA and KIAS, Seoul, Korea 02455}%
\email{sjoh@math.berkeley.edu}%


\begin{abstract}
We introduce a general method for understanding the late time tail for solutions to wave equations on asymptotically flat spacetimes with odd space dimensions. In particular, for a large class of equations, we prove that the precise late time tail is determined by the limits of higher radiation field at future null infinity.

In the setting of stationary linear equations, we recover and generalize the Price law decay rates. In particular, in addition to reproving known results on $(3+1)$-dimensional black holes, this allows one to obtain the sharp decay rate for the wave equation on higher dimensional black hole spacetimes, which exhibits an anomalous rate due to subtle cancellations. More interesting, our method goes beyond the stationary linear case and applies to both equations on \textbf{dynamical} background and \textbf{nonlinear} equations. In this case, our results can be used to show that in general there is a correction to the Price law rates.

\end{abstract}
\maketitle

\setcounter{tocdepth}{1}
\tableofcontents
\section{Introduction} \label{sec:intro}

In this paper, we develop a general method for understanding the late time tail for solutions to wave equations on asymptotically flat spacetimes with odd space dimensions, which is \textbf{applicable to nonlinear problems on dynamical backgrounds}. In addition to its inherent interest, such information is crucial for studying problems involving the interaction of waves with a spatially localized object. Indeed, our motivation for developing this method comes from the strong cosmic censorship conjecture  in general relativity --- recent works \cite{DL, LO.interior, Sbierski.Teukolsky} suggest that a lower bound on the decay rate of waves is crucial for proving that the interior of generic dynamical black holes is singular. As a particular case, our method recovers and refines the well-known \emph{Price's law} asymptotics for linear problems on stationary backgrounds \cite{AAGPrice, HintzPriceLaw, Price}. Moreover, we also show that the late time tails are in general \textbf{\emph{different}} from the linear stationary case in the presence of nonlinearity and/or a dynamical background.


Taking the point of view of wave equations on general asymptotically conic manifolds \cite{GHS} or wave equations with (critical) inverse square potentials at infinity \cite{Gajic.inverse.square}, the wave equation on an odd space dimensional asymptotically flat spacetime can be viewed as exceptional. More precisely, after separating into spherical harmonics $\phi = \sum_{\ell} \phi_{(\ell)}$ and denoting $\Phi_{(\ell)} = r^{\f{d-1}2} \phi_{(\ell)}$, the wave equation in the usual polar coordinate $(t,r,\th)$ takes the form
\begin{equation}\label{eq:intro.1D.potential}
\Big(-\rd_t^2  + \rd_r^2 - \f {\alp}{r^2} \Big) \Phi_{(\ell)} = 0,\qquad \alp = \left( \ell + \tfrac{d-3}{2} \right) \left(\ell + \tfrac{d-1}{2} \right) .
\end{equation}
For the one-dimensional wave equation \eqref{eq:intro.1D.potential}, in the generic case $\alp \neq k(k+1)$ for $k \in \mathbb Z_{\geq 0}$, a generic solution attains an $O(t^{-(1+\sqrt{1+4\alp})})$ late time tail on spatially compact set.
The wave equations in even space dimension $d$ fall into the generic case, where already on Minkowski space the $\ell$-th spherical harmonic mode generically attains an $O(t^{-1+\sqrt{1+4\alp}}) = O(t^{-(d-1+2\ell)})$ tail. This tail is moreover expected to persist under perturbations; see the recent interesting works \cite{Gajic.inverse.square, Hintz.linear.waves} in the case of stationary perturbations. However, the case for odd space dimensions $d\geq 3$ is different: indeed in this case $\alp = k(k+1)$ for $k = \ell + \f{d-3}2$, which leads to important cancellations causing a faster decay rate.

On the exact Minkowski space in odd space dimensions $d\geq 3$, the strong Huygens principle holds. Thus compactly supported data to the linear wave equation give rise to solutions with \emph{no} late time tails. However, the strong Huygens principle is very unstable (see \cite{Gunther,Mathisson}). Indeed, Price's law-type decay shows that a long-range perturbation of the Minkowskian wave operator may create a nontrivial late time tail for solutions. Notice however that even for Price's law-type asymptotics on stationary spacetimes, we have an $O(t^{-d-2\ell})$ tail, which is still faster than the generic case. Indeed, as we will see in various examples, in general there are different types of subtle cancellations in odd space dimensions which determine the actual late time tail. Nonetheless, \textbf{our main theorem proves that there is a simple algorithmic way to determine the late time tail in general} by computing the higher radiation fields (see \eqref{eq:intro.higher.radiation} below) and their corresponding Minkowskian contribution to late time tails.

We now give a rough statement of our general result (see Theorem~\ref{thm:intro.main}). 
Let $d\geq 3$ be an odd integer. Consider an asymptotically flat $(d+1)$-dimensional Lorentzian manifold $(\calM, \bfg)$ (either on $\mathbb R^{d+1}$ or in a black hole exterior or obstacle problem setting, see Section~\ref{sec:global.assumptions}).  Introduce $(\tau,r,\theta = \{\th^A\}_{A=1}^{d-1})$ coordinates, where $r$ is a radial function (with $r\to \infty$ towards the asymptotically flat end) and $\tau$ is a time function which is null near infinity and timelike on a finite $r$ region. (In a black hole setting, one should think the constant-$\tau$ hypersurfaces are horizon penetrating.) In the large-$r$ region, we will also write $u = \tau$ to emphasize that the time function is null. Consider the following wave equation:
\begin{equation}\label{eq:intro.eq}
\mathcal P\phi := (\bfg^{-1})^{\alp \bt} \bfD_{\alp} \rd_{\bt} \phi + \bfB^{\alp} \rd_{\alp} \phi  + V \phi = \calN(p,\phi) + f,
\end{equation}
where the following assumptions hold:
\begin{enumerate}[label=(\Alph*)]
\item $\bfg$, $\bfB$ and $V$ are \emph{asymptotically flat} as $r\to \infty$ and \emph{ultimately stationary} as $\tau\to \infty$, but are allowed to be possibly dynamical. \label{item:intro.assumption.A}
\item $\calN$ is a nonlinearity which is at least quadratic and is allowed to be quasilinear. In dimension $d=3$, assume also that the classical null condition \cite{Chr.86, Kla.86} holds.
\item $f$ is an inhomogeneity which decays sufficiently fast as $r\to \infty$ or $\tau \to \infty$.
\end{enumerate}
In order to give a slightly simpler first version of the theorem, we will make an additional assumption. It will be weakened in the main theorem of the paper, but will create a number of additional cases to consider.
\begin{enumerate}[label=(\Alph*),start=4]
\item $\bfg$, $\bfB$, $V$ and $f$ are conformally regular in the sense that they all admit an expansion near null infinity of the form $\bfg = \displaystyle\sum_j r^{-j} \mathring{\bfg}_j(u,\th)$, etc. \label{item:intro.assumption.D}
 \end{enumerate}

There are three ingredients necessary for describing the late time tail.
\begin{enumerate}
\item We prove that $\phi$ admits an expansion into higher radiation fields for $r \gtrsim u$:
\begin{equation}\label{eq:intro.higher.radiation}
r^{\f{d-1}2} \phi(u,r,\theta) = \rPhi_0(u,\theta) + r^{-1} \rPhi_1(u,\theta) + r^{-2} \rPhi_2(u,\theta) + \ldots .
\end{equation}
The \textbf{time-decay rate} is determined as follows. For each $j, \ell \in \mathbb Z_{\geq 0}$, define $\rPhi_{(\leq \ell)j}$ to be the projection of $\rPhi_{j}$ to spherical harmonics with degree $\leq \ell$. Suppose $J_{\mathfrak f} \in \mathbb N$ satisfy
\begin{align}
\lim_{u\to \infty} \rPhi_{(\leq j - \f{d-1}2)j}(u,\theta) =0 \hbox{ when }0\leq j \leq J_{\mathfrak f}-1, \label{eq:intro.Jf.def.1}
\end{align}
The time-decay rate of $\phi$ is a finite-$r$ region is then $\tau^{-J_{\mathfrak f} - \f{d-1}2}$, with the leading-order contribution proportional to $\lim_{u\to \infty} \rPhi_{(\leq J_{\mathfrak f} - \f{d-1}2)J_{\mathfrak f}}(u,\theta)$.

We emphasize that in \eqref{eq:intro.Jf.def.1}, it is the only the angular modes up to $j -\f{d-1}2$ that contribute to late time tails. This is related to the fact (see Lemma~\ref{lem:med.corrected}) that for the Minkowskian wave equation, given characteristic data $\phi(0,r,\theta) = r^{-\f{d-1}2-j} \rPhi_{j}(u,\theta)$, no late time tail would be generated if $\rPhi_{j}$ is supported on $\geq j-\f{d-3}2$ spherical modes. In particular, we could assume without loss of generality that $J_{\mathfrak f} \geq \f{d-1}2$ (for otherwise the condition \eqref{eq:intro.Jf.def.1} is vacuous).
\item In the large $r$ region, say $\{r \geq \tau^{1-\de'}\}$, $\phi$ can be approximated by the \textbf{Minkowskian profile} $\varphi^{\bfm}_{J_{\mathfrak f}}[\rPhi_{J_{\mathfrak f}}(\infty)]$, which we now define. Let $\rPhi_{(\leq J_{\mathfrak f}-\f{d-1}2)J_{\mathfrak f}}(u,\theta)$ be as in \eqref{eq:intro.higher.radiation} above. Define $\varphi^{\bfm}_{J_{\mathfrak f}}[\rPhi_{J_{\mathfrak f}}(\infty)]$ to be the solution to the \underline{Minkowskian} wave equation $\Box_\bfm \varphi^{\bfm}_{J_{\mathfrak f}}[\rPhi_{J_{\mathfrak f}}(\infty,\theta)]= 0$ with the following characteristic initial data on $\{u=0\}$
\begin{equation}
\varphi^{\bfm}_{j}[\rPhi_{J_{\mathfrak f}}(\infty)](r,\th) = \begin{cases}
\lim_{u\to \infty}\rPhi_{(\leq J_{\mathfrak f}-\f{d-1}2)J_{\mathfrak f}}(u,\theta) r^{-J_{\mathfrak f}-\f{d-1}2} & \hbox{ if } r \geq 1 \\
0 & \hbox{ if } r \leq \f 12.
\end{cases}
\end{equation}

\item \label{item:intro.assumption.3} The \textbf{spatial profile} of the late time tails in the region $\{r \leq \tau^{1-\de'}\}$ is determined by the \emph{stationary operator}. Consider the linear operator $\calP$ in \eqref{eq:intro.eq} and define $\calP_0$ to be linear operator obtained from $\calP$ by removing all the terms involving $\rd_\tau$ derivatives. (For instance, if $\calP = \Box_{\bfm} + V$, then $^{(\tau)}\calP_0 = \Delta_{\bfe} + V$, where $\Delta_{\bfe}$ is the Euclidean Laplacian; see \eqref{eq:intro.Box.in.terms.of.P0.T}, \eqref{eq:intro.calP.def}.) Assume there is a limit $^{(\infty)}\calP_0 = \lim_{\tau\to \infty} {}^{(\tau)}\calP_0$ which is invertible, and denote this inverse by $^{(\infty)}\calR_0$. 

The precise late time asymptotics in $\{r \leq \tau^{1-\de'}\}$ is then given by $\tau^{-J_{\mathfrak f}-\f{d-1}2}  (1 - {}^{(\infty)}\calR_0 {}^{(\infty)}\calP_0 1)$ with a constant that is proportional to the limit $\rPhi_{(0)J_{\mathfrak f}}(\infty)$.
\end{enumerate}

We are now ready to state our main theorem informally. In order to simplify the exposition, our theorem is formulated \emph{conditionally}. We will prove the sharp decay bounds and precise asymptotics under the assumption that a solution $\phi$ already satisfies some \emph{weak decay bounds}. The weak decay bounds that we assume (see \ref{hyp:sol} in Section~\ref{subsec:assumptions}) in Theorem~\ref{thm:intro.main} are quite standard, and are known to hold in a variety of (possibly dynamical and nonlinear) settings; see Remark~\ref{rem:sol} and examples in Section~\ref{sec:examples}. Separating out the weak decay bounds allows us to focus on the main new aspects the problem.

\begin{theorem}[Main theorem with conformally regular coefficients, informal version]\label{thm:intro.main}
Consider the equation \eqref{eq:intro.eq} with the assumptions \ref{item:intro.assumption.A}--\ref{item:intro.assumption.D} all hold. Let $\phi$ be a solution to \eqref{eq:intro.eq} arising from compactly supported data and assume that it satisfies some weak decay bounds. Suppose $J_{\mathfrak f}$ satisfies \eqref{eq:intro.Jf.def.1}.

Then, the following holds:
\begin{enumerate}
\item (Upper bound) The solution $\phi$ satisfies the upper bound
\begin{equation}\label{eq:intro.main.upper.bound}
|\phi|(\tau,r,\theta) \ls \min\{\tau^{-J_{\mathfrak f} - \f{d-1}2}, \tau^{-J_{\mathfrak f}} r^{-\f{d-1}2} \}.
\end{equation}
\item (Precise late time asymptotics) Let $\rPhi_{(\leq J_{\mathfrak f}-\f{d-1}2)J_{\mathfrak f}}(u,\theta)$ be as in \eqref{eq:intro.higher.radiation} above.

Then there exist constants $c_{J_{\mathfrak f},d} \neq 0$ and $0<\de' \ll \de_{\mathfrak f}$ such that the following holds:
\begin{enumerate}
\item In $\{r \gtrsim \tau^{1-\de'}\}$,
\begin{equation}\label{eq:intro.main.precise.1}
\Big| \phi - \varphi^{\bfm}_{J_{\mathfrak f}}[\rPhi_{(\leq J_{\mathfrak f} - \f{d-1}2)J_{\mathfrak f}}(\infty)] \Big|(u,r,\theta) \ls r^{-\f{d-1}2} \tau^{-J_{\mathfrak f}-\de_{\mathfrak f}}.
\end{equation}
\item In $\{r \ls \tau^{1-\de'}\}$,
\begin{equation}\label{eq:intro.main.precise.2}
\Big| \phi(\tau,r,\theta) - c_{J_{\mathfrak f},d} \rPhi_{(0)J_{\mathfrak f}}(\infty) \tau^{-J_{\mathfrak f}-\f{d-1}2} (1 - {}^{(\infty)}\calR_0 {}^{(\infty)}\calP_0 1) \Big| \ls \tau^{-J_{\mathfrak f}-\f{d-1}2-\de_{\mathfrak f}}.
\end{equation}
\end{enumerate}
\end{enumerate}
\end{theorem}

We refer the reader to Section~\ref{subsec:assumptions} for the precise assumptions we need, and to Main Theorem~\ref{thm:upper} and Main Theorem~\ref{thm:lower} in Section~\ref{subsec:mainthm} for a precise statement.

\begin{remark}[Decay rate and final asymptotic charges along null infinity]
As indicated in Theorem~\ref{thm:intro.main}, the late-time asymptotics of a solution is determined by a finite collection of numbers $\rPhi_{(\leq J_{\mathfrak f} - \f{d-1}2)J_{\mathfrak f}}(\infty)$ which can be thought of as ``final asymptotic charges'' at null infinity. In this special case of $d=3$ and $J_{\mathfrak f} = 2$, these charges are already known to be relevant to late-time asymptotics  in the physics literature \cite{BR.dynamical, Poisson.decay}, and have also played an important role in our previous works on late-time asymptotics in spherically symmetric, but possibly dynamical and nonlinear, settings \cite{LO1,LO.instab,LO.exterior}.

In the stationary setting such as on stationary black holes, these asymptotic charges can be rephrased in terms of the initial data \cite[Section~1.6]{AAG2018}, but as we will see in \cite{LO.part2}, it is more natural to compute these asymptotic charges along null infinity in dynamical and/or nonlinear settings.
\end{remark}

\begin{remark}[Failure of peeling and the assumption \ref{item:intro.assumption.D}]\label{rmk:peeling}
Motivated by the failure of peeling for spacetimes arising in general relativity \cite{CHRISTODOULOU2002,Friedrich.peeling,dGlK2022,KehI,KehII,KehIII}, our main theorem also allows for the failure of the assumption \ref{item:intro.assumption.D}. In this case, we assume the coefficients of the equations to merely have a polyhomogeneous expansion in $r^{-1}$ near null infinity (i.e., the expansion may include $r^{-j} \log^k r$ terms), in which case the statement of the theorem (as well as the proof) requires suitable modifications; see Section~\ref{sec:intro.log} and Section~\ref{sec:main.theorem} for details.
\end{remark}

\begin{remark}[Contribution from data and the inhomogeneous term]
In Theorem~\ref{thm:intro.main}, we have chosen to consider the case where $f$ (in \eqref{eq:intro.eq}) and the data decay sufficiently fast so that they do not contribute to the late time asymptotics. In our main theorem, we also allow for the case where their contribution dominates. The proof is only requires mild modifications; See Section~\ref{sec:assumption.data.inho.sol}.
\end{remark}

\begin{remark}[Higher derivative estimates]
Our result also gives higher derivative estimates, as well as higher ``vector field bounds'' (see \eqref{eq:intro.vector.field}). In particular, if we denote by $\bfT$ the vector field corresponding to the asymptotic stationarity in the $\tau \to \infty$ limit, we show that every $\bfT$ derivative gives an additional power of $\tau^{-1}$ of decay.
\end{remark}

\begin{remark}[Improved decay for higher angular modes on spherically symmetric background] \label{rmk:intro.higher.ell}
If $\bfg$, $\bfB$, $V$ and $f$ are spherically symmetric, then we can consider solutions $\phi$ to \eqref{eq:intro.eq} supported on spherical harmonics of degree $\geq \ell$, i.e., $\phi = \phi_{(\geq \ell)}$. In this case, $J_{\mathfrak f} \geq \f{d-1}2 + \ell$ and moreover the upper bound \eqref{eq:intro.main.upper.bound} can be improved to
\begin{equation}
\min\{ \tau^{-J_{\mathfrak f}-\f{d-1}2-\ell} r^\ell, \tau^{-J_{\mathfrak f}} r^{-\f{d-1}2}\}.
\end{equation}
Additionally, in this case \eqref{eq:intro.main.precise.2} can be replaced by
\begin{equation}
\Big| \phi(\tau,r,\theta) - c_{J_{\mathfrak f},d,\ell} \rPhi_{J_{\mathfrak f}}(\infty) \tau^{-J_{\mathfrak f}-\f{d-1}2-\ell} \Big(r^\ell Y_{(\ell)} - \calR_0 \calP_0 (r^\ell Y_{(\ell)}) \Big) \Big| \ls \tau^{-J_{\mathfrak f}-\f{d-1}2-\ell-\de_{\mathfrak f}} r^\ell
\end{equation}
for some eigenfunction $Y_{(\ell)} = Y_{(\ell)}(\th)$ of the Laplacian on the unit round sphere satisfying $-\mathring{\slashed{\Delta}} Y_{(\ell)} = \ell (\ell+d-2) Y_{(\ell)}$.

See Main Theorem~\ref{thm:upper-sphsymm} and Main Theorem~\ref{thm:lower-sphsymm} for precise statements.
\end{remark}

\begin{remark}[Sharp asymptotics for generic data]\label{rmk:genericity}
For specific settings of interest, we would like to obtain the sharp asymptotics for solutions arising from \emph{generic} initial data. On the other hand, Theorem~\ref{thm:intro.main} gives the asymptotics in terms of $J_{\mathfrak f}$ and the limit $\lim_{u\to \infty} \Phi_{(\leq J_{\mathfrak f} - \f{d-1}2)J_{\mathfrak f}}(u,\theta)$. Thus, in order to prove sharp asymptotics for generic solutions, we need to show that for an appropriate problem-dependent $J_{\mathfrak f}$, the limit $\lim_{u\to \infty} \Phi_{(\leq J_{\mathfrak f} - \f{d-1}2)J_{\mathfrak f}}(u,\theta) \neq 0$. This can in turn be achieved for many problems of interest --- including dynamical and nonlinear settings --- on a case by case basis. See Section~\ref{sec:intro.consequences} for examples.
\end{remark}

\begin{remark}[Lower bounds on the spatial profile]
To obtain a lower bound from Theorem~\ref{thm:intro.main}, we also need a lower bound on $\psi = 1 - {}^{(\infty)}\calR_0 {}^{(\infty)}\calP_0 1$, which can be established in problem-specific manner. There are a few cases where this is particularly simple: (1) if $V\equiv 0$ in \eqref{eq:intro.eq}, then $\psi \equiv 1$; (2)  if $\calP_0$ is elliptic (but $V$ not necessarily vanishing), then an averaged lower bound can be established using unique continuation type results \cite{LM.unique.continuation}.
\end{remark}

\begin{remark}[Systems of wave equations]\label{rmk:systems}
For expositional purposes, we have only chosen to state Theorem~\ref{thm:intro.main} (as well as its precise version in Main Theorem~\ref{thm:upper} and Main Theorem~\ref{thm:lower}) for scalar equations. However, the theorem and its proof apply equally well for systems of (possibly nonlinear) wave equations, where $\phi: \calM \to \mathbb R^N$ for any $N \in \mathbb N$. In that case, \eqref{eq:intro.Jf.def.1} is required to hold componentwise, and the conclusions \eqref{eq:intro.main.upper.bound}, \eqref{eq:intro.main.precise.1} and \eqref{eq:intro.main.precise.2} are also to be understood componentwise.
\end{remark}

The remainder of the introduction will be structured as follows. In \textbf{Section~\ref{sec:intro.consequences}}, we discuss some corollarites of the main theorem, including Price's law for linear stationary equations. We will also highlight some consequences in the non-stationary and nonlinear settings which will be further discussed in \cite{LO.part2}. In \textbf{Section~\ref{sec:ideas}}, we sketch the main ideas of the proof. In \textbf{Section~\ref{sec:related.works}}, we discuss some related works.

Finally, in \textbf{Section~\ref{sec:outline}}, we will give an outline of the remainder of the paper.

\subsection{Corollaries of the main theorem}\label{sec:intro.consequences}

In this subsection, we consider some immediate applications of Theorem~\ref{thm:intro.main}. First, in \textbf{Section~\ref{sec:intro.general.upper.bound}}, we derive some upper bound results which hold for a very general class of equations. These are in general sharp and are often achieved on dynamical backgrounds. For \emph{stationary} linear equations, however, there is an improved upper bound and the decay rate is at least as fast as that given by Price's law. This can be viewed as a consequence of the conservative structure for the higher radiation fields \eqref{eq:intro.higher.radiation} in the stationary setting. We will discuss the improved decay in \textbf{Section~\ref{sec:intro.price.upper.bound}} (see also Section~\ref{sec:price}).

As can be seen from the statement of Theorem~\ref{thm:intro.main}, in addition to obtaining upper bounds, we can determine the precise late time tails. In specific settings of interest, the determination of the precise late time tails reduces to an (often simple) analysis of the higher radiation fields in \eqref{eq:intro.higher.radiation}. We will defer to \cite{LO.part2} for a thorough discussion of various settings on a case-by-case basis, but we will highlight some results in \textbf{Section~\ref{sec:intro.anomalous.etc}}.

\subsubsection{General upper bounds}\label{sec:intro.general.upper.bound}

The first consequences of Theorem~\ref{thm:intro.main} are general upper bounds that hold for a large class of equations. We first consider the case (for both linear and nonlinear equations) where the coefficients are conformally regular as in Theorem~\ref{thm:intro.main}. Notice that the condition \eqref{eq:intro.Jf.def.1} is vacuous when $J_{\mathfrak f} \leq \f{d-1}2$ (because of the projection to $\leq j - \f{d-1}2$ spherical modes). In particular, we can apply Theorem~\ref{thm:intro.main} with $J_{\mathfrak f} = \f{d-1}2$ to obtain
\begin{theorem}[Minimal decay rate of $d-1$ with conformally regular coefficients] \label{thm:decay.least}
For the class of equations with conformally regular coefficients as in Theorem~\ref{thm:intro.main}, at least the following decay rate holds:
$$|\phi |(\tau,r,\th) \ls \min\{\tau^{-(d-1)}, r^{-\f{d-1}2} \tau^{-\f{d-1}2}\}.$$
\end{theorem}
In the general linear (but still possibly dynamical) case, Hintz\footnote{The work \cite{Hintz.linear.waves} is restricted to non-trapping settings without horizons. For the more general case, see the forthcoming \cite{HV23a}.} \cite{Hintz.linear.waves} proved a similar upper bound, up to arbitrarily small loss.

The decay rate in Theorem~\ref{thm:decay.least} is in general sharp (see \cite{LO.part2} for some examples). Notice that this also coincide with the sharp decay rate in even space dimensions. Here, it is important to assume that the coefficients are conformally regular; otherwise, the decay rate could be much slower! We again refer the reader to \cite{LO.part2} for examples.

If we impose better $r$-decay for the coefficients in linear homogeneous equations $\mathcal P \phi = 0$ (for $\mathcal P$ in \eqref{eq:intro.eq}), then we also obtain faster decay for the solution. This can be viewed as saying that stronger asymptotically flat conditions imply that the solutions decay more like their Minkowskian counterparts (which are super-polynomial by the strong Huygens principle). More precisely, the following theorem holds:
\begin{theorem}[Improved decay for linear equations with improved $r$-decay for the coefficients]\label{thm:decay.least.2}
Suppose there exists $j_0 \in \mathbb N\cup \{0\}$ such that $\bfg^{-1} - \bfm = O(r^{-1-j_0})$, $\bfB = O(r^{-2-j_0})$ and $V = O(r^{-3-j_0})$ (with suitable consistent higher derivative bounds and improved decay for certain components of $\bfg$ and $\bfB$; see Section~\ref{sec:main.theorem}). Then at least the following decay rate holds:
$$|\phi|(\tau,r,\th) \ls  \min\{\tau^{-(d-1)}, \tau^{-(j_0+2)-\f{d-1}{2}}, r^{-\f{d-1}2} \tau^{-\f{d-1}2}, r^{-\f{d-1}2} \tau^{-(j_0+2)} \}.$$
\end{theorem}
This can be easily inferred because in the recurrence equations for the higher radiation fields (see \eqref{eq:recurrence-jk}), the $\rd_u \Phi_{j}$ does not involve any contribution from $\bfg^{-1} - \bfm$, $\bfB$ or $V$ when $j \leq j_0 +1$. Thus, we can apply Theorem~\ref{thm:intro.main} with $J_{\mathfrak f} = \max\{j_0+2, \f{d-1}2\}$, which gives the result in Theorem~\ref{thm:decay.least.2}.

In the special case where $d=3$ and $j_0=0$, this recovers the result of \cite{MTT}; when $d=3$ and $j_0 \geq 1$, this recovers the result of \cite{szL.linear}. Both of these results apply to dynamical spacetimes.

Like Theorem~\ref{thm:decay.least}, the decay rate in Theorem~\ref{thm:decay.least.2} is in general sharp (see \cite{LO.part2} for some examples).

\subsubsection{Stationary equations and Price's law upper bounds}\label{sec:intro.price.upper.bound}

If we consider linear equations which are \textbf{stationary}, the decay rate is always faster than that given by Theorem~\ref{thm:decay.least}. This one power of additional decay arises from a cancellation in the recurrence equations satisfied by the higher radiation fields.
\begin{theorem}[Price's law upper bound] \label{thm:price}
Consider the linear equation
\begin{equation}\label{eq:intro.linear}
\mathcal P\phi := (\bfg^{-1})^{\alp \bt} \bfD_{\alp} \rd_{\bt} \phi + \bfB^{\alp} \rd_{\alp} \phi  + V \phi = 0,
\end{equation}
where $\bfg^{-1}$, $\bfB$ and $V$ are \textbf{stationary} for large $r$. Then the following holds:
\begin{enumerate}
\item Any solution $\phi$ to \eqref{eq:intro.linear} satisfies the upper bound
$$|\phi |(\tau,r,\th) \ls \min\{\tau^{-d}, r^{-\f{d-1}2} \tau^{-\f{d+1}2}\}.$$
\item If $\bfg^{-1}$, $\bfB$ and $V$ are also \textbf{spherically symmetric}, then any solution $\phi$ to \eqref{eq:intro.linear} which is \textbf{supported on spherical harmonics of degree $\geq \ell$} satisfies the upper bound
$$|\phi |(\tau,r,\th) \ls \min\{\tau^{-d-\ell}, r^{-\f{d-1}2} \tau^{-\f{d+1}2-\ell}\}.$$
\end{enumerate}
\end{theorem}

Theorem~\ref{thm:price} will be proven in Section~\ref{sec:price}. Note that in the particular case $d=3$, part (1) of the theorem recovers the result of \cite{Ta}.



Let us emphasize that 
Theorem~\ref{thm:price} 
does \underline{not} require the coefficients to be conformally regular, in stark contrast to the result in Theorem~\ref{thm:decay.least}.

\subsubsection{Anomalous cancellations, non-stationary and nonlinear effects}\label{sec:intro.anomalous.etc}

In general, the upper bounds proven in Section~\ref{sec:intro.general.upper.bound} and the upper bounds in the stationary linear setting proven in Section~\ref{sec:intro.price.upper.bound} are sharp. In the stationary linear case, the upper bounds in Theorem~\ref{thm:price} are in particular saturated in the very well-studied case of $(3+1)$-dimensional stationary black hole spacetimes, as has been proven Angelopoulos--Aretakis--Gajic \cite{AAG2018,AAGKerr,AAGPrice} and Hintz \cite{HintzPriceLaw} (see also Section~\ref{sec:intro.price.refereces}).

In a companion paper \cite{LO.part2}, we will introduce a method to construct specific perturbations of the higher radiation field, which, when combined with Theorem~\ref{thm:intro.main} (see also Main Theorem~\ref{thm:lower} or Main Theorem~\ref{thm:lower-sphsymm}), will in turn allow us to make statements about the precise decay rates of \emph{generic} solutions in a wide variety of settings. Our construction in \cite{LO.part2} can be viewed as an extension of our previous works \cite{LO1, LO.instab, LO.exterior} in spherical symmetry in $3+1$ dimensions. As a particular case, this also reproves the sharp asymptotics on $(3+1)$-dimensional stationary black hole spacetimes.

On the other hand, unlike in \cite{AAG2018,AAGKerr,AAGPrice,HintzPriceLaw}, where stationarity is used in a crucial way, Theorem~\ref{thm:intro.main} also allows us to probe non-stationary and nonlinear settings. We will defer to \cite{LO.part2} for a thorough discussion of the generic decay rates in various settings. Here, we will only highlight a few phenomena that can be captured by the method.
\begin{enumerate}
    \item (Anomalous cancellations) There are linear stationary settings where the decay rate is faster than that given in Theorem~\ref{thm:price}. Our method can capture the algebraic cancellations present in the higher radiation field and derive the sharp decay rate in these cases.
    \begin{enumerate}
        \item In the stationary linear case, if the coefficients have improved $r$-decay (similar to Theorem~\ref{thm:decay.least.2}), then the solution also exhibits improved decay. Moreover, in the stationary case, the decay rate is faster than the general non-stationary case: for $j_0$ as in Theorem~\ref{thm:decay.least.2}, one would have a decay rate of $\min \{\tau^{-d-j_{0}}, r^{-\f{d-1}2}\tau^{-\f{d-1}2-j_0}\}$ in the stationary case. (See \cite{Morgan} for this result when $d=3$.)
        \item More surprisingly, there can be additional cancellations beyond that in point (a) above. For instance, using our main theorem and an additional analysis of the structure of the higher radiation field in the specific case of $(d+1)$-dimensional Schwarzschild or Reissner--Nordstr\"om spacetimes with $d \geq 5$, we show that the decay rate is faster than those predictly in (a). More precisely, because these spacetimes have improved $r$-decay with $j_0 = d-3$, considerations as in (a) above gives an upper bound of $\min\{\tau^{-2d+3}, r^{-\f{d-1}2} \tau^{-\f{3d}2+\f 52}\}$. However, the decay rate is always faster on these special spacetimes, and the spherically symmetric mode decays with an exact rate of $\min\{\tau^{-3d+5}, r^{-\f{d-1}2} \tau^{-\f{5d-9}2}\}$ on a finite-$r$ region. In fact, for Reissner--Nordstr\"om in $(5+1)$-dimensions, there could even be further cancellations for a specific choice of black hole parameters, giving rise to an even faster decay rate.
    \end{enumerate}
    \item (Non-stationary effects) We show that in general, the decay rates are slower on non-stationary spacetimes, even in some cases where the non-stationary spacetimes settle down to stationarity very rapidly. In particular, as we mentioned above, there are non-stationary examples which saturate the decay rates given in Theorems~\ref{thm:decay.least} and \ref{thm:decay.least.2}.

    It is interesting to note that in the most studied case of $(3+1)$-dimensional black hole spacetimes, the decay rates on stationary or non-stationary spacetimes actually coincide. This coincidence occurs because the upper bounds in Theorem~\ref{thm:decay.least.2} and Theorem~\ref{thm:price} agree.

    However, the above case should be thought of as an accident, and in general, the decay rate is indeed slower on non-stationary spacetimes. As specific examples, we consider (a) dynamical spherically symmetric $(3+1)$-dimensional black holes settling down to Schwarzschild or Reissner--Nordstr\"om (Example~\ref{ex:Schwarzschild.dynamical}) and (b) dynamical higher-dimensional ($d+1$ dimensions with $d\geq 5$) black holes (without symmetry assumptions) settling down to Schwarzschild or Reissner--Nordstr\"om (Example~\ref{ex:higher.d.BH.dynamical}). In case (a), for generic dynamical black holes, generic solutions supported on spherical harmonics of order $\geq \ell$ (for $\ell \geq 1$) decay with a rate of $\min\{\tau^{-2-2\ell}, r^{-1} \tau^{-1-2\ell}\}$, in contrast to the Price law rate of $\min\{\tau^{-3-2\ell}, r^{-1} \tau^{-2-2\ell}\}$. In case (b), as long as the Bondi mass is non-constant, generic solutions decay with a rate of $\min \{\tau^{-\f{3(d-1)}2}, r^{-\f{d-1}2} \tau^{-(d-1)}\}$, which is much slower than the rate discussed in point 1(b) above.
    \item (Nonlinear effects) Our main theorem also applies to a wide range of nonlinear models (including possibly quasilinear ones). In \cite{LO.part2}, we will discuss the sharp asymptotics on various models on a case by case basis, including the examples in Section~\ref{sec:examples}.

    It is often of particular physical interest to consider small-data solutions to nonlinear equations in $(3+1)$ dimensions satisfying the null condition (see Example~\ref{ex:null.condition}). In this case, the general decay rate is no better than $O(\min\{r^{-1}\tau^{-1}, \tau^{-2} \})$ (as is given by Theorem~\ref{thm:decay.least}); see \cite{LO.part2} for explicit examples saturating the upper bound. However, our method also captures additional decay in specific settings. For instance, in the case of constant-coefficient null forms, the decay rate is at least $O(\min\{r^{-1}\tau^{-2}, \tau^{-3} \})$. In particular, small data solutions to the wave map system obey these bounds, but there could be (provable) further improvements depending on the curvature of the target manifold.
\end{enumerate}

As we mentioned earlier, one of our motivations to study the precise tails for nonlinear wave equations comes from the strong cosmic censorship conjecture in general relativity. The conjecture in particular predicts that the interior of generic black holes should be ``singular'' and inextendible in a suitably regular manner. Recent progress  \cite{D1,D2,D3,DL,DafShl,Gleeson,GSNS,LO.instab,LO.interior,LO.exterior,LOSR,LS,MZ.interior,McN,Sbierski.Teukolsky} suggests that the structure of the singularity in the interior of dynamical vacuum black holes is intimately tied to the rate at which the black hole settles down to Kerr.

Since the Einstein equations are nonlinear, and the black hole background in this setting is dynamical, our work suggests that one should take into account the non-stationary and nonlinear contribution to the decay rate. In particular, thinking of dimension $d =7$ or the higher spherical modes $\ell \geq 2$ in dimension $d=3$ as models for higher spin equations, one may expect that generic dynamical black holes settle down to Kerr with a slower --- but more stable --- rate of $\tau^{-6}$ instead of the rate $\tau^{-7}$ predicted by Price's law. We refer the reader to \cite{LO.part2} for further discussions in relation to the strong cosmic censorship conjecture.

\subsection{Ideas of the proof}\label{sec:ideas}

\subsubsection{Simple model problem}

We will first explain our method for a simple model problem (see \eqref{eq:intro.5d.ex} below), which captures some of the difficulties, but is simplified enough for us to sketch the full proof. After studying the model problem, we will discuss in Sections~\ref{sec:intro.further} and \ref{sec:intro.log} further ideas that are used for our more general result.

In what follows, we will use the smooth cutoff functions
\begin{equation}\label{eq:intro.cutoff}
\chi_{>1}(s) \equiv \begin{cases} 0 & s\leq \f 12 \\ 1 & s \geq 1\end{cases},\quad \chi_{>\lambda}(\cdot) := \chi_{>1}(\lambda^{-1}( \cdot)),\quad \chi_{<\lambda} := 1-\chi_{>2\lambda}.
\end{equation}

To set up model problem, we consider the Minkowski spacetime $(\mathbb R^{d+1},\bfm)$ and use the standard radial coordinates $(t,r,\th)$, where $\th \in \mathbb S^{d-1}$. Fix $R_{\far} \geq 1$. Define $u = t - r + 3 R_{\far}$ and let $\tau$ be a smooth ``time'' function\footnote{We note carefully that $\tau$ is not timelike, but we will abuse the terminology for simplicity. See also the discussion in Remark~\ref{rem:Sigma-1-causality}.} such that
\begin{equation*}
	\tau = t - \chi_{>4R_{\far}}(r) (r - 3 R_{\far}).
\end{equation*}

Denote the coordinate partial derivatives in the $(u,r,\th)$  coordinate system by $(\rd_u, \rd_r, \rd_{\th^A})$. Denote $\bfT = \rd_u$ (i.e., $\bfT = \rd_{x^0}$ in  $(x^0,x^1,\ldots,x^d)$ coordinates).

A crucial ingredient for our approach is \emph{vector field regularity}: we denote
\begin{equation}\label{eq:intro.vector.field}
h = O_{\bfGmm}(B) \quad \iff \quad \Big| (\langle \tau \rangle \bfT)^{I_\tau} (\langle r \rangle \rd_r)^{I_r} \bfOmg^I h \Big| \ls_{I_\tau, I_r, I} B,
\end{equation}
where $\bfOmg$ denote the rotations in Minkowski spacetime. In the introduction, we will not keep track of the number of vector field derivatives at all. We only note that each step will require at worst a loss of finitely many vector field derivatives.

Take as our model problem a linear wave equation on $\mathbb R^{d+1}$ with a potential:
\begin{equation}\label{eq:intro.5d.ex}
\Box_\bfm \phi = V\phi,
\end{equation}
where $V$ is globally small, satisfies the asymptotically flat bound $V = O_{\bfGmm}(\langle r \rangle^{-3})$, and admits an expansion near null infinity for all $J \in \mathbb N$:
\begin{equation}\label{eq:intro.5d.ex.V}
V = \sum_{j=3}^J r^{-j} \rV_j(u,\theta) + O_{\bfGmm}(r^{-J-1}),\quad \rV_j(u,\theta) = O_{\bfGmm}(1).
\end{equation}

Consider a solution $\phi$ to \eqref{eq:intro.5d.ex} with smooth and compactly support data on $\{\tau = 1\}$. We will assume that the solution $\phi$ already has some decay with very high vector field regularity, e.g.,
\begin{equation}\label{eq:intro.assumed.bound}
\phi = O_{\bfGmm}(\min\{ \tau^{-\f{d}2}, r^{-\f{d-1}2} \tau^{-\f 12}\}),
\end{equation}
which can be established with ``standard vector field method;'' see for instance \cite{jOjS2023}.

Let $\nB = \f{d-1}2$. In the proof, we will justify an expansion of the form
\begin{equation}\label{eq:intro.idea.expansion}
r^{\nB}\phi = \rPhi_0(u,\theta)+ r^{-1} \rPhi_1(u,\theta) + \ldots + r^{-J_{\mathfrak f}} \rPhi_{J_{\mathfrak f}}(u,\theta) + \rho_{J_{\mathfrak f}+1}(u,r,\theta) \quad \hbox{ in }\calM_\wave\cup \calM_\med,
\end{equation}
where $J_{\mathfrak f}$ is defined such that (see \eqref{eq:intro.Jf.def.1})
\begin{equation}\label{eq:intro.idea.Jf.def}
\lim_{u\to \infty} \rPhi_{(\leq j - \nB)j}(u,\theta) =0 \hbox{ when }0\leq j \leq J_{\mathfrak f}-1.
\end{equation}
By definition, $J_{\mathfrak f} \geq \nB$, but in general, the value of $J_{\mathfrak f}$ could be larger depending on the potential $V$ (as well as the particular solution in question).

Corresponding to Theorem~\ref{thm:intro.main}, we first show that upper bound
$\phi = O_{\bfGmm}(\min\{ \tau^{-J_{\mathfrak f}-\nB},\tau^{-J_{\mathfrak f}} r^{-\nB}\})$ and then give a detailed description of the late time asymptotics of $\phi$, i.e., $\phi\simeq \varphi_{J_{\mathfrak f}}^{\bfm}[\rPhi_{(\leq J_{\mathfrak f}-\f{d-1}2)J_{\mathfrak f}}(\infty)]$ when $r\geq \tau^{1-\de'}$ and $\phi \simeq c_{J_{\mathfrak f},d} \rPhi_{(\leq J_{\mathfrak f}-\f{d-1}2)J_{\mathfrak f}}(\infty)\tau^{-J_{\mathfrak f}-\nB}(1 - ^{(\infty)}\calR_0{}^{(\infty)}\calP_0 1)$ when $r \leq \tau^{1-\de'}$.

\subsubsection{Main structure of the proof}\label{sec:ideas.structure}

Our proof is \textbf{completely based on arguments in physical space} and it is structured in an \textbf{iteration} argument. The iteration (cf.~\cite{DRPL, MTT}) allows us to start with the assumed decay rate and improve the rate in each iteration until we arrive at the sharp rate. In each iteration, we divide the spacetime into three regions:
$$\calM_{\wave} = \{ r \geq 4u \},\quad \calM_{\med} = \{ R_{\far}\leq r \leq 400 u \},\quad \calM_{\near} = \{ r\leq 2R_{\far}\}.$$

Here is a brief outline of the strategy in each region and the tools involved in the iteration:
\begin{enumerate}
\item In $\calM_{\wave}$, the key is to justify the expansion \eqref{eq:intro.idea.expansion} and to prove decay estimates for the higher radiation fields $\rPhi_j$ and the error $\rho_J$. In particular, we use the the conformally regular vector field $\bfK = r^2 \rd_r$ to single out the terms in the expansion of $r^\nB \phi$. Here are the important ingredients for the analysis:
\begin{enumerate}
\item The higher radiation fields $\rPhi_j$ satisfy \emph{recurrence equations} (see \eqref{eq:intro.recurrence.orig} and \eqref{eq:recurrence-jk}). A particularly important structural property in the recurrence equations is the \emph{Newman--Penrose cancellation} (see discussions in Step~3 of Section~\ref{sec:ideas.wave}), which is crucial for analyzing the low angular modes of both the higher radiation fields and the error terms.
\item For the high angular modes, we use the \emph{Angelopoulos--Aretakis--Gajic positivity}. The key point here is that in the energy identities for the Dafermos--Rodnianski $r^p$ estimates for $\bfK^j \rho_J$, certain bulk terms are non-negative after projection to suitable high angular modes.
\end{enumerate}
\item In $\calM_{\med}$, we view the equation as a perturbation of the Minkowskian wave equation. Here, due to the \emph{strong Huygens principle} (!) of the flat wave equation, the main contribution comes from the wave zone; we carry out explicit Minkowskian calculations to capture this main contribution. In particular, the explicit computation shows that some terms in the expansion in \eqref{eq:intro.idea.expansion} gives no contribution. The error terms in this region are bounded using simple Minkowskian estimates.
\item In $\calM_{\near}$, we use that $\bfT$ gives extra decay (by the vector field bound \eqref{eq:intro.vector.field}) to treat the equation as $\calP_0\phi = \mathrm{error}$ after localizing into $\calM_{\near}\cup \calM_{\med}$ (where $\calP_0$ is defined in assumption \ref{item:intro.assumption.3} of Theorem~\ref{thm:intro.main}). We then use sharp weighted elliptic estimates associated to $\calP_0$ to control $\phi$.
\end{enumerate}
The analysis in $\calM_{\wave}$ and $\calM_{\med}$ are new, while the main idea for the analysis in $\calM_{\near}$ is borrowed from \cite{MTT, AAG2020, AAGPrice}. In particular, while the analysis in $\calM_{\wave}$ shares some elements in \cite{AAGPrice}, it is important for non-stationary and nonlinear settings that our scheme avoids inverting $\bfT$ derivatives; this is in turn achieved by being able to estimate $\rPhi_j$ directly for $j \geq \f{d+1}2$.

\subsubsection{Wave zone}\label{sec:ideas.wave}

As mentioned above, the key to understanding the wave zone is to justify, control and use an expansion of the form \eqref{eq:intro.idea.expansion}. In the process, we need to bound the higher radiation fields $\rPhi_j$, $j=0,\ldots, J_{\frkf}$, which are defined to be solutions to the \emph{recurrence equations}:
\begin{equation}\label{eq:intro.recurrence.orig}
\rd_u \rPhi_{j} = - \f 1{2j}\Big((j-1)j - \f{(d-1)(d-3)}4 + \rslap \Big) \rPhi_{j-1} + \f 1{2j} \sum_{j' + j'' = j+1} \rV_{j'} \rPhi_{j''}.
\end{equation}
In the analysis, we will justify the expansion, control the higher radiation fields, and bound the error terms. This will be carried out in an iteration argument: in each step of the iteration, we add a term in the expansion, as well as improve the decay rates that we prove.

\pfstep{Step~1: Initial expansion} To start the analysis in the wave zone, we first prove an expansion (recall $\nB = \f{d-1}2$)
\begin{equation}\label{eq:intro.initial.expansion}
r^{\nB} \phi = \rPhi_0 + \rho_1,\quad \Phi_0(u,\theta) = \lim_{r\to \infty} r^{\nB} \phi(u,r,\theta), \quad \rPhi_0 = O_{\bfGmm}(u^{-\f 13}),\quad \rho_1 = O_{\bfGmm}(r^{-\f 13}).
\end{equation}
This can be justified using the Dafermos--Rodnianski $r^p$ method \cite{DRNM}, which provides integral identities for showing that (for $p \in (0,2)$)
$$\int_{\mathcal C_u} r^p (\rd_r (r^{\nB} \phi))^2 \, \ud \mathring{\slashed{\sigma}}\ud r \ls 1.$$
With this, it is easy to justify the existence of the limit for the Friedlander radiation field $\Phi_0(u,\theta)$ (see also \cite[Theorem~1.5]{gM2016}). Moreover, $\rPhi_0 = O_{\bfGmm}(u^{-\f 12})$ by \eqref{eq:intro.assumed.bound}, which is better than needed in \eqref{eq:intro.initial.expansion}. Finally, when using the $r^p$ method again, we can bound
$$\int_{\mathcal C_u} r^p (\rd_r \rho_1)^2 \, \ud \mathring{\slashed{\sigma}}\ud r  = \int_{\mathcal C_u} r^p (\rd_r (r^{\nB} \phi))^2 \, \ud \mathring{\slashed{\sigma}}\ud r \ls 1,$$
for all $p \in (0,2)$, where $\calC_u$ is constant-$u$ null cone (cf.~\eqref{eq:intro.rp} below). This implies, after choosing $p = \f 53$ and losing vector field regularity, that
$$\rho_1 := r^\nB\phi - \Phi_0 = O_{\bfGmm}(r^{-\f{p-1}2}) = O_{\bfGmm}(r^{-\f 13}).$$

\pfstep{Step~2: Iteration argument in the wave zone} For the iteration argument, we start with the following assumption:
\begin{align}
\phi = &\: O_{\bfGmm}(u^{-\alp}) \quad \hbox{in } \calM_\med, \label{eq:intro.wave.interation.1}\\
r^{\nB} \phi = &\: \sum_{j=0}^{J-1} r^{-j}\rPhi_j + \rho_J \quad \hbox{in } \calM_\med \cup \calM_\wave, \label{eq:intro.wave.interation.2}\\
\rPhi_{(\leq j-\nB)j} = &\:  O_{\bfGmm}(u^{j-\alp+\nB}),\quad j = 0,1,\ldots,J-1, \label{eq:intro.wave.interation.3}\\
\rPhi_{(\geq j-\nB+1)j} = &\: O_{\bfGmm}(u^{j-\alp+\nB+1}),\quad j = 0,1,\ldots,J-1, \label{eq:intro.wave.interation.4}\\
\rho_J = &\: O_{\bfGmm}(r^{-\alp+\nB}), \label{eq:intro.wave.interation.5}
\end{align}
where $\rPhi_j$ is defined by the recurrence equation \eqref{eq:intro.recurrence}, and we have chosen $J$ so that $\alp +\f 23 = J + \nB$. In the initial step, we take $J= 1$, $\alp = \nB + \f 13$ so that \eqref{eq:intro.wave.interation.1} and \eqref{eq:intro.wave.interation.2}--\eqref{eq:intro.wave.interation.5} follow from \eqref{eq:intro.assumed.bound} and \eqref{eq:intro.initial.expansion} respectively.

Assume $J \leq J_{\mathfrak f} - 1$. Our goal in this analysis, to be achieved in the steps below, is to improve \eqref{eq:intro.wave.interation.2}--\eqref{eq:intro.wave.interation.5} to replace $J$ by $J+1$ and $\alp$ by $\alp+1$. In other words, our goal is to show that
\begin{align}\label{eq:intro.improved.higher.radiation.goal.J-1}
\rPhi_{(\leq j-\nB)j} = O_{\bfGmm}(u^{j-\alp+\nB-1}),\quad \rPhi_{(\geq j-\nB+1)j} =  O_{\bfGmm}(u^{j-\alp+\nB}),\quad j = 0,1,\ldots,J-1,
\end{align}
and that for $\rho_{J+1}$ defined by $\rho_{J} = r^{-J} \rPhi_{J} + \rho_{J+1}$, where $\rPhi_J$ is defined by the recurrence equation \eqref{eq:intro.recurrence}, it holds that
\begin{align}
\rPhi_{(\leq J-\nB)J} = O_{\bfGmm}(u^{J-\alp+\nB-1})& = O_{\bfGmm}(u^{-\f 13}),\quad \rPhi_{(\geq J-\nB+1)J} = O_{\bfGmm}(u^{J-\alp+\nB}) = O_{\bfGmm}(u^{\f 23}), \label{eq:intro.wave.rho.improved.1}\\
& \rho_{J+1} = O_{\bfGmm}(r^{-\alp+\nB-1}).\label{eq:intro.wave.rho.improved.2}
\end{align}

\pfstep{Step~3: Improvement for the higher radiation fields} We begin with the estimates for $\rPhi_j$. First, we show that
\begin{equation}\label{eq:intro.expansion.improved.1}
\rPhi_j = O_{\bfGmm}(u^{j-\alp+\nB}), \quad j = 0, 1,\ldots, J-1,
\end{equation}
without needing to project to particular spherical modes, which then implies the estimate for the higher angular modes in \eqref{eq:intro.improved.higher.radiation.goal.J-1}. To achieve \eqref{eq:intro.expansion.improved.1}, we observe that $\bfK^{J-1} r^{-j} = (r^2 \rd_r)^{J-1} r^{-j} = 0$ when $j =0,1,\ldots, J-2$. Thus, in $\calM_\med \cap \calM_\wave$ (where $u \sim r$),
$$\rPhi_{J-1} = \f{(-1)^{J-1}}{(J-1)!} \bfK^{J-1} (r^{-J+1} \rPhi_{J-1})= \f{(-1)^{J-1}}{(J-1)!} \bfK^{J-1}(r^{\nB}\phi - \rho_J).$$
We then use the bound $\bfK^{J-1}(r^{\nB}\phi) = O_{\bfGmm}(r^{\nB+J-1} u^{-\alp})$ in $\calM_{\med}$ and $\bfK^{J-1} \rho_J = O_{\bfGmm}(r^{-\alp+\nB+J-1})$ in $\calM_\wave$ (since each $\bfK$ derivative loses one power of $r$ by \eqref{eq:intro.vector.field}). Combining these bounds and remembering that $\rPhi_{J-1}$ is independent of $r$, we obtain \eqref{eq:intro.expansion.improved.1} for $j=J-1$. We then induct in decreasing $j$. For example, for $j=J-2$, we write
$$\rPhi_{J-2} = \f{(-1)^{J-2}}{(J-2)!} \bfK^{J-2} (r^{-J+2} \rPhi_{J-2})= \f{(-1)^{J-2}}{(J-2)!} \bfK^{J-2}(r^{\nB}\phi - \rPhi_{J-1} - \rho_J),$$
and then argue as before, now using also the bound \eqref{eq:intro.expansion.improved.1} for $j= J-1$ that we have just obtained.

The next step is to obtain \eqref{eq:intro.improved.higher.radiation.goal.J-1} for low spherical harmonics. For this we use the recurrence equation \eqref{eq:intro.recurrence.orig} (cf.~\eqref{eq:recurrence-jk}). Plugging the estimates \eqref{eq:intro.5d.ex} and \eqref{eq:intro.expansion.improved.1} into the recurrence equations \eqref{eq:intro.recurrence.orig}, we obtain
\begin{equation}\label{eq:intro.recurrence}
\rd_u \rPhi_{j} = - \f 1{2j}\Big((j-1)j - \f{(d-1)(d-3)}4 + \rslap \Big) \rPhi_{j-1} + O_{\bfGmm}(u^{j-\alp+\nB-2}).
\end{equation}
One important structure in the recurrence equations is what we call the \emph{Newman--Penrose cancellation} (since it relates to the existence of the conserved Newman--Penrose constants \cite{NP1965, NPB1968}), which allows us to prove additional decay. This is the fact that $\Big((j-1)j - \f{(d-1)(d-3)}4 + \rslap\Big) \mathbb S_{(j-\nB)} = 0$, where $\mathbb S_{(j-\nB)}$ is the projection to the $\ell = j-\nB$ spherical harmonics. Thus, since $\lim_{u\to \infty} \rPhi_{(j-\nB)j}(u) = 0$ for all $j = 0,1,\ldots,J-1$ by \eqref{eq:intro.wave.interation.3}, we can integrate from $u=\infty$ to obtain
\begin{equation}
\rd_u \rPhi_{(j-\nB)j} =  O_{\bfGmm}(u^{j-\alp+\nB-2}) \implies \rPhi_{(j-\nB)j} = O_{\bfGmm}(u^{j-\alp+\nB-1}),\quad j=0,1,\ldots,J-1.
\end{equation}
This is our desired improved bound for $\ell = j-\nB$. Now, we can induct in decreasing $j$ to obtain improved bounds for all $\ell \leq j -\nB$. For instance, for $\ell = j-\nB-1$, we have
\begin{equation}
\rd_u \rPhi_{(j-\nB-1)j} = O_{\bfGmm}(1) \rPhi_{(j-\nB-1)j-1} + O_{\bfGmm}(u^{j-\alp+\nB-2}) = O_{\bfGmm}(u^{j-\alp+\nB-2}), \quad j = 0,1,\ldots,J-1,
\end{equation}
where the second equality is obtained using the bound for $\rPhi_{(j-\nB-1)j-1}$ that we have just achieved. As before, we integrate from $u = \infty$ to bound $\rPhi_{(j-\nB-1)j}$. Continuing in this manner, we thus obtain the desired improvement:
\begin{equation}\label{eq:intro.expansion.improved.2}
\rPhi_{(\leq j-\nB)j} =  O_{\bfGmm}(u^{j-\alp+\nB-1}),\quad j = 0,1,\ldots,J-1.
\end{equation}
Combining \eqref{eq:intro.expansion.improved.1} and \eqref{eq:intro.expansion.improved.2} yields \eqref{eq:intro.improved.higher.radiation.goal.J-1}.

Finally, we turn to the estimates for $\rPhi_{J}$, i.e., the bounds in \eqref{eq:intro.wave.rho.improved.1}. The bounds for the low spherical harmonics $\rPhi_{(\leq J- \nB)J}$ can be derived in a similar manner as that for $\rPhi_{(\leq j - \nB) j}$ when $j = 0, \ldots, J-1$. Indeed, we first consider $\ell = J-\nB$, when the Newman--Penrose cancellation is present, to obtain
\begin{equation}\label{eq:intro.improved.eqn.low.modes}
\rd_u \rPhi_{(J-\nB)J} = O_{\bfGmm}(u^{J-\alp+\nB-2}).
\end{equation}
Since $J \leq J_{\frkf} -1$, it holds that $\lim_{u\to \infty}\rPhi_{(J-\nB)J}(u) = 0$ and thus integrating \eqref{eq:intro.improved.eqn.low.modes} gives the bound for $\rPhi_{(J-\nB)J}$ in \eqref{eq:intro.wave.rho.improved.1}. For $\ell \leq J - \nB - 1$, we use the improved bound $\rPhi_{(\leq J-\nB-1)J-1} = O_{\bfGmm}(u^{J-\alp+\nB-2})$ obtained in \eqref{eq:intro.expansion.improved.2} above so that \eqref{eq:intro.recurrence} also implies
\begin{equation}\label{eq:intro.improved.eqn.low.modes.2}
\rd_u \rPhi_{(\leq J-\nB-1)J} = O_{\bfGmm}(u^{J-\alp+\nB-2}).
\end{equation}
Integrating in $u$ as before, we obtain the estimate in \eqref{eq:intro.wave.rho.improved.1} for $\rPhi_{(\leq J-\nB)J}$.

For the $\ell \geq J-\nB+1$ angular modes, \eqref{eq:intro.recurrence.orig} only gives  $\rd_u \rPhi_{(\geq J-\nB+1) J} = O_{\bfGmm}(u^{J-\alp+\nB-1})$. However, this is already sufficient as we only need the weaker bound $\rPhi_{(\geq J-\nB+1)J} = O_{\bfGmm}(u^{J-\alp+\nB})$ in \eqref{eq:intro.wave.rho.improved.1}.

\pfstep{Step~4: Estimates for the remainder term} 
We now turn to the estimate for $\rho_{J+1}$ in \eqref{eq:intro.wave.rho.improved.2}. To this end, we consider the equation satisfied by $\rho_J$. Define the (Minkowskian) operator $Q_0$ by $Q_0(r^{\nB} \phi) = r^{\nB}\Box_{\bfm} \phi$, where $\nB = \f{d-1}2$. In the Bondi--Sachs type $(u,r,\theta)$ coordinates, this is given explicitly by
$$Q_0 = - 2\rd_u \rd_r + \rd_r^2 - \f{(d-1)(d-3)}{4}\f 1{r^2} + \f 1{r^2} \rslap.$$
(In particular, the conjugation by $r^{\nB}$ removes the $\rd_u$ term in the equation.) Using \eqref{eq:intro.wave.interation.3}, \eqref{eq:intro.wave.interation.5} and the improved estimate \eqref{eq:intro.expansion.improved.1} obtained above (cf.~Lemma~\ref{lem:exp-error-wave}), we deduce that
\begin{equation}\label{eq:intro.Q0.rhoJ}
Q_0 \rho_J = -E_{J;\Box} + r^{-J-1} \mathring{e}_{J+1}(u,\theta)  + O_{\bfGmm}(r^{-J-2}u^{-\f 13})\quad \hbox{ in }\calM_\wave,
\end{equation}
where
\begin{equation}\label{eq:intro.EJBox}
E_{J;\Box}:=r^{-J-1} \Big[ ((J-1) J + (\nu_{\Box}-1)\nu_{\Box} + \rslap)  \rPhi_{J-1} \Big],\quad \hbox{and} \quad \mathring{e}_{J+1} = O_{\bfGmm}(u^{-\f 43}).
\end{equation}

Let $\bfK = r^2 \rd_r$ in $(u,r,\theta)$ coordinates as before. Define the operators $Q_j$ ($j \in \mathbb N$) inductively by requiring $Q_j \bfK^j = (\bfK + 2r)^j Q_0$ and define $Q_{(\ell)j} = Q_j \mathbb S_{(\ell)}$, where $\mathbb S_{(\ell)}$ is the projection to spherical harmonics of degree $\ell$. Direct computation shows that (see Lemma~\ref{lem:Qj})
\begin{equation}\label{eq:intro.Qjl}
Q_{(\ell)j} =  - 2 \rd_{u} \rd_{r} + \rd_{r}^{2} - \frac{2j}{r} \rd_{r} + \Big( \big(j + \frac{d-1}{2}\big)\big(j - \frac{d-3}{2} \big) - \ell (\ell + d-2) \Big) \frac{1}{r^{2}}.
\end{equation}

We will separately treat two different cases, namely that of the low angular modes $\ell \leq J-\nB$ and of the high angular modes $\ell \geq J - \nB+1$. (Note that the $\ell \leq J-\nB$ case could be empty.) For the low angular modes, we view \eqref{eq:intro.Qjl} (with $j=\ell+\nB-1$) as a transport equation and control $\rho_{(\ell)j}$ via transport estimates (Step~4(a)). For the high angular modes, we use the \emph{Angelopoulos--Aretakis--Gajic positivity}, which shows that when using the Dafermos--Rodnianski $r^p$ method for $\bfK^J \rho_{(\geq J-\nB+1)J}$, the bulk term has a good sign (Step~4(b)). Interestingly, the two regimes are exactly complementary!

\pfstep{Step~4(a): Estimates for the remainder term for low angular modes} For the angular modes $\ell \leq J-\nB$, we start with \eqref{eq:intro.Q0.rhoJ} but explicitly put in a correction term $r^{-J} \rPhi_J$. Due to the recurrence equation \eqref{eq:intro.recurrence.orig}, we know that the $r^{-J}\rPhi_J$ term removes the contribution $r^{-J-1} \mathring{e}_{J+1}(u,\theta)$ on the right-hand side of \eqref{eq:intro.Q0.rhoJ} at the expense of a term $E_{J+1;\Box}$, defined as in \eqref{eq:intro.EJBox} but with $J$ replaced by $J+1$, i.e., we have, for $\rho_{J+1} = \rho_J - r^{-J} \rPhi_J$,
\begin{equation}\label{eq:intro.Q0.rhoJ.low.with.EJ+1}
Q_0 \rho_{J+1}  = Q_0 (\rho_J - r^{-J} \rPhi_J) =  -E_{J+1;\Box}+ O_{\bfGmm}(r^{-J-2} u^{-\f 13}),
\end{equation}
where $E_{J+1;\Box}$ is defined similarly as \eqref{eq:intro.EJBox}, but with
\begin{equation}\label{eq:intro.EJ+1Box}
E_{J+1;\Box}:=r^{-J-2} \Big[ (J(J+1) + (\nu_{\Box}-1)\nu_{\Box} + \rslap)  \rPhi_{J} \Big].
\end{equation}
Since we have already achieved the improvement of the $\leq J-\nB$ angular modes of $\rPhi_{J}$ (see \eqref{eq:intro.wave.rho.improved.1}), we control the $E_{J+1;\Box}$ term in \eqref{eq:intro.Q0.rhoJ.low.with.EJ+1} and obtain
\begin{equation}\label{eq:intro.Q0.rhoJ.low.removed}
Q_0 \rho_{(\ell)J+1}  = O_{\bfGmm}(r^{-J-2} u^{-\f 13}),\quad \hbox{when $\ell \leq J-\nB$}.
\end{equation}
For each $\ell \leq J - \nB$, we commute \eqref{eq:intro.Q0.rhoJ.low.removed} with $\bfK^{\ell+\nB-1}$ so as to remove the $\f 1{r^2}$ potential term in \eqref{eq:intro.Qjl}. We then obtain
\begin{equation}\label{eq:Qrho.low.commuted}
Q_{(\ell)\ell+\nB-1}(\bfK^{\ell+\nB-1} \rho_{(\ell)J+1}) = O_{\bfGmm}(r^{-J-3+\ell+\nB} u^{-\f 13}).
\end{equation}
To derive \eqref{eq:Qrho.low.commuted}, we used as before that for each action of $\bfK + 2r$, the estimates on the right-hand side of \eqref{eq:intro.Q0.rhoJ.low.removed} is at worst worsened by one power of $r$ (using the vector field bounds \eqref{eq:intro.vector.field} which control $r\rd_r$ derivatives).
Recalling \eqref{eq:intro.Qjl}, it follows that \eqref{eq:Qrho.low.commuted} implies
\begin{equation}\label{eq:Qrho.low.commuted.transport}
\begin{split}
&\: r^{2 (\ell+\nB-1)} (-2\rd_u + \rd_r)(r^{-2 (\ell+\nB-1)} \rd_r (\bfK^{\ell+\nB-1} \rho_{(\ell)J+1})) \\
= &\:   \Big( -2 \rd_u \rd_r + \rd_r^2 - \f{2 (\ell+\nB-1)}r \rd_r \Big) (\bfK^{\ell+\nB-1} \rho_{(\ell)J+1}) \\
= &\: Q_{(\ell)\ell+\nB-1}(\bfK^{\ell+\nB-1} \rho_{(\ell)J+1})= O_{\bfGmm}(r^{-J-3+\ell+\nB} u^{-\f 13}).
\end{split}
\end{equation}
Integrating equation \eqref{eq:Qrho.low.commuted.transport} along the integral curves of $-2\rd_u + \rd_r$ (see Section~\ref{subsec:char}), we obtain
\begin{equation}\label{eq:intro.rho.low.improved}
 \rd_r  \bfK^{\ell+\nB-1} \rho_{(\ell)J+1} = O_{\bfGmm} (r^{-J-3+\ell+\nB} u^{\f 23}) = O_{\bfGmm} (r^{-J-\f 73+\ell+\nB}) \quad \hbox{in }\calM_\wave,
 \end{equation}
where we have used that $u \ls r$ in the wave zone.

We then obtain estimates for $\rho_{(\ell)J+1}$ using \eqref{eq:intro.rho.low.improved} by integrating from $r=\infty$ as follows. By the assumption \eqref{eq:intro.wave.interation.5} on $\rho_J$, for any $j \leq \ell + \nB-1 \leq J-1$, we have $\bfK^{j} \rho_{(\ell)J}(u,r,\theta) =O_{\bfGmm}(r^{-J+j +\f 23}) \xrightarrow{r\to \infty} 0$. Clearly, we also have $\bfK^{j}(r^{-J} \rPhi_J) \xrightarrow{r\to \infty} 0$ (since $\rPhi_J$ is independent of $r$). Hence, $\lim_{r\to \infty} \bfK^{j} \rho_{(\ell)J+1}(u,r,\theta) =0$ for $j \leq \ell + \nB-1 \leq J-1$. We can therefore use \eqref{eq:intro.rho.low.improved} to deduce that $\rho_{(\ell)J+1} = O_{\bfGmm}(r^{-J-\f 13}) = O_{\bfGmm}(r^{-\alp-1+\nB})$, which gives the desired estimates for $\rho_{(\leq J-\nB)J+1}$ in \eqref{eq:intro.wave.rho.improved.2}.

\pfstep{Step~4(b): Estimates for the remainder term for high angular modes} To deal with the high spherical harmonics, we use the Dafermos--Rodnianski $r^p$ method \cite{DRNM}, and the \emph{Angelopoulos--Aretakis--Gajic positivity} observed in \cite{AAG2020, AAGPrice}. The key estimate is as follows: given an equation $Q_j \Psi = F$ (on Minkowski) and for $-4j< p < 2$, we have the following estimate \emph{after projecting to spherical modes $\ell \geq j - \nB+1$}:
\begin{equation}\label{eq:intro.rp}
\begin{split}
&\: \Big( \int_{r \geq 2R} r^p \Big(\rd_r \Psi_{(\geq j-\nB+1)} \Big)^2(U,r) \, \ud r \Big)^{\f 12} \\
\ls &\: \int_1^{U} \Big( \int_{\max\{u, R\}}^\infty r^p F_{(\geq j - \nB+1)}^2(u,r) \, \ud r \Big)^{\f 12} \ud u + \Big( \int_1^U \int_{R}^{u} r^{p+1} F_{(\geq j - \nB+1)}^2(u,r)  \, \ud r \ud u \Big)^{\f 12} \\
&\: + \mbox{boundary terms at $u=1$ and $r \sim R$}.
\end{split}
\end{equation}
In the derivation of \eqref{eq:intro.rp}, it is important to observe that the borderline terms in $Q_j$ have a \emph{good} sign after projecting to $\ell  \geq j-\nB+1$; we call this the Angelopoulos--Aretakis--Gajic positivity \cite{AAG2020, AAGPrice}. (Note that the Angelopoulos--Aretakis--Gajic positivity is the only way that we use the $r^p$ method; in particular we do not need to perform any iterations with a hierarchy of $r^p$ weighted energies as in \cite{AAG2020, DRNM}.)

We apply \eqref{eq:intro.rp} for $\Psi = \bfK^J\rho_J$. Note that to use \eqref{eq:intro.rp} we need to bound $Q_J (\bfK^J \rho_J)$ in both $\calM_\wave$ and $\calM_\med$. In $\calM_{\wave}$, we start with \eqref{eq:intro.Q0.rhoJ} and note that the bad term $E_{J;\Box}$ (which is bad both for $r$-decay and total decay) takes the form of $r^{-J-1} \times [\hbox{function of $(u,\th)$}]$. Since $Q_J \bfK^J = (\bfK + 2r)^J Q_0$, and that $(\bfK + 2r)^Jr^{-J-1} = 0$, we deduce from \eqref{eq:intro.Q0.rhoJ} that
\begin{equation}\label{eq:intro.QJ.high.wave}
Q_J(\bfK^J \rho_{J}) =
O_{\bfGmm}(r^{-2} u^{J-\alp+\nB-1}) = O_{\bfGmm}(r^{-2} u^{-\f 13}) \quad \hbox{ in }\calM_\wave,
\end{equation}
where, as in Step~4(a), we used that each $\bfK+2r$ loses one power of $r$.

In $\calM_{\med}$, the bound \eqref{eq:intro.Q0.rhoJ} is not sufficient, as we need an estimate with the same total decay but with better $u$-decay. Instead, simply using $V = O_{\bfGmm}(\brk{r}^{-3})$ and the assumed bound \eqref{eq:intro.wave.interation.1}, we have
$$Q_0 \Phi = O_{\bfGmm}(r^{-3+\nB} u^{-\alp}) ,\quad \hbox{in $\calM_{\med}$}.$$
Then, notice that $(\bfK+2r)^J$ annihilates all monomials $r^{-j}$ up to $r^{-J-1}$ so that $\bfK^J \rho_J = \bfK^J \Phi$. Hence,
\begin{equation}\label{eq:intro.QJ.high.med}
Q_J(\bfK^J \rho_{J}) = O_{\bfGmm}(r^{-3+J+\nB} u^{-\alp}) \quad  \hbox{ in }\calM_\med.
\end{equation}


In order to bound $\bfK^J \rho_{(\geq J-\nB+1)J}$ using \eqref{eq:intro.QJ.high.wave}--\eqref{eq:intro.QJ.high.med}, we apply the $r^p$ estimate \eqref{eq:intro.rp} with a fixed $p \in (\f 53,2)$. For the error term in $\calM_\wave$, we estimate
\begin{equation}\label{eq:intro.rp.error.wave}
\int_1^{U} \Big( \int_{\calC_u \cap \{r\gtrsim u\}} r^p (\hbox{error})^2 \, \ud \mathring{\slashed{\sigma}}\ud r \Big)^{\f 12} \ud u \ls \int_1^U \Big( \int_{r\gtrsim u} r^p u^{-\f 23} r^{-4} \, \ud r \Big)^{\f 12} \ud u \ls \int_1^U u^{\f p2 -\f {11}6} \, \ud u \ls U^{\f p2 - \f 56}.
\end{equation}
For the error term in $\calM_\med$, we note that $p+1-6+2J+2\nB > -1$ so we bound
\begin{equation}\label{eq:intro.rp.error.med}
\begin{split}
\Big( \int_1^U \int_{\calC_u \cap \{r\ls u\}} r^{p+1} (\hbox{error})^2 \, \ud \mathring{\slashed{\sigma}} \ud r \ud u \Big)^{\f 12} \ls &\: \Big( \int_1^U \int_{r\ls u} r^{p+1} u^{-2\alp} r^{-6+2J+2\nB} \, \ud r \ud u \Big)^{\f 12}  \\
\ls &\: \Big( \int_1^U u^{p+1-6+\f 43+1} \, \ud u   \Big)^{\f 12} = \Big( \int_1^U u^{p-\f 83} \, \ud u   \Big)^{\f 12}  \ls U^{\f p2 - \f 56}.
\end{split}
\end{equation}
Using the $r^p$ estimate \eqref{eq:intro.rp} with the error bounds \eqref{eq:intro.rp.error.wave} and \eqref{eq:intro.rp.error.med}, we obtain
\begin{equation}\label{eq:intro.rp.final}
\Big( \int_{\calC_U}  r^p \Big(\rd_r (\bfK^J \rho_{(\geq J-\nB+1)J}) \Big)^2 \, \ud \mathring{\slashed{\sigma}}\ud r \Big)^{\f 12} \ls U^{\f p2 - \f 56}.
\end{equation}
In particular, since $p>1$, \eqref{eq:intro.rp.final} implies that $\rPhi'_{(\leq J-\nB+1),J}(u,\theta) = \f{(-1)^J}{J!}\lim_{r\to \infty} \bfK^J \rho_{(\geq J-\nB+1)J}(u,r,\theta)$ exists
and such that
\begin{equation}\label{eq:intro.rho'.est}
 \rho_{(\geq J-\nB+1)J}(u,r,\theta) - r^{-J} \rPhi'_{(\leq J-\nB+1),J}(u,\theta)  = O_{\bfGmm}(r^{-\f{p-1}2-J} u^{\f p2 - \f 56}) = O_{\bfGmm}(r^{-J-\f 13}).
\end{equation}

It can then be shown that $\rPhi'_{(\leq J-\nB+1)J}$ satisfies \eqref{eq:intro.recurrence.orig} (with $j=J$ and projected to $\ell \leq J-\nB+1$) and thus it coincides with $\rPhi_{(\leq J-\nB+1)J}$. Hence, the left-hand of \eqref{eq:intro.rho'.est} is $\rho_{(\geq J-\nB+1)J+1}$, and the right-hand side of \eqref{eq:intro.rho'.est} is the bound we wished to establish.

\subsubsection{Intermediate zone} \label{sec:intro.med}

In this region, we begin with
\begin{align}
\phi = &\: O_{\bfGmm}(\tau^{-\alp}) \quad \hbox{in } \calM_\med \cup \calM_\near , \label{eq:intro.med.assumption.1}\\
r^{\nB} \phi = &\: \sum_{j=0}^{J} r^{-j}\Phi_j + \rho_{J+1} \quad \hbox{in } \calM_\wave, \label{eq:intro.med.assumption.2}\\
\rPhi_{(\leq j-\nB)j} = &\:  O_{\bfGmm}(u^{j-\alp+\nB-1}), \, \rPhi_{(\geq j-\nB)j} = O_{\bfGmm}(u^{j-\alp+\nB})\quad 0\leq j \leq J, \label{eq:intro.med.assumption.3}\\
\rho_{J+1} = &\: O_{\bfGmm}(r^{-\alp+\nB-1}). \label{eq:intro.med.assumption.4}
\end{align}
One can think that using Section~\ref{sec:ideas.wave}, there is already an improved bound in $\calM_{\wave}$, and the goal is to obtain a bound
\begin{equation}\label{eq:intro.med.improved}
\phi = O_{\bfGmm}(u^{-\alp-1+\nB} r^{-\nB}) \quad \hbox{in } \calM_\med.
\end{equation}
The reader will note that \eqref{eq:intro.med.improved} is not an improvement of \eqref{eq:intro.med.assumption.1} in the regime $r \ll u$. Nonetheless, in Section~\ref{sec:intro.near}, we will improve this estimate by trading $r^{-\nB}$ for $\tau^{-\nB}$ when we treat the near-intermediate region with elliptic estimates.

In order to prove \eqref{eq:intro.med.improved}, we use the assumptions \eqref{eq:intro.med.assumption.1}--\eqref{eq:intro.med.assumption.4} and the wave equation \eqref{eq:intro.5d.ex} to derive the following:
\begin{equation}\label{eq:intro.med.wave.eqn}
\Box_{\bfm} \phi = O_{\bfGmm}(r^{-3}u^{-\alp}) \quad \hbox{in }\calM_\near\cup \calM_\med \cup \calM_\wave.
\end{equation}
The idea is to treat the right-hand side of \eqref{eq:intro.med.wave.eqn} perturbatively so that the main contribution to $\phi$ comes from the input in $\calM_\wave$, where we have already obtained an improvement.

Suppose we want to bound $\phi$ at some $(U,R,\Theta) \in \calM_\med$. Take $U_0 = \eta_1 U$ (for fixed small $\eta_1 > 0$ independent of $U$), and consider the function $\chi_{>U_0}(u) \phi$ (see \eqref{eq:intro.cutoff}). Consider the wave equation for $\Box_{\bfm} (\chi_{>U_0}(u) \phi)$, which has the following two types of contributions:
\begin{enumerate}
\item contribution from differentiating $\chi_{>U_0}(u)$, and
\item contribution from inhomogeneous terms coming from \eqref{eq:intro.med.wave.eqn},
\end{enumerate}

(1) contains the main contribution and (2) are treated as error terms. Importantly, we use the \emph{strong Huygens principle} associated with $\Box_{\bfm}$ to show that the main term (1) \textbf{only has contribution from the wave zone $\calM_{\wave}$}. Thus, to handle (1), we use the expansion and estimates in \eqref{eq:intro.med.assumption.2}--\eqref{eq:intro.med.assumption.4}. In turn, there are two types of terms: those involving $\rPhi_{j}$ and those involving $\rho_{J+1}$. For the $\rPhi_{j}$ terms, we explicitly \emph{compute} the contribution by a Minkowskian computation (see Section~\ref{sec:Minkowski.wave}). Importantly, the computation shows that $\rPhi_{(\geq j-\nB+1) j}$ (for $0\leq j \leq J-1$) does not contribute to the late time tails at all (see Lemma~\ref{lem:med.corrected})! Therefore, even though the decay rate for the high angular modes is not improved in \eqref{eq:intro.med.assumption.3}, we can nonetheless obtain an improvement in $\calM_{\med}$.

To bound the error terms from (2) requires only just Minkowskian estimates. We use a sharp (Minkowskian) integrated local decay estimate \cite{MetTat} when $d \geq 5$ and use the positivity of the fundamental solution together with integration by characteristic when $d = 3$, cf.~\cite{MTT}.

\subsubsection{Near and intermediate zone} \label{sec:intro.near}

In the near-intermediate zone $\calM_\near \cup \calM_\med$, we start with the estimates \eqref{eq:intro.med.assumption.1} and \eqref{eq:intro.med.improved}, which gives
\begin{equation}\label{eq:intro.near.assumption}
\phi = O_{\bfGmm}(\tau^{-\alp-1+\nB} r^{-\nB})\quad\hbox{in $\calM_{\med} \cup \calM_{\near}$}.
\end{equation}
This is the desired total power of decay, but it is unsatisfactory in the region $r \ll \tau$. To deal with this, we use (a sequence of) weighted elliptic estimates to exchange $r$ decay for $\tau$ decay. This strategy is standard by now; as far as we are aware it first appeared in \cite{MTT} and the more refined sequence of weighted elliptic estimates was introduced in \cite{AAGPrice}.

More precisely, starting with the assumption \eqref{eq:intro.near.assumption},
our goal is to trade all the $r$-decay for $\tau$-decay and to obtain the following bound:
\begin{equation}\label{eq:intro.near.improved}
\phi = O_{\bfGmm}(\tau^{-\alp-1}) \quad \hbox{in } \calM_\med \cup \calM_\near.
\end{equation}

We denote by $\urd_\tau$, $\urd_r$ the coordinate partial derivatives in the $(\tau, r,\th^A)$ coordinate system. Then $\Box_{\bfm}$ takes the form
\begin{equation}\label{eq:intro.Box.in.terms.of.P0.T}
\begin{split}
\Box_{\bfm} = &\: O_{\bfGmm}(1) \bfT^2 + O_{\bfGmm}(1) \urd_r \bfT+ \urd_r^2 +\f 2r(-(\chi'_{>2R}(r) r + \chi_{>2R}) \bfT + \urd_r) + r^{-2} \rslap.
\end{split}
\end{equation}
Define
\begin{equation}\label{eq:intro.calP.def}
\mathcal P_0 = \urd_r^2 +\f 2r \urd_r + r^{-2} \rslap - V.
\end{equation}
Because we are establishing vector field bounds, every derivative of $\bfT$ gives an extra power of $\tau$ decay. As a result, using \eqref{eq:intro.Box.in.terms.of.P0.T} and \eqref{eq:intro.calP.def}, the equation \eqref{eq:intro.5d.ex} implies that $\mathcal P_0 \phi$ is perturbative.
\begin{equation}\label{eq:intro.med.P0.first.bound}
\mathcal P_0 (\chi_{<\tau}(r) \phi) = O_{\bfGmm}(\tau^{-\alp-2+\nB} \brk{r}^{-\nB-1}),
\end{equation}
where $\chi_{<\tau}(r)$ is a cutoff localizing to the near-intermediate zone.

The key estimate we will prove for $\calP_0$ is that for $-\f d2 \leq \gamma \leq \f d2 - 2$,
\begin{equation}\label{eq:intro.elliptic.pointwise}
\calP_0 (\chi_{<\tau}(r) \psi) = O_{\bfGmm}(\tau^{\bt} \brk{r}^{-\f{d+2\gamma}2-2-\sigma}) \quad \implies \quad  \chi_{<\tau}(r)\psi = \begin{cases}
O_{\bfGmm}(\tau^\bt \brk{r}^{-\f{d+2\gamma}2}) & \hbox{ if }\sigma >0, \\
O_{\bfGmm}(\tau^\bt \log \tau \brk{r}^{-\f{d+2\gamma}2}) & \hbox{ if }\sigma =0, \\
O_{\bfGmm}(\tau^{\bt-\sigma} \brk{r}^{-\f{d+2\gamma}2}) & \hbox{ if }\sigma <0.
\end{cases}
\end{equation}
Such an estimate is particularly simple for the model problem \eqref{eq:intro.5d.ex}: it can be derived by combining a family of weighted elliptic estimates with Sobolev embedding, first on Euclidean space, and then perturbatively to absorb the potential term; see Section~\ref{subsec:stat-est}.

Using the bound \eqref{eq:intro.med.P0.first.bound} and applying \eqref{eq:intro.elliptic.pointwise} with $(\beta, \sigma,\gamma) = (-\alp-2+\nB, -\f 12, -1)$, we obtain $\phi = \chi_{<\tau}(r) \phi = O_{\bfGmm}(\tau^{-\alp-\f 32 + \nB} r^{-\nB+\f 12})$ when $r < \tau$. In particular, we have traded $\f 12$ power of $r$-decay in \eqref{eq:intro.near.assumption} for $\f 12$ power of $\tau$-decay.

We can then repeat the above procedure. With the improved bound on $\phi$, we get $\mathcal P_0 (\chi_{<\tau}(r) \phi) = O_{\bfGmm}(\tau^{-\alp-\f 52+\nB} \brk{r}^{-\nB-\f 12})$. Using \eqref{eq:intro.elliptic.pointwise} with $(\beta, \sigma,\gamma) = (-\alp-\f 52+\nB, -\f 12, -\f 32)$, this then gives the further improved estimate $\phi = O_{\bfGmm}(\tau^{-\alp-2+\nB} r^{-\nB+1})$. Continuing this way we arrive at the desired estimate \eqref{eq:intro.near.improved} (and we cannot continue further because we need $\gamma \geq -\f d2$).

\subsubsection{End of iteration and proof of upper bound}\label{sec:intro.end.of.iteration} The iteration stops when we get to $J_{\mathfrak f}$. At that point, since it is possible that $\lim_{u\to \infty}\rPhi_{(\leq J_{\mathfrak f} - \nB)J_{\mathfrak f}}(u,r,\th) \neq 0$, we cannot integrate \eqref{eq:intro.improved.eqn.low.modes} (with $J=J_{\mathfrak f}$) from $u=\infty$. Instead, integrating from $u=1$, we obtain
\begin{equation}\label{eq:end.of.iteration}
\rPhi_{(\leq J_{\mathfrak f} - \nB)J_{\mathfrak f}} = O_{\bfGmm}(1).
\end{equation}
In fact, one obtains the slightly more precise estimate
\begin{equation}\label{eq:end.of.iteration.precise}
\rPhi_{(\leq J_{\mathfrak f} - \nB)J_{\mathfrak f}} = \rPhi_{(\leq J_{\mathfrak f} - \nB)J_{\mathfrak f}}(\infty) + O_{\bfGmm}(u^{-\f 13}),\quad \rPhi_{(\leq J_{\mathfrak f} - \nB)J_{\mathfrak f}}(\infty) = O_{\bfGmm}(1).
\end{equation}

Since the bound \eqref{eq:end.of.iteration} for the higher radiation field is weaker, the estimates for low spherical harmonic parts of the remainder $\rho_{(\leq J_{\frkf}-\nB) J_{\frkf}+1}$ also become weaker. Indeed, using \eqref{eq:end.of.iteration}, we only obtain that $\mathbb S_{(\leq J_{\frkf}-\nB)} E_{J_{\frkf}+1;\Box} = O_{\bfGmm}(r^{-J-2})$ (see \eqref{eq:intro.EJ+1Box}) so that by \eqref{eq:intro.Q0.rhoJ.low.with.EJ+1} (with $J = J_{\frkf}$), we have
$$Q_0 \rho_{(\leq J_{\frkf}-\nB) J_{\frkf}+1} = O_{\bfGmm}(r^{-J_{\frkf}-2}).$$
(compare this with \eqref{eq:intro.Q0.rhoJ.low.removed}). Repeating the integration along characteristic argument in Step~4(a) of Section~\ref{sec:ideas.wave} (including commuting with $\bfK^{\ell+\nB-1}$ to control $\bfK^{\ell+\nB-1}\rho_{(\ell)J_{\frkf}+1}$ and then integrating successively from $r=\infty$), we obtain
\begin{equation}\label{eq:end.of.iteration.rho}
\rho_{(\leq J_{\frkf}-\nB) J_{\frkf}+1} = O_{\bfGmm}(r^{-J_{\frkf}-1} u).
\end{equation}
(Notice that this is not strong enough to imply the analogous $O_{\bfGmm}(r^{-J-\f 13})$ estimates when $J< J_{\frkf}$.) On the other hand, for the higher angular modes, the argument remain unchanged and one could still obtain the stronger estimates \eqref{eq:intro.rp.final} and \eqref{eq:intro.rho'.est}. In particular, putting the above estimates together gives the upper bound $\phi = O_{\bfGmm}( \tau^{-J_{\mathfrak f}} r^{-\nB})$ in $\calM_{\wave}$.

Repeating the arguments in Sections~\ref{sec:intro.med} and \ref{sec:intro.near}, but now only with the weaker input \eqref{eq:end.of.iteration} and \eqref{eq:end.of.iteration.rho}, we obtain the desired upper bound $\phi = O_{\bfGmm}(\min\{\tau^{-J_{\mathfrak f}-\nB}, \tau^{-J_{\mathfrak f}} r^{-\nB}\})$.

\subsubsection{Precise tails}\label{sec:intro.precise.tails} The above argument gives the sharp upper bound in \eqref{eq:intro.main.upper.bound}. We need an extra argument to obtain the precise tail in \eqref{eq:intro.main.precise.1}--\eqref{eq:intro.main.precise.2}. See Section~\ref{sec:lower}. The main term \eqref{eq:end.of.iteration.precise} in $\calM_{\wave}$ will become the source of the precise tail.

\begin{enumerate}
\item In $\calM_\wave \cup \calM_\med$, we refine the analysis above to show that $\phi$ is well-approximated by the Minkowskian profile $\varphi^{\bfm}_{J_{\frkf}}[\rPhi_{J_{\mathfrak f}}(\infty)]$, precisely that
\begin{equation}\label{eq:intro.precise.wave.med}
\Big| \phi - \varphi^{\bfm}_{J_{\frkf}} [\rPhi_{(\leq J_{\frkf - \nB})J_{\mathfrak f}}(\infty)] \Big|(u,r,\th) = O_{\bfGmm}(u^{-J_{\mathfrak f} - \de_{\mathfrak f}} r^{-\nB}) \quad \hbox{in $\calM_{\med}\cup \calM_{\wave}$}
\end{equation}
for some $\de_{\mathfrak f} >0$. 

     In order to prove \eqref{eq:intro.precise.wave.med}, say at a point $(U,R,\Theta)$, we define (recall \eqref{eq:intro.cutoff})
     \begin{equation}\label{eq:intro.varphif}
    \varphi_{\frkf} = \Box_{\bfm}^{-1} \left(-2 \chi_{>U_{0}}'(u) r^{-\nu_{\Box}} \rd_{r} \left(\sum_{j=0}^{J_{\mathfrak f}}  \rPhi_{j}(u) r^{-j} \right)\right),
    \end{equation}
    where $U_0 = U^{1-\bt}$ for some well-chosen $\bt >0$.

    Using explicit computations on Minkowski spacetime together with the decay estimates \eqref{eq:intro.wave.rho.improved.1} (for $J \leq J_{\frkf}-1$) and \eqref{eq:end.of.iteration.precise}, we see that the main contribution from \eqref{eq:intro.varphif} are those corresponding to the non-decay terms when $j = J_{\frkf}$, i.e.,
    $$\Big| \varphi_{\frkf} - \varphi^{\bfm}_{J_{\frkf}} [\rPhi_{(\leq J_{\frkf - \nB})J_{\mathfrak f}}(\infty)] \Big|(u,r,\th) = O_{\bfGmm}(u^{-J_{\mathfrak f} - \de_{\mathfrak f}} r^{-\nB}) \quad \hbox{in $\calM_{\med}\cup \calM_{\wave}$}.$$

    To prove \eqref{eq:intro.precise.wave.med}, it thus remains to show that $\varphi_{\frkf}$ is indeed a good approximation of $\phi$. For this, we use the equation for $\Box_{\bfm} (\chi_{>U_0}(u)\phi - \varphi_{\frkf})$:
\begin{equation}\label{eq:intro.Box.diff}
   \Box_{\bfm} (\chi_{>U_0}(u)\phi - \varphi_{\frkf}) = -2\chi_{> U_0}'(u) r^{-\nu_{\Box}} \rd_{r} \rho_{J_{\frkf}+1} + O_{\bfGmm}(\min\{ r^{-3-\nB} u^{-J_{\frkf}}, r^{-3} u^{-J_{\frkf}-\nB} \}),
    \end{equation}
    where we used the upper bounds in Section~\ref{sec:intro.end.of.iteration} above. By \eqref{eq:end.of.iteration.rho}, $2\chi_{> U_0}'(u) r^{-\nu_{\Box}} \rd_{r} \rho_{J_{\frkf}+1}$ is bounded by $O_{\bfGmm}(r^{-\nB-J_{\frkf}-2})$. This is to be contrasted with the decay in the main terms in \eqref{eq:intro.varphif}, i.e., $\Box_{\bfm} \varphi_{\frkf} = O_{\bfGmm}(r^{-\nB-J_{\frkf}-1}u^{-1})$. The key point here is to exploit the extra $r$-decay and to break the scaling by imposing $U_0 = U^{1-\bt}$. By the strong Huygens principle of $\Box_{\bfm}$, $(\chi_{>U_0}(u)\phi - \varphi_{\frkf})(U,R,\Theta)$ only depends on $-2\chi_{> U_0}'(u) r^{-\nu_{\Box}} \rd_{r} \rho_{J_{\frkf}+1}$ in the region where $r \sim U_0$. Thus the better $r$ decay implies that the contribution of $-2\chi_{> U_0}'(u) r^{-\nu_{\Box}} \rd_{r} \rho_{J_{\frkf}+1}$ to  $(\chi_{>U_0}(u)\phi - \varphi_{\frkf})(U,R,\Theta)$ is better than $\varphi_{\frkf}$.

    In order to carry out the above analysis for \eqref{eq:intro.Box.diff}, we also need to choose $\bt>0$ sufficiently small so that the contribution from the $O_{\bfGmm}(\min\{ r^{-3-\nB} u^{-J_{\frkf}}, r^{-3} u^{-J_{\frkf}-\nB} \})$ term in \eqref{eq:intro.Box.diff} remains perturbative. In this process of controlling the perturbative terms, we use some of the ideas that we have outlined in Sections~\ref{sec:ideas.wave} and \ref{sec:intro.med} above for proving upper bounds.

\item We then turn to $\calM_\near$. First note that the Minkowskian profile (which can be explicitly computed, see Proposition~\ref{prop:med.explicit.precise}) satisfies
\begin{equation}\label{eq:intro.varphim.in.near}
\varphi^{\bfm}_{J_{\frkf}}[\rPhi_{(\leq J_{\frkf}-\nB)J_{\mathfrak f}}](\tau,r,\th) = c_{d,J_{\mathfrak f}} \tau^{-J_{\mathfrak f} - \nB} + O_{\bfGmm}(\tau^{-J_{\mathfrak f}-\nB-\de_m})\quad \hbox{when $r \leq \tau^{1-\de_m}$}
\end{equation}
for some (explicit) constant $c_{d,J_{\mathfrak f}}$. Thus We need to propagate this estimate

Fix some scale $\tau_0 \simeq \tau^{1-\de'}$. We cut off $\phi$ by $\chi_{<\tau_0}(r)$ and show that it is close to the desired asymptotic spatial profile $c_{d,J_{\mathfrak f}} \tau^{-J_{\mathfrak f} - \nB} (1 - ^{(\infty)}\calR_0 ^{(\infty)}\calP_0 1)$ using an elliptic estimate. The profile has two properties:
\begin{enumerate}[label=(\alph*),ref=\alph*]
\item \label{item:elliptic.profile.prop.1} $^{(\infty)}\calP_0 (1 - ^{(\infty)}\calR_0 ^{(\infty)}\calP_0 1) = 0$ since $^{(\infty)}\calR_0$ is the inverse of $^{(\infty)}\calP_0$ by definition.
\item \label{item:elliptic.profile.prop.2} $^{(\infty)}\calR_0 ^{(\infty)}\calP_0 1 = O_{\bfGmm}(r^{-1})$ (with even better bounds if $d \geq 5$) by elliptic estimates associated with $^{(\infty)}\calR_0$, see Proposition~\ref{cor:near-pointwise}.
\end{enumerate}

Writing $\Box_{\bfm}$ in terms of $^{(\infty)}\calP_0$ using \eqref{eq:intro.Box.in.terms.of.P0.T}--\eqref{eq:intro.calP.def}, and use the sharp upper bound we have already obtained to derive $^{(\infty)}\calP_0 \phi = O_{\bfGmm}(\tau^{-J_{\frkf}-\nB-1}r^{-1})$. Thus, using also (\ref{item:elliptic.profile.prop.1}) above, we have
\begin{equation}\label{eq:elliptic.profile.extra.decay}
    \begin{split}
        &\: \calP_0\Big(\chi_{<\tau_0}(r) (\phi - c_{d,J_{\mathfrak f}} \tau^{-J_{\mathfrak f} - \nB} (1 - ^{(\infty)}\calR_0 ^{(\infty)}\calP_0 1)\Big) \\
        = &\: [\calP_0, \chi_{<\tau_0}(r)](\phi - c_{d,J_{\mathfrak f}} \tau^{-J_{\mathfrak f} - \nB} (1 - ^{(\infty)}\calR_0 ^{(\infty)}\calP_0 1) + O_{\bfGmm}(\tau^{-J_{\frkf}-\nB-1}r^{-1}).
    \end{split}
\end{equation}
The main term involving the commutator $[\calP_0, \chi_{<\tau_0}(r)]$ is supported in $\calM_\med$. Combining the estimate \eqref{eq:intro.precise.wave.med} with \eqref{eq:intro.varphim.in.near}, and using property (\ref{item:elliptic.profile.prop.2}) above, we know that
$$\phi - c_{d,J_{\mathfrak f}} \tau^{-J_{\mathfrak f} - \nB} (1 - ^{(\infty)}\calR_0 ^{(\infty)}\calP_0 1)= O_{\bfGmm}(\tau^{-J_{\mathfrak f} - \nB} r^{-2}) \quad \hbox{on $\calM_{\med}$}$$
Thus we see that the right-hand side of \eqref{eq:elliptic.profile.extra.decay} has improved decay and we can conclude the argument using elliptic estimates as in Section~\ref{sec:intro.near}.

\end{enumerate}

\subsubsection{Proof in the general setting}\label{sec:intro.further}

Most of the ideas we discussed for the model problem \eqref{eq:intro.5d.ex} can be adapted to the more general case in Theorem~\ref{thm:intro.main}. We discuss a sequence of generalizations that can be made and indication the modifications that are needed. The more delicate issue of adding logarithmic terms in the expansion in the wave zone will be treated separately in Section~\ref{sec:intro.log}.

\begin{enumerate}
\item \textbf{Large ultimately stationary perturbations of Minkowski.} While it was convenient to assume that $V$ is small in the model problem \eqref{eq:intro.5d.ex}, it is only in the bound \eqref{eq:intro.elliptic.pointwise} in the near zone that we have explicitly used smallness. In general, it suffices to assume that $\lim_{u\to \infty} V(u,r,\th) = V(r,\th)$ for some stationary asymptocially flat potential (with suitable estimate) such that the elliptic estimates \eqref{eq:intro.elliptic.pointwise} hold. In fact, we can start with a very weak invertibility statement on $\calP_0$ (with potentially a loss of finitely many derivatives) in \eqref{eq:intro.calP.def}; see \ref{hyp:ult-stat} in Section~\ref{sec:global.assumptions}. This can be viewed as an assumption on the lack of eigenvalue/resonance at zero energy (see Remark~\ref{rmk:no.eigenvalue.resonance.zero.energy}).

At the same time, it is also possible to relax the decay rate assumption \eqref{eq:intro.5d.ex.V} at infinity. It suffices to require that $V = O_{\bfGmm}(r^{-2-\de_c})$ in $\calM_\med\cup \calM_\near$ and
\begin{equation}\label{eq:intro.5d.ex.V.weaker}
V = \sum_{j=2}^{J_{c}} r^{-j} \rV_j(u,\theta) + O_{\bfGmm}(r^{-J_{c}-\eta_c} u^{J_{c}-2 + \eta_c - \de_c}),\quad \rV_j(u,\theta) = O_{\bfGmm}(u^{j-2-\de_c}) \quad \hbox{in }\calM_\wave
\end{equation}
for some $\de_c >0$. In this case, each iteration step may improve the decay rate only by a power $\de_0 \in (0,1)$, increasing the total number of iteration steps needed. Moreover, because of the weaker improvement in each step, in the iteration in the wave zone, we do not necessarily extend the expansion \eqref{eq:intro.wave.interation.2} by one one order in each iteration. We refer the reader to Section~\ref{sec:wave} for details.

\item \textbf{General asymptotically flat equations.} We also go beyond \eqref{eq:intro.5d.ex} and allow for more general variable coefficient equations:
\begin{equation}\label{eq:intro.linear.more.general.AF}
(\bfg^{-1})^{\alp \bt} \bfD_{\alp} \rd_{\bt} \phi + \bfB^{\alp} \rd_{\alp} \phi  + V \phi = 0,
\end{equation}
where $(\bfg^{-1})$, $\bfB$ and $V$
\begin{enumerate}
\item converge to stationary coefficients as $\tau \to \infty$,
\item satisfy $\bfg^{-1} - \bfm = O_{\bfGmm}(\brk{r}^{-\de_c})$, $\bfB = O_{\bfGmm}(\brk{r}^{-1-\de_c})$, $V = O_{\bfGmm}(\brk{r}^{-\de_c})$ in $\calM_\med\cup \calM_\near$,
\item admit as expansion in $r$ near null infinity but with decay rates consistent with \eqref{eq:intro.5d.ex.V.weaker}; see Section~\ref{sec:assumption.far.away}.
\end{enumerate}

In such a general setting, we will \emph{assume} that $\calP_0$ is invertible, allowing us to still carry out the analysis in the near-intermediate zone (see Section~\ref{sec:intro.near}). In fact, $\calP_0$ needs not even be elliptic, which in particular allows us to consider black hole settings; see Remark~\ref{rmk:P0.elliptic}. The analysis in $\calM_{\med}$ and $\calM_{\wave}$ is essentially unchanged, after suitably localizing to the asymptotically flat region.

\item \textbf{Weaker decay assumptions on the solution.} The above discussion starts with the assumption \eqref{eq:intro.assumed.bound}, but a close inspection of the argument shows that for any linear equation we can start with a much weaker decay assumption, e.g., $\phi = O_{\bfGmm}(\min\{u^{-\alp}, u^{-\alp+\f{d-1}2} r^{-\f{d-1}2}\})$ even possibly $\alp <0$, which is even easier to achieve. In other words, our approach furnishes a different proof of \eqref{eq:intro.assumed.bound}. (Our approach moreover uses slightly fewer assumptions than some existing works. For instance, compared to \cite{gM2016}, we do not need the extra good commutating vector $K$, and compared to \cite{MTT} we allow for weaker decay assumptions of the coefficients more in line with \cite{jOjS2023}, etc.)

\item \textbf{Nonlinear equations.} Our approach applies to nonlinear (including quasilinear) equations as well, as long as we assume sufficient decay of the solution so that the nonlinearity is perturbative. We only consider smooth nonlinearities with the following requirements:
\begin{enumerate}
\item In $(3+1)$ dimensional, we impose the classical \emph{null condition} in $(3+1)$ dimensions; null conditions are not necessary in higher dimensions. See Section~\ref{sec:nonlinearity.assumptions} for the precise assumptions. The null condition is assumed in $(3+1)$ dimensions in part to ensure that there is a regular expansion near null infinity.
\item Another requirement is that $\phi$ has sufficient decay to ensure that the nonlinearity is perturbative. We introduce a(n equation-dependent) parameter $\alp_{\calN}$, where if
\begin{equation}\label{eq:intro.phi.alp.decay}
\phi = O_{\bfGmm}(\min\{\tau^{-\alp}, \tau^{-\alp+\nB} r^{-\nB}\})
\end{equation} (with an expansion near null infinity), then the nonlinear terms are perturbative and can be treated just in the linear terms in \eqref{eq:intro.linear.more.general.AF}. We then impose this decay as part of the weak decay bound assumed in Theorem~\ref{thm:intro.main}. (We remark that this assumed bound is typically much weaker than that obtained from standard small-data global existence results. For instance, for nonlinearities satisfying the null condition in $(3+1)$-dimensions, typical proofs of small data global existence gives \eqref{eq:intro.phi.alp.decay} with $\alp = 1$, while we only need $\alp >0$ as an input to Theorem~\ref{thm:intro.main}.)
\end{enumerate}
Under these assumptions, the nonlinearity poses no additional difficulty. In particular, since we allow for a loss of derivatives, all the nonlinear terms (including the quasilinear ones) can be treated as error terms.

\item \textbf{Non-compactly supported data and inhomogeneous equations.} It is easy to adapt our method to allow for non-compactly supported data and inhomogeneous equations. When the initial data and the inhomogeneous terms decay sufficiently rapidly, then the above argument can be carried out without modification.

If the decay of the data and/or the inhomogeneity is slower, but there is nonetheless an expansion near null infinity, then a slightly more careful argument also gives a sharp upper bound of the solution, which could now be determined by the data or the inhomogeneous terms. See Section~\ref{sec:assumption.data.inho.sol} for details on the assumptions. Moreover, the same argument would allow us to consider coefficients satisfying weaker assumptions (say, as compared to \eqref{eq:intro.5d.ex.V}) so that the upper bound is determined by these coefficients.

We point out that in the particular case where the decay is determined by the data, inhomogeneous terms or the coefficients, we need a further idea in our analysis in the wave zone. In particular, in the last step of the iteration (see Section~\ref{sec:intro.end.of.iteration}), we need to prove additional decay for the remainder $\rho$, which is achieved by introducing a cutoff in spacetime so as to use the strong Huygens principle for the flat wave operator. We refer the reader to Case~4 of Proposition~\ref{prop:wave-main} for details.
\end{enumerate}

\subsubsection{Logarithmic terms}\label{sec:intro.log}

A technically more challenging aspect is that we also allow for logarithmic terms in the expansion near infinity, either coming from initial data or coming from the coefficients in the equation. (As mentioned above, this is natural from the point of view of general relativity, which such logarithmic terms in the expansion of the metric arise in physical settings.)

To make the ideas transparent, we consider a equation as in \eqref{eq:intro.5d.ex} ($\Box_{\bfm}\phi = V\phi$), but instead of \eqref{eq:intro.5d.ex.V}, we add in a \emph{single} term logarithmic term $r^{-3} \log (\tfrac ru) \rV_{3,1}(u,\th)$ in the expansion of $V$, i.e., for every $J \geq 4$,
\begin{equation}\label{eq:intro.5d.ex.V.log}
V = r^{-3} \rV_{3,0}(u,\theta) + r^{-3} \log (\tfrac ru) \rV_{3,1}(u,\th) + \sum_{j=4}^J  r^{-j} \rV_{j,0}(u,\theta) + O_{\bfGmm}(r^{-J-1}),\quad \rV_j(u,\theta) = O_{\bfGmm}(1).
\end{equation}
(In our theorem, we also allow for $u$-growth in $\rV_j$ and in the error term as in \eqref{eq:intro.5d.ex.V.weaker}; we omit the discussion here.)

We first note that the logarithmic terms are easier to treat in $\calM_\med$ and $\calM_\near$. In $\calM_\med$ the main new ingredient is to also carry out the Minkowski computation when logarithmic terms are present in the expansion near null infinity. These terms them give rise to $\log \tau$ losses in the final decay rate. In $\calM_\near$, the argument in fact remains unchanged except at the end of the iteration, where one also keeps track of the $\log \tau$ powers in the final decay rates.

The main new difficulties thus lie in the logarithmic terms in $\calM_\wave$. Here are a few observations that allow the proof to be extended to include these terms.
\begin{enumerate}
\item (Modified expansion) When the logarithmic terms are present, we define our expansion and the remainder $\rho_{J}$ in a way that captures these terms. Instead of \eqref{eq:intro.wave.interation.2}, we define
\begin{equation}\label{eq:intro.log.modified.remainder}
 r^{\nB}\phi = \sum_{j=0}^{J-1} r^{-j} \rPhi_{j, 0} + \chi_{>4}\left( \tfrac{r}{u} \right)  \sum_{j=0}^{J-1} \sum_{k =1}^{K_{j}} r^{-j} \log^{k} (\tfrac{r}{u}) \rPhi_{j, k} + \rho_J,\quad \hbox{in $\calM_{\wave} \cup \calM_{\med}$}.
\end{equation}
Define $J_{\mathfrak f}$ so that it satisfies (as opposed to \eqref{eq:intro.idea.Jf.def})
\begin{align}
\lim_{u\to \infty} \Phi_{(\leq j - \nB)j,0}(u,\theta)=0 \hbox{ when }0\leq j \leq J_{\mathfrak f}-1,\quad \hbox{and} \label{eq:intro.idea.Jf.def.log.1} \\
\lim_{u\to \infty} \Phi_{j,k}(u,\theta)=0 \hbox{ when }0\leq j \leq J_{\mathfrak f}-1, \, k\geq 1. \label{eq:intro.idea.Jf.def.log.2}
\end{align}
Notice that we made the expansion in powers of $\log (\tfrac{r}{u})$ (instead of powers of $\log r$). Thus these terms capture the $\log r$ growth when $r \gg u$, but they become bounded as $r \sim u$, where we introduce a smooth cutoff. With this choice, the upper bounds for $\rPhi_{j,k}$ and $\rho_{J}$ are essentially unchanged in the iterations (with $\rPhi_{j,k}$ satisfying the same estimates as $\rPhi_{j,0}$ when $k\geq 1$); the only change would arise in the end of the iteration, see point (4) below. (Expanding in $\log r$ instead would incur additional powers of $\log u$ in the estimates.)

In the presence of logarithmic terms, we need, in addition to \eqref{eq:intro.log.modified.remainder}, a different expansion in terms of $r^{-j} \log^k r$ (instead of $r^{-k} \log^k (\tfrac ru)$ above) with what we call the renormalized higher radiation fields $\rcPhi_{j,k}$ (see Section~\ref{sec:renormalized.higher.radiation.fields}):
\begin{equation}\label{eq:intro.log.modified.remainder.2}
 r^{\nB}\phi = \sum_{j=0}^{J-1} r^{-j} \rcPhi_{j, 0} + \chi_{>4}\left( \tfrac{r}{u} \right)  \sum_{j=0}^{J-1} \sum_{k =1}^{K_{j}} r^{-j} \log^{k} r \rcPhi_{j, k} + \rho_J,\quad \hbox{in $\calM_{\wave} \cup \calM_{\med}$}.
\end{equation}
While $\rPhi_{j,k}$ are useful in carrying out the iteration, in the final step, it is the renormalized $\rcPhi_{J_{\frkf}, k}$ that achieve finite limits $\rcPhi_{J_{\frkf}, k}(\infty) = \lim_{u\to \infty}\rcPhi_{J_{\frkf}, k}(u,\theta)$.

\item \label{item:log.radiation.improved}(Improvement for the higher radiation field for the logarithmic terms) One important ingredient in the analysis of the higher radiation fields is the Newman--Penrose cancellation, which allowed us to show that the low spherical harmonics $\rPhi_{(\leq j-\nB)j}$ decays better (by one power in the model example); see Step~3 in Section~\ref{sec:ideas.wave}. A similar improvement also occur for the higher radiation field associated with the logarithmic terms (for all spherical harmonics).

%

In the setting of the new model, we carry out a similar iteration in $\calM_{\wave}$ as Step~2 of Section~\ref{sec:ideas.wave}, except that in addition to \eqref{eq:intro.wave.interation.1}--\eqref{eq:intro.wave.interation.5} (where now $\rPhi_{j} = \rPhi_{j,0}$), we have the extra assumption
\begin{equation}\label{eq:intro.wave.interation.log}
\rPhi_{j,k} =   O_{\bfGmm}(u^{j-\alp+\nB}),\quad j = 0,1,\ldots,J-1,\quad k\geq 1.
\end{equation}

When $J \leq J_{\frkf} -1$, our goal, in addition to \eqref{eq:intro.improved.higher.radiation.goal.J-1}--\eqref{eq:intro.wave.rho.improved.2}, is to improve \eqref{eq:intro.wave.interation.log} as well as to get the estimate of one more term in the expansion, i.e., we wish to establish
\begin{equation}\label{eq:intro.wave.interation.log.improved}
\rPhi_{j,k} =   O_{\bfGmm}(u^{j-\alp+\nB-1}),\quad j = 0,1,\ldots,J,\quad k\geq 1.
\end{equation}
Unlike for the $k=0$ case, this can be handled without separating into high and low spherical harmonics. There a \emph{hierarchical} structure in the recurrence equations so that the term $\rPhi_{j,k}$ (with $k\geq 1$) --- without needing to project to low spherical harmonics --- exhibits improved decay. This is ultimately due to the fact that on Minkowski spacetimes, these logarithmic terms cannot be generated if they were no present in the data. The analogue of \eqref{eq:intro.recurrence.orig} in this case now reads:
\begin{equation}\label{eq:intro.recurrence.with.log}
\begin{split}
& \rd_{u} \rPhi_{j, k}- (k+1) u^{-1} \rPhi_{j, k+1} \\
	&= \frac{k+1}{j} \left( \rd_{u} \rPhi_{j, k+1} - (k+2) u^{-1} \rPhi_{j, k+2} \right)
- \frac{1}{2j}\Big( (j-1)j - \frac{(d-1)(d-3)}{4} + \rslap \Big) \rPhi_{j-1, k}   \\
&\peq
+ \frac{1}{2j} (j-1)(k+1) \rPhi_{j-1, k+1} - \frac{1}{2j}(k+2)(k+1) \rPhi_{j-1, k+2} \\
& \peq + \frac{1}{2j} \sum_{\substack{j' + j'' = j+1 \\ k' + k'' = k}} \rV_{j',k'} \rPhi_{j'',k''}.
\end{split}
\end{equation}
By our assumptions, the last term is $O_{\bfGmm}(u^{j-\alp+\nB-2})$ for $j \leq J$ (cf.~\eqref{eq:intro.recurrence}).

This allows for an induction according to the ordering that $(j_1,k_1) < (j_2,k_2)$ if $j_1 < j_2$ or if $j_1 = j_2$ and $k_1 > k_2$. Since $r^{\nB}\phi$ is bounded, $\rPhi_{0,k} = 0$ for $k\geq 1$ (see \eqref{eq:intro.log.modified.remainder}), logarithmic terms only start occurring when $j\geq 2$. Now the equation \eqref{eq:intro.recurrence.with.log} takes the following form when $j=2,3$:
\begin{align}
\rd_u \rPhi_{2,1} = &\: O_{\bfGmm}(u^{-\alp+\nB}),\quad \rd_u \rPhi_{2,k} = 0,\quad k \geq 2 \\
\rd_u \rPhi_{3,1} = &\: 2u^{-1} \rPhi_{3,2} -\f 16 \Big(6 - \f{(d-1)(d-3)}{4} + \rslap \Big)\rPhi_{2,1}+ O_{\bfGmm}(u^{1-\alp+\nB}),\\
\rd_u \rPhi_{3,2} = &\:  O_{\bfGmm}(u^{1-\alp+\nB}),\quad \rd_u \rPhi_{3,k} = 0,\quad k \geq 3.
\end{align}
The key structure is that we can start with $\rPhi_{2,1}$, then $\rPhi_{3,2}$, and then $\rPhi_{3,1}$ so that in the process we either see terms that decay sufficiently, or terms that already controlled in the previous step. This structure continues for higher $j$'s and and at the end we obtain \eqref{eq:intro.wave.interation.log.improved}.

The $J = J_{\frkf}$ case is also similar (cf.~Section~\ref{sec:intro.end.of.iteration}), except we may not have $\lim_{u\to \infty} \rPhi_{J_{\frkf},k}(u) = 0$ so that the $u^{-1}$ factor in \eqref{eq:intro.recurrence.with.log} would lead to logarithmic growth. It is then more appropriate to consider $\rcPhi_{J_{\frkf},f}$ (see \eqref{eq:intro.log.modified.remainder.2}) for which we can prove to have finite limits as $u\to \infty$.

Finally, it is useful to note that in the induction procedure one can a priori bound the top degree $K_j$ of $\log^k (\tfrac ru)$ for each $j$ (with $\rPhi_{j,k} = 0$ for $k \geq j+1$ in this example).

\item (Iterations in the presence of logarithmic terms) We now discuss the changes needed when there are logarithmic terms. The estimates for the low angular modes (i.e., $\ell \leq J-\nB$) can mostly be treated similarly as before (see Step~4(a) in Section~\ref{sec:ideas.wave}). The only difference is that there would be extra logarithmic factors, e.g., in \eqref{eq:intro.Q0.rhoJ.low.removed}, we only have $O_{\bfGmm} (r^{-J-3+\ell+\nB} u^{-\f 13} \log^{K}(\tfrac ru))$ (for some fixed $K$). As a result, in \eqref{eq:intro.rho.low.improved}, we only obtain $O_{\bfGmm} (r^{-J-3+\ell+\nB} u^{\f 23} \log^{K}(\tfrac ru))$, which this still be controlled by $O_{\bfGmm} (r^{-J-\f 73+\ell+\nB})$ in $\calM_{\wave}$ since $u \ls r$.

For the high angular modes (i.e., $\ell \geq J-\nB+1$), we need to be more careful with the logarithmic terms. Instead of \eqref{eq:intro.Q0.rhoJ}--\eqref{eq:intro.EJBox}, $E_{J; \Box}$ now takes the form
\begin{equation}\label{eq:intro.bad.EJBox.with.log}
E_{J; \Box} = \sum_{k=0}^{K_{J-1}} \left(\rd_{r}^{2} - \nu_{\Box} (\nu_{\Box}-1) r^{-2} + r^{-2} \rslap \right) \left(r^{-J+1} \log^{k} (\tfrac{r}{u}) \rPhi_{J-1, k} \right),
\end{equation}
which includes some terms $r^{-J-1} \log^k(\tfrac ru) \rPhi_{J-1,k}$ with $k \geq 1$. These terms are not annihilated by $(\bfK + 2r)^J$ (a property that we relied on crucially previously in Step~4(b) in Section~\ref{sec:ideas.wave}). This gives $Q_0 (\bfK^J \rho_J) = O_{\bfGmm}(r^{-1} u^{-\f 43} \log^{K_J} (\tfrac ru))$, where the right-hand side, despite having sufficient total decay, has an $r$-decay rate that is too slow to apply \eqref{eq:intro.rp.error.wave} with $p>1$.

Thus, for higher angular modes, instead of $\rho_{J}$, we control
\begin{equation}\label{eq:intro.crho.def}
\crho_{J} = \rho_{J} - \chi_{>4}(\tfrac r u) \sum_{k=1}^{K_J} r^{-J} \log^k(\tfrac ru) \rPhi_{J,k}.
\end{equation}
The introduction of $\chi_{>4}(\tfrac r u) \sum_{k=1}^{K_J} r^{-J} \log^k(\tfrac ru) \rPhi_{J,k}$ removes the worst terms in \eqref{eq:intro.bad.EJBox.with.log} at the expense of terms of the form
\begin{equation}\label{eq:intro.new.terms.from.crho}
r^{-J-2} \log^k(\tfrac ru) \rPhi_{J,k},\quad 1\leq k \leq K_J - 1.
\end{equation}
Altogether, using also that the $\rPhi_{J,k}$ terms satisfy improved decay (by point \eqref{item:log.radiation.improved} above), we have $Q_0 (\bfK^J \crho_J) = O_{\bfGmm}(r^{-2} u^{-\f 13} \log^{K_J} (\tfrac ru))$ in $\calM_{\wave}$, which is similar to \eqref{eq:intro.QJ.high.wave} before (except for some extra logarithmic factors). Notice moreover that both in \eqref{eq:intro.log.modified.remainder} and in \eqref{eq:intro.crho.def}, the higher radiation field with logarithms are only introduced for $r \gtrsim u$. Thus, they only contribute to $\calM_{\med}$ when $r$ and $u$ are comparable and so that the logarithms are bounded. As a result, the exact same argument as in Step~4(b) in Section~\ref{sec:ideas.wave} also give the bound \eqref{eq:intro.QJ.high.med}. The rest of the argument follows as before.

Let us remark that there is also a fundamental reason for which we need to consider $\crho_{J}$ as in \eqref{eq:intro.crho.def}. Similar to Step~4(b) in Section~\ref{sec:ideas.wave}, we need to take the limit $\f{(-1)^J}{J!}\lim_{r\to \infty} \bfK^J \crho_{(\geq J-\nB+1)J}(u,r,\theta)$. It is necessary to consider $\crho_{J}$ instead of $\rho_{J}$, for otherwise such a limit does not exist.

Finally, let us comment on an additional issue to be dealt with for the higher angular modes in the final step of the iteration (cf.~Section~\ref{sec:intro.end.of.iteration}). At this stage, it is possible that $\lim_{u\to \infty} \rcPhi_{(\geq J_{\frkf}-\nB+1)J_{\mathfrak f},k}(u, \theta) \neq 0$. (In particular, unlike in previous steps of the iteration, they do not exhibit improved decay.) As a result, when taking into account the terms \eqref{eq:intro.new.terms.from.crho}, we could have addition terms on the right-hand side of $Q_0(\bfK^J \rho_J)$ of size $O_{\bfGmm}(r^{-2} \log (\tfrac{r}{u}) \log^{K_{J}-1} r)$, which is quite a bit weaker than $O_{\bfGmm}(r^{-2} u^{-\f 13} \log^{K_J}(\tfrac ru))$. As a result, we are not able to prove the estimate \eqref{eq:intro.rho'.est}, but
\begin{equation}
\rho_{(\geq J_{\frkf} -\nB+1)J_{\frkf}+1}= \crho_{(\geq J_{\frkf}-\nB+1)J_{\frkf}}(u,r,\theta) - r^{-J_{\frkf}} \rPhi'_{(\leq J_{\frkf}-\nB+1),J_{\frkf}}(u,\theta)  = O_{\bfGmm}(r^{-\f{p-1}2-J_{\frkf}} u^{\f {p-1}2}).
\end{equation}
This is no better than the main terms in total decay, but has stronger $r$-decay (since $p>1$). This is thus similar to the estimates for the lower angular modes in the final step in Section~\ref{sec:intro.end.of.iteration} (see \eqref{eq:end.of.iteration.rho}), and is sufficient for the argument in Section~\ref{sec:intro.precise.tails} to obtain the precise tails.

\end{enumerate}

\subsection{Related works}\label{sec:related.works}

We discuss some related works in both the mathematics and the physics literature. Note that some of the works already discussed earlier are not mentioned again here.

\subsubsection{Wave equations on black hole spacetimes}

Motivated by the black hole stability problem, the boundedness and decay for solutions to wave equations have been extensively studied. In the Schwarzschild case, boundedness was proven in the classical works \cite{KW,rW1979}, and quantitative decay estimates were achieved in \cite{BSt, DRS,MMTT}. In the rotating Kerr case, boundedness \cite{DRK}, integrated local energy decay \cite{AB, DRSub2, DRL, TT} and pointwise decay \cite{DRL, DRSub2} were first proven when $|a| \ll M$. The Kerr case with the full subextremal range of parameters  $|a|< M$ was finally settled in \cite{DRSR}. See also \cite{DRSR2, Schlue, modestability} for related results.

Additionally, we refer the reader to \cite{ABB.Maxwell,AB.Maxwell,Blue.Maxwell,Ma.Maxwell,P.Maxwell,ST.Maxwell} for results on the Maxwell equations on black hole spacetimes, and to \cite{DHRT2,hLmT2018,LukNull,P.MBI} for nonlinear model problems.

Finally, the linearized gravitational perturbations on black hole spacetimes was first treated in the result of Dafermos--Holzegel--Rodnianski \cite{DHR}. There has been a flurry of recent activities associated to the problem: various extensions and alternative approaches for linearized gravitational perturbations on both Schwarzschild and rotating Kerr have been achieved \cite{ABBM, Ben22a, Ben22b, DHR2, HHV, HKW, Johnson, Ma.Spin2, SRTdC1, SRTdC2}. (See also results on perturbations of charged black holes in \cite{Giorgi1, Giorgi2, Giorgi3}.) More recently, the full nonlinear stability of asymptotically flat black holes have been proven both in the Schwarzschild case \cite{DHRT, KS} and in the Kerr case with $|a|\ll M$ \cite{GKS, KS.Kerr}.


\subsubsection{Price's law tails on black hole spacetimes}\label{sec:intro.price.refereces}
The sharp decay rate for wave equation was first derived heuristically by Price in the seminal \cite{Price}. There is a huge physics literature related to Price's law decay, see \cite{Barack, BurkoOri, CLSY1995, GPPI, GPPII, Leaver, Leaver.erratum, MarsaChoptuik, Poisson.decay} and the references therein.

In terms of rigorous mathematical results, the first proof of Price's law as an \textbf{upper bound} (up to a small loss) was proven \cite{DRPL} in a spherically symmetric, but nonlinear, setting. The sharp upper bound was later proven for the linear wave equation on Schwarzschild without symmetry assumptions \cite{DSS1,DSS2}. For more general stationary asymptotically flat spacetimes, the Price law upper bound was proven by Tataru \cite{Ta}; an alternative proof which also applies to non-stationary spacetimes was later given by Metcalfe--Tataru--Tohanuanu \cite{MTT}. See \cite{szL.linear, Morgan, MoWu} for related upper bound results.

The first \textbf{lower bound} result in the spirit of Price's law was achieved in our \cite{LO.instab} for Reissner--Nordstr\"om. (Although the lower bound was only proven on the event horizon in an averaged sense, it was nonetheless sufficient for the instability of the Cauchy horizon.) \textbf{Sharp pointwise asymptotics} was achieved independently by Angelopoulos--Aretakis--Gajic \cite{AAG2018,AAGPrice,AAGKerr} and by Hintz \cite{HintzPriceLaw}. Remarkably, \cite{AAGKerr} also understood precise mode couplings in Kerr, and resolved the conflicting numerology in the physics literature \cite{BO99, BK14, Hod99, KLPA, ZKB14}. Both the Angelopoulos--Aretakis--Gajic approach and the Hintz approach generalize to higher spin equations on a subextremal Kerr black hole; see \cite{syMlZ2021, syMlZ2022, syMlZ2022.2, pM2023}.

\subsubsection{General upper bound results}

In part motivated by the study of wave equations on asymptotically flat black hole spacetimes, there is a flurry of works on decay of wave equations which apply in more general asymptotically flat setting. We refer the reader to \cite{DRNM, gM2016, MTT, jOjS2023, Ta} for various approaches in obtaining pointwise upper bounds in general settings. For integrated local energy decay estimates in general settings, see \cite{MST}. For works on trapping in black hole spacetimes and beyond, see \cite{Hintz.trapping, jWmZ2011}. See also \cite{BVW2015, BVW2018, pHaV2023, Schlag.pointwise}.

For results on related but potentially different classes of equations, we refer the readers to \cite{dBjGRjlM2022, oCwSwSsT2008, Gajic.inverse.square, GajVan, Hintz.linear.waves, eS2022}.

\subsubsection{Related results on late time tails for linear equations on black holes} There are many works concerning late time tails of other related models on black hole spacetimes. We refer the readers to \cite{AAG.log} for results on next-to-leading order tails on black hole spacetimes (see also \cite{BVW2018} for general polyhomogeneous expansions for asymptotically flat spacetimes), to \cite{AAG2020, Gajic.EK} (see also \cite{Are1, Are2, Are3, CGZ, dGcW2021, GW}) for results on extremal black holes, to \cite{Gajic.inverse.square, mVdM2022} for results on charged scalar field and to \cite{PasShlRotVan, ShlRotVan} on results for the Klein--Gordon equation on Schwarzschild spacetime.

\subsubsection{Anomalous, dynamical and nonlinear tails in the physics literature}

Many of the phenomena discussed in Section~\ref{sec:intro.anomalous.etc} and follow from our main theorem have been observed and studied in the physics. The anomalously faster decay rate for linear wave in higher dimensional Schwarzschild black holes was discussed in \cite{vCsjYsDjL2003}. This was further analyzed in \cite{BCR.higher.d} within the context of a large class of wave equations that exhibit similar anomalous decay.

For modified linear tails on dynamical background, it was indicated in \cite{BR.dynamical} that on a spherically symmetric but dynamical background, scalar waves supported on mode $\ell$ decay with a tail of $t^{-3}$ if $\ell =0$ and $t^{-2-2\ell}$ if $\ell \geq 1$; in particular, the decay rate is different from the stationary case. This was already apparent in the numerics in \cite{GPPII} though it was not noticed by the authors. A similar phenomenon for higher $\ell$ was observed heuristically and numerically for the $\ell$-equivariant Einstein--wave map system \cite{BCRZ.WM}.

In a highly related but slightly different directions, there are heuristic and numerical works which studying nonlinear tails, for instance for power-type nonlinearity \cite{Szpak.I, Szpak.et.al.II}, for the Skyrme model in a neighborhood of the Skyrmion in spherical symmetry \cite{Bizon.Skyrmion}, for the Yang--Mills system in spherical symmetry \cite{BCR.YM} and for dispersive solutions to the Einstein--scalar field system in spherical symmetry \cite{BCR.SGSF, BCR.SGSF.reply, Szpak.comment}.

\subsubsection{Sharp decay estimates for nonlinear wave equations}

There are some recent works established strong upper bounds for nonlinear wave equations on the Minkowski space $\mathbb R^{3+1}$ \cite{sjDsyMyMxyY2022, szL2022.2, szL2022,  szLmT2021, mT2022}, as well as variations on more general asymptotically flat spacetimes \cite{Looi.2022.3, Looi.2022.4}, using the methods from \cite{AAGPrice, MTT}. Our present work in particular shows that the decay rates for the power-type nonlinearity in \cite{szL2022.2, szL2022} are sharp. However, for nonlinear equation satisfying the classical null condition, the sharp decay rates are more subtle; see \cite{LO.part2}.

As far as we are aware, there are very few rigorous results concerning lower bounds (or sharp asymptotics) in nonlinear settings except when the late time tails are determined by the propagation of an initial tail \cite{sjDsyMyMxyY2022}. The only result we are aware of is our paper \cite{LO1}, where we proved a sharp lower bound for a class of dispersive solutions to the Einstein--scalar field system in spherical symmetry.

\subsection{Organization of the paper}\label{sec:outline}

The remainder of the paper will be organized as follows.

In \textbf{Section~\ref{sec:main.theorem}}, we will introduce the general assumptions for the class of equations we consider and give the precise statements of the main theorem. In \textbf{Section~\ref{sec:examples}}, we discuss some examples for which our main theorem applies.

In \textbf{Section~\ref{sec:price}}, we will prove Price's law upper bound for stationary linear equations (see Theorem~\ref{thm:price}), assuming the main theorems in Section~\ref{sec:main.theorem}.

In the next two sections, we introduce some basic tools. In \textbf{Section~\ref{sec:vf}}, we introduce tools that fall into the general category of ``vector field method''. In \textbf{Section~\ref{sec:Minkowski.wave}}, we carry out some useful computations for solutions to the wave equation on exact Minkowski spacetime.

We then turn to the main arguments of the paper. As indicated in Section~\ref{sec:ideas.structure}, our analysis will be different in various regions of spacetimes: the analysis in the near-intermediate region is treated in \textbf{Section~\ref{sec:near}}, that in the intermediate region is treated in \textbf{Section~\ref{sec:med}}, and that in the wave zone is treated in \textbf{Section~\ref{sec:wave}}.

Putting together the analysis in the previous three sections, we prove our main theorems on upper bounds and on sharp asymptotics in \textbf{Section~\ref{sec:upper}} and \textbf{Section~\ref{sec:lower}}, respectively.

Finally, in \textbf{Section~\ref{sex:ext}}, we turn to the exterior region and extend our results to initial data on asymptotically flat initial data.

In \textbf{Appendix~\ref{sec:app}}, we give a list of notations and parameters used in the paper.

\subsection*{Acknowledgements} J.~Luk is partially supported by a Terman fellowship and the NSF grant DMS-2304445. S.-J.~Oh is partially supported by a Sloan Research Fellowship and an NSF CAREER grant DMS-1945615.

\section{Assumptions and the main theorems}\label{sec:main.theorem}
\subsection{Assumptions} \label{subsec:assumptions}
Let $d$ be an odd integer that is at least $3$. In this article, we shall consider the following scalar problem on a $(d+1)$-dimensional spacetime $(\calM, \bfg)$:
\begin{equation} \label{eq:main-eq}
	\calP \phi = \calN + f,
\end{equation}
where $\calN$ is a nonlinearity whose properties will be discussed in \ref{hyp:nonlin}--\ref{hyp:nonlin-null} below. The linear part of the equation, $\calP \phi$, is defined as
\begin{equation}\label{eq:P.def}
	\calP \phi := (\bfg^{-1})^{\alp \bt} \bfD_{\alp} \rd_{\bt} \phi + \bfB^{\alp} \rd_{\alp} \phi  + V \phi,
\end{equation}
where $\bfg^{-1}$ is the inverse metric, $\bfD$ is the Levi-Civita connection of $\bfg$, $\bfB$ is a smooth vector field, $V$ is a smooth real-valued function on $\calM$. The solution $\phi$ and the forcing term $f$ are also real-valued functions on $\calM$.

\subsubsection{List of parameters}
Our assumptions will be quantified using a number of parameters. We list them here for the convenience of the reader (the terminologies used here shall be clarified below):
\begin{itemize}
\item $R_{\far} > 0$: the radius used in the definition of $\calM_{\far}$ and $\calM_{\near}$,
\item $M_{c} \in \bbZ_{\geq 0}$: Klainerman vector field regularity for the coefficients,
\item $\dlt_{c} \in (0, 1]$: a small parameter in all the decay assumptions for the coefficients,
\item $J_{c} \in \bbZ_{\geq 2}$: (essentially) the maximum $j$ in the expansion of the coefficients in $r^{-j} \log^{k}(\tfrac{r}{u})$ in $\calM_{\wave}$,
\item $K_{c} \in \bbZ_{\geq 0}$: (essentially) the maximum $k$ in the expansion of the coefficients in $r^{-j} \log^{k}(\tfrac{r}{u})$ in $\calM_{\wave}$,
\item $\eta_{c} \in (0, 1]$: a small parameter in the decay assumptions for the remainders in $\calM_{\wave}$ for the coefficients,
\item $\alp_{\calN} \in \bbR$: minimal decay rate for the solution that ensure that the nonlinearity has suitable decay properties,
\item $M_{0} \in \bbZ_{\geq 0}$: Klainerman vector field regularity for the data and the solution,
\item $A_{0} \geq 0$: initially assumed size of the solution,
\item $\alp_{0} \in \bbR$: initially assumed $\tau$-decay rate for the solution,
\item $\nu_{0} \in \bbR$: $r^{\nu_{0}-\frac{d-1}{2}}$ is the initially assumed $r$-decay rate for the solution in $\calM_{\wave}$,
\item $D \geq 0$: size of the data,
\item $\alp_{d} > \frac{d-1}{2}$: decay rate for the data,
\item $\dlt_{d} \in (0, 1]$: a small parameter in all the decay assumptions for the data,
\item $J_{d} = \lceil \alp_{d} \rceil - \frac{d-1}{2}$: (essentially) the maximum $j$ in the expansion of the data in $r^{-j} \log^{k}(\tfrac{r}{u})$ in $\calM_{\wave}$,
\item $K_{d} \in \bbZ_{\geq 0}$: (essentially) the maximum $k$ in the expansion of the data in $r^{-j} \log^{k}(\tfrac{r}{u})$ in $\calM_{\wave}$,
\item $\eta_{d} = \alp_{d} + 1 -J_{d} - \frac{d-1}{2} \in (0, 1]$: a small parameter in the decay assumptions for the remainder terms in $\calM_{\wave}$ for the data.
\end{itemize}

\subsubsection{Global assumptions}\label{sec:global.assumptions}
We begin with the main global assumptions on the Lorentzian manifold $(\calM,\bfg)$. For the sake of simplicity, we assume in this paper that $\calM$ admits a single coordinate system. We will also restrict ourselves to a setting when $\calM$ can be thought of as only the future of an asymptotically flat $\{x^0 = 0\}$ hypersurface. While the scope of our method is more general, such an assumption already covers many cases of interest, such as the Minkowski spacetime, black hole exteriors, obstacle problems and perturbations thereof.
\begin{enumerate}[label=(G\arabic*)]
\item \label{hyp:topology} {\it Topology of $\calM$.} $\calM$ is a manifold (possibly with boundary) diffeomorphic to $(0,\infty) \times (\mathbb R^d \setminus \calK)$, where $\calK$ is either $\0$ or is a compact subset of $B_{R_{\far}} \subset \mathbb R^d$ with $C^{\infty}$ boundary $\rd \calK$ such that $B_{R_{\far}} \setminus \calK$ is nonempty. Here, $B_{R_{\far}}$ denotes the ball of radius $R_{\far}$ in $\bbR^{d}$ centered at the origin. Without loss of generality, we assume $R_{\far} \geq 1$.
\end{enumerate}


When the boundary $\rd \calM \neq \0$, we make the following assumption about the boundary:
\begin{enumerate}[resume*]
\item \label{hyp:bdry} {\it The boundary $\rd \calM$ of $\calM$.}
Assume that each component of $\rd \calM$ is either spacelike or null or timelike.
\end{enumerate}

We introduce globally defined rectangular coordinates $(x^{0}, x^{1}, \ldots, x^{d})$ using the diffeomorphism $\calM \to (0,\infty) \times (\mathbb R^{d} \setminus \calK)$. We will use $\rd_{\mu} = \rd_{x^\mu}$ to denote the coordinate vector field in this coordinate system.

With respect to the globally defined coordinates $(x^{0}, x^{1}, \ldots, x^{d})$, define the smooth vector field
\begin{equation}\label{eq:bfT.def}
	\bfT = \rd_{x^{0}} \quad \hbox{ on } \calM,
\end{equation}
as well as the level hypersurfaces
\begin{equation*}
	\calS_{t_{0}} = \set{p \in \calM : x^{0}(p) = t_{0}}.
\end{equation*}
By definition,
\begin{equation*}
	\bfT(x^{0}) = 1 \hbox{ on } \calM.
\end{equation*}
Moreover, by \ref{hyp:topology}, we see that $\rd \calM$ is diffeomorphic to $\bbR \times \rd \calK$. If $\rd \calM \neq \0$, then $\bfT$ is tangent to $\rd \calM$.


As our final global assumption on $(\calM, \bfg)$, we require the following causality conditions to hold:
\begin{enumerate}[resume*]
\item \label{hyp:T-tau} {\it Causality conditions.}
Assume that $\ud x^{0}$ is uniformly timelike on the initial hypersurface, in that $-C_{c} < \bfg^{-1}(\ud x^{0}, \ud x^{0}) |_{\calS_1} < - C_{c}^{-1}$. Moreover, for every $t > 1$, the level hypersurface $\calS_{t}$ lies in the future domain of dependence of $\calS_{1}$, in the sense that every past inextendible curve starting from a point on $\calS_{t}$ must intersect either $\calS_{1}$ or $\rd_{(\mathrm{t})} \calM$, where $\rd_{(\mathrm{t})} \calM$ is the union of all timelike components of $\rd \calM$ (see \ref{hyp:bdry}).
\end{enumerate}

\subsubsection{Alternative coordinate functions and the decomposition of $\calM$}\label{sec:decomposition}

Before we proceed to the remaining assumptions we impose, we first introduce a decomposition of $\calM$ into different regions.

Using the global rectangular coordinates $(x^{0}, x^{1}, \ldots, x^{d})$ introduced in Section~\ref{sec:global.assumptions}, we define the functions
\begin{equation} \label{eq:far-r-u-tht}
	r = \Big( \sum_{a=1}^{d} (x^{a})^{2} \Big)^{\frac{1}{2}}, \qquad u = x^{0} - r + 3 R_{\far}, \qquad \tht = \frac{x}{r} \in \bbS^{d-1}.
\end{equation}
For $R_{\far}\geq 1$ as in \ref{hyp:topology} and $u$ as above, introduce a globally defined smooth time function $\tau$
\begin{equation}\label{eq:tau.def}
\tau = x^{0} - \chi_{>4R_{\far}}(r) (r - 3 R_{\far}).
\end{equation}
With the above definitions, note that
\begin{equation*}
	\bfT \tau = 1 \hbox{ on } \calM, \quad \hbox{ and }
	\tau = \begin{cases}
	x^{0} & \hbox{ when } r \leq 2R_{\far}, \\
	u & \hbox{ when } r \geq 4R_{\far}.
	\end{cases}
\end{equation*}
For each $\tau_{0} \in \mathbb R$, define
\begin{equation}\label{eq:Sigma.def}
	\Sgm_{\tau_{0}} = \set{p \in \calM : \tau(p) = \tau_{0}}, \quad \rd_{(\mathrm{t})} \Sgm_{\tau_{0}} = \rd_{(\mathrm{t})} \calM \cap \Sgm_{\tau_{0}}.
\end{equation}

As indicated above, we will use $(\rd_{0},\rd_{1},\ldots,\rd_{d})$ to denote coordinate vector fields in the global coordinate system $(x^{0}, x^{1},\ldots, x^{d})$. From now on, we also use $(\rd_{u}, \rd_{r}, \rd_{\th^A})$ to denote the coordinate vector fields in the $(u,r,\th^A)$ coordinates system, after choosing $d-1$ coordinate functions $\th^A$ to form a local coordinate system on $\mathbb S^{d-1}$. Finally, it will also be convenient to introduce the notation that $(\urd_{\tau}, \urd_i = \urd_{x^i})$ (for $i=1,\ldots d$) denotes the coordinate vector fields in the $(\tau, x^{1}, \ldots, x^{d})$, as well as the notation $(\urd_{\tau}, \urd_r, \urd_{\th^A})$ in the $(\tau,r,\th^{A})$ coordinates. (Note that the coordinate vector fields $\urd_{\tau}$ indeed coincide so that the notation is unambiguous.)

We decompose $\calM$ into the following non-disjoint pieces
\begin{equation*}
	\calM = \calM_{\near} \cup \calM_{\med} \cup \calM_{\wave} \cup \calM_{\ext},
\end{equation*}
where
\begin{align*}
	\calM_{\near} &= \set{(u,r,\theta) \in \calM: r\leq 2R_{\far}, \, \tau \geq 1}, \\
	\calM_{\med} &= \set{(u,r,\theta) \in \calM: R_{\far} \leq r \leq 100 \eta_{0}^{-1} u, \, \tau \geq 1}, \\
	\calM_{\wave} &= \set{(u,r,\theta) \in \calM: r \geq \eta_{0}^{-1} u, \, \tau \geq 1}, \\
	\calM_{\ext} &= \set{(u,r,\theta) \in \calM: r \geq R_{\far}, \, x^{0} \geq 0, \, \tau \leq 2},
\end{align*}
where $\eta_{0} = \f 14$. We also introduce the notation that
\begin{equation}
	\calM_{\far} = \calM_{\med}\cup \calM_{\wave}.
\end{equation}

The crux of our analysis will involve the initial value problem for \eqref{eq:P.def} with initial data posed on the asymptotically null hypersurface $\Sigma_{1}$. Our main theorems (Main Theorem~\ref{thm:upper}--Main Theorem~\ref{thm:lower-sphsymm}) in Section~\ref{subsec:mainthm} will also be stated for such initial data. (Note that this includes as a special case compactly supported initial data posed on the asymptotically flat hypersurface $\calS_{1}$.)
In the remainder of Section~\ref{subsec:assumptions}, we will thus restrict ourselves to the assumptions in the region to the future of $\Sigma_{1}$, i.e., $\calM_{\near} \cup \calM_{\far}$. Later, in Section~\ref{sec:Cauchy}, we will state the assumptions in $\calM_{\ext}$ and state a more general theorem allowing for (non-compactly supported) data on $\calS_{1}$.

\subsubsection{Assumptions on $\bfg$, $\bfB$ and $V$}\label{sec:assumption.far.away}
We consider the assumptions on $\bfg$, $\bfB$ and $V$ in the region $\calM_{\far}\cup \calM_{\near}$. In order to precisely formulate these assumptions, we need to discuss ways to measure the size of the coefficients $\bfg^{-1}$, $\bfB$, $V$ and their derivatives.



Define the (Klainerman) vector fields by
\begin{equation}\label{eq:Klainerman}
\bfS := u \rd_u + r \rd_r,\quad \bfOmg_{ab} := x^a \rd_{x^b} - x^b \rd_{x^a}.
\end{equation}
Given a smooth function $a$ on (some subset of) $\calM_{\far} \cup \calM_{\near}$, a nonnegative integer $N$ and a fixed function $m(\tau,r)$, we define
\begin{align}
	a = O_{\bfGmm}^{N}(m(\tau, r)) \quad \impmi \quad \abs{(\tau \bfT)^{I_{\bfT}} (\brk{r} \rd_{r})^{I_{r}} \bfOmg_{ab}^{I_{\bfOmg_{ab}}} a} \aleq m(\tau, r) \quad \hbox{ for all } \abs{I} \leq N. \label{eq:O-Gmm-far-scalar}
\end{align}
In $\calM_{\far}$, we will also sometimes write $m(u,r)$ instead of $m(\tau,r)$. We remark that \eqref{eq:O-Gmm-far-scalar} is equivalent to
\begin{align}
	a = O_{\bfGmm}^{N}(m(\tau, r)) \quad \impmi \quad \abs{(\tau \bfT)^{I_{\bfT}} (\brk{r} \urd_{x^i})^{I_{x}} a} \aleq m(\tau, r) \quad \hbox{ for all } \abs{I} \leq N, \label{eq:O-Gmm-far-scalar.2}
\end{align}
where, as defined in Section~\ref{sec:decomposition}, $\urd_{x^i}$ denotes the coordinate derivatives in the $(\tau,x^1,\ldots,x^d)$ coordinate system.

In case $\bfa$ is a smooth contravariant tensor field, we replace the usual derivatives by Lie derivatives and define
\begin{align}
	\bfa \in O_{\bfGmm}^{N}(m(\tau, r)) \quad \impmi \quad \abs{\calL_{\tau\bfT}^{I_{\bfT}} \calL_{\brk{r} \rd_{r}}^{I_{r}} \calL_{\bfOmg_{ab}}^{I_{\bfOmg_{ab}}} \bfa} \aleq m(\tau, r) \quad \hbox{ for all } \abs{I} \leq N, \label{eq:O-Gmm-far}
\end{align}
where the absolute value of a contravariant tensor of rank $s$ is defined with respect to the coordinate vectors associated with $(x^{0}, x^{1}, \ldots, x^{d})$, i.e.,
\begin{equation} \label{eq:tensor-abs-near}
	\abs{\bfa}^{2} = \sum_{\alp_{j} \in \set{0, 1, \ldots, d}} (\bfa^{\alp_{1} \ldots \alp_{s}})^{2}.
\end{equation}

Here, the regularity index $N$ shall be referred to as the order of \emph{vector field regularity} of $a$.

\begin{remark} \label{rem:vf-reg}
The notion of vector field regularity is motivated by the following properties of a solution $\phi$ to the linear wave equation $\Box_{\bfm} \phi = 0$ with control over $\bfT^{I_{\bfT}} \bfS^{I_{\bfS}} \bfOmg_{jk}^{I_{\bfOmg_{jk}}} \phi$:
\begin{enumerate}
\item in $\set{r \ageq u}$, a derivative tangential to the outgoing null cones (i.e., $\rd_{r}$, $\rd_{\tht^{A}}$ in Bondi--Sachs-type coordinates) gains a power of $r^{-1}$, while a general derivative only gains $u^{-1}$;
\item in $\set{u \ageq r}$, a general derivative only gains $\brk{r}^{-1}$, while the stationary vector field $\rd_{u}$ (which equals $\rd_{x^{0}}$ in the rectangular coordinates) gains a power of $u^{-1}$.
\end{enumerate}
This statement is essentially the classical observation of Klainerman--Sideris \cite{KlSi} (see also Section~\ref{subsec:kl-sid}). We also note that $\bfS$ and $\bfOmg$ obey good commutation properties with the Minkowskian wave operator $\Box_{\bfm}$ (see Lemma~\ref{lem:comm-Qj}).

We also remark that inherent to the definition of $O_{\bfGmm}^{N}$ is the implicit statement that $\bfT$ is an almost stationary vector field in our generalized setting.
\end{remark}

We are now ready to state the assumptions on $\bfg$, $\bfB$ and $V$.  In $\calM_{\near}$, we simply assume the following:

\begin{enumerate}[label=($\bfg\bfB V$\arabic*)]
\item \label{hyp:T-almost-stat}
{\it Boundedness of $\bfg^{-1}$, $\bfB$ and $V$ in $\calM_{\near}$.} In $\calM_{\near}$, we assume that
\begin{align*}
	\bfg^{-1} = O_{\bfGmm}^{M_{c}}(1), \quad
	\bfB = O_{\bfGmm}^{M_{c}}(1), \quad
	V = O_{\bfGmm}^{M_{c}}(1).
\end{align*}
\end{enumerate}
In $\calM_{\med}$, we make similar assumptions, except for requiring decay in $r$:
\begin{enumerate}[resume*]
\item \label{hyp:med} {\it Asymptotics of $\bfg^{-1}$, $\bfB$ and $V$ in $\calM_{\med}$.} In $\calM_{\med}$, the coefficients $\bfg^{-1}$, $\bfB$ and $V$ have asymptotics of the form
\begin{align}
	\bfg^{-1} - \bfm^{-1} &\in O_{\bfGmm}^{M_{c}}(r^{-\dlt_{c}}) \label{eq:med-gab}, \\
	\bfB &\in O_{\bfGmm}^{M_{c}}(r^{-1-\dlt_{c}}) \label{eq:med-B}, \\
	V &\in O_{\bfGmm}^{M_{c}}(r^{-2-\dlt_{c}}) \label{eq:med-V}.
\end{align}
\end{enumerate}

On the other hand, in $\calM_{\wave}$ we need different assumptions on different components of $\bfg^{-1}$ and $\bfB$.
\begin{remark}[Heuristic principle for the assumptions in $\calM_{\wave}$]\label{rem:wave-heuristic}
The ensuing list of assumptions in $\calM_{\wave}$ is long mostly because,
in order to define higher radiation fields (see Section~\ref{subsec:recurrence-formal} below), we need to keep track of the expansion in $r^{-j} \log^{k}(\tfrac{r}{u})$ of each coefficient. However, our assumptions in all the cases are guided by a simple heuristic principle:
\begin{quote}
replacing $\rd_{u}$ by $u^{-1}$ and $\rd_{r}, r^{-1} \rd_{\tht^{A}}$ by $r^{-1}$, each term in $\calP$ must be $O(u^{-\alp+\nu} r^{-\nu})$ with $\nu > 1$ and $\alp > 2$.
\end{quote}
See also Remark~\ref{rem:wave-heuristic-met} below.
\end{remark}

Before we turn to the conditions in $\calM_{\wave}$, let us remind the reader that, the (inverse) Minkowski metric $\bfm^{-1}$ in the Bondi--Sachs coordinates $(u, r, \tht)$ takes the form
\begin{equation}\label{eq:m-1}
\bfm^{-1} = - \rd_{u} \otimes \rd_{r} - \rd_{r} \otimes \rd_{u} + \rd_{r} \otimes \rd_{r} + r^{-2} \rsgmm^{-1},
\end{equation}
while the Minkowski metric $\bfm$ itself takes the form
\begin{equation*}
\bfm = - \ud u \otimes \ud u - \ud u \otimes \ud r - \ud r \otimes \ud u + r^2 \rsgmm.
\end{equation*}
Here, and below, $\rsgmm$ and $\rsgmm^{-1}$ denote that standard unit round metric on $\mathbb S^{d-1}$ and its inverse, respectively.

The precise list of assumptions in $\calM_{\wave}$ are as follows.
\begin{enumerate}[resume*]
\item \label{hyp:wave-g} {\it Asymptotics of $\bfg^{-1}$ in $\calM_{\wave}$.} In $\calM_{\wave}$, the metric difference $\bfg^{-1} - \bfm^{-1}$ admits an expansion of the following form\footnote{The notation $\rh_{j, k}$ is consistent with our notation $\bfh = \bfg^{-1} - \bfm^{-1}$ below; see \eqref{eq:bfh-def}.}.
For $0 < \eta_{c} \leq 1$,
\begin{align}
	(\bfg^{-1})^{uu} &=
	\sum_{j=2}^{J_{c}} \sum_{k=0}^{K_{c}} r^{-j} \log^{k} (\tfrac{r}{u}) \rh_{j, k}^{uu}(u, \tht)
	+ \rem_{J_{c}+\eta_{c}}[\bfh^{uu}],   \label{eq:wave-g-uu-phg} \\
	(\bfg^{-1})^{u r} &= - 1
	+ \sum_{j=1}^{J_{c}-1} \sum_{k=0}^{K_{c}} r^{-j} \log^{k} (\tfrac{r}{u}) \rh_{j, k}^{ur}(u, \tht)
	+ \rem_{J_{c}-1+\eta_{c}}[\bfh^{ur}],   \label{eq:wave-g-ur-phg} \\
	r (\bfg^{-1})^{u A} &=
	\sum_{j=1}^{J_{c}-1} \sum_{k=0}^{K_{c}} r^{-j} \log^{k} (\tfrac{r}{u}) \rsh_{j, k}^{uA}(u, \tht)
	+ \rem_{J_{c}-1+\eta_{c}}[r \bfh^{uA}],   \label{eq:wave-g-uA-phg} \\
	(\bfg^{-1})^{r r} &= 1
	+ \sum_{j=0}^{J_{c}-2} \sum_{k=0}^{K_{c}} r^{-j} \log^{k} (\tfrac{r}{u}) \rh_{j, k}^{rr}(u, \tht)
	+ \rem_{J_{c}-2+\eta_{c}}[\bfh^{rr}],   \label{eq:wave-g-rr-phg} \\
	r (\bfg^{-1})^{r A} &=
	\sum_{j=0}^{J_{c}-2} \sum_{k=0}^{K_{c}} r^{-j} \log^{k} (\tfrac{r}{u}) \rsh_{j, k}^{r A}(u, \tht)
	+ \rem_{J_{c}-2+\eta_{c}}[r\bfh^{rA}],   \label{eq:wave-g-rA-phg} \\
	r^{2} (\bfg^{-1})^{AB} &= (\rsgmm^{-1})^{AB} + \rsh_{0, 0}^{AB}(u, \tht)
	+ \sum_{j=1}^{J_{c}-2} \sum_{k=0}^{K_{c}} r^{-j} \log^{k} (\tfrac{r}{u}) \rsh_{j, k}^{AB}(u, \tht)
	+ \rem_{J_{c}-2+\eta_{c}}[r^{2} \bfh^{AB}],   \label{eq:wave-g-AB-phg}
\end{align}
where
\begin{align}
	\rh_{j, k}^{uu}, \,
	\rh_{j, k}^{ur}, \,
	\rsh_{j, k}^{uA}, \,
	\rh_{j, k}^{rr}, \,
	\rsh_{j, k}^{rA}, \,
	\rsh_{j, k}^{AB}
	 &= O_{\bfGmm}^{M_{c}}(u^{j-\dlt_{c}}), \label{eq:wave-gjk-phg}
\end{align}
whenever the object on the left-hand side is defined. Moreover, the remainders satisfy
\begin{align}
	\rem_{J_{c}+\eta_{c}}[\bfh^{uu}]
	 &= O_{\bfGmm}^{M_{c}}(r^{-(J_{c}+\eta_{c})} u^{J_{c}+\eta_{c}-\dlt_{c}}), \label{eq:wave-grem-uu-phg} \\
	\rem_{J_{c}-1+\eta_{c}}[\bfh^{ur}], \rem_{J_{c}-1+\eta_{c}}[r \bfh^{uA}]
	 &= O_{\bfGmm}^{M_{c}}(r^{-(J_{c}-1+\eta_{c})}u^{J_{c}-1+\eta_{c}-\dlt_{c}}), \label{eq:wave-grem-u-phg}  \\
	\rem_{J_{c}-2+\eta_{c}}[\bfh^{rr}],
	\rem_{J_{c}-2+\eta_{c}}[r \bfh^{rA}],
	\rem_{J_{c}-2+\eta_{c}}[r^{2} \bfh^{AB}]
	 &= O_{\bfGmm}^{M_{c}}(r^{-(J_{c}-2+\eta_{c})}u^{J_{c}-2+\eta_{c}-\dlt_{c}}). \label{eq:wave-grem-phg}
\end{align}
\end{enumerate}

\begin{remark}\label{rmk:gAB.log}
In \eqref{eq:wave-g-AB-phg} above, while we have allowed for a $j=0$ term, we have imposed the condition that $k=0$ when $j=0$. This can be relaxed to include $k \geq 1$ terms in the expansion, but we need to impose an additional condition that $\det(((\rsgmm^{-1})^{AB} + \sum_{k=0}^{K_{c}} \log^{k}(\tfrac ru) \mathring{\slashed{\bfh}}^{AB}_{0,k})_{A,B=1}^{d-1})$ is a function of $(u,\theta)$ alone (i.e., has no log terms).
See Remark~\ref{rmk:angular.cancellations} for the relevance of this condition.
\end{remark}

\begin{enumerate}[resume*]
\item \label{hyp:wave-B} {\it Asymptotics of $\bfB$ in $\calM_{\wave}$.} The vector field $\bfB$ admits an expansion of the form
\begin{align}
	\bfB^{u} &= \sum_{j=2}^{J_{c}} \sum_{k=0}^{K_{c}} r^{-j} \log^{k} (\tfrac{r}{u}) \rbfB_{j, k}^{u}(u, \tht) + \rem_{J_{c}+\eta_{c}}[\bfB^{u}], \label{eq:wave-B-u-phg} \\
	\bfB^{r} &= \sum_{j=1}^{J_{c}-1} \sum_{k=0}^{K_{c}} r^{-j} \log^{k} (\tfrac{r}{u}) \rbfB_{j, k}^{r}(u, \tht) + \rem_{J_{c}-1+\eta_{c}}[\bfB^{r}], \label{eq:wave-B-r-phg} \\
	r \bfB^{A} &= \sum_{j=1}^{J_{c}-1} \sum_{k=0}^{K_{c}} r^{-j} \log^{k} (\tfrac{r}{u}) \rsbfB_{j, k}^{A}(u, \tht) + \rem_{J_{c}-1+\eta_{c}}[r \bfB^{A}], \label{eq:wave-B-A-phg}
\end{align}
where
\begin{align}
	\rbfB_{j, k}^{u}(u, \tht), \,
	\rbfB_{j, k}^{u}(u, \tht), \,
	\rsbfB_{j, k}^{A}(u, \tht)
	 &= O_{\bfGmm}^{M_{c}}(u^{j-1-\dlt_{c}}), \label{eq:wave-Bjk-phg}
\end{align}
whenever $\rbfB_{j, k}^{\alp}$ is defined, and the remainders satisfy
\begin{align}
	\rem_{J_{c}+\eta_{c}}[\bfB^{u}] &= O_{\bfGmm}^{M_{c}}(r^{-(J_{c}+\eta_{c})}u^{J_{c}-1+\eta_{c}-\dlt_{c}}). \label{eq:wave-Brem-u-phg} \\
	\rem_{J_{c}-1+\eta_{c}}[\bfB^{r}], \rem_{J_{c}-1+\eta_{c}}[r \bfB^{A}] &= O_{\bfGmm}^{M_{c}}(r^{-(J_{c}-1+\eta_{c})}u^{J_{c}-2+\eta_{c}-\dlt_{c}}). \label{eq:wave-Brem-phg}
\end{align}
\item \label{hyp:wave-V} {\it Asymptotics of $V$ in $\calM_{\wave}$.}
The function $V$ admits an expansion of the form
\begin{align}
	V &= \sum_{j=2}^{J_{c}} \sum_{k=0}^{K_{c}} r^{-j} \log^{k} (\tfrac{r}{u}) \rV_{j, k}(u, \tht) + \rem_{J_{c}+\eta_{c}}[V], \label{eq:wave-V-phg}
\end{align}
\begin{align}
	\rV_{j, k}(u, \tht) &\in O_{\bfGmm}^{M_{c}}(u^{j-2-\dlt_{c}}), \label{eq:wave-Vj}\\
	\rem_{J_{c}+\eta_{c}}[V] &\in O_{\bfGmm}^{M_{c}}(r^{-(J_{c}+\eta_{c})}u^{J_{c}-2+\eta_{c}-\dlt_{c}}). \label{eq:wave-Vrem-phg}
\end{align}
\end{enumerate}

Some remarks concerning \ref{hyp:wave-g}--\ref{hyp:wave-V} are in order.
\begin{remark} \label{rem:wave-heuristic-met}
Note that the coefficients $(\bfg^{-1} - \bfm^{-1})^{u \alp}$ for $\alp = u, r, A$ and $\bfB^{u}$ are assumed to decay better in $r$ than the rest. These stronger assumptions compensate the worse $r$-decay of the terms in $\calP \phi$ involving the ``bad derivative'' $\rd_{u}$ (i.e., worst-decaying in $r$); recall Remark~\ref{rem:wave-heuristic}. 
\end{remark}

\begin{remark}[Bondi--Sachs-type condition] \label{rem:bondi}
We note in particular that the assumptions \eqref{eq:wave-g-uu-phg} and \eqref{eq:wave-g-uA-phg} for $(\bfg^{-1})^{uu} = \bfg^{-1}(\ud u, \ud u)$ and $(\bfg^{-1})^{uA} = \bfg^{-1}(\ud u, \ud \tht^{A})$, respectively, may be thought of as approximations of the \emph{Bondi--Sachs conditions}\footnote{The full Bondi--Sachs coordinate system would consist of $(u, \check{r}, \tht)$, where $(u, \tht)$ satisfy \eqref{eq:bondi-sachs} and $\check{r}$ is the \emph{areal coordinate} $\check{r} = \left(- \frac{\det \bfg}{\det \rsgmm}\right)^{\frac{1}{2(d-1)}}$, where $\rsgmm = \rsgmm(\tht)$ is the metric on the unit round sphere $\bbS^{d-1}$. By \ref{hyp:med} and \ref{hyp:wave-g}, $\check{r} = r(1+O_{\bfGmm}^{M_{c}}(r^{-\dlt_{c}}))$ in $\calM_{\far}$.}, which are \cite{BBM1962,Sachs1962}:
\begin{equation} \label{eq:bondi-sachs}
	(\bfg^{-1})^{uu} = 0, \quad (\bfg^{-1})^{u A} = 0.
\end{equation}
where $\rsgmm =\rsgmm(\tht)$ is the metric on the unit round sphere $\bbS^{d-1}$. The first equation says that $u$ is an eikonal function, and the second equation says that the shift vector of the $r$-foliation of each constant-$u$ hypersurface is zero. One expects in principle that our decay assumptions imposed above imply that one can introduce a change of coordinates to $(u^{\ast}, r, \tht^{\ast})$ so that the actual Bondi--Sachs condition is satisfied. However, in the proof of our theorem, we will not require the actual Bondi--Sachs condition.
\end{remark}

\subsection{Stationary estimate}

We now formulate an important assumption, which we call the \emph{stationary estimate} (following similar terminology in \cite{MST, MTT}). This can be viewed as a dynamical analogue of the absence of zero eigenvalues and resonances at zero energy; see Remark~\ref{rmk:no.eigenvalue.resonance.zero.energy}.

Before introducing the assumption on stationary estimate, we need an assumption on the boundary condition of $\phi$ on the timelike components of $\rd \calM$ (see \eqref{hyp:bdry}).
\begin{enumerate}[label=(SE\arabic*)]
\item \label{hyp:bdry.cond} {\it Boundary conditions.}
On the union of all timelike components of $\rd \calM$, denoted by $\rd_{(\mathrm{t})} \calM$, assume the Dirichlet boundary condition
\begin{equation*}
\left. \phi \right|_{\rd_{(\mathrm{t})} \calM \cap \calS_{t}} = 0 \hbox{ for all } t \geq 1.
\end{equation*}
\end{enumerate}

\begin{remark}
If $\rd_{(\mathrm{t})} \calM \neq \0$, the condition \ref{hyp:bdry.cond} would guarantee the well-posedness of the initial-boundary value problem on $\cup_{t \geq 1} \calS_{t}$ with initial conditions on $\calS_{1}$. We shall \emph{not} directly need such a well-posedness result, however, since one of our initial assumptions is the existence of a solution $\phi$ with appropriate a-priori (vector field regularity) bounds; see \ref{hyp:sol} below.
\end{remark}

\begin{remark}[Other boundary conditions]
We assume the Dirichlet boundary condition \ref{hyp:bdry.cond} for simplicity. (More specifically, its technical advantage is that it is independent of the metric $\bfg$ unlike, say, the Neumann boundary condition.) With appropriate modifications of the uniqueness part of \ref{hyp:ult-stat} or \ref{hyp:ult-stat'} below, our results may be extended to other boundary conditions as well; we leave the precise formulation to the interested reader.
\end{remark}

To formulate the stationary estimate, we introduce the quotient space $\Sgm_{\infty} := \calM / \sim$, where $p \sim q$ if and only if $p$ and $q$ lie on an integral curve of $\bfT$. Since this quotient space is clearly diffeomorphic to $\Sgm_{\tau}$ for any $\tau$ (see \eqref{eq:Sigma.def}), we may equip $\Sgm_{\infty}$ with the smooth coordinates $x$ and view
\begin{equation*}
	\Sgm_{\infty} = \set{x \in \bbR^{n} : (\tau, x) \in \Sgm_{\tau} \hbox{ for some } \tau}.
\end{equation*}
For any $\tau \in \bbR \cup \set{\infty}$ and a subset $U$ of $\Sgm_{\tau}$ with $C^{\infty}$ boundary, we define the spaces $H^{s}(U)$ and $L^{2}(U)$ using the coordinates $(x^{1}, \ldots, x^{d})$ on $\Sgm_{\tau}$. Moreover, define
\begin{align*}
H^{s}_{comp}(\Sgm_{\tau}) &= \set{f \in H^{s}(\Sgm_{\tau}) : f = 0 \hbox{ outside } \set{r \leq R} \hbox{ for some } R > 0}, \\
H^{s}_{loc}(\Sgm_{\tau}) &= \set{f \in L^{1}_{loc}(\Sgm_{\tau}) : f \in H^{s}(\Sgm_{\tau} \cap \set{r \leq R}) \hbox{ for all } R > 0}.
\end{align*}

\begin{enumerate}[resume*]
\item \label{hyp:ult-stat}
{\it Ultimate stationarity of $(\bfg, \bfB, V)$ and stationary estimate.}
There exists an asymptotically flat metric ${}^{(\infty)} \bfg$, vector field ${}^{(\infty)} \bfB$ and a real-valued function ${}^{(\infty)} V$ on $\calM_{\far} \cup \calM_{\near}$ such that $\calL_{\bfT} ({}^{(\infty)}\bfg) = 0$, $\calL_{\bfT} ({}^{(\infty)} \bfB) = 0$, $\bfT ({}^{(\infty) }V) = 0$ and
\begin{equation}\label{eq:T-almost-stat-limit}
	\abs{\bfg^{-1} - {}^{(\infty)}\bfg^{-1}}_{C^{M_{c}}(\Sgm_{\tau})}
	+ \abs{\bfB - {}^{(\infty)}\bfB}_{C^{M_{c}}(\Sgm_{\tau})}
	+ \abs{V - {}^{(\infty)} V}_{C^{M_{c}}(\Sgm_{\tau})} \leq C \tau^{-\dlt_{c}},
\end{equation}
and for the $\bfT$ derivatives,
\begin{align}
	\calL_{\bfT}\bfg^{-1} = O_{\bfGmm}^{M_{c}-1}(\tau^{-1-\de_c}), \quad
	\calL_{\bfT}\bfB = O_{\bfGmm}^{M_{c}-1}(\tau^{-1-\de_c}), \quad
	\bfT V = O_{\bfGmm}^{M_{c}-1}(\tau^{-1-\de_c}). \label{eq:T-almost-stat-2}
\end{align}
Let
\begin{equation}\label{eq:Pinfty.def}
	{}^{(\infty)}\calP \phi := ({}^{(\infty)}\bfg^{-1})^{\alp \bt} {}^{(\infty)}\bfD_{\alp} \rd_{\bt} \phi + {}^{(\infty)}\bfB^{\alp} \rd_{\alp} \phi  + {}^{(\infty)}V \phi,
\end{equation}
where ${}^{(\infty)}\bfD_{\alp}$ is the Levi-Civita connection of ${}^{(\infty)}\bfg^{-1}$. Moreover, define ${}^{(\infty)} \calP_{0}$, which will be referred to as the \emph{stationary operator}, by dropping $\rd_{\tau}$ from the expression for ${}^{(\infty)} \calP$ in the coordinate system $(\tau, x^{1}, \ldots, x^{d})$. We may view ${}^{(\infty)} \calP_{0}$ as a linear operator on the space of functions on $\Sgm_{\infty}$. We assume the following two properties of ${}^{(\infty)} \calP_{0}$:
\begin{enumerate}
\item (Existence) There exists a linear right-inverse ${}^{(\infty)} \calR_{0} : H^{s_{c}}_{comp}(\Sgm_{\infty}) \to L^{2}_{loc}(\Sgm_{\infty})$ of ${}^{(\infty)} \calP_{0}$ such that for every $0 \leq s \leq M_{c}-1$ and $R, R' > 0$ there exists $C_{s, R, R'}$ so that if $f \in H^{s+s_{c}}(\Sgm_{\infty})$ and $\supp f \subseteq \set{r \leq R'}$, then
\begin{equation} \label{eq:zero-res-est}
	\nrm{{}^{(\infty)} \calR_{0} f}_{H^{s+2}(\Sgm_{\infty} \cap \set{r \leq R})}
	\leq C_{s, R, R'} \nrm{f}_{H^{s+s_{c}}(\Sgm_{\infty})}.
\end{equation}

\item (Uniqueness) For $\psi \in C^{M_{c}-s_{c}-\lfloor \frac{d}{2} \rfloor}(\br{\Sgm_{\infty}})$,
\begin{equation*}
	{}^{(\infty)} \calP_{0} \psi = 0, \quad \left. \psi \right|_{\rd_{(\mathrm{t})} \Sgm_{\infty}} = 0, \quad
	\abs{\psi} = O(\brk{r}^{-(d-2)}) \quad \imp \quad \psi = 0.
\end{equation*}
\end{enumerate}
\end{enumerate}

A few remarks are in order concerning \ref{hyp:ult-stat}.
\begin{remark}[\ref{hyp:ult-stat} as no eigenvalue/resonance at zero energy]\label{rmk:no.eigenvalue.resonance.zero.energy}
We view \ref{hyp:ult-stat} as a dynamical generalization of the \emph{no-eigenvalue/resonance-at-zero-energy condition}. Indeed, when $(\bfg, \bfB, V)$ is stationary (i.e., $\calL_{\bfT} \bfg = 0$, $\calL_{\bfT} \bfB = 0$ and $\bfT V = 0$), we may take $(\bfg, \bfB, V) = ({}^{(\infty)} \bfg, {}^{(\infty)} \bfB, {}^{(\infty)} V)$ and ${}^{(\infty)} \calP_{0}$ is the operator obtained by (formally) taking the Fourier transform in $t$ of $\calP$ and evaluating at zero frequency. Then (b) is what is usually called the no-eigenvalue/resonance-at-zero-energy condition.
\end{remark}

\begin{remark}[Condition for ellipticity of ${}^{(\infty)} \calP_{0}$ and the exponent $s_{c}$]\label{rmk:P0.elliptic}
Observe that ${}^{(\infty)} \calP_{0}$ is elliptic if and only if $\bfT$ is timelike with respect to ${}^{(\infty)} \bfg$ on the entire $\Sgm_{(\infty)}$. In this case, by Fredholm alternative and elliptic regularity, (a) with $s_{c} = 0$ would follow from (b). See discussions in Example~\ref{ex:classical}.

However, there are interesting cases when $\bfT$ is \emph{not} everywhere timelike yet (a) and (b) hold. For instance, this is the case of the Kerr spacetime; see \cite{Ta}. In fact, the argument of \cite{Ta} suggests that (a) and (b) would follow from an integrated local energy decay estimate, possibly with derivative losses, on a stationary spacetime. In this case, we generally have $s_{c} > 0$.
\end{remark}

\begin{remark}[Further technical remarks]
Observe that $H^{s}$ may be defined with respect to $\left. \bfg \right|_{\Sgm_{\tau}}$, $\left. {}^{(\infty)} \bfg \right|_{\Sgm_{\tau}}$, or in coordinates $(x^{1}, \ldots x^{d})$; they all lead to equivalent definitions.

In (2), note that $O(r^{-(d-2)})$ is the spatial decay of the Newtonian potential. We also remark that the regularity $C^{M_{c}-s_{c}-\lfloor \frac{d}{2} \rfloor}(\br{\Sgm_{\infty}})$ is the expected regularity of ${}^{(\infty)} \calR_{0} {}^{(\infty)} \calP_{0} \psi$ for $\psi \in C^{\infty}_{c}(\Sgm_{\infty})$ via (1) and Sobolev embedding; see the proof of Proposition~\ref{prop:stat-est} in Section~\ref{subsec:stat-est} for the appearance of this regularity.
\end{remark}

We also give an alternative to \ref{hyp:ult-stat}, which does not require that $\bfg$ settles down to a stationary metric, but is still sufficient for the sharp upper bound theorem, i.e., Main Theorem~\ref{thm:upper}. (Note, however, that the sharp asymptotics theorem, i.e., Main Theorem~\ref{thm:lower}, requires the assumption \ref{hyp:ult-stat}.)
\begin{enumerate}[label=\ref{hyp:ult-stat}']
\item \label{hyp:ult-stat'}
For every $\tau \geq \tau_{0}$, define ${}^{(\tau)} \calP_{0}$ by dropping $\rd_{\tau}$ from the expression for $\calP$ in the coordinate system $(\tau, x^{1}, \ldots, x^{d})$. We may view ${}^{(\tau)} \calP_{0}$ as a linear operator on the space of functions on $\Sgm_{\tau}$. We assume that, for some $\tau_{1} \geq \tau_{0}$, the following two properties of ${}^{(\tau)} \calP_{0}$ for all $\tau \geq \tau_{1}$:
\begin{enumerate}
\item (Existence) There exists a linear right-inverse ${}^{(\tau)} \calR_{0} : H^{s_{c}}_{comp}(\Sgm_{\tau_{0}}) \to L^{2}_{loc}(\Sgm_{\tau_{0}})$ of ${}^{(\tau)} \calP_{0}$ such that, for every $0 \leq s \leq M_{c}-1$ and $R, R' > 0$, there exists $C_{s, R, R'}$ \emph{independent} of $\tau$ so that if $f \in H^{s+s_{c}}(\Sgm_{\tau})$ and $\supp f \subseteq \set{r \leq R'}$, then
\begin{equation} \label{eq:zero-res-est'} \tag{\ref{eq:zero-res-est}'}
	\nrm{{}^{(\tau)} \calR_{0} f}_{H^{s+2}(\Sgm_{\tau} \cap \set{r \leq R})}
	\leq C_{s, R, R'} \nrm{f}_{H^{s+s_{c}}(\Sgm_{\tau})}.
\end{equation}

\item (Uniqueness) For $\psi \in C^{M_{c}-s_{c}-\lfloor \frac{d}{2} \rfloor}(\br{\Sgm_{\tau}})$,
\begin{equation*}
	{}^{(\tau)} \calP_{0} \psi = 0, \quad \left. \psi \right|_{\rd_{(\mathrm{t})} \Sgm_{\tau}} = 0, \quad
	\abs{\psi} = O(r^{-(d-2)}) \quad \imp \quad \psi = 0.
\end{equation*}
\end{enumerate}

\end{enumerate}

\begin{remark}
We note that when ${}^{(\tau)} \calP_{0}$ is elliptic, or equivalently, if $\bfT$ is everywhere timelike, we have $s_{c} = 0$ and, by using the techniques in Section~\ref{subsec:stat-est}, it is possible to show that \ref{hyp:ult-stat'} is weaker than \ref{hyp:ult-stat} (after restricting to sufficiently large $\tau$'s) . Such an implication may hold in general, but we do not know of a proof that does not rely on specific features of the proof of \eqref{eq:zero-res-est} for ${}^{(\infty)} \calP_{0}$, which is why we separated \ref{hyp:ult-stat'} from \ref{hyp:ult-stat}.
\end{remark}

\subsubsection{Assumptions on nonlinearity}\label{sec:nonlinearity.assumptions}
The method developed in this paper is applicable to a wide class of nonlinearities on asymptotically flat spacetimes of odd space dimensions. Our formulation  includes many cases of interest (smooth quasilinear nonlinearity with bounded coefficients). 

\begin{enumerate}[label=($\calN$\arabic*)]
\item \label{hyp:nonlin} {\it Form of the nonlinearity.} We consider a general quasilinear nonlinearity $\calN$ of the form
\begin{align*}
	\calN(p; \phi(\cdot))
	=
	& \quasi^{\mu \nu} (p, \phi, \ud \phi) \bfD_{\mu} \rd_{\nu} \phi
	+ \semi(p, \phi, \ud \phi),
\end{align*}
where
\begin{align*}
	\quasi : & \set{(p, z, \bfxi) : p \in \calM, \, z \in \bbR, \, \bfxi \in T^{\ast}_{p} \calM} \to (T \calM)^{\otimes 2}, \quad \quasi(p, z, \bfxi) \in (T_{p} \calM)^{\otimes 2} \\
	\semi : & \set{(p, z, \bfxi) : p \in \calM, \, z \in \bbR, \, \bfxi \in T^{\ast}_{p} \calM} \to \bbR,
\end{align*}
are smooth.
\end{enumerate}

By the smoothness of $\quasi$ and $\semi$, we mean the following. Consider $\calT : \set{(p, z, \bfxi) : p \in \calM, \, z \in \bbR, \, \bfxi \in T^{\ast}_{p} \calM} \to (T \calM)^{\otimes r}$ with $\calT(p, z, \bfxi) \in (T_{p} \calM)^{\otimes r}$ ($r$ is called the \emph{rank} of $\calT$). Note that $\quasi$ and $\semi$ are particular instances of such an object. We say that $\calT$ is \emph{smooth} if for each $p \in \calM$ and for each local coordinate system $(y^{0}, \ldots, y^{d})$ defined in a neighborhood of $p$, we have
\begin{equation*}
\calT = \calT^{\alp_{1} \ldots \alp_{r}}(y^{0}, \ldots y^{d}, z, \eta_{0}, \ldots, \eta_{d}) \rd_{y^{\alp_{1}}} \ldots \rd_{y^{\alp_{r}}}
\end{equation*}
with each component $\calT^{\alp_{1} \ldots \alp_{r}}(y^{0}, \ldots y^{d}, z, \eta_{0}, \ldots, \eta_{d})$ smooth, where $(\eta_{0}, \ldots, \eta_{d})$ is the coordinate system on $T^{\ast}_{p} \calM$ defined with respect to the basis $(\ud y^{0}, \ldots, \ud y^{d})$ (i.e., $\bfxi = \eta_{\bt} \ud y^{\bt}$), and $(\rd_{y^{0}}, \ldots, \rd_{y^{d}})$ is the standard coordinate basis for $T_{p} \calM$. Observe that if $\phi'$ is a smooth function and $\bfomg$ is a smooth $1$-form on $\calM$, then $\calT(\cdot, \phi', \bfomg)$ defines a smooth contravariant $r$-tensor field on $\calM$.

Before proceeding to the assumptions on $\quasi$ and $\semi$, we need to discuss ways to differentiate objects of the form $\calT$. Fix a local coordinate system $(y, z, \eta)$ as above. We define $\rd_{z} \calT : \set{(p, z, \bfxi) : p \in \calM, \, z \in \bbR, \, \bfxi \in T^{\ast}_{p} \calM} \to (T \calM)^{\otimes r}$ and $\rd_{\bfxi} \calT : \set{(p, z, \bfxi) : p \in \calM, \, z \in \bbR, \, \bfxi \in T^{\ast}_{p} \calM} \to (T \calM)^{\otimes (r+1)}$ in local coordinates as
\begin{align*}
	\rd_{z} \calT &= \rd_{z} \calT^{\alp_{1} \ldots \alp_{r}} \rd_{y^{\alp_{1}}} \ldots \rd_{y^{\alp_{r}}}, \\
	\rd_{\bfxi} \calT &= \rd_{\eta_{\bt}} \calT^{\alp_{1} \ldots \alp_{r}} \rd_{y^{\alp_{1}}} \ldots \rd_{y^{\alp_{r}}} \rd_{y^{\bt}}.
\end{align*}
Observe that $\rd_{z} \calT$ and $\rd_{\bfxi} \calT$ are objects of the same type as $\calT$ with ranks $r$ and $r+1$, respectively.


Given a vector field $\bfX$, we define the \emph{Lie derivative} by the local coordinate expression
\begin{align*}
	\calL_{\bfX} \calT^{\alp_{1} \ldots \alp_{r}}
	&= \bfX^{\bt} \rd_{y^{\bt}} \calT^{\alp_{1} \ldots \alp_{r}}(y, z, \eta)
	- \rd_{\eta^{\bt'}} \calT^{\alp_{1} \ldots \alp_{r}}(y, z, \eta) (\rd_{\bt'} \bfX^{\bt})\eta_{\bt}  \\
	&\peq - \calT^{\bt \ldots \alp_{r}}(y, z, \eta) \rd_{y^{\bt}} \bfX^{\alp_{1}}
	- \ldots
	- \calT^{\alp_{1} \ldots \bt}(y, z, \eta) \rd_{y^{\bt}} \bfX^{\alp_{r}}.
\end{align*}
Such definitions allow us to have the following \emph{chain rules}:
\begin{itemize}
\item for any smooth functions $\phi$ and $\phi'$,
\begin{equation} \label{eq:nonlin-chain-rule-X}
	\calL_{\bfX}(\calT(x, \phi', \ud \phi))
	= (\calL_{\bfX} \calT)(x, \phi', \ud \phi) + (\rd_{z} \calT)(x, \phi', \ud \phi) \bfX \phi' + (\rd_{\bfxi} \calT)(x, \phi', \ud \phi) (\ud \bfX \phi),
\end{equation}
where $(\rd_{\bfxi} \calT)(x, \phi', \ud \phi) (\ud \bfX \phi) = \rd_{\eta_{\bt}} \calT(x, \phi', \ud \phi) \rd_{y^{\bt}} \bfX \phi$ in local coordinates; and
\item for any one-parameter families of smooth functions $\phi_{h}$ and $\phi'_{h}$ defined for $h \in (-\dlt_{h}, \dlt_{h})$ for some $\dlt_{h} > 0$,
\begin{equation} \label{eq:nonlin-chain-rule-z}
	\left. \frac{\ud}{\ud h} (\calT(x, \phi_{h}', \ud \phi_{h})) \right|_{h = 0}
	= (\rd_{z} \calT)(x, \phi'_{0}, \ud \phi_{0})  \dlt \phi' + (\rd_{\bfxi} \calT)(x, \phi'_{0}, \ud \phi_{0}) (\ud  \dlt \phi),
\end{equation}
where $\dlt \phi = \left. \frac{\ud}{\ud h} \phi_{h} \right|_{h = 0}$ and $\dlt \phi' = \left. \frac{\ud}{\ud h} \phi'_{h} \right|_{h = 0}$.
\end{itemize}
The former is a consequence of the definitions above. The latter follows from the former by adding a dummy variable $h$, viewing $\calT$ as defined on $(-\dlt_{h}, \dlt_{h}) \times \calM$, and then expressing $\frac{\ud}{\ud h} = \calL_{\rd_{h}}$ (the vector field $\rd_{h}$ is well-defined due to the product structure).

Using the notions just introduced, we may extend the $O_{\bfGmm}$ notation in the following way. Let $(x^{0}, x^{1}, \ldots, x^{d})$ be the global coordinate system given in Section~\ref{sec:global.assumptions}, and define $\tau$, $u$, $r$ as in \eqref{eq:far-r-u-tht}--\eqref{eq:tau.def}. We write
$\calT = O_{\bfGmm}^{M}(m(\tau, r))$ if there exists $C > 0$ such that
\begin{equation*}
	\sup_{p \in \calM_{\far}, \, z \in \bbR, \, \bfxi \in T_{p}^{\ast} M} \abs{\calZ_{a_{1}} \ldots \calZ_{a_{M'}} \calT(p, z, \bfxi)} \leq C m(\tau, r) \quad \hbox{ for all } 0 \leq M' \leq M,
\end{equation*}
where each $\calZ_{a_{m}}$ is an operation of the form $\calZ = \rd_{z}$, $\calZ = \rd_{\bfxi}$ or $\calZ = \calL_{\bfGmm}$ with $\bfGmm \in \set{\tau \bfT, r \rd_{r}, \bfOmg}$ (cf.~\eqref{eq:O-Gmm-far-scalar}), and $\abs{\cdot}$ is defined as in \eqref{eq:tensor-abs-near}.

We are now ready to state the remaining assumptions for $\calN$.
We begin with a basic definition.
\begin{definition} [Minimum degree of $\calN$] \label{def:nonliear-deg}
Let $\calN$ be as in \ref{hyp:nonlin}. The \emph{minimum degree} of $\calN$, denoted by $n_{\calN}$, is the least integer such that
\begin{align*}
	\rd_{z}^{I_{1}} \rd_{\bfxi}^{I_{2}} \quasi (\cdot, 0, 0) &\equiv 0 \quad \hbox{ for all } I_{1} + \abs{I_{2}} < n_{\calN}-1, \\
	\rd_{z}^{I_{1}} \rd_{\bfxi}^{I_{2}} \semi (\cdot, 0, 0) &\equiv 0 \quad \hbox{ for all } I_{1} + \abs{I_{2}} < n_{\calN},
\end{align*}
where $I_{1} \in \bbZ_{\geq 0}$ and $I_{2}$ is a multi-index.

\end{definition}
\begin{remark}\label{rem:nonlinearity=0}
We note that, according to Definition~\ref{def:nonliear-deg}, $n_{\calN} = +\infty$ if $\calN = 0$.
\end{remark}
Our next assumption says that $\calN$ is indeed nonlinear.
\begin{enumerate}[resume*]
\item \label{hyp:nonlin-degree} {\it Nonlinearity is at least quadratic.}
Assume $n_{\calN} \geq 2$.
\end{enumerate}

The next three assumptions essentially say that the coefficients of $\calN$ and their derivatives are bounded, and moreover admit expansions in $r^{-j} \log^{k}(\tfrac{r}{u})$ up to some orders (essentially $J_{c}$) in $\calM_{\wave}$.
\begin{enumerate}[resume*]
\item \label{hyp:nonlin-near} {\it Assumptions on the nonlinearity in $\calM_{\near}$.}
In $\calM_{\near}$, we assume that
\begin{align*}
\quasi &= O_{\bfGmm}^{M_{c}}(1), \quad \semi = O_{\bfGmm}^{M_{c}}(1).
\end{align*}

\item \label{hyp:nonlin-med} {\it Assumptions on the nonlinearity in $\calM_{\med}$.} In $\calM_{\med}$, we assume that
\begin{align*}
\quasi &= O_{\bfGmm}^{M_{c}}(1), \quad \semi = O_{\bfGmm}^{M_{c}}(1).
\end{align*}

\item \label{hyp:nonlin-wave} {\it Assumptions on the nonlinearity in $\calM_{\wave}$.}
In $\calM_{\wave}$, with respect to the Bondi--Sachs-type coordinates $(u, r, \tht)$,
\begin{align*}
\quasi^{\mu \nu} &= \sum_{j = 0}^{J_{c}-\frac{d-1}{2}} \sum_{k=0}^{K_{c}} \rquasi_{j, k}^{\mu \nu} r^{-j} \log^{k} (\tfrac{r}{u}) + \rem_{J_{c}-\frac{d-1}{2} +\eta_{c}}[\quasi^{\mu \nu}], \\
\semi &= \sum_{j = 0}^{J_{c}- \frac{d-1}{2}} \sum_{k=0}^{K_{c}} \rsemi_{j, k} r^{-j} \log^{k} (\tfrac{r}{u}) + \rem_{J_{c}-\frac{d-1}{2} + \eta_{c}}[\semi],
\end{align*}
with
\begin{align*}
\rquasi_{j, k} &= O_{\bfGmm}^{M_{c}}(u^{j}), \\
\rem_{J_{c}-\frac{d-1}{2}+\eta_{c}}[\quasi] &= O_{\bfGmm}^{M_{c}}(r^{-J_{c}+\frac{d-1}{2}-\eta_{c}} u^{J_{c}-\frac{d-1}{2}+\eta_{c}}), \\
\rsemi_{j, k} &= O_{\bfGmm}^{M_{c}}(u^{j}), \\
\rem_{J_{c}-\frac{d-1}{2}+\eta_{c}}[\semi] &= O_{\bfGmm}^{M_{c}}(r^{-J_{c}+\frac{d-1}{2}-\eta_{c}} u^{J_{c}-\frac{d-1}{2}+\eta_{c}}),
\end{align*}
where $\rquasi_{j, k}^{\mu \nu} = \rquasi_{j, k}^{\mu \nu}(u,\theta,z,\bfxi)$ and $\rsemi_{j, k} = \rsemi_{j, k}(u,\theta,z,\bfxi)$ are both independent of $r$.
\end{enumerate}

We note that there is no need to use the same $J_{c}$, $\eta_{c}$, and $M_{c}$ here as in \ref{hyp:T-almost-stat}--\ref{hyp:wave-V}, \ref{hyp:ult-stat} and \ref{hyp:ult-stat'}, but we do this to avoid introducing more parameters.

Finally, we need to make assumptions to ensure that $\calN$ has acceptable decay properties. Heuristically, we want the linearization $D \calN(\phi) \psi = \left. \frac{\ud}{\ud \eps} \calN(\phi+\eps \psi) \right|_{\eps = 0}$ of $\calN$ to obey the same bounds as the linear coefficients. This means that we want (cf.~\ref{hyp:T-almost-stat}, \ref{hyp:med} and Remark~\ref{rem:wave-heuristic}), schematically and for some $\dlt > 0$,
\begin{equation*}
	D \calN(\phi) \scheq \begin{cases}
	O(\tau^{-\dlt}) \hbox{ in $\calM_{\near}$, after replacing $\urd_{\tau}$ by $\tau^{-1}$ and $\urd_{x^{j}}$ by $1$}, \\
	O(r^{-2-\dlt}) \hbox{ in $\calM_{\med}$, after replacing $\rd_{u}$ by $u^{-1}$ and $\rd_{r}, r^{-1} \rd_{\tht^{A}}$ by $r^{-1}$}, \\
	O(u^{-2+\nu-\dlt}r^{-\nu}) \hbox{ for some $\nu > 1$ in $\calM_{\wave}$, after replacing $\rd_{u}$ by $u^{-1}$ and $\rd_{r}, r^{-1} \rd_{\tht^{A}}$ by $r^{-1}$},
	\end{cases}
\end{equation*}
where we remind the reader that $\tau \aeq u$ in $\calM_{\med} \cup \calM_{\wave}$.

Since we expect the solution $\phi$ to decay like $r^{-\frac{d-1}{2}}$ as $r \to \infty$ on each fixed $u$-hypersurface, we make the following assumption -- which is essentially the \emph{classical null condition} \cite{Chr.86, Kla.86}  -- to ensure the desired $r$-decay rate in $\calM_{\wave}$:
\begin{enumerate}[resume*]
\item \label{hyp:nonlin-null} {\it Null condition when $d = 3$.}
When $d = 3$, we furthermore assume the following \emph{null conditions} in $\calM_{\wave}$: With respect to the Bondi--Sachs-type coordinates $(u, r, \tht)$,
\begin{align*}
\rd_{\xi_{u}} \rquasi_{0, k}^{uu}  (p, 0, 0) &=0, \quad
\rd_{z} \rquasi_{0, k}^{uu} (p, 0, 0) =0,  \\
\rd_{\xi_{u}} \rd_{\xi_{u}} \rsemi_{0, k}  (p, 0, 0) &=0, \quad
\rd_{z} \rd_{\xi_{u}}  \rsemi_{0, k}  (p, 0, 0) =0, \quad
\rd_{z}^{2}  \rsemi_{0, k} (p, 0, 0) = 0,
\end{align*}
for all $0 \leq k \leq K_{c}$.
\end{enumerate}
Observe that when $d \geq 5$, a quadratic nonlinearity would already have a sufficient $r$-decay rate in $\calM_{\wave}$, so no null condition is necessary.

Towards ensuring the appropriate decay condition in $\tau$ (and $u$), we introduce a definition:
\begin{definition} \label{def:alp-N-min}
Let $\calN$ be as in \ref{hyp:nonlin}--\ref{hyp:nonlin-wave}. If $\calN \neq 0$, we define the \emph{minimal decay exponent} $\alp_{\calN}$ \emph{for $\calN$} to be the infimum of all admissible decay exponents $\alp_{\calN}'$ for $\calN$, whose precise definition is found in Definition~\ref{def:alp-N} in Section~\ref{subsec:nonlin-est}. If $\calN = 0$, we define $\alp_{\calN} = -\infty$.
\end{definition}

We refrain from giving the precise definition of an admissible decay exponent here, since it is rather long. Nevertheless, the underlying idea may be explained as follows. According to the standard vector field method (see Remark~\ref{rem:sol}), as well as the results of this paper, the solution $\phi$ is expected to obey bounds of the form
\begin{equation} \label{eq:alp-N-phi-decay}
	\phi = \begin{cases}
	O_{\bfGmm}^{M}(A \tau^{-\alp}) \hbox{ in } \calM_{\near}, \\
	O_{\bfGmm}^{M}(A u^{-\alp}) \hbox{ in } \calM_{\med}, \\
	O_{\bfGmm}^{M}(A u^{-\alp + \frac{d-1}{2}} r^{-\frac{d-1}{2}}) \hbox{ in } \calM_{\wave},
	\end{cases}
\end{equation}
for some $M \in \bbZ_{\geq 0}$, $A \geq 0$ and $\alp \in \bbR$. Roughly speaking, $\alp_{\calN}'$ is an admissible decay exponent for $\calN$ if for any $\phi$ satisfying \eqref{eq:alp-N-phi-decay} with $\alp > \alp_{\calN}'$, the linearization of $\calN$ around $\phi$ exhibits appropriate decay in $\tau$ (which is equal to $u$ in $\calM_{\far} = \calM_{\med} \cup \calM_{\wave}$) as discussed above. Moreover, an admissible decay exponent $\alp_{\calN}'$ is also easily computed in practice thanks to the following result:

\begin{proposition} \label{prop:alp-N}
Let $\calN$ satisfy the assumptions \ref{hyp:nonlin}--\ref{hyp:nonlin-null}.
\begin{enumerate}
\item In $\calM_{\far} = \calM_{\med} \cup \calM_{\wave}$, let $\calN$ be a homogeneous term (in $\phi, \ud \phi$) of the form
\begin{equation*}
\calN(\phi) = \begin{cases} c^{\mu \nu; \alp_{1} \ldots \alp_{n_{1}}}(p) \phi^{n_{0}} \rd_{\alp_{1}} \phi \ldots \rd_{\alp_{n_{1}}} \phi \bfD_{\rd_{\mu}} \rd_{\nu} \phi; \hbox{ or } \\
c^{\alp_{1} \ldots \alp_{n_{1}}}(p) \phi^{n_{0}} \rd_{\alp_{1}} \phi \ldots \rd_{\alp_{n_{1}}} \phi,
\end{cases}
\end{equation*}
where, in each case,
\begin{align*}
	c &= O_{\bfGmm}^{M_{c}}(r^{-\bt}) \quad \hbox{ in } \calM_{\med}, \\
	c &= \sum_{j=0}^{J_{c}-\frac{d-1}{2}} \sum_{k=0}^{K_{c}} \mathring{c}_{j, k} r^{-j} \log^{k} (\tfrac{r}{u}) + \rem_{J_{c}-\frac{d-1}{2}+\eta_{c}}[c] \quad \hbox{ in } \calM_{\wave}, \\
	\mathring{c}_{j, k} &= O_{\bfGmm}^{M_{c}}(u^{j-\bt}) , \\
	\rem_{J_{c}-\frac{d-1}{2}+\eta_{c}}[c] &=  O_{\bfGmm}^{M_{c}}(r^{-J_{c}+\frac{d-1}{2}-\eta_{c}} u^{J_{c}-\frac{d-1}{2}+\eta_{c}-\bt}) \quad \hbox{ in } \calM_{\wave},
\end{align*}
for some $\bt \geq 0$. In both cases, $\alp_{\calN}' = \max\set{0, \frac{2-o-\bt}{n-1}}$ is admissible, where
\begin{align*}
	o &= \hbox{(total number of derivatives in the homogeneous term)}, \\
	n &= \hbox{(total number of factors of $\phi$ in the homogeneous term)}.
\end{align*}
If $\calN$ is a sum of homogeneous terms of the above form, then the maximum of admissible decay exponents of the homogeneous terms is admissible.

\item In general, any $\calN$ satisfying \ref{hyp:nonlin}--\ref{hyp:nonlin-null} may be Taylor expanded in $\phi, \ud \phi$ to arbitrary order, where each term in the expansion is of the form considered in (1) in $\calM_{\far}$. The supremum $\alp_{\calN}'$ of the nonnegative admissible decay exponents of all Taylor coefficients (e.g., that given in (1)) is an admissible decay exponent for $\calN$. Moreover, it satisfies $0 \leq \alp_{\calN}' \leq \frac{d-1}{2}$.

\end{enumerate}
\end{proposition}

We defer the proof of this proposition until Section~\ref{subsec:nonlin-est}, after we precisely formulate the notion of an admissible decay exponent. Nonetheless, we encourage the reader to compare the behavior of (the linearization of) a homogeneous nonlinearity under \eqref{eq:alp-N-phi-decay} and the schematic requirement for $D \calN(\phi)$ above to see the ideas behind our definition and Proposition~\ref{prop:alp-N}. Observe also that $\alp > \alp_{\calN}' \geq 0$ would ensure, thanks to \ref{hyp:nonlin-degree} and \ref{hyp:nonlin-near}, that an adequate $\tau$-decay \emph{always} holds in $\calM_{\near}$; this explains why we only need to consider the behavior of $\calN$ in $\calM_{\far}$ in Proposition~\ref{prop:alp-N}.

\subsubsection{Assumptions on the data, the inhomogeneous terms and the solution}\label{sec:assumption.data.inho.sol}
Finally, we state our assumptions on the data and the solution. Let $\alp_{0} \in \bbR$, $\nu_{0} \in [0, \frac{1}{2})$, $\alp_{d} > \frac{d-1}{2}$, $J_{d} := \lceil \alp_{d} \rceil - \frac{d-1}{2}$, $K_{d} \in \bbZ_{\geq 0}$, $M_{0} \in \bbZ_{\geq 0}$, $A_{0} \geq 0$ and $D \geq 0$.

The following is our assumption on the initial data.
\begin{enumerate}[label=(D${}_{\Sigma_1}$)]
\item \label{hyp:id}
	{\it Assumptions on the initial data.} On $\Sgm_{1} \cap (\calM_{\near} \cup \calM_{\med})$, assume that
\begin{align}
\abs{(\tau \bfT)^{I_{\bfT}} \rd_{x^{1}}^{I_{x^{1}}} \ldots \rd_{x^{d}}^{I_{x^{d}}} \phi} &\leq D & & \hbox{ on } \Sgm_{1} \cap \calM_{\near} \quad \hbox{ for all } \abs{I} \leq M_{0}, \label{eq:near-id} \\
\abs{(u \rd_{u})^{I_{u}} (r \rd_{r})^{I_{r}} \bfOmg_{jk}^{I^{\bfOmg_{jk}}} \phi} &\leq D & & \hbox{ on } \Sgm_{1} \cap \calM_{\med} \quad \hbox{ for all } \abs{I} \leq M_{0}. \label{eq:med-id}
\end{align}
On $\Sgm_{1} \cap \calM_{\far}$, for $J_{d} = \lceil \alp_{d} \rceil - \frac{d-1}{2}$ and some $0 \leq K_{d} < +\infty$, we assume that $\phi$ admits the expansion
\begin{equation} \label{eq:wave-id-exp}
	\phi(1, r, \tht) = r^{-\frac{d-1}{2}} \left[ \sum_{j=0}^{J_{d}-1} \sum_{k=0}^{K_{d}} r^{-j} \log^{k} r {}^{(\Sgm_{1})}\rPhi_{j, k}(\tht) + {}^{(\Sgm_{1})} \rho_{J_{d}}(r, \tht) \right],
\end{equation}
where
\begin{align}
	\abs{\bfOmg_{jk}^{I_{\bfOmg_{jk}}} {}^{(\Sgm_{1})} \rPhi_{j, k}} &\leq D \quad \hbox{ for all } \abs{I} \leq M_{0},  \label{eq:wave-id-jk-phg} \\
	\abs{(r \rd_{r})^{I_{r}} \bfOmg_{jk}^{I_{\bfOmg_{jk}}} {}^{(\Sgm_{1})} \rho_{J_{d}}} &\leq D r^{-\alp_{d}+\frac{d-1}{2}} \quad \hbox{ for all } \abs{I} \leq M_{0}. \label{eq:wave-id-rem}
\end{align}
\end{enumerate}

\begin{remark} \label{rem:Sigma-1-causality}
We note explicitly that under our assumptions on $\bfg$ in \ref{hyp:T-almost-stat}, \ref{hyp:med} and \ref{hyp:wave-g}, $\Sigma_{1}$ need not be spacelike (or even acausal). Nonetheless, this will not be an issue for the main theorems in Section~\ref{subsec:mainthm} because we are not solving the Cauchy problem, but instead we will be proving properties of solutions that are already assumed to exist and obey the bounds in \ref{hyp:sol} below.

Note also that if we instead take the point of view of solving the Cauchy problem for data on $\calS_1$, then the assumption \ref{hyp:id} can be proven to hold; see Section~\ref{sec:Cauchy}.
\end{remark}

The following is our assumption on the inhomogeneous term.
\begin{enumerate}[label=(F)]
\item \label{hyp:forcing}
{\it Assumptions on the forcing term.}
In $\calM_{\near} \cup \calM_{\med}$, assume that
\begin{align}
f &= O_{\bfGmm}^{M_{0}}(D \tau^{-\alp_{d}}) \hbox{ in } \calM_{\near}, \label{eq:near-f} \\
f &= O_{\bfGmm}^{M_{0}}(D u^{-\alp_{d}} r^{-2 - \dlt_{d}})
 \hbox{ in } \calM_{\med}. \label{eq:med-f}
\end{align}
In $\calM_{\far}$, for $J_{d} = \lceil \alp_{d} \rceil - \frac{d-1}{2}$ and $\alp_{d}$, $K_{d}$ as in \ref{hyp:id}, we assume that $f$ admits the expansion
\begin{equation} \label{eq:wave-f-exp}
	f(u, r, \tht) = \sum_{j=2}^{J_{d}} \sum_{k=0}^{K_{d}} r^{-j-\frac{d-1}{2}} \log^{k} (\tfrac{r}{u}) \rf_{j+\frac{d-1}{2}, k}(u, \tht) + \rem_{\alp_{d}+1}[f](u, r, \tht),
\end{equation}
where
\begin{align}
	\rf_{j+\f{d-1}2, k} &= O_{\bfGmm}^{M_{0}}(D u^{j-2-\alp_{d} - \dlt_{d} + \frac{d-1}{2}}) \hbox{ in } \calM_{\wave},  \label{eq:wave-fjk-phg} \\
	\rem_{\alp_{d}+1}[f] &= O_{\bfGmm}^{M_{0}}(D r^{-\alp_{d}-1} u^{-1-\dlt_{d}}) \hbox{ in } \calM_{\wave}. \label{eq:wave-f-rem}
\end{align}
and we omit the sum in \eqref{eq:wave-f-exp} if $J_{d}  < 2$ (or equivalently, $\alp_{d} \leq \tfrac{d-1}{2} + 1$).
\end{enumerate}

Finally, we impose an a priori assumption on the solution, which is a weak decay statement with high vector field regularity.
\begin{enumerate}[label=(S)]
\item \label{hyp:sol}
	{\it Klainerman vector field regularity of $\phi$.} Assume, for $\nu_{0} \in [0, \frac{1}{2})$,
\begin{align}
	\phi &= O_{\bfGmm}^{M_{0}}(A_{0} \tau^{-\alp_{0}}) \hbox{ in } \calM_{\near}, \label{eq:sol-near} \\
	\phi &= O_{\bfGmm}^{M_{0}}(A_{0} u^{-\alp_{0}}) \hbox{ in } \calM_{\med}, \label{eq:sol-med} \\
	\phi &= O_{\bfGmm}^{M_{0}}(A_{0} r^{\nu_{0} - \frac{d-1}{2}} u^{-\alp_{0} - \nu_{0} + \frac{d-1}{2}} ) \hbox{ in } \calM_{\wave}. \label{eq:sol-wave}
\end{align}
\end{enumerate}

\begin{remark}[On \ref{hyp:sol}] \label{rem:sol}
Note that our assumption \ref{hyp:sol} may be recovered from many existing methods.
For instance, via Klainerman--Sideris and Klainerman's Sobolev-type bounds (see Sections~\ref{subsec:kl-sid} and \ref{subsec:kl-sob}), we may deduce \eqref{eq:sol-near}--\eqref{eq:sol-wave} (with $\nu_{0} = 0$) from the following bounds involving the standard Klainerman commuting vector fields:
\begin{align}
\abs{(\tau \bfT)^{I_{\bfT}} \rd_{x^{1}}^{I_{x^{1}}} \ldots \rd_{x^{d}}^{I_{x^{d}}} \phi} &\leq A_{0} \brk{\tau}^{-\alp_{0}} & & \hbox{ in } \calM_{\near} \cap \set{\tau \geq 1} \hbox{ for all } \abs{I} \leq M_{0}, \label{eq:sol-vf-near} \\
\abs{\bfT^{I_{\bfT}} \bfS^{I_{\bfS}} \bfOmg^{I_{\bfOmg}} \phi} &\leq A_{0} \brk{u}^{-\alp_{0}+\frac{d-1}{2}} r^{-\frac{d-1}{2}} & & \hbox{ in } \calM_{\far} \cap \set{\tau \geq 1} \hbox{ for all } \abs{I} \leq M_{0}, \label{eq:sol-vf-far}
\end{align}
where $C = C(d)$.  Another, more general, route is to use the powerful philosophy schematically expressed as:
\begin{equation*}
\boxed{\hbox{uniform boundedness of energy}} + \boxed{\hbox{integrated local energy decay}} + \boxed{\hbox{asymptotic flatness}} \imp \boxed{\hbox{stronger decay}}.
\end{equation*}
Among the many different manifestations of this philosophy, bounds of the form \ref{hyp:sol} follow either from $r^{p}$-method of Dafermos--Rodnanski \cite{DRNM}, Schlue \cite{Schlue} and Moschidis \cite{gM2016}, or the methods of Metcalfe--Tataru--Tohaneanu \cite{MTT} and Oliver--Sterbenz \cite{jOjS2023} that directly involve the vector fields $\bfT$, $\bfS$ and $\bfOmg_{jk}$ (see also \cite{MMTT, MetTat, TT} for different realizations of this philosophy). Note also that vector fields bounds of the form \ref{hyp:sol} are usually a byproduct of the proof of \emph{nonlinear} stability theorems via vector field method -- this fact underlies the nonlinear applications of our theorem; see Section~\ref{sec:nonlinear.ex} and \cite{LO.part2}.

We also refer the reader to the recent works of Hintz, Hintz--Vasy \cite{Hintz.linear.waves, pHaV2023}, which provides an alternative approach to obtaining the kind of bounds in \ref{hyp:sol}.

It must be emphasized that the black box assumption \ref{hyp:sol} is the key reason why we were able to formulate our main theorems concerning late time tails in the current generality without worrying about the (often times extremely) delicate behavior of waves in an $r$-bounded region such as trapping, superradiance, resonance and so on. These issues would rather come up in the \emph{proof} of \ref{hyp:sol} using the above-mentioned ideas.
\end{remark}

\begin{remark} We make some further comments on the above assumptions.
\begin{itemize}

\item The assumption $\alp_{d} > \frac{d-1}{2}$ is placed to ensure the existence of the Friedlander radiation field; see Proposition~\ref{prop:wave-0} below.

\item In \eqref{eq:wave-id-rem}, \eqref{eq:med-f}, \eqref{eq:wave-fjk-phg} and \eqref{eq:wave-f-rem}, $\dlt_{d}$ may be removed by assuming appropriate summability on dyadic regions, but we shall not pursue such an optimization here.
\end{itemize}
\end{remark}

\subsection{Higher radiation fields and recurrence equations} \label{subsec:recurrence-formal}
The central thesis of this paper, made precise in the main theorems stated in Section~\ref{subsec:mainthm} below, is that the late time tail of each individual solution $\phi$ to $\calP \phi = f$ is dictated by the coefficients of its expansion along future null infinity -- which we refer to as \emph{higher radiation fields} -- that are governed by a system of recurrence equations. Algebraic properties of these recurrence equations, in turn, determine the late time tail of \emph{generic} solutions.

The goal of this subsection is to give a formal discussion of the expansion along future null infinity, higher radiation fields and the recurrence equations they satisfy. We begin by considering the conjugated variable $\Phi = r^{\frac{d-1}{2}} \phi$, which solves the equation
\begin{equation} \label{eq:Phi-eq-0}
	 Q_{0} \Phi = - \bfh^{\alp \bt} \nb_{\alp} \rd_{\bt} \Phi - \bfC^{\alp} \rd_{\alp} \Phi - W \Phi + r^{\frac{d-1}{2}} \calN(r^{-\frac{d-1}{2}} \Phi) + r^{\frac{d-1}{2}} f,
\end{equation}
where $Q_{0}$ is the conjugated d'Alembertian on the Minkowski spacetime, i.e.,
\begin{equation*}
	Q_{0} = r^{\frac{d-1}{2}} \Box_{\bfm} r^{-\frac{d-1}{2}} = - 2 \rd_{u} \rd_{r} + \rd_{r}^{2} - \frac{(d-1)(d-3)}{4} \frac{1}{r^{2}} + \frac{1}{r^{2}} \rslap;
\end{equation*}
$\nb$ is the Minkowskian covariant derivative, i.e., the covariant derivative associated with $(\calM_{\far}, \bfm = - (\ud u)^{2} + 2 \ud u \cdot \ud r + r^{2} \rsgmm)$; and
\begin{align}
	\bfh^{\alp \bt} &= (\bfg^{-1})^{\alp \bt} - (\bfm^{-1})^{\alp \bt}, \label{eq:bfh-def} \\
	\bfC^{\alp} &= - \frac{d-1}{r} \Big[(\bfg^{-1})^{r \alp} - (\bfm^{-1})^{r\alp}\Big] + \left( {}^{(\bfg)} \Gmm_{\mu \nu}^{\alp} - {}^{(\bfm)} \Gmm^{\alp}_{\mu \nu} \right) (\bfg^{-1})^{\mu \nu} + \bfB^{\alp}, \label{eq:bfC-def} \\
	W &=r^{\frac{d-1}{2}} \Box_{\bfg} r^{-\frac{d-1}{2}} + \frac{(d-1)(d-3)}{4} \frac{1}{r^{2}} - \frac{d-1}{2r} \bfB^{r}  + V. \label{eq:W-def}
\end{align}
Note that the conjugation weight $r^{-\frac{d-1}{2}}$ is precisely the sharp decay rate of solutions to $\Box_{\bfm} \psi = 0$ in the wave zone. In the present context, the key reason for introducing the conjugated variable $\Phi$ is to remove the dangerous term $(\bfm^{-1})^{AB} {}^{(\bfm)} \Gmm_{AB}^{u} \rd_{u} = (d-1) r^{-1} \rd_{u}$; see Section~\ref{subsec:conj-wave} below for details.


We are now ready to introduce the formal expansion along future null infinity. By \emph{future null infinity}, we mean the ideal boundary $\calI^{+} = \set{(u, r = \infty, \tht) : u \geq 1, \, \tht \in \bbS^{d-1}}$ with respect to the Bondi--Sachs-type coordinates $(u, r, \tht)$ on $\calM_{\far}$. Consider, for now, the following \emph{formal polyhomogeneous expansion of $\Phi$ along $\calI^{+}$}:
\begin{equation}\label{eq:formal.expansion}
	\Phi(u, r, \tht) = \rPhi_{0}(u, \tht) + \sum_{j=1}^{\infty} \sum_{k=0}^{\infty} r^{-j} \log^{k} (\tfrac{r}{u}) \rPhi_{j, k}(u, \tht),
\end{equation}
where we further assume that \emph{for each $j \geq 1$, $\rPhi_{j, k}$ is zero except for finitely many $k$'s} (soon, this assumption shall be refined so as to be compatible with the recurrence equations).

The initial summand $\rPhi_{0}(u, \tht)$ -- which corresponds to $(j, k) = (0, 0)$ -- is formally the \emph{Friedlander radiation field}, i.e.,
\begin{equation*}
\rPhi_{0}(u, \tht) = \lim_{r \to \infty} r^{\frac{d-1}{2}} \phi(u, r, \tht).
\end{equation*}
The Friedlander radiation field exists under our assumptions; see Proposition~\ref{prop:wave-0} below.
The other coefficients $\rPhi_{j, k}$, which we shall call \emph{higher radiation fields}, are determined by \emph{recurrence equations} obtained by plugging $\Phi$ into \eqref{eq:Phi-eq-0}, computing a (formal) polyhomogeneous expansion, and then requiring each coefficient in front of $r^{-j} \log^{k} r$ to be zero, in the increasing lexicographic order for $\set{(j, -k) : j \geq 1, k \geq 0}$. More precisely, the left-hand side of \eqref{eq:Phi-eq-0}, which we shall refer to as the \emph{Minkowskian contribution}, admits the following formal expansion:
\begin{align*}
& Q_{0} \left( \rPhi_{0}(u, \tht) + \sum_{j=1}^{\infty} \sum_{k=0}^{\infty} r^{-j} \log^{k} (\tfrac{r}{u}) \rPhi_{j, k}(u, \tht) \right) \\
&= \sum_{j=0}^{\infty} \sum_{k=0}^{\infty} r^{-j-1} \log^{k} (\tfrac{r}{u}) \\
&\peq \phantom{\sum_{j=0}^{\infty} \sum_{k=0}^{\infty}}
\times \Big[ 2 j \rd_{u} \rPhi_{j, k}
- 2 (k+1) \rd_{u} \rPhi_{j, k+1}
- 2 j (k+1) u^{-1} \rPhi_{j, k+1}
+ 2(k+2)(k+1) u^{-1} \rPhi_{j, k+2} \\
&\peq \phantom{\sum_{j=0}^{\infty} \sum_{k=0}^{\infty} \times \Big[}
+ \Big( (j-1)j + \frac{(d-1)(d-3)}{4} + \rslap \Big) \rPhi_{j-1, k}
- (2j-1) (k+1)  \rPhi_{j-1, k+1} \\
&\peq \phantom{\sum_{j=0}^{\infty} \sum_{k=0}^{\infty} \times \Big[}
+ (k+2)(k+1) \rPhi_{j-1, k+2} \Big],
\end{align*}
where we use the convention that $\rPhi_{0, 0} = \rPhi_{0}$ and $\rPhi_{0, k} = 0$ for $k \geq 1$.
This identity follows readily from \eqref{eq:monomial-du}--\eqref{eq:monomial-drr} in Section~\ref{subsec:recurrence}.
Thanks to the assumptions \ref{hyp:wave-g}--\ref{hyp:wave-V}, \ref{hyp:nonlin}, \ref{hyp:nonlin-wave}--\ref{hyp:nonlin-null} and \ref{hyp:forcing}, the right-hand side of \eqref{eq:Phi-eq-0} admits a formal expansion in the wave zone, truncated to a finite order determined by $d$, $J_{c}$, $K_{c}$, $J_{d} = \lceil \alp_{d} \rceil - \frac{d-1}{2}$, $K_{d}$ and the structure of the nonlinearity $\calN$:
\begin{equation*}
	\hbox{(RHS of \eqref{eq:Phi-eq-0})}
	= \sum_{j=1}^{\min\set{J_{c}, J_{d}}-1} \sum_{k=0}^{\infty} r^{-j-1} \log^{k} (\tfrac{r}{u}) \, \rF_{j, k} + \ldots,
\end{equation*}
where each $\rF_{j, k}$ is an explicitly computable coefficient from $(\rPhi_{j', k'})_{j' \leq j-1, k' \leq k+2}$, as well as $\rmet_{j, k}$, $\rbfB_{j, k}$, $\rV_{j, k}$, $\calN$ and $\rf_{j, k}$. By identifying the coefficients of $r^{-j-1} \log^{k} r$ on both sides for $1 \leq j \leq \min\set{J_{c}, J_{d}} -1$, we obtain the following \emph{recurrence equations}:
\begin{equation} \label{eq:recurrence-jk}
\begin{aligned}
	\rd_{u} \rPhi_{j, k}&=
	(k+1) u^{-1} \rPhi_{j, k+1}
	+ \frac{k+1}{j} \left( \rd_{u} \rPhi_{j, k+1} - (k+2) u^{-1} \rPhi_{j, k+2} \right)
\\
&\peq
- \frac{1}{2j}\Big( (j-1)j - \frac{(d-1)(d-3)}{4} + \rslap \Big) \rPhi_{j-1, k}   \\
&\peq
+ \frac{1}{2j} (j-1)(k+1) \rPhi_{j-1, k+1} - \frac{1}{2j}(k+2)(k+1) \rPhi_{j-1, k+2}
+ \frac{1}{2j} \rF_{j, k},
\end{aligned}
\end{equation}
where we use the convention that $\rPhi_{j, k} = 0$ for $j < 0$ and for $j = 0$ but $k > 0$.
In order to use the equations \eqref{eq:recurrence-jk} to recursively determine $\rPhi_{j, k}$ -- which would justify the name -- we need the following two important properties of $\rF_{j, k}$, which follow from \ref{hyp:wave-g}--\ref{hyp:wave-V}, \ref{hyp:nonlin}, \ref{hyp:nonlin-wave}--\ref{hyp:nonlin-null}, and \ref{hyp:forcing}:
\begin{enumerate}
\item $\rF_{j, k}$ is determined from $\rPhi_{j', k'}$ with only $j' \leq j-1$;
\item If $\rPhi_{j', k'} = 0$ for all $j' \leq j-1$ and $k' > K_{j-1}$, then $\rF_{j, k} = 0$ for all $k$ greater than some finite number $K$ that depends on $d$, $K_{c}$, $K_{d}$, $j$ and $K_{j-1}$.
\end{enumerate}
For a verification of these points, we refer the reader to Section~\ref{subsec:recurrence} below. To emphasize the first point, we shall write $\rF_{j, k} = \rF_{j, k}((\rPhi_{j', k'})_{j' \leq j-1})$. The second point allows us to recursively define numbers $K_{j} = K_{j}(d, K_{c}, K_{d}, j, K_{j-1})$ starting with $K_{0} = 0$, such that the assumption
\begin{equation*}
	\rPhi_{j, k} = 0 \quad \hbox{ for } k > K_{j}
\end{equation*}
is preserved by \eqref{eq:recurrence-jk}; see Lemma~\ref{lem:Kj-def} below. In particular, we may drop the terms $\rd_{u} \rPhi_{j, K_{j}+1}$, $u^{-1} \rPhi_{j, K_{j}+1}$, and $u^{-1} \rPhi_{j, K_{j}+2}$ on the RHS's of these equations, so that \eqref{eq:recurrence-jk} can be solved recursively to determine $\rPhi_{j, k}$ in the order
\begin{equation*}
\rPhi_{1, K_{1}} \to \ldots \rPhi_{1, 0} \to \rPhi_{2, K_{2}} \to \ldots \rPhi_{2, 0} \to \ldots \to \rPhi_{\min\set{J_{c}, J_{d}} - 1, 0}.
\end{equation*}

We shall see later in Section~\ref{subsec:recurrence} that, as a consequence of our assumptions, every term on the right-hand side of \eqref{eq:recurrence-jk} contributes a term with an improved rate $O_{\bfGmm}^{M'}(u^{j-1-\dlt_{0}+\nu_{\Box}})$ for some $M'$ and $\dlt_{0} > 0$, except for the Minkowskian contribution
\begin{equation*}
- \frac{1}{2j}\Big( (j-1)j - \frac{(d-1)(d-3)}{4} + \rslap \Big) \rPhi_{j-1, 0}
\end{equation*}
in the case $k = 0$. To quantify the effect of $-\rslap$ in this term, we are motivated to decompose $\rPhi_{j, k}$ into the eigenfunctions of $-\rslap$, which are nothing but the standard spherical harmonics on the round unit sphere $\bbS^{d-1}$ defined with respect to the angular variables $\tht$.

We write $\bbS_{(\ell)}$ for the projection to the eigenspace corresponding to the $\ell$-th eigenvalue of $-\rslap$. Hence, we have
\begin{equation} \label{eq:S-ell-def}
- \rslap \bbS_{(\ell)} = \ell(\ell + d - 2) \bbS_{(\ell)}.
\end{equation}
We also introduce the shorthands
\begin{equation*}
	\bbS_{(\geq \ell)} = \sum_{\ell' \geq \ell} \bbS_{(\ell)}, \quad
	\rPhi_{(\ell) j, k} = \bbS_{(\ell)} \rPhi_{(\ell) j, k}, \quad
	\rPhi_{(\geq \ell) j, k} = \bbS_{(\geq \ell)} \rPhi_{(\ell) j, k}.
\end{equation*}
Since the Minkowskian part $Q_{0}$ commutes with $\bbS_{(\ell)}$, we have the following projected recurrence equations:
\begin{align}
	\rd_{u} \rPhi_{(\ell) j, k}&=
	(k+1) u^{-1} \rPhi_{(\ell) j, k+1}
	+ \frac{k+1}{j} \left( \rd_{u} \rPhi_{(\ell) j, k+1} - (k+2) u^{-1} \rPhi_{(\ell) j, k+2} \right)
\label{eq:recurrence-jk-ell} \\
&\peq
- \frac{1}{2j}  \left( j - \frac{d-1}{2} - \ell \right) \left( j + \frac{d-1}{2} + \ell - 1\right)  \rPhi_{(\ell) j-1, k}  \notag \\
&\peq
+ \frac{1}{2j} (j-1)(k+1) \rPhi_{(\ell) j-1, k+1} - \frac{1}{2j}(k+2)(k+1) \rPhi_{(\ell) j-1, k+2}
+ \frac{1}{2j} \bbS_{(\ell)} \rF_{j, k}, \notag
\end{align}
where we simplified the coefficient in front of $\rPhi_{(\ell) j-1, k}$ using the algebraic identity
\begin{equation*}
(j-1)j - \frac{(d-1)(d-3)}{4} - \ell(\ell+d-2) = \left( j - \frac{d-1}{2} - \ell \right) \left( j + \frac{d-1}{2} + \ell - 1\right).
\end{equation*}
%
A crucial consequence of $d$ being odd is that $\frac{d-1}{2} + \ell$ is always an integer, so that the main term
\begin{equation*}
- \frac{1}{2j} \left( j - \frac{d-1}{2} - \ell \right) \left( j + \frac{d-1}{2} + \ell - 1\right) \rPhi_{(\ell) j-1, k}
\end{equation*}
vanishes for $j = \frac{d-1}{2} + \ell$. We shall refer to this cancellation as the \emph{Newman--Penrose cancellation}, since it is the key cancellation behind the conservation of the Newman--Penrose constants $I_{(\ell)} = \rPhi_{(\ell) \frac{d-1}{2}+\ell, 0}$ for $\Box_{\bfm} \phi = 0$ on the Minkowski spacetime.

\subsection{Renormalized higher radiation fields}\label{sec:renormalized.higher.radiation.fields}

In addition to the higher radiation fields $\rPhi_{j,k}$, we also introduce the \emph{renormalized higher radiation fields} $\rcPhi_{j,k}$, which are defined so that for each fixed $j$, the following holds:
\begin{equation}\label{eq:change.rPhi.rcPhi}
	\sum_{k=0}^{K_{j}} \rPhi_{j, k}(u, \theta) \log^{k}(\tfrac{r}{u}) = \sum_{k=0}^{K_{j}} \rcPhi_{j, k}(u, \theta) \log^{k} r.
\end{equation}
We also use the convention, consistent with the above, that $\rPhi_0 = \rPhi_{0,0} = \rcPhi_{0,0}$. Alternatively, and equivalently, $\rcPhi_{j,k}$ can be defined in terms of $\rPhi_{j,k}$ by
\begin{equation} \label{eq:rcPhi-rPhi}
	\rcPhi_{j, k}(u, \theta) = \sum_{K=k}^{K_{j}} \binom{K}{k} (-1)^{K-k}\rPhi_{j, K}(u, \theta) \log^{K-k} u.
\end{equation}
Indeed, by the binomial theorem,
\begin{align*}
	&\sum_{k=0}^{K_{j}} r^{-j} \log^{k} r \left(\sum_{K=k}^{K_{j}} \binom{K}{k} (-1)^{K-k}\rPhi_{j, K}(u, \theta) \log^{K-k} u \right) \\
	&= \sum_{K=0}^{K_{j}} r^{-j} \sum_{k = 0}^{K} \binom{K}{k} (\log^k r) (- \log u)^{K-k}   \rcPhi_{j, K}(u, \theta)
	= \sum_{K=0}^{K_{j}} r^{-j} \log^{K} (\tfrac{r}{u}) \rPhi_{j, K}(u, \theta).
\end{align*}

In the paper, we will need to switch back and forth between $\rPhi_{j,k}$ and $\rcPhi_{j,k}$. Most of the time, $\rPhi_{j,k}$ are the most convenient for carrying out the estimates in the proof. On the other hand, the renormalized higher radiation fields $\rcPhi_{j,k}$ are important because their limits (as $u\to \infty$) determine the precise asymptotics of the solutions.

As in Section~\ref{subsec:recurrence-formal}, we use the following notation for projection to the $\ell$-th eigenspace of $-\rslap$:
$$\rcPhi_{(\ell)j,k} = \mathbb S_{(\ell)} \rcPhi_{j,k},\quad \rcPhi_{(\geq \ell)j,k} = \mathbb S_{(\geq \ell)} \rcPhi_{j,k},\quad \rcPhi_{(\leq \ell)j,k} = \mathbb S_{(\leq \ell)} \rcPhi_{j,k}.$$

Following the derivation of \eqref{eq:recurrence-jk}, but writing in terms of $\rcPhi$ instead of $\rPhi$, we obtain
\begin{equation}\label{eq:recurrence-jk.rcPhi}
\begin{split}
	\rd_{u} \rcPhi_{j, k}&=
	 \frac{k+1}{j} \rd_{u} \rcPhi_{j, k+1}
- \frac{1}{2j}\Big( (j-1)j - \frac{(d-1)(d-3)}{4} + \rslap \Big) \rcPhi_{j-1, k}  \\
&\peq
+ \frac{1}{2j} (j-1)(k+1) \rcPhi_{j-1, k+1} - \frac{1}{2j}(k+2)(k+1) \rcPhi_{j-1, k+2}
+ \frac{1}{2j} \rcF_{j, k}, \quad j\geq 1, \, k \geq 0,
\end{split}
\end{equation}
where, similar to the case for $\rF_{j,k}$ above, $\rcF_{j,k}$ is explicitly computable by comparing coefficients in
\begin{equation*}
	\hbox{(RHS of \eqref{eq:Phi-eq-0})}
	= \sum_{j=1}^{\min\set{J_{c}, J_{d}}-1} \sum_{k=0}^{\infty} r^{-j-1} \log^{k} r \, \rcF_{j, k} + \cdots.
\end{equation*}

Concerning the renormalized higher radiation fields, we have the following lemma. This lemma will be useful in practice to analyze the higher radiation fields; see \cite{LO.part2}.

\begin{lemma}\label{lem:radiation.field}
    Let $\calP$ satisfy the assumptions \ref{hyp:topology}--\ref{hyp:T-tau},  \ref{hyp:T-almost-stat}--\ref{hyp:wave-V}, \ref{hyp:bdry.cond}, as well as either \ref{hyp:ult-stat} or \ref{hyp:ult-stat'}. In case $\calN \neq 0$, assume that $\calN$ satisfies \ref{hyp:nonlin}--\ref{hyp:nonlin-null}.
Let $\phi$, $f$ solve the main equation \eqref{eq:main-eq} and satisfy \ref{hyp:id}, \ref{hyp:forcing}, and \ref{hyp:sol} with some $\alp_{0} \in \bbR$ and $\alp_{d} > \alp_{0}$. When $\calN \neq 0$, let $\alp_{\calN}$ be its minimal decay exponent and assume also that for some $\de_{0} >0$,
\begin{equation} 
	\alp_{0} > \alp_{\calN} + 2 \dlt_{0}.
\end{equation}
From $\rPhi_{0}$, define $\rPhi_{j, k}(u, \tht)$ for $0 \leq j \leq \min\set{J_{c}, J_{d}} - 1$, $0 \leq k \leq K_{j}$ using the formal recurrence equations \eqref{eq:recurrence-jk}, and define $\rcPhi_{j,k}(u,\tht)$ using \eqref{eq:rcPhi-rPhi}.

Then the following holds:
\begin{enumerate}
    \item The radiation field decays:
    \begin{equation}
        \lim_{u\to \infty} \rPhi_{0,0}(u,\th) = 0.
    \end{equation}
    \item There exists a sequence $0 = K_{0} \leq K_{1} \leq \dots \leq K_{\min\{J_{c},J_{d}\}-1}$ such that $\rPhi_{j,k} = \rcPhi_{j,k} = 0$ if $k > K_j$.
    \item For each $1\leq J \leq \min \{J_{c},J_d\} - 1$, if
    $$\lim_{u\to \infty} \rPhi_{(\leq j-\f{d-1}2)j,0}(u,\theta) =0, \quad \lim_{u\to \infty} \rPhi_{j,k}(u,\theta) =0,\quad 0\leq j \leq J-1,\quad 1 \leq k \leq K_{J-1},$$
    then the following limits are well-defined:
    $$\rcPhi_{J,k}(\infty) = \lim_{u\to \infty} \rcPhi_{J,k}(u,\cdot),\quad 1\leq k \leq K_J,\quad \rcPhi_{(\leq J-\f{d-1}2)J,0}(\infty) = \lim_{u\to \infty} \rcPhi_{(\leq J-\f{d-1}2)J,0}(u, \cdot).$$
\end{enumerate}
\end{lemma}

Lemma~\ref{lem:radiation.field} will be proven in Section~\ref{sec:lem.radiation.field}.

\subsection{Statement of the main theorems} \label{subsec:mainthm}
We are now ready to state the central theorems of this paper, of which the results stated in Section~\ref{sec:intro} are consequences.

\subsubsection{Main upper bound theorem}

\begin{maintheorem} [Main upper bound theorem] \label{thm:upper}
Let $\calP$ satisfy the assumptions \ref{hyp:topology}--\ref{hyp:T-tau},  \ref{hyp:T-almost-stat}--\ref{hyp:wave-V}, \ref{hyp:bdry.cond}, as well as either \ref{hyp:ult-stat} or \ref{hyp:ult-stat'}. In case $\calN \neq 0$, assume that $\calN$ satisfies \ref{hyp:nonlin}--\ref{hyp:nonlin-null}.
Let $\phi$, $f$ solve the main equation \eqref{eq:main-eq} and satisfy \ref{hyp:id}, \ref{hyp:forcing}, and \ref{hyp:sol} with some $\alp_{0} \in \bbR$ and $\alp_{d} > \alp_{0}$. When $\calN \neq 0$, let $\alp_{\calN}$ be its minimal decay exponent and assume also that for some $\de_{0} >0$,
\begin{equation} \label{eq:upper-hyp-alp0}
	\alp_{0} > \alp_{\calN} + 2 \dlt_{0}.
\end{equation}
From $\rPhi_{0}$, define $\rPhi_{j, k}(u, \tht)$ for $0 \leq j \leq \min\set{J_{c}, J_{d}} - 1$, $0 \leq k \leq K_{j}$ using the formal recurrence equations \eqref{eq:recurrence-jk}. Suppose $J_{\mathfrak f} \in \mathbb N$, $1 \leq J_{\mathfrak f} \leq \min\set{J_{c}, J_{d}}$ satisfies
\begin{equation}\label{eq:Jf.def}
\lim_{u\to \infty} \rcPhi_{(\leq j - \frac{d-1}{2}) j, 0}(u) =0,\quad \lim_{u\to \infty} \rcPhi_{j, k}(u) =0 ,\quad \forall 0\leq j \leq J_{\mathfrak f}-1,\, \forall 1\leq k \leq K_j,
\end{equation}
for $K_j$ as in Lemma~\ref{lem:radiation.field}.

Then there exist $m_{0} \in \mathbb N$ large, $a_{0} \in (0,1)$ small, both depending only on $d$, $\de_0$, $\de_{c}$, $\de_{d}$, $\eta_{c}$, $\alp_{0}$, $\alp_{d}$, $J_{c}$, $J_{d}$, $J_{\mathfrak f}$, $K_{c}$, $K_{d}$, $s_{c}$, such that if $M_{0}, M_{c} \geq m_0$, then the solution $\phi$ obeys the following upper bounds in $\calM$:
\begin{align}
	\phi &= O_{\bfGmm}^{\lfloor a_{0} M_{0} \rfloor}(A_{\mathfrak f} \min\{ \tau^{-\alp_{\mathfrak f}} ,\tau^{-\alp_{\mathfrak f}+\nB} r^{-\nu_{\Box}} \} \log^{K_{\mathfrak{f}}} \tau ), \label{eq:thm1.main.decay}
\end{align}
where the decay rate $\alp_{\mathfrak f}$ is given by
\begin{align}\label{eq:alp.def}
	\alp_{\mathfrak f} &= \min \set*{J_{\mathfrak f}+\nu_{\Box}, J_{c}-1+\eta_{c}+\nu_{\Box}, \alp_{d}},
\end{align}
the power in the logarithm (see Lemma~\ref{lem:radiation.field}.(2)) is given by
\begin{equation}\label{eq:Kf.def}
    K_{\mathfrak{f}} := \begin{cases}
        K_{J_{\mathfrak f}} & \hbox{when $\alp_{\mathfrak f} = J_{\mathfrak f} + \nu_{\Box}$} \\
        0 & \hbox{otherwise}
    \end{cases},
\end{equation}
the amplitude $A_{\mathfrak f}$ can be chosen
\begin{equation} \label{eq:Af}
\begin{aligned}
A_{\mathfrak f}&= D+A + A^{(2)}(A) \quad \hbox{ if } \calN \neq 0 \\
A_{\mathfrak f} &= D+ A \quad \hbox{ if } \calN = 0,
\end{aligned}\end{equation}
where $A^{(2)}(A) \geq 0$ is nondecreasing function depending on $\calN$ satisfying $A^{(2)}(A) = O(A^{2})$ as $A \to 0$,
and the dependence of the implicit constants in \eqref{eq:thm1.main.decay} on $\phi$ and $f$ is only through $\de_0, J_{\mathfrak f}, \de_{d}, \alp_d, M_0$.
\end{maintheorem}

\subsubsection{Main asymptotics theorem}
Using Main Theorem~\ref{thm:upper} as a starting point, we may now state our \emph{main asymptotics theorem}. To this end, we need to describe three additional notions: the \emph{final asymptotic charges} $\frkL_{k}$, the \emph{Minkowskian profile} $\varphi^{\bfm[1]}_{(0)J_{\mathfrak f},k}$, and the \emph{asymptotic spatial profile} $\psi$.

In the setting of Main Theorem~\ref{thm:upper}, suppose in addition that $J_{\mathfrak f} \leq \min \set{J_{c}, J_{d}}-1$, so that $\rPhi_{J_{\mathfrak f}, k}$ for $k = 0, \ldots, K_{J_{\mathfrak f}}$ are also well-defined via \eqref{eq:recurrence-jk}. According to Lemma~\ref{lem:radiation.field}, the following limits are well-defined:
\begin{align}\label{eq:rcPhi.limit.sec.2}
\rcPhi_{(\leq J_{\mathfrak f} - \frac{d-1}{2}) J_{\mathfrak f}, 0}(\infty) := \lim_{u\to \infty} \rcPhi_{(\leq J_{\mathfrak f} - \frac{d-1}{2}) J_{\mathfrak f}, 0}(u, \cdot),\quad \rcPhi_{J_{\mathfrak f}, k}(\infty) := \lim_{u \to \infty} \rcPhi_{J_{\mathfrak f}, k}(u, \cdot) \quad \hbox{ for } 1 \leq k \leq K_{J_{\mathfrak f}}.
\end{align}
Our main asymptotics theorem says that, in fact, the leading order asymptotics of the late time tail of $\phi$ are determined by \eqref{eq:rcPhi.limit.sec.2}.

Given the functions $\rcPhi_{(\leq J_{\mathfrak f}+\f{d-1}2)J_{\mathfrak f},0}(\infty)$ and $\{\rcPhi_{J_{\mathfrak f},k}(\infty)\}_{k=1}^{K_{J_{\mathfrak f}}}$ as above, the late-time asymptotics when $r\geq \tau^{1-\de_{a}}$ are given by the \textbf{Minkowskian profiles}, which are defined as follows:
\begin{itemize}
\item for $u< 0$, $\varphi^{\bfm[1]}_{J_{\mathfrak f},0}[\rcPhi_{(\leq J_{\mathfrak f}+\f{d-1}2)J_{\mathfrak f},0}(\infty)]= \varphi^{\bfm[1]}_{J_{\mathfrak f},k}[\rcPhi_{J_{\mathfrak f},k}(\infty)] = 0$, and
\item for $u\geq 0$, $\varphi^{\bfm[1]}_{J_{\mathfrak f},0}[\rcPhi_{(\leq J_{\mathfrak f}+\f{d-1}2)J_{\mathfrak f},0}(\infty)]$ and $\varphi^{\bfm[1]}_{J_{\mathfrak f},k}[\rcPhi_{J_{\mathfrak f},k}(\infty)]$ solves the wave equation on \emph{Minkowski} space with characteristic data on $\{u=1\}$ given by
\begin{equation}\label{eq:varphi.m.data.summed.1}
\varphi^{\bfm[1]}_{J_{\mathfrak f},0}[\rcPhi_{(\leq J_{\mathfrak f}+\f{d-1}2)J_{\mathfrak f},0}(\infty)](0,r,\th) = \begin{cases}
\rcPhi_{(\leq J_{\mathfrak f}+\f{d-1}2)J_{\mathfrak f},0}(\infty) r^{-J_v-\nB} & \hbox{ if } r \geq 1 \\
0 & \hbox{ if } r \leq \f 12,
\end{cases}
\end{equation}
and
\begin{equation}\label{eq:varphi.m.data.summed.2}
\varphi^{\bfm[1]}_{J_{\mathfrak f},k}[\rcPhi_{J_{\mathfrak f},k}(\infty)](0,r,\th) = \begin{cases}
 r^{-j-\nB}\log^k (\tfrac ra) & \hbox{ if } r \geq 1 \\
0 & \hbox{ if } r \leq \f 12,
\end{cases}
\quad 1\leq k \leq K_{J_{\mathfrak f}}.
\end{equation}
\end{itemize}

When $r\leq \tau^{1-\de_{a}}$, the asymptotics are given by the final asymptotic charges and the asymptotic spatial profiles. The \textbf{final asymptotic charges} are defined by
\begin{align}
\frkL_{k} := \rcPhi_{(0) J_{\mathfrak f}, k}(\infty) \quad \hbox{ for } 0 \leq k \leq K_{J_{\mathfrak f}}. \label{eq:frkL-def}
\end{align}
To define the asymptotic spatial profiles, we need to assume \ref{hyp:ult-stat} (instead of \ref{hyp:ult-stat'}). The \textbf{asymptotic spatial profiles} are then defined as follows: we take $\psi = \psi(x)$ (satisfying  $\bfT \psi = 0$) to be defined by
\begin{equation}\label{eq:psi.def}
	\psi = \begin{cases}
	1 - {}^{(\infty)} \calR_{0} {}^{(\infty)} \calP_{0} 1 & \hbox{ if } \rd_{(\mathrm{t})} \calM = \0, \\
	1 - \chi_{< R_{\far}} - {}^{(\infty)} \calR_{0} {}^{(\infty)} \calP_{0} (1- \chi_{< R_{\far}}) & \hbox{ if } \rd_{(\mathrm{t})} \calM \neq \0.
	\end{cases}
\end{equation}
Here, we are abusing the notation a bit to write ${}^{(\infty)} \calR_{0}$ for an extension of ${}^{(\infty)} \calR_{0}$ in \ref{hyp:ult-stat} 
to handle non-compactly supported functions. Its construction, as well as the fact that $\psi$ is well-defined, will be proven in Section~\ref{subsec:stat-est}.

Observe that $\psi$ is a solution to ${}^{(\infty)} \calP_{0} \psi = 0$ with $\left. \psi \right|_{\rd \Sgm_{\tau}} = 0$ (if $\rd_{(\mathrm{t})} \calM \neq \0$) and $\psi - 1 = o(1)$ as $r \to \infty$, which is unique if $\psi - 1$ belongs to a suitable class (namely, $\psi - 1\in \ell^{1} \calH^{s, -\frac{d}{2}}(\Sgm_{\tau})$ for $s > \frac{d}{2}$ with the notation in Section~\ref{sec:near}).

\begin{maintheorem} [Main asymptotics theorem] \label{thm:lower}
Let $\calP$ satisfy the assumptions \ref{hyp:topology}--\ref{hyp:T-tau},  \ref{hyp:T-almost-stat}--\ref{hyp:wave-V}, \ref{hyp:bdry.cond}, as well as \ref{hyp:ult-stat}.
In case $\calN \neq 0$, assume that $\calN$ satisfies \ref{hyp:nonlin}--\ref{hyp:nonlin-null}. 

Let $\phi$, $f$ solve the main equation \eqref{eq:main-eq} and satisfy \ref{hyp:id}, \ref{hyp:forcing}, and \ref{hyp:sol} with some $\alp_{0} \in \bbR$ and $\alp_{d} > \alp_{0}$. When $\calN \neq 0$, assume also that \eqref{eq:upper-hyp-alp0} holds for $\alp_{\calN}$ being the minimal decay exponent for $\calN$.

From $\rPhi_{0}$, define $\rPhi_{j, k}(u, \tht)$ for $0 \leq j \leq \min\set{J_{c}, J_{d}} - 1$, $0 \leq k \leq K_{j}$ using the formal recurrence equations \eqref{eq:recurrence-jk}. Assume that \eqref{eq:Jf.def} holds for some $J_{\mathfrak f}\in \mathbb N$, $1\leq J_{\mathfrak f} \leq \min\{J_{c}, J_{d}\}-1$. Assume, in addition, that
\begin{equation} \label{eq:lower-hyp-alpf}
J_{c}-1+\eta_{c}+\nu_{\Box},\, \alp_d > J_{\mathfrak f} + \nB.
\end{equation}
Define\footnote{Note that this definition is consistent with \eqref{eq:alp.def} and \eqref{eq:Kf.def} thanks to \eqref{eq:lower-hyp-alpf}.} $\alp_{\mathfrak f} = J_{\mathfrak f} + \nB$ and $K_{\mathfrak f} = K_{J_{\mathfrak f}}$.

Then there exist $m'_{0} \in \mathbb N$ large, $a'_{0}  \in (0,1)$ small, $\de_{\mathfrak f} \in (0,1)$ small, all depending only on $d$, $\de_0$, $\de_{c}$, $\de_{d}$, $\eta_{c}$, $\alp_{0}$, $\alp_{d}$, $J_{c}$, $J_{d}$, $J_{\mathfrak f}$, $K_{c}$, $K_{d}$, $s_{c}$, such that if $M_0, M_{c} \geq m_0'$, then the solution obeys the following sharp asymptotics estimates whenever $\de_a \in (0,\f{2\de_{\mathfrak f}}{d-1})$:
\begin{enumerate}
\item (Sharp asymptotics\footnote{Note that when $r\geq \tau^{1-\de_a}$ and $0< \de_a \ll \de_{\mathfrak f}$, the error term $A_{\mathfrak f} r^{-\f{d-1}2} \tau^{-\alp_{\mathfrak f}+\f{d-1}2-\de_{\mathfrak f}}$ is indeed smaller than the main term.} when $r \geq \tau^{1-\de_a}$)
\begin{equation} \label{eq:asymptotics.wave.zone}
\begin{split}
\phi(\tau,r,\theta) = &\: \varphi^{\bfm[1]}_{J_{\mathfrak f},0}[\rcPhi_{(\leq J_{\mathfrak f}+\f{d-1}2)J_{\mathfrak f},0}(\infty)](\tau,r,\theta) + \sum_{k=1}^{K_{J_{\mathfrak f}}} \varphi^{\bfm[1]}_{J_{\mathfrak f},k}[\rcPhi_{J_{\mathfrak f},k}(\infty)](\tau,r,\theta) \\
&\: + O_{\bfGmm}^{\lfloor a'_{0} M_{0} \rfloor}(A'_{\mathfrak f} r^{-\f{d-1}2} \tau^{-\alp_{\mathfrak f}+\f{d-1}2-\de_{\mathfrak f}})  \quad \hbox{ in } \set{ r \geq R_{\far}},
\end{split}
\end{equation}
where $\varphi^{\bfm[1]}_{J_{\mathfrak f},0}[\rcPhi_{(\leq J_{\mathfrak f}+\f{d-1}2)J_{\mathfrak f},0}(\infty)]$ and $\varphi^{\bfm[1]}_{J_{\mathfrak f},k}[\rcPhi_{J_{\mathfrak f},k}(\infty)]$ are given by \eqref{eq:varphi.m.data.summed.1}--\eqref{eq:varphi.m.data.summed.2}.
\item (Sharp asymptotics when $r\leq \tau^{1-\de_a}$) Let $c_{k,k'} = \mathfrak b^{(J_{\mathfrak f},k,0)}_{k-k'} \f{\Big( \f{d-3}2 \Big)!}{(d-2)!} 2^{J_{\mathfrak f}+\nB}$, with $\mathfrak b^{(J_{\mathfrak f},k,0)}_{k'}$ given in Lemma~\ref{lem:med.explicit.1}. Then
\begin{equation} \label{eq:asymptotics}
\phi(\tau,x)
= \sum_{k,k'=0}^{K_{J_{\mathfrak f}}} c_{k,k'} \frkL_{k} \tau^{-\alp_{\mathfrak f}} (\log^{k'} (\tfrac \tau 2)) \psi(x)
+ O_{\bfGmm}^{\lfloor a'_{0} M_{0} \rfloor}(A'_{\mathfrak f} \tau^{-\alp_{\mathfrak f}-\f{\dlt_{a}}4}) \quad \hbox{ in } \set{r \leq \tau^{1-\dlt_{a}}},
\end{equation}
where $\mathfrak L_k$ and $\psi$ are given by \eqref{eq:frkL-def} and \eqref{eq:psi.def}, respectively.
\end{enumerate}
In both \eqref{eq:asymptotics.wave.zone} and \eqref{eq:asymptotics},
the amplitude $A'_{\mathfrak f}$ can be chosen so that
\begin{equation} 
\begin{aligned}
A'_{\mathfrak f}&= D+A + A^{(2)}(A) \quad \hbox{ if } \calN \neq 0 \\
A'_{\mathfrak f} &= D+ A \quad \hbox{ if } \calN = 0,
\end{aligned}\end{equation}
where $A^{(2)}(A) \geq 0$ is nondecreasing function depending on $\calN$ satisfying $A^{(2)}(A) = O(A^{2})$ as $A \to 0$.
The dependence of the implicit constants in \eqref{eq:asymptotics.wave.zone} and \eqref{eq:asymptotics} on $\phi$ and $f$ is only through $\de_0, J_{\mathfrak f}, \de_{d}, \alp_d, M_0$. Moreover, the implicit constant in \eqref{eq:asymptotics} depends on $\de_a$.
\end{maintheorem}

\begin{remark}[Bounds and asymptotics for the Minkowskian profile]
The Minkowskian profiles are further investigated in Section~\ref{sec:Minkowski.wave}. See \eqref{eq:med.explicit.expanded} for their explicit form. We note the following properties:
\begin{enumerate}
\item $\varphi^{\bfm[1]}_{J_{\mathfrak f},0}[\rcPhi_{(\leq J_{\mathfrak f}+\f{d-1}2)J_{\mathfrak f},0}(\infty)]$, $\varphi^{\bfm[1]}_{J_{\mathfrak f},k}[\rcPhi_{J_{\mathfrak f},k}(\infty)]$ obey the upper bounds (see Propositions~\ref{prop:phi.ell.j.k.upper}, \ref{prop:radiation.field})
\begin{align*}
|\varphi^{\bfm[1]}_{J_{\mathfrak f},0}[\rcPhi_{(\leq J_{\mathfrak f}+\f{d-1}2)J_{\mathfrak f},0}(\infty)]|\ls &\: \min \{r^{-\nB} \tau^{-J_{\mathfrak f}},\tau^{-J_{\mathfrak f}-\nB} \}, \\
|\varphi^{\bfm[1]}_{J_{\mathfrak f},k}[\rcPhi_{J_{\mathfrak f},k}(\infty)]| \ls &\: \min\{r^{-\nB} \tau^{-J_{\mathfrak f}},\tau^{-J_{\mathfrak f}-\nB} \}\log^k \tau,
\end{align*}
consistent with the upper bounds in Main Theorem~\ref{thm:upper}.
\item Near null infinity, $\varphi^{\bfm[1]}_{J_{\mathfrak f},0}[\rcPhi_{(\leq J_{\mathfrak f}+\f{d-1}2)J_{\mathfrak f},0}(\infty)]$ and $\varphi^{\bfm[1]}_{J_{\mathfrak f},k}[\rcPhi_{J_{\mathfrak f},k}(\infty)]$ are approximated by the corresponding radiation fields, which are computed in Proposition~\ref{prop:radiation.field}.
\item In the region $\{r \ls \tau^{1-\de_a}\}$, $\varphi^{\bfm[1]}_{J_{\mathfrak f},k}[\rcPhi_{(\leq J_{\mathfrak f}+\f{d-1}2)J_{\mathfrak f},k}(\infty)] \simeq \tau^{-J_{\mathfrak f}-\nB} \log^k \tau$ for any $k \in \mathbb Z_{\geq 0}$; see Proposition~\ref{prop:med.explicit.precise}. In particular, the estimates in \eqref{eq:asymptotics.wave.zone} and \eqref{eq:asymptotics} are consistent when $r \approx \tau^{1-\de_a}$.
\end{enumerate}
\end{remark}

\begin{remark}[Basic properties of $c_{k,k'}$]
It is easy to check that the constants $c_{k,k'}$ satisfy the following properties:
\begin{itemize}
\item Whenever $k<k'$, we have $c_{k,k'} = 0$.
\item For each $0\leq k \leq K_{J_{\mathfrak f}}$, $c_{k,k} =0$ if $J_{\mathfrak f} < \f{d-1}2$ and $c_{k,k} \neq 0$ if $J_{\mathfrak f} \geq \f{d-1}2$.
\item For each $1\leq k \leq K_{J_{\mathfrak f}}$, $c_{k,k-1}\neq 0$.
\end{itemize}
\end{remark}

\begin{remark}[Some special cases]
From the properties of $c_{k,i}$ above, we deduce the following asymptotics statements.
\begin{enumerate}
\item If $K_{J_{\mathfrak f}} = 0$ and $\mathfrak L_0 \neq 0$, then we must have $J_{\mathfrak f} \geq \f{d-1}2$ (by the definition of $J_{\mathfrak f}$) and thus
$$\phi = c_{0,0} \mathfrak L_0 \tau^{-\alp_{\mathfrak f}} \psi(\tau,x) + O_{\bfGmm}^{\lfloor a_0 M_{0} \rfloor}(A_{\mathfrak f} \tau^{-\alp_{\mathfrak f}-\dlt_{\mathfrak f}}),$$
where the leading coefficient $c_{0,0}\neq 0$.
\item  Let
$$\widehat{K}_{J_{\mathfrak f}} := \max\{ k: \mathfrak L_k \neq 0 \}.$$
Suppose $\widehat{K}_{J_{\mathfrak f}} \geq 1$. Then
$$\phi = \begin{cases}
c_{\widehat{K}_{J_{\mathfrak f}},\widehat{K}_{J_{\mathfrak f}}-1} \frkL_{\widehat{K}_{J_{\mathfrak f}}}  \tau^{-\alp_{\mathfrak f}} (\log^{\widehat{K}_{J_{\mathfrak f}}-1} \tfrac \tau 2) \psi(\tau, x)
+ O_{\bfGmm}^{\lfloor a_0 M_{0} \rfloor}(A_{\mathfrak f} \tau^{-\alp_{\mathfrak f}} \log^{\widehat{K}_{J_{\mathfrak f}}-2} \tfrac \tau 2) & \hbox{ if } J_{\mathfrak f} < \f{d-1}2 \hbox{ and } \widehat{K}_{J_{\mathfrak f}} \geq 2 \\
c_{\widehat{K}_{J_{\mathfrak f}},\widehat{K}_{J_{\mathfrak f}}-1} \frkL_{\widehat{K}_{J_{\mathfrak f}}}  \tau^{-\alp_{\mathfrak f}}  \psi(\tau, x)
+ O_{\bfGmm}^{\lfloor a_0 M_{0} \rfloor}(A_{\mathfrak f} \tau^{-\alp_{\mathfrak f} - \de_{\mathfrak f}}) & \hbox{ if } J_{\mathfrak f} < \f{d-1}2 \hbox{ and } \widehat{K}_{J_{\mathfrak f}} = 1 \\
c_{\widehat{K}_{J_{\mathfrak f}},\widehat{K}_{J_{\mathfrak f}}-1} \frkL_{\widehat{K}_{J_{\mathfrak f}}}  \tau^{-\alp_{\mathfrak f}} (\log^{\widehat{K}_{J_{\mathfrak f}}} \tfrac \tau 2)\psi(\tau, x)
+ O_{\bfGmm}^{\lfloor a_0 M_{0} \rfloor}(A_{\mathfrak f} \tau^{-\alp_{\mathfrak f}} \log^{\widehat{K}_{J_{\mathfrak f}}-1} \tfrac \tau 2) & \hbox{ if } J_{\mathfrak f} \geq \f{d-1}2
\end{cases},$$
where the leading coefficient $c_{\widehat{K}_{J_{\mathfrak f}},\widehat{K}_{J_{\mathfrak f}}-1} \neq 0$.
\end{enumerate}
\end{remark}

%
%
%


\subsubsection{The case of spherically symmetric spacetimes}
Main Theorem~\ref{thm:lower} shows that Main Theorem~\ref{thm:upper} is sharp when the first non-vanishing non-corrected coefficient has a \emph{non-zero spherical mean} (i.e., $\frkL_{k} := \rcPhi_{(0) J_{\mathfrak f}, k}(\infty)  \neq 0$). While $(\rcPhi_{(0) J_{\mathfrak f}, k}(\infty))_{k=0, \ldots, K_{J_{\mathfrak f}}} \neq 0$ is generic in many linear and nonlinear cases of interest, it is nevertheless interesting to ask what governs the late time tails when $\rcPhi_{(0) J_{\mathfrak f}, k}(\infty) = 0$ for $k=0, \ldots, K_{J_{\mathfrak f}}$ yet $\rPhi_{(\ell) J_{\mathfrak f}, k'}(\infty) \neq 0$ for some $\ell > 0$ and $0 \leq k' \leq K_{J_{\mathfrak f}}$. Intuitively speaking, the reason why our results are not sharp in this case is because of ``angular mode coupling,'' i.e., the possible contribution of $\rcPhi_{(\ell) J_{\mathfrak f}, k}(\infty)$ to $\bbS_{(\ell')} \phi$ for $\ell' \neq \ell$.

However, in the special case of linear equations on spherically symmetric spacetimes, there is a complete decoupling among different spherical modes, and our methods can be applied to obtain sharp results. The following results generalizes the results in \cite{AAG2018, DSS1, DSS2, HintzPriceLaw} and includes linear equations on dynamical, but still spherically symmetric, spacetimes. (To obtain sharp results on a non-spherically symmetric background in the case when $\rcPhi_{(0) J_{\mathfrak f}, k}(\infty) = 0$ for $k=0, \ldots, K_{J_{\mathfrak f}}$ yet $\rPhi_{(\ell) J_{\mathfrak f}, k'}(\infty) \neq 0$ for some $\ell > 0$ and $0 \leq k' \leq K_{J_{\mathfrak f}}$, it is likely that more specific assumptions must be made on the angular dependency of the coefficients $\bfg^{-1}$, $\bfB$, $V$ and $\calN$ than those in this paper. We shall leave this task to a future investigation.)

We start with an improved upper bound for the late time tail of solutions supported in high spherical harmonics.
\begin{maintheorem} [Upper bound for linear equations on spherically symmetric spacetimes] \label{thm:upper-sphsymm}
Let $\calP$ satisfy the assumptions \ref{hyp:topology}--\ref{hyp:T-tau},  \ref{hyp:T-almost-stat}--\ref{hyp:wave-V}, \ref{hyp:bdry.cond}, as well as either \ref{hyp:ult-stat} or \ref{hyp:ult-stat'}. Assume that $\calN = 0$, and also that $\calP$ is spherically symmetric. Let $\phi$, $f$ solve the main equation \eqref{eq:main-eq} and satisfy \ref{hyp:id}, \ref{hyp:forcing}, and \ref{hyp:sol} with some $\alp_{0} \in \bbR$ and $\alp_{d} > \alp_{0}$. Assume also that, for some $\ell \in \bbZ_{\geq 0}$,
\begin{equation*}
	\phi = \bbS_{(\geq \ell)} \phi, \quad f = \bbS_{(\geq \ell)} f.
\end{equation*}
From $\rPhi_{0}$, define $\rPhi_{j, k}(u, \tht)$ for $0 \leq j \leq \min\set{J_{c}, J_{d}} - 1$, $0 \leq k \leq K_{j}$ using the formal recurrence equations \eqref{eq:recurrence-jk}. Suppose $J_{\mathfrak f}$ satisfies \eqref{eq:Jf.def} as in Main Theorem~\ref{thm:upper}.

Then there exist $m_{0,\ell} \in \mathbb N$ large, $a_{0,\ell} \in (0,1)$ small, both depending only on $d$, $\ell$, $\de_{0}$, $\de_{c}$, $\de_{d}$, $\eta_{c}$, $\alp_{0}$, $\alp_{d}$, $J_{c}$, $J_{d}$, $J_{\mathfrak f}$, $K_{c}$, $K_{d}$, $s_{c}$, such that if $M_{0}, M_{c} \geq m_{0,\ell}$, then the solution $\phi$ obeys the following upper bounds in $\calM$:
\begin{align}
	\phi &= O_{\bfGmm}^{\lfloor a_{0,\ell} M_{0} \rfloor}((D+A_0) \min\{ \tau^{-\alp_{\mathfrak f}-\ell} \brk{r}^{\ell} ,\tau^{-\alp_{\mathfrak f}+\nB} r^{-\nu_{\Box}} \} \log^{K_{\mathfrak f}} \tau ), \label{eq:thm3.main.decay}
\end{align}
where the decay rate $\alp_{\mathfrak f}$ is given by \eqref{eq:alp.def}, the power $K_{\mathfrak f}$ in the logarithm is given by \eqref{eq:Kf.def},
and the dependence of the implicit constants in \eqref{eq:thm3.main.decay} on $\phi$ and $f$ is only through $\de_0, J_{\mathfrak f}, \de_{d}, \alp_d, M_0$.
\end{maintheorem}

For initial data supported on a fixed spherical harmonic, we also have a refined asymptotics theorem. To state the refined asymptotic theorem, we need slight modifications of the \emph{final asymptotic charge} and the \emph{asymptotic spatial profiles}. (The Minkowskian profiles are still defined as in \eqref{eq:varphi.m.data.summed.1}--\eqref{eq:varphi.m.data.summed.2}.) For the final asymptotic charge, define
\begin{align}
	\frkL_{(\ell) k} &= \rcPhi_{(\ell) J_{\mathfrak f}, k}(\infty) \quad \hbox{ for } 1 \leq k \leq K_{J_{\mathfrak f}}, \label{eq:frkL-ell-k-def} \\
	\frkL_{(\ell) 0} &= \begin{cases}
				\rcPhi_{(\ell) J_{\mathfrak f}, 0}(\infty) & \hbox{ for } \ell \leq J_{\mathfrak f} - \frac{d-1}{2}, \\
				0 & \hbox{ otherwise.}
				\end{cases} 						\label{eq:frkL-ell-0-def}
\end{align}
For the asymptotic spatial profiles (cf.~\eqref{eq:psi.def}), define
\begin{equation}\label{eq:psi.def.ell}
	\psi(x) = \begin{cases}
	r^\ell Y_{(\ell)} - {}^{(\infty)} \calR_{0} {}^{(\infty)} \calP_{0} (r^\ell Y_{(\ell)}) & \hbox{ if } \rd_{(\mathrm{t})} \calM = \0, \\
	r^\ell Y_{(\ell)} - \chi_{< R_{\far}} - {}^{(\infty)} \calR_{0} {}^{(\infty)} \calP_{0} (r^\ell Y_{(\ell)} - \chi_{< R_{\far}}) & \hbox{ if } \rd_{(\mathrm{t})} \calM \neq \0,
	\end{cases}
\end{equation}
where $Y_{(\ell)}$ is the spherical harmonic on which the initial data are supported.

\begin{maintheorem} [Asymptotics for linear equations on spherically symmetric spacetimes] \label{thm:lower-sphsymm}
Let $\calP$ satisfy the assumptions \ref{hyp:topology}--\ref{hyp:T-tau},  \ref{hyp:T-almost-stat}--\ref{hyp:wave-V}, \ref{hyp:bdry.cond}--\ref{hyp:ult-stat}. Assume that $\calN = 0$, and also that $\calP$ is spherically symmetric. Let $\phi$, $f$ solve the main equation \eqref{eq:main-eq} and satisfy \ref{hyp:id}, \ref{hyp:forcing}, and \ref{hyp:sol} with some $\alp_{0} \in \bbR$ and $\alp_{d} > \alp_{0}$. Assume also that, for some $\ell \in \bbZ_{\geq 0}$,
\begin{equation*}
	\phi = \bbS_{(\ell)} \phi, \quad f = \bbS_{(\ell)} f.
\end{equation*}
Assume in addition that \eqref{eq:lower-hyp-alpf} holds.

Define $\rPhi_{j, k}(u, \tht)$ as in Main Theorem~\ref{thm:upper}. Assume that \eqref{eq:Jf.def} holds for some $J_{\mathfrak f}\in \mathbb N$, $1\leq J_{\mathfrak f} \leq \min\{J_{c}, J_{d}\}-1$. Assume, in addition, that \eqref{eq:lower-hyp-alpf} holds and define $\alp_{\mathfrak f} = J_{\mathfrak f} + \nB$ and $K_{\mathfrak f} = K_{J_{\mathfrak f}}$.

Then the following holds:
\begin{enumerate}
\item The estimate \eqref{eq:asymptotics.wave.zone} holds in $\{r \geq R_{\far}\}$.
\item Let $c^{(\ell)}_{k,k'}=\mathfrak b^{(j,k,\ell)}_{k-k'}   \f{\left( \f{d-3}2+\ell \right)!}{(d-2+2\ell)!} 2^{J_{\mathfrak f}+\nB+\ell}$, where $\mathfrak b^{(J_{\mathfrak f},k,\ell)}_{k'}$ are given in Lemma~\ref{lem:med.explicit.1}. There exist $m'_{0,\ell} \in \mathbb N$ large, $a'_{0,\ell}  \in (0,1)$ small, $\de_{f,\ell}  \in (0,1)$ small, with all three of them depending only on $d$, $\ell$, $\de_{0}$, $\de_{c}$, $\de_{d}$, $\eta_{c}$, $\alp_{0}$, $\alp_{d}$, $J_{c}$, $J_{d}$, $J_{\mathfrak f}$, $K_{c}$, $K_{d}$, $s_{c}$, such that if $M_{0} \geq m'_{0,\ell}$, then the following holds for every $\de_a \in (0,\f{2\de_{\mathfrak f}}{d-1})$:
\begin{equation} \label{eq:asymptotics-sphsymm}
\begin{split}
\phi(\tau,x)
= &\: \sum_{k,k'=0}^{K_{J_{\mathfrak f}}} c^{(\ell)}_{k,k'} \frkL_{(\ell)k} \tau^{-\alp_{\mathfrak f}-\ell} (\log^{k'} (\tfrac \tau 2)) \psi(x)\\
&\: + O_{\bfGmm}^{\lfloor a'_{0,\ell} M_{0} \rfloor}((D+A_0) \tau^{-\alp_{\mathfrak f}-\ell-\f{\dlt_{a}}2} \brk{r}^\ell) \quad \hbox{ in } \set{r \leq \tau^{1-\dlt_{a}}},
\end{split}
\end{equation}
where $\alp_{\mathfrak f} = J_{\mathfrak f} + \nB$, and $\mathfrak L_{(\ell)k}$, $\mathfrak L_{(\ell)0}$ and $\psi$ are given by \eqref{eq:frkL-ell-k-def}, \eqref{eq:frkL-ell-0-def} and \eqref{eq:psi.def.ell}, respectively.
\end{enumerate}
The dependence of the implicit constants in \eqref{eq:asymptotics-sphsymm} on $\phi$ and $f$ is only through $\de_0, J_{\mathfrak f}, \de_{d}, \alp_d, M_0$. Moreover, the implicit constant in \eqref{eq:asymptotics-sphsymm} depends on $\de_a$.

\end{maintheorem}

\begin{remark}[Minkowskian profiles for higher angular modes]
We refer the reader to Section~\ref{sec:Minkowski.wave} for the explicit form of the Minkowskian profiles. We only note for now that for higher $\ell$'s,
 $\varphi^{\bfm[1]}_{J_{\mathfrak f},0}[\rcPhi_{(\ell)J_{\mathfrak f},0}(\infty)]$, $\varphi^{\bfm[1]}_{J_{\mathfrak f},k}[\rcPhi_{(\ell)J_{\mathfrak f},k}(\infty)]$ obey the upper bounds (see Propositions~\ref{prop:phi.ell.j.k.upper}, \ref{prop:radiation.field})
\begin{align*}
|\varphi^{\bfm[1]}_{J_{\mathfrak f},0}[\rcPhi_{(\leq J_{\mathfrak f}+\f{d-1}2)J_{\mathfrak f},0}(\infty)]|\ls &\: \min \{r^{-\nB} \tau^{-J_{\mathfrak f}},r^\ell \tau^{-J_{\mathfrak f}-\nB-\ell} \}, \\
|\varphi^{\bfm[1]}_{J_{\mathfrak f},k}[\rcPhi_{J_{\mathfrak f},k}(\infty)]| \ls &\: \min\{r^{-\nB} \tau^{-J_{\mathfrak f}},r^\ell \tau^{-J_{\mathfrak f}-\nB-\ell} \}\log^k \tau,
\end{align*}
consistent with the upper bounds in Main Theorem~\ref{thm:upper-sphsymm}.
\end{remark}

\subsection{Statements of main theorems starting from Cauchy data}\label{sec:Cauchy}

In this final subsection of the section, we consider an extension of our main theorems to include the case where the initial data are posed on the asymptotically flat hypersurface $\calS_{1}$ (as opposed to an asymptotically null hypersurface). The main result is Main Theorem~\ref{thm:Cauchy}, which will be given Section~\ref{sec:thm.Cauchy}. We will describe the assumptions need in $\calM_{\ext}$ in Section~\ref{sec:E-assumptions} and then discuss some preliminaries in Section~\ref{sec:recurrence.ext} and Section~\ref{sec:ext.prelim}. The main analytic ingredient will be the exterior stability theorem given in Section~\ref{sec:ext.stab.thm}. Our main theorem (Main Theorem~\ref{thm:Cauchy}) will then be a consequence of the exterior stability; see  Section~\ref{sec:thm.Cauchy}.

These considerations are somewhat separate from those described in the subsections above. Indeed, by the finite speed of propagation, one can localize the analysis to this region. In particular, in this region, one needs not deal with issues of trapping, superradiance, etc. As a result, we also only need fewer assumptions in this region. (Notice in particular that no analogues of \ref{hyp:bdry.cond}, \ref{hyp:ult-stat} and \ref{hyp:sol} are needed in this region.)

\subsubsection{Assumptions in the exterior region}\label{sec:E-assumptions}

Let $\mbrk{u}$ be a smooth even function such that
\begin{equation}\label{eq:def.mbrku}
\mbrk{u} = \begin{cases}
\abs{u} & \hbox{when $\abs{u} \geq 1$},\\
\frac{1}{2} & \hbox{when $\abs{u} \leq \frac{1}{2}$},
\end{cases}
\end{equation}
and nondecreasing for $u > 0$.

In $\calM_{\ext}$, introduce the following vector field bounds (cf.~\eqref{eq:O-Gmm-far-scalar}, \eqref{eq:O-Gmm-far}).
Given a smooth function $a$ on (some subset of) $\calM_{\ext}$, a nonnegative integer $N$ and a fixed function $m(u,r)$, we define
\begin{align}
	a = O_{\bfGmm}^{N}(m(u, r)) \quad \impmi \quad \abs{(\mbrk{u} \bfT)^{I_{\bfT}} (\brk{r} \rd_{r})^{I_{r}} \bfOmg_{ab}^{I_{\bfOmg_{ab}}} a} \aleq m(u, r) \quad \hbox{ for all } \abs{I} \leq N. \label{eq:O-Gmm-ext-scalar}
\end{align}
In case $\bfa$ is a contravariant tensor field, we replace the usual derivatives by Lie derivatives and define
\begin{align}
	\bfa \in O_{\bfGmm}^{N}(m(u, r)) \quad \impmi \quad \abs{\calL_{\mbrk{u}\bfT}^{I_{\bfT}} \calL_{\brk{r} \rd_{r}}^{I_{r}} \calL_{\bfOmg_{ab}}^{I_{\bfOmg_{ab}}} \bfa} \aleq m(u, r) \quad \hbox{ for all } \abs{I} \leq N, \label{eq:O-Gmm-ext}
\end{align}
where as in \eqref{eq:tensor-abs-near}, the absolute value of a contravariant tensor of rank $s$ is defined with respect to the coordinate vectors associated with $(x^{0}, x^{1}, \ldots, x^{d})$.

Using the notations in \eqref{eq:O-Gmm-ext-scalar}, \eqref{eq:O-Gmm-ext}, we now introduce the assumptions in $\calM_{\ext}$.

\begin{enumerate}[label=(E-$\bfg$)]
\item \label{hyp:ext-g} {\it Asymptotics of $\bfg^{-1}$ in $\calM_{\ext}$.} In $\calM_{\ext}$, the metric difference $\bfg^{-1} - \bfm^{-1}$ admits an expansion of the following form.
For $J_{c} \in \bbZ_{\geq 2}$ and $0 < \eta_{c} \leq 1$,
\begin{align}
	(\bfg^{-1})^{u u} &= \sum_{j=2}^{J_{c}} \sum_{k=0}^{K_{c}} r^{-j} \log^{k} (\tfrac{r}{\mbrk{u}}) \rh_{j, k}^{uu}(u, \tht)
	+ \rem_{J_{c}+\eta_{c}}[\bfh^{uu}],   \label{eq:ext-g-uu-phg} \\
	(\bfg^{-1})^{u r} &= - 1
	+ \sum_{j=1}^{J_{c}-1} \sum_{k=0}^{K_{c}} r^{-j} \log^{k} (\tfrac{r}{\mbrk{u}}) \rh_{j, k}^{ur}(u, \tht)
	+ \rem_{J_{c}-1+\eta_{c}}[\bfh^{ur}],   \label{eq:ext-g-ur-phg} \\
	r (\bfg^{-1})^{u A} &=
	\sum_{j=1}^{J_{c}-1} \sum_{k=0}^{K_{c}} r^{-j} \log^{k} (\tfrac{r}{\mbrk{u}}) \rsh_{j, k}^{uA}(u, \tht)
	+ \rem_{J_{c}-1+\eta_{c}}[r \bfh^{uA}],   \label{eq:ext-g-uA-phg} \\
	(\bfg^{-1})^{r r} &= 1 
	+ \sum_{j=1}^{J_{c}-2} \sum_{k=0}^{K_{c}} r^{-j} \log^{k} (\tfrac{r}{\mbrk{u}}) \rh_{j, k}^{rr}(u, \tht)
	+ \rem_{J_{c}-2+\eta_{c}}[\bfh^{rr}],   \label{eq:ext-g-rr-phg} \\
	r (\bfg^{-1})^{r A} &= 
	\sum_{j=1}^{J_{c}-2} \sum_{k=0}^{K_{c}} r^{-j} \log^{k} (\tfrac{r}{\mbrk{u}}) \rsh_{j, k}^{r A}(u, \tht)
	+ \rem_{J_{c}-2+\eta_{c}}[r\bfh^{rA}],   \label{eq:ext-g-rA-phg} \\
	r^{2} (\bfg^{-1})^{AB} &= \rsgmm^{AB} 
	+ \sum_{j=1}^{J_{c}-2} \sum_{k=0}^{K_{c}} r^{-j} \log^{k} (\tfrac{r}{\mbrk{u}}) \rsh_{j, k}^{AB}(u, \tht)
	+ \rem_{J_{c}-2+\eta_{c}}[r^{2} \bfh^{AB}],   \label{eq:ext-g-AB-phg}
\end{align}
where, whenever the object (on the left-hand side below) is defined,
\begin{align}
	\rh_{j, k}^{uu}, \,
	\rh_{j, k}^{ur}, \,
	\rsh_{j, k}^{uA}, \,
	\rh_{j, k}^{rr}, \,
	\rsh_{j, k}^{rA}, \,
	\rsh_{j, k}^{AB}
	 &= O_{\bfGmm}^{M_{c}}(A_{c} \mbrk{u}^{j-\dlt_{c}}), \label{eq:ext-gjk-phg}
\end{align}
with all implicit constants in $O_{\bfGmm}^{M_{c}}$ equal to $1$.
Moreover, the remainders satisfy
\begin{align}
	\rem_{J_{c}+\eta_{c}}[\bfh^{uu}]
	 &= O_{\bfGmm}^{M_{c}}(A_{c} r^{-(J_{c}+\eta_{c})} \mbrk{u}^{J_{c}+\eta_{c}-\dlt_{c}}), \label{eq:ext-grem-uu-phg} \\
	\rem_{J_{c}-1+\eta_{c}}[\bfh^{ur}], \rem_{J_{c}-1+\eta_{c}}[r \bfh^{uA}]
	 &= O_{\bfGmm}^{M_{c}}(A_{c} r^{-(J_{c}-1+\eta_{c})} \mbrk{u}^{J_{c}-1+\eta_{c}-\dlt_{c}}), \label{eq:ext-grem-u-phg}  \\
	\rem_{J_{c}-2+\eta_{c}}[\bfh^{rr}],
	\rem_{J_{c}-2+\eta_{c}}[r \bfh^{rA}],
	\rem_{J_{c}-2+\eta_{c}}[r^{2} \bfh^{AB}]
	 &= O_{\bfGmm}^{M_{c}}(A_{c} r^{-(J_{c}-2+\eta_{c})} \mbrk{u}^{J_{c}-2+\eta_{c}-\dlt_{c}}), \label{eq:ext-grem-phg}
\end{align}
with all implicit constants in $O_{\bfGmm}^{M_{c}}$ equal to $1$.
\end{enumerate}
\begin{enumerate}[label=(E-$\bfB$)]
\item \label{hyp:ext-B} {\it Asymptotics of $\bfB$ in $\calM_{\ext}$.} The vector field $\bfB$ admits an expansion of the form. For $J_{c} \in \bbZ_{\geq 2}$ and $0 < \eta_{c} \leq 1$,
\begin{align}
	\bfB^{u} &= \sum_{j=2}^{J_{c}} \sum_{k=0}^{K_{c}} r^{-j} \log^{k} (\tfrac{r}{\mbrk{u}}) \rbfB_{j, k}^{u}(u, \tht) + \rem_{J_{c}+\eta_{c}}[\bfB^{u}], \label{eq:ext-B-u-phg} \\
	\bfB^{r} &= \sum_{j=1}^{J_{c}-1} \sum_{k=0}^{K_{c}} r^{-j} \log^{k} (\tfrac{r}{\mbrk{u}}) \rbfB_{j, k}^{r}(u, \tht) + \rem_{J_{c}-1+\eta_{c}}[\bfB^{r}], \label{eq:ext-B-r-phg} \\
	r \bfB^{A} &= \sum_{j=1}^{J_{c}-1} \sum_{k=0}^{K_{c}} r^{-j} \log^{k} (\tfrac{r}{\mbrk{u}}) \rsbfB_{j, k}^{A}(u, \tht) + \rem_{J_{c}-1+\eta_{c}}[r \bfB^{A}], \label{eq:ext-B-A-phg}
\end{align}
where
\begin{align}
	\rbfB_{j, k}^{u}(u, \tht), \,
	\rbfB_{j, k}^{u}(u, \tht), \,
	\rsbfB_{j, k}^{A}(u, \tht)
	 &= O_{\bfGmm}^{M_{c}}(A_{c} \mbrk{u}^{j-1-\dlt_{c}}), \label{eq:ext-Bjk-phg}
\end{align}
whenever $\rbfB_{j, k}^{\alp}$ is defined, and the remainders satisfy
\begin{align}
	\rem_{J_{c}+\eta_{c}}[\bfB^{u}] &= O_{\bfGmm}^{M_{c}}(A_{c} r^{-(J_{c}+\eta_{c})}\mbrk{u}^{J_{c}-1+\eta_{c}-\dlt_{c}}). \label{eq:ext-Brem-u-phg} \\
	\rem_{J_{c}-1+\eta_{c}}[\bfB^{r}], \rem_{J_{c}-1+\eta_{c}}[r \bfB^{A}] &= O_{\bfGmm}^{M_{c}}(A_{c} r^{-(J_{c}-1+\eta_{c})}\mbrk{u}^{J_{c}-2+\eta_{c}-\dlt_{c}}), \label{eq:ext-Brem-phg}
\end{align}
with all implicit constants in $O_{\bfGmm}^{M_{c}}$ equal to $1$.
\end{enumerate}
\begin{enumerate}[label=(E-$V$)]
\item \label{hyp:ext-V} {\it Asymptotics of $V$ in $\calM_{\ext}$.}
The function $V$ admits an expansion of the form. For $J_{c} \in \bbZ_{\geq 2}$ and $0 < \eta_{c} \leq 1$,
\begin{align}
	V &= \sum_{j=2}^{J_{c}} \sum_{k=0}^{K_{c}} r^{-j} \log^{k} (\tfrac{r}{\mbrk{u}}) \rV_{j, k}(u, \tht) + \rem_{J_{c}+\eta_{c}}[V], \label{eq:ext-V-phg}
\end{align}
\begin{align}
	\rV_{j, k}(u, \tht) &\in O_{\bfGmm}^{M_{c}}(A_{c} \mbrk{u}^{j-2-\dlt_{c}}), \label{eq:ext-Vj}\\
	\rem_{J_{c}+\eta_{c}}[V] &\in O_{\bfGmm}^{M_{c}}(A_{c} r^{-(J_{c}+\eta_{c})} \mbrk{u}^{J_{c}-2+\eta_{c}-\dlt_{c}}), \label{eq:ext-Vrem-phg}
\end{align}
with all implicit constants in $O_{\bfGmm}^{M_{c}}$ equal to $1$.
\end{enumerate}
\begin{enumerate}[label=(E-$\calN$)]
\item \label{hyp:ext-N} {\it Assumptions on the nonlinearity in $\calM_{\ext}$.} Assume that the nonlinearity satisfies \ref{hyp:nonlin}--\ref{hyp:nonlin-degree}, as well as the null condition \ref{hyp:nonlin-null} when $d = 3$.
Assume, in addition, that in $\calM_{\ext}$,
\begin{align}
\quasi^{\mu \nu} &= \sum_{j = 0}^{J_{c}-\frac{d-1}{2}} \sum_{k=0}^{K_{c}} \rquasi_{j, k}^{\mu \nu} r^{-j} \log^{k} (\tfrac{r}{\mbrk{u}}) + \rem_{J_{c}-\frac{d-1}{2} +\eta_{c}}[\quasi^{\mu \nu}], \label{eq:ext-quasi-phg} \\
\semi &= \sum_{j = 0}^{J_{c}- \frac{d-1}{2}} \sum_{k=0}^{K_{c}} \rsemi_{j, k} r^{-j} \log^{k} (\tfrac{r}{\mbrk{u}}) + \rem_{J_{c}-\frac{d-1}{2} + \eta_{c}}[\semi], \label{eq:ext-semi-phg}
\end{align}
with
\begin{align}
\rquasi_{j, k} &= O_{\bfGmm}^{M_{c}}(\mbrk{u}^{j}), \label{eq:ext-quasijk}\\
\rem_{J_{c}-\frac{d-1}{2}+\eta_{c}}[\quasi] &= O_{\bfGmm}^{M_{c}}(r^{-J_{c}+\frac{d-1}{2}-\eta_{c}} \mbrk{u}^{J_{c}-\frac{d-1}{2}+\eta_{c}}), \label{eq:ext-quasirem-phg} \\
\rsemi_{j, k} &= O_{\bfGmm}^{M_{c}}(\mbrk{u}^{j}), \label{eq:ext-semijk}\\
\rem_{J_{c}-\frac{d-1}{2}+\eta_{c}}[\semi] &= O_{\bfGmm}^{M_{c}}(r^{-J_{c}+\frac{d-1}{2}-\eta_{c}} \mbrk{u}^{J_{c}-\frac{d-1}{2}+\eta_{c}}). \label{eq:ext-semirem-phg}
\end{align}
\end{enumerate}
\begin{enumerate}[label=(E-F)]
\item \label{hyp:ext-forcing}
{\it Assumptions on the forcing term.}
In $\calM_{\ext}$, for $J_{d} := \lceil \alp_{d} \rceil - \nu_{\Box}$ and $\eta_{d} := \alp_{d} - (J_{d} - 1 + \nu_{\Box})$, we assume that $f$ admits the expansion
\begin{equation} \label{eq:ext-f-exp}
	f(u, r, \tht) = \sum_{j=2}^{J_{d}} \sum_{k=0}^{K_{d}} r^{-j-\nu_{\Box}} \log^{k} (\tfrac{r}{\mbrk{u}}) \rf_{j+\f{d-1}2, k}(u, \tht) + \rem_{\alp_{d}+1}[f](u, r, \tht),
\end{equation}
where
\begin{align}
	\rf_{j+\f{d-1}2, k} &= O_{\bfGmm}^{M_{0}}(D \mbrk{u}^{(j-2)-\alp_{d} - \dlt_{d}+\nu_{\Box}}) \hbox{ in } \calM_{\ext},  \label{eq:ext-fjk-phg} \\
	\rem_{\alp_{d}+1}[f] &= O_{\bfGmm}^{M_{0}}(D r^{-1-\alp_{d}} \mbrk{u}^{-1-\dlt_{d}}) \hbox{ in } \calM_{\ext}, \label{eq:ext-f-rem}
\end{align}
and we omit the sum in \eqref{eq:ext-f-exp} if $J_{d}  < 2$.
\end{enumerate}
\begin{enumerate}[label=(E-D${}_{\calS_{t_{0}}}$)]
\item \label{hyp:ext-id}
	{\it Assumptions on the Cauchy data.} On $\set{x^{0} = t_{0}, \, r > R_{0}} \subseteq \calS_{t_{0}}$, assume that the jet of $\phi$ determined by the Cauchy data $(\phi_{0}, \dot{\phi}_{0})$ and the equation $\calP \phi = \calN + f$ obeys
\begin{equation} \label{eq:ext-id}
\sum_{I_{u} + I_{r} + \abs{I_{\bfOmg}} \leq M_{0}} \int_{\set{x^{0} = t_{0}, \, r \geq R_{0}}} \mbrk{u}^{2 (\alp_{d} - \nu_{\Box})} \abs{(\mbrk{u} \rd_{u})^{I_{u}} (r \rd_{r})^{I_{r}} \bfOmg^{I_{\bfOmg}} \phi}^{2} \, \ud \sgm \leq D^{2}.
\end{equation}
\end{enumerate}

Some remarks concerning these assumptions are in order.

\begin{remark}[Slightly strengthened assumptions near null infinity]
We have imposed slightly stronger assumptions on $\bfg$ in $\calM_{\ext}$ as compared to $\calM_{\far}$. More precisely, in \eqref{eq:ext-g-rr-phg}, \eqref{eq:ext-g-rA-phg} and \eqref{eq:ext-g-AB-phg}, we do {\bf not} allow for any $j=0$ terms in the expansion, in contrast to \eqref{eq:wave-g-rr-phg}, \eqref{eq:wave-g-rA-phg} and \eqref{eq:wave-g-AB-phg}. This is to ensure that constant $x^0$ hypersurfaces are spacelike for the weighted energy estimates in Section~\ref{sec:EE}.

We remark that these assumptions can be relaxed in the following ways. 
\begin{enumerate}
\item In \eqref{eq:ext-g-rr-phg}, \eqref{eq:ext-g-rA-phg} and \eqref{eq:ext-g-AB-phg}, it is easy to allow for $\rh_{0,0}^{rr}$, $\rsh_{0,0}^{rA}$ and $\rsh_{0, 0}^{AB}$ to be non-zero, but in addition to \eqref{eq:ext-gjk-phg}, these coefficients need to be small and satisfy 
$$	\rh_{0, 0}^{rr}, \,
	\rsh_{0, 0}^{rA}, \,
	\rsh_{0, 0}^{AB}
	 = O_{\bfGmm}^{M_{c}}(\ep_{e}A_{c} \mbrk{u}^{-\dlt_{c}})$$
	 for $\ep_{e\ast}>0$ sufficiently small. This is then sufficient for the weighted energy estimates in Section~\ref{sec:EE} without any modifications (see the conditions \eqref{eq:en-ext-coeff} needed for the energy estimates).
\item It appears to be in principle possible to even removing the smallness condition for allow for $\rh_{0,0}^{rr}, \rsh_{0,0}^{rA}, \rsh_{0, 0}^{AB}$, as well $\rh_{0,k}^{rr}$, $\rsh_{0,k}^{rA}$ terms $k\geq 1$ terms. This will however require modifications in the energy estimates; see Remark~\ref{rmk:ext.improved.coeff}.
\end{enumerate}

\end{remark}

\begin{remark}[No a priori assumptions on the solution]
In contrast to the assumption \ref{hyp:sol} in $\calM_{\far} \cup \calM_{\near}$, in the region $\calM_{\ext}$, we do not impose a priori assumptions on the solution. In particular since we do not need to deal with issues of trapping, superradiance, etc., we will carry out the analysis in $\calM_{\ext}$ to \emph{prove} the analogues of the estimates in \ref{hyp:sol}.
\end{remark}

\subsubsection{Recurrence equations in $\calM_{\ext}$}\label{sec:recurrence.ext}

In $\calM_{\ext}$, for $J = \min\{J_{c},J_{d}\}$ (with $J_{c}$, $J_{d}$ as in \ref{hyp:ext-g}--\ref{hyp:ext-forcing}), we introduce the expansion
\begin{equation}\label{eq:ext-formal.expansion}
r^{\nB} \phi = \rPhi_{0}(u, \tht) + \sum_{j=1}^{J-1} \sum_{k=0}^{\infty} r^{-j} \log^{k} (\tfrac{r}{\mbrk{u}}) \rPhi_{j, k}(u, \tht) + \rho_{J}.
\end{equation}
Recalling \eqref{eq:def.mbrku}, $\rPhi_{j,k}$ agree with the definition in \eqref{eq:formal.expansion} when $u\geq 1$. On the other hand, when $u\leq 1$, the higher radiation fields obey the following slightly modified \textbf{recurrence equations} (cf.~\eqref{eq:recurrence-jk}), which can be derived in a similar manner as in Section~\ref{subsec:recurrence-formal}:
\begin{equation} \label{eq:ext-recurrence-jk}
\begin{aligned}
	\rd_{u} \rPhi_{j, k}&=
	(k+1) \f{\rd_{u} \mbrk{u}}{u} \rPhi_{j, k+1}
	+ \frac{k+1}{j} \left( \rd_{u} \rPhi_{j, k+1} - (k+2) \f{\rd_{u} \mbrk{u}}{u} \rPhi_{j, k+2} \right)
\\
&\peq
- \frac{1}{2j}\Big( (j-1)j - \frac{(d-1)(d-3)}{4} + \rslap \Big) \rPhi_{j-1, k}   \\
&\peq
+ \frac{1}{2j} (j-1)(k+1) \rPhi_{j-1, k+1} - \frac{1}{2j}(k+2)(k+1) \rPhi_{j-1, k+2}
+ \frac{1}{2j} \rF_{j, k}.
\end{aligned}
\end{equation}
As in Lemma~\ref{lem:radiation.field}, there exists a sequence $K_{0} \leq K_{1} \leq K_{2} \leq \ldots \leq K_{\min\set{J_{c}, J_{d}}-1}$, depending only on $d$, $K_{c}$, $K_{d}$, $\calN$ and $\min\set{J_{c}, J_{d}}-1$, such that
$$\rPhi_{j,k} = \rcPhi_{j,k} = 0, \quad \hbox{if $k > K_j$}.$$

\subsubsection{Further preliminaries for the exterior stability theorem}\label{sec:ext.prelim}

Before we formulate the exterior stability theorem in Section~\ref{sec:ext.stab.thm}, we need to introduce two additional preliminary notions.

First, we define the minimal decay exponent for $\calN$ to ensure appropriate decay for the nonlinear terms. This is similar to Definition~\ref{def:alp-N-min}. As in Definition~\ref{def:alp-N-min}, the precise definition of admissible decay exponents will be postponed; see  Definition~\ref{def:ext-alp-N}.
\begin{definition}[Minimal decay exponent] \label{def:ext-alp-N-min}
In $\calM_{\ext}$, define the \emph{minimal decay exponent} $\alp_{\calN}$ \emph{for $\calN$} to be the infimum of all admissible decay exponents $\alp_{\calN}'$ for $\calN$ in $\calM_{\ext}$. If $\calN = 0$, we define $\alp_{\calN} = -\infty$.
\end{definition}

The second notion we need is a ``hyperboloidal'' foliation asymptotic to $u$, i.e., a foliation by \emph{spacelike} hypersurfaces that asymptote to constant $u$-hypersurfaces. This will allow us to localize (essentially) to $\calM_{\ext}$. More precisely, given $A_{h}, \dlt_{h} > 0$, we introduce
\begin{equation} \label{eq:ext-uh}
	u_{h} := u - \mbrk{u} \gmm_{h}(r), \quad \gmm_{h}(r) := 2 A_{h} \int_{r}^{\infty} (r')^{-1-\dlt_{h}} \, \ud r'.
\end{equation}
In $\calM_{\ext}$, $u_{h}$ will be comparable to $u$, and with appropriate choices of $A_{h}$ and $\dlt_{h}$, each level-$u_{h}$ hypersurface is spacelike (equivalently, $\ud u_{h}$ is timelike) and asymptotically null as $r \to \infty$.

\subsubsection{Exterior stability theorem}\label{sec:ext.stab.thm}

We are now ready to formulate the exterior stability theorem. 

\begin{theorem}[Exterior stability] \label{thm:ext-stab}
Let $\calP$ satisfy the assumptions \ref{hyp:ext-g}, \ref{hyp:ext-B} and \ref{hyp:ext-V}. In case $\calN \neq 0$, assume that $\calN$ satisfies \ref{hyp:ext-N} and let $\alp_{\calN}$ be its minimal decay exponent in Definition~\ref{def:ext-alp-N-min}; if $\calN = 0$, set $\alp_{\calN} = - \infty$. Define $u_{h}$ as in \eqref{eq:ext-uh} with $A_{h} \geq 2 \sup_{\calM_{\ext}} \frac{r^{1+\dlt_{h}}}{\mbrk{u}}\abs{(\bfg^{-1})^{uu}}$ and $0 < \dlt_{h} \leq \dlt_{c}$ (to ensure $\ud u_{h}$ is timelike for $r$ large) and introduce the function
\begin{equation}\label{eq:Uh.def}
	U_{h}(t, R) := (t - R + 3R_{\far}) - \mbrk{(t-R + 3R_{\far})} \gmm_{h}(R),
\end{equation}
which is the value of $u_{h}$ when $(x^{0}, r) = (t, R)$. Then there exist $R_{e\ast} > 0$ and $\eps_{e\ast}(\cdot) > 0$ depending on $A_{h}$, $\dlt_{h}$, $\dlt_{c}$ and the implicit constants in \ref{hyp:ext-g}--\ref{hyp:ext-N}; $m_{e\ast}, C_{e\ast} \in \bbZ_{\geq 0}$ depending only on $d$; and $m_{e\ast}', C_{e\ast}' \in \bbZ_{\geq 0}$, depending on $d$, $\dlt_{c}$, $\dlt_{d}$, $J_{c}$ and $J_{d}$ such that the following holds.
\begin{enumerate}
\item $\ud u_{h}$ is timelike (hence each level-$u_{h}$ hypersurface is spacelike) with respect to $\bfg$ in the region $\calM_{\ext} \cap \set{r \geq R_{e\ast}}$. Moreover, $\f 34 < \f{\mbrk{u_h}}{\mbrk{u}} < \f 54$.
\item Given a forcing term $f$ on $\calM_{\ext}$ and Cauchy data $(\phi_{0}, \dot{\phi}_{0})$ on $\set{x^{0} = t_{0}, \, r > R_{0}}$ satisfying \ref{hyp:ext-forcing}--\ref{hyp:ext-id} with
\begin{equation*}
R_{0} \geq R_{e\ast}, \quad U_{h}(t_{0}, R_{0}) \leq \frac{3}{2}, \quad \alp_{d} > \max\set{\nu_{\Box}, \alp_{\calN}}, \quad D \leq \eps_{e\ast}(R_{0}), \quad m_{e\ast} \leq M_{0} \leq M_{c},
\end{equation*}
there exists a unique solution $\phi$ to \eqref{eq:main-eq} with Cauchy data $(\phi, \bfn_{\calS_{t_{0}}} \phi) |_{\set{x^{0} = t_{0}, \, r \geq R_{0}}} = (\phi_{0}, \dot{\phi}_{0})$ in $\set{x^{0} \geq t_{0}, \, u_{h} \leq U_{h}(t_{0}, R_{0})}$ satisfying
\begin{equation} \label{eq:ext-stab-en}
\phi = O_{\bfGmm}^{M_{0}-C_{e\ast}}(D \mbrk{u}^{-\alp+\nu_{\Box}} r^{-\nu_{\Box}}),
\end{equation}
where $\alp$ is defined by
\begin{equation}\label{eq:ext-alp.def}
\alp = \alp_d = J_{d}-1+\eta_{d}+\nu_{\Box},
\end{equation}
and the implicit constant depends on $d$, $A_{h}$, $\dlt_{h}$, $\dlt_{c}$, $\dlt_{d}$, $\alp$, $\alp_{\calN}$ and the implicit constants in \ref{hyp:ext-g}--\ref{hyp:ext-N}.

\item If, in addition, we have
\begin{equation*}
	m_{e\ast}' \leq M_{0} \leq M_{c},
\end{equation*}
then the solution $\phi$ in (2) has the following structure in $\set{x^{0} \geq t_{0}, \, u_{h} \leq U_{h}(R_{0})}$:
\begin{equation} \label{eq:ext-stab-exp}
	\phi = r^{-\nu_{\Box}} \left( \sum_{j = 0}^{\min\set{J_{d}, J_{c}}-1} r^{-j} \log^{k} (\tfrac{r}{\mbrk{u}}) \rPhi_{j, k}(u, \tht) + \rho_{\min\set{J_{d}, J_{c}}}(u, r, \tht) \right),
\end{equation}
where
\begin{align}
	\rPhi_{j, k} &= O_{\bfGmm}^{M_{0}-C_{e\ast}'}(D \mbrk{u}^{j - \min\set{J_{c}-1+\eta_{c}, J_{d}-1+\eta_{d}}}), \label{eq:ext-stab-rphijk} \\
	\rho_{\min\set{J_{d}, J_{c}}} &= O_{\bfGmm}^{M_{0}-C_{e\ast}'}(D r^{-\min\set{J_{c}-1+\eta_{c}, J_{d}-1+\eta_{d}}} \log^{K_{\min\set{J_{d}, J_{c}}}}(\tfrac r{\mbrk{u}}))\log r. \label{eq:ext-stab-rem}
\end{align}
where the implicit constants depend on $d$, $A_{c}$, $A_{h}$, $\dlt_{h}$, $\dlt_{c}$, $J_{c}$, $J_{d}$, $\eta_{c}$, $\eta_{d}$ and the implicit constants in \ref{hyp:ext-N}, and $C_{e\ast}'$ depends on $d, \dlt_{c}, \dlt_{0}, J_{c}$ and $J_{d}$. When $\eta_{c},\,\eta_{d} <1$,  we can moreover omit the factor $\log^{K_{\min\set{J_{d}, J_{c}}}}(\tfrac r{\mbrk{u}}) \log r$ in \eqref{eq:ext-stab-rem}.
\end{enumerate}

\end{theorem}
Some remarks concerning the theorem are in order.
\begin{remark}[Asymptotics of $\rPhi_{j, k}$ as $u \to -\infty$]
Observe that, under our hypotheses, $\rPhi_{j, k} \to 0$ as $u \to - \infty$ for $j \leq \min\set{J_{d}, J_{c}} - 1$. If instead of \eqref{eq:ext-id} we assume that the Cauchy data has a polyhomogeneous expansion in $r$, then these limits may be nontrivial. However, a complete result in this regard would require another set of assumptions for the coefficients, nonlinearity and the forcing term near $i^{0}$ (expansion in terms of $\frac{1}{(-u)}$ and $\frac{1}{r}$). We will refrain from considering this setting in this paper. We refer to \cite{HintzVasyMinkStab} for related ideas.
\end{remark}
\begin{remark}
The statements for $\phi$ concern a subregion of $\calM_{\ext}$ where $r$ is bounded from below by a sufficiently large number $R_{0}$. In practice, this lower bound can be ensured by either fixing $t_{0}$ and taking $R_{0}$ large (which is the case in Main Theorem~\ref{thm:Cauchy}), or fixing $U_{h}(t_{0}, R_{0})$ and making $t_{0}$ large (which is useful when we wish to construct the solution essentially up to some finite $u$).
\end{remark}

\subsubsection{Main theorem for initial data on asymptotically flat hypersurface}\label{sec:thm.Cauchy}

We now state our main theorem for initial data prescribed on the asymptotically flat hypersurface $\calS_{1}$. In this case, we prove that when the condition \ref{hyp:id} on $\Sigma_{1}$ is not assumed, it can be derived as a consequence of the estimates in $\calM_{\ext}$ established in Theorem~\ref{thm:ext-stab}. As a result, Main Theorem~\ref{thm:upper}, Main Theorem~\ref{thm:lower}, Main Theorem~\ref{thm:upper-sphsymm} and Main Theorem~\ref{thm:lower-sphsymm} are all extend to hold with initial data $\calS_{1}$.

The main result is given in Theorem~\ref{thm:Cauchy} below. We note that due to the application of Theorem~\ref{thm:ext-stab}, we will need a smallness condition which amounts to taking $R_{\far}$ sufficiently large (so that the coefficients are small) and taking the data (and inhomogeneous terms) to be small when $r \geq R_{\far}$.

\begin{maintheorem}\label{thm:Cauchy}
Suppose that other than the assumption \ref{hyp:id}, all the assumptions of Main Theorem~\ref{thm:upper} hold. Make the following assumptions in $\calM_{\ext}$:
\begin{enumerate}
\item Instead of \ref{hyp:id}, impose Cauchy data on $\calS_{1}$ which satisfy the assumption \ref{hyp:ext-id} with $t_0 = 1$ and $R_{0} = 3 R_{\far} - \f12$. 
\item In $\calM_{\ext}$, let $\calP$ satisfy the assumptions \ref{hyp:ext-g}, \ref{hyp:ext-B} and \ref{hyp:ext-V}. In case $\calN \neq 0$, assume that $\calN$ satisfies \ref{hyp:ext-N} and let $\alp_{\calN}$ be its minimal decay exponent in Definition~\ref{def:ext-alp-N-min}; if $\calN = 0$, set $\alp_{\calN} = - \infty$. 
\item Assume that the forcing term $f$ on $\calM_{\ext}$ satisfies \ref{hyp:ext-forcing}.
\end{enumerate}
Let $A_{h}$ be a constant satisfying $A_{h} \geq 2 \sup_{\calM_{\ext}} \frac{r^{1+\dlt_{h}}}{\mbrk{u}}\abs{(\bfg^{-1})^{uu}}$ and $0 < \dlt_{h} \leq \dlt_{c}$. Let $R_{e\ast} >0$ and $m_{e\ast}', C_{e\ast}' \in \bbZ_{\geq 0}$ be the constants as in Theorem~\ref{thm:ext-stab} (which depend on $A_{h}$, $\de_{h}$, $d$, $\dlt_{c}$, $\dlt_{d}$, $J_{c}$, $J_{d}$ and constants in \ref{hyp:ext-g}, \ref{hyp:ext-B} and \ref{hyp:ext-V}). Then the following hold:
\begin{enumerate}
\item Assuming \begin{equation*}
3 R_{\far} - \f12\geq R_{e\ast}, \quad \alp_{d} > \max\set{\nu_{\Box}, \alp_{\calN}}, \quad D \leq \eps_{e\ast}(R_{0}), \quad m'_{e\ast} \leq M_{0} \leq M_{c}',
\end{equation*}
then the condition \ref{hyp:id} holds (with $M$ in \ref{hyp:id} replaced by $M_{0} - C_{e\ast}'$ in the conclusion of part (3) of Theorem~\ref{thm:ext-stab}.).
\item As a result of (1), the conclusion of Main Theorem~\ref{thm:upper} also holds. In this setting, the higher radiation fields $\rPhi_{j,k}$ are to be computed with the formal recurrence equations \eqref{eq:ext-recurrence-jk} for all $u \in \mathbb R$, with $\lim_{u\to -\infty} \rPhi_{j,k}(u,\th) = 0$ for all $0\leq k \leq K_j$, $0\leq j \leq J_{\frkf}$.
\item Similar statements as in part (2) also hold for Main Theorem~\ref{thm:lower}, Main Theorem~\ref{thm:upper-sphsymm} and Main Theorem~\ref{thm:lower-sphsymm}.
\end{enumerate}
\end{maintheorem}

It is easy to check that Theorem~\ref{thm:Cauchy} follows immediately from an application of Theorem~\ref{thm:ext-stab} with $t_0 = 1$ and $R_{0} =3 R_{\far} - \f12$. Indeed, to check the theorem, it suffices to prove that \ref{hyp:id} holds. The needed bounds on $\Sigma_{1} \cap \{u_h \leq U_h(1, 3 R_{\far} - \f12) \}$ is given by Theorem~\ref{thm:ext-stab}. On the other hand, note that the region $\Sigma_{1} \setminus \{u_h \leq U_h(1, 3 R_{\far} - \f12) \}$ has finite-$r$ range since by $\f 34 < \f{\mbrk{u_h}}{\mbrk{u}} < \f 54$, we know that $\{ u = 1\} \subset \{u_h \leq \frac{5}{4} \} \subset \{u_h \leq U_h(1, 3 R_{\far} - \f12) \}$ (for $R_{\far}$ large enough). Therefore, the needed estimates on $\Sigma_{1} \setminus \{u_h \leq U_h(1, 3 R_{\far} - \f12) \}$ is already a consequence of \ref{hyp:sol}.

\section{Examples}\label{sec:examples}

In this section, we give some examples for which our main theorems apply. These examples serve to illustrate our assumptions.

We have three different groups of examples. In \textbf{Section~\ref{sec:classical.ex}}, we consider classical examples of wave equations on $\mathbb R^{d+1}$, including the obstacle problem. In \textbf{Section~\ref{sec:black.holes.ex}}, we give examples of linear wave equations on black hole backgrounds. Finally, in \textbf{Section~\ref{sec:nonlinear.ex}}, we give examples of nonlinear (and possibly quasilinear) wave equations on Minkowski spacetime.

In what follows, we will describe the examples and briefly go through the proof that all assumptions are satisfied. We will only consider initial data satisfying the assumption \ref{hyp:id}. In the examples below, we also set the inhomogeneous terms $f \equiv 0$ so that \ref{hyp:forcing} is vacuous (though one could  add in inhomogeneous terms for most of these equations). We will not discuss \ref{hyp:id} and \ref{hyp:forcing} further below.

\subsection{Classical examples}\label{sec:classical.ex}

\begin{example}\label{ex:classical}(Wave equations on Minkowski spacetime with lower order terms)
    Consider the following wave equation on $((0,\infty)\times \mathbb R^{d},\bfm)$ with lower order terms:
    \begin{equation}\label{eq.wave.with.lower.order}
    (\bfm^{-1})^{\alp\bt} \bfD_\alp \rd_\bt \phi + \bfB^\alp \rd_\alp \phi + V\phi =0,
    \end{equation}
    where $\bfB$ and $V$ are stationary, i.e., $\calL_{\bfT} \bfB = 0$, $\bfT V = 0$, where $\bfT = \rd_t$ in the $(t,x^1,\ldots,x^d)$ coordinate system.

    In this case, we claim that our assumptions can be reduced to the following three conditions:
    \begin{enumerate}
    	\item $\bfB$ and $V$ satisfy the decay assumptions in \ref{hyp:T-almost-stat}, \ref{hyp:med}, \ref{hyp:wave-B}, \ref{hyp:wave-V}, \ref{hyp:ext-B} and \ref{hyp:ext-V}.
	\item The elliptic operator $\calP_0 = \Delta_{\bfe} + \bfB^\alp \rd_\alp + V$ satisfies the uniqueness statement in part (b) of \ref{hyp:ult-stat}.
	\item For any $C^\infty_c$ initial data and any $M \in \mathbb N$, there exists $A_M$ such that the solution $\phi$ to \eqref{eq.wave.with.lower.order} satisfies
	\begin{equation}
	\phi = O_{\bfGmm}^M(A_M \brk{r}^{-1}).
	\end{equation}
    \end{enumerate}
    One very special case for which (2)--(3) hold is when $\bfB$ and $V$ satisfy a global $C^1$ smallness assumption:
    $$\sum_{i+|I| \leq 1} \sup_{x\in \mathbb R^{d}} \Big( \brk{r} |\calL_{\brk{r} \rd_r}^i \calL_{\bfOmg}^I \bfB| + \brk{r}^2 | (\brk{r} \rd_r)^i \bfOmg^I V| \Big) \leq \ep_{d},$$
    for a suitably small constant $\ep_d$ depending only on the dimension $d$, where $\bfOmg$ are the standard rotation vector fields (see \eqref{eq:Klainerman}).

    More generally, it is possible for conditions (2) and (3) to hold for large $\bfB$ and $V$. The condition (2) is often called the \emph{no-eigenvalue/resonance-at-zero-energy condition}. We remark that the condition (3) is well-known to follow as a consequence if energy boundedness and an integrated local energy decay estimate (see Lemma~\ref{lem:iled-flat}) hold. For \eqref{eq.wave.with.lower.order}, these estimates are in turn known to be guaranteed by certain spectral properties of $\calP_0$. We refer the reader to \cite{MST} for such results.

    We now show that that our assumptions can indeed be reduced to the conditions (1)--(3) above. Since we are considering a linear homogeneous equation with stationary coefficients on the Minkowskian spacetime, the assumptions \ref{hyp:topology}--\ref{hyp:T-tau}, \ref{hyp:wave-g}, \ref{hyp:bdry.cond}, \ref{hyp:nonlin}--\ref{hyp:nonlin-null} are vacuous. On the other hand, the assumptions \ref{hyp:T-almost-stat}, \ref{hyp:med}, \ref{hyp:wave-B}, \ref{hyp:wave-V}, \ref{hyp:ext-B}, \ref{hyp:ext-V} and \ref{hyp:sol} are immediate consequence of the conditions (1) and (3) above.

    It thus remains to check \ref{hyp:ult-stat}. Since the operator is stationary, we only need to check the conditions (a) and (b) in \ref{hyp:ult-stat} for $\calP_0 = \calP_0^{(\infty)}$. The condition (b) is assumed (in condition (2)); we claim that it also implies condition (a) by Fredholm alternative. Indeed, by the decay assumptions of $\bfB$ and $V$, it is easy to show the following elliptic estimates for any $s \in \mathbb Z_{\geq 0}$:
    $$\| \phi \|_{H^{s+2}}  \ls \| \calP_0 \phi\|_{H^s} + \|\phi \|_{L^2(\brk{r}^{-2} \ud x)}.$$
    Since for any $s \in \mathbb Z_{\geq 0}$, $L^2(\brk{r}^{-2} \ud x)$ embeds compactly into $H^{s+2}$, standard Fredholm alternative shows that $\calP_0$ is invertible with an inverse $\calR_0$ if and only if the kernel of $\calP_0$ is trivial in $H^{s+2}$. Moreover, in this case, the following estimate holds.
    \begin{equation}\label{eq:zero-res-est.in.basic.case}
    \| \calR_0 \phi \|_{H^{s+2}}  \ls \| \phi\|_{H^s} .
    \end{equation}
    Noticing that \eqref{eq:zero-res-est.in.basic.case} is stronger than the estimate \eqref{eq:zero-res-est} we need, it remains to rule out the possibility of a nontrivial kernel of $\calP_0$ in $H^{s+2}$. Suppose for the sake of contradiction that there is $\phi_0 \in H^{s+2} \setminus \{0\}$ such that $\calP_0 \phi_0 = 0$. Then standard elliptic theory gives that $|\phi_0| = O(\brk{r}^{-(d-2)})$ which then contradicts condition (2) above.

\end{example}

\begin{example}\label{ex:product}(Product Lorentzian manifold)
    Consider a product metric $\bfg = -\ud t^2 + \bfg_0$ on $(0\infty)_t\times \mathbb R^d$, where $(\mathbb R^d,\bfg_0)$ is a smooth asymptotically flat Riemannian manifold. Assume that there exists $R_0>0$ such that the Riemannian metric $\bfg_0$ satisfies the following condition\footnote{We use the notation $\overbullet{\bfh}$ here so that it is consistent with Section~\ref{sec:renormalized.higher.radiation.fields}; this expansion is more natural in view of the stationarity assumption. The same comment applies to Proposition~\ref{prop:longrange}. See also Section~\ref{sec:price}.}:
    \begin{enumerate}
    \item $$\bfg_0^{-1}(r,\th) = \bfe^{-1} + r^{-1} \overbullet{\bfh}_{\mathrm{sym}}(\th) + \sum_{j=2}^{J_{c}}\sum_{k=0}^{K_{c}} r^{-j} \log r \overbullet{\bfh}_{j,k}(\th) + \rem_{J_{c}+\eta_{c}}[\bfh] \quad \hbox{for $|x| \geq R_0$}, $$\label{hyp:product.ex.1}
    \end{enumerate}
    where $\bfe$ is the Euclidean metric on $\mathbb R^d$, $\bfh_{\mathrm{sym}}$ is spherically symmetric, i.e., in the coordinates \eqref{eq:far-r-u-tht}, $\bfh_{\mathrm{sym}} = \mathfrak a \rd_r\otimes \rd_r + \mathfrak b r^{-2} \mathring{\slashed{\bfgmm}}^{-1}$ for some constants $\mathfrak a, \mathfrak b\in \mathbb R$, and the remainder satisfies the following bounds in the coordinate system $(x^1, \ldots, x^d)$:
    $$ \rem_{J_{c}+\eta_{c}}[\bfh] = O_{\bfGmm}^{M_{c}} (r^{-(J_{c}+\eta_{c})}).$$


    Consider the linear wave equation
    \begin{equation}\label{eq:wave.on.product.mfld}
    \Box_{\bfg} \phi = 0
    \end{equation}
    on $(\calM = (0,\infty)_t \times \mathbb R^d, \bfg = -\ud t^2 + \bfg_0)$. Assume, in addition to \eqref{hyp:product.ex.1}, that the following conditions hold:
    \begin{enumerate}\setcounter{enumi}{1}
    \item The elliptic operator $\calP_0 = \Delta_{\bfg_0}$ satisfies the uniqueness statement in part (b) of \ref{hyp:ult-stat}, where $\Delta_{\bfg_0}$ denotes the Laplace--Beltrami operator associated to the Riemmann metric $\bfg_0$.\label{hyp:product.ex.2}
	\item \label{hyp:product.ex.3} For any $C^\infty_c$ initial data and any $M \in \mathbb N$, there exists $A_M$ such that the solution $\phi$ to \eqref{eq:wave.on.product.mfld} satisfies
	\begin{equation}
	\phi = O_{\bfGmm}^M(A_M \brk{r}^{-1}).
	\end{equation}
    \end{enumerate}

    We claim that if \eqref{hyp:product.ex.1}, \eqref{hyp:product.ex.2}, \eqref{hyp:product.ex.3} above hold, then all the assumptions of our main theorems hold and our main theorems are applicable. Indeed, it can be seen that except for \ref{hyp:T-almost-stat}, \ref{hyp:med}, \ref{hyp:wave-g} and \ref{hyp:ext-g} (concerning the estimates for $\bfg$), the remaining assumptions can be checked in a manner similar to the Example~\ref{sec:classical.ex}, including using Fredholm alternative; we omit the details.

    It thus remains to check that the conditions \ref{hyp:T-almost-stat}, \ref{hyp:med}, \ref{hyp:wave-g} and \ref{hyp:ext-g} follow from the condition \eqref{hyp:product.ex.1}. First notice that if we naively choose $u = t-r$, $r = |x|$ and $\th = \f xr$, then the conditions \eqref{eq:wave-g-uu-phg} and \eqref{eq:ext-g-uu-phg} fail and the expansion has an additional $j=1$ term. In order to guarantee these conditions, we need to introduce a different coordinate system in $\calM_\wave$. This relies on the $r^{-1}$ part in \eqref{hyp:product.ex.1} being both stationary and spherically symmetric. We prove this as a consequence of a more general fact in Proposition~\ref{prop:longrange}.

    Again as in Example~\ref{ex:classical}, (2) and (3) hold if $\bfg_0$ is globally close to the flat metric. More generally, (3) would follow from energy boundedness and integrated local energy decay estimates. In the case the $\bfg$ is non-trapping, there are results which reduce these estimates to spectral conditions; we again refer the reader to \cite{MST} for details.
\end{example}

In Example~\ref{ex:product} above, we have referred to Proposition~\ref{prop:longrange}, which allows us to introduce a change of variables to remove the $r^{-1}$ part of $(\bfg^{-1})^{uu}$.
\begin{proposition}\label{prop:longrange}
(Renormalization of the $r^{-1}$-term in the metric in the stationary case) %
Assume that $(\bfg, \bfB, V)$ is stationary, in the sense that $\calL_{\rd_{u}} \bfg = 0$, $\calL_{\rd_{u}} \bfB = 0$ and $\rd_{u} V = 0$ in $\calM_{\wave} \cup \calM_{\ext}$, and satisfies \ref{hyp:med}--\ref{hyp:wave-V} in $\calM_{\far}$ and \ref{hyp:ext-g} except for \eqref{eq:wave-g-uu-phg} and \eqref{eq:ext-g-uu-phg}. Instead, the $(\bfg^{-1})^{uu}$ component admits the following expansion
\begin{equation} \label{eq:wave-g-uu-phg-longrange} \tag{\ref{eq:wave-g-uu-phg}$'$}
(\bfg^{-1})^{uu} = \rch_{1, 0}^{uu} r^{-1}  + \sum_{j = 2}^{J_{c}} \sum_{k=0}^{K_{c}} \rch_{j, k}^{uu} (\tht) r^{-j} \log^{k} r + \rem_{J_{c}+\eta_{c}}[\bfh^{uu}](r, \tht),
\end{equation}
where $\rch_{1, 0}^{uu}$ is \emph{independent} of $u$ and $\tht$, the assumptions \eqref{eq:wave-gjk-phg} and \eqref{eq:ext-gjk-phg} on $\rch_{j,k}^{u,u}$ hold, and the assumptions \eqref{eq:wave-grem-uu-phg}, \eqref{eq:ext-grem-uu-phg} on the remainder hold. Define, in $\calM_{\far}$,
\begin{equation} \label{eq:u-ast-longrange}
	u^{\ast} = u + \rcu^{\ast}_{0, 1} \log r, \quad
	\hbox{ where } \rcu^{\ast}_{0, 1} = -\frac{1}{2} \rch_{1, 0}^{uu}.
\end{equation}
Then with respect to the coordinates $(u^{\ast}, r, \tht)$, we have $\rd_{u^{\ast}} = \rd_{u}$ and $(\bfg, \bfB, V)$ satisfies \ref{hyp:med}--\ref{hyp:wave-V} and \ref{hyp:ext-g} with the same $J_{c}$, $K_{c}$, $M_{c}$, $\dlt_{c}$ and $\eta_{c}$.
\end{proposition}
\begin{proof}
Proposition~\ref{prop:longrange} is a consequence of the following computations:
\begin{align*}
\ud u^{\ast} &= \ud u + \rcu^{\ast}_{0, 1}(\tht) r^{-1} \ud r, \\
(\bfg^{-1})^{u^{\ast} u^{\ast}} &=
(\bfg^{-1})^{uu} + 2 (\bfg^{-1})^{u r} \rcu^{\ast}_{0, 1} r^{-1} + (\bfg^{-1})^{r r} (\rcu^{\ast}_{0, 1} r^{-1})^{2}, \\
(\bfg^{-1})^{u^{\ast} r} &= (\bfg^{-1})^{ur} + (\bfg^{-1})^{rr} \rcu^{\ast}_{0, 1} r^{-1},  \\
(\bfg^{-1})^{u^{\ast} A} &= (\bfg^{-1})^{u A} + (\bfg^{-1})^{r A} \rcu^{\ast}_{0, 1} r^{-1}, \\
\bfB^{u^{\ast}} &= \bfB^{u} + \bfB^{r} \rcu^{\ast}_{0, 1} r^{-1}.
\end{align*}
In particular, observe that \eqref{eq:u-ast-longrange} ensures that $(\rch_{1, 0}^{uu} + 2 (\bfm^{-1})^{u r} \rcu^{\ast}_{0, 1}) r^{-1}$ in $(\bfg^{-1})^{u^{\ast} u^{\ast}}$ cancels. To verify that the higher order vector field bounds hold, note that
\begin{align*}
\rd_{u^{\ast}} = \rd_{u}, \quad
\rd_{r} = \rd_{r} - \rcu_{0, 1}^{\ast} r^{-1} \rd_{u}, \quad
\rd_{\tht^{A}} = \rd_{\tht^{A}},
\end{align*}
where the partial derivatives on the left- and right-hand sides are computed in the $(u^{\ast}, r, \tht)$ and the $(u, r, \tht)$ coordinates, respectively.
\end{proof}

\begin{example}\label{ex:obstacle}(Obstacle problems)
    Our assumptions are set up so that we can consider analogues of Example~\ref{ex:classical} and Example~\ref{ex:product}, but with an obstacle.

    Let $\mathcal O \subset \mathbb R^d$ be a bounded open set with a smooth boundary. Consider the following obstacle problem on $(0,\infty)_t \times (\mathbb R^d \setminus \mathcal O)$:
    \begin{equation}
    (\bfg^{-1})^{\alp \bt} \bfD_{\alp} \rd_{\bt} \phi + \bfB^{\alp} \rd_{\alp} \phi  + V \phi = 0,\quad \phi|_{\mathbb R\times \mathcal O} = 0,
    \end{equation}
    where $\bfg$, $\bfB$ and $V$ are as in Example~\ref{ex:classical} and Example~\ref{ex:product}, i.e., they are stationary and obey the conditions (1)--(3) in Example~\ref{ex:classical} and Example~\ref{ex:product}, except that now condition (2) is to be understood with a nontrivial boundary. Noticing that our main theorems allow for this type of domains and boundary conditions (see \ref{hyp:bdry} and \ref{hyp:bdry.cond}), it follows from similar considerations as before that our main theorems apply to this setting.

    (We note that for the classical case of $\Box_{\bfm} \phi = 0$ with $\phi|_{\mathbb R\times \mathcal O} = 0$, then in fact there is \emph{no} tail, i.e., the solutions decay faster than any polynomial, as long as the hypothesis \ref{hyp:sol} and the uniqueness statement in part (b) of \ref{hyp:ult-stat}. This is a very classical and well-studied problem. In full generality \ref{hyp:sol} is difficult to verify and may not hold. However, in many cases in fact exponential decay is known; see for instance \cite{pdLcsMrsP1963, LaxPhillipsBook, rbM1979, cM1975}.)

\end{example}

\begin{example}(Dynamical examples)
Our assumptions are exactly designed so that exact stationarity is not needed. In particular, all of Examples~\ref{ex:classical}, \ref{ex:product}, \ref{ex:obstacle} can be replaced by equations with \emph{dynamical} coefficients which settle down to these stationary models. We only formulate the dynamical version of Example~\ref{ex:classical}, but Examples~\ref{ex:product} and \ref{ex:obstacle} can be treated analogously.

Consider again the following equation from Example~\ref{ex:classical}:
 \begin{equation}\label{eq.wave.with.lower.order.dynamical}
    (\bfm^{-1})^{\alp\bt} \bfD_\alp \rd_\bt \phi + \bfB^\alp \rd_\alp \phi + V\phi =0,
    \end{equation}
but no longer assume that the coefficients are stationary. Instead assume that
\begin{enumerate}
\item $\bfB$ and $V$ satisfy the decay assumptions in \ref{hyp:T-almost-stat}, \ref{hyp:med}, \ref{hyp:wave-B} and \ref{hyp:wave-V} and that they approach stationarity as $\tau \to \infty$ in the sense that \eqref{eq:T-almost-stat-limit} and \eqref{eq:T-almost-stat-2} hold for some limiting $^{(\infty)} \bfB$ and $^{(\infty)}V$.
\item For the stationary limits $^{(\infty)} \bfB$ and $^{(\infty)}V$ as in \eqref{eq:T-almost-stat-limit}, the elliptic operator $\calP_0 = \Delta_{\bfe} + \bfB^\alp \rd_\alp + V$ satisfies the uniqueness statement in part (b) of \ref{hyp:ult-stat}.
\item For any $C^\infty_c$ initial data and any $M \in \mathbb N$, there exists $A_M$ such that the solution $\phi$ to \eqref{eq.wave.with.lower.order.dynamical} satisfies
	\begin{equation}
	\phi = O_{\bfGmm}^M(A_M \brk{r}^{-1}).
	\end{equation}
\end{enumerate}

We reiterate (see Section~\ref{sec:intro.anomalous.etc}) that in the dynamical setting, the Price law decay rates in Theorem~\ref{thm:price} need not hold. Indeed, even in the case where the coefficients are conformally regular near null infinity, one can construct examples for which the weaker decay estimates in Theorem~\ref{thm:decay.least} are sharp.

\end{example}


%

\subsection{Black hole spacetimes}\label{sec:black.holes.ex}

Next, we discuss some examples of black hole spacetimes for which our theorems apply. These include well-known explicit stationary Kerr--Newman solutions to the Einstein--Maxwell equations. We note that in $(3+1)$ dimensions, the precise decay rates on these explicit stationary black hole spacetimes have been very well-studied, see the discussions in Section~\ref{sec:related.works}. Our theorems also apply to dynamical black holes; see Example~\ref{ex:Kerr.dynamical}--Example~\ref{ex:higher.d.BH.dynamical}.

    We first consider the linear wave equation on exterior region of the Schwarzschild black hole. Even though it will follow as a subcase of Example~\ref{ex:Kerr-Newman}, it is useful to first consider this algebraically simpler situation.
\begin{example}(Schwarzschild black holes)\label{ex:Schwarzschild}
	Let $M>0$. Define $\calM_{\mathrm{Schw}} = \mathbb R_{t}\times (2M,\infty)_r \times \mathbb S^2_{(\vartheta,\varphi)}$. Define the Schwarzschild metric $\bfg_{0,0,M}$ on $\calM_{\mathrm{Schw}}$ by
    \begin{equation}
    	\bfg_{0,0,M} = -\Big(1-\f{2M}r \Big) \, \ud t\otimes \ud t + \Big( 1-\f{2M}r \Big)^{-1} \, \ud r\otimes \ud r + r^2 \rsgmm,
    \end{equation}
    where as before $\rsgmm$ is the round metric on $\mathbb S^2$.
    We consider the linear scalar wave on Schwarzschild
    \begin{equation}
    \Box_{\bfg_{0,0,M}} \phi = 0.
    \end{equation}

    In order to verify the assumptions of our main theorems, we need to introduce two changes of coordinates. Introduce a new coordinate function $t^*$ by
    \begin{equation}
    	t^*(t,r) = t+ \chi_{<3M}(r)2M \log (r-2M).
    \end{equation}
    We note that for $r \leq 3M$, the metric in the $(t^*,r,\vartheta,\varphi)$ coordinates takes the form
    \begin{equation}
    	\bfg_{0,0,M} = -\Big(1-\f{2M}r \Big) \, \ud t^* \otimes \ud t^* +   \f{2M}{r} (\ud t^*\otimes \ud r + \ud r^*\otimes \ud t) + \Big( 1 + \f{2M}r \Big) \ud r \otimes \ud r + r^2 \rsgmm.
    \end{equation}
    In this coordinate system, we can extend the metric smoothly up to $r=2M$. We introduce a further change and define $t_*$ by
    \begin{equation}\label{eq:t_*.Schwarzschild}
    	t_*(t,r) = t^*(t,r) - \chi_{>12M(r)}2M\log (r-2M),
    \end{equation}
    and define $u = t_* -r$. 
    When $r \geq 12M$, a similar computation as above shows that the metric takes the form
    \begin{equation}\label{eq:Schw.large.r}
    	\begin{split}
    		\bfg_{0,0,M} = &\: -\Big(1-\f{2M}r \Big) \, \ud t_* \otimes \ud t_* -   \f{2M}{r} (\ud t^*\otimes \ud r + \ud r\otimes \ud t^*) + \Big( 1 + \f{2M}r \Big) \ud r \otimes \ud r + r^2 \rsgmm \\
		= &\: -\Big(1-\f{2M}r \Big) \, \ud u \otimes \ud u - (\ud u \otimes \ud r + \ud r \otimes \ud u) + r^2 \rsgmm.
	\end{split}
    \end{equation}

    Define now the manifold with boundary $\calM_{0,0,M}$  by $\calM_{0,0,M}= (0,\infty)_{t_*} \times [2M, \infty)_r \times S^2_{(\vartheta,\varphi)}$. Introduce a diffeomorphism $\calM_{0,0,M} \to (0,\infty) \times (\mathbb R^3 \setminus B(0,2M))$ via $x^0 = t_*$, $x^1 = r \sin \vartheta \cos \varphi$, $x^2 = r \sin \vartheta \sin \varphi$, $x^3 = r \cos \vartheta$. Let $\bfg_{0,0M}$ be the smooth extension of the metric above to the manifold with boundary $\calM_{0,0,M}$.

    We can now check that $(\calM_{0,0,M}, \bfg_{0,0M})$ obeys all the assumptions. First, $\calM_{0,0,M}$ satisfies \ref{hyp:topology} (for $R_{\far} = \max\{3M, 1\}$, say) with the diffeomorphism defined above. The boundary $\rd M = \{ r = 2M\}$ is a null hypersurface so that \ref{hyp:bdry} satisfied. Moreover, the causality conditions in \ref{hyp:T-tau} hold; indeed, $\bfg^{-1}(\ud t, \ud t) = \begin{cases} -(1+\f{2M}r) & r\leq 3M \\ -(1-\f{2M}r)^{-1} & r \geq 6M \end{cases}$ so that $-C_c < \bfg^{-1}(\ud t, \ud t) < - C_c^{-1}$.

    Regarding the necessary bounds on $\bfg$, the assumptions \ref{hyp:T-almost-stat} and \ref{hyp:med} can be easily read off. The assumption \ref{hyp:wave-g} and \ref{hyp:ext-g} also hold; notice that the change of variables \eqref{eq:t_*.Schwarzschild} is designed exactly so that \eqref{eq:wave-g-uu-phg} and \eqref{eq:ext-g-uu-phg} are satisfied. Indeed, if we were to take $u^* = t^* -r$, then there would be an additional $r^{-1}$ term in \eqref{eq:wave-g-uu-phg}. Nonetheless, using \eqref{eq:Schw.large.r}, one computes that
    \begin{equation}
    	\bfg_{0,0,M}^{-1} = \Big(1-\f{2M}r \Big) \rd_r \otimes \rd_r + \rd_r \otimes \rd_u + \rd_u \otimes \rd_r + r^{-2} \rsgmm^{-1}
    \end{equation}
    so that the exact Bondi--Sachs condition \eqref{eq:bondi-sachs} holds. In particular, the conditions \ref{hyp:wave-g} and \ref{hyp:ext-g} hold. (Let us note that in view of the spherical symmetry and stationarity of $\bfg_{0,0,M}$, we know a change of variables can be introduced to remove the $r^{-1}$ term in $(\bfg^{-1})^{uu}$ using Proposition~\ref{prop:longrange}. The choice of $t_*$ gives an explicit such change of variables; a closer look at the proof of Proposition~\ref{prop:longrange} shows that it is in fact the same choice.)

    Finally, since the boundary is null, we need not impose boundary conditions, i.e., \ref{hyp:bdry.cond} is vacuous. As for the condition \ref{hyp:ult-stat}, since the metric is stationary, \eqref{eq:T-almost-stat-limit}--\eqref{eq:T-almost-stat-2} are vacuous. For the stationary estimate, observe that $\calP_0$ takes the form
    \begin{equation}\label{eq:calP0.Schwarzschild}
    \calP_0 \phi = \f 1{r^2} \rd_r \Big( (r^2-2Mr)\rd_r\phi \Big) + \f 1{r^2} \rslap \phi.
    \end{equation}
    This is a degenerate elliptic equation with a degeneration at $r = 2M$, but its invertibility properties follows as a consequence of the results in \cite{AAGKerr}.
\end{example}

\begin{example}(Kerr--Newman black holes in the full sub-extremal range)\label{ex:Kerr-Newman}
	For $(a,Q,M)$ satisfying $a^2 + Q^2 < M^2$ and $M>0$, we first consider the manifold $\calM_{KN} = \mathbb R_t \times (r_+,\infty)_r \times \mathbb S^2_{(\vartheta,\varphi^*)}$, where $r_+ = M + \sqrt{M^2 - (a^2+Q^2)}$ is the larger root of $r^2+a^2+Q^2-2Mr$.

	In what is known as the Boyer--Lindquist coordinates, the Kerr--Newman metric on $\calM_{KN}$ takes the following form.
	\begin{equation}\label{BL}
\begin{split}
\bfg_{a,Q,M}=&-(1-\frac {2Mr-Q^2}{\Sigma})\, \ud t\otimes \ud t+\frac{\Sigma}{\Delta}\, \ud r\otimes \ud r+\Sigma\, \ud \vartheta\otimes \ud \vartheta\\
&+ R^2\sin^2\vartheta\, \ud \varphi\otimes \ud \varphi-\frac{a(2Mr -Q^2)\sin^2\vartheta}{\Sigma}(\ud \varphi\otimes \ud t+\ud t\otimes \ud \varphi),
\end{split}
\end{equation}
where
$$\Sigma=r^2+a^2\cos^2\vartheta,\quad R^2=r^2+a^2+\frac{a^2(2Mr-Q^2)\sin^2\vartheta}{\Sigma},\quad\Delta=r^2+a^2+Q^2-2Mr$$
and $t\in \mathbb R$, $r\in (r_+,\infty)$, $\vartheta\in (0,\pi)$ and $\varphi\in [0,2\pi)$.

We consider the linear scalar wave equation on $(\calM_{KN}, \bfg_{a,Q,M})$:
\begin{equation}
\Box_{g_{a,Q,M}} \phi = 0.
\end{equation}

First, introduce the Kerr star coordinates, defined by
$$t^* = t + \chi_{<3M}(r) \bar{t}(r),\quad \varphi^* = \varphi +  \bar{\varphi}(r),\quad
\hbox{where }
\f{\ud \bar{t}}{\ud r} = \f{2Mr-Q^2}{\Delta},\quad \f{\ud \bar{\varphi}}{\ud r} = \f{a}{\Delta},$$
and the equations involving $\varphi^*$, $\bar{\varphi}$ are understood mod $2\pi$.
When $r \leq 3M$, in the Kerr star coordinates, the metric takes the form
\begin{equation}\label{eq:Kerr.star}
\begin{split}
\bfg_{a,Q,M}=&\:-(1-\frac {2Mr-Q^2}{\Sigma})\, \ud t^* \otimes \ud t^* +\Big( 1+ \f{2Mr-Q^2}{\Sigma}\Big)\, \ud r\otimes \ud r + \Sigma\, \ud \vartheta\otimes \ud \vartheta\\
&\:+ R^2\sin^2\vartheta\, \ud \varphi^*\otimes \ud \varphi^*- \frac{a(2Mr -Q^2)\sin^2\vartheta}{\Sigma}(\ud \varphi^*\otimes \ud t^*+\ud t^*\otimes \ud \varphi^*) \\
&\: +\f{2Mr-Q^2}{\Sigma} (\ud t^*\otimes \ud r + \ud r \otimes \ud t^*) - \f{a(2Mr-Q^2+\Sigma)\sin^2\vartheta}{\Sigma} (\ud \varphi^*\otimes \ud r + \ud r \otimes \ud \varphi^*),
\end{split}
\end{equation}
and in particular extends smoothly to $r = r_+$.
As in Example~\ref{sec:black.holes.ex}, $t^*$ is not a good coordinate near infinity. We thus introduce another change and define
$$t_* = t^* - \chi_{>12M}(r) 2M \log(r - 2M).$$

We now consider $(\calM_{a,Q,M}, \bfg_{a,Q,M})$, where $\calM_{a,Q,M} = (0,\infty)_{t_*} \times [r_+,\infty)_r\times \mathbb S^2_{(\vartheta,\varphi^*)}$ and $\bfg_{a,Q,M}$ is the extension of the metric above up to the boundary. The coordinate functions induces a natural diffeomorphism into $(0,\infty) \times (\mathbb R^3 \setminus B(0,r_+))$ via $x^0 = t_*$, $x^1 = r \sin \vartheta \cos \varphi^*$, $x^2 = r \sin \vartheta \sin \varphi^*$, $x^3 = r \cos \vartheta$.

We now check that $(\calM_{a,Q,M}, \bfg_{a,Q,M})$ satisfies all the assumptions of the main theorems. $\calM_{a,Q,M}$ by definition satisfies the topological assumption \ref{hyp:topology}. Using \eqref{eq:Kerr.star}, one computes that $\{r = r_-\}$ is s null hypersurface so that \ref{hyp:bdry} satisfied. One also checks that \ref{hyp:T-tau} holds; in particular, for $r \leq 3M$, we have $\bfg_{a,Q,M}^{-1}(\ud t_*,\ud t_*) = -1-\f{2Mr-Q^2}{\Sigma}$.

For the condition \ref{hyp:ult-stat}, note that the operator $\calP_0$ in the Kerr case is given by
$$\calP_0 \phi = \f 1{r^2}\Big(\rd_r (\Delta \rd_r \phi) + 2a\rd_r \rd_{\varphi^*}\phi + \rslap \phi\Big),$$
where as before $\rslap$ is the Laplacian corresponding to the round metric on $\mathbb S^{2}$.
The invertibility of $\calP_0$ and the needed estimate follows from \cite[Section~7.1]{AAGKerr}. The Kerr--Newman case does not seem to be explicitly in the literature, but follows along the same lines as \cite{AAGKerr}.
One does expect that the needed estimates for $\calP_0$ are weaker than the integrated local decay estimates since it concerns only $0$ time frequency \cite{Ta}. In turn, the integrated local decay estimates are proven by Civin \cite{Civin} in the full subextremal range of Kerr--Newman black holes. This followed the work of Dafermos--Rodnianski--Shlapentokh-Rothman \cite{DRSR} who achieved these estimates in the full subextremal range of Kerr black holes; see also \cite{AB, BSt, DRS, DRSub2, DRL, MMTT, TT}.

Finally, the condition \ref{hyp:sol} is highly nontrivial: in the case of Kerr (in the full subextremal range), this follows from the work of Dafermos--Rodnianski--Shlapentokh-Rothman \cite{DRSR}. The work \cite{DRSR} does not explicitly give higher vector field bounds as required by \ref{hyp:sol}. However, such bounds are known to follow from the energy boundedness and integrated local energy decay estimates after using suitably defined $\bfS$ and $\bfOmg$ vector fields as commutators; see for instance \cite{MTT}.

As a result, both Main Theorem~\ref{thm:upper} and Main Theorem~\ref{thm:lower} apply to the Kerr--Newman black hole spacetimes in the full subextremal range.
\end{example}

\begin{example}\label{ex:higher.D}(Higher dimensional Schwarzschild/Reissner--Nordstr\"om black holes)
The Schwarzschild spacetime in Example~\ref{ex:Schwarzschild} has higher dimensional analogues. Let $d\geq 5$ be odd and $M>0$. First consider the following metric
    \begin{equation}
    	\bfg^{(d+1)}_{Q,M} = -\Big(1-\f{2M}{r^{d-2}} + \f{Q^2}{r^{2(d-2)}}\Big) \, \ud t\otimes \ud t + \Big( 1-\f{2M}{r^{d-2}}+ \f{Q^2}{r^{2(d-2)}} \Big)^{-1} \, \ud r\otimes \ud r + r^2 \rsgmm
    \end{equation}
on the manifold $\mathbb R_t\times ((2M)^{\f 1{d-2}},\infty)_r \times \mathbb S^{d-1}$.

We introduce a change of coordinate similar to Example~\ref{ex:Schwarzschild} and Example~\ref{ex:Kerr-Newman} so that in the new coordinate system, the metric extends smooth to the boundary and that the causality condition \ref{hyp:T-tau} holds. More precisely, define
$$t^*(t,r) = t + \chi_{<(3M)^{\f 1{d-2}}}(r)\bar{t}(r),\quad \f{\ud \bar{t}}{ \ud r} = \f{2 M r^{d-2} - Q^2}{r^{2(d-2)} - 2M r^{d-2} + Q^2}.$$
In terms of $(t^*,r)$, when $r \in ((2M)^{\f 1{d-2}}, (3M)^{\f 1{d-2}}]$, the metric takes the form
\begin{equation}
\begin{split}
\bfg^{(d+1)}_{Q,M} = &\: -\Big(1-\f{2M}{r^{d-2}} + \f{Q^2}{r^{2(d-2)}}\Big) \, \ud t^* \otimes \ud t^* +   \Big( \f{2M}{r^{d-2}}- \f{Q^2}{r^{2(d-2)}} \Big) (\ud t^*\otimes \ud r + \ud r\otimes \ud t^*) \\
&\: + \Big( 1 + \f{2M}{r^{d-2}} - \f{Q^2}{r^{2(d-2)}}\Big) \ud r \otimes \ud r + r^2 \rsgmm.
\end{split}
\end{equation}
It is then easy to check that the metric can be extended to the manifold with boundary $\calM_{Q,M}^{(d+1)} = \mathbb R_{t^*} \times ((2M)^{\f 1{d-2}},\infty)_r \times \mathbb S^{d-1}$ and that the conditions \ref{hyp:topology}--\ref{hyp:T-tau}, \ref{hyp:T-almost-stat}--\ref{hyp:wave-g} all hold.
(Note in particular that when $d \geq 5$, the decay is already sufficiently fast as $r \to \infty$ for \eqref{eq:wave-g-uu-phg} to hold, without requiring the additional change of variable near infinity as in Example~\ref{ex:Schwarzschild} or Example~\ref{ex:Kerr-Newman}.)

For the wave equation on higher dimensional Schwarzschild, energy boundedness and integrated local energy decay of solutions \cite{pLjM2012, Schlue} as well as pointwise decay estimates \cite{Schlue} are known. After combining with vector field commutations with $\bfT$, $\bfS$ and $\bfOmg$, this is in principle sufficient to obtain the needed estimate \ref{hyp:sol}. Finally, for the stationary estimate \ref{hyp:ult-stat}, note that the operator $\calP_0$ assumes the form
\begin{equation}\label{eq:calP0.Schwarzschild.higher.d}
\calP_0 \phi = \f 1{r^{d-1}}\Big( \rd_r \Big(r^{d-1} - 2Mr + \f{Q^2}{r^{d-3}}\Big)\rd_r \phi\Big)  + \f 1{r^2} \mathring{\slashed{\Delta}} \phi.
\end{equation}
Degenerate elliptic estimates for this operator $\calP_0$ do not seem to be in the literature explicitly, but it can be treated in a very similar manner as \eqref{eq:calP0.Schwarzschild}. In particular, the red-shift effect that is crucially used in \cite{AAGKerr} is also present in \eqref{eq:calP0.Schwarzschild.higher.d}.

Finally, we remark that these spacetimes are spherically symmetric and thus Main Theorem~\ref{thm:upper-sphsymm} and Main Theorem~\ref{thm:lower-sphsymm} apply.

\end{example}

\begin{example}\label{ex:Kerr.dynamical}(Dynamical black holes settling down to subextremal Kerr--Newman black holes)
    Our assumptions allow us to black holes spacetimes which are \emph{dynamical}, and settle down to a subextremal Kerr--Newman black hole. Fix $(a,Q,M)$ such that $a^2 + Q^2 < M^2$ and $M>0$ and let $\calM_{a,Q,M} = (0,\infty)_t \times [r_+,\infty)_r\times \mathbb S^2_{(\vartheta,\varphi^*)}$ as in Example~\ref{ex:Kerr-Newman}. Then let $\bfg$ be a metric on $\calM_{a,Q,M}$ such that the following all hold:
    \begin{enumerate}
    \item $\bfg$ is chosen so that the boundary $\rd \calM$ is either everywhere null or everywhere spacelike. Assume also that the causality conditions \ref{hyp:T-tau} are satisfied.
    \item $\bfg$ satisfies all the assumptions \ref{hyp:T-almost-stat}, \ref{hyp:med}, \ref{hyp:wave-g} and \ref{hyp:ext-g}.
    \item $\bfg$ settles down to $\bfg_{a,Q,M}$ in the $(t_*,r,\vartheta,\varphi^*)$ coordinates in Example~\ref{ex:Kerr-Newman} in the sense that \eqref{eq:T-almost-stat-limit}--\eqref{eq:T-almost-stat-2} hold with $^{(\infty)}\bfg = \bfg_{a,Q,M}$.
    \end{enumerate}

\end{example}

\begin{example}\label{ex:Schwarzschild.dynamical}(Dynamical spherically symmetric black holes settling down to Schwarzschild or Reissner--Nordstr\"om black holes)
    As a particular subcase of Example~\ref{ex:Kerr.dynamical}, we can restrict to spherically symmetric black holes which settle down to Schwarzschild or Reissner--Nordstr\"om black holes. This in particular furnishes spherically symmetric examples for which Main Theorem~\ref{thm:upper-sphsymm} and Main Theorem~\ref{thm:lower-sphsymm} apply.

	One of the motivations in separating out the spherically symmetric examples from the general considerations in Example~\ref{ex:Kerr.dynamical} is that for a \emph{generic} subclass of \emph{dynamical}, spherically symmetric black hole spacetimes settling down to Schwarzschild or Reissner--Nordstr\"om, the $\ell \geq 1$ modes of generic solutions decay (in a compact-$r$ region) only with a rate of $O(t_*^{-2\ell-2})$, which is slower than the Price law rate of  $O(t_*^{-2\ell-3})$ in the stationary case. See \cite{LO.part2} for details.

%
\end{example}

\begin{example}\label{ex:higher.d.BH.dynamical}(Dynamical black holes settling down to higher dimensional Schwarzschild or Reissner--Nordstr\"om black holes)
    In a similar manner as in Example~\ref{ex:Kerr.dynamical}, our assumptions allow for dynamical black hole spacetimes higher dimensions ($d\geq 5$) settling down to Schwarzschild or Reissner--Nordstr\"om black holes considered in Example~\ref{ex:higher.D}. This is similar to Example~\ref{ex:Kerr.dynamical} and we omit the details.
\end{example}

\begin{remark}
The reader may notice that we have not discussed higher dimensional rotating black holes, i.e., the Myers--Perry black holes. This is because even boundedness of solutions to the wave equation (cf.~\ref{hyp:sol}) is largely unexplored in this setting. (See, however, \cite{pLjMsTmT2015}.)
\end{remark}

\subsection{Nonlinear examples}\label{sec:nonlinear.ex}

We give some examples of nonlinear equations. We will focus on nonlinear models on the Minkowski spacetime $(\mathbb R^{d+1}, \bfm)$, but in view of the generality of our main theorems, it is easy to consider a wide variety of nonlinear models on different asymptotically flat backgrounds (including in particular black hole backgrounds).

\begin{example}(Power nonlinearity)\label{ex:power.nonlinearity}
Consider the following nonlinear wave equation with power nonlinearity on $((0,\infty)_t\times \mathbb R^{d},\bfm)$ with $d \geq 3$ odd:
    \begin{equation}\label{eq:power.law}
        \Box_{\bfm} \phi = \pm \phi^p,
    \end{equation}
    where either the $+$ or $-$ sign can be taken. We require the nonlinearity to be smooth and at least quadratic, i.e., $p \in \mathbb N$, $p \geq 2$. In the case $d =3$, we additionally require $p \geq 3$.

    In general, solutions to \eqref{eq:power.law} need not be global in time. However, it is known that for sufficiently small and localized data, the solution remains regular globally in time \cite{GLS97,fJ1979,LS96}. (The $d\geq 9$ case is strictly speaking not in these papers, but since we only consider smooth nonlinearity with $p\geq 2$, this is much easier in higher dimensions.) The higher vector field bounds needed in \ref{hyp:sol} do not directly follow from these works, but in principle one should be able to prove it with $(\alp_0,\nu_0) = (1,0)$, say, by combining the $r^p$ method \cite{DRNM, gM2016} with commuting vector fields. From now on, we consider solutions which are global and satisfy \ref{hyp:sol} with $(\alp_0,\nu_0) = (1,0)$.

    The assumptions \ref{hyp:topology}--\ref{hyp:T-tau}, \ref{hyp:T-almost-stat}--\ref{hyp:wave-V}, \ref{hyp:bdry.cond}--\ref{hyp:ult-stat}, \ref{hyp:ext-g}, \ref{hyp:ext-B} and \ref{hyp:ext-V} obviously hold, since they just correspond to Minkowski spacetime.

    We now check the assumptions \ref{hyp:nonlin}--\ref{hyp:nonlin-null} on the nonlinearity, The nonlinearity is a smooth nonlinear function of $\phi$. Thus, \ref{hyp:nonlin} is satisfied. By Definition~\ref{def:nonliear-deg}, $n_{\calN} = p \geq 2$ and thus \ref{hyp:nonlin-degree} holds.

    We next verify the conditions \ref{hyp:nonlin-near}--\ref{hyp:nonlin-wave}. Notice that $\mathcal F(z) = z^p$ is not a bounded function. However, since $\phi$ is bounded (recall that we assume \ref{hyp:sol} is satisfied with $(\alp_0,\nu_0) = (1,0)$), say, with $|\phi| \leq A$, we can redefine the nonlinearity by setting $\widetilde{\mathcal F}$ to be a smooth function satisfy
    \begin{equation}
    	\widetilde{\mathcal F}(z) = \begin{cases}
	z^p & \hbox{when $|z| \leq A$}\\
	0 & \hbox{when $|z| \geq 2A$}
	\end{cases}.
    \end{equation}
Notice that the same function $\phi$ solves the equation with the truncated nonlinearity $\widetilde{\calF}$:
    $$\Box_{\bfm} \phi = \widetilde{\mathcal F}(\phi).$$
With this redefinition, the bounds \ref{hyp:nonlin-near} and \ref{hyp:nonlin-med} obviously hold. It is easy to see that \ref{hyp:nonlin-wave} also holds with the one-term expansion $\calF = \mathring{\calF}_{0,0}$.

	Finally, in this setting, the condition \ref{hyp:nonlin-null} corresponds exactly to our requirement that $p \geq 3$ when $d = 3$.

\end{example}

\begin{example}(Nonlinear wave equations with quadratic nonlinearities, with a null condition when $d=3$)\label{ex:null.condition}
    For this example, we allow for systems of equations (which are applicable; see Remark~\ref{rmk:systems}).

    Consider a system of nonlinear wave equations on $((0,\infty)_t\times \mathbb R^{d}, \bfm)$ for $\phi:(0,\infty)_t\times \mathbb R^{d} \to \mathbb R^N$ (with components denoted by $\phi^I$, $I=1,\ldots, N$) of the form
    \begin{equation}\label{eq:general.null.condition}
        \Box_{\bfm} \phi^I = \frkc^I_s(x)  B^{I,\mu\nu}_{LP}(\phi) \rd_\mu \phi^L \rd_\nu\phi^P + \frkc^I_q(x) D^{I,\alp\mu\nu}_{L}(\phi) \rd^2_{\alp\mu} \phi^I \rd_\nu \phi^L
    \end{equation}
    where the coefficients $\frkc^I_s, \,\frkc^I_q:\mathbb R^{d+1} \to \mathbb R^N$ satisfying vector field bounds up to sufficiently high order in  $\calM_{\near} \cup \calM_{\med}$, i.e.,
    \begin{align*}
    \frkc^I_s , \,\frkc^I_q= O_{\bfGmm}^{M_{c}}(1) \quad \hbox{in $\calM_{\near} \cup \calM_{\med}$}.
    \end{align*}
    Moreover, in $\calM_{\wave}$, $\frkc^I_s$ and $\frkc^I_q$ admit an expansion
    \begin{align*}
    \frkc^I_s =  &\: \sum_{j=0}^{J_c-\f{d-1}2} \sum_{k=0}^{K_{c}} r^{-j} \log^k(\tfrac r{\mbrk{u}}) (\mathring{\frkc}^I_s)_{j, k}(u,\th) + \rem_{J_{c}-\frac{d-1}{2}+\eta_{c}}[\frkc^I_s], \\
    \frkc^I_q = &\: \sum_{j=0}^{J_c-\f{d-1}2} \sum_{k=0}^{K_{c}} r^{-j} \log^k(\tfrac r{\mbrk{u}}) (\mathring{\frkc}^I_q)_{j, k}(u,\th) + \rem_{J_{c}-\frac{d-1}{2}+\eta_{c}}[\frkc^I_q],
    \end{align*}
    where
    \begin{align*}
	(\mathring{\frkc}^I_s)_{j, k},\, (\mathring{\frkc}^I_q)_{j, k} &= O_{\bfGmm}^{M_{c}}(u^{j}) , \\
	\rem_{J_{c}-\frac{d-1}{2}+\eta_{c}}[\frkc^I_s],\, \rem_{J_{c}-\frac{d-1}{2}+\eta_{c}}[\frkc^I_q] &=  O_{\bfGmm}^{M_{c}}(r^{-J_{c}+\frac{d-1}{2}-\eta_{c}} u^{J_{c}-\frac{d-1}{2}+\eta_{c}}) \quad \hbox{ in } \calM_{\wave},
	\end{align*}
    In the case $d = 3$, assume in addition that the nonlinear terms verify the classical null condition, i.e.,
    \begin{equation}\label{eq:classical.null}
        B^{I,\mu\nu} \xi_{\mu}\xi_{\nu} = 0,\quad D^{I,\alp\mu\nu}_{L} \xi_\alp \xi_\mu \xi_\nu=0, \quad \hbox{whenever $\bfm^{\alp\bt} \xi_\alp \xi_\bt = 0$}.
    \end{equation}
    In all of the above, repeated lower case Greek letters are summed over $0,1,\ldots, d$, while repeated upper case Latin letters are summed over $1,\ldots, N$.

	We consider the case where $\phi^I$ is a given global solution which satisfies the bounds \eqref{eq:sol-near}--\eqref{eq:sol-wave} in \ref{hyp:sol} for some $\alp_0 >0$ and some sufficiently large $M_0$. In particular, we assume that $\phi$ and its first and second derivatives are globally bounded (which will be useful in verifying \ref{hyp:nonlin-near}--\ref{hyp:nonlin-wave} below). In the small data case when $d=3$, \ref{hyp:sol} is classical for equation of the type \eqref{eq:general.null.condition}, which was proven independently by Christodoulou \cite{Chr.86} and Klainerman \cite{Kla.86}. The work \cite{Chr.86} gives slightly stronger estimates so that \ref{hyp:sol} holds with $(\alp_0,\nu_0) = (2,0)$. In higher dimensions, the dispersion is stronger and it is easier to obtain \ref{hyp:sol} for small-data solutions; indeed, this can be achieved even without the null condition. In addition to the small-data solutions, there are cases of large data for which \ref{hyp:sol} holds for the solution; see \cite{sA2009, sMlPpY2019, swY2015}.

    As in Example~\ref{ex:power.nonlinearity}, the assumptions \ref{hyp:topology}--\ref{hyp:T-tau}, \ref{hyp:T-almost-stat}--\ref{hyp:wave-V}, \ref{hyp:bdry.cond}--\ref{hyp:ult-stat}, \ref{hyp:ext-g}, \ref{hyp:ext-B} and \ref{hyp:ext-V} obviously hold, as we are considering a (system of) nonlinear wave equation(s) on Minkowski spacetimes.

    For the assumptions regarding the nonlinearity, \ref{hyp:nonlin}--\ref{hyp:nonlin-degree} are obvious. For \ref{hyp:nonlin-near}--\ref{hyp:nonlin-wave} and \ref{hyp:ext-N}, we need to redefine the nonlinearities as bounded functions as in Example~\ref{ex:power.nonlinearity}, and this can be achieved using that the boundedness of $\phi$ and its first and second derivatives (which follows from \ref{hyp:sol} for $\alp_0 >0$). The null condition \ref{hyp:nonlin-null}, which is required when $d=3$, follows as a consequence of \eqref{eq:classical.null}.

\end{example}

\begin{example}(Wave maps)
Consider the wave map system for a map $\phi: ((0,\infty)_t\times \mathbb R^{d},\bfm) \to (\mathcal N^D, \frkh)$, where $\mathcal N^{D}$ is a $D$-dimensional manifold and $\frkh$ is a Riemannian metric on $\mathcal N$. Let $\{y^I\}_{I=1,\ldots D}$ be local coordinates in some open set $U \subset \mathcal N^{D}$. Then, when $\phi$ takes value in $U$, the wave map system takes the form
$$\Box_{\bfm}\phi^I = -\sum_{J,K=1}^D \Gamma_{JK}^I(\phi) (\bfm^{-1})^{\alp\bt} \rd_\alp \phi^J \rd_\bt \phi^K,$$
where $\Gamma^{I}_{JK}$ are the Christoffel symbols for the metric $\frkh$.

    At least when the image of the wave map stays in one coordinate patch of the target, the wave map equations form a semilinear system on Minkowski spacetime which satisfies the null condition. Hence, it satisfies the assumptions of our main theorem by Example~\ref{ex:null.condition}.
\end{example}

\begin{example}(Hyperbolic vanishing mean curvature equation)
Consider the following equation for $\phi: ((0,\infty)_t\times \mathbb R^{d},\bfm) \to \mathbb R$:
    \begin{equation} \label{eq:hvmc}
	\Box_{\bfm} \phi = - \frac{q(\phi, q(\phi, \phi))}{2 (1 + q(\phi, \phi))},
\end{equation}
where $q$ is the null form given by $q(\phi, \psi) = (\bfm^{-1})^{\alp \bt} \rd_{\alp} \phi \rd_{\bt} \psi$. This does not fall into the class considered in Example~\ref{ex:null.condition}, but because it is better than cubic, it is in fact easier and at least in the small data regime, all the estimates in Example~\ref{ex:null.condition} still hold and our main theorem applies.
\end{example}

\section{Price's law cancellation: Conservative structure of the recurrence equations in the stationary case (Proof of Theorem~\ref{thm:price})} \label{sec:price}

In this section, we give and prove a precise version of Theorem~\ref{thm:price} (given below as Theorem~\ref{thm:price.precise}), which gives the Price's law upper bound in the stationary setting. In light of Main Theorem~\ref{thm:upper}, the Price's law upper bound can be viewed as a consequence of a conservative structure in the recurrence equations for the higher radiation fields, which holds because of stationarity. We emphasize again that this applies even in the presence of logarithmic terms in the expansion near null infinity.

\begin{theorem}[Price's law]\label{thm:price.precise}
Consider the linear equation $\calP \phi = 0$, with $\calP$ given by \eqref{eq:P.def}. Let $\calP$ satisfy the assumptions \ref{hyp:topology}--\ref{hyp:T-tau},  \ref{hyp:T-almost-stat}--\ref{hyp:wave-V}, \ref{hyp:bdry.cond}, as well as either \ref{hyp:ult-stat} or \ref{hyp:ult-stat'}. Assume also that $J_{c}$ is sufficiently large. \textbf{Assume in addition that all the coefficients are \emph{stationary} in $\calM_\wave$}, i.e.,
\begin{equation}\label{eq:stationary.Mwave.price}
\calL_{\bfT} \bfg^{-1},\, \calL_{\bfT} B,\quad \bfT V = 0\quad \hbox{in }\calM_\wave.
\end{equation}
Then for every solution $\phi$ to $\calP \phi = 0$ satisfying assumptions \ref{hyp:sol} and \ref{hyp:id}, and \textbf{in addition arise from compactly supported initial data}, the following upper bound holds:
\begin{equation}
|\phi|\ls \min\{\tau^{-d}, r^{-\f{d-1}2} \tau^{-\f{d+1}2}\}.
\end{equation}
\end{theorem}
\begin{proof}
\pfstep{Step~0: Recurrence equations in the stationary case and schematic notation} Consider the renormalized higher radiation fields $\rcPhi_{j,k}$ as defined in Section~\ref{sec:renormalized.higher.radiation.fields}. Recall that $\rcPhi_{j,k}$ satisfies the recurrence equations \eqref{eq:recurrence-jk.rcPhi}. In this setting, the equations are linear and inhomogeneous, and that the coefficients are stationary. It thus follows that the term $\rcF_{j,k}$ in \eqref{eq:recurrence-jk.rcPhi} is given schematically as
\begin{equation}\label{eq:price.F.notation}
\rcF_{j,k} = \rcF_{j,k}((\rcPhi_{j',k'})_{j'\leq j-2}, (\rd_u \rcPhi_{j',k'}, \rd_u^2 \rcPhi_{j',k'})_{j'\leq j-1}),
\end{equation}
where in \eqref{eq:price.F.notation}, and for the remainder of the proof, we use this notation to mean that $\rcF_{j,k}$ is a \underline{linear} function of $(\rcPhi_{j',k'})_{j'\leq j-2}$, $(\rd_u \rcPhi_{j',k'}, \rd_u^2 \rcPhi_{j',k'})_{j'\leq j-1}$, possibly acted on by a finite order pseudo-differential \underline{operator on $\mathbb S^{d-1}$}, and with \underline{stationary coefficients}, i.e., \eqref{eq:price.F.notation} means that
$$\rcF_{j,k} =  \sum_{k'} \Big( \sum_{j'=0}^{j-2}\mathfrak f_{j,k,(1)}^{j',k'} \mathcal O^{j',k'}_{(1)}\rcPhi_{j',k'} + \sum_{j'=0}^{j-1}(\mathfrak f_{j,k,(2)}^{j',k'} \mathcal O^{j',k'}_{(2)} \rd_u \rcPhi_{j',k'} + \mathfrak f_{j,k,(3)}^{j',k'} \mathcal O^{j',k'}_{(3)}\rd_u^2\rcPhi_{j',k'}) \Big)$$
with $u$-independent functions $\mathfrak f$ and with angular operators $\mathcal O$.

\pfstep{Step~1: The main induction argument} Fix $d$. We prove by induction in $j$ the following two statements (where we continue to use the schematic notation as in \eqref{eq:price.F.notation}):
\begin{enumerate}
\item For $1\leq j\leq \f{d+1}{2}$, $k\geq 1$, $\rcPhi_{j-1,k}$ can be expressed as a linear combination of derivatives of $\rcPhi_{j',k'}$ in the following manner:
\begin{equation}\label{eq:stat-cancellation.BA1}
\rcPhi_{j-1,k} = \overline{\mathfrak H}_{j-1,k}((\rcPhi_{j',k'}, \rd_u \rcPhi_{j',k'})_{j'\leq j-2}).
\end{equation}
\item For $2 \leq j \leq \f{d-1}2$, $\rcPhi_{j-2,0}$ can be written as a total $\rd_u$ derivative in the following manner:
\begin{equation}\label{eq:stat-cancellation.BA2}
\rcPhi_{j-2,0} = \rd_u \Big(\mathfrak H_{j-2}((\rcPhi_{j',k'}, \rd_u \rcPhi_{j',k'})_{j'\leq j-1}) \Big).
\end{equation}
\end{enumerate}

\pfstep{Step~1(a): The base case} We take $j=1$ to be the base case. The statement \eqref{eq:stat-cancellation.BA1} holds because $\rcPhi_{0,k}= 0$ for $k\geq 1$ (due to the finiteness of the radiation field), while the statement \eqref{eq:stat-cancellation.BA2} is vacuous.

\pfstep{Step~1(b): The induction step} Assume that $\exists j_* \in \mathbb N$ with $1\leq j_* \leq \f{d-3}2$ such that \eqref{eq:stat-cancellation.BA1} and \eqref{eq:stat-cancellation.BA2} hold for all $1\leq j \leq j_*$. We will prove \eqref{eq:stat-cancellation.BA1} and \eqref{eq:stat-cancellation.BA2} hold for $j$ replaced by $j_*+1$.

Define
\begin{equation}\label{eq:urcF.def}
\underline{\rcF}_{j,k} = (j-1)(k+1) \rcPhi_{j-1, k+1} - (k+2)(k+1) \rcPhi_{j-1, k+2}
+ \rcF_{j, k}.
\end{equation}
We claim that
\begin{equation}\label{eq:Price.key.claim}
\underline{\rcF}_{j_*,k} = \rd_u \Big( \underline{\rcH}_{j_*,k}(( \rcPhi_{j',k'}, \rd_u \rcPhi_{j',k'})_{j'\leq j_*-1}) \Big),\quad k\geq 0.
\end{equation}
This is because
\begin{itemize}
\item By the induction hypothesis \eqref{eq:stat-cancellation.BA1} (since $k+1,\,k+2\geq 1$),
$$\rcPhi_{j_*-1, k+1},\,\rcPhi_{j_*-1, k+2} = G \Big((\rcPhi_{j',k'})_{j'\leq j_*-2}, (\rd_u \rcPhi_{j',k'}, \rd_u^2 \rcPhi_{j',k'})_{j'\leq j_*-1} \Big).$$
Thus, using also \eqref{eq:price.F.notation}, all terms in \eqref{eq:urcF.def} depend on $(\rcPhi_{j',k'})_{j'\leq j_*-2}$ and $(\rd_u \rcPhi_{j',k'}, \rd_u^2 \rcPhi_{j',k'})_{j'\leq j_*-1}$.
\item Since all the coefficients are stationary, we can write terms involving $\rd_u \rcPhi_{j',k'}$ or $\rd_u^2 \rcPhi_{j',k'}$ as a total $\rd_u$ derivative, for instance by writing $\rch_{2,k}^{uu} \rd^2_u \rcPhi_{0,0} = \rd_{u} (\rch_{2,k}^{uu} \rd_u \rcPhi_{0,0})$, etc.
\item For the $(\rcPhi_{j',k'})_{j'\leq j_*-2}$ terms, we use \eqref{eq:stat-cancellation.BA1} iteratively to write them in terms of $(\rcPhi_{j',0})_{j'\leq j_*-2}$ as follows. Since by \eqref{eq:stat-cancellation.BA1}, $(\rcPhi_{j',k'})_{j'\leq j_*-2,k'\geq 1}$ can be written in terms of $(\rcPhi_{j',k'})_{j'\leq j_*-3}$, we see that $(\rcPhi_{j',k'})_{j'\leq j_*-2}$ can be written in terms of $(\rcPhi_{j',0})_{j'\leq j_*-2}$ and $(\rcPhi_{j',k'})_{j'\leq j_*-3,k'\geq 1}$. Then we use \eqref{eq:stat-cancellation.BA1} again to write these terms as $(\rcPhi_{j',0})_{j'\leq j_*-2}$ and $(\rcPhi_{j',k'})_{j'\leq j_*-4,k'\geq 1}$. This procedure has to stop since $\rcPhi_{0,k} =0$ if $k\geq 1$ by the finiteness of the radiation field.

Now that we have written $(\rcPhi_{j',k'})_{j'\leq j_*-2}$ in terms of $(\rcPhi_{j',0})_{j'\leq j_*-2}$, we then use the induction hypothesis \eqref{eq:stat-cancellation.BA2} to write $\rcPhi_{j',0}$ as a $\rd_u$ derivative of $(\rcPhi_{j',k'},\rd_u \rcPhi_{j',k'})_{j'\leq j_*-1}$. Then, arguing as in the above point, we use the stationary of the coefficients to write the corresponding terms as a total $\rd_u$ derivative.
\end{itemize}
We have thus established \eqref{eq:Price.key.claim}, which is the key claim to carry out the induction for \eqref{eq:stat-cancellation.BA2}.

We now prove \eqref{eq:stat-cancellation.BA2}. Apply \eqref{eq:recurrence-jk.rcPhi} with $j=j_*$, $k\geq 1$, and use \eqref{eq:urcF.def} and \eqref{eq:Price.key.claim} to get
\begin{equation}\label{eq:recurrence-jk.stat.simplified}
\begin{split}
	\rd_{u} \rcPhi_{j_*, k}&=
	 \frac{k+1}{j} \rd_{u} \rcPhi_{j_*, k+1}
- \frac{1}{2j_*}\Big( (j_*-1)j_* - \frac{(d-1)(d-3)}{4} + \rslap \Big) \rcPhi_{j_*-1, k}
+ \frac{1}{2j_*} \rd_u \underline{\rcH}_{j_*,k}.
\end{split}
\end{equation}

When $k\geq 1$, we use the induction hypothesis corresponding to \eqref{eq:stat-cancellation.BA1}, to write
$ \frac{1}{2j_*}\Big( (j_*-1)j_* - \frac{(d-1)(d-3)}{4} + \rslap \Big) \rcPhi_{j_*-1, k} $ as a $\rd_u$ derivative of $(\rcPhi_{j',k'},\rd_u \rcPhi_{j',k'})_{j'\leq j_*-1}$.
Therefore, using \eqref{eq:recurrence-jk.stat.simplified}, there is some $\underline{\underline{\rcH}}_{j, k} = \underline{\underline{\rcH}}_{j, k}((\rcPhi_{j',k'},\rd_u \rcPhi_{j',k'})_{j'\leq j_*-1})$ such that
\begin{equation}\label{eq:Price.rcPhik.rewrite.0}
\rd_{u} \Big(\rcPhi_{j_*, k} - \frac{k+1}{j_*}  \rcPhi_{j_*, k+1} -   \f 1{2j_*} \underline{\underline{\rcH}}_{j_*, k} \Big) =0.
\end{equation}
Since the data are compactly supported, \eqref{eq:Price.rcPhik.rewrite.0} implies that $\rcPhi_{j_*, k}-
	 \frac{k+1}{j_*}  \rcPhi_{j_*, k+1} -   \f 1{2j_*} \underline{\underline{\rcH}}_{j_*, k}$ vanishes identically, $\forall u$, $\forall k\geq 1$. Using also that there exists $K\geq 0$ such that $\rcPhi_{j_*, k} = 0$ for all $k > K$ (see Lemma~\ref{lem:Kj-def}), we obtain
	 \begin{equation}\label{eq:Price.rcPhik.rewrite}
	 \rcPhi_{j_*,k} = \sum_{k'=0}^{K-k}  \f {(k+k')!}{2j_*^{k'+1}k!} \underline{\underline{\rcH}}_{j_*, k+k'} =: \overline{\mathfrak H}_{j_*,k}.
	 \end{equation}
	 We have thus proven \eqref{eq:stat-cancellation.BA1} for $j = j_*+1$.

To prove \eqref{eq:stat-cancellation.BA2}, we now apply \eqref{eq:recurrence-jk.rcPhi} with $(j,k)=(j_*,0)$. Notice that $(j_*-1)(j_*-2) - \frac{(d-1)(d-3)}{4} + \rslap$ is invertible since $j_* \leq \f{d-3}2$; we denote its inverse by $\mathcal O$. Thus,
\begin{equation}
\begin{split}
	\rcPhi_{j_*-1, 0} &=
	   \mathcal O \rd_{u} \Big( - 2j\rcPhi_{j_*, 0} + 2 \rcPhi_{j_*, 1} + \underline{\rcH}_{j_*, 0}\Big) \\
	   &= \mathcal O \rd_{u} \Big( - 2j_*\rcPhi_{j_*, 0} + 2 \sum_{k'=0}^{K-1}  \f {(k'+1)!}{2j^{k'+1}} \underline{\underline{\rcH}}_{j_*, k'+1} + \underline{\rcH}_{j_*, 0}\Big) =: \rd_u \mathfrak H_{j_*-1},
\end{split}
\end{equation}
where we have used \eqref{eq:Price.rcPhik.rewrite} for the second equality. This proves \eqref{eq:stat-cancellation.BA2} for $j = j_*+1$.

By the induction above, we have proved that \eqref{eq:stat-cancellation.BA1} and \eqref{eq:stat-cancellation.BA2} both hold for $j\leq \f{d-1}2$. Finally, notice that we can repeat the inductive argument above to show that \eqref{eq:stat-cancellation.BA1} also holds for $j = \f{d+1}2$. (On the other hand, in the proof of \eqref{eq:stat-cancellation.BA2} above, we explicitly used $j_* +1 \leq \f{d-1}2$. In fact, in general \eqref{eq:stat-cancellation.BA2} does not hold for $j = \f{d+1}2$. This is related to the Newman--Penrose cancellation in Step~2 below.)

\pfstep{Step~2: The Newman--Penrose cancellation} We now use the Newman--Penrose cancellation. More precisely, consider \eqref{eq:recurrence-jk.rcPhi} with $(j,k) = (\f{d-1}2,0)$, and then project to the zeroth spherical harmonic to obtain
\begin{equation}\label{eq:recurrence-nuBox0.stat}
\begin{split}
	\rd_{u} \rcPhi_{(0)\f{d-1}2, 0}=&\:
	 \frac{2}{d-1} \rd_{u} \rcPhi_{(0)\f{d-1}2, 1}
+ \frac{d-3}{2(d-1)}  \rcPhi_{(0)\f{d-3}{2}, 1} - \frac{2}{d-1} \rcPhi_{(0)\f{d-3}{2}, 2}
+ \frac{1}{d-1} \rcF_{(0)\f{d-1}2, 0}.
\end{split}
\end{equation}
The cancellation is that $(j-1)j-\f{(d-1)(d-3)}{4} =0$ so that there is no $\rcPhi_{(0)\f{d-3}2, 0}$ term no the right-hand side.

Using \eqref{eq:stat-cancellation.BA1} and \eqref{eq:stat-cancellation.BA2}, we can derive \eqref{eq:urcF.def} and \eqref{eq:Price.key.claim} for $j = \f{d-1}{2}$ with the same proof. Moreover, by \eqref{eq:stat-cancellation.BA1}, we can write $\rcPhi_{(0)\f{d-1}2, 1}$. in terms of $(\rcPhi_{j',k'}, \rd_u \rcPhi_{j',k'})_{j'\leq \f{d-3}2}$. Therefore, \eqref{eq:recurrence-nuBox0.stat} can be rewritten as
\begin{equation}\label{eq:NP.ode}
\rd_{u} \rcPhi_{(0)\f{d-1}2, 0}= \rd_u \mathfrak H_{NP}((\rcPhi_{j',k'}, \rd_u \rcPhi_{j',k'})_{j'\leq \f{d-3}2}).
\end{equation}

Using the compact support of the initial data, \eqref{eq:NP.ode} implies a conservation law of the form
\begin{equation}\label{eq:NP}
\rcPhi_{(0)\f{d-1}2, 0} = \mathfrak H_{NP}((\rcPhi_{j',k'}, \rd_u \rcPhi_{j',k'})_{j'\leq \f{d-3}2}).
\end{equation}

\pfstep{Step~3: Proof of the improved decay} Our goal is to apply Main Theorem~\ref{thm:upper} with $J_{\mathfrak f} = \f{d+1}2$. To this end, we need to show that \eqref{eq:Jf.def} holds for $J_{\mathfrak f} = \f{d+1}2$.

Assume for the sake of contradction that \eqref{eq:Jf.def} fails for $J_{\mathfrak f} = \f{d+1}2$. We know by Lemma~\ref{lem:radiation.field} that $\lim_{u\to \infty} \rcPhi_{0}(u,\th) = \lim_{u\to \infty} \rPhi_{0}(u,\th) = 0$. Combining these two statements, there must be a \underline{smallest} $J_* \leq \f{d+1}2 - 1 = \f{d-1}2$ such that $\lim_{u\to \infty} \rcPhi_{J_*,k}(u,\th) \neq 0$ for some $1\leq k \leq K_{J_*}$ or $\lim_{u\to \infty} \rcPhi_{(\leq J_* - \f{d-1}2)J_*,0}(u,\th) \neq 0$. We now analyze these two possibilities: 
\begin{enumerate}
\item Suppose that $\lim_{u\to \infty} \rcPhi_{J_*,k}(u,\th) \neq 0$ for some $k\geq 1$, and $J_*$ is the minimal integer for which this happens. This is impossible because this would contradict \eqref{eq:stat-cancellation.BA1}.
\item Suppose $\lim_{u\to \infty} \rcPhi_{(\leq J_* - \f{d-1}2)J_*,0}(u,\th) \neq 0$ with $J_*$ being the minimal integer that this happens. For the projection to $\leq J_* - \f{d-1}2$ spherical harmonic modes to be nontrivial, we necessarily need $J_* \geq \f{d-1}2$. Since we also have $J_* \leq \f{d-1}2$, it follows that $J_* = \f{d-1}2$. Hence, $\lim_{u\to \infty} \rcPhi_{(\leq J_*- \f{d-1}2)J_*,0}(u,\th)  = \lim_{u\to \infty} \rcPhi_{(0)\f{d-1}2,0}(u,\th) \neq 0$. On the other hand, by the minimality of $J_*$, we have $\rcPhi_{j',k'}, \rd_u \rcPhi_{j',k'} \to 0$ as $u\to \infty$ for all $j' \leq \f{d-3}2$. These two facts together contradict \eqref{eq:NP}. 
\end{enumerate}
It thus follows that \eqref{eq:Jf.def} holds for $J_{\mathfrak f} = \f{d+1}2$. The proof of the theorem can therefore be concluded after using Main Theorem~\ref{thm:upper} with $J_{\mathfrak f} = \f{d+1}2$. \qedhere

\end{proof}

\begin{remark}[Effects of dynamic perturbations]
It is important to note that the argument in the proof of Theorem~\ref{thm:price.precise} relies heavily on the stationarity assumption. In particular, we repeatedly make use of facts such as $\mathring{\mathcal W}_{j_1,k_2} \rd_u \rPsi_{j_2,k_2} = \rd_u (\mathring{\mathcal W}_{j_1,k_2} \rPsi_{j_2,k_2}) $. Indeed, as indicated in Section~\ref{sec:intro.anomalous.etc} (see also \cite{LO.part2}), when dynamic perturbations are present --- even when those perturbations decay very rapidly --- the decay rate is generically slower.
\end{remark}

The proof of Theorem~\ref{thm:price.precise} shows that in fact initial compact support is not needed, and one only needs suitable ``initial charges'' to vanish. We summarize this in the corollary below.
\begin{corollary}
Suppose all the assumptions for $\calP$ in Theorem~\ref{thm:price.precise} (including, importantly, \eqref{eq:stationary.Mwave.price}) hold.

Suppose $\phi$ is a solution to $\calP\phi = 0$ satisfying assumptions  \ref{hyp:sol} and \ref{hyp:id} (which is not necessarily compactly supported) such that the following holds for the initial data on $\Sigma_1$:
$$\Big( \rcPhi_{j, k} -   \overline{\mathfrak H}_{j,k} \Big)(1,\theta)= 0,\, \forall 0\leq j \leq \f{d-1}2, k \geq 1 \quad  \Big(\rcPhi_{(0)\f{d-1}2, 0} - \mathfrak H_{NP}\Big)(1,\th) = 0.$$
and
$$ \Big(\rcPhi_{(0)\f{d-1}2, 0} - \mathfrak H_{NP}\Big)(1,\th) = 0$$
where $ \overline{\mathfrak H}_{j,k} =  \overline{\mathfrak H}_{j,k}((\rcPhi_{j',k'}, \rd_u \rcPhi_{j',k'})_{j'\leq j-1})$ is as in \eqref{eq:Price.rcPhik.rewrite.0} and \eqref{eq:Price.rcPhik.rewrite}, and $\mathfrak H_{NP} = \mathfrak H_{NP}((\rcPhi_{j',k'}, \rd_u \rcPhi_{j',k'})_{j'\leq \f{d-3}2})$ is as in \eqref{eq:NP.ode}.

Then the upper bound in Theorem~\ref{thm:price.precise} holds.
\end{corollary}

\section{Tools related to the vector field method} \label{sec:vf}
In this section, we collect tools for the proofs of our main theorems that may be (loosely) grouped under the heading ``vector field method.''
\subsection{Convenient expressions for $\calP$ in $\calM_{\far}$} \label{subsec:conj-wave}
In this subsection, we discuss ways to express $\calP$ in $\calM_{\far}$ that are convenient for our analysis in $\calM_{\med}$ and $\calM_{\wave}$.

\subsubsection{$\calP$ as a perturbation of $\Box_{\bfm}$ in $\calM_{\far}$}
In $\calM_{\far}$, it is convenient to write $\calP$ as a perturbation of the Minkowskian d'Alembertian $\Box_{\bfm} = - \rd_{x^{0}}^{2} + \sum_{j=1}^{d} \rd_{x^{j}}^{2}$. We write
\begin{equation} \label{eq:main-eq-far}
	\calP = \Box_{\bfm} + (\bfg^{-1} - \bfm^{-1})^{\alp \bt} \nb_{\alp} \rd_{\bt} + (\bfA^{\alp} + \bfB^{\alp}) \rd_{\alp} + V,
\end{equation}
where $\nb_{\alp}$ is the covariant derivative associated with the reference Minkowski metric $\bfm = - (\ud x^{0})^{2} + \sum_{j=1}^{d} (\ud x^{j})^{2}$, and $\bfA^{\alp}$ is given by
\begin{equation} \label{eq:bfA-def}
\bfA^{\alp} = \left( {}^{(\bfg)}\Gmm^{\alp}_{\mu \nu} - {}^{(\bfm)} \Gmm^{\alp}_{\mu \nu} \right) (\bfg^{-1})^{\mu \nu},
\end{equation}
where ${}^{(\bfg)} \Gmm$ and ${}^{(\bfm)} \Gmm$ are the Christoffel symbols associated with $\bfg$ and $\bfm$, respectively\footnote{Of course, with respect to the rectangular coordinates we have ${}^{(\bfm)} \Gmm^{\alp}_{\mu \nu} = 0$ and thus $\bfA^{\alp} = {}^{(\bfg)}\Gmm^{\alp}_{\mu \nu} (\bfg^{-1})^{\mu \nu}$. Nevertheless, \eqref{eq:bfA-def} has the advantage of being coordinate-invariant, i.e., $\bfA^{\alp}$ as in \eqref{eq:bfA-def} is a vector field.}.

From \ref{hyp:wave-g} and Koszul's formula
\begin{equation*}
	{}^{(\bfg)} \Gmm^{\alp}_{\mu \nu} = \frac{1}{2} (\bfg^{-1})^{\alp \bt} \left(\rd_{\mu} \bfg_{\nu \bt} + \rd_{\nu} \bfg_{\bt \mu} - \rd_{\bt} \bfg_{\mu \nu} \right),
\end{equation*}
it is straightforward to verify that $\bfA$ enjoys the same properties as $\bfB$ (except for the loss of one order of $\bfGmm$-regularity) in $\calM_{\med}$:
\begin{lemma} \label{lem:bfA-med}
In $\calM_{\med}$, we have
\begin{align}
	\bfA^{\alp} = O_{\bfGmm}^{M_{c}-1}(r^{-1-\dlt_{c}}). \label{eq:med-bfA}
\end{align}
\end{lemma}
We omit the proof. We postpone the discussion of the region $\calM_{\wave}$ until Lemma~\ref{lem:conj-wave} below.

\subsubsection{Conjugation of $\calP$ in $\calM_{\far}$}
We now specialize to the Bondi--Sachs-type coordinate system $(u, r, \tht)$ introduced in Section~\ref{subsec:assumptions}. Recall that the non-zero Christoffel symbols associated with $\bfm$ in the $(u, r, \tht)$-coordinates are:
\begin{equation} \label{eq:bondi-Gmm}
\begin{aligned}
	{}^{(\bfm)} \Gmm^{u}_{AB}
	&= r \rsgmm_{AB}, \\
	{}^{(\bfm)} \Gmm^{r}_{AB}
	&= - r \rsgmm_{AB}, \\
	{}^{(\bfm)} \Gmm^{B}_{rA} = {}^{(\bfm)} \Gmm^{B}_{A r}
	&= \frac{1}{r} \bfdlt^{B}_{A}, \\
	{}^{(\bfm)} \Gmm^{C}_{AB} &= {}^{(\rsgmm)} \Gmm^{C}_{AB}
\end{aligned}
\end{equation}
Note, in particular, the appearance of the term $(\bfm^{-1})^{AB} {}^{(\bfm)} \Gmm_{AB}^{u} \rd_{u} = (d-1) r^{-1} \rd_{u}$ in $\Box_{\bfm}$, which is of similar strength with $(\bfm^{-1})^{ru} \rd_{r} \rd_{u} = - 2 \rd_{r} \rd_{u}$. To remove this first-order term, we are motivated to conjugate the equation with the weight $r^{\frac{d-1}{2}}$. Here, observe that the exponent $\frac{d-1}{2}$ is precisely the sharp uniform decay rate of solutions to the linear wave equation on $\bbR^{d+1}$. As this number shall be used frequently, we introduce the shorthand
\begin{equation*}
\nu_{\Box} = \frac{d-1}{2}.
\end{equation*}
\begin{lemma} \label{lem:conj-wave}
Define $\Phi = r^{\nu_{\Box}} \phi$. Then the conjugated operator $\calQ \Phi := r^{\nu_{\Box}} \calP (r^{-\nu_{\Box}} \Phi)$ takes the form
\begin{equation} \label{eq:conj-wave}
	 \calQ \Phi = Q_{0} \Phi + \bfh^{\alp \bt} \nb_{\alp} \rd_{\bt} \Phi + \bfC^{\alp} \rd_{\alp} \Phi+ W \Phi,
\end{equation}
where
\begin{equation*}
	Q_{0} = - 2 \rd_{u} \rd_{r} + \rd_{r}^{2} - \frac{(d-1)(d-3)}{4} \frac{1}{r^{2}} + \frac{1}{r^{2}} \rslap,
\end{equation*}
and $\bfh$, $\bfC$, $W$ are defined by \eqref{eq:bfh-def}, \eqref{eq:bfC-def} and \eqref{eq:W-def}, respectively.
\end{lemma}
We remark that $Q_{0}$ is the conjugated operator corresponding to $\Box_{\bfm}$, i.e.,
\begin{equation*}
	\Box_{\bfm} (r^{-\nu_{\Box}} \Psi) = r^{-\nu_{\Box}} Q_{0} \Psi.
\end{equation*}
The meaning of the subscript $0$ in $Q_{0}$ will become clear in Section~\ref{subsec:aag} below.
\begin{proof}[Proof of Lemma~\ref{lem:conj-wave}]
Recalling the definition of $\calP$ in \eqref{eq:P.def}, we compute as follows:
\begin{equation}
\begin{split}
	& r^{\nB} \calP (r^{-\nB} \Phi) \\
	&= (\bfg^{-1})^{\alp \bt} \bfD_{\alp} \rd_{\bt} \Phi + r^{\nB} (\Box_{\bfg} r^{-\nB})\Phi + 2 \nB (\bfg^{-1})^{\alp \bt} (\rd_{\bt} \log r) \rd_{\alp} \Phi \\
	&\peq  + \bfB^{\alp} \rd_{\alp} \Phi + \nB \bfB^{\alp} (\rd_{\alp} \log r) \Phi + V \Phi \\
	&=   (\bfm^{-1})^{\alp\bt} \nb_{\alp} \rd_{\bt} \Phi + r^{\nB} (\Box_{\bfm} r^{-\nB})\Phi - \f{d-1}r (\bfm^{-1})^{\alp r} \rd_\alp \Phi  \\
	&\peq + \Big[ (\bfg^{-1})^{\alp \bt} - (\bfm^{-1})^{\alp \bt} \Big] \nb_{\alp} \rd_\bt \Phi + (\bfg^{-1})^{\alp \bt} (\bfD_\alp - \nb_\alp) \rd_\bt \Phi  - \f{d-1}r \Big[ (\bfg^{-1})^{\alp r} - (\bfm^{-1})^{\alp r} \Big]  \rd_{\alp} \Phi \\
	&\peq  + \bfB^{\alp} \rd_{\alp} \Phi - \f{d-1}{2r} \bfB^{r} \Phi + r^{\nB} \Big[(\Box_{\bfg} r^{-\nB}) -  (\Box_{\bfm} r^{-\nB}) \Big] \Phi + V \Phi.
\end{split}
\end{equation}
To obtain the conclusion, notice that (by \eqref{eq:m-1})
$$Q_0\Phi = (\bfm^{-1})^{\alp\bt} \nb_{\alp} \rd_{\bt} \Phi + r^{\nB} (\Box_{\bfm} r^{-\nB})\Phi - \f{d-1}r (\bfm^{-1})^{\alp r} \rd_\alp \Phi,$$
and that the remaining terms are exactly given by \eqref{eq:conj-wave}, \eqref{eq:bfh-def}, \eqref{eq:bfC-def} and \eqref{eq:W-def} (where we have noted that $r^{\nB}(\Box_{\bfm} r^{-\nB}) = -\frac{(d-1)(d-3)}{4} \frac{1}{r^{2}}$). \qedhere
\end{proof}

We note that by \ref{hyp:med}, \ref{hyp:wave-g}, \ref{hyp:wave-B} and \ref{hyp:wave-V}, it follows that $\bfh$, $\bfC$ and $W$ satisfy the following properties.

\begin{lemma} \label{lem:conj-wave-med}
Let $\bfh$, $\bfC$, $W$ be defined by \eqref{eq:bfh-def}, \eqref{eq:bfC-def} and \eqref{eq:W-def}, respectively. In $\calM_{\med}$, we have
\begin{align}
	\bfh &\in O_{\bfGmm}^{M_{c}}(r^{-\dlt_{c}}) \label{eq:med-hab}, \\
	\bfC &\in O_{\bfGmm}^{M_{c}-1}(r^{-1-\dlt_{c}}) \label{eq:med-C}, \\
	W &\in O_{\bfGmm}^{M_{c}-1}(r^{-2-\dlt_{c}}) \label{eq:med-W}.
\end{align}
Moreover, for $M \leq M_{c}$ and any function $\rho = O_{\bfGmm}^{M}(A_\rho r^{-\alp_\rho+\nB} u^{\bt_\rho} )$, the following holds
\begin{equation}\label{eq:conj-wave-med-main}
\bfh^{\alp\bt} \nabla_\alp \rd_\bt\rho + \bfC^\alp \rd_\alp \rho + W \rho= O_{\bfGmm}^{M-2} (A_\rho r^{-2-\de_c} r^{-\alp_\rho+\nB} u^{\bt_\rho} ).
\end{equation}
\end{lemma}

\begin{lemma} \label{lem:conj-wave-wave}
Let $\bfh$, $\bfC$, $W$ be defined by \eqref{eq:bfh-def}, \eqref{eq:bfC-def} and \eqref{eq:W-def}, respectively.
In $\calM_{\wave}$, the components of $\bfh$ with respect to $(u, r, \tht)$ admit expansions of the form
\begin{align}
	\bfh^{uu} &=
	\sum_{j=2}^{J_{c}} \sum_{k=0}^{K_{c}} r^{-j} \log^{k} (\tfrac{r}{u}) \rh_{j, k}^{uu}(u, \tht)
	+ \rem_{J_{c}+\eta_{c}}[\bfh^{uu}],   \label{eq:wave-h-uu-phg} \\
	\bfh^{u r} &=
	\sum_{j=1}^{J_{c}-1} \sum_{k=0}^{K_{c}} r^{-j} \log^{k} (\tfrac{r}{u}) \rh_{j, k}^{ur}(u, \tht)
	+ \rem_{J_{c}-1+\eta_{c}}[\bfh^{ur}],   \label{eq:wave-h-ur-phg} \\
	r \bfh^{u A} &=
	\sum_{j=1}^{J_{c}-1} \sum_{k=0}^{K_{c}} r^{-j} \log^{k} (\tfrac{r}{u}) \rsh_{j, k}^{uA}(u, \tht)
	+ \rem_{J_{c}-1+\eta_{c}}[r \bfh^{uA}],   \label{eq:wave-h-uA-phg} \\
	\bfh^{r r} &= \sum_{j=0}^{J_{c}-2} \sum_{k=0}^{K_{c}} r^{-j} \log^{k} (\tfrac{r}{u}) \rh_{j, k}^{rr}(u, \tht)
	+ \rem_{J_{c}-2+\eta_{c}}[\bfh^{rr}],   \label{eq:wave-h-rr-phg} \\
	r \bfh^{r A} &=
	\sum_{j=0}^{J_{c}-2} \sum_{k=0}^{K_{c}} r^{-j} \log^{k} (\tfrac{r}{u}) \rsh_{j, k}^{r A}(u, \tht)
	+ \rem_{J_{c}-2+\eta_{c}}[r\bfh^{rA}],   \label{eq:wave-h-rA-phg} \\
	r^{2} \bfh^{AB} &=
	\sum_{j=0}^{J_{c}-2} \sum_{k=0}^{K_{c}} r^{-j} \log^{k} (\tfrac{r}{u}) \rsh_{j, k}^{AB}(u, \tht)
	+ \rem_{J_{c}-2+\eta_{c}}[r^{2} \bfh^{AB}],   \label{eq:wave-h-AB-phg}
\end{align}
where
\begin{align}
	\rh_{j, k}^{uu}, \,
	\rh_{j, k}^{ur}, \,
	\rsh_{j, k}^{uA}, \,
	\rh_{j, k}^{rr}, \,
	\rsh_{j, k}^{rA}, \,
	\rsh_{j, k}^{AB}
	 &= O_{\bfGmm}^{M_{c}}(u^{j-\dlt_{c}}), \label{eq:wave-hjk-phg}
\end{align}
whenever the object on the left-hand side is defined, and
the remainders satisfy
\begin{align}
	\rem_{J_{c}+\eta_{c}}[\bfh^{uu}]
	 &= O_{\bfGmm}^{M_{c}}(r^{-(J_{c}+\eta_{c})} u^{J_{c}+\eta_{c}-\dlt_{c}}), \label{eq:wave-hrem-uu-phg} \\
	\rem_{J_{c}-1+\eta_{c}}[\bfh^{ur}], \rem_{J_{c}-1+\eta_{c}}[r \bfh^{uA}]
	 &= O_{\bfGmm}^{M_{c}}(r^{-(J_{c}-1+\eta_{c})}u^{J_{c}-1+\eta_{c}-\dlt_{c}}), \label{eq:wave-hrem-u-phg}  \\
	\rem_{J_{c}-2+\eta_{c}}[\bfh^{rr}],
	\rem_{J_{c}-2+\eta_{c}}[r \bfh^{rA}],
	\rem_{J_{c}-2+\eta_{c}}[r^{2} \bfh^{AB}]
	 &= O_{\bfGmm}^{M_{c}}(r^{-(J_{c}-2+\eta_{c})}u^{J_{c}-2+\eta_{c}-\dlt_{c}}). \label{eq:wave-hrem-phg}
\end{align}
In $\calM_{\wave}$, components of $\bfC$ with respect to $(u, r, \tht)$ admit expansions of the form
\begin{align}
	\bfC^{u} &= \sum_{j=1}^{J_{c}} \sum_{k=0}^{K_{c}'} r^{-j} \log^{k} (\tfrac{r}{u}) \rbfC_{j, k}^{u}(u, \tht) + \rem_{J_{c}+\eta_{c}}[\bfC^{u}], \label{eq:wave-C-u-phg} \\
	\bfC^{r} &= \sum_{j=1}^{J_{c}-1} \sum_{k=0}^{K_{c}'} r^{-j} \log^{k} (\tfrac{r}{u}) \rbfC_{j, k}^{r}(u, \tht) + \rem_{J_{c}-1+\eta_{c}}[\bfC^{r}], \label{eq:wave-C-r-phg} \\
	r \bfC^{A} &= \sum_{j=1}^{J_{c}-1} \sum_{k=0}^{K_{c}'} r^{-j} \log^{k} (\tfrac{r}{u}) \rsbfC_{j, k}^{A}(u, \tht) + \rem_{J_{c}-1+\eta_{c}}[r \bfC^{A}], \label{eq:wave-C-A-phg}
\end{align}
for some $K_{c}'$ depending only on $d$, $J_{c}$ and $K_{c}$, where
\begin{align}
	\rbfC_{j, k}^{u}, \rbfC_{j, k}^{r}, \rsbfC_{j, k}^{A}
	 &= O_{\bfGmm}^{M_{c}-1}(u^{j-1-\dlt_{c}}), \label{eq:wave-Cjk-phg}
\end{align}
whenever the object on the left-hand side is defined, and the remainders satisfy
\begin{align}
	\rem_{J_{c}+\eta_{c}}[\bfC^{u}] &= O_{\bfGmm}^{M_{c}-1}(r^{-(J_{c}+\eta_{c})}u^{J_{c}-1+\eta_{c}-\dlt_{c}}). \label{eq:wave-Crem-u-phg} \\
	\rem_{J_{c}-1+\eta_{c}}[\bfC^{r}], \rem_{J_{c}-1+\eta_{c}}[r \bfC^{A}] &= O_{\bfGmm}^{M_{c}-1}(r^{-(J_{c}-1+\eta_{c})}u^{J_{c}-2+\eta_{c}-\dlt_{c}}). \label{eq:wave-Crem-phg}
\end{align}
Finally, in $\calM_{\wave}$, $W$ admits an expansion of the form
\begin{align}
	W &= \sum_{j=2}^{J_{c}} \sum_{k=0}^{K_{c}'} r^{-j} \log^{k} (\tfrac{r}{u}) \rW_{j, k}(u, \tht) + \rem_{J_{c}+\eta_{c}}[W], \label{eq:wave-W-phg}
\end{align}
for some $K_{c}'$ depending only on $d$, $J_{c}$ and $K_{c}$, where
\begin{align}
	\rW_{j, k}(u, \tht) &\in O_{\bfGmm}^{M_{c}-1}(u^{j-2-\dlt_{c}}), \label{eq:wave-Wjk-phg}\\
	\rem_{J_{c}+\eta_{c}}[W] &\in O_{\bfGmm}^{M_{c}-1}(r^{-(J_{c}+\eta_{c})}u^{J_{c}-2+\eta_{c}-\dlt_{c}}). \label{eq:wave-Wrem-phg}
\end{align}

Moreover, for $M \leq M_{c}$ and any function $\rho = O_{\bfGmm}^{M}(A_\rho r^{-\alp_\rho+\nB} u^{\bt_\rho} )$, it holds that
\begin{equation}\label{eq:conj-wave-wave-main}
\bfh^{\alp\bt} \nabla_\alp \rd_\bt\rho + \bfC^\alp \rd_\alp \rho + W \rho= O_{\bfGmm}^{M-2} (A_\rho r^{-2} u^{-\de_{c}} r^{-\alp_\rho+\nB} u^{\bt_\rho} \log^{K_{c}'} (\tfrac ru)).
\end{equation}

\end{lemma}
\begin{proof}[Proof of Lemma~\ref{lem:conj-wave-med} and Lemma~\ref{lem:conj-wave-wave}]
The estimates for $\bfh$, $\bfC$ and $W$ in both lemmas are straightforward to verify from our hypotheses (namely, \ref{hyp:med}, \ref{hyp:wave-g}, \ref{hyp:wave-B} and \ref{hyp:wave-V}) and the definitions of $\bfh$, $\bfC$ and $W$ (namely, \eqref{eq:bfh-def}, \eqref{eq:bfC-def} and \eqref{eq:W-def}). Hence, we shall only sketch the main points and omit the details of the proof. In Lemma~\ref{lem:conj-wave-med}, the loss of one vector field regularity in \eqref{eq:med-C} and \eqref{eq:med-W} arise from the appearance of $\bfg^{-1} \rd \bfg$ through ${}^{(\bfg)} \Gmm$ in \eqref{eq:bfC-def} and \eqref{eq:W-def}. There is a similar loss of one vector field regularity in \eqref{eq:wave-Crem-u-phg}, \eqref{eq:wave-Crem-phg} and \eqref{eq:wave-Wrem-phg} in Lemma~\ref{lem:conj-wave-wave}. Moreover, the possibly higher powers of $\log (\tfrac{r}{u})$ appears in the expansion for $\bfC$ and $W$ (with $K_{c}'$ instead of $K_{c}$) because of the appearance of  $\bfg$

Finally, \eqref{eq:conj-wave-med-main} is an easy consequence of the estimates for $\bfh$, $\bfC$ and $W$ in Lemma~\ref{lem:conj-wave-med}. For \eqref{eq:conj-wave-wave-main}, using the estimates above, all terms obey the acceptable estimates except we only have the slightly weaker estimate for the terms
$$\bfh^{\alp\bt} {}^{(\bfm)}\Gamma^u_{\alp\bt} \rd_u \rho  ,\, \bfC^u \rd_u \rho  = O_{\bfGmm}^{M-2} (A_\rho r^{-1} u^{-1-\de_{c}} r^{-\alp_\rho+\nB} u^{\bt_\rho} \log^{K_{c}'} (\tfrac ru)).$$
To conclude \eqref{eq:conj-wave-wave-main}, we thus need a cancellation in the term
\begin{equation}\label{eq:cancellation.for.Cu}
({}^{(\bfm)}\Gamma^u_{\mu\nu} \bfh^{\mu\nu} - \bfC^u) \rd_u \rho.
\end{equation}
The only uncontrollable contribution from $\bfC^u$ comes from $\left( {}^{(\bfg)} \Gmm_{\mu \nu}^{u} - {}^{(\bfm)} \Gmm^{u}_{\mu \nu} \right) (\bfg^{-1})^{\mu \nu}$. We rewrite
\begin{equation*}
\begin{split}
&\: {}^{(\bfm)}\Gamma^u_{\mu\nu} \bfh^{\mu\nu} + \left( {}^{(\bfg)} \Gmm_{\mu \nu}^{u} - {}^{(\bfm)} \Gmm^{u}_{\mu \nu} \right) (\bfg^{-1})^{\mu \nu}
=  - {}^{(\bfm)}\Gamma^u_{\mu\nu} \bfm^{\mu\nu} + {}^{(\bfg)} \Gmm_{\mu \nu}^{u} (\bfg^{-1})^{\mu \nu}.
\end{split}
\end{equation*}
By \eqref{eq:m-1}, we compute
$${}^{(\bfm)}\Gamma^u_{\mu\nu} \bfm^{\mu\nu} = \f{1}{\sqrt{-\det \bfm}}\rd_r  \Big((\bfm^{-1})^{ur}\sqrt{-\det \bfm} \Big) = -\f{d-1}{r}.$$
To compute ${}^{(\bfg)} \Gmm_{\mu \nu}^{u} (\bfg^{-1})^{\mu \nu}$, we use the assumption \ref{hyp:wave-g} to compute
\begin{equation}\label{eq:det.expansion}
\begin{split}
\det \bfg^{-1} = &\: -r^{-2(d-1)} \det \Big( ((\rsgmm^{-1})^{AB} + \rsh_{1, 0}^{AB})_{A,B=1}^{d-1} \Big)(u,\th) \\
&\: + \sum_{j=1}^{J_{c}-2} \sum_{k=0}^{(d+1)K_{c}} r^{-2(d-1)-j}\log^k(\tfrac ru) \mathring{d}_{j,k}(u,\th) + r^{-2(d-1)} \rem_{J_{c}-2+\eta_{c}}[\det](u,r,\th),
\end{split}
\end{equation}
where $\mathring{d}_{j,k} = O_{\bfGmm}^{M_{c}}(u^{j-\de_{c}})$ and $\rem_{J_{c}+\eta_{c}}[\det] = (r^{-(J_{c}-2+\eta_{c})}u^{J_{c}-2+\eta_{c}-\de_{c}})$.
(Notice in particular that the $j=0$ terms in $(\bfg^{-1})^{rr}$ and $r(\bfg^{-1})^{rA}$ do not contribute to $\det \bfg^{-1}$ in the lowest order.)

Using \eqref{eq:wave-g-uu-phg}, \eqref{eq:wave-g-ur-phg}, \eqref{eq:wave-g-uA-phg} and \eqref{eq:det.expansion}, it follows that
\begin{equation*}
\begin{split}
&\: {}^{(\bfg)} \Gmm_{\mu \nu}^{u} (\bfg^{-1})^{\mu \nu} = \f{1}{\sqrt{-\det \bfg}}\rd_\alp  \Big((\bfg^{-1})^{u\alp}\sqrt{-\det \bfg} \Big)  =  -\f{d-1}{r} + O_{\bfGmm}^{M_{c}}(r^{-2} u^{1-\de_{c}}).
\end{split}
\end{equation*}
Thus the main terms in \eqref{eq:cancellation.for.Cu} cancel and the bound \eqref{eq:conj-wave-wave-main} holds. \qedhere
%
\end{proof}

\begin{remark}\label{rmk:angular.cancellations}
Recall from Remark~\ref{rmk:gAB.log} that we could allow for logarithmic terms in $\log^k (\tfrac ru)\rsh_{0,k}^{AB}$ in the expansion \eqref{eq:wave-g-AB-phg}, provided we assume that $\det \Big( ((\rsgmm^{-1})^{AB} + \sum_{k=0}^{K_{c}} \log^k(\tfrac ru) \rsh_{0,k}^{AB})_{A,B=1}^{d-1} \Big)$ is a function of $(u,\th)$ alone. In this case, the first term on the right-hand side of \eqref{eq:det.expansion} now becomes
$$-r^{-2(d-1)} \det \Big( ((\rsgmm^{-1})^{AB} + \sum_{k=0}^{K_{c}} \log^k(\tfrac ru)\rsh_{0,k}^{AB})_{A,B=1}^{d-1} \Big)(u,\th).$$
Importantly, since $\det \Big( ((\rsgmm^{-1})^{AB} + \sum_{k=0}^{K_{c}}\log^k(\tfrac ru) \rsh_{0,k}^{AB})_{A,B=1}^{d-1} \Big)(u,\th)$ is a function of $(u,\th)$ alone, the remainder of the proof proceeds similarly as before.
\end{remark}

\subsection{Vector field commutators}

We will use the following (standard) commuting vector fields: the \emph{translation vector fields} $\bfT_{\alp} = \rd_{\alp}$, the \emph{scaling vector field} $\bfS = u \rd_{u} + r \rd_{r}$ (in $(u,r,\th)$ coordinates), the \emph{rotation vector fields} $\bfOmg_{a b} = x^{a} \rd_{b} - x^{b} \rd_{a}$ and the \emph{Lorentz boosts} $\bfL_{a} = (x^{0} + 3 R_{\far}) \rd_{a} + x^{a} \rd_{0}$ (in the $(x^{0}, \ldots, x^{d})$ coordinates). In $\calM_{\far}$, we use $\bfT = \bfT_0$, $\bfS$ and $\bfOmg$ as commuting vector fields, while later in $\calM_{\ext}$, we will use all of the commuting vector fields introduced above. As is well-known, they have simple commuting properties with $\Box_{\bfm}$:
\begin{lemma} \label{lem:ext-vf-comm}
For $\bfGmm \in \set{\bfT_\alp, \bfS, \bfOmg_{ab}, \bfL_a}$, define
\begin{equation*}
	c_{\bfT} = 0, \quad c_{\bfS} = 2, \quad c_{\bfOmg} = 0, \quad c_{\bfL} = 0.
\end{equation*}
Then we have
\begin{equation*}
	(\bfGmm + c_{\bfGmm}) \Box_{\bfm} = \Box_{\bfm} \bfGmm.
\end{equation*}
\end{lemma}
We omit the short proof. These vector fields are related to $u \rd_{u}$ and $r \rd_{r}$ (used in our definition of $O_{\bfGmm}$) in the following way:
\begin{lemma} \label{lem:ext-vf}
In the $(u, r, \tht)$ coordinates, we have, for $x^{0} \geq 0$, $r > 0$ and $u \leq 2$,
\begin{align*}
	r \rd_{r} &= O_{\bfGmm}^{\infty}(1) \bfS + \sum_{a} O_{\bfGmm}^{\infty}(1) \bfL_{a}, \\
	u \rd_{u} &= O_{\bfGmm}^{\infty}(1) \bfS + \sum_{a} O_{\bfGmm}^{\infty}(1) \bfL_{a}, \\
	\rd_{r} &= \sum_{\alp} O_{\bfGmm}^{\infty}(1) \bfT_{\alp}, \\
	\rd_{u} &= \sum_{\alp }O_{\bfGmm}^{\infty}(1) \bfT_{\alp}.
\end{align*}
Conversely, for $x^{0} \geq 0$, $r > 0$ and $u \leq 2$
\begin{align*}
	\bfS &= u \rd_{u} + r \rd_{r}, \\
	\bfL_{a} &= O_{\bfGmm}^{\infty}(1) u \rd_{u} +  O_{\bfGmm}^{\infty}(1) r \rd_{r}  - \sum_{b} O_{\bfGmm}^{\infty}(1) \bfOmg_{a b}, \\
	\bfT_{\alp} &= O_{\bfGmm}^{\infty}(1) \rd_{u} +  O_{\bfGmm}^{\infty}(1) \rd_{r}  + \sum_{b} O_{\bfGmm}^{\infty}(\tfrac{1}{r}) \bfOmg_{a b}.
\end{align*}
\end{lemma}
\begin{proof}
The expression $\bfS = u \rd_{u} + r \rd_{r}$ holds by definition. Recall that, $u = x^{0} - r + 3 R_{\far}$ (by \eqref{eq:far-r-u-tht}), and that in the rectangular coordinates, we have $\bfL_{a} = x^{a} \rd_{x^{0}} + (x^{0} + 3 R_{\far}) \rd_{x^{a}}$ and $\bfOmg_{a b} = x^{a} \rd_{x^{b}} - x^{b} \rd_{x^{a}}$. It follows that, in the $(u, r, \tht)$ coordinates, we have
\begin{align}
	\bfT_{0} &= \rd_{u}, \quad
	\bfT_{a} = - \frac{x^{a}}{r} \rd_{u} + \frac{x^{a}}{r} \rd_{r} + \frac{x^{b}}{r^{2}} \bfOmg_{a b}, \label{eq:trans2vfs} \\
	\bfL_{a} &= x^{a} \rd_{u} +  \frac{(u + r) x^{a}}{r} (\rd_{r} - \rd_{u}) - \frac{(u + r) x^{b}}{r^{2}} \bfOmg_{a b}. \label{eq:boost2vfs}
\end{align}
The coefficients in front of $\rd_{u}$, $\rd_{r}$ and $\frac{1}{r} \bfOmg_{ab}$ in \eqref{eq:trans2vfs} and $u \rd_{u}$, $r \rd_{r}$ and $\bfOmg_{ab}$ in \eqref{eq:boost2vfs} are $O_{\bfGmm}^{\infty}(1)$, as desired.

To proceed, observe from \eqref{eq:boost2vfs}
\begin{equation*}
	\frac{x^{a}}{r} \bfL_{a}
	= (u + r) \rd_{r} - u \rd_{u}.
\end{equation*}
Using $\bfS = u \rd_{u} + r \rd_{r}$ and solving for $u \rd_{u}$ and $r \rd_{r}$, we obtain
\begin{equation*}
	r \rd_{r} = \frac{r}{u+2r}\bfS + \frac{x^{a}}{u + 2r} \bfL_{a}, \quad u \rd_{u} = \frac{u + r}{u+2r}\bfS - \frac{x^{a}}{u + 2r} \bfL_{a}.
\end{equation*}
Observe that the coefficients in front of $\bfS$ and $\bfL_{a}$ are $O_{\bfGmm}^{\infty}(1)$. Finally, note also that
\begin{equation*}
	\rd_{r} = \bfT_{0} + \frac{x^{a}}{r} \bfT_{a}, \quad
	\rd_{u} = \bfT_{0}.
\end{equation*}
where the coefficients in front of $\bfT_{\alp}$ are clearly $O_{\bfGmm}^{\infty}(1)$.\qedhere
\end{proof}

\subsection{Vector field bounds and elliptic estimates} \label{subsec:kl-sid}

In this subsection, we show the derivation of $O_{\bfGmm}(\cdot)$ bounds (see \eqref{eq:O-Gmm-far-scalar}, \eqref{eq:O-Gmm-far}) on Minkowski spacetime using $\bfT$, $\bfS$ and $\bfOmg$ as commutators. By Lemma~\ref{lem:ext-vf-comm}, if we used all the commuting vector fields in $\set{\bfT_\alp, \bfS, \bfOmg_{ab}, \bfL_a}$, the $O_{\bfGmm}(\cdot)$ bounds in $\calM_{\ext}$ are immediate. The key point here is that the subset $\bfT = \bfT_{0}$, $\bfS$ and $\bfOmg$ already suffices. This argument involves a spacetime elliptic estimates, which can be viewed as a spacetime version of the ideas in the work of Klainerman--Sideris \cite{KlSi}.

\begin{lemma} \label{lem:kl-sid-med}
Fix $(U,R)$ such that $1 \leq R \leq \f{1}{27} U$. For $\alp \in [\f 14,\f 12]$, define
\begin{equation}\label{eq:CUR.def.small.R}
\begin{split}
^{(\alp)}{C}_{U}^{R} = &\: \{(u,r,\theta):  (1-\alp)U \leq u \leq (1+\alp)U,\, (1-\alp)R \leq r \leq (1+\alp)R\}.
\end{split}
\end{equation}
Then, for every $s \in \mathbb Z_{\geq 0}$ and $\f 14 \leq \alp' < \alp \leq \f 12$,
\begin{equation} \label{eq:kl-sid-med}
	\sum_{\abs{I} + i \leq s+2} \nrm{r^{\abs{I}}\rd^{I} (u \bfT)^i h}_{L^{2}(^{(\alp)}C_{U}^{R})}
	\aleq \nrm{\bfGmm^{\leq s+2} h}_{L^{2}(^{(\alp')}{C}_{U}^{R})}
	+ \sum_{\abs{I}+ i \leq s} R^{\abs{I} + 2} \nrm{\rd^{I} (u\bfT)^i \Box_{\bfm} h}_{L^{2}(^{(\alp')}{C}_{U}^{R})} ,
\end{equation}
where the implicit constant depends only on $d$, $s$, $\alp$, $\alp'$, and $\bfGmm$ runs over $\bfGmm \in \{\bfOmg, \bfS\}$.
\end{lemma}

\begin{proof}
The key step (Step~1) will be to realize that a suitable linear combination of $\bfS^2$ and $\Box_{\bfm}$ is elliptic, and use elliptic estimates to gain $u\rd_u$ and $r\rd_r$ control.

\pfstep{Step~1: The radial derivatives in the $s=0$ case} In Bondi coordinates, $\bfS = u\rd_u + r \rd_r$. We compute
\begin{equation}\label{eq:S.elliptic.1}
\bfS^2 h - \bfS h = u^2 \rd_u^2 h + r^2 \rd_r^2 h + 2 ur \rd^2_{ur} h .
\end{equation}
On the other hand,
\begin{equation}\label{eq:S.elliptic.2}
-2 \rd^2_{ur} h + \rd_r^2 h + \f{d-1}r \rd_r h - \f{d-1}r \rd_u h = \Box_{\bfm} h + O_{\bfGmm}^{\infty}(r^{-2}) \sum_{|I|= 2} \bfOmg^I h
\end{equation}
Taking  $\mbox{\eqref{eq:S.elliptic.2}} \times r^2 + 9\times \mbox{\eqref{eq:S.elliptic.1}}$, we obtain
\begin{equation}\label{eq:S.elliptic.3}
\begin{split}
&\: u^2 \rd_u^2 h + 10 r^2 \rd_r^2 h + 9(d-1)r (\rd_r - \rd_u) h + 2 (ur-9r^2) \rd^2_{ur} h\\
= &\:  \bfS^2 h - \bfS h + r^2 \Box_{\bfm} h + O_{\bfGmm}^{\infty}(1) \sum_{|I| = 2} \bfOmg^I h.
\end{split}
\end{equation}
The key point here is that the operator on the left-hand side of \eqref{eq:S.elliptic.3} is elliptic. We take the second order terms on the left-hand side and observe that
\begin{equation}\label{eq:S.elliptic.4}
\begin{split}
&\: \Big( u^2 \rd_u^2 h + 10 r^2 \rd_r^2 h + 2 (ur-9r^2) \rd^2_{ur} h \Big)^2 \\
= &\: \underbrace{u^4 (\rd_u^2 h)^2 }_{=:I} + \underbrace{100 r^4 (\rd_r^2 h)^2}_{=:II} + \underbrace{4 (ur-9r^2)^2 (\rd^2_{ur} h)^2 }_{=:III} \\
&\: + \underbrace{20 u^2 r^2 \rd_u^2 h \rd_r^2 h }_{=:IV} + \underbrace{4 u^2 (ur-9r^2)\rd_u^2 h \rd^2_{ur} h }_{=:V} + \underbrace{40 r^2 (ur-2r^2) \rd_r^2 h \rd^2_{ur} h}_{=:VI}.
\end{split}
\end{equation}
The terms $I$, $II$, $III$ in \eqref{eq:S.elliptic.4} are non-negative. The term $IV$ can be written as a non-negative term plus lower order terms:
\begin{equation}\label{eq:S.elliptic.5}
\begin{split}
&\: 20 u^2 r^2 \rd_u^2 h \rd_r^2 h \\
= &\:  20 u^2 r^2 (\rd^2_{ur} h)^2 + 20 \rd_u (u^2 r^2 \rd_u h \rd_r^2 h) - 20 \rd_r (u^2 r^2 \rd_u h \rd_{ur}^2 h) + 40 u^2 r \rd_u h \rd_{ur}^2 h - 40 u r^2 \rd_u h \rd_r^2 h \\
= &\:  20 u^2 r^2 (\rd^2_{ur} h)^2 + \mbox{acceptable error terms},
\end{split}
\end{equation}
where, from now on, we say that a term is an ``acceptable error'' if it is of the form $X_1 h X_2 h$, $X_1 h X_2 X_3 h$, $\rd_u (u X_1 h X_2 X_3 h)$ or $\rd_r (r X_1 h X_2 X_3 h)$, where $X_i \in \{ r \rd_r, r \rd_u, u \rd_u\}$. For $V$, $VI$ in \eqref{eq:S.elliptic.4}, notice that
\begin{align}
| V| = &\: (\f 43 u |\rd_u^2 h|)(3 r(u-9r) \rd^2_{ur} h|)   \leq \f 89 u^4 (\rd_u^2 h)^2 + \f 92 (ur-9r^2)^2 (\rd^2_{ur} h)^2, \label{eq:S.elliptic.6} \\
|VI| = &\: (10 r^2 |\rd_r^2 h|)(4 (ur-9r^2)|\rd^2_{ur}h| )\leq 50 r^2 (\rd_r^2 h)^2 + 8 (ur-9r^2)^2 (\rd^2_{ur} h)^2. \label{eq:S.elliptic.7}
\end{align}
Combining \eqref{eq:S.elliptic.3}--\eqref{eq:S.elliptic.7}, we thus obtain that when $r \leq \f u9$,
\begin{equation}\label{eq:S.elliptic.8}
\begin{split}
\f 19 u^4 (\rd_u^2 h)^2 + 50 r^4 (\rd_r^2 h)^2 + \f{23}2 u^2 r^2 (\rd^2_{ur} h)^2 \leq \Big( \mbox{RHS of \eqref{eq:S.elliptic.3}} \Big)^2 + \mbox{acceptable error terms}.
\end{split}
\end{equation}
(Here, we noted that the first-order terms on the left-hand side of \eqref{eq:S.elliptic.3} are also acceptable errors.)

Let $\chi$ be a cutoff function $\equiv 1$ on $^{(\alp')}C^R_U$ and supported in $^{(\alp)}C^R_U$. To lighten the notation, we write $^{(\alp')}C = {}^{(\alp')}C^R_U$, $^{(\alp)}C = {}^{(\alp)}C^R_U$ for the rest of the proof. We choose $\chi$ such that $|\rd_r^i \rd_u^j \chi| \ls R^{-i} U^{-j}$. Considering $\int \chi^2 \mbox{\eqref{eq:S.elliptic.8}}$ (with the Minkowskian volume form $r^{d-1}\ud\rssgm\ud r \ud u$, which is notationally suppressed), we obtain
\begin{equation}\label{eq:S.elliptic.9}
\begin{split}
&\: \| \chi u^2 \rd_u^2 h \|_{L^2(^{(\alp)}C)}^2 + \| \chi r^2 \rd_r^2 h \|_{L^2(^{(\alp)}C)}^2 + \| \chi ru \rd^2_{ur} h \|_{L^2(^{(\alp)}C)}^2 \\
\ls &\:  \| \bfGmm^{\leq 2} h \|_{L^{2}(^{(\alp)}C)}^2 + R^2 \| \Box_\bfm h \|_{L^{2}(^{(\alp)}C)}^2 + \int \chi^2 \times \mbox{acceptable error terms},
\end{split}
\end{equation}
where we have used that $r \leq \f u9$ in $^{(\alp)}C$ (by \eqref{eq:CUR.def.small.R}). The acceptable error terms can be treated with integration by parts, e.g.,
\begin{equation}\label{eq:S.elliptic.10}
\begin{split}
\Big| \int \chi^2 u^2 r \rd_u h \rd_{ur}^2 h \Big| = &\: \Big| \f 12 \int \rd_r(\chi^2  r^{d}) u^2 (\rd_u h)^2\, \ud \rssgm \, \ud r \, \ud u \Big| \\
\leq &\:  \Big| \f 14 \int \rd^3_{uur}(\chi^2  r^{d} u^2) h^2 \, \ud \rssgm \, \ud r \, \ud u \Big| + |\f 12 \int \rd_r(\chi^2 r^d) u^2 h \rd_u^2 h \, \ud \rssgm \, \ud r \, \ud u \Big| \\
\ls &\: \| h\|_{L^{2}(^{(\alp)}C)}^2 + \| h \|_{L^{2}(^{(\alp)}C)} \|\chi u^2 \rd_u^2 h \|_{L^{2}(^{(\alp)}C)}.
\end{split}
\end{equation}
All other error terms can be treated in a similar manner. We omit the details but just note that
\begin{equation}\label{eq:S.elliptic.11}
\begin{split}
\int \chi^2 \times \mbox{acceptable error terms}
\ls  \| h\|_{L^{2}(^{(\alp)}C)}^2 + \| h \|_{L^{2}(^{(\alp)}C)} (\mbox{LHS of \eqref{eq:S.elliptic.9}})^{\f 12}.
\end{split}
\end{equation}

Plugging \eqref{eq:S.elliptic.11} back into \eqref{eq:S.elliptic.9}, we have thus proved that
\begin{equation}\label{eq:S.elliptic.12}
\| \chi u^2 \rd_u^2 h \|_{L^2(^{(\alp)}C)} + \| \chi r^2 \rd_r^2 h \|_{L^2(^{(\alp)}C)} + \| \chi ru \rd^2_{ur} h \|_{L^2(^{(\alp)}C)} \ls \mbox{RHS of \eqref{eq:kl-sid-med}}.
\end{equation}

\pfstep{Step~2: $s=0$ for all derivatives} We now control (up to second order) derivatives of $h$ which are not already controlled in \eqref{eq:S.elliptic.12}. First, note that
\begin{equation}\label{eq:S.elliptic.13}
\| \chi \bfOmg^{(\leq 2)} h \|_{L^2(^{(\alp)}C)} \ls \mbox{RHS of \eqref{eq:kl-sid-med}}.
\end{equation}
This is trivially true because the right-hand side include the terms with rotations. Integrating by parts, and treating the lower order terms that arise as in \eqref{eq:S.elliptic.10}, we can then control the mixed radial and angular derivatives using \eqref{eq:S.elliptic.12} and \eqref{eq:S.elliptic.13} so that
\begin{equation}\label{eq:S.elliptic.14}
\| \chi u \rd_u \bfOmg h \|_{L^2(^{(\alp)}C)} + \| \chi r \rd_r \bfOmg h \|_{L^2(^{(\alp)}C)} \ls \mbox{RHS of \eqref{eq:kl-sid-med}}.
\end{equation}

With \eqref{eq:S.elliptic.12}--\eqref{eq:S.elliptic.14}, we have thus controlled all second derivatives. For zeroth order derivative and first order angular derivative, we can use \eqref{eq:S.elliptic.13}. It thus remains only to control the first order $u\rd_u$ or $r\rd_r$ derivatives. This can be achieved by integrations by parts, e.g.,
$$\int \chi^2 u^2 (\rd_u h)^2 \ls U \| h \|_{L^2({^{(\alp)}C})} \| \chi \rd_u h \|_{L^2({^{(\alp)}C})} + U^2 \| h \|_{L^2({^{(\alp)}C})} \| \rd_u^2 h \|_{L^2({^{(\alp)}C})} \ls \mbox{RHS of \eqref{eq:kl-sid-med}},$$
where we have used the bounds \eqref{eq:S.elliptic.12}--\eqref{eq:S.elliptic.13} already obtained. The $r\rd_r$ derivative can be treated similarly.

\pfstep{Step~3: Higher order estimates} We now carry out an induction in $s$. Suppose $\exists s_0 \in \mathbb Z_{\geq 0}$ such that \eqref{eq:kl-sid-med} holds for $s\leq s_0$.

Introduce a new scale $\alp'' $ with $\alp' < \alp'' < \alp$. Since \eqref{eq:kl-sid-med} holds for arbitrary regular functions, we use the induction hypothesis to obtain
\begin{equation}\label{eq:kl-sid-higher.with.Gmm}
\begin{split}
	&\: \sum_{\substack{\abs{I} + i \leq s_0+2}} \nrm{r^{\abs{I}}\rd^{I} (u \bfT)^{i} \bfGmm h}_{L^{2}(^{(\alp'')}C)} \\
	\aleq &\:   \nrm{\bfGmm^{\leq s_0+3} h}_{L^{2}(^{(\alp')}{C})}
	+ \sum_{\abs{I}+i \leq s_0 } R^{\abs{I} + 2} \nrm{\rd^{I} (u\bfT)^i \Box_{\bfm} (\bfGmm h)}_{L^{2}(^{(\alp')}{C})} \\
	\ls &\: \nrm{\bfGmm^{\leq s_0+3} h}_{L^{2}(^{(\alp')}{C})}
	+ \sum_{\abs{I}+i \leq s_0 +1} R^{\abs{I} + 2} \nrm{\rd^I (u\bfT)^i \Box_{\bfm} h}_{L^{2}(^{(\alp')}{C})},
\end{split}
\end{equation}
where in the last step we used $[\Box_{\bfm}, \bfGmm] = c_{\bfGmm} \Box_{\bfm}$ and bound $\bfGmm$ trivially by coordinates derivatives.

Now, we start with the $s=0$ case of \eqref{eq:kl-sid-med} (applied to all $r^{\abs{I'}}\rd^{I'} (u \bfT)^{i'} h$ with $|I'|+i' \leq s_0+2$) and compute as follows:
\begin{equation}\label{eq:kl-sid-higher}
\begin{split}
&\:  \sum_{\substack{\abs{I} + i \leq s_0+3}} \nrm{r^{\abs{I}}\rd^{I} (u \bfT)^{i} h}_{L^{2}(^{(\alp)}C)} \\
\ls &\: \sum_{\substack{\abs{I} + i \leq s_0+1}} \nrm{\bfGmm^{\leq 2} (r^{\abs{I}}\rd^{I} (u \bfT)^{i} h)}_{L^{2}(^{(\alp'')}C)} + \sum_{|I|+i\leq s_0+1} R^2\nrm{\Box_{\bfm} (r^{\abs{I}}\rd^{I} (u \bfT)^{i} h)}_{L^{2}(^{(\alp'')}C)} \\
\ls &\: \sum_{\substack{\abs{I} + i \leq s_0+2}} \nrm{r^{\abs{I}}\rd^{I} (u \bfT)^{i} \bfGmm h}_{L^{2}(^{(\alp'')}C)} + \sum_{|I|+i\leq s_0+1} R^2\nrm{r^{\abs{I}}\rd^{I} (u \bfT)^{i} \Box_{\bfm} h}_{L^{2}(^{(\alp'')}C)} \\
&\: + \sum_{\substack{\abs{I} + i \leq s_0+2}} \nrm{r^{\abs{I}}\rd^{I} (u \bfT)^{i} h}_{L^{2}(^{(\alp'')}C)} \\
\ls &\: \sum_{\substack{\abs{I} + i \leq s_0+2}} \nrm{r^{\abs{I}}\rd^{I} (u \bfT)^{i} \bfGmm h}_{L^{2}(^{(\alp')}C)} + \sum_{|I|+i\leq s_0+1} R^{|I|+2}\nrm{\rd^{I} (u \bfT)^{i} \Box_{\bfm} h}_{L^{2}(^{(\alp)}C)}  + \nrm{\bfGmm^{\leq s_0+2} h}_{L^{2}(^{(\alp')}{C})}.
\end{split}
\end{equation}
where in the second step we commuted the derivatives at the expense of the last commutator term, and in the last step we controlled the commutator term using  the induction hypothesis.

Combining \eqref{eq:kl-sid-higher.with.Gmm} and \eqref{eq:kl-sid-higher}, we obtain \eqref{eq:kl-sid-med} for $=s_0+1$, which completes the induction. \qedhere

\end{proof}

We need an analogue of Lemma~\ref{lem:kl-sid-med} in the complementary region $U\ls R$. The estimate here turns out to be slightly simpler; in particular, the exact constant in $U\ls R$ plays no role. We will only sketch the derivative of the elliptic operator in this case; the rest of the proof follows as in Lemma~\ref{lem:kl-sid-med}.
\begin{lemma} \label{lem:kl-sid-wave}
Suppose $1\leq U \aleq R$. For $\alp \in [\f 14, \f 12]$, define $^{(\alp)}{C}_{U}^{R}$ as in \eqref{eq:CUR.def.small.R}. Denote $\bar{\rd} \in \{\rd_r, \f 1r \mathring{\slashed{\nabla}}\}$. Then the following estimate holds for all $s \in \mathbb Z_{\geq 0}$ and $\f 14 \leq \alp' < \alp \leq \f 12$:
\begin{equation*}
\sum_{\abs{I} +i \leq s+2}R^{\abs{I}} \nrm{(u\rd_u)^i \bar{\rd}^I h}_{L^{2}(^{(\alp')}C_U^R)}
	\aleq  \nrm{\bfGmm^{\leq s+2} h}_{L^{2}(^{(\alp)}C_U^R)}
	+ \sum_{\abs{I} +i \leq s} U R^{|J|+1} \nrm{(u\rd_u)^i \bar{\rd}^J \Box_{\bfm} h}_{L^{2}(^{(\alp)}C_{U}^{R})},
\end{equation*}
where the implicit constant depends only on $d$, $s$, $\alp$, $\alp'$, and $\bfGmm$ runs over $\bfGmm \in \{\bfOmg, \bfS\}$.
\end{lemma}
\begin{proof}
We compute further with \eqref{eq:S.elliptic.1} to obtain
\begin{equation}\label{eq:S.elliptic.1.alt}
\begin{split}
\bfS^2 h - \bfS h = u^2 \rd_u^2 h + r^2 \rd_r^2 h + 2 ur \rd^2_{ur} h = u^2 (\rd_u - \f 12 \rd_r)^2 + \f 14 (u+2r)^2 \rd_r^2 + u(u+2r) (\rd_u - \f 12 \rd_r) \rd_r.
\end{split}
\end{equation}
This is because
$$u^2 (\rd_u - \f 12 \rd_r)^2 = u^2 \rd^2_u - u^2 \rd^2_{ur} + \f 14 u^2\rd_r^2, \quad \f 14 (u+2r)^2 \rd_r^2 = \f 14 u^2 \rd_r^2 + ur \rd_r^2 + r^2 \rd_r^2.$$


Taking $\mbox{\eqref{eq:S.elliptic.1.alt}} + \f 12 u ( u+ 2r) \times \mbox{\eqref{eq:S.elliptic.2}}$, we obtain
\begin{equation}\label{eq:S.elliptic.total.alt}
\begin{split}
&\: u^2 (\rd_u - \f 12 \rd_r)^2 h+ \f 14 (u+2r)^2 \rd_r^2 h + \f{(d-1)u(u+2r)}{2r} (\rd_r h -\rd_u h) \\
= &\: \bfS^2 h - \bfS h + \f 12 u(u+2r) \Box_{\bfm} h + O_{\bfGmm}^{\infty}(1) \sum_{|I| = 2} \bfOmg^I h.
\end{split}
\end{equation}

The operator on the left-hand side of \eqref{eq:S.elliptic.total.alt} is obviously elliptic (as there are no cross terms $\rd_r (\rd_u - \f 12 \rd_r) h$). Therefore, the remainder of the proof proceeds as in Lemma~\ref{lem:kl-sid-med}. \qedhere

\end{proof}


\subsection{Klainerman's Sobolev-type lemmas} \label{subsec:kl-sob}

In this subsection, we upgrade the $L^2$ estimates in Section~\ref{subsec:kl-sid} to obtain $L^2$--$L^\infty$ type weighted Sobolev result.

For the following lemma, we work in general $\mathbb R^{d'}$; we denote by $\rd$ the coordinate derivatives with respect to $x = (x^{1}, \ldots, x^{d})$ on $\bbR^{d}$.
\begin{lemma} \label{lem:rescaled-sob}
Let $d'\in \mathbb N$ be odd, $\mathfrak R >0$ and $c>1$. Define $B_{\frkR},\,B_{\frkR}^+\subset \mathbb R^{d'}$ by $B_{\frkR} = \{x \in \mathbb R^{d'}: |x|<{\frkR}\}$ and $B_{\frkR}^+ = \{ x \in \mathbb R^d: |x|< {\frkR},\, x^d \geq 0\}$. Then the following rescaled Sobolev embedding inequalities hold:
\begin{align}
	\| h\|_{L^\infty(B_{\frkR})} \aleq {\frkR}^{-\frac{d}{2}} \nrm{(\frkR \rd)^{(\leq \f{d+1}2)}h}_{L^{2}(B_{c{\frkR}})}, \label{eq:rescaled-sob} \\
	\| h\|_{L^\infty(B_{\frkR}^+)} \aleq {\frkR}^{-\frac{d}{2}} \nrm{(\frkR \rd)^{(\leq \f{d+1}2)}h}_{L^{2}(B_{c{\frkR}}^+)}, \label{eq:rescaled-sob.half}
\end{align}
where the implicit constants depend only on $c$ and $d$.
\end{lemma}
\begin{proof}
We claim that for any Schwartz function $\frkh$, the following holds:
\begin{equation}\label{eq:Sobolev.simple.2}
\| \frkh\|_{L^\i(\mathbb R^{d'})}  \ls   \| \frkh \|_{L^2(\mathbb R^{d'})}^{\f 1{d'+1}}\| \rd^{\f{d'+1}2} \frkh \|_{L^2(\mathbb R^{d'})}^{\f {d'}{d'+1}},\quad \| \frkh\|_{L^\i(\mathbb R_+^{d'})}  \ls   \| \frkh \|_{L^2(\mathbb R_+^{d'})}^{\f 1{d'+1}}\| \rd^{\f{d'+1}2} \frkh \|_{L^2(\mathbb R_+^{d'})}^{\f {d'}{d'+1}},
\end{equation}
where $\mathbb R_+^{d'} = \{ x\in \mathbb R^{d'}: x^{d'} \geq 0\}$.

We will prove the first inequality in \eqref{eq:Sobolev.simple.2}. Without loss of generality, assume $h \not\equiv 0$ (for otherwise this is obvious). By standard Sobolev embedding,
\begin{equation}\label{eq:Sobolev.simple}
\| \frkh\|_{L^\i(\mathbb R^{d'})} \ls \| \frkh \|_{L^2(\mathbb R^{d'})} + \| \rd^{\f{d+1}2} \frkh \|_{L^2(\mathbb R^{d'})}.
\end{equation}
Applying \eqref{eq:Sobolev.simple} to $\frkh_\lambda(x) = \frkh(\lambda x)$ for all $\lambda >0$, we obtain
\begin{equation}
\| \frkh\|_{L^\i(\mathbb R^{d'})} = \| \frkh_\lambda \|_{L^\i(\mathbb R^{d'})} \ls \| \frkh_\lambda \|_{L^2(\mathbb R^{d'})} + \| \rd^{\lfloor \f d2 \rfloor +1} \frkh_\lambda \|_{L^2(\mathbb R^{d'})} = \lambda^{-\f {d'}2} \| \frkh \|_{L^2(\mathbb R^{d'})} + \lambda^{\f 12} \| \rd^{\f{d'+1}2} \frkh_\lambda \|_{L^2(\mathbb R^{d'})}.
\end{equation}
Choosing $\lambda = \| \frkh\|_{L^2(\mathbb R^{d'})}^{\f 2{d'+1}} \| \rd^{\f{d'+1}2} \frkh \|_{L^2(\mathbb R^{d'})}^{-\f 2{d'+1}}$, we obtain the first inequality in \eqref{eq:Sobolev.simple}. The second inequality is similar, except that instead of \eqref{eq:Sobolev.simple}, we start with the standard Sobolev embedding on a half-space.

To prove \eqref{eq:rescaled-sob} (resp.~\eqref{eq:rescaled-sob.half}), we apply the first (resp.~the second) estimate in \eqref{eq:Sobolev.simple.2} to $\frkh = \chi h$, where $\chi \geq 0$ is a suitably defined smooth cutoff function which is $\equiv 1$ in $B_{\frkR}$ (resp.~$B_{\frkR}^+$) and is supported in $B_{c\frkR}$ (resp.~$B_{c\frkR}^+$). \qedhere
\end{proof}


\begin{lemma} \label{lem:kl-sob}
Let $C^R_U = {}^{(\f 12)}C^R_U$ (see \eqref{eq:CUR.def.small.R}).
\begin{equation}
\begin{split}
&\: \sum_{I_u + I_r + |I| \leq s} |(u\rd_u)^{I_u} (r\rd_r)^{I_r} \bfOmg^{I} h|(U,R,\Theta)\\
\ls &\: U^{-\f 12} R^{-\f d2}\Big( \| \bfGmm^{\leq s+\f{d+3}2} h \|_{L^2(C^R_U)} + R \min\{R,U\} \sum_{I_u + I_r + |I| \leq s + \f{d+3}2} \| (u\rd_u)^{I_u} (r\rd_r)^{I_r} \bfOmg_{ab}^{I_{ab}} \Box_{\bfm} h \|_{L^2(C^R_U)} \Big).
\end{split}
\end{equation}
\end{lemma}
\begin{proof}
We apply \eqref{eq:rescaled-sob} in Lemma~\ref{lem:rescaled-sob} twice, first with $d'=d$ for each fixed $u$ and then with $d'=1$ for the $u$-dimension. 
Hence, we obtain
\begin{equation}
\sum_{I_u + I_r + |I| \leq s} |(u\rd_u)^{I_u} (r\rd_r)^{I_r} \bfOmg^{I} h|(U,R,\Theta) \ls R^{-\f d2} U^{-\f 12} \sum_{I_u + I_r + |I| \leq s+ \f{d+1}2} \| (u\rd_u)^{I_u} (r\rd_r)^{I_r} \bfOmg^{I} h \|_{L^2({}^{(\f 14)}C^R_U)}.
\end{equation}
Now applying the estimates in Lemmas~\ref{lem:kl-sid-med} and \ref{lem:kl-sid-wave} yields the desired result. \qedhere
\end{proof}

\subsection{Angelopoulos--Aretakis--Gajic commutation identities} \label{subsec:aag}

In this section, we derive (Minkowskian) commutation identities for $Q_{0}$ and the commutation vector fields
\begin{equation*}
\bfK = r^{2} \rd_{r},\quad r\rd_r,\quad \bfS = r \rd_{r} + u\rd_{u},\quad \bfOmg_{jk},
\end{equation*}
where $\bfOmg_{jk}$ are the usual rotations on Minkowski spacetime. The commutations with $\bfK$ will be treated in Section~\ref{sec:aaq.K}, while all the other commutations will be computed in Section~\ref{sec:aaq.Gmm}.

The use of $\bfK$ as a commutator comes from the work of Angelopoulos--Aretakis--Gajic. This idea is also very natural in view of our method of using expansions in powers and logarithms of $r^{-1}$ in the Bondi--Sachs-type coordinate system $(u, r, \tht)$, since $\bfK$ can also be viewed as the coordinate vector field $\rd_{z}$ in the coordinate system $(u, z = r^{-1}, \tht)$.

\subsubsection{Commutation identities for $\bfK$}\label{sec:aaq.K}

Recall that we denoted the conjugated d'Alembertian on $\bbR^{d+1}$ by
\begin{align}\label{eq:Q0-def}
	Q_{0} = - 2 \rd_{u} \rd_{r} + \rd_{r}^{2} - \frac{(d-1)(d-3)}{4} \frac{1}{r^{2}} + \frac{1}{r^{2}} \rslap.
\end{align}
For $j \geq 0$, we define $Q_{j+1}$ by the relation
\begin{equation} \label{eq:Qj-def}
	(\bfK+2r) Q_{j} = Q_{j+1} \bfK.
\end{equation}
A straightforward computation gives the following expression for $Q_{j}$.
\begin{lemma} \label{lem:Qj}
For each nonnegative integer $j$, we have
\begin{equation} \label{eq:Qj}
	Q_{j} = - 2 \rd_{u} \rd_{r} + \rd_{r}^{2} - \frac{2j}{r} \rd_{r} + \left(j + \frac{d-1}{2}\right)\left(j - \frac{d-3}{2}\right) \frac{1}{r^{2}} + \frac{1}{r^{2}} \rslap.
\end{equation}
\end{lemma}
\begin{proof}
Note the basic identities
\begin{align}\label{eq:K.basic}
[\bfK, \rd_{u}] = 0, \quad
[\bfK, \rd_{r}] = -2 r^{-1} \bfK, \quad
[\bfK, r^\ell] = \ell r^{\ell+1}, \quad
[\bfK, \rslap] = 0.
\end{align}
From this, it is easy to deduce that
\begin{align}
[\bfK, \rd_u \rd_r] = &\:  \rd_u [\bfK, \rd_r] = -2 r^{-1} \rd_u \bfK, \\
[\bfK, \rd_r^2] = &\: -2 r^{-1} \bfK \rd_r -2 \rd_r (r^{-1} \bfK) = -2 r^{-1} \rd_r \bfK + 4 r^{-2} \bfK - 2 r^{-1} \rd_r \bfK + 2 r^{-2} \bfK \notag \\
= &\: -4r^{-1} \rd_r \bfK + 6 r^{-2} \bfK, \\
[\bfK, r^{-1} \rd_r] = &\: - \rd_r - 2 r^{-2} \bfK = -3 r^{-2} \bfK.
\end{align}

We now prove the identity \eqref{eq:Qj} by induction in $j$. The base case holds by definition (see \eqref{eq:Q0-def}). Assume that \eqref{eq:Qj} holds for some $j\geq 0$.

Then, by \eqref{eq:K.basic}, we have
\begin{equation}\label{eq:K.commute.almost}
\begin{split}
[\bfK, Q_j] = &\: 4 r^{-1} \rd_u \bfK - 4 r^{-1} \rd_r \bfK + 6 r^{-2} \bfK + 6jr^{-2} \bfK \\
&\: - 2 \left(j + \frac{d-1}{2}\right)\left(j - \frac{d-3}{2}\right) r^{-1} -2 r^{-1} \rslap \\
= &\: - 2 r Q_j + 2 r \rd_r^2 -4j r^{-2} \bfK - 4 r^{-1} \rd_r \bfK + (6j + 6)r^{-2} \bfK \\
= &\: - 2 r Q_j + 2 r^{-1} \rd_r \bfK - 4 r^{-2} \bfK -4j r^{-2} \bfK - 4 r^{-1} \rd_r \bfK + (6j + 6)r^{-2} \bfK \\
= &\: - 2 r Q_j - 2 r^{-1} \rd_r \bfK + (2j + 2)r^{-2} \bfK.
\end{split}
\end{equation}
Using \eqref{eq:K.commute.almost}, and noting that $-2j - 2 = -2(j+1)$ and
\begin{equation*}
\begin{split}
&\:  \left(j + \frac{d-1}{2}\right)\left(j - \frac{d-3}{2}\right) + (2j + 2) \\
= &\: j^2 + j - \f{(d-1)(d-3)}{4} + 2j + 2 = j^2 + 3 j  - \f{(d-1)(d-3)}{4} + 2 \\
= &\: \left(j + 1+ \frac{d-1}{2}\right)\left(j + 1- \frac{d-3}{2}\right),
\end{split}
\end{equation*}
we obtain \eqref{eq:Qj} for $j+1$ in place of $j$, thus completing the induction. \qedhere
\end{proof}

\subsubsection{Commutation identities for $\protect\bfGmm \in \protect\set{r \rd_{r}, \protect\bfS, \protect\bfOmg}$}\label{sec:aaq.Gmm}

We begin with the following simple commutation identities.
\begin{lemma} \label{lem:comm-Qj}
For each nonnegative integer $j$, we have
\begin{align}
	[\bfOmg, \bfK] &= 0, \label{eq:comm-K-Omg}\\
	[\bfS, \bfK] &= \bfK, \label{eq:comm-K-S}\\
	[r \rd_{r}, \bfK] &= \bfK, \label{eq:comm-K-rdr} \\
	[\bfOmg, Q_{j}] &= 0, \label{eq:comm-Qj-Omg}\\
	[\bfS, Q_{j}] &= - 2 Q_{j}. \label{eq:comm-Qj-S}
\end{align}
\end{lemma}
\begin{proof}
The identities \eqref{eq:comm-K-Omg} and \eqref{eq:comm-Qj-Omg} are obvious, and identities \eqref{eq:comm-K-S} and \eqref{eq:comm-Qj-S} follow from the homogeneity properties of $\bfK$ and $Q_{j}$, respectively (i.e., $\bfK(\phi(\lmb^{-1} \cdot)) = \lmb (\bfK \phi)(\lmb^{-1} \cdot)$ and $Q_{j}(\phi(\lmb^{-1} \cdot)) = \lmb^{-2} (Q_{j} \phi)(\lmb^{-1} \cdot)$ for all $\lmb > 0$, respectively). Then \eqref{eq:comm-K-rdr} follows since $[\bfS - r \rd_{r}, \bfK] = [u \rd_{u}, r^{2} \rd_{r}] = 0$.  \qedhere
\end{proof}


The following is the main result concerning the commutation of $Q_j$ (recall \eqref{eq:Qj-def}, \eqref{eq:Qj}) and $\bfGmm$:
\begin{proposition}
For $\bfGmm \in \set{r \rd_{r}, \bfS, \bfOmg}$, define $c_{\bfGmm}$ by
\begin{equation*}
	c_{\bfS} = 2, \quad c_{r \rd_{r}} = 1, \quad c_{\bfOmg} = 0.
\end{equation*}
Define also the multi-index notation that
\begin{equation*}
	\bfGmm^{I} = (r\rd_r)^{I_{r\rd_r}} \bfS^{I_{\bfS}} \Pi_{j,k:j<k} \bfOmg_{jk}^{I_{\bfOmg_{jk}}},\quad I = (I_{r\rd_r},I_{\bfS},(I_{\bfOmg_{jk}})_{j,k:j<k}).
\end{equation*}

Then
\begin{equation} \label{eq:comm-Qj}
\begin{aligned}
	(\bfGmm + c_{\bfGmm})^{I} Q_{j} - Q_{j} \bfGmm^{I} &= I_{r \rd_{r}} \left( - r^{-1} \rd_{r} \bfGmm^{I} - r^{-2} \rslap \bfGmm^{I-(1, \ldots)} \right) \\
	&\peq + O_{\bfGmm}^{\infty} (r^{-1}) \rd_{r} \bfGmm^{(\leq \abs{I}-1)}
	+ O_{\bfGmm}^{\infty} (r^{-2}) \rslap \bfGmm^{(\leq \abs{I}-2)}
	+ O_{\bfGmm}^{\infty} (r^{-2}) \bfGmm^{(\leq \abs{I}-1)}.
\end{aligned}
\end{equation}
Moreover, in the case $j = \ell + \nB - 1$, after projecting to the $\ell$-th spherical harmonics, we have, for $\bfGmm \in \{ r \rd_{r}, \bfS\}$,
\begin{equation} \label{eq:comm-Qj-ell}
\begin{aligned}
	(\bfGmm + c_{\bfGmm})^{I}  Q_{\ell+\nu_{\Box}-1} \bbS_{(\ell)}  - Q_{\ell+\nu_{\Box}-1} \bfGmm^{I} \bbS_{(\ell) }
	&= - I_{r \rd_{r}} r^{-1} \rd_{r} \bfGmm^{I} \bbS_{(\ell)}
	+ O_{\bfGmm}^{\infty} (r^{-1}) \rd_{r} \bfGmm^{(\leq \abs{I}-1)} \bbS_{(\ell)} \\
	&\peq + O_{\bfGmm}^{\infty} (r^{-2}) \bfGmm^{(\leq \abs{I}-1)} \bbS_{(\ell)}.
\end{aligned}
\end{equation}
\end{proposition}
\begin{proof}
We prove \eqref{eq:comm-Qj} by induction. The base case ($|I|=0$) is obvious. The induction in $I_{\bfOmg_{jk}}$ and $I_{\bfS}$ are also easy and follows from \eqref{eq:comm-Qj-Omg} and \eqref{eq:comm-Qj-S}, respectively.

To complete the proof of \eqref{eq:comm-Qj}, it thus remains to compute the commutator $[r \rd_{r}, Q_{j}]$:
\begin{equation} \label{eq:comm-Qj-rdr}
\begin{split}
[r \rd_{r}, Q_{j}]
&= \left[ r \rd_{r}, - 2 \rd_{u} \rd_{r} + \rd_{r}^{2} - \frac{2j}{r} \rd_{r} + \left(j + \frac{d-1}{2}\right)\left(j - \frac{d-3}{2}\right) \frac{1}{r^{2}} + \frac{1}{r^{2}} \rslap \right] \\
&= 2 \rd_{u} \rd_{r} - 2 \rd_{r}^{2} + 2 \frac{2j}{r} \rd_{r} - 2\left(j + \frac{d-1}{2}\right)\left(j - \frac{d-3}{2}\right) \frac{1}{r^{2}} - 2 \frac{1}{r^{2}} \rslap \\
&= - Q_{j} - \rd_{r}^{2} + \frac{2j}{r} \rd_{r} - \left(j + \frac{d-1}{2}\right)\left(j - \frac{d-3}{2}\right) \frac{1}{r^{2}} - \frac{1}{r^{2}} \rslap.
\end{split}
\end{equation}

We rewrite \eqref{eq:comm-Qj-rdr} as
\begin{align}\label{eq:comm-Qj-rdr-1}
	(r \rd_{r} + 1) Q_{j} - Q_{j} r \rd_{r} = - r^{-1} \rd_{r} r \rd_{r} - r^{-2} \rslap + (2j+1) r^{-2} r \rd_{r}  - (j + \nu_{\Box})(j-\nu_{\Box}+1) r^{-2},
\end{align}
from which it is clear that we can induct to obtain \eqref{eq:comm-Qj}.

Finally, to prove \eqref{eq:comm-Qj-ell}, notice that when $j = \ell + \nB - 1$, after acting $\bbS_{(\ell)} $ on \eqref{eq:comm-Qj-rdr-1}, the angular Laplacian cancels with the inverse square potential term so that
\begin{align*}
	(r \rd_{r} + 1) Q_{\ell+\nu_{\Box}-1} \bbS_{(\ell)}- Q_{\ell+\nu_{\Box}-1} r \rd_{r} \bbS_{(\ell)} = - r^{-1} \rd_{r} r \rd_{r} \bbS_{(\ell)} + (2 \ell+2 \nu_{\Box}-1) r^{-2} r \rd_{r} \bbS_{(\ell)}.
\end{align*}
Using this in the induction, we obtain \eqref{eq:comm-Qj-ell}. \qedhere


%
%
%
%

\end{proof}

\begin{remark}
It is important to note that the main term $ - r^{-1} \rd_{r} r \rd_{r} - r^{-2} \rslap$ on the right-hand side of \eqref{eq:comm-Qj} contributes to a good bulk term in the $r^{p} \rd_{r}$ multiplier argument in Lemma~\ref{lem:Qj-rp} below. This has been observed in \cite{DRNM, gM2016, Schlue}.
\end{remark}

\subsection{The $r^{p}$ estimates in $\calM_{\far}$} \label{subsec:rp}




In this subsection, we recall the Dafermos--Rodnianski $r^p$ identities associated to the (Minkowskian) operators $Q_j$ (see the work \cite[Proposition~6.5]{AAG2020} of Angelopoulos--Aretakis--Gajic), as well as a version for higher derivatives.

To state the $r^p$ identities, introduce the following notation for subsets of $\calM_{\far}$:
\begin{align*}
	\calC_{U} &= \set{(u, r, \tht) \in \calM_{\far} : u = U}, \quad
	\ul{\calC}_{V} = \set{(u, r, \tht) \in \calM_{\far} : u + 2r = V}, \\
	\calD_{U_{0}}^{U} &= \set{(u, r, \tht) \in \calM_{\far} : U_{0} \leq u \leq U}.
\end{align*}

\begin{lemma} \label{lem:Qj-rp-id}
Consider the inhomogeneous problem
\begin{equation}\label{eq:QjPsi=G}
Q_{j} \Psi = G \quad \hbox{ in } \calM_{\far}.
\end{equation}
Let $R_{1} \geq 2R_{\far}$. For any $p \in \bbR$, we have
\begin{equation} \label{eq:Qj-rp-id}
\begin{aligned}
& 2 \int_{\calC_{U}} \chi_{>R_{1}} r^{p} (\rd_{r} \Psi)^{2}  \, \ud r \ud \rssgm(\tht) \\
&+\frac{1}{2} \lim_{V \to \infty} \int_{\ul{\calC}_{V} \cap \calD_{U_{0}}^{U}} r^{p-2} \left( \abs{\rsnb \Psi}^{2} - (j-\nu_{\Box}+1)(j+\nu_{\Box}) \Psi^{2} \right) \, \ud u  \ud \rssgm(\tht) \\
&+ \f 12\iint_{\calD_{U_{0}}^{U}} \chi_{>R_{1}} (p + 4j) r^{p-1} (\rd_{v} \Psi)^{2} \, \ud u \ud r  \ud \rssgm(\tht) \\
&+ \frac{1}{2} \iint_{\calD_{U_{0}}^{U}} \chi_{>R_{1}} (2-p) r^{p-3} \left(\abs{\rsnb \Psi}^{2} - (j-\nu_{\Box}+1)(j+\nu_{\Box}) \Psi^{2} \right) \, \ud u \ud r  \ud \rssgm(\tht) \\
&= 2 \int_{\calC_{U_{0}}} \chi_{>R_{1}} r^{p} (\rd_{r} \Psi)^{2}  \, \ud r \ud \rssgm(\tht)
- \iint_{\calD_{U_{0}}^{U}} \chi_{>R_{1}} G r^{p} \rd_{r} \Psi \, \ud u \ud r  \ud \rssgm(\tht) \\
&\peq
+ \iint_{\calD_{U_{0}}^{U}} \chi_{>R_{1}}' \left(- r^{p} (\rd_{r} \Psi)^{2} + \frac{1}{2} r^{p-2} \left(\abs{\rsnb \Psi}^{2} - (j-\nu_{\Box}+1)(j+\nu_{\Box}) \Psi^{2}\right) \right)\, \ud u \ud r  \ud \rssgm(\tht).
\end{aligned}
\end{equation}
\end{lemma}

\begin{remark}[Angelopoulos--Aretakis--Gajic positivity] \label{rem:aag-pos}
We shall apply Lemma~\ref{lem:Qj-rp-id} with $- 4j \leq p <  2$. Note that the last term on the left-hand side is, in general, of an indefinite sign. However, if we apply the projection $\bbS_{(\geq \ell)}$ to the equation and consider $\Psi_{(\geq \ell)}$, then
\begin{align*}
&\int_{\bbS^{d-1}} \left(\abs{\rsnb \Psi_{(\geq \ell)}}^{2} - (j-\nu_{\Box}+1)(j+\nu_{\Box}) \Psi_{(\geq \ell)}^{2} \right) \, \ud \rssgm(\tht) \\
&\geq \int_{\bbS^{d-1}} \left(\ell (\ell+d-2) - (j-\nu_{\Box}+1)(j+\nu_{\Box}) \right) \Psi_{(\geq \ell)}^{2}  \, \ud \rssgm(\tht) \\
&= \int_{\bbS^{d-1}} (\ell - j + \nu_{\Box} - 1)(\ell + j + \nu_{\Box})  \Psi_{(\geq \ell)}^{2}  \, \ud \rssgm(\tht).
\end{align*}
Hence, this term is nonnegative if
\begin{equation*}
\ell \geq j - \nu_{\Box} + 1.
\end{equation*}
This fact was observed by Angelopolous--Aretakis--Gajic \cite{AAG2020, AAGPrice}.
\end{remark}
\begin{proof}
Multiplying both sides of \eqref{eq:QjPsi=G} by $- r^{p} \chi_{>R_{1}} \rd_{r} \Psi$ and using the expression in \eqref{eq:Qj}, we obtain
\begin{align*}
& (\rd_{u} - \f 12 \rd_{r}) \left(2 \chi_{>R_{1}} r^{p} (\rd_{r} \Psi)^{2} \right)
+ \rd_{r} \left(\frac{1}{2} \chi_{>R_{1}} r^{p-2} \abs{\rsnb \Psi}^{2} - \frac{(j-\nu_{\Box}+1)(j+\nu_{\Box})}{2} \chi_{>R_{1}} r^{p-2} \Psi^{2} \right) + \rsdiv(\ldots) \\
& + \f 12 (p + 4j) \chi_{>R_{1}} r^{p-1} (\rd_{r} \Psi)^{2} + \frac{1}{2} (2-p) \chi_{>R_{1}} r^{p-3} \left(\abs{\rsnb \Psi}^{2} - (j-\nu_{\Box}+1)(j+\nu_{\Box}) \Psi^{2} \right) \\
&= - \chi_{>R_{1}} G r^{p} \rd_{r} \Psi - \chi_{>R_{1}}' r^{p} (\rd_{r} \Psi)^{2} + \frac{1}{2} \chi_{>R_{1}}' r^{p-2} \left(\abs{\rsnb \Psi}^{2} - (j-\nu_{\Box}+1)(j+\nu_{\Box}) \Psi^{2}\right),
\end{align*}
where $\rsdiv$ is the angular divergence defined with respect to $\ud \rssgm(\tht)$ on $\bbS^{d-1}$. The lemma now follows by integrating this identity on $\calD_{U_{0}}^{U}$ with respect to the coordinate volume form $\ud u \ud r  \ud \rssgm(\tht)$. \qedhere
\end{proof}

The following is a corollary of Lemma~\ref{lem:Qj-rp-id}, which is in a more convenient form for our use.
\begin{lemma} \label{lem:Qj-rp}
Let $R_{1} \geq 2R_{\far}$. Consider the inhomogeneous problem
\begin{equation*}
Q_{j} \bfGmm^{I} \Psi + I_{r \rd_{r}} \left(r^{-1} \rd_{r} \bfGmm^{I} + r^{-2} \rslap \bfGmm^{I-(1, 0, \ldots)} \right) \Psi = G_{1} + G_{2} \quad \hbox{ in } \calM_{\far}.
\end{equation*}
where
\begin{equation*}
	\Psi = \bbS_{(\geq j - \nu_{\Box} + 1)} \Psi.
\end{equation*}
Then for any $- 4 j < p < 2$, we have
\begin{equation} \label{eq:Qj-rp}
\begin{aligned}
& \sup_{u \in [U_{0}, U]} \left(\int_{\calC_{u}} \chi_{>R_{1}} r^{p} (\rd_{r} \bfGmm^{I} \Psi)^{2}  \, \ud r \ud \rssgm(\tht) \right)^{\frac{1}{2}} \\
&+ \left( \iint_{\calD_{U_{0}}^{U}} \chi_{>R_{1}} r^{p-1} \left((\rd_{r} \bfGmm^{I} \Psi)^{2} + r^{-2} \abs{\rsnb \bfGmm^{I} \Psi}^{2} + r^{-2} (\bfGmm^{I} \Psi)^{2} \right)\, \ud u \ud r \ud \rssgm(\tht) \right)^{\frac{1}{2}} \\
&\aleq_{p, j, I_{r \rd_{r}}} \left( \int_{\calC_{U_{0}}} \chi_{>R_{1}} r^{p} (\rd_{r} \bfGmm^{I} \Psi)^{2}  \, \ud r \ud \rssgm(\tht) \right)^{\frac{1}{2}} \\
&\phantom{\aleq_{p, j, I_{r \rd_{r}}}}
+ \left( \iint_{\calD_{U_{0}}^{U} \cap \set{\frac{1}{2}R_{1} < r < R_{1}}} R_{1}^{p-1} \left(\abs{\rd_{r} \bfGmm^{I} \Psi}^{2} + r^{-2} \abs{\rsnb \bfGmm^{I} \Psi}^{2} + r^{-2} (\bfGmm^{I} \Psi)^{2}\right)\, \ud u \ud r \ud \rssgm(\tht) \right)^{\frac{1}{2}} \\
&\phantom{\aleq_{p, j, I_{r \rd_{r}}}}
+ \int_{U_{0}}^{U} \left( \int_{\calC_{u} \cap \set{r \geq \frac{1}{2} R_{1}}} r^{p} G_{1}^{2} \, \ud r \ud \rssgm(\tht) \right)^{\frac{1}{2}} \ud u
+ \left(\iint_{\calD_{U_{0}}^{U} \cap \set{r \geq \frac{1}{2} R_{1}}} r^{p+1} G_{2}^{2} \, \ud u \ud r \ud \rssgm(\tht)\right)^{\frac{1}{2}} \\
&\phantom{\aleq_{p, j, I_{r \rd_{r}}}} + \left(I_{r \rd_{r}} \iint_{\calD_{U_{0}}^{U}} \chi_{>R_{1}} \frac{(2-p)^{2}}{2} r^{p-3} \abs{\rsnb \bfGmm^{I-(1,0,\ldots)}\Psi}^{2} \, \ud u \ud r \ud \rssgm(\tht) \right)^{\frac{1}{2}} \\
&\phantom{\aleq_{p, j, I_{r \rd_{r}}}} +\liminf_{R \to \infty} \abs*{I_{r \rd_{r}} \iint_{\calD_{U_{0}}^{U} \cap \set{\frac{1}{2} R \leq r \leq R}} R^{-1} r^{p-1} (\rd_{r} + (2-p) r^{-1}) \abs{\rsnb \bfGmm^{I-(1,0,\ldots)}\Psi}^{2} \, \ud u \ud r \ud \rssgm(\tht)}^{\frac{1}{2}}
\end{aligned}
\end{equation}
\end{lemma}
Observe that we have the freedom to split the right-hand side of the equation into $G_{1} + G_{2}$, for each of which we may look for a different control.
\begin{proof}
We begin with the case $I_{r \rd_{r}} = 0$. Observe that, in view of $\Psi = \bbS_{(\geq j - \nu_{\Box}+1)} \Psi$ and Remark~\ref{rem:aag-pos}, each term on the left-hand side of \eqref{eq:Qj-rp-id} is nonnegative. By a straightforward argument involving taking the supremum in $U$, and applying Cauchy--Schwarz to the contributions of $G_{1}$ and $G_{2}$ on the right-hand side, we arrive at
\begin{align*}
& \sup_{u \in [U_{0}, U]} \left(\int_{\calC_{u}} \chi_{>R_{1}} r^{p} (\rd_{r} \bfGmm^{I} \Psi)^{2}  \, \ud r \ud \rssgm(\tht) \right)^{\frac{1}{2}} \\
&+ \left( \iint_{\calD_{U_{0}}^{U}} \chi_{>R_{1}} (p + 4j) r^{p-1} (\rd_{r} \bfGmm^{I} \Psi)^{2} \, \ud u \ud r \ud \rssgm(\tht) \right)^{\frac{1}{2}} \\
&+ \left( \iint_{\calD_{U_{0}}^{U}} \chi_{>R_{1}} r^{p-3} (2-p) \left( \abs{\rsnb \bfGmm^{I} \Psi}^{2} - (j-\nu_{\Box}+1)(j+\nu_{\Box}) (\bfGmm^{I} \Psi)^{2} \right) \, \ud u \ud r \ud \rssgm(\tht) \right)^{\frac{1}{2}} \\
&\aleq \left( \int_{\calC_{U_{0}}} \chi_{>R_{1}} r^{p} (\rd_{r} \bfGmm^{I} \Psi)^{2}  \, \ud r \ud \rssgm(\tht) \right)^{\frac{1}{2}} \\
&\peq
+ \int_{U_{0}}^{U} \left( \int_{\calC_{u} \set{r \geq \frac{1}{2} R_{1}}} r^{p} G_{1}^{2} \, \ud r \ud \rssgm(\tht) \right)^{\frac{1}{2}} \ud u
+ \left(\iint_{\calD_{U_{0}}^{U} \cap \set{r \geq \frac{1}{2} R_{1}}} r^{p+1} G_{2}^{2} \, \ud u \ud r \ud \rssgm(\tht)\right)^{\frac{1}{2}} \\
&\peq
+ \left( \iint_{\calD_{U_{0}}^{U} \cap \set{\frac{1}{2}R_{1} < r < R_{1}}} r^{p-1} \left(\abs{\rd_{r} \bfGmm^{I} \Psi}^{2} + r^{-2} \abs{\rsnb \bfGmm^{I} \Psi}^{2} + (j-\nu_{\Box}+1)(j+\nu_{\Box}) r^{-2} (\bfGmm^{I} \Psi)^{2}\right)\, \ud u \ud r \ud \rssgm(\tht) \right)^{\frac{1}{2}}.
\end{align*}

To complete the proof of \eqref{eq:Qj-rp} when $I_{r \rd_{r}} =0$, note that, since $p < 2$, Lemma~\ref{lem:hardy-radial} below (with $r_{0} = R_{1}$) implies
\begin{align*}
\left( \iint_{\calD_{U_{0}}^{U}} \chi_{>R_{1}} r^{p-3} (\bfGmm^{I} \Psi)^{2} \, \ud u \ud r \ud \rssgm(\tht) \right)^{\frac{1}{2}}
&\aleq_{p}
\left( \iint_{\calD_{U_{0}}^{U}} \chi_{>R_{1}} r^{p-1} (\rd_{r} (\bfGmm^{I} \Psi))^{2} \, \ud u \ud r \ud \rssgm(\tht) \right)^{\frac{1}{2}} \\
&\phantom{\aleq_{p}} + \left( \iint_{\calD_{U_{0}}^{U} \cap \set{\frac{1}{2}R_{1} < r < R_{1}}} r^{p-3} (\bfGmm^{I} \Psi)^{2} \, \ud u \ud r \ud \rssgm(\tht) \right)^{\frac{1}{2}}.
\end{align*}
Combined with the preceding inequality, the desired estimate follows.

Finally, we treat the case when $I_{r \rd_{r}} \neq 0$. We apply Lemma~\ref{lem:Qj-rp-id} to $Q_{j} \bfGmm^{I} \Psi = G_{0} + G_{1} + G_{2}$, where
\begin{equation*}
G_{0} = - I_{r \rd_{r}} (r^{-1} \rd_{r} \bfGmm^{I} \Psi + r^{-2} \rslap \bfGmm^{I-(1, 0, \ldots)} \Psi),
\end{equation*}
and proceed as before. To treat the new term
\begin{equation*}
\bfe_{G_{0}} := \iint_{\calD_{U_{0}}^{U}} \chi_{>R_{1}} G_{0} r^{p} \rd_{r} \bfGmm^{I} \Psi \, \ud u \ud r \ud \rssgm(\tht),
\end{equation*}
we use the following identity: For $R' > R_{1}$, we have
\begin{equation} \label{eq:Qj-rp-I>0}
\begin{aligned}
&- \iint_{\calD_{U_{0}}^{U} \cap \set{r \leq R'}} \chi_{>R_{1}} \left( r^{-1} \rd_{r} \bfGmm^{I} \Psi + r^{-2} \rslap \bfGmm^{I-(1, 0, \ldots)} \Psi \right) r^{p} \rd_{r} \bfGmm^{I} \Psi \, \ud u \ud r \ud \rssgm(\tht) \\
& = - \iint_{\calD_{U_{0}}^{U} \cap \set{r \leq R'}} \chi_{>R_{1}} r^{p-1} (\rd_{r} \bfGmm^{I} \Psi)^{2} + r^{p-3} \abs{\rsnb \bfGmm^{I}\Psi}^{2}  \, \ud u \ud r \ud \rssgm(\tht) \\
&\peq +\left. \int \left[ r^{p-1} (\rsnb^{A} \bfGmm^{I-(1, 0, \ldots)} \Psi) \rsnb_{A} \bfGmm^{I} \Psi + \tfrac{2-p}{2} r^{p-2} \abs{\rsnb \bfGmm^{I-(1,0,\ldots)}\Psi}^{2}\right] \ud u \ud \rssgm(\tht)\right|_{r=R'} \\
&\peq + \frac{(p-2)^{2}}{2}\iint_{\calD_{U_{0}}^{U} \cap \set{r \leq R'}} \chi_{>R_{1}} r^{p-3} \abs{\rsnb \bfGmm^{I-(1,0,\ldots)}\Psi}^{2} \, \ud u \ud r \ud \rssgm(\tht) \\
&\peq + \iint_{\calD_{U_{0}}^{U}  \cap \set{r \leq R'}} \left(\tfrac 12 \chi_{>R_{1}}'' r^{p-1} + \tfrac{2p-3}{2}\chi_{>R_{1}}' r^{p-2}\right) \abs{\rsnb \bfGmm^{I-(1, 0, \ldots)} \Psi}^{2} \, \ud u \ud r \ud \rssgm(\tht).
\end{aligned}
\end{equation}
We omit the proof of this identity, which just consists of several applications of integration by parts in $r$. To arrive at $\bfe_{G_{0}}$, we average this identity in $R'$ over $R' \in [\frac{1}{2} R, R]$, let  $R = R_{n}$ with a sequence $R_{n} \to \infty$ and take the limit as $n \to \infty$ and multiply by $-I_{r \rd_{r}}$. The contribution of the first term on the right-hand side of \eqref{eq:Qj-rp-I>0} has a favorable sign. The remaining terms contribute to the last three lines in \eqref{eq:Qj-rp} (where $\liminf_{R \to \infty}$ arises since the choice of the sequence $R_{n}$ is arbitrary). \qedhere
\end{proof}

\subsection{Integration along Minkowskian radial characteristics} \label{subsec:char}
For low spherical harmonics in $\calM_{\wave}$, we shall use the following simple integration along characteristics lemma.
\begin{lemma} \label{lem:Qj-char}
Consider the inhomogeneous problem (written in the $(u, r, \tht)$-coordinates)
\begin{equation*}
	(Q_{\ell + \nu_{\Box} - 1} - s r^{-1} \rd_{r})\Psi_{(\ell)} = G_{(\ell)} \quad \hbox{ in } \calM_{\far},
\end{equation*}
where $s \in \bbR$. Then
\begin{equation}\label{eq:Qj-char}
\begin{aligned}
\rd_{r} \Psi_{(\ell)}(U, R, \Tht) &= R^{2(\ell + \nu_{\Box} - 1)+s} \left. r^{-2(\ell + \nu_{\Box} - 1)-s} \rd_{r} \Psi_{(\ell)}\right|_{(u, r, \tht) = (U_{0}, R + \frac{U-U_{0}}{2}, \Tht)} \\
&\peq - \frac{1}{2} R^{2(\ell + \nu_{\Box} - 1)+s} \int_{U_{0}}^{U} \left. r^{-2(\ell + \nu_{\Box} - 1)-s} G_{(\ell)} \right|_{(u, r, \tht) = (\sgm, R +\frac{U-\sgm}{2}, \Tht)} \,\ud \sgm.
\end{aligned}\end{equation}
\end{lemma}
\begin{proof}
Write $j = \ell + \nu_{\Box} - 1$. We have
\begin{equation*}
	( - 2 \rd_{u} + \rd_{r}) \rd_{r} \Psi_{(\ell)} - (2 j + s) r^{-1} \rd_{r} \Psi_{(\ell)} = G_{(\ell)},
\end{equation*}
which implies
\begin{equation*}
	( - 2 \rd_{u} + \rd_{r})(r^{-2j-s} \rd_{r} \Psi_{(\ell)}) = r^{-2j-s} G_{(\ell)}.
\end{equation*}
Now the lemma follows by an integration along the integral curve of $\rd_u - \f 12 \rd_r$. \qedhere
\end{proof}

\begin{remark} \label{rem:Qj-char-supp}
Observe that if $(U, V, \Tht) \in \calM_{\wave}$, then the integration path $\set{(u, V, \Tht) \in \calM_{\far} : U_{0} \leq u \leq U}$ also lies in $\calM_{\wave}$.
\end{remark} 


\subsection{Hardy inequality} \label{subsec:hardy}
Here we state the standard one-dimensional Hardy inequality.
\begin{lemma} \label{lem:hardy-radial}
For $p \neq 1$,
\begin{equation} \label{eq:hardy-radial}
\frac{(p-1)^{2}}{4} \int_{r_{0}}^{r_{1}} h^{2} r^{p-2} \, \ud r
\leq \int_{r_{0}}^{r_{1}} (\rd_{r} h)^{2} r^{p} \, \ud r + \frac{p-1}{2} \left.r^{p-1} h^{2}\right|_{r_{0}}^{r_{1}}.
\end{equation}
\end{lemma}
\begin{proof}
For $c \in \bbR$ to be determined below, we compute
\begin{align*}
	r^{p} (\rd_{r} h + c r^{-1} h)^{2}
	&= r^{p} (\rd_{r} h)^{2} + 2 c r^{p-1} h \rd_{r} h + c^{2} r^{p-2} h^{2}.
\end{align*}
Integrating this identity on $[r_{0}, r_{1}]$, writing $2 h \rd_{r} h = \rd_{r} h^{2}$ and integrating by parts, we obtain
\begin{align*}
	\int_{r_{0}}^{r_{1}}r^{p} (\rd_{r} h + c r^{-1} h)^{2} \, \ud r
	= \int_{r_{0}}^{r_{1}} r^{p} (\rd_{r} h)^{2} \, \ud r - \int_{r_{0}}^{r_{1}}((p-1)-c) c r^{p-2} h^{2} \, \ud r + \left. c r^{p-1} h^{2} \right|_{r_{0}}^{r_{1}}.
\end{align*}
Note that the left-hand side is nonnegative. Choosing $c = \frac{p-1}{2}$, which optimizes the expression $((p-1)-c) c$, and rearranging terms, we arrive at \eqref{eq:hardy-radial}. \qedhere
\end{proof}

\subsection{Tools and results for handling the nonlinearity} \label{subsec:nonlin-est}
The main goal of this subsection is to precisely formulate the notion of an admissible decay exponent for $\calN$ (see Definition~\ref{def:alp-N}), which was cited in Definition~\ref{def:alp-N-min} and Proposition~\ref{prop:alp-N}. This notion is designed to streamline the treatment of the nonlinearity in Sections~\ref{sec:near}, \ref{sec:med} and \ref{sec:wave} below. Using this definition, we prove Proposition~\ref{prop:alp-N}, which justifies the use of Definition~\ref{def:alp-N} in applications.

\subsubsection{Convenient expressions for $\calN$ and its difference in $\calM_{\far}$} In order to state the definition of an admissible decay exponent, we need to introduce several other definitions. We begin by rewriting the nonlinearity in $\calM_{\far}$ as we did for the linear terms before. We write
\begin{equation} \label{eq:nonlin-far}
	\calN(\phi) = \quasi^{\mu \nu}(p, \phi, \ud \phi) \nb_{\mu} \rd_{\nu} \phi + \csemi(p, \phi, \ud \phi),
\end{equation}
where
\begin{align*}
	\csemi(p, z, \bfxi) = \semi(p, z, \bfxi) + \quasi^{\mu \nu}(p, z, \bfxi)  \left( {}^{(\bfg)}\bfGmm^{\alp}_{\mu \nu}(p) - {}^{(\bfm)} \bfGmm^{\alp}_{\mu \nu}(p) \right) \bfxi_{\alp}.
\end{align*}
As before, the following lemmas are straightforward to verify using \ref{hyp:nonlin-med} and \ref{hyp:nonlin-wave} (the proofs are omitted):
\begin{lemma} \label{lem:csemi-med}
In $\calM_{\med}$, we have
\begin{align}
	\csemi = O_{\bfGmm}^{M_{c}-1}(1).
\end{align}
\end{lemma}

\begin{lemma} \label{lem:conj-nonlin-wave}
In $\calM_{\wave}$, we have
\begin{align}
\quasi^{\mu \nu} &= \sum_{j = 0}^{J_{c} - \nu_{\Box}} \sum_{k=0}^{K_{c}} \rquasi_{j, k}^{\mu \nu} r^{-j} \log^{k} (\tfrac{r}{u}) + \rem_{J_{c}-\nu_{\Box} +\eta_{c}}[\quasi^{\mu \nu}], \label{eq:wave-quasi-phg} \\
\csemi &= \sum_{j = 0}^{J_{c} - \nu_{\Box}} \sum_{k=0}^{(d+1) K_{c}} \rcsemi_{j, k} r^{-j} \log^{k} (\tfrac{r}{u}) + \rem_{J_{c}-\nu_{\Box} + \eta_{c}}[\csemi], \label{eq:wave-csemi-phg}
\end{align}
with
\begin{align}
\rquasi_{j, k} &= O_{\bfGmm}^{M_{c}}(u^{j}), \label{eq:wave-quasi-jk-phg} \\
\rem_{J_{c}-\nu_{\Box}+\eta_{c}}[\quasi] &= O_{\bfGmm}^{M_{c}}(r^{-J_{c}+\nu_{\Box}-\eta_{c}} u^{J_{c}-\nu_{\Box}+\eta_{c}}), \label{eq:wave-quasirem-phg} \\
\rcsemi_{j, k} &= O_{\bfGmm}^{M_{c}-1}(u^{j}), \label{eq:wave-csemi-jk-phg} \\
\rem_{J_{c}-\nu_{\Box}+\eta_{c}}[\csemi] &= O_{\bfGmm}^{M_{c}-1}(r^{-J_{c}+\nu_{\Box}-\eta_{c}} u^{J_{c}-\nu_{\Box}+\eta_{c}}). \label{eq:wave-csemirem-pgh}
\end{align}
\end{lemma}

We also introduce a set of notation that expresses the difference $\calN(\phi+ \psi) - \calN(\phi)$ (after conjugation by $r^{\nu_{\Box}}$) as almost linear terms in $\Psi = r^{\nu_{\Box}} \psi$. By the fundamental theorem of calculus, we may write $\calN(\phi+ \psi) - \calN(\phi) = \int_{0}^{1} \tfrac{\ud}{\ud h} \calN(\phi + h \psi) \, \ud h$. We may compute the integrand using \eqref{eq:nonlin-chain-rule-z}. Using this computation, we write out $\calN(\phi + \psi) - \calN(\phi)$ and define $\bfh_{\calN}^{\mu \nu}(\phi, \psi)$, $\bfC_{\calN}^{\mu}(\phi, \psi)$ and $W_{\calN}(\phi, \psi)$ (cf.~\eqref{eq:conj-wave}) as follows:
\begin{equation}\label{eq:calN.expansion}
\begin{split}
&\calN(\phi + \psi) - \calN(\phi) \\
&= \nb_{\mu} \rd_{\nu} \psi \int_{0}^{1} \quasi^{\mu \nu}(p, \phi + h \psi, \ud (\phi + h \psi)) \,\ud h  \\
&\peq +\rd_{\bt} \psi \int_{0}^{1} \left( \rd_{\bfxi_{\bt}} \quasi^{\mu \nu}(p, \phi + h \psi, \ud (\phi + h \psi)) \nb_{\mu} \rd_{\nu} (\phi+ h \psi) + \rd_{\bfxi_{\bt}} \csemi(p, \phi+h \psi, \ud (\phi + h \psi)) \right) \,\ud h  \\
&\peq +\psi \int_{0}^{1} \left( \rd_{z} \quasi^{\mu \nu}(p, \phi + h \psi, \ud (\phi + h \psi)) \nb_{\mu} \rd_{\nu} (\phi+ h \psi) + \rd_{z} \csemi(p, \phi+h \psi, \ud (\phi + h \psi)) \right) \,\ud h \\
&=:
 r^{-\nu_{\Box}} \bfh_{\calN}^{\mu \nu}(\phi; \psi) \nb_{\mu} \rd_{\nu} \Psi
+r^{-\nu_{\Box}}  \bfC_{\calN}^{\bt}(\phi; \psi) \rd_{\bt} \Psi
+r^{-\nu_{\Box}}  W_{\calN}(\phi; \psi) \Psi,
\end{split}
\end{equation}
where $\Psi = r^{\nu_{\Box}} \psi$.

\subsubsection{Formal expansion of $\calN$ in the wave zone}
Given $J \in \bbN$, $K \in \bbZ_{\geq 0}^{J}$, and functions $\rPhi_{j, k} = \rPhi_{j, k}(u, \tht)$ for $0 \leq j \leq J$, $0 \leq k \leq K$, $u \geq 1$ and $\tht \in \bbS^{d-1}$, define, in $\calM_{\wave}$,
\begin{equation} \label{eq:Phi<J-formal}
	\Phi_{<J} = \sum_{j=0}^{J-1} \sum_{k=0}^{K} r^{-j} \log^{k}(\tfrac{r}{u}) \rPhi_{j, k}(u, \tht).
\end{equation}
Consider the expression $\calN(r^{-\nu_{\Box}} \Phi_{<J})$ in $\calM_{\wave}$. Using \ref{hyp:nonlin-wave} and Taylor expansion in $(\phi, \rd_{u} \phi, \rd_{r} \phi, r^{-1} \rd_{\tht^{A}} \phi)$, we may obtain a formal expansion of the form
\begin{equation} \label{eq:rNjk}
	\calN(r^{-\nu_{\Box}} \Phi_{<J}) = \sum_{j=0}^{\infty} \sum_{k=0}^{\infty} \rN^{<J}_{j, k}(u, \tht) r^{-j-1} \log^{k}(\tfrac{r}{u}) + E_{J; \nonlinear, \rem}
\end{equation}
where
\begin{equation} \label{eq:E-J-nonlin-rem}
\begin{aligned}
	E_{J; \nonlinear, \rem}
	&= r^{\nu_{\Box}} \rem_{J_{c}-\nu_{\Box}+\eta_{c}}[\quasi^{\mu \nu}](u, r, \tht, r^{-\nu_{\Box}}\Phi_{<J}, \ud (r^{-\nu_{\Box}}\Phi_{<J})) \rd_{\mu} \rd_{\nu} (r^{-\nu_{\Box}}\Phi_{<J}) \\
	&\peq
	+ r^{\nu_{\Box}} \rem_{J_{c}-\nu_{\Box}+\eta_{c}}[\csemi](u, r, \tht, r^{-\nu_{\Box}}\Phi_{<J}, \ud (r^{-\nu_{\Box}}\Phi_{<J})),
\end{aligned}
\end{equation}
and we take \eqref{eq:rNjk} as the defining equation for the coefficients $\rN^{<J}_{j, k} = \rN^{<J}_{j, k}(u, \tht)$. Observe that while the sum in \eqref{eq:rNjk} is in general infinite, each coefficient $\rN^{<J}_{j, k}$ involves at most a finite number of multilinear expressions in $\rPhi_{j, k}$, thanks to \ref{hyp:nonlin}, \ref{hyp:nonlin-degree}, and \ref{hyp:nonlin-wave} (see also the proof of Lemma~\ref{lem:rNjk} below).

We record as a lemma the following algebraic properties of $\rN^{<J}_{j, k}$, which follow from our hypotheses on $\calN$.
\begin{lemma} \label{lem:rNjk}
Let $\calN$ satisfy \ref{hyp:nonlin}, \ref{hyp:nonlin-degree}, and \ref{hyp:nonlin-wave}, as well as \ref{hyp:nonlin-null} if $d = 3$.
\begin{enumerate}
\item $\rN^{<J}_{j, k}$ depends only on $\rPhi_{j', k'}$ with $j' \leq j-1$.
\item For every $J \in \bbN$, $K \in \bbZ_{\geq 0}$  and $\rPhi_{<J}$ as in \eqref{eq:Phi<J-formal}, if $\rPhi_{j', k'} = 0$ for all $0 \leq j' \leq j-1$ and $k' > K'$ for some $K' \in \bbZ_{\geq 0}$, then $\rN^{<J}_{j, k} = 0$ for all $k > K_{c} + (\frac{j+1+\nu_{\Box}}{\nu_{\Box}}) K' $. \end{enumerate}
\end{lemma}
This result will be useful in the rigorous justification of the properties of $\rF_{j, k}$ assumed in Section~\ref{subsec:recurrence-formal}; see Lemmas~\ref{lem:recurrence-forcing-basic} and \ref{lem:recurrence-forcing} below.

\begin{proof}
We need to look a bit closer into the structure of $\rN^{<J}_{j, k}$. Fix $\rPhi_{<J}$ as in \eqref{eq:Phi<J-formal}. Then $\rN^{<J}_{j, k}$ consists of contributions from the following terms:
\begin{itemize}
\item (Quasilinear) With $n = n_{0} + n_{r} + n_{u} + n_{\tht}+1$ (total degree in $(\phi, \ud \phi)$) and $\phi_{i} = r^{-\nu_{\Box}} r^{-j_{i}} \log^{k_{i}} (\tfrac{r}{u}) \rPhi_{j_{i}, k_{i}}$,
\begin{align*}
\hbox{(i)~}&r^{\nu_{\Box}} r^{-n_{\tht}} \mathring{\slashed{c}}_{\rquasi_{j_{0}, k_{0}}}^{A_{1} \ldots A_{n_{\tht}}}(u, \tht) r^{-j_{0}} \log^{k_{0}}(\tfrac{r}{u}) \left(\tprod_{i=1}^{n_{0}} \phi_{i} \right)  \\
&\quad \times \left(\tprod_{i=1}^{n_{u}}\rd_{u} \phi_{n_{0}+i}\right) \left(\tprod_{i=1}^{n_{r}}\rd_{r} \phi_{n_{0}+n_{u}+i}\right) \left(\tprod_{i=1}^{n_{\tht}} \rd_{\tht^{A_{i}}} \phi_{n_{0}+n_{u}+n_{r}+i}\right) (\rd_{u}, \rd_{r}, r^{-1}) (\rd_{u}, \rd_{r}) \phi_{n}, \\
\hbox{(ii)~}&r^{\nu_{\Box}} r^{-n_{\tht}-1} \mathring{\slashed{c}}_{\rquasi_{j_{0}, k_{0}}}^{A; A_{1} \ldots A_{n_{\tht}}}(u, \tht) r^{-j_{0}} \log^{k_{0}}(\tfrac{r}{u}) \left(\tprod_{i=1}^{n_{0}} \phi_{i} \right)  \\
&\quad \times \left(\tprod_{i=1}^{n_{u}}\rd_{u} \phi_{n_{0}+i}\right) \left(\tprod_{i=1}^{n_{r}}\rd_{r} \phi_{n_{0}+n_{u}+i}\right) \left(\tprod_{i=1}^{n_{\tht}} \rd_{\tht^{A_{i}}} \phi_{n_{0}+n_{u}+n_{r}+i}\right) (\rd_{u}, \rd_{r}, r^{-1}) \rd_{\tht^{A}} \phi_{n}, \hbox{ or }\\
\hbox{(iii)~}&r^{\nu_{\Box}}r^{-n_{\tht}-2} \mathring{\slashed{c}}_{\rquasi_{j_{0}, k_{0}}}^{AB; A_{1} \ldots A_{n_{\tht}}}(u, \tht) r^{-j_{0}} \log^{k_{0}}(\tfrac{r}{u}) \left(\tprod_{i=1}^{n_{0}} \phi_{i} \right)  \\
&\quad \times \left(\tprod_{i=1}^{n_{u}}\rd_{u} \phi_{n_{0}+i}\right) \left(\tprod_{i=1}^{n_{r}}\rd_{r} \phi_{n_{0}+n_{u}+i}\right) \left(\tprod_{i=1}^{n_{\tht}} \rd_{\tht^{A_{i}}} \phi_{n_{0}+n_{u}+n_{r}+i}\right) \rsnb_{\tht^{A}} \rd_{\tht^{B}} \phi_{n},
\end{align*}
where
\begin{gather*}
	- \nu_{\Box} + j_{0} + \nu_{\Box} n + j_{1} + \ldots + j_{n} + o_{g} = j+1, \\
	k \leq k_{0} + k_{1} + \ldots k_{n} \leq k + n_{r} + n_{u} + 2,
\end{gather*}
and $o_{g}$ is the total number of $\rd_{r}$, $r^{-1} \rd_{\tht^{A}}$ or $r^{-1}$ (i.e., good derivatives) arising in that term; hence, $o_{g} = n_{r} + n_{\tht} + 2 - (\hbox{number of $\rd_{u}$'s falling on $\phi_{n}$})$. We note that multiplication of $\phi_{n}$ by $r^{-1}$ in terms of the form (i) arises due to the nontrivial Christoffel symbols in the $(u, r, \tht)$ coordinates; see \eqref{eq:bondi-Gmm}.

\item (Semilinear) With $n = n_{0} + n_{r} + n_{u} + n_{\tht}$ (total degree in $(\phi, \ud \phi)$) and $\phi_{i} = r^{-\nu_{\Box}} r^{-j_{i}} \log^{k_{i}} (\tfrac{r}{u}) \rPhi_{j_{i}, k_{i}}$,
\begin{align*}
& r^{\nu_{\Box}} r^{-n_{\tht}} \mathring{\slashed{c}}_{\rcsemi_{j_{0}, k_{0}}}^{A_{1} \ldots A_{n_{\tht}}}(u, \tht) r^{-j_{0}} \log^{k_{0}}(\tfrac{r}{u}) \left(\tprod_{i=1}^{n_{0}} \phi_{i} \right)  \\
&\quad \times \left(\tprod_{i=1}^{n_{u}}\rd_{u} \phi_{n_{0}+i}\right) \left(\tprod_{i=1}^{n_{r}}\rd_{r} \phi_{n_{0}+n_{u}+i}\right) \left(\tprod_{i=1}^{n_{\tht}} \rd_{\tht^{A_{i}}} \phi_{n_{0}+n_{u}+n_{r}+i}\right)
\end{align*}where
\begin{gather*}
	- \nu_{\Box} + j_{0} + \nu_{\Box} n + j_{1} + \ldots + j_{n} + o_{g} = j+1, \\
	k \leq k_{0} + k_{1} + \ldots k_{n} \leq k + n_{r}+n_{u},
\end{gather*}
where $o_{g} = n_{r} + n_{\tht}$.
\end{itemize}
In order to arrive at the above conclusion, we have simply counted the possible powers of $r^{-1}$ and $\log (\tfrac{r}{u})$ in each expression; note that we get a range for $\sum_{i \geq 0} k_{i}$ since a derivative may decrease the power of $\log^{k_{i}} (\tfrac{r}{u})$. In both cases, note that $\nu_{\Box} (n-1) \leq j+1$, which implies that $\rN^{<J}_{j, k}$ consists of finitely many contributions of the above forms.

We may now quickly prove (1) and (2). In both cases, observe that $j_{1}+j_{2}+\ldots+j_{n} = j+1- \nu_{\Box}(n-1) - j_{0} - o_{g}$, where $o_{g} \geq 0$ and $j_{i} \geq 0$ ($i \geq 1$) for obvious reasons, $j_{0} \geq 0$ by \ref{hyp:nonlin-wave}, and $n \geq 2$ by \ref{hyp:nonlin-degree}. If $d \geq 5$, then $\nu_{\Box} \geq 2$ so it follows that the RHS is at most $j-1$, which proves (1). If $d = 3$, then either $o_{g} \geq 1$ or $n \geq 3$ by \ref{hyp:nonlin-null}, so the RHS is again at most $j-1$ and (1) follows. To prove (2), we note that, under the assumption of (2), $\rN^{<J}_{j, k}$ can be nontrivial only if $k_{i} \leq K'$ ($i \geq 1$) for each contribution of one of the above forms, hence only if $k \leq k_{0} + k_{1} + \ldots k_{n} \leq K_{c} + \frac{j+1+\nu_{\Box}}{\nu_{\Box}} K'$ (where we used $k_{0} \leq K_{c}$ from \ref{hyp:nonlin-wave}). \qedhere

\end{proof}
\subsubsection*{Precise definition of an admissible decay exponent and the proof of Proposition~\ref{prop:alp-N}} We now define the notion of an \emph{admissible decay exponent} for $\calN$ (see the beginning of this subsection for the notation):
\begin{definition} [Admissible decay exponent] \label{def:alp-N}
We say that $\alp_{\calN}'$ is an \emph{admissible decay exponent} for $\calN$ if there exists a nonnegative nondecreasing function $A_{\calN} = A_{\calN}(A)$ with $A_{\calN} = O(A^{2})$ as $A \to 0$ such that, for every $\dlt_{0} > 0$, nonnegative integer $M \leq M_{c}$, and $A \geq 0$, the following holds:

\begin{enumerate}
\item (Estimates in $\calM_{\near}$) For $\phi$ satisfying
\begin{equation*}
	\phi = O_{\bfGmm}^{M}(A\tau^{-\alp}) \quad \hbox{ in } \calM_{\near}
\end{equation*}
with $\alp \geq \alp_{\calN}' + \dlt_{0}$, we have
\begin{align*}
	\calN(\phi) &= O_{\bfGmm}^{M-2}(A_{\calN}(A) \tau^{-\dlt_{0}-\alp}) \quad \hbox{ in } \calM_{\near}.
\end{align*}

\item (Estimates in $\calM_{\med}$) For $\phi$ and $\psi$ satisfying
\begin{equation*}
	\phi, \psi = O_{\bfGmm}^{M}(A u^{-\alp})  \quad \hbox{ in } \calM_{\med}
\end{equation*}
with $\alp \geq \alp_{\calN}' + \dlt_{0}$, we have, in $\calM_{\med}$,
\begin{equation} \label{eq:alp-N-med}
\begin{aligned}
	\bfh_{\calN}(\phi; \psi) &= O_{\bfGmm}^{M-1}(\tfrac{1}{A} A_{\calN}(A) r^{-\dlt_{0}}), \\
	\bfC_{\calN}(\phi; \psi) &= O_{\bfGmm}^{M-1}(\tfrac{1}{A} A_{\calN}(A) r^{-1-\dlt_{0}}), \\
	W_{\calN}(\phi; \psi) &= O_{\bfGmm}^{M-1}(\tfrac{1}{A} A_{\calN}(A) r^{-2-\dlt_{0}}).
\end{aligned}
\end{equation}

\item (Estimates in $\calM_{\wave}$)
For $\phi$ and $\psi$ satisfying
\begin{equation*}
	\phi, \psi = O_{\bfGmm}^{M}(A u^{-\alp+\nu_{\Box}} r^{-\nu_{\Box}})  \quad \hbox{ in } \calM_{\wave}
\end{equation*}
with $\alp \geq \alp_{\calN}' + \dlt_{0}$, we have, in $\calM_{\wave}$,
\begin{equation} \label{eq:alp-N-wave}
\begin{aligned}
	\bfh_{\calN}(\phi; \psi) &= O_{\bfGmm}^{M-1}(\tfrac{1}{A} A_{\calN}(A) u^{-\dlt_{0}+\nu_{\Box}} r^{-\nu_{\Box}}), \\
	\bfC_{\calN}(\phi; \psi) &= O_{\bfGmm}^{M-1}(\tfrac{1}{A} A_{\calN}(A) u^{-1-\dlt_{0}+\nu_{\Box}} r^{-\nu_{\Box}}), \\
	W_{\calN}(\phi; \psi) &= O_{\bfGmm}^{M-1}(\tfrac{1}{A} A_{\calN}(A) u^{-2-\dlt_{0}+\nu_{\Box}} r^{-\nu_{\Box}}).
\end{aligned}
\end{equation}
Furthermore, if $d = 3$, then for the same $\phi$ and $\psi$ we also have, in $\calM_{\wave}$,
\begin{equation} \label{eq:alp-N-wave-3d}
\begin{aligned}
	\bfh_{\calN}^{uu}(\phi; \psi) &= O_{\bfGmm}^{M-1}(\tfrac{1}{A} A_{\calN}(A) u^{-\dlt_{0}+2} r^{-2}), \\
	\bfC_{\calN}^{u}(\phi; \psi) &= O_{\bfGmm}^{M-1}(\tfrac{1}{A} A_{\calN}(A) u^{1-\dlt_{0}} r^{-2}), \\
	W_{\calN}(\phi; \psi) &= O_{\bfGmm}^{M-1}(\tfrac{1}{A} A_{\calN}(A) u^{-\dlt_{0}} r^{-2}).
\end{aligned}
\end{equation}

\item (Additional estimates in $\calM_{\wave}$ related to higher radiation fields) For every $J \in \bbZ_{\geq 0}$, $\vec{K} \in (\bbZ_{\geq 0})^{J}$ and $\rPhi_{j, k}$ for $0 \leq j \leq J$, $0 \leq k \leq K_{j}$ satisfying
\begin{equation*}
	\rPhi_{j, k} =  O_{\bfGmm}^{M}(A u^{j-\alp+\nu_{\Box}})
\end{equation*}
with $\alp \geq \alp_{\calN}' + \dlt_{0}$, we have
\begin{align}
	\rN^{<J}_{j, k} &= O_{\bfGmm}^{M-2}(C_{j, k} A_{\calN}(A) u^{j-1-\alp-\dlt_{0}+\nu_{\Box}}), \label{eq:alp-N-rNjk<J} \\
	E_{J; \nonlinear, \rem}
	&= O_{\bfGmm}^{M-2}(A_{\calN}(A) r^{-J_{c}-\eta_{c}} u^{J_{c}-2+\eta_{c}-\alp-\dlt_{0}+\nu_{\Box}}). \label{eq:alp-N-rem}
\end{align}
\end{enumerate}
\end{definition}

We are now ready to verify Proposition~\ref{prop:alp-N}, which is rather straightforward from Definition~\ref{def:alp-N}. 
\begin{proof}[Proof of Proposition~\ref{prop:alp-N}]
We first prove (1). For concreteness, assume that
\begin{equation*}
\calN(\phi) = c^{\alp_{1} \ldots \alp_{n_{1}}}(p) \phi^{n_{0}} \rd_{\alp_{1}} \phi \ldots \rd_{\alp_{n_{1}}} \phi
\end{equation*}
and let $\alp_{\calN}' = \max\set{0, \frac{2 - o - \bt}{n - 1}}$, with $c$, $o$ and $n$ as in the statement of Proposition~\ref{prop:alp-N}. The other case $\calN(\phi) = c^{\mu \nu; \alp_{1} \ldots \alp_{n_{1}}}(p) \phi^{n_{0}} \rd_{\alp_{1}} \phi \ldots \rd_{\alp_{n_{1}}} \phi \bfD_{\mu} \rd_{\nu} \phi$ can be treated similarly and shall be omitted (in fact, the proof is somewhat simpler in this case, since $o \geq 2$ and thus $\alp_{\calN}' = 0$).

Given $\alp \geq \alp_{\calN}' + \dlt_{0}$, we will verify Definition~\ref{def:alp-N}.(1)--(4) with $A_{\calN}(A) = A^{n}$. First, Definition~\ref{def:alp-N}.(1) is clear from \ref{hyp:nonlin}, \ref{hyp:nonlin-degree} and the fact that $\alp \geq \alp_{\calN}' + \dlt_{0} \geq \dlt_{0}$. Next, Definition~\ref{def:alp-N}.(2) and (3) are verified writing out $\bfh_{\calN}(\phi, \psi)$, $\bfC_{\calN}(\phi, \psi)$, $W_{\calN}(\phi, \psi)$, which is a straightforward computation given the simple form of $\calN$. Indeed, we have $\bfh_{\calN}(\phi, \psi) = 0$ and
\begin{align*}
	\bfC_{\calN}^{\mu}(\phi, \psi) \rd_{\mu} &= c^{\alp_{1} \ldots \alp_{n_{1}}} \sum_{i=1}^{n_{1}} \left[\int_{0}^{1} (\phi+h\psi)^{n_{0}} \prod_{i' : i' \neq i} \rd_{\alp_{i'}} (\phi+h\psi) \, \ud h \right] \rd_{\alp_{i}}, \\
	W_{\calN}(\phi, \psi) &= c^{\alp_{1} \ldots \alp_{n_{1}}} n_{0}\int_{0}^{1} (\phi+h\psi)^{n_{0}-1} \rd_{\alp_{1}} (\phi+h\psi) \ldots \rd_{\alp_{n_{1}}} (\phi+h\psi) \, \ud h,
\end{align*}
where $\bfC_{\calN} = 0$ if $n_{1} = 0$.
If $\phi, \psi = O_{\bfGmm}^{M}(A u^{-\alp})$ in $\calM_{\med}$, it follows that
\begin{align*}
	\bfC_{\calN}(\phi, \psi) &= O_{\bfGmm}^{M-1}(n_{1} A^{n-1} u^{-(n-1) \alp} r^{-(o-1)-\bt}) = O_{\bfGmm}^{M-1}(A^{n-1} r^{-1-(n-1) \dlt_{0}}), \\
	W_{\calN}(\phi, \psi) &= O_{\bfGmm}^{M-1}(A^{n-1} u^{-(n-1) \alp} r^{-o-\bt}) = O_{\bfGmm}^{M-1}(A^{n-1} r^{-2-(n-1) \dlt_{0}}),
\end{align*}
in $\calM_{\med}$, which proves Definition~\ref{def:alp-N}.(2). Similarly, if $\phi, \psi = O_{\bfGmm}^{M}(A u^{-\alp+\nu_{\Box}} r^{-\nu_{\Box}})$ in $\calM_{\wave}$, then
\begin{align*}
	\bfC_{\calN}(\phi, \psi) &= O_{\bfGmm}^{M-1}(n_{1} A^{n-1} u^{-(o-1)-\bt-(n-1)(\alp-\nu_{\Box})} r^{-(n-1) \nu_{\Box}}) = O_{\bfGmm}^{M-1}(A^{n-1} u^{-1-(n-1) \dlt_{0} + (n-1) \nu_{\Box}} r^{-(n-1) \nu_{\Box}}) , \\
	W_{\calN}(\phi, \psi) &= O_{\bfGmm}^{M-1}(n_{1} A^{n-1} u^{-o-\bt-(n-1)(\alp-\nu_{\Box})} r^{-(n-1) \nu_{\Box}}) = O_{\bfGmm}^{M-1}(A^{n-1} u^{-2-(n-1) \dlt_{0} + (n-1) \nu_{\Box}} r^{-(n-1) \nu_{\Box}}),
\end{align*}
in $\calM_{\wave}$, which proves \eqref{eq:alp-N-wave}. When $d = 3$, \ref{hyp:nonlin-null} implies a gain of $u r^{-1}$ compared to the above estimate, which proves \eqref{eq:alp-N-wave-3d}. Finally, to verify Definition~\ref{def:alp-N}.(4), observe that
\begin{align*}
	\mathring{\csemi}_{j, k} &= \mathring{c}^{\alp_{1} \ldots \alp_{n_{1}}}_{j, k}(p) \phi^{n_{0}} \rd_{\alp_{1}} \phi \ldots \rd_{\alp_{n_{1}}} \phi, \\
	\rem_{J_{c} - \nu_{\Box} + \eta_{c}}[\csemi] &= \rem_{J_{c} - \nu_{\Box} + \eta_{c}}[c^{\alp_{1} \ldots \alp_{n_{1}}}_{j, k}] \phi^{n_{0}} \rd_{\alp_{1}} \phi \ldots \rd_{\alp_{n_{1}}} \phi.
\end{align*}
Then \eqref{eq:alp-N-rNjk<J} follows from the structure of $\rN^{<J}_{j, k}$ uncovered in the proof of Lemma~\ref{lem:rNjk}. Moreover,
\begin{align*}
	E_{J; \nonlinear, \rem}
	&= r^{\nu_{\Box}} \rem_{J_{c}-\nu_{\Box}+\eta_{c}}[c^{\alp_{1} \ldots \alp_{n_{1}}}] (r^{-\nu_{\Box}}\Phi_{<J})^{n_{0}} \rd_{\alp_{1}} (r^{-\nu_{\Box}}\Phi_{<J}) \ldots \rd_{\alp_{n_{1}}} (r^{-\nu_{\Box}}\Phi_{<J}) \\
	&= O_{\bfGmm}^{M-1}(A^{n} r^{\nu_{\Box}} r^{-J_{c} + \nu_{\Box} - \eta_{c}} u^{J_{c}-\nu_{\Box}+\eta_{c}-\bt} u^{-o} u^{-n(\alp - \nu_{\Box})} r^{-n \nu_{\Box}} ) \\
	&= O_{\bfGmm}^{M-1}(A^{n} r^{-J_{c} - \eta_{c} - (n-2) \nu_{\Box}} u^{J_{c}-2+\eta_{c}-\alp-(n-1)\dlt_{0}+(n-1)\nu_{\Box}}),
\end{align*}
which proves \eqref{eq:alp-N-rem} since $n \geq 2$.

Next, we prove (2). For each monomial in the Taylor expansion of $\calN$ in $(\phi, \ud \phi)$, which is of the form considered in (1), we may associate $n$ and $o$. Note that the higher the $n$, the smaller the exponent $\max\set{0, \frac{2 - o - \bt}{n-1}}$; hence, the supremum $\alp_{\calN}'$ of $\max\set{0, \frac{2 - o - \bt}{n-1}}$ for all monomials in this expansion is well-defined. In fact, $\alp_{\calN}'$ is determined by the first nontrivial Taylor monomials with $o=0$ and $o = 1$. That $0 \leq \alp_{\calN}' \leq \nu_{\Box}$ is obvious when $d > 3$, and follows from \ref{hyp:nonlin-null} when $d = 3$. That $\alp_{\calN}'$ is an admissible decay exponent for $\calN$ follows from an argument similar to (1) applied to the Taylor remainder formula; we omit the details. \qedhere
\end{proof}

%
%
%
%

\section{Solutions to the wave equation on Minkowski spacetime}\label{sec:Minkowski.wave}

In this section, we carry out some explicit computations on Minkowski spacetime.
\begin{definition}\label{def:varphi.m.summed}
Let $j,k\in \mathbb N\cup \{0\}$ and $a \geq 1$. For every sufficiently regular $h:\mathbb S^{d-1} \to \mathbb R$, define the function $\varphi_{j,k}^{\bfm[a]}[h](u,r,\theta)$ such that
\begin{itemize}
\item for $u< 0$, $\varphi^{\bfm[a]}_{j,k}[h](u,r,\th) = 0$, and
\item for $u\geq 0$, $\varphi^{\bfm[a]}_{j,k}[h](u,r,\theta)$ solves $\Box_{\bfm} (\varphi^{\bfm[a]}_{j,k}[h])(u,r,\theta) = 0$
with characteristic data
\begin{equation}\label{eq:varphi.m.data.summed}
\varphi^{\bfm[a]}_{j,k}[h](0,r,\th) = \begin{cases}
h(\th) r^{-j-\nB}\log^k (\tfrac ra) & \hbox{ if } r \geq 1 \\
0 & \hbox{ if } r \leq \f 12.
\end{cases}
\end{equation}
\end{itemize}
\end{definition}

When carrying out the computations, it will be convenient to decompose into spherical harmonics. We thus make the following definition.
\begin{definition}\label{def:varphi.m}
Let $j,k\in \mathbb N\cup \{0\}$ and $a \geq 1$. Define the function $\varphi^{\bfm[a]}_{(\ell)j,k}(u,r)$ such that
\begin{itemize}
\item for $u< 0$, $\varphi^{\bfm[a]}_{(\ell)j,k}(u,r) = 0$, and
\item for $u\geq 0$, $\varphi^{\bfm[a]}_{(\ell)j,k}(u,r)$ is the unique solution to the linear wave equation on Minkowski spacetime for the $\ell$-th spherically symmetric mode, i.e.,
\begin{equation}\label{eq:wave.eqn.for.phi.m.ell}
\Big( -2 \partial_u \partial_r + \partial_r^2 + \f{d-1}r(\partial_r - \partial_u) - \f {(d-\ell-2)\ell}{r^2} \Big)\varphi^{\bfm[a]}_{(\ell)j,k}=0 ,
\end{equation}
with the characteristic data $\varphi^{\bfm[a]}_{(\ell)j,k}(r,0)$ which is smooth and
\begin{equation}\label{eq:varphi.m.data}
\varphi^{\bfm[a]}_{(\ell)j,k}(r,0) = \begin{cases}
r^{-j-\nB}\log^k (\tfrac ra) & \hbox{ if } r \geq 1 \\
0 & \hbox{ if } r \leq \f 12.
\end{cases}
\end{equation}
\end{itemize}
\end{definition}

\begin{remark}
Notice that there is an ambiguity in the definition of $\varphi^{\bfm[a]}_{(\ell)j,k}$ in Definition~\ref{def:varphi.m} since  $\varphi^{\bfm[a]}_{(\ell)j,k}(r,0)$ is not fixed by \eqref{eq:varphi.m.data} when $r \in (\f 12, 1)$. Nonetheless, by the strong Huygens principle, this does not affect the solution $\varphi^{\bfm[a]}_{(\ell)j,k}(u,r)$ when $u\geq 2$, which is the only relevant region for our decay estimates below. Similar comments apply to Definition~\ref{def:varphi.m.summed}.
\end{remark}

\begin{lemma}\label{lem:med.explicit}
When $u \geq 2$, $\varphi^{\bfm[a]}_{(\ell)j,k}(u,r)$ is given explicitly by
\begin{equation*}
\begin{split}
&\: r^\nB \varphi^{\bfm[a]}_{(\ell)j,k}(u,r) \\
= &\: \int_{0}^{r} r_{\f{d-3}2+\ell}^{-2} \ldots \int_0^{r_2} r_1^{-2} \int_0^{r_1} \Big(\f{r'}{r'+\f u2}\Big)^{d-3+2\ell} \rd_r \bfK^{\f{d-3}2 + \ell} \Big( (r'+\f u2)^{-j} \log^k (\f{r'}a + \f u{2a})\Big) \, \ud r' \, \ud r_1 \, \ldots \, \ud r_{\f{d-3}2+\ell}.
\end{split}
\end{equation*}
\end{lemma}
\begin{proof}
Define $Q_{(\ell)j} = Q_j \mathbb S_{(\ell)}$. By \eqref{eq:wave.eqn.for.phi.m.ell} and the facts $r^{\nB} \Box_{\bfm}  = Q_0 r^{\nB}$ and $(\bfK + 2r) Q_j = Q_{j+1} \bfK$, we have
\begin{equation}\label{eq:QKrnphi.sum}
\begin{split}
 Q_{(\ell)\f{d-3}2 + \ell} \bfK^{\f{d-3}2 + \ell}(r^\nB \varphi^{\bfm[a]}_{(\ell)j,k}) = 0
\end{split}
\end{equation}

Recall that
$Q_{j} = -2 \rd_u \rd_r + \rd^2_r - \f{2j}r \rd_r + ( j + \f{d-1}2 ) ( j - \f{d-3}2 ) \f 1{r^2} - \f 1{r^2} \mathring{\slashed{\Delta}}$
and that $\mathring{\slashed{\Delta}}Y_{(\ell)} = - (d+\ell-2) \ell Y_{(\ell)}$. Thus, \eqref{eq:QKrnphi.sum} is equivalent to
\begin{equation}
\Big( -2\rd_u + \rd_r -\f{2 (\f{d-3}2 + \ell)}r \Big) \rd_r \bfK^{\f{d-3}2 + \ell}(r^\nB \varphi_{(\ell)}) = 0,
\end{equation}
or equivalently,
\begin{equation}
(-2\rd_u + \rd_r)(r^{-d+3 - 2\ell} \rd_r \bfK^{\f{d-3}2 + \ell}(r^\nB \varphi_{(\ell)})) = 0.
\end{equation}
Using the data condition \eqref{eq:varphi.m.data} and integrating along the integral curves of $-2\rd_u + \rd_r$ yields
\begin{equation}
\begin{split}
 r^{-d+3 - 2\ell} \rd_r \bfK^{\f{d-3}2 + \ell}(r^\nB \varphi_{(\ell)})(u,r)
= &\:   (r + \f u2)^{-d+3 - 2\ell} \rd_r \bfK^{\f{d-3}2 + \ell} \Big( (r+\f u2)^{-j} \log^k (\f ra + \f u{2a})\Big).
\end{split}
\end{equation}

Integrating repeatedly in $r$ (starting from $r=0$, and noting that all contributions from $r=0$ vanish), we obtain the desired explicit formula. \qedhere
\end{proof}

We will give some estimates for $\varphi^{\bfm[a]}_{(\ell)j,k}$ in Lemma~\ref{lem:med.corrected}--Proposition~\ref{prop:med.explicit.precise} below. First, we note an important cancellation: when $k=0$ and $\ell \geq j-\nB +1$, then $\varphi^{\bfm}_{(\ell)j,k}(u,r)$ does not contribute to late time asymptotics:
\begin{lemma}\label{lem:med.corrected}
If $k=0$ and $\ell \geq j-\nB +1$, then
$$\varphi^{\bfm[a]}_{(\ell)j,k}(u,r) = 0$$
whenever $u\geq 2$.
\end{lemma}
\begin{proof}
We compute that
$$\rd_r \bfK^{\f{d-3}2 + \ell}  r^{-j} = (-j+\f{d-3}2+\ell)\ldots(-j+1)(-j) r^{-j+\f{d-3}2+\ell},$$
which vanishes when $\ell \geq j-\nB +1$.
The desired conclusion hence follows from Lemma~\ref{lem:med.explicit}. \qedhere

\end{proof}

For more general values of $k$ and $\ell$, $\varphi^{\bfm}_{(\ell)j,k}(u,r)$ does not vanish. To compute them, we begin with the following lemma.
\begin{lemma}\label{lem:med.explicit.1}
$$\rd_r \bfK^{\f{d-3}2 + \ell}  (r^{-j} \log^k r) = \sum_{k' = 0}^k \mathfrak b_{k'}^{(j,k,\ell)} r^{-j+\f{d-5}2 + \ell} \log^{k-k'} r,$$
where the constants $\mathfrak b_{k'}^{(j,k,\ell)}$ are given by
$$\mathfrak b_{k'}^{(j,k,\ell)} = \begin{cases}
0 & \mathrm{if \,} \nB + \ell < k' \leq k \\
(-j)(-j+1) \ldots (-j+\nB+\ell) & \mathrm{if \,} k'=0, \\
\displaystyle\sum_{ -j \leq n_1 < n_2<\ldots <n_{k'} \leq -j+i+1 } \f{(-j)(-j+1) \ldots (-j+\nB+\ell)}{n_1 \times \ldots \times n_{k'}} \f{k!}{(k-k')!} & \mathrm{if \,} 1\leq k' \leq \min\{\nB + \ell,k\}.
\end{cases}$$
\end{lemma}
\begin{proof}
Note that
$$\bfK (r^{-j} \log^{k} r) = -j r^{-j+1} \log^{k} r + k r^{-j+1} \log^{k-1} r.$$
Therefore, inductively, for any $(i,j,k) \in (\mathbb Z_{\geq 0})^3$, we have
$$\bfK^i (r^{-j} \log^{k} r) = \sum_{k'=0}^k \mathfrak b_{i,k'}^{(j,k,\ell)} r^{-j+i} \log^{k-k'} r,$$
where the constants $\mathfrak b_{i,k'}^{(j,k,\ell)}$ are given by
$$\mathfrak b_{i,k'}^{(j,k,\ell)} = \begin{cases}
0 & \mathrm{if \,} 1\leq i< k' \leq k \\
1 & \mathrm{if \,} i=0 \\
(-j)(-j+1) \ldots (-j+i) & \mathrm{if \,} k'=0,\,i\geq 1 \\
\displaystyle\sum_{ -j \leq n_1 < n_2<\ldots <n_{k'} \leq -j+i } \f{(-j)(-j+1) \ldots (-j+i)}{n_1 \times \ldots \times n_{k'}} \f{k!}{(k-k')!} & \mathrm{if \,} 1\leq k' \leq \min\{i,k\}.
\end{cases}$$
In particular, $\partial_r \bfK^i (r^{-j} \log^{k} r) = \sum_{k'=0}^k \mathfrak b^{(j,k,\ell)}_{i+1,k'} r^{-j+i-1} \log^{k-k'} r.$ Thus, setting $\mathfrak b^{(j,k,\ell)}_{k'} = \mathfrak b^{(j,k,\ell)}_{\f{d-3}2 + \ell+1,k'}$ yields the desired result. \qedhere
\end{proof}

To write down $\varphi^{\bfm[a]}_{(\ell)j,k}$ in a more convenient form, it will be useful to introduce the following notations:
\begin{definition}\label{def:notations.for.Minkowski}
\begin{enumerate}
\item (Averaging operator $\mathfrak I_s$) For any $s \in \mathbb N$, define an operator $\mathfrak I_s: C^\infty([0,\infty)) \to C^\infty([0,\infty))$ by
\begin{equation}
\mathfrak I_s[f](r):=r^{-s} \int_0^r f(r') (r')^{s-1}\, \ud r'.
\end{equation}
\item (Functions $\mathfrak f_u^{(p_1,p_2)[a]}$) For any $p_1,p_2 \in \mathbb Z_{\geq 0}$ and any $u \geq 2$, define  $\mathfrak f_u^{(p_1,p_2)[a]} \in C^\infty([0,\infty))$ by
\begin{equation}\label{eq:def.fu}
\mathfrak f_u^{(p_1,p_2)[a]}(r) = (r+\tfrac u2)^{-p_1} \log^{p_2} (\f{2r+u}{2a}).
\end{equation}
\end{enumerate}
\end{definition}

Combining Lemma~\ref{lem:med.explicit} and Lemma~\ref{lem:med.explicit.1}, we obtain the following proposition:
\begin{proposition}
When $u \geq 2$, $\varphi^{\bfm[a]}_{(\ell)j,k}(u,r)$ is given explicitly by
\begin{equation}\label{eq:med.explicit.expanded}
\begin{split}
\varphi^{\bfm[a]}_{(\ell)j,k}(u,r)
= &\: \sum_{k'=0}^k  \mathfrak b_{k'}^{(j,k,\ell)} r^{\ell} \mathfrak I_{\f{d-1}2+\ell}[ \mathfrak I_{\f{d+1}2+\ell}[\cdot\,\cdot\,\cdot \mathfrak I_{d-2+2\ell}[\mathfrak f_{u}^{(j + \f{d-1}2+\ell, k-k')[a]}]\ldots]](r),
\end{split}
\end{equation}

where $\mathfrak I_s$ and $\mathfrak f_u^{(p_1,p_2)[a]}$ are as in Definition~\ref{def:notations.for.Minkowski} and the constants $\mathfrak b^{(j,k,\ell)}_{k'}$ are as in Lemma~\ref{lem:med.explicit.1}.
\end{proposition}

\subsection{Sharp upper bound for $\varphi^{\bfm[a]}_{(\ell)j,k}$ in the intermediate zone}

%

\begin{proposition}\label{prop:phi.ell.j.k.upper}
For any $(I_u,I_r) \in (\mathbb N\cup \{0\})^2$, $\varphi^{\bfm[a]}_{(\ell)j,k}$ obeys the following bound when $u \gtrsim_{d,j,k,I_u,I_r} \max\{a,1\}$
\begin{equation}\label{eq:phi.ell.j.k.upper}
|\rd_u^{I_u} \rd_r^{I_r} \varphi^{\bfm[a]}_{(\ell)j,k}(u,r)|
\ls_{d,j,k,,I_u,I_r} u^{-I_u} r^{-I_r} \brk{\ell}^{I_u +I_r} u^{-j-\nB} \log^{k} (\tfrac ua) \min\{ r^\ell (\tfrac u2)^{-\ell}, 1 \},
\end{equation}
where the implicit constant depends on $d,j,k,I_u,I_r$, but importantly, it is independent of $\ell$.
\end{proposition}
\begin{proof}
In the proof, we will take $u$ large in terms of $d,j,k,I_u,I_r$ and allow all implicit constants to depend on $d,j,k,I_u,I_r$ without further comments.

\pfstep{Step~0: Preliminary estimates} We start with two basic estimates for objects in Definition~\ref{def:notations.for.Minkowski}. First, for the operator $I_s$, by H\"older's inequality, we have
\begin{equation}\label{eq:I_s.pointwise}
\mathfrak I_s[f](r) \leq \Big( \sup_{r' \in [0,r]} |f(r')| \Big) r^{-s} \int_0^r (r')^{s-1} \, \ud r = \f 1 s \sup_{r' \in [0,r]} |f(r')|.
\end{equation}
Second, since $x^{-p_1} \log^{p_2} (\tfrac{x}{a})$ is decreasing for $x \geq a e^{\f{p_2}{p_1}}$, by the definition \eqref{eq:def.fu}, we have
\begin{equation}\label{eq:fu.pointwise}
\mathfrak f_u^{(p_1,p_2)[a]}(r) \leq (\tfrac u2)^{-p_1} \log^{p_2} (\tfrac{u}{2a}),\quad \hbox{if $u \geq 2ae^{\f{p_2}{p_1}}$}.
\end{equation}

\pfstep{Step~1: Proof of the upper bound \eqref{eq:phi.ell.j.k.upper} when $I_u = I_r = 0$} To make the computation more transparent, we first consider the special case $I_u=I_r=0$. Recall the definition of $\mathfrak b_{k'}^{(j,k,\ell)}$ in Lemma~\ref{lem:med.explicit.1}. Note that all the negative factors in the definition can be bounded by a constant depending on $d$, $j$ and $k$, independently of $\ell$. Thus to understand the dependence on $\ell$, we only need to bound the positive factors and obtain a rough bound as follows:
\begin{equation}\label{eq:QKrnphi.sum.constants.2}
|\mathfrak b^{(j,k,\ell)}_{k'}| \ls_{d,j,k} (\tfrac{d-1}{2}+\ell)!.
\end{equation}

We now bound the integral in \eqref{eq:med.explicit.expanded}. We will give two different bounds, which will be useful in the regions $\{r\leq \f u2\}$ and $\{r \geq \f u2\}$, respectively. For the first bound, we use \eqref{eq:I_s.pointwise} repeatedly and the use the pointwise bound \eqref{eq:fu.pointwise} to obtain
\begin{equation}\label{eq:med.explicit.r.integral.1.pre}
\mathfrak I_{\f{d-1}2+\ell}[\mathfrak I_{\f{d+1}2+\ell}[\cdot\,\cdot\,\cdot\, \mathfrak I_{d-2+2\ell}[\mathfrak f_{u}^{(j + \f{d-1}2+\ell, k-k')[a]}]\ldots]](r) \leq \Big(\f u2\Big)^{-j-\nB-\ell} \log^{k-k'} \Big(\f u{2a}\Big) \f {(\f{d-3}2+\ell)!}{(d-2+2\ell)!}.
\end{equation}
Plugging \eqref{eq:med.explicit.r.integral.1.pre} back into \eqref{eq:med.explicit.expanded} and using \eqref{eq:QKrnphi.sum.constants.2}, we obtain
\begin{equation}\label{eq:med.explicit.r.integral.1}
\begin{split}
|\varphi^{\bfm[a]}_{(\ell)j,k}|(u,r) \ls_{d,j,k} \Big(\f u2\Big)^{-j-\nB-\ell} \log^{k-k'} \Big(\f u{2a}\Big) \f {(\f{d-3}2+\ell)!(\f{d-1}2+\ell)!}{(d-2+2\ell)!} r^{\ell}.
\end{split}
\end{equation}

For the second bound, we first use that $\Big(\f{r}{r+\f u2}\Big)^{\ell} \leq 1$, which implies
\begin{equation}\label{eq:med.explicit.r.integral.2.pre}
\begin{split}
\mathfrak I_{\f{d-1}2+\ell}[\mathfrak I_{\f{d+1}2+\ell}[\ldots \mathfrak I_{d-2+2\ell}[\mathfrak f_{u}^{(j + \f{d-1}2+\ell, k-k')[a]}]\ldots]](r)
\leq r^{-\ell} \mathfrak I_{\f{d-1}2}[\mathfrak I_{\f{d+1}2}[\ldots \mathfrak I_{d-2+\ell}[\mathfrak f_{u}^{(j + \f{d-1}2, k-k')[a]}]\ldots]](r).
\end{split}
\end{equation}
Hence, arguing as in \eqref{eq:med.explicit.r.integral.1.pre} and \eqref{eq:med.explicit.r.integral.1}, we obtain
\begin{equation}\label{eq:med.explicit.r.integral.2}
|\varphi^{\bfm[a]}_{(\ell)j,k}|(u,r) \ls_{d,j,k} \Big(\f u2\Big)^{-j-\nB} \log^{k-k'} \Big(\f u{2a} \Big) \f {(\f{d-3}2+\ell)!}{(d-2+\ell)!}.
\end{equation}

Taking the minimum of the bounds in \eqref{eq:med.explicit.r.integral.1} and \eqref{eq:med.explicit.r.integral.2}, we thus obtain
\begin{equation}\label{eq:med.explicit.sharp.bound}
\begin{split}
 |\varphi^{\bfm[a]}_{(\ell)j,k}|(u,r)
\ls_{d,j,k} &\: \Big(\f u2 \Big)^{-j-\nB} \log^{k} \Big(\f u{2a}\Big) \min \Big\{ \f{(\f{d-1}2+\ell)!(\f{d-3}2+\ell)!}{(d-2+2\ell)!} r^\ell \Big(\f u2\Big)^{-\ell}, \f{(\f{d-1}2+\ell)!(\f{d-3}2)!}{(d-2+\ell)!} \Big\} \\
\leq &\: u^{-j-\nB} \log^{k} (\tfrac ua) \min \Big\{ r^\ell u^{-\ell}, 1 \Big\},
\end{split}
\end{equation}
where in the last line we used $\f{( \f{d-1}2+\ell)!(\f{d-3}2+\ell)!}{(d-2+2\ell)!} \leq 1$ and $\f{(\f{d-1}2+\ell)! (\f{d-3}2)! }{(d-2+\ell)!} \leq 1$, which can be obtained by comparison with suitable binomial coefficients.

This completes the proof of the estimates for the pointwise bound of $\varphi^{\bfm[a]}_{(\ell)j,k}$.

\pfstep{Step~2: Higher derivative estimates} It remains to show that a similar bound holds with finitely many $\rd_u$ and $\rd_r$. First, note that it suffices to show
\begin{equation}\label{eq:phi.ell.j.k.upper.alt}
|\rd_u^{I_u} \rd_r^{I_r} (r^{-\ell} \varphi^{\bfm[a]}_{(\ell)j,k})(u,r)|
\ls_{d,j,k,,I_u,I_r} u^{-I_u} r^{-I_r-\ell} \brk{\ell}^{I_u +I_r} u^{-j-\nB} \log^{k} (\tfrac ua) \min\{ r^\ell (\tfrac u2)^{-\ell}, 1 \},
\end{equation}

Next, to note that the derivative in $u$ is easy to compute since \eqref{eq:med.explicit.expanded} depends on $u$ only via $\mathfrak f_u$. For the derivative in $r$, we note the following simple identity for the operator $\mathfrak I_s$ that can be established with integration by parts (see, for instance, \cite[Lemma~2.6]{LOY}):
\begin{equation}\label{eq:dr.Is}
\rd_r \mathfrak I_s[f](r) = \mathfrak I_{s+1}[\rd_r f](r), \quad \forall s \in \mathbb N.
\end{equation}
We observe that if we take $\rd_u$ or $\rd_r$ derivative of $\mathfrak f_{u}^{(j + \f{d-1}2+\ell, k-k')[a]}$, then each derivative would bring an additional factor of $(r + \f u2)^{-1}$, possibly also removing one logarithmic factor. Taking into account the dependence on $\ell$ (because $\ell$ appears in the power of $(r + \f u2)$), we have
\begin{equation}\label{eq:dr.du.fu}
\begin{split}
|\rd_u^{I_u} \rd_r^{I_r} \mathfrak f_{u}^{(j + \f{d-1}2+\ell, k-k')[a]}| &\: \ls_{k,I_u,I_r} \f{(j + \f{d-3}2+\ell+I_u+I_r)!}{(j + \f{d-3}2+\ell)!} \mathfrak f_{u}^{(j + \f{d-1}2+\ell+I_u+I_r, k-k')[a]} \\
&\: \ls_{d,j,k,I_u,I_r} \langle \ell \rangle^{I_u+I_r} \mathfrak f_{u}^{(j + \f{d-1}2+\ell+I_u+I_r, k-k')[a]}.
\end{split}
\end{equation}
We differentiate the expression in \eqref{eq:med.explicit.expanded} using \eqref{eq:dr.Is} and \eqref{eq:dr.du.fu}, and use the bound for $\mathfrak b_{k'}^{j,k}$ in \eqref{eq:QKrnphi.sum.constants.2} to obtain
\begin{equation}\label{eq:du.dr.varphi.m}
\begin{split}
&\: \Big| \rd_r^{I_r} \rd_u^{I_u} (r^{-\ell} \varphi^{\bfm[a]}_{j,k}) (u,r) \Big| \\
&\: \quad \ls_{d,j,k,I_u,I_r} (\tfrac{d-1}2+\ell)! \langle \ell \rangle^{I_u+I_r}  \mathfrak I_{\f{d-1}2+\ell+I_r}[ \mathfrak I_{\f{d+1}2+\ell+I_r}[\cdot\,\cdot\,\cdot\, \mathfrak I_{d-2+2\ell+I_r}[\mathfrak f_{u}^{(j + \f{d-1}2+\ell+I_u+I_r, k)[a]}]\ldots]](r)
\end{split}
\end{equation}
Starting with \eqref{eq:du.dr.varphi.m}, we use $(\f{r}{r+\f u2})^{I_r} \leq 1$ to shift the powers, and then argue as in \eqref{eq:med.explicit.r.integral.1.pre} to obtain
\begin{equation}\label{eq:du.dr.varphi.m.1}
\begin{split}
&\: \Big| \rd_r^{I_r} \rd_u^{I_u} (r^{-\ell} \varphi^{\bfm[a]}_{j,k}) (u,r) \Big| \\
&\: \quad \ls_{d,j,k,I_u,I_r} (\tfrac{d-1}2+\ell)! \langle \ell \rangle^{I_u+I_r} r^{-I_r} \mathfrak I_{\f{d-1}2+\ell}[ \mathfrak I_{\f{d+1}2+\ell}[\cdot\,\cdot\,\cdot\, \mathfrak I_{d-2+2\ell}[\mathfrak f_{u}^{(j + \f{d-1}2+\ell+I_u, k)[a]}]\ldots]](r)\\
&\: \quad \ls_{d,j,k,I_u,I_r} (\tfrac{d-1}2+\ell)! \langle \ell \rangle^{I_u+I_r} r^{-I_r} \f{(\f{d-3}2+\ell)!}{(d-2+2\ell)!} \Big(\f u2\Big)^{-j-\nB-\ell-I_u-I_r} \log^{k} \Big(\f u{2a}\Big) \\
&\: \quad \ls_{d,j,k,I_u,I_r} \langle \ell \rangle^{I_u+I_r} r^{-I_r} \Big(\f u2\Big)^{-j-\nB-\ell-I_u} \log^{k} \Big(\f u{2a}\Big).
\end{split}
\end{equation}
Alternatively, we start with \eqref{eq:du.dr.varphi.m} and use $(\f{r}{r+\f u2})^{I_r+\ell} \leq 1$ to shift $\ell$ additional powers so as to obtain
\begin{equation}\label{eq:du.dr.varphi.m.2}
\begin{split}
&\: \Big| \rd_r^{I_r} \rd_u^{I_u} (r^{-\ell} \varphi^{\bfm[a]}_{j,k}) (u,r) \Big| \\
&\: \quad \ls_{d,j,k,I_u,I_r} (\tfrac{d-1}2+\ell)! \langle \ell \rangle^{I_u+I_r} r^{-I_r-\ell} \mathfrak I_{\f{d-1}2}[ \mathfrak I_{\f{d+1}2}[\cdot\,\cdot\,\cdot\, \mathfrak I_{d-2+\ell}[\mathfrak f_{u}^{(j + \f{d-1}2+I_u, k)[a]}]\ldots]](r)\\
&\: \quad \ls_{d,j,k,I_u,I_r} (\tfrac{d-1}2+\ell)! \langle \ell \rangle^{I_u+I_r} r^{-I_r-\ell} \f{(\f{d-3}2)!}{(d-2+\ell)!} \Big(\f u2\Big)^{-j-\nB-I_u} \log^{k} \Big(\f u{2a}\Big) \\
&\: \quad \ls_{d,j,k,I_u,I_r} \langle \ell \rangle^{I_u+I_r} r^{-\ell-I_r} \Big(\f u2\Big)^{-j-\nB-I_u} \log^{k} \Big(\f u{2a}\Big).
\end{split}
\end{equation}
Taking the minimum of \eqref{eq:du.dr.varphi.m.1} and \eqref{eq:du.dr.varphi.m.2} gives \eqref{eq:phi.ell.j.k.upper.alt}. \qedhere

\end{proof}

\subsection{Precise asymptotics for $\varphi^{\bfm[1]}_{(\ell) j,k}$ when $r \leq u^{1-\de_m}$}

We now turn to the estimate for the precise asymptotics for $\varphi^{\bfm[1]}_{(\ell)j,k}$. (For later applications, we will not need to track the exact dependence on $\ell$ as we did in Proposition~\ref{prop:phi.ell.j.k.upper}.)
\begin{proposition}[Precise asymptotics for the incoming radiation on Minkowski]\label{prop:med.explicit.precise}
For every $\de_m \in (0,1)$, the following bound holds when $r \leq u^{1-\de_m}$:
\begin{equation}
\begin{split}
&\: \qquad  \Bigg| (u\rd_u)^{I_u} (r\rd_r)^{I_r} \Big(\varphi^{\bfm[1]}_{(\ell)j,k} - \sum_{k'=0}^k \mathfrak b^{(j,k,\ell)}_{k'}   \f{\Big( \f{d-3}2+\ell \Big)!}{(d-2+2\ell)!} r^\ell \Big( \f{u}2 \Big)^{-j-\nB-\ell}  \log^{k-k'} (\tfrac u2) \Big) \Bigg| \\
&\: \ls_{d,j,k,\ell,I_u,I_r,\de_m} r^\ell u^{-j-\nB-\ell-\de_m} \log^k u,
\end{split}
\end{equation}
where $\{\mathfrak b^{(j,k,\ell)}_{k'}\}_{k'=0}^k$ are as in Lemma~\ref{lem:med.explicit.1}.

\end{proposition}
\begin{proof}
We will allow all implicit constants to depend on $d,j,k,\ell,I_u,I_r,\de_m$ without further comments. Our goal will be to extract the main term in the asymptotics for the expression \eqref{eq:med.explicit.expanded} (only when $a =1$). Writing $\mathfrak f_u^{(p_1,p_2)} = \mathfrak f_u^{(p_1,p_2)[1]}$, we first note that $\rd_u^{I_u} \rd_r^{I_r} \mathfrak f_u^{(j+\f{d-1}2+\ell,k-k')} = 2^{I_r} \rd_u^{I_u+I_r} \mathfrak f_u^{(j+\f{d-1}2+\ell,k-k')}$.
Hence, for $r \leq u^{1-\de_m}$, we have
\begin{equation}\label{eq:du.dr.fu.small.r}
\begin{split}
\rd_u^{I_u} \rd_r^{I_r} \mathfrak f_u^{(j+\f{d-1}2+\ell,k-k')} = &\: 2^{I_r} \rd_u^{I_u+I_r} \mathfrak f_u^{(j+\f{d-1}2+\ell,k-k')}(r) \\
 = &\: 2^{I_r} \rd_u^{I_u+I_r} \mathfrak f_u^{(j+\f{d-1}2+\ell,k-k')}(0) + O(u^{-j-\nB-\ell-I_u-I_r} (\tfrac ru)) \\
 = &\: 2^{I_r} \rd_u^{I_u+I_r} \Big( \Big( \f{u}2 \Big)^{-j-\nB-\ell}  \log^{k-k'} (\tfrac u2)\Big) + O(u^{-j-\nB-\ell-I_u-I_r-\de_m}).
\end{split}
\end{equation}
Now differentiating \eqref{eq:med.explicit.expanded} using \eqref{eq:dr.Is}, substituting in \eqref{eq:du.dr.fu.small.r}, and using
\begin{equation}
\mathfrak I_{\f{d-1}2+\ell+I_r}[ \mathfrak I_{\f{d+1}2+\ell+I_r}[\cdot\,\cdot\,\cdot\, \mathfrak I_{d-2+2\ell+I_r}[1]\ldots]](r) = \f{(\tfrac{d-3}2+\ell+I_r)!}{(d-2+2\ell+I_r)!},
\end{equation}
it follows that
\begin{equation}
\begin{split}
&\: \rd_u^{I_u} \rd_r^{I_r} (r^{-\ell} \varphi^{\bfm[1]}_{(\ell)j,k})(u,r) \\
= &\: \sum_{k'=0}^k  \mathfrak b_{k'}^{(j,k,\ell)} \rd_u^{I_u} \rd_r^{I_r} \Big(\mathfrak I_{\f{d-1}2+\ell}[ \mathfrak I_{\f{d+1}2+\ell}[\cdot\,\cdot\,\cdot \mathfrak I_{d-2+2\ell}[\mathfrak f_{u}^{(j + \f{d-1}2+\ell, k-k')}]\ldots]](r) \Big) \\
= &\: \sum_{k'=0}^k  \mathfrak b_{k'}^{(j,k,\ell)} 2^{I_r} \mathfrak I_{\f{d-1}2+\ell+I_r}[ \mathfrak I_{\f{d+1}2+\ell+I_r}[\cdot\,\cdot\,\cdot\, \mathfrak I_{d-2+2\ell+I_r}[\rd_u^{I_u+I_r} \mathfrak f_u^{(j+\f{d-1}2+1,k-k')}]\ldots]](r) \\
&\: + O(u^{-j-\nB-\ell-I_u-I_r-\de_m})\\
= &\: \sum_{k'=0}^k  \mathfrak b_{k'}^{(j,k,\ell)} \f{(\tfrac{d-3}2+\ell+I_r)!}{(d-2+2\ell+I_r)!} \rd_u^{I_u+I_r} \Big( \Big( \f{u}2 \Big)^{-j-\nB-\ell}  \log^{k-k'} (\tfrac u2)\Big) 2^{I_r} + O(u^{-j-\nB-\ell-I_u-I_r-\de_m}).
\end{split}
\end{equation}
Using again $r \leq u^{1-\de_m}$, this implies
\begin{equation}
\begin{split}
&\: u^{I_u} r^{I_r} \rd_u^{I_u} \rd_r^{I_r} (r^{-\ell} \varphi^{\bfm[1]}_{(\ell)j,k})(u,r) \\
= &\: \begin{cases}
\sum_{k'=0}^k  \mathfrak b_{k'}^{(j,k,\ell)} \f{(\tfrac{d-3}2+\ell)!}{(d-2+2\ell)!} (u\rd_u)^{I_u} \Big( \Big( \f{u}2 \Big)^{-j-\nB-\ell}  \log^{k-k'} (\tfrac u2)\Big)  + O(u^{-j-\nB-\ell-\de_m}) & \hbox{ if }I_r = 0 \\
 O(u^{-j-\nB-\ell-(I_r+1)\de_m}) & \hbox{ if }I_r \geq 1,
\end{cases}
\end{split}
\end{equation}
which immediately implies the desired conclusion. \qedhere

\end{proof}

\subsection{Upper bound for $\varphi^{\bfm[1]}_{(\ell) j,k}$ in the wave zone and precise asymptotics for the radiation field}

\begin{proposition}\label{prop:radiation.field}
\begin{enumerate}
\item There exist constants $\{\mathfrak d_{k'}\}_{k'=0}^k$ depending on $d$, $j$, $k$, $\ell$ and $k'$ (whose exact values can be inferred from the proof) such that
\begin{equation}\label{eq:radiation.field.expansion}
\lim_{r\to \infty} r^{\nB} \varphi^{\bfm[1]}_{(\ell) j,k}(u,r) = \sum_{k'=0}^k \mathfrak d_{k'} u^{-j-1} \log^{k-k'} u, \quad u \geq 2.
\end{equation}
Moreover, $\mathfrak d_{0} = 0$ if and only if $\ell \geq j - \nB+1$, and $\mathfrak d_{1}\neq 0$ when $\mathfrak d_0 = 0$.
\item $\varphi^{\bfm[1]}_{(\ell) j,k}$ satisfies the upper bound
\begin{equation}\label{eq:explicit.wave.upper}
\Big|\rd_u^{I_u} \rd_r^{I_r} \varphi^{\bfm[1]}_{(\ell) j,k} \Big|(u,r) \ls_{d,j,k,I_u,I_r} u^{-I_u} r^{-I_r} \brk{\ell}^{I_u+I_r} r^{-\nB} u^{-j} \log^k (\tfrac u2).
\end{equation}
\item $(\rd_{r} + \frac{\nB}{r}) \varphi^{\bfm[1]}_{(\ell) j,k}$ satisfies the upper bound
\begin{equation}\label{eq:explicit.wave.dr.upper}
\Big|\rd_u^{I_u} \rd_r^{I_r} (\rd_{r} + \tfrac{\nB}{r}) \varphi^{\bfm[1]}_{(\ell) j,k} \Big|(u,r) \ls_{d,j,k,I_u,I_r} u^{-I_u} r^{-I_r} \brk{\ell}^{I_u+I_r+1} r^{-\nB-2} u^{-j} \log^k (\tfrac u2).
\end{equation}
\end{enumerate}
\end{proposition}
In \eqref{eq:explicit.wave.dr.upper}, notice that $(\rd_{r} + \frac{\nu_{\Box}}{r})$ acting on $\varphi^{\bfm[1]}_{(\ell) j, k}$ improves the upper bound by $\brk{\ell} r^{-2}$ instead of $\brk{\ell} r^{-1}$ as $\rd_{r}$ would. This behavior is consistent with the power of the leading order term in $r$ revealed in \eqref{eq:radiation.field.expansion}.
\begin{proof}
\pfstep{Step~1: Proof of \eqref{eq:radiation.field.expansion}} Our goal will be to take the $r\to \infty$ limit in the expression \eqref{eq:med.explicit.expanded}. We will take this limit for each term in the expression.

Recalling the notation in \eqref{eq:def.fu}, define $\mathfrak H_{(p_3)}^{(p_1,p_2)}(u,r) = r^{p_3}\mathfrak f_u^{(p_1,p_2)}$, where from now on we write $\mathfrak f_u^{(p_1,p_2)} = \mathfrak f_u^{(p_1,p_2)[1]}$. By Fubini's theorem,
\begin{equation}\label{eq:varphi.explicit.radiation.1}
\begin{split}
&\: r^{\nB+\ell} \mathfrak I_{\f{d-1}2+\ell}[ \mathfrak I_{\f{d+1}2+\ell}[\cdot\,\cdot\,\cdot \mathfrak I_{d-2+2\ell}[\mathfrak f_{u}^{(j + \f{d-1}2+\ell, k-k')}]\ldots]](r)\\
= &\: \int_{0}^{r} r_{\f{d-3}2+\ell}^{-2} \ldots \int_0^{r_2} r_1^{-2} \int_0^{r_1} \Big(\f{r'}{r'+\f u2}\Big)^{d-3+2\ell}  (r'+\f u2)^{-j+\f{d-5}2 + \ell}\log^{k-k'} (r' +\f u 2) \, \ud r' \, \ud r_1 \, \ldots \, \ud r_{\f{d-3}2+\ell} \\
= &\: \int_0^r  \int_{r'}^r r_1^{-2} \ldots \int_{r_{\f{d-3}2+\ell-2}}^r r_{\f{d-3}2+\ell-1 }^{-2}\Big( \int_{r_{\f{d-3}2+\ell-1}}^r r_{\f{d-3}2+\ell}^{-2} \ud r_{\f{d-3}2+\ell}\Big) \ud r_{\f{d-3}2+\ell-1} \ldots \ud r_1 \mathfrak H_{(d-3+2\ell)}^{(j+\f{d-1}2+\ell,k-k')}(u,r') \, \ud r' \\
= &\: \f{1}{(\f{d-1}2 +\ell)!} \int_0^r (r')^{-\f{d-5}2-\ell} \mathfrak H_{(d-3+2\ell)}^{(j+\f{d-1}2+\ell,k-k')}(u,r') \, \ud r' +o_{r\to \infty}(1),
\end{split}
\end{equation}
where we have noted that in all the integrals in $r_1,\ldots, r_{\f{d-3}2+\ell}$, the boundary terms at $r_i = r$ give rise to a contribution which is $o(1)$ as $r\to \infty$. Now, to evaluate the integral in \eqref{eq:varphi.explicit.radiation.1}, we compute
\begin{equation}\label{eq:varphi.explicit.radiation.2}
\begin{split}
&\: \int_0^r (r')^{-\f{d-5}2-\ell} \mathfrak H_{(d-3+2\ell)}^{(j+\f{d-1}2+\ell,k-k')}(u,r') \, \ud r' \\
= &\: \int_0^{r} (r')^{\f{d-1}2 +\ell}  (r'+\f u2)^{-j-\f{d+1}2-\ell}\log^{k-k'} (r' +\f u 2) \, \ud r' \\
= &\: \int_{\f u2}^{\infty} (s-\f u2)^{\f{d-1}2 +\ell}  (s)^{-j-\f{d+1}2-\ell}\log^{k-k'} s \, \ud s + o_{r\to \infty}(1) \\
= &\: \sum_{k'' = 0}^{k-k'} {k-k' \choose k''} (\tfrac u2)^{-j} \log^{k''} (\tfrac u2) \int_{1}^{\infty} (\sigma-1)^{\f{d-1}2 +\ell}  (\sigma)^{-j-\f{d+1}2-\ell} \log^{k-k'-k''} (\sigma) \, \ud \sigma + o_{r\to \infty}(1).
\end{split}
\end{equation}
where we have used the substitution $s = r'+\f u2$ and then $\sigma = \f{2s}{u}$.

To compute the constants that arise in \eqref{eq:varphi.explicit.radiation.2}, we introduce the change of variable $t = \f{1}{\sigma}$ and obtain 
\begin{equation}\label{eq:varphi.explicit.radiation.3}
\begin{split}
&\:  \int_{1}^{\infty} (\sigma-1)^{\f{d-3}2 +\ell}  (\sigma)^{-j-\f{d+1}2-\ell} \log^{k-k'-k''} (\sigma) \, \ud \sigma \\
= &\: (-1)^{k-k'-k''}\int_0^1 (1-t)^{\f{d-3}2+\ell} t^j \log^{k-k'-k''} t \, \ud t \\
= &\: (-1)^{k-k'-k''} \sum_{i=0}^{\f{d-3}2+\ell} {\f{d-3}2+\ell \choose i} (-1)^i \int_0^1 t^{i+j} \log^{k-k'-k''} t \, \ud t \\
= &\: \sum_{i=0}^{\f{d-3}2+\ell} (-1)^{k-k'-k''+i} {\f{d-3}2+\ell \choose i} (-1)^{k-k'-k''} (i+j+1)^{-k+k'+k''-1} (k-k'-k'')! .
\end{split}
\end{equation}

Thus, using the expression in \eqref{eq:med.explicit.expanded} and combining \eqref{eq:varphi.explicit.radiation.1}--\eqref{eq:varphi.explicit.radiation.3}, we obtain
\begin{align}
&\: \lim_{r\to \infty} r^{\nB} \varphi^{\bfm[1]}_{(\ell)j,k}(u,r) \notag\\
= &\: \f{1}{(\f{d-1}2 +\ell)!}\sum_{k'=0}^k \mathfrak b_{k'}^{(j,k,\ell)} \sum_{k''=0}^{k-k'} \f{(k-k')!}{k''!} \sum_{i=0}^{\f{d-3}2+\ell} (-1)^i {\f{d-3}2+\ell \choose i} (i+j+1)^{-k+k'+k''-1} \log^{k''} (\tfrac u2) \label{eq:Mink.RF.1}\\
= &\: \f{\mathfrak b_{0}^{(j,k,\ell)}}{(\f{d-1}2 +\ell)!} \sum_{i=0}^{\f{d-3}2+\ell} (-1)^i {\f{d-3}2+\ell \choose i} (i+j+1)^{-1} \log^{k} (\tfrac u2) \label{eq:Mink.RF.2}\\
&\: + \f{1}{(\f{d-1}2 +\ell)!}\Bigg( \sum_{i=0}^{\f{d-3}2+\ell} (-1)^i {\f{d-3}2+\ell \choose i} \Big( \f{\mathfrak b_{0}^{(j,k,\ell)} k}{ (i+j+1)^{2}} + \f{\mathfrak b_{1}^{(j,k,\ell)} }{i+j+1} \Big) \Bigg) \log^{k-1} (\tfrac u2) + \ldots. \label{eq:Mink.RF.3}
\end{align}
Now \eqref{eq:radiation.field.expansion} follows from \eqref{eq:Mink.RF.1}.

Reading off the coefficient in \eqref{eq:Mink.RF.2} and plugging in value of $\mathfrak b_{0}^{(j,k,\ell)}$ from the Lemma~\ref{lem:med.explicit.1}, we see that
\begin{equation}\label{eq:d1.check.nonzero}
\begin{split}
\mathfrak d_0 = &\: \f{\mathfrak b_{0}^{(j,k,\ell)}}{(\f{d-1}2 +\ell)!} \sum_{i=0}^{\f{d-3}2+\ell} (-1)^i {\f{d-3}2+\ell \choose i} (i+j+1)^{-1} \\
= &\: \f{(-j)(-j+1) \ldots (-j+\nB+\ell)}{(\f{d-1}2 +\ell)!} \int_0^1 (1-t)^{\f{d-3}2+\ell} t^j \, \ud t.
\end{split}
\end{equation}
It follows that $\mathfrak d_{0} = 0$ if and only if $\ell \geq j - \nB+1$. Now when $\mathfrak d_{0} = 0$, \eqref{eq:Mink.RF.3} shows that
\begin{equation}
\mathfrak d_1 = \f{\mathfrak b_{1}^{(j,k,\ell)}}{(\f{d-1}2 +\ell)!} \sum_{i=0}^{\f{d-3}2+\ell} (-1)^i {\f{d-3}2+\ell \choose i}   (i+j+1)^{-1},
\end{equation}
which is $\neq 0$ by a similar computation as \eqref{eq:d1.check.nonzero} using the expression for $\mathfrak b_{1}^{(j,k,\ell)}$ from the Lemma~\ref{lem:med.explicit.1}.

\pfstep{Step~2: Proof of \eqref{eq:explicit.wave.upper}} We will allow all implicit constants to depend on $d,j,k,I_u,I_r$ without further comments, and only track the dependence on $\ell$.

Before we proceed, we give a bound on \eqref{eq:varphi.explicit.radiation.3}. Simple calculus shows that $\sup_{t\in [0,1]} |t \log^k t| \leq e^{-k} k^k$ (since the function is maximized at $t=e^{-k}$). Therefore, since $j\geq 0$, using the second line in \eqref{eq:varphi.explicit.radiation.3}, we have
\begin{equation}\label{eq:silly.calculus}
\Big| \int_{1}^{\infty} (\sigma-1)^{\f{d-3}2 +\ell}  (\sigma)^{-j-\f{d+1}2-\ell} \log^{k-k'-k''} (\sigma) \, \ud \sigma  \Big| \leq e^{-k} k^k.
\end{equation}

Returning to \eqref{eq:explicit.wave.upper}, we use \eqref{eq:du.dr.varphi.m} to obtain
\begin{equation}\label{eq:varphi.explicit.radiation.derivatives.1}
\begin{split}
&\: \Big| r^{\nB+\ell} \rd_u^{I_u} \rd_r^{I_r} (r^{-\ell} \varphi^{\bfm[1]}_{(\ell)j,k})\Big|(u,r) \\
\ls &\: (\tfrac{d-1}2+\ell)! \brk{\ell}^{I_u+I_r} r^{\nB} \mathfrak I_{\f{d-1}2+\ell+I_r}[ \mathfrak I_{\f{d+1}2+\ell+I_r}[\cdot\,\cdot\,\cdot \mathfrak I_{d-2+2\ell+I_r}[\mathfrak f_{u}^{(j + \f{d-1}2+\ell+I_u+I_r, k-k')}]\ldots]](r).
\end{split}
\end{equation}
where in the last line we used $(\f{r}{r+\f u2})^{I_r} \leq 1$.

On the other hand, arguing as in \eqref{eq:varphi.explicit.radiation.1} with Fubini's theorem (but only aiming for an easier upper bound without needing to extract the main term),
\begin{equation}\label{eq:varphi.explicit.radiation.derivatives.2}
\begin{split}
&\: r^{\nB} \mathfrak I_{\f{d-1}2+\ell+I_r}[ \mathfrak I_{\f{d+1}2+\ell+I_r}[\cdot\,\cdot\,\cdot \mathfrak I_{d-2+2\ell+I_r}[\mathfrak f_{u}^{(j + \f{d-1}2+\ell+I_u+I_r, k-k')}]\ldots]](r) \\
\ls &\: r^{-I_r} \int_{0}^{r} r_{\f{d-3}2+\ell}^{-2} \ldots \int_0^{r_2} r_1^{-2} \int_0^{r_1} \mathfrak H_{(d-3+2\ell+I_r)}^{(j+\f{d-1}2+\ell+I_u+I_r,k-k')}(u,r') \, \ud r' \, \ud r_1 \, \ldots \, \ud r_{\f{d-3}2+\ell} \\
\ls &\: \f{r^{-I_r}}{(\f{d-1}2 +\ell)!} \int_0^r (r')^{-\f{d-5}2-\ell} \mathfrak H_{(d-3+2\ell+I_r)}^{(j+\f{d-1}2+\ell+I_u+I_r,k-k')}(u,r') \, \ud r' .
\end{split}
\end{equation}
Using $(\f r{r+\f u2})^{I_r} \leq 1$, we have $\mathfrak H_{(d-3+2\ell+I_r)}^{(j+\f{d-1}2+\ell+I_u+I_r,k-k')}(u,r') \leq \mathfrak H_{(d-3+2\ell)}^{(j+\f{d-1}2+\ell+I_u,k-k')}(u,r')$. Thus, using \eqref{eq:varphi.explicit.radiation.derivatives.1}, \eqref{eq:varphi.explicit.radiation.derivatives.2} together with the bounds \eqref{eq:varphi.explicit.radiation.2}, \eqref{eq:varphi.explicit.radiation.3} and \eqref{eq:silly.calculus} to obtain
\begin{equation} \label{eq:explicit.wave.upper.key}
\begin{split}
 \Big| r^{\nB+\ell} \rd_u^{I_u} \rd_r^{I_r} (r^{-\ell} \varphi^{\bfm[1]}_{(\ell)j,k})\Big|(u,r)
\ls  \brk{\ell}^{I_u+I_r} r^{-I_r} u^{-j-I_u} \log^k (\tfrac u2),
\end{split}
\end{equation}
which implies \eqref{eq:explicit.wave.upper}.

\pfstep{Step~3: Proof of \eqref{eq:explicit.wave.dr.upper}}
Using \eqref{eq:med.explicit.expanded} and the fact that $(\rd_{r} + \tfrac{\nB}{r}) r^{\ell} \frkI_{\frac{d-1}{2} + \ell}[f] = r^{\ell-1} f$, we have
\begin{equation} \label{eq:med.explicit.expanded.dr}
(\rd_{r} + \tfrac{\nB}{r})\varphi^{\bfm[1]}_{(\ell)j,k}(u,r)
=  \sum_{k'=0}^k  \mathfrak b_{k'}^{(j,k,\ell)} r^{\ell-1} \mathfrak I_{\f{d+1}2+\ell}[\cdot\,\cdot\,\cdot \mathfrak I_{d-2+2\ell}[\mathfrak f_{u}^{(j + \f{d-1}2+\ell, k-k')}]\ldots](r).
\end{equation}
Repeating the proof of \eqref{eq:explicit.wave.upper.key}, but using \eqref{eq:med.explicit.expanded.dr} in place of \eqref{eq:med.explicit.expanded}, we obtain
\begin{equation} \label{eq:explicit.wave.dr.upper.key}
\begin{split}
 \Big| r^{\nB+\ell+1} \rd_u^{I_u} \rd_r^{I_r} (r^{-(\ell-1)} (\rd_{r} + \tfrac{\nB}{r}) \varphi^{\bfm[1]}_{(\ell)j,k})\Big|(u,r)
\ls  \brk{\ell}^{I_u+I_r+1} r^{-I_r} u^{-j-I_u} \log^k (\tfrac u2),
\end{split}
\end{equation}
which implies \eqref{eq:explicit.wave.dr.upper}
\qedhere

\end{proof}

\subsection{An auxiliary estimate}
We will need the following auxiliary estimate later. The proof is very similar to Propositions~\ref{prop:phi.ell.j.k.upper} and \ref{prop:radiation.field}.
\begin{lemma}
If $u' \leq \frac{u}{2}$ and $u \gtrsim \max\{a,1\}$ is sufficiently large, then
\begin{equation}
\begin{split}
	&\: \Big| \rd_u^{I_u} \rd_r^{I_r} \Big( \varphi^{\bfm [a]}_{(\ell) j, k}(u-u', r) - \varphi^{\bfm [a]}_{(\ell) j, k}(u, r) \Big) \Big| \\
	&\: \quad \ls_{d,j,k,I_u,I_r}  \brk{\ell}^{I_u+I_r+1} u' \min\{ u^{-j-\nB-1}, u^{-j-1}r^{-\nB}\} \log^{k} (\tfrac ua) \min\{r^\ell (\tfrac u2)^{-\ell} , 1\},\label{eq:profile.move.around} 
	\end{split}
	\end{equation}
\end{lemma}
\begin{proof}

Let $u' \in [2,\frac{u}2]$. We compute the derivatives using \eqref{eq:dr.du.fu}, and then use the mean value theorem to obtain
\begin{equation*}
\begin{split}
&\: \Big|\rd_r^{I_r} \rd_u^{I_u} \Big( \mathfrak f_{u-u'}^{(j+\f{d-1}2+\ell,k-k')}(r) - \mathfrak f_{u}^{(j+\f{d-1}2+\ell,k-k')}(r) \Big) \Big| \ls \brk{\ell}^{I_u+I_r+1} u' \mathfrak f_u^{(j+\f{d-1}2+\ell+1+I_u+I_r,k-k')}(r).
\end{split}
\end{equation*}
We then argue as in the proofs of Propositions~\ref{prop:phi.ell.j.k.upper} and \ref{prop:radiation.field} to obtain the desired estimates. \qedhere
\end{proof}

\section{Analysis in the near-intermediate zone: stationary estimate} \label{sec:near}


\subsection{Statement of the main iteration lemma}

The following is the main iteration lemma we prove in the near-intermediate zone.

\begin{proposition} \label{prop:near-med}
Let $0 < \dlt_{0} \leq \min\set{\dlt_{c}, \dlt_{d}, \f 12}$. In case $\calN \neq 0$, assume that
\begin{equation} \label{eq:near-med-alp}
\alp_{0} \geq \alp_{\calN} + 2 \dlt_{0},
\end{equation}
where $\alp_{0}$ and $\alp_{\calN}$ are from \ref{hyp:sol} and Definition~\ref{def:alp-N}, respectively. %
Assume that, for some $M \in \bbZ_{\geq 0}$ (with $M\leq \min\{M_0, M_c\}$), $A > 0$, $\alp \geq \alp_0$, and $K \in \mathbb Z_{\geq 0}$, the following estimates hold:
\begin{align}
\phi &= O_{\bfGmm}^{M}(A \tau^{-\alp}) \quad \hbox{ in } \calM_{\near} \cup \calM_{\med}, \label{eq:near-med-ass-near} \\
\phi &= O_{\bfGmm}^{M}(A \tau^{-\alp-\dlt_{0}+\nu_{\Box}} r^{-\nu_{\Box}} \log^{K}\tau) \quad \hbox{ in } \calM_{\med}. \label{eq:near-med-ass-med}
\end{align}
Then, the following estimate holds:
\begin{align*}
\phi &= O_{\bfGmm}^{M'}\left( A' \max\{\tau^{-\alp-\dlt_{0}} \log^K \tau, \tau^{-\alp_{d}}\} \right) \quad \hbox{ in } \calM_{\near} \cup \calM_{\med},
\end{align*}
for $M' \leq\f{M}{2(\lfloor \f d2 \rfloor + s_c)} - \f{5d}2$, and $A'$ can be chosen as
\begin{equation} \label{eq:A'}
\begin{aligned}
A' &= D+A + A^{(2)}(A) \quad \hbox{ if } \calN \neq 0 \\
A' &= D+ A \quad \hbox{ if } \calN = 0,
\end{aligned}\end{equation}
where $A^{(2)}(A) \geq 0$ is nondecreasing function depending on $\calN$ satisfying $A^{(2)}(A) = O(A^{2})$ as $A \to 0$.
\end{proposition}

We also state an improved version of the iteration lemma for higher angular modes for \textbf{linear} equations when $\calP$ is \textbf{spherically symmetric}.
\begin{proposition} \label{prop:near-med-sphsymm}
Let $0 < \dlt_{0} \leq \min\set{\dlt_{c}, \dlt_{d},\f 12}$. Assume $\calN = 0$, and that $\calP$ is spherically symmetric.
Assume that, for some $M \in \bbZ_{\geq 0}$ (with $M\leq \min \{M_0,M_c\}$), $A > 0$, $\alp \in \mathbb R$, and $K \in \mathbb Z_{\geq 0}$, the following estimates hold:
\begin{align*}
\phi_{(\geq \ell)} &= O_{\bfGmm}^{M}(A \brk{r}^{\ell} \tau^{-\alp-\ell}) \quad \hbox{ in } \calM_{\near} \cup \calM_{\med},
\end{align*}
\begin{equation*}
\phi_{(\geq \ell)} = O_{\bfGmm}^{M}(A u^{-\alp-\dlt_{0}+\nu_{\Box}} r^{-\nu_{\Box}} \log^{K}\tau) \quad \hbox{ in } \calM_{\med}.
\end{equation*}

Then the following estimate holds for $M' \leq\f{M}{2(\lfloor \f d2 \rfloor + s_c)} - \f{5d}2 - 5\ell$:
\begin{align*}
\phi_{(\geq \ell)} &= O_{\bfGmm}^{M'}\left( (D+A) \brk{r}^{\ell} \tau^{-\ell} \max\{\tau^{-\alp-\dlt_{0}} \log^K \tau, \tau^{-\alp_{d}}\} \right) \quad \hbox{ in } \calM_{\near} \cup \calM_{\med}.
\end{align*}
\end{proposition}

\subsection{Extension of the stationary estimate} \label{subsec:stat-est} 
The goal of this subsection is to extend the right-inverse ${}^{(\tau)} \calR_{0}$ in \ref{hyp:ult-stat} ($\tau = \infty$) or \ref{hyp:ult-stat'} to inputs which do not necessarily have compact support. The proof below resembles the argument in \cite{MTT-maxwell, MST}. In what follows, {\bf we take $\tau$ to be either (1) $\infty$ if \ref{hyp:ult-stat} holds, or (2) sufficiently large if \ref{hyp:ult-stat'} holds}.

 We begin by defining the relevant weighted $L^{2}$-based Sobolev spaces. In view of the condition \ref{hyp:topology}, we identify $\Sgm_{\tau}$ with $\bbR^{d} \setminus \calK$. We denote by $x = (x^{1}, \ldots, x^{d})$ the standard Cartesian coordinates on $\bbR^{d}$, and by $\urd = (\rd_{x^{1}}, \ldots \rd_{x^{d}})$ the associated coordinate derivatives.

\begin{definition} \label{def:w-soblev}
For $1 \leq p \leq \infty$, $s \in \bbZ_{\geq 0}$ and $\gmm \in \bbR$, define the norms
\begin{align*}
	\nrm{h}_{\ell_{r}^{p} L^{2}(\brk{r}^{2 \gmm} \, \ud x)} &= \nrm*{R^{\gmm}\nrm*{h }_{L^{2}(A_{R})}}_{\ell_{r}^{p}\set{R \in 2^{\bbZ_{\geq 0}}}}, \\
	\nrm{h}_{\ell_{r}^{p} \calH^{s, \gmm}} &= \sum_{\alp : \abs{\alp} \leq s} \nrm*{(\brk{r} \urd)^{\alp} h}_{\ell_{r}^{p} L^{2}(\brk{r}^{2 \gmm} \, \ud x)}.
\end{align*}
The corresponding function spaces $\ell_{r}^{p} L^{2}(\brk{r}^{2 \gmm} \, \ud x)$ and $\ell_{r}^{p} \calH^{s, \gmm} = \ell_{r}^{p} \calH^{s, \gmm}(\Sgm_{\tau})$ consist of the distributions on $\Sgm_{\tau}$ with finite respective norms. When $p = 2$, we simply write $L^{2}(\brk{r}^{2 \gmm} \, \ud x) = \ell_{r}^{2} L^{2}(\brk{r}^{2 \gmm} \, \ud x)$ and $\calH^{s, \gmm} = \ell_{r}^{2} \calH^{s, \gmm}$.
\end{definition}
Observe the obvious embeddings
\begin{equation*}
	\ell_{r}^{p_{1}} \calH^{s, \gmm} \subseteq \ell_{r}^{p_{2}} \calH^{s, \gmm} \hbox{ for } p_{1} < p_{2}, \qquad
	\ell_{r}^{p} \calH^{s, \gmm_{1}} \subseteq \ell_{r}^{p} \calH^{s, \gmm_{2}} \hbox{ for } \gmm_{1} < \gmm_{2}, \qquad
	\ell_{r}^{p} \calH^{s_{2}, \gmm} \subseteq \ell_{r}^{p} \calH^{s_{1}, \gmm} \hbox{ for } s_{1} < s_{2}.
\end{equation*}

The main result of this subsection is as follows.
\begin{proposition} \label{prop:stat-est}
The operator ${}^{(\tau)} \calR_{0}$ extends to a bounded linear operator
\begin{equation*}
{}^{(\tau)} \calR_{0} : \ell_{r}^{1} \calH^{s_{c}, -\frac{d}{2}+2} \to \ell_{r}^{\infty} \calH^{2, -\frac{d}{2}},
\end{equation*}
such that the following properties hold:
\begin{enumerate}
\item (right-inverse of ${}^{(\tau)} \calP_{0}$) ${}^{(\tau)} \calP_{0} {}^{(\tau)} \calR_{0} h = h$  for all $h \in \ell_{r}^{1} \calH^{s_{c}, -\frac{d}{2}+2}$.
\item (left-inverse of ${}^{(\tau)} \calP_{0}$) ${}^{(\tau)} \calR_{0} {}^{(\tau)} \calP_{0} \varphi = \varphi$  for all $\varphi \in \ell_{r}^{1} \calH^{s_{c}+2, -\frac{d}{2}}$.
\item (higher regularities and weights) For every $0 \leq s \leq M_{c}-C$ and $\gmm \in [-\frac{d}{2}, \frac{d}{2}-2]$, we have
\begin{align}\label{eq:stat-est.main.1}
	\nrm{{}^{(\tau)} \calR_{0} h}_{\ell_{r}^{\infty} \calH^{s+2, \gmm}}
	&\aleq_{s} \nrm{h}_{\ell_{r}^{1} \calH^{s+s_{c}, \gmm+2}}.
\end{align}
\end{enumerate}
\end{proposition}

The range of $\gmm$ is dictated by the mapping properties of $\Delta_{\bfe}$ on $\bbR^{d}$; see Lemma~\ref{lem:stat-est-lap} below. We note that Statement~(2) in Proposition~\ref{prop:stat-est} implies a version of Liouville's theorem for ${}^{(\tau)} \calP$, i.e., any regular entire solution $\varphi$ to ${}^{(\tau)} \calP \varphi = 0$ satisfying $\varphi \in \ell^{1}_{r} \calH^{s_{c}+2, -\frac{d}{2}}$ (to which $1$ fails to belong just logarithmically) must be trivial. Also, \eqref{eq:stat-est.main.1} holds with $\ell^{\infty}_{r}$ and $\ell^{1}_{r}$ replaced by $\ell^{2}_{r}$ if $-\frac{d}{2} < \gmm < \frac{d}{2} - 2$ (as in the case of $\Delta_{\bfe}$ on $\bbR^{d}$), but we will not need this fact.

When ${}^{(\tau)} \calP_{0}$ is spherically symmetric, we have the following refinement of Proposition~\ref{prop:stat-est}.
\begin{proposition} \label{prop:stat-est.2}
Assume that ${}^{(\tau)} \calP_{0}$ and ${}^{(\tau)} \calR_{0}$ are spherically symmetric (as in Main Theorem~\ref{thm:upper-sphsymm} or \ref{thm:lower-sphsymm}). Then ${}^{(\tau)} \calP_{0}$ and ${}^{(\tau)} \calR_{0}$ commute with $\bbS_{(\ell)}$ for any $\ell \geq 0$. Moreover, the operator ${}^{(\tau)} \calR_{0}$ extends to a bounded linear operator
\begin{equation*}
{}^{(\tau)} \calR_{0} :  \bbS_{(\geq \ell)} \ell_{r}^{1} \calH^{s_{c}, -\frac{d}{2}-\ell+2} \to  \bbS_{(\geq \ell)} \ell_{r}^{\infty} \calH^{2, -\frac{d}{2}-\ell},
\end{equation*}
such that the following properties hold:
\begin{enumerate}
\item (right-inverse of ${}^{(\tau)} \calP_{0}$) ${}^{(\tau)} \calP_{0} {}^{(\tau)} \calR_{0} h = h$  for all $h \in  \bbS_{(\geq \ell)} \ell_{r}^{1} \calH^{s_{c}, -\frac{d}{2}-\ell+2}$.
\item (left-inverse of ${}^{(\tau)} \calP_{0}$) ${}^{(\tau)} \calR_{0} {}^{(\tau)} \calP_{0} \varphi = \varphi$  for all $\varphi \in  \bbS_{(\geq \ell)} \ell_{r}^{1} \calH^{s_{c}+2, -\frac{d}{2}-\ell}$.
\item (higher regularities and weights) For every $0 \leq s \leq M_{c}-C$ and $\gmm \in [-\frac{d}{2}-\ell, \frac{d}{2}-\ell-2]$, we have
\begin{align}\label{eq:stat-est.main.2}
	\nrm{{}^{(\tau)} \calR_{0}  \bbS_{(\geq \ell)} h}_{\ell_{r}^{\infty} \calH^{s+2, \gmm}}
	&\aleq_{s} \nrm{ \bbS_{(\geq \ell)} h}_{\ell_{r}^{1} \calH^{s+s_{c}, \gmm+2}}.
\end{align}
\end{enumerate}
\end{proposition}

\begin{remark}
It will be clear from the proof below that, when ${}^{(\tau)} \calP_{0}$ and ${}^{(\tau)} \calR_{0}$ are spherically symmetric, Proposition~\ref{prop:stat-est.2} still holds if the decay assumption on $\psi$ in (2) of \ref{hyp:ult-stat} and \ref{hyp:ult-stat'} is modified to $O(r^{-(d+\ell-2)})$.
\end{remark}

The underlying idea of the proof of Propositions~\ref{prop:stat-est} and \ref{prop:stat-est.2} is that since ${}^{(\tau)} \calP_{0}$ asymptotes to $\Delta_{\bfe}$ as $x \to \infty$, the mapping properties of $\Delta_{\bfe}$ can be transferred. Accordingly, we begin by stating the mapping properties of $\Delta_{\bfe}$ among $\ell^{p} \calH^{s, \gmm}$'s.

\begin{lemma} \label{lem:stat-est-lap}
Let $\ell \in \bbZ_{\geq 0}$. For any $s \in \bbZ_{\geq 0}$ and $- \frac{d}{2} - \ell < \gmm < \frac{d}{2} + \ell - 2$, we have
\begin{align}\label{eq:stat-est-lap}
	\nrm{(-\Delta_{\bfe})^{-1} \bbS_{(\geq \ell)} h}_{\calH^{s+2, \gmm}}
	\aleq_{d, \ell, s, \gmm} \nrm{\bbS_{(\geq \ell)} h}_{\calH^{s, \gmm}}.
\end{align}
For the endpoints cases, we have
\begin{align*}
	\nrm{(-\Delta_{\bfe})^{-1} \bbS_{(\geq \ell)} h}_{\ell^{\infty}_{r} \calH^{s+2, -\frac{d}{2}-\ell}}
	\aleq_{d, \ell, s} &\: \nrm{\bbS_{(\geq \ell)} h}_{\ell^{1}_{r} \calH^{s, -\frac{d}{2}-\ell+2}}, \\
	\nrm{(-\Delta_{\bfe})^{-1} \bbS_{(\geq \ell)} h}_{\ell^{\infty}_{r} \calH^{s+2, \frac{d}{2}+\ell-2}}
	\aleq_{d, \ell, s} &\: \nrm{\bbS_{(\geq \ell)} h}_{\ell^{1}_{r} \calH^{s, \frac{d}{2}+\ell}}.
\end{align*}
\end{lemma}
\begin{proof}

\pfstep{Step~1(a): Lowest order estimate (non-endpoint cases)}
We claim that, for $- \frac{d}{2} - \ell < \gmm < \frac{d}{2} + \ell - 2$,
\begin{equation} \label{eq:stat-est-lap-nonendpoint}
	\nrm{(-\Delta_{\bfe})^{-1} \bbS_{(\geq \ell)} h}_{L^{2}(r^{2 \gmm} \, \ud x)}
	\aleq_{d, \ell} \nrm{r^{2} \bbS_{(\geq \ell)} h}_{L^{2}(r^{2 \gmm} \, \ud x)}.
\end{equation}
To prove this, we use an energy argument. Let $\varphi \in C^{\infty}_{c}(\bbR^{d})$ satisfy $\varphi = \bbS_{(\geq \ell)} \varphi$. We compute
\begin{align*}
	\int \left[ - \Delta_{\bfe} \varphi \, r^{2 \gmm + 2} \varphi \right] \, \ud x
	&= - \int \left(\rd_{r} (r^{d-1} \rd_{r} \varphi) + r^{d-3} \rslap \varphi \right) r^{2 \gmm + 2} \varphi \, \ud r \ud \rssgm \\
	&= \int \left[r^{2\gmm + d + 1} (\rd_{r} \varphi)^{2} + (2 \gmm + 2) r^{2 \gmm + d} \varphi \rd_{r} \varphi + r^{2 \gmm + d-1} \abs{\rsnb \varphi}^{2} \right] \, \ud r \ud \rssgm \\
	&= \int \left[r^{2\gmm + d + 1} (\rd_{r} \varphi)^{2} - (\gmm + 1)(2 \gmm + d) r^{2 \gmm + d-1} \varphi^{2} + r^{2 \gmm + d-1} \abs{\rsnb \varphi}^{2} \right] \, \ud r \ud \rssgm.
\end{align*}
Using Lemma~\ref{lem:hardy-radial} (Hardy inequality) with $(r_{0}, r_{1}) = (0, \infty)$ and $p = 2 \gmm + d + 1$, as well as the Poincar\'e inequality
\begin{equation*}
\int \abs{\rsnb \varphi}^{2} \, \ud \rssgm \geq \ell (\ell + d - 2) \int \varphi^{2} \, \ud \rssgm,
\end{equation*}
which holds since $\varphi = \bbS_{(\geq \ell)} \varphi$, we have
\begin{align*}
	\int \left[ - \Delta_{\bfe} \varphi \, r^{2 \gmm + 2} \varphi \right] \, \ud x
	&\geq \int \left( \frac{1}{4} (2 \gmm + d)^{2} - (\gmm + 1)(2 \gmm + d) + \ell (\ell + d - 2) \right) r^{2 \gmm + d-1} \varphi^{2} \, \ud r \ud \rssgm \\
	&\geq (-\gmm + \tfrac{d}{2} + \ell - 2) (\gmm + \tfrac{d}{2} + \ell) \int r^{2 \gmm + d-1} \varphi^{2} \, \ud r \ud \rssgm.
\end{align*}
In the application of Lemma~\ref{lem:hardy-radial}, we need to take care to ensure that the boundary term at $r = 0$ vanishes. We begin by observing that if $\varphi = \bbS_{(\geq \ell)} \varphi$, then $\varphi = O(r^{\ell})$ as $r \to 0$. Inspecting Lemma~\ref{lem:hardy-radial}, we see that we need
\begin{equation*}
p - 1 + 2\ell = 2 \gmm + d + 1 - 1 + 2\ell > 0
\end{equation*}
to ensure that the boundary term at $r = 0$ vanishes. This condition is equivalent to $\gmm > - \frac{d}{2} - \ell$, which is one of the assumptions for $\gmm$. Observe furthermore that the assumptions on $\gmm$ ensures that the $(-\gmm + \tfrac{d}{2} + \ell - 2) (\gmm + \tfrac{d}{2} + \ell) > 0$. Hence, we have proved
\begin{equation*}
	\nrm{\varphi}_{L^{2}(r^{2 \gmm} \ud x)} \aleq_{\gmm, d, \ell} \nrm{r^{2} (-\Delta_{\bfe}) \varphi}_{L^{2}(r^{2 \gmm} \ud x)}.
\end{equation*}
By duality and the fact that $-\Delta_{\bfe}$ is self-adjoint with respect to $L^{2}(\ud x)$, it follows that $-\Delta_{\bfe}$ is invertible between the weighted $L^{2}$ spaces defined with the above norms. In conclusion, \eqref{eq:stat-est-lap-nonendpoint} follows.

\pfstep{Step~1(b): Lowest order estimates (endpoint cases)}
For the endpoint case, we only need to consider a single spherical harmonic. For each $\ell$, we prove
\begin{align}
	\nrm{(-\Delta_{\bfe})^{-1} \bbS_{(\ell)} h}_{\ell^{\infty}_{r} L^{2}(r^{-d-2\ell} \, \ud x)}
	\aleq_{d, \ell} \nrm{r^{2} \bbS_{(\ell)} h}_{\ell^{1}_{r} L^{2}(r^{-d-2\ell} \, \ud x)}, \label{eq:stat-est-lap-endpoint} \\
	\nrm{(-\Delta_{\bfe})^{-1} \bbS_{(\ell)} h}_{\ell^{\infty}_{r} L^{2}(r^{d+2\ell-4} \, \ud x)}
	\aleq_{d, \ell} \nrm{r^{2} \bbS_{(\ell)} h}_{\ell^{1}_{r} L^{2}(r^{d+2\ell-4} \, \ud x)}. \label{eq:stat-est-lap-endpoint'}
\end{align}
By duality, it suffices to prove one of the two bounds; we focus on \eqref{eq:stat-est-lap-endpoint}.

Suppose $\varphi \in C^\infty_c(\mathbb R^d)$ with $\varphi = \mathbb S_{(\ell)} \varphi$ and such that
$$\Delta_{\bfe} \varphi = r^{-d+1} \rd_r (r^{d-1} \rd_r \varphi)  - r^{-2} \ell(\ell + d -2) \varphi = h$$
(for some $h \in C^\infty_c(\mathbb R^d)$ with $h = \mathbb S_{(\ell)} h$).

Notice that
\begin{equation}
\begin{split}
\rd_r (r^{d+2\ell-1} \rd_r (r^{-\ell} \varphi)) =&\: \rd_r (r^{d+\ell-1} \rd_r\varphi) - \ell \rd_r (r^{d+\ell-2}  \varphi) \\
= &\: r^{\ell} \rd_r (r^{d-1} \rd_r \varphi) - \ell (\ell + d - 2) r^{d+\ell-3} \varphi = r^{d+\ell-1} h,
\end{split}
\end{equation}
from which we deduce (by integrating twice)
\begin{equation}\label{eq:varphi.endpoint}
 \varphi(r) = - r^\ell \int_r^\infty (r')^{-d-2\ell+1} \int_0^{r'} (r'')^{d+\ell-1} h(r'') \,\ud r''\, \ud r'.
\end{equation}

By the Cauchy--Schwarz inequality, we have
\begin{equation*}
\begin{split}
\int_0^{r'} (r'')^{d+\ell-1} h(r'') \,\ud r'' \leq &\: \Big(\int_0^{r'} (r'')^{2d+4\ell-5} \, \ud r'' \Big)^{\f 12} \Big( \int_0^{r'} (r'')^{3-2\ell} h^2(r'') \, \ud r'' \Big)^{\f 12} \\
= &\: \f{(r')^{d+2\ell-4}}{\sqrt{d+2\ell-4}} \Big( \int_0^{r'} (r'')^{3-2\ell} h^2(r'') \, \ud r'' \Big)^{\f 12},
\end{split}
\end{equation*}
so that using \eqref{eq:varphi.endpoint} we obtain the pointwise estimate
$$\sup_{r\geq 0} \Big( r^{-\ell} |\varphi(r)| \Big) \ls \nrm{r^{2} \bbS_{(\ell)} h}_{\ell^{1}_{r} L^{2}(r^{-d-2\ell} \, \ud x)}.$$
Integrating over a dyadic $r$-interval, and recalling $-\varphi = (-\Delta_{\bfe})^{-1} \mathbb S_{(\ell)} h$, we obtain \eqref{eq:stat-est-lap-endpoint}.

\pfstep{Step~2(a): Weighted elliptic estimates and bounds up to second order (non-endpoint cases)}
To obtain up to second order bounds, we apply a standard weighted $L^2$-based elliptic estimates. We continue to take $\Delta_{\bfe} \varphi = h$ and assume both $\varphi$ and $h$ to be supported on spherical modes $\geq \ell$. We compute
\begin{equation}\label{eq:stat-est-elliptic.1}
\begin{split}
h^2  r^{2\gamma+d+3} = &\: (\Delta_{\bfe} \varphi)^2 r^{2\gamma+d+3} \\
= &\: (\rd_r(r^{d-1} \rd_r\varphi))^2 r^{2\gamma-d+5} + 2 (\rd_r(r^{d-1} \rd_r\varphi)) \mathring{\slashed{\Delta}} \varphi r^{2\gamma+2} + (\mathring{\slashed{\Delta}} \varphi)^2 r^{2\gamma+d-1} \\
=&\: (\rd_r(r^{d-1} \rd_r\varphi))^2 r^{2\gamma-d+5} + (\mathring{\slashed{\Delta}} \varphi)^2 r^{2\gamma+d-1} + 2 |\mathring{\slashed{\nabla}}\rd_r \varphi|^2 r^{2\gamma+d+1}\\
&\: - 2(\gamma+1) (2\gamma + d) |\mathring{\slashed{\nabla}} \varphi|^2 r^{2\gamma+d-1} + \rd_r (\ldots)  + \mathring{\slashed{\mathrm{div}}}(\ldots).
\end{split}
\end{equation}
We further manipulate the first two terms in \eqref{eq:stat-est-elliptic.1}. For the first term,
\begin{equation}\label{eq:stat-est-elliptic.2}
\begin{split}
&\: (\rd_r(r^{d-1} \rd_r\varphi))^2 r^{2\gamma-d+5} \\
 = &\: (r^{d-2} \rd_r (r\rd_r \varphi) + (d-2) r^{d-2} \rd_r \varphi)^2 r^{2\gamma-d+5} \\
 = &\: ((r\rd_r)^2 \varphi)^2 r^{d+2\gamma-1} + (d-2)^2 (r \rd_r \varphi)^2 r^{d+2\gamma-1} + (d-2) \rd_r [(r\rd_r\varphi)^2 r^{d+2\gamma}] \\
 &\: - (d-2) (d+2\gamma) (r\rd_r\varphi)^2 r^{d+2\gamma-1} \\
 = &\: ((r\rd_r)^2 \varphi)^2 r^{d+2\gamma-1} - 2(\gamma+1) (d-2) (r \rd_r \varphi)^2 r^{d+2\gamma-1} + (d-2) \rd_r [(r\rd_r\varphi)^2 r^{d+2\gamma}] \\
  = &\: [((r\rd_r)^2 \varphi)^2 + 2(\gamma+1)(d-2)\varphi ((r\rd_r)^2\varphi)] r^{d+2\gamma-1} + \rd_r (\ldots).
 \end{split}
 \end{equation}
For the second term in \eqref{eq:stat-est-elliptic.1}, we use that the unit round sphere has Gauss curvature $=1$ to obtain
\begin{equation}\label{eq:stat-est-elliptic.3}
\begin{split}
(\mathring{\slashed{\Delta}} \varphi)^2 r^{2\gamma+d-1} = [|\mathring{\slashed{\nabla}}{}^2 \varphi|^2 + |\mathring{\slashed{\nabla}} \varphi|^2 + \mathring{\slashed{\mathrm{div}}}(\ldots)]r^{2\gamma+d-1}.
\end{split}
\end{equation}
Combining \eqref{eq:stat-est-elliptic.1}--\eqref{eq:stat-est-elliptic.3}, and integrating over $\mathbb R^d$ with the volume form $\ud r \ud \rssgm$, we obtain
\begin{equation}
\begin{split}
&\: \int \Big[ ((r\rd_r)^2 \varphi)^2 + |\mathring{\slashed{\nabla}}{}^2 \varphi|^2 + |\mathring{\slashed{\nabla}} \varphi|^2 + 2 r^2 |\mathring{\slashed{\nabla}}\rd_r \varphi|^2 \Big] r^{2\gamma+d-1} \, \ud r \ud \rssgm \\
\leq &\: 2 |\gamma + 1| \int \Big( (d-2) |\varphi| | (r \rd_r )^2 \varphi| + |2\gamma+d| |\mathring{\slashed{\nabla}} \varphi|^2 \Big) \, \ud r \ud \rssgm + \int  h^2 r^{2\gamma+d+3} \, \ud r \ud \rssgm \\
\leq &\: \f 12  \int \Big[ ((r\rd_r)^2 \varphi)^2 + |\mathring{\slashed{\nabla}}{}^2 \varphi|^2 \Big] r^{2\gamma+d-1} \, \ud r \ud \rssgm + C \| \varphi \|_{L^2(r^{2\gmm} \, \ud x)}^2 + \| r^2 h \|_{L^2(r^{2\gamma}\, \ud x)}^2.
\end{split}
\end{equation}
The first term on the right-hand side can be absorbed to the left-hand side. After that, use \eqref{eq:stat-est-lap-nonendpoint} so that we obtain
\begin{equation}\label{eq:elliptic.2nd.only}
\nrm{(r \urd)^{(2)} (-\Delta_{\bfe})^{-1} \bbS_{(\geq \ell)} h}_{L^{2}(r^{2 \gmm} \, \ud x)}
		\aleq_{d, \ell} \nrm{r^{2}  \bbS_{(\geq \ell)} h}_{L^{2}(r^{2 \gmm} \, \ud x)},
\end{equation}

Combining \eqref{eq:elliptic.2nd.only} with \eqref{eq:stat-est-lap-nonendpoint}, we control both zeroth and second derivatives. The first derivative estimates can be derived as a consequence after suitably integrating by parts. Thus we have
\begin{equation}\label{eq:elliptic.up.to.2.bad.r}
\nrm{(r \urd)^{(\leq 2)} (-\Delta_{\bfe})^{-1} \bbS_{(\geq \ell)} h}_{L^{2}(r^{2 \gmm} \, \ud x)}
		\aleq_{d, \ell} \nrm{r^{2}  \bbS_{(\geq \ell)} h}_{L^{2}(r^{2 \gmm} \, \ud x)}.
\end{equation}
Now combining \eqref{eq:elliptic.up.to.2.bad.r} with its $\gamma = 0$ version, we obtain
\begin{equation}\label{eq:elliptic.up.to.2.better.r}
\nrm{(r \urd)^{(\leq 2)} (-\Delta_{\bfe})^{-1} \bbS_{(\geq \ell)} h}_{L^{2}(\brk{r}^{2 \gmm} \, \ud x)}
		\aleq_{d, \ell} \nrm{r^{2}  \bbS_{(\geq \ell)} h}_{L^{2}(\brk{r}^{2 \gmm} \, \ud x)}.
\end{equation}

Finally, notice that integrating by parts in Cartesian coordinates gives
\begin{equation}
\| \Delta_{\bfe} \varphi \|_{L^2(\ud x)}^2 =  \sum_{i,j=1}^d \int \urd_i \urd_i \varphi \urd_j \urd_j \varphi \, \ud x = \sum_{i,j=1}^d \int \urd_i \urd_j \varphi \urd_i \urd_j \varphi \, \ud x
\end{equation}
so that
\begin{equation}\label{eq:elliptic.up.to.2.small.r}
\nrm{\urd^{(2)} (-\Delta_{\bfe})^{-1} \bbS_{(\geq \ell)} h}_{L^{2}(\ud x)}
		\aleq_{d, \ell} \nrm{\bbS_{(\geq \ell)} h}_{L^{2}(\ud x)}.
\end{equation}

Combining \eqref{eq:elliptic.up.to.2.better.r} and \eqref{eq:elliptic.up.to.2.small.r}, and controlling the first derivatives in $r\leq 1$ using the Hardy inequality in Lemma~\ref{lem:hardy-radial}, we obtain
\begin{equation}\label{eq:elliptic.up.to.2}
\nrm{(\brk{r} \urd)^{(\leq 2)} (-\Delta_{\bfe})^{-1} \bbS_{(\geq \ell)} h}_{L^{2}(\brk{r}^{2 \gmm} \, \ud x)}
		\aleq_{d, \ell} \nrm{\brk{r}^{2}  \bbS_{(\geq \ell)} h}_{L^{2}(\brk{r}^{2 \gmm} \, \ud x)}.
\end{equation}

\pfstep{Step~2(b): Weighted elliptic estimates and bounds up to 2nd order (endpoint cases)} To treat the endpoint cases, we argue as in Step~2(a) except for extra cutoff functions $\chi(r)$ such that either (1) $\chi \equiv 1$ when $\{2^k \leq r \leq 2^{k+1}\}$ and $\mathrm{supp}(\chi)\subset \{2^{k-1} \leq r \leq 2^{k+2}\}$, or (2) $\chi \equiv 1$ when $\{r\leq 2\}$ and $\mathrm{supp}(\chi)\subset \{ r \leq 4\}$.

\pfstep{Step~3: Commutation and higher order estimates} We will only prove in detail the higher order estimates for the non-endpoint cases. The commutation procedure (assuming the lower order bounds) is similar for the endpoint cases; we omit the details.

To handle the commutator term, we need to be slightly careful with the choice of commutators to avoid borderline $r$-weights. Notice that
$$[\bfOmg,\Delta_{\bfe}] = 0,\quad [r \urd_r, \Delta_{\bfe}] = -2\Delta_{\bfe}. $$
Iterating, we obtain
$$|[(r\urd_r)^k {\bf \Omg}^I, \Delta_{\bfe}] \varphi | \ls |(r\urd)^{(\leq k-1)} {\bf \Omg}^I \Delta_{\bfe} \varphi |.$$
As a result,
$$|\Delta_{\bfe} (r\urd_r)^k {\bf \Omg}^I \varphi| \leq |(r\urd)^{k} {\bf \Omg}^I \Delta_{\bfe} \varphi| +  |[(r\urd_r)^k {\bf \Omg}^I, \Delta_{\bfe}] \varphi | \ls |(r\urd)^{(\leq k)} {\bf \Omg}^I \Delta_{\bfe} \varphi|.  $$
Suppose now that $\varphi = \mathbb S_{(\geq \ell)} \varphi$, it follows that $(r\urd_r)^k {\bf \Omg}^I \varphi = \mathbb S_{(\geq \ell)} (r\urd_r)^k {\bf \Omg}^I \varphi$ (since $\mathring{\slashed\Delta} {\bf \Omg}^I Y_{(\ell)} = -\ell(d+\ell-2) Y_{(\ell)}$ for any spherical harmonic $Y_{(\ell)}$). We can therefore apply the estimate \eqref{eq:elliptic.up.to.2} to obtain
\begin{equation}\label{eq:elliptic.commuted.large.r}
\nrm{(\brk{r} \urd)^{(\leq 2)} (\brk{r}\urd_r)^k {\bf \Omg}^I \bbS_{(\geq \ell)} \varphi}_{L^{2}(\brk{r}^{2 \gmm} \, \ud x)}
		\aleq_{d, \ell} \nrm{r^{2} (\brk{r}\urd)^{(\leq k)} {\bf \Omg}^I \bbS_{(\geq \ell)} h}_{L^{2}(\brk{r}^{2 \gmm} \, \ud x)},
\end{equation}
which implies the desired estimate except for the weights for $r\geq 1$.

To obtain the sharp estimates when $r \leq 1$, we commute with $\urd_i$, noting that
$$[\urd_i, \Delta_{\bfe}] = 0.$$
Hence, upon $s$ commutations of Cartesian derivatives and using \eqref{eq:elliptic.up.to.2}, we obtain
\begin{equation}\label{eq:elliptic.commuted.small.r}
\nrm{(\brk{r} \urd)^{(\leq 2)} \urd^{(s)} \bbS_{(\geq \ell)} \varphi}_{L^{2}(\brk{r}^{2 \gmm} \, \ud x)}
		\aleq_{d, \ell} \nrm{r^{2} \urd^{(s)} \bbS_{(\geq \ell)} h}_{L^{2}(\brk{r}^{2 \gmm} \, \ud x)},
\end{equation}

Using \eqref{eq:elliptic.commuted.large.r} for $r\geq 1$ and \eqref{eq:elliptic.commuted.small.r} for $r\leq 1$, we obtain the final estimate \eqref{eq:stat-est-lap}.
 \qedhere
\end{proof}

Next, we form
\begin{equation}\label{eq:P0far.def}
	\calP^{\far}_{0} = \Delta_{\bfe} + \chi_{>R_{0}}(r) \left(\left( (\bfg^{-1})^{j k} - (\bfm^{-1})^{j k} \right) \nb_{j} \rd_{k} + \bfB^{j} \rd_{j} + V \right),
\end{equation}
in the coordinate system $(\tau, x)$, which has the property that $\calP^{\far}_{0} = {}^{(\tau)} \calP_{0}$ in $\set{r > R_{0}}$ and is a small perturbation of $\Delta_{\bfe}$ thanks to \ref{hyp:med}--\ref{hyp:wave-V}; see Lemma~\ref{lem:P-far} below. The following lemma captures quantitatively the fact that $\calP^{\far}_{0}$ is close to $\Delta_{\bfe}$:
\begin{lemma} \label{lem:P-far}
For any $s\in \mathbb N\cup \{0\}$, $s\leq M_c$ and any $\gmm\in \mathbb R$, we have
\begin{align}
\nrm{(\calP^{\far}_{0} - \Delta_{\bfe}) \varphi}_{\calH^{s, \gmm+2}}
\aleq R_{0}^{-\dlt_{c}} \nrm{\varphi}_{\calH^{s, \gmm}}. \label{eq:P-far.1}
\end{align}
Moreover,
\begin{align}
\nrm{(\calP^{\far}_{0} - \Delta_{\bfe}) \varphi}_{\ell^{1} \calH^{s, -\frac{d}{2}+2}}
&\aleq R_{0}^{-\dlt_{c}} \nrm{\varphi}_{\ell^{\infty} \calH^{s, -\frac{d}{2}}}, \label{eq:P-far.2} \\
\nrm{(\calP^{\far}_{0} - \Delta_{\bfe}) \varphi}_{\ell^{1} \calH^{s, \frac{d}{2}}}
&\aleq R_{0}^{-\dlt_{c}} \nrm{\varphi}_{\ell^{\infty} \calH^{s, \frac{d}{2}-2}}. \label{eq:P-far.3}
\end{align}
\end{lemma}

\begin{proof}
By the assumptions \ref{hyp:med}--\ref{hyp:wave-V}, we have the pointwise estimate
\begin{equation}\label{eq:P-far.pointwise}
|(\brk{r}\urd)^k (\calP^{\far}_{0} - \Delta_{\bfe}) \varphi| \ls \brk{r}^{-\de_c} |(\brk{r}\urd)^{(\leq k +2)} \varphi|.
\end{equation}
Since by definition, $\calP^{\far}_{0} - \Delta_{\bfe} = 0$ when $r\leq R_0$, the bound \eqref{eq:P-far.1} is immediate from \eqref{eq:P-far.pointwise}.
Furthermore, since $\de_c > 0$, the $\brk{r}^{-\de_c}$ allows us to sum over $r$ dyadically so that \eqref{eq:P-far.2} and \eqref{eq:P-far.3} also follow from \eqref{eq:P-far.pointwise}. \qedhere
\end{proof}

We are now ready to prove Propositions~\ref{prop:stat-est} and \ref{prop:stat-est.2}.
\begin{proof}[Proof of Propositions~\ref{prop:stat-est} and \ref{prop:stat-est.2}]
To simplify the notation, we shall omit ${}^{(\tau)}$ and simply write $\calP_{0} = {}^{(\tau)} \calP_{0}$, $\calR_{0} = {}^{(\tau)} \calR_{0}$ etc.

\pfstep{Step~1: Invertibility of $\mathcal P_0^{\mathrm{far}}$} We begin by introducing a large parameter $R_{0}$ such that $R_{0} \gg R_{\far}$ and defining $\calP_{0}^{\far}$ as in \eqref{eq:P0far.def}. If $R_0$ sufficiently large, $\mathcal P_0^{\mathrm{far}}$ is a perturbation of $\Delta_{\bfe}$, we expect that it would inherit all the invertibility properties of $\Delta_{\bfe}$ as proven in Lemma~\ref{lem:stat-est-lap}. In particular, we can obtain estimates with all the desired weights for $\calP_{0}^{\far}$.

The aim of Step~1 is to make the preceding paragraph precise. For the remainder of Step~1, \textbf{take $\gamma$ as in the range in Lemma~\ref{lem:stat-est-lap}}. Suppose $\mathcal P_0^{\mathrm{far}} \widetilde \varphi = \widetilde h$. Writing $\mathcal P_0^{\mathrm{far}} = \Delta_{\bfe} + (\mathcal P_0^{\mathrm{far}} - \Delta_{\bfe})$, we have, by Lemma~\ref{lem:stat-est-lap} and Lemma~\ref{lem:P-far}, that
\begin{equation}
\begin{split}
&\:  \|(\brk{r}\urd)^{(\leq s+ 2)} \widetilde \varphi \|_{\ell^\infty_r L^2(\brk{r}^{2\gmm}\, \ud x)} \\
\ls &\: \|\brk{r}^2 (\brk{r}\urd)^{(\leq s)}\widetilde  h\|_{\ell^1_r L^2(\brk{r}^{2\gmm}\, \ud x)} + \| \brk{r}^2 (\brk{r}\urd)^{(\leq s)}(\mathcal P_0^{\mathrm{far}} - \Delta_{\bfe}) \widetilde \varphi\|_{\ell^1_r L^2(\brk{r}^{2\gmm} \, \ud x)} \\
\ls &\: \| \brk{r}^2 (\brk{r}\urd)^{(\leq s)} \widetilde h\|_{\ell^1_r L^2(\brk{r}^{2\gmm}\, \ud x)} + R_0^{-\de_c} \|(\brk{r}\urd)^{(\leq s+2)} \widetilde \varphi\|_{\ell^\infty_r L^2(\brk{r}^{2\gmm}\, \ud x)}.
\end{split}
\end{equation}

For $R_0$ sufficiently large, we can absorb the last term to the left-hand side to obtain
\begin{equation}\label{eq:est.P0far-1}
\|(\brk{r}\urd)^{(\leq s+ 2)}  \widetilde \varphi \|_{\ell^\infty_r L^2(\brk{r}^{2\gmm}\, \ud x)} \ls \| \brk{r}^2 (\brk{r}\urd)^{(\leq s)} \widetilde h\|_{\ell^1_r L^2(\brk{r}^{2\gmm}\, \ud x)} = \| \brk{r}^2 (r\urd)^{(\leq s)}(\mathcal P_0^{\mathrm{\far}} \widetilde \varphi) \|_{\ell^1_r L^2(\brk{r}^{2\gmm}\, \ud x)} ,
\end{equation}
which in particular implies the invertibility of $\mathcal P_0^{\mathrm{\far}}$ in the spaces above.

\pfstep{Step~2: Construction of a right-inverse $\td{\calR}_{0}$ with improved weights for the input $h$}
Consider the equation $\mathcal P_0 \varphi = h$. By \ref{hyp:ult-stat} or \ref{hyp:ult-stat'}, we already have a right-inverse $\calR_{0}$ for $\calP_{0}$ which is well-defined when $h \in H^{s_{c}}_{comp}$. The goal of this step is to construct a right-inverse $\td{\calR}_{0}$ for $\calP_{0}$ for $h$ in suitable weighted Sobolev spaces. Here, we will crucially use the fact that $\calP_{0} = \calP_{0}^{\far}$ in $\set{r > R_{0}}$, for which estimates with the desired weights have been proved in Step~1. (We note that, pedantically speaking, at this stage $\td{\calR}_{0}$ and $\calR_{0}$ may be completely different on the common domain; however, in Step~4, we shall show that $\td{\calR}_{0}$ is an extension of $\calR_{0}$.)

Before proceeding, let us define a version of $(\mathcal P_0^\far)^{-1}$ adapted to our geometric setting. Recall that by \ref{hyp:topology}, our domain is $\mathbb R^d\setminus \mathcal K$, where either $\mathcal K = \emptyset$ or $\mathcal K \subset B_{R_{\mathrm{far}}}$ is compact with a $C^\infty$ boundary. In the case where $\mathcal K \neq \emptyset$, let $\mathfrak R$ denotes the restriction operator, i.e., $\mathfrak R h = h_{|\mathbb R^d\setminus \calK}$. Fix (arbitrarily) a continuous bounded linear extension operator $\mathfrak E: H^s(\mathbb R^d\setminus \mathcal K) \to H^s(\mathbb R^d)$ such that $\mathfrak R\circ\mathfrak E = I$. Define $(\bar{\calP}^{\far}_{0})^{-1} = \mathfrak R \circ (\calP^{\far}_{0})^{-1}\circ \mathfrak E$. In the case where $\mathcal K = \emptyset$, we simply denote $(\bar{\calP}^{\far}_{0})^{-1} = (\calP^{\far}_{0})^{-1}$.

Let $h \in \ell^{1} \calH^{s_{c}, -\frac{d}{2}+2}$. Our goal is to find $\varphi$ that is obtained by applying linear operators on $h$ and furthermore solves $\calP_{0} \varphi = h$. To do this, we write $\varphi = (\bar\calP^{\far}_{0})^{-1} h + (\varphi - (\bar\calP^{\far}_{0})^{-1} h)$, where we note that $(\bar\calP^{\far}_{0})^{-1} h$ is well-defined thanks to Step~1. Observe that
\begin{equation*}
	\calP_{0}(\varphi - (\bar\calP^{\far}_{0})^{-1} h) = h - \calP_{0} (\bar\calP^{\far}_{0})^{-1} h =: h^{\near}
\end{equation*}
and also that for any $\gmm$ as in the proposition,
\begin{equation}\label{eq:hnear.est}
	\supp h^{\near} \subseteq \set{r \leq R_{0}}, \quad
	\nrm{h^{\near}}_{H^{s}}
	\aleq \nrm{\urd^{(\leq 2)}(\bar{\calP}_{0}^{\far})^{-1} h}_{H^{s}(B_{R_{0}})} \aleq \nrm{h}_{\ell^{1}_{r} \calH^{s, \gmm+2}}.
\end{equation}
In particular, $\calR_{0}$ may now be applied to $h^{\near}$, and we are led to the definition
\begin{equation*}
	\td{\calR}_{0} : \ell^{1} \calH^{s_{c}, -\frac{d}{2}+2} \to L^{2}_{loc}, \quad
	\td{\calR}_{0} h := (\bar\calP^{\far}_{0})^{-1} h + \calR_{0} (h - \calP_{0} (\bar\calP^{\far}_{0})^{-1} h).
\end{equation*}
For each constituents of $\td{\calR}_{0}$, we have the following bounds. For any $s \in \bbZ_{\geq 0}$ and $\gmm$ as in the proposition, by Step~1,
\begin{equation} \label{eq:stationary.h.improved.0}
	\nrm{(\bar\calP^{\far}_{0})^{-1} h}_{\ell^{\infty} \calH^{s+2, \gmm}} \aleq \nrm{h}_{\ell^{1} \calH^{s, \gmm+2}},
\end{equation}
and for any $R' > R_{0}$, $0 \leq s \leq M_{c}-1$ and $\gmm$ as in the proposition, by \eqref{eq:zero-res-est} or \eqref{eq:zero-res-est'} we have
\begin{equation}\label{eq:stationary.h.improved}
	\nrm{\calR_{0} (h - \calP_{0} (\bar\calP^{\far}_{0})^{-1} h)}_{H^{s+2}(B_{<10 R'})}
	\aleq_{R'} \nrm{h}_{\ell^{1}_{r} \calH^{s+s_{c}, \gmm}}.
\end{equation}

\pfstep{Step~3: Estimates for $\td{\calR}_{0}$ and $\calR_{0}$ with improved weights for the output} Our next aim is to improve the weights in the spaces in the range of $\td{\calR}_{0}$ and $\calR_{0}$. We will treat both cases simultaneously. For $\td{\calR}_{0}$, let $\gmm$ be as in the statement of the proposition, and $h \in \ell^{1} \calH^{\gmm+2}$ so that $\varphi = \td{\calR}_{0} h$ is well-defined. For $\calR_{0}$, we simply take $h \in H^{s+s_{c}}_{comp}$ and let $\varphi = \calR_{0} h$. In both cases, we have $\calP_{0} \varphi = h$. The main idea for this step is again to exploit $\calP_{0} = \calP_{0}^{\far}$ in $\set{r > R_{0}}$, but this time we decompose $\varphi$.

For $R' > R_0$ to be chosen later, define the decomposition
\begin{equation}\label{eq:stationary.varphi.decomp}
\varphi = \chi_{>R'} \varphi + (1-\chi_{>R'}) \varphi.
\end{equation}
Notice that $\calP_{0} = \bar{\calP}^{\far}_{0}$ on the support of $\chi_{>R'}$. Therefore,
$$\bar{\calP}^{\far}_{0} (\chi_{>R'} \varphi) = \calP_{0} (\chi_{>R'} \varphi) = \chi_{>R'} h + [ \calP_{0}, \chi_{>R'}] \varphi.$$
Since $\mathrm{supp}(\chi_{>R'} \varphi)\subset \{r \geq R'\}$, we can extend $\chi_{>R'} \varphi$ trivially and view it as a function on $\mathbb R^d$ so that we can use the estimate \eqref{eq:est.P0far-1} (with $\widetilde\varphi = \chi_{>R'} \varphi$) to deduce that
\begin{equation}\label{eq:elliptic.largeR}
\begin{aligned}
\nrm{\chi_{>R'} \varphi}_{\ell^{\infty}_{r} \calH^{s+2, \gmm}}
&\aleq \nrm{\chi_{>R'} h}_{\ell^{1} \calH^{s, \gmm+2}} + \nrm{[\calP_{0}, \chi_{>R'}] \varphi}_{\ell^{1} \calH^{s, \gmm+2}} \\
&\aleq \nrm{\chi_{>R'} h}_{\ell^{1} \calH^{s, \gmm+2}} + \nrm{\varphi}_{H^{s+1}(B_{< 10 R'})} \\
\end{aligned}
\end{equation}
where in the second inequality, we used the fact that $[\calP_{0}, \chi_{>R'}] \varphi$ is supported in $B_{10 R'}$ and that $[\calP_{0}, \chi_{>R'}]$ is order $1$. On the other hand, since $1-\chi_{>R'}$ is a smooth cutoff to $B_{R'}$, we obviously have
\begin{equation}\label{eq:elliptic.smallR}
\nrm{(1-\chi_{>R'}) \varphi}_{\ell^{\infty}_{r} \calH^{s+2, \gmm}}
\aleq \nrm{\varphi}_{H^{s+2}(B_{< 10 R'})}.
\end{equation}
In both \eqref{eq:elliptic.largeR} and \eqref{eq:elliptic.smallR}, note that $\nrm{\chi_{>R'} h}_{\ell^{1} \calH^{s, \gmm+2}}$ is obviously bounded above by $\nrm{h}_{\ell_{r}^{1} \calH^{s+s_{c}, \gmm+2}}$ for $\td{\calR}_{0}$ and $\nrm{h}_{H^{s+s_{c}}}$ for $\calR_{0}$, whereas $\nrm{\varphi}_{H^{s+2}(B_{< 10 R'})}$ can be bounded using Step~2 for $\td{\calR}_{0}$ and by the $H^{s+2}(\Sgm_{\tau} \cap \set{r \leq R'})$ bound in \eqref{eq:zero-res-est} or \eqref{eq:zero-res-est'} for $\calR_{0}$. In sum, for $0 \leq s \leq M_{c} - 1$,
\begin{equation}\label{eq:elliptic.final.large.r}
\nrm{\varphi}_{\ell^{\infty}_{r} \calH^{s+2, \gmm}} \aleq
\begin{cases}
\nrm{h}_{\calH^{s+s_{c}, \gmm+2}} & \hbox{ when } \varphi = \td{\calR}_{0} h, \\
\nrm{h}_{H^{s+s_{c}}}& \hbox{ when } \varphi = \calR_{0} h,
\end{cases}
\end{equation}
where the implicit constant also depends on $\supp h$ in the case of $\calR_{0}$.

\pfstep{Step~4: Completion of the proof}
We have shown that (a) and (c) in the proposition hold with $\td{\calR}_{0}$. It remains to establish (b) and also that $\td{\calR}_{0}$ is an extension of $\calR_{0}$. For any $\varphi \in C^{\infty}_{c}$, note that $\calP_{0} \varphi \in C^{M_{c}}_{c} \subseteq H^{M_{c}}_{comp}$ and hence, by Step~3 (with $s = M_{c} - 1$), $\td{\calR}_{0} \calP_{0} \varphi$ is well-defined, belong to $\ell^{\infty} \calH^{M_{c}+1-s_{c}, \frac{d}{2}-2}$ and satisfies $\calP_{0} (\varphi - \td{\calR}_{0} \calP_{0} \varphi) = 0$. We note that, by Lemma~\ref{lem:rescaled-sob} (see also \ref{eq:GN.spatial} in the proof of Corollary~\ref{cor:near-pointwise} below), we have $\td{\calR}_{0} \calP_{0} \varphi \in C^{M_{c}-s_{c}-\lfloor \frac{d}{2} \rfloor}$ and
\begin{equation*}
\abs{(\brk{r} \urd)^{(\leq M_{c}-s_{c}-\lfloor \frac{d}{2} \rfloor)} \td{\calR}_{0} \calP_{0} \varphi}
\aleq \brk{r}^{-(d-2)}.
\end{equation*}
Hence, by the uniqueness assertion in \ref{hyp:ult-stat} or \ref{hyp:ult-stat'}, we have
\begin{equation} \label{eq:stat-est-unique}
\td{\calR}_{0} \calP_{0} \varphi = \varphi \quad \hbox{ for all $\varphi \in C^{\infty}_{c}$.}
\end{equation}
By the density of $C^{\infty}_{c}$ in $\ell^{1} \calH^{s_{c}+2, -\frac{d}{2}-\ell}$ (with $\ell =0$ for Proposition~\ref{prop:stat-est}), (b) follows.

It remains to prove that $\td{\calR}_{0} = \calR_{0}$ on $H^{s_{c}}_{comp}$. Given $h \in H^{s_{c}}_{comp}$ with $\supp h \subseteq \set{r < R_{1}}$ for some $R_{1} > 0$, consider an approximating sequence $h_{n} \in C^{\infty}_{c}$ with $\supp h_{n} \subseteq \set{r < R_{1}}$ such that $h_{n} \to h$ in $H^{s_{c}}$. Choosing $\gmm > -\frac{d}{2}$ and using \eqref{eq:elliptic.final.large.r}, observe that $\calR_{0} h_{n} \in \ell^{\infty} \calH^{s_{c}, \gmm} \subseteq \ell^{1} \calH^{s_{c}, -\frac{d}{2}}$. Hence, $\td{\calR}_{0} h_{n} = \td{\calR}_{0} \calP_{0} \calR_{0} h_{n} = \calR_{0} h_{n}$. By taking the limit, it follows that $\td{\calR}_{0} h = \calR_{0} h$, i.e., $\td{\calR}_{0} = \calR_{0}$ on $H^{s_{c}}_{comp}$.  \qedhere
\end{proof}

To conclude this subsection, we derive pointwise estimates on the solution $\varphi$ to ${}^{(\tau)} \calP_{0} \varphi = h$ for later.

\begin{corollary}\label{cor:near-pointwise}
Let $\varphi \in H^{s_{c}+2}_{comp}$. Then for $-\f d2\leq \gamma \leq \f d2-2$,
\begin{equation*}
	\nrm{\brk{r}^{\frac{d+2\gmm}{2}} (\brk{r} \urd)^{(\leq s)} \varphi}_{L^{\infty}}
	\aleq \nrm{{}^{(\tau)} \calP_{0} \varphi}_{\ell^{1}_{r} \calH^{s+s_{c}+\lfloor \frac{d}{2} \rfloor - 1, \gmm+2}}.
\end{equation*}
Suppose that $\mathcal P_0$ and $\mathcal R_{0}$ are spherically symmetric as in Proposition~\ref{prop:stat-est.2}. Then for $-\f d2 - \ell \leq \gmm \leq \f d2+\ell-1$,
\begin{equation*}
	\nrm{\brk{r}^{\frac{d+2\gmm}{2}} (\brk{r} \urd)^{(\leq s)} \bbS_{(\geq \ell)} \varphi}_{L^{\infty}}
	\aleq \nrm{{}^{(\tau)} \calP_{0} \bbS_{(\geq \ell)} \varphi}_{\ell^{1}_{r} \calH^{s+s_{c}+\lfloor \frac{d}{2} \rfloor - 1, \gmm+2}}.
\end{equation*}
\end{corollary}
\begin{proof}
The desired estimate follows directly from Proposition~\ref{prop:stat-est} and the following Gagliardo--Nirenberg estimate:
\begin{equation}\label{eq:GN.spatial}
\| \brk{r}^{\f{d+2\gamma}2} h \|_{L^\infty} \ls \| (\brk{r} \urd)^{(\leq \lfloor \f d2\rfloor + 1)} h \|_{\ell^\infty_r L^2(\brk{r}^{2\gamma} \, \ud x)},
\end{equation}
which in turn follows from Lemma~\ref{lem:rescaled-sob}.

The spherically symmetric improvement can be proved similarly, except for also using Proposition~\ref{prop:stat-est.2}. \qedhere
\end{proof}

\subsection{Proof of the iteration lemmas}

\begin{proof} [Proof of Proposition~\ref{prop:near-med}]

We will derive the desired estimate in an induction argument, where in each step we trade some $r$-decay for some $\tau$-decay. For each fixed $\tau$, take $\tau_0 \leq \tau$. (In Step~4 of the proof, we will fix $\tau_0 = \tau$. However, it is useful to carry out Steps~1--3 with a more general parameter $\tau_0$, as the more general estimates established will be used later in the proof of Proposition~\ref{prop:sharp.near}.) Notice this in particular implies that $\mathrm{supp}(\chi_{<\tau_0}) \subset \calM_{\med}$ so that we may use the estimates in Corollary~\ref{cor:near-pointwise}.

\pfstep{Step~1: Setting up the induction argument} We proceed by induction (on $j$) to prove in $\calM_{\near} \cup \calM_{\med}$ that
\begin{equation}\label{eq:med.induction.hyp}
\sum_{0\leq s \leq M - (2i+5j+5) (\lfloor \f d2 \rfloor + s_c)} |(\brk{r} \urd)^s (\tau\bfT)^i \phi| \ls A' \max\{\tau^{-\alp-\dlt_{0}} \log^K \tau, \tau^{-\alp_{d}}\} \tau^{\nu_\Box-\f j2} \brk{r}^{-\nu_\Box+\f j2}.
\end{equation}
for all $j \in \{0,1, \ldots, 2\nu_\Box\}$.  The $j = 2\nu_\Box$ case then gives the desired estimate stated in the proposition. (Note that the restriction $M' \leq \f{M}{2(\lfloor \f d2 \rfloor + s_c)} - \f{5d}2$ in the statement of the proposition comes from the worst case scenario where we take all derivatives to be $\bfT$ derivatives.)

%

To establish \eqref{eq:med.induction.hyp}, first note that the base case, i.e., the $j=0$ case, holds by assumption: this is true in $\calM_{\med}$ by \eqref{eq:near-med-ass-med} and $\de_0>0$, and this is true in $\calM_{\near}$ by \eqref{eq:near-med-ass-near} by $\de_0\leq 1$.

\textbf{For the remainder of the proof, we assume that \eqref{eq:med.induction.hyp} holds for some $j \in \{0,1, \ldots, 2\nu_\Box-1\}$}, and show that \eqref{eq:med.induction.hyp} holds for $j$ replaced by $j+1$.

\pfstep{Step~2: The equation for $\calP_{0} \bfT^i \phi$}
Starting with the equation $\calP\phi = f + \calN(\phi)$, commuting with $\bfT^i$, isolating $\calP_{0}$ from $\calP$, and inserting the cutoff $\chi_{<\tau_0} (r)$, we have
\begin{equation}\label{eq:P0.cutoff}
\begin{split}
\calP_{0} (\chi_{<\tau_0} (r) \bfT^{i} \phi) &= - [\chi_{<\tau_0}(r), \calP_{0}] \bfT^{i} \phi \\
&\peq  + \chi_{<\tau_0}(r) (\calP_{0} - \calP) (\bfT^{i} \phi) \\
&\peq + \chi_{<\tau_0}(r) \bfT^{i} f \\
&\peq + \chi_{<\tau_0}(r) \bfT^{i} \calN(\phi) \\
&\peq + \chi_{<\tau_0}(r) [\calP_{0}, \bfT^i] \phi \\
&=: e_{1} + \ldots + e_{5}.
\end{split}
\end{equation}

\pfstep{Step~3: The main estimates} We now control each term in \eqref{eq:P0.cutoff} under our induction hypothesis.

\textbf{The term $e_1$.} Since each derivative of $\chi_{<\tau_0}(r)$ gives $\tau_0^{-1}$, we have
\begin{equation}\label{eq:P0.chi.commute}
e_1 = - [\chi_{<\tau_0}(r), \calP_{0}] \bfT^{i} \phi = O^{M_c}(\tau_0^{-1}) \rd \bfT^i \phi + O^{M_c}(\tau_0^{-2}) \bfT^i \phi.
\end{equation}
Moreover, $e_1$ is localized to $r \aeq \tau_0$. Therefore, by the assumption on the input in \eqref{eq:near-med-ass-med}, we have
\begin{equation}\label{eq:med.induction.e1}
\begin{split}
|(\brk{r} \urd)^s e_1| \ls &\: A' \tau^{-\alp-\de_0+\nB-i}\tau_0^{-1} \brk{r}^{-\nB-1} \log^{K}\tau \\
 \ls &\: A' \tau^{-\alp-\de_0+\nB-i}\tau_0^{-\f j2-1} \brk{r}^{-\nB+\f j2-1} \log^{K}\tau
\end{split}
\end{equation}
whenever $s+i+1\leq M$.

\textbf{The term $e_2$.} For this term, we use that each term in $\calP_{0} - \calP$ has at least one $\bfT$ derivative, i.e.,
\begin{equation*}
\begin{split}
e_2=&\: \chi_{<\tau_0}(r) (\calP_{0} - \calP) (\bfT^{i} \phi) \\
=&\:  O_{\bfGmm}^{M_{c}}(1) \rd \bfT^{i+1} \phi +  \chi_{<\tau_0}(r) O_{\bfGmm}^{M_{c}}(\brk{r}^{-1})  \bfT^{i+1} \phi \\
=:&\: e_{2,1} + e_{2,2}.
\end{split}
\end{equation*}
We now estimate these terms using the induction hypothesis. Take $s \in \mathbb Z$ such that $0\leq s \leq M - (2i+5(j+1)+5) (\lfloor \f d2 \rfloor + s_c )$. Since $M-(2i+5(j+1)+5) (\lfloor \f d2 \rfloor + s_c ) \leq M- (2(i+2)+5j+5) (\lfloor \f d2 \rfloor + s_c )$, we can use the induction hypothesis to obtain
\begin{equation}\label{eq:med.induction.e2}
\begin{split}
|(\brk{r} \urd)^s e_{2,1}| + |(\brk{r} \urd)^s e_{2,2}| \ls A' \tau^{-i-1} \max\{\tau^{-\alp-\dlt_{0}} \log^K \tau, \tau^{-\alp_{d}}\} \tau^{\nu_\Box-\f j2} \brk{r}^{-\nu_{\Box}+\f j2-1}.
\end{split}
\end{equation}

%

\textbf{The term $e_3$.} By \eqref{eq:near-f} and \eqref{eq:med-f}, whenever $0\leq s +i \leq M_0$,
\begin{equation}\label{eq:med.induction.e3}
|(\brk{r} \urd)^s e_3| = | (\brk{r} \urd)^s(\chi_{<\tau_0}(r) \bfT^{i} f) | \ls D \tau^{-\alp_d-i} \brk{r}^{-2-\de_{d}} \ls A' \tau^{-\alp_d-i}\tau_0^{\nB-\f j2-\f 12}\brk{r}^{-\nB+ \f j 2-\f 32-\de_0},
\end{equation}
where in the last inequality, we simply used $\nB-\f j2-\f 12 \geq 0$ and $\brk{r} \ls \tau_0$ on the support of $\chi_{<\tau_0}$, and that $\de_0\leq \de_d$. 

\textbf{The term $e_4$.} Now we turn to the nonlinearity. Using \eqref{eq:near-med-alp}, \eqref{eq:sol-near}, \eqref{eq:sol-med}, \eqref{eq:near-med-ass-near} and the fact $\alp \geq \alp_0$, we apply Definition~\ref{def:alp-N} with $\alp_{\calN}' = \alp_{\calN} + \f{\de_0}2$ to obtain
\begin{equation*}
\calN(\phi) = O_{\bfGmm}^{M-2}(A' \tau^{-\alp-\f{3\de_0}2} r^{-2}) \quad \hbox{in }\calM_{\med} \cup \calM_{\near}.
\end{equation*}
Hence, differentiating by $\bfT^i$ and $(\brk{r}\urd)^s$, and using $\tau^{-1} \ls \brk{r}^{-1}$ (twice) and that $j \leq 2\nB -1$, we obtain
\begin{equation}\label{eq:med.induction.e4}
|(\brk{r} \urd)^s e_4| =  |(\brk{r} \urd)^s \bfT^{i} \calN(\phi)| \ls A' \tau^{-\alp- \f{3\de_0}2-i} \brk{r}^{-2} \ls A' \tau^{-\alp-\f{3\de_0}2-i+\nB-\f j2 -\f 12} \brk{r}^{-\nB+ \f j 2-\f 32}
\end{equation}
whenever $s+i\leq M-2$.



\textbf{The term $e_5$.} For $e_{5}$, we introduce another induction in $i$ within the induction-in-$j$ argument. (Thus, $j$ should still be considered fixed.) First, for the base case $i=0$, we have $e_5 = 0$, which obviously satisfies the desired bound. Fix some $i\geq 1$ and assume that $\bfT^{i'}\phi$ obeys the desired bound \eqref{eq:med.induction.hyp} for $j$ replaced by $j+1$ and for all $0\leq i' \leq i-1$:
\begin{equation}\label{eq:med.induction.hyp.2}
\sum_{0\leq s \leq M - (2i'+5j+10) (\lfloor \f d2 \rfloor + s_c)} |(\brk{r} \urd)^s (\tau\bfT)^{i'} \phi| \ls A' \max\{\tau^{-\alp-\dlt_{0}} \log^K \tau, \tau^{-\alp_{d}}\} \tau^{\nu_\Box-\f{j+1}2} \brk{r}^{-\nu_\Box+\f{j+1}2}.
\end{equation}

We estimate the commutator $[\calP_0, \bfT^i]$ using \eqref{eq:T-almost-stat-2} so as to obtain
\begin{equation}\label{eq:e5.123.def}
\begin{split}
e_5 = &\: \chi_{<\tau_0}(r) \sum_{i' = 0}^{i-1}O_{\bfGmm}^{M_{c}-(i-i')} (\tau^{-(i-i')-\de_c} \brk{r}^{-2-\dlt_{c}}) (\brk{r} \urd)^{(\leq 2)} \bfT^{i'} \phi  \\
&\: + \chi_{<\tau_0}(r) \sum_{i' = 0}^{i-1}O_{\bfGmm}^{M_{c}-(i-i')} (\tau^{-(i-i')} \brk{r}^{-1-\dlt_{c}}) (\brk{r} \urd)^{(\leq 1)} \bfT^{i'+1} \phi  \\
&\: + \chi_{<\tau_0}(r) \sum_{i' = 0}^{i-1}O_{\bfGmm}^{M_{c}-(i-i')} (\tau^{-(i-i')} \brk{r}^{-\dlt_{c}})  \bfT^{i'+2} \phi =: e_{5,1} + e_{5,2} + e_{5,3}.
\end{split}
\end{equation}

Take $s \in \mathbb Z$ with $0\leq s \leq M - (2i+5j+10) (\lfloor \f d2 \rfloor + s_c)$.

We first handle $e_{5,1}$. Noting that $\lfloor \f d2 \rfloor+s_c \geq 1$, we see that for every $i' \leq i -1$, it holds that $s+2\leq M - (2(i'+1)+5j+10) (\lfloor \f d2 \rfloor + s_c )+2 \leq M - (2i'+5j+10) (\lfloor \f d2 \rfloor + s_c)$ so that \eqref{eq:med.induction.hyp.2} gives
\begin{equation}\label{eq:med.induction.e51}
\begin{split}
|(\brk{r} \urd)^s e_{5,1}| \ls &\: A' \sum_{i' = 0}^{i-1} \tau^{-(i-i')-\de_c} \brk{r}^{-2-\dlt_{c}} \max\{\tau^{-\alp-\dlt_{0}} \log^K \tau, \tau^{-\alp_{d}}\} \tau^{\nu_\Box-i'-\f{j+1}2} \brk{r}^{-\nu_\Box+\f{j+1}2} \\
\ls &\: A' \max\{\tau^{-\alp-\dlt_{0}} \log^K \tau, \tau^{-\alp_{d}}\} \tau^{\nB-i-\f{j+1}2-\de_0} \brk{r}^{-\nu_\Box+\f{j+1}2-2-\de_0},
\end{split}
\end{equation}
where we have used $\de_0 \leq \de_c$.

For $e_{5,2}$ and $e_{5,3}$, we note respectively that if $i' \leq i-1$, then
\begin{align*}
s +1\leq M - (2(i'+1)+5j+10) (\lfloor \f d2 \rfloor + s_c) +1 \leq M - (2(i'+1)+5j+5) (\lfloor \f d2 \rfloor + s_c), \\
s\leq M - (2(i'+1)+5j+10) (\lfloor \f d2 \rfloor + s_c ) \leq M - (2(i'+2)+5j+5) (\lfloor \f d2 \rfloor + s_c ),
\end{align*}
so that we can use \eqref{eq:med.induction.hyp} and $\brk{r} \ls \tau$ to obtain
\begin{equation}\label{eq:med.induction.e523}
\begin{split}
|(\brk{r} \urd)^s e_{5,2}| + |(\brk{r} \urd)^s e_{5,3}| \ls &\: A' \sum_{i' = 0}^{i-1} \tau^{-(i-i')} \brk{r}^{-1-\dlt_{c}} \max\{\tau^{-\alp-\dlt_{0}} \log^K \tau, \tau^{-\alp_{d}}\} \tau^{\nu_\Box-i'-\f j2} \brk{r}^{-\nu_\Box+ \f j2} \\
\ls &\: A' \max\{\tau^{-\alp-\dlt_{0}} \log^K \tau, \tau^{-\alp_{d}}\} \tau^{\nB-i-\f j2-1} \brk{r}^{-\nu_\Box+\f j2-1-\de_0}.
\end{split}
\end{equation}

\pfstep{Step~4: Applying Corollary~\ref{cor:near-pointwise}} We continue to work within the induction-in-$j$ argument introduced in Step~1 and the additional induction-in-$i$ argument introduced for the $e_5$ term in Step~3. At this point, we fix $\tau_0 = \tau$. Combining the estimates \eqref{eq:med.induction.e1}, \eqref{eq:med.induction.e2}, \eqref{eq:med.induction.e3}, \eqref{eq:med.induction.e4}, \eqref{eq:med.induction.e51} and \eqref{eq:med.induction.e523}, we obtain
\begin{equation}\label{eq:med.induction.goal.Li}
\begin{split}
&\:  |(\brk{r} \urd)^s \calP_{0} (\chi_{<\tau_0} (r)\bfT^i \phi)| \\
\ls &\: \max\{A' \max\{\tau^{-\alp-\dlt_{0}} \log^K \tau, \tau^{-\alp_{d}}\} \tau^{\nB-i-\f j2-1} \brk{r}^{-\nB+\f j2-1}, \\
&\: \phantom{\max\{} A' \max\{\tau^{-\alp-\dlt_{0}} \log^K \tau,
\tau^{-\alp_{d}}\} \tau^{\nB-i-\f j2-\f 12} \brk{r}^{-\nB+\f j2-\f 32-\de_0}, \\
&\: \phantom{\max\{} A' \tau^{-(\alp+\f{3\de_0}2)+\nB-i-\f j2-\f 12} \brk{r}^{-\nB+\f j2-\f 32}\}
\end{split}
\end{equation}
for $0 \leq s \leq M - (2i+5j+9) (\lfloor \f d2 \rfloor + s_c)$.

Notice now that
\begin{equation}\label{eq:near.how.to.integrate}
\begin{split}
&\: \| \chi_{<\tau_0} \brk{r}^{-\nB+\f j2-1} \|_{\ell^{1}_{r} \calH^{0,-\f j2+1}} \ls \sum_{k: 2^k \leq \tau_0} \Big(\int_{2^k}^{2^{k+1}} (r^{-\nB+\f j2-1})^2 r^{(d-1)-2j+2} \, \ud r \Big)^{1/2} \\
= &\: \sum_{k: 2^k \leq \tau_0} \Big( \int_{2^k}^{2^{k+1}} \, \ud r \Big)^{1/2} \ls \tau_0^{\f 12} \ls \tau^{\f 12}.
\end{split}
\end{equation}
Similarly, we have
\begin{equation}\label{eq:near.how.to.integrate.2}
\| \chi_{<\tau_0} \brk{r}^{-\nB+\f j2-\f 32-\de_0} \|_{\ell^{1}_{r} \calH^{0,-\f j2+1}} \ls 1,\quad \| \chi_{<\tau_0} \brk{r}^{-\nB+\f j2-\f 32} \|_{\ell^{1}_{r} \calH^{0,-\f j2+1}}\ls \log \tau .
\end{equation}

Therefore, integrating the pointwise estimate in \eqref{eq:med.induction.goal.Li} implies that for $\gamma = -\f j2-1$, we have
\begin{equation}\label{eq:med.induction.goal.L2}
 \nrm{\calP_{0} (\chi_{<\tau_0} (r)\bfT^i \phi) }_{\ell^{1}_{r} \calH^{s+\lfloor \f d2 \rfloor + s_{c}, \gmm+2}} \ls A' \max\{\tau^{-\alp-\dlt_{0}} \log^K \tau,
\tau^{-\alp_{d}}\}\tau^{\nB-i-\f{j+1}2},
\end{equation}
whenever $0\leq s \leq M - (2i+5j+10) (\lfloor \f d2 \rfloor + s_c )$. (Notice that to handle the last term in \eqref{eq:med.induction.goal.Li}, where the $r$-integation gives a logarithmic divergence, we have used $\tau^{-\f{\de_0}2} \log \tau$.)

By Corollary~\ref{cor:near-pointwise} with $\gamma = -\f j2-1$, we obtain
\begin{equation*}
\sum_{0\leq s \leq M - (2i'+5j+10) (\lfloor \f d2 \rfloor + s_c )} |(\brk{r} \urd)^s (\tau\bfT)^{i'} \phi| \ls A' \max\{\tau^{-\alp-\dlt_{0}} \log^K \tau,
\tau^{-\alp_{d}}\} \tau^{\nu_\Box-\f{j+1}2} \brk{r}^{-\nu_\Box+\f{j+1}2}.
\end{equation*}
(Note that $\gamma = -\f j2-1 \in [-\f d2, \f d2-2]$ for the range of $j$ under consideration so that Corollary~\ref{cor:near-pointwise} is applicable.) We have thus concluded the induction step, first for the induction-in-$i$ argument and then for the induction-in-$j$ argument. This concludes the proof. \qedhere
\end{proof}

\begin{proof} [Proof of Proposition~\ref{prop:near-med-sphsymm}]
We perform a similar induction as in \eqref{eq:med.induction.hyp}, except that we prove the estimate for $j \in \{0,1,\ldots, 2\nu_\Box,\ldots, 2\nu_\Box + 2\ell \}$. This is possible is precisely because we have a larger allowable range of $\gamma$ in Corollary~\ref{cor:near-pointwise} (which is in turn due to the fact that we have access to a longer hierarchy of stationary estimates). \qedhere

\end{proof}

\section{Analysis in the intermediate zone: analysis based on Minkowskian computations} \label{sec:med}

In this section, we perform the analysis in the intermediate zone. Our main iteration lemma takes as input the expansion in the wave zone to improve the decay estimate in the intermediate zone. The main estimates of this section can be found in \textbf{Section~\ref{sec:iteration.intermediate}}.

In  \textbf{Section~\ref{sec:Huygens}}, we prove some estimates for the inhomogeneous wave equation \emph{on Minkowski}. In \textbf{Section~\ref{subsec:huygens-corr}}, we show how the inhomogeneous wave equation on Minkowski relates to the explicit computations in Section~\ref{sec:Minkowski.wave}. In \textbf{Section~\ref{sec:med.error}}, we estimate some error terms in preparation of the proof of the main iteration lemma. Finally, in \textbf{Section~\ref{sec:iteration.intermediate.proof}}, we combine all the above tools to prove the main iteration lemma.

\subsection{Statement of the main iteration lemma}\label{sec:iteration.intermediate}

The following is the main iteration lemma in the intermediate zone.
\begin{proposition}\label{prop:med}
Let $0< \de_{0} \leq \min\{\de_c,\de_d,\f 12\}$ and $\alp \in \mathbb R$. In the case $\calN \neq 0$, assume that
\begin{equation} \label{eq:alpN-med}
\alp_{0} \geq \alp_{\calN} + 2 \dlt_{0},
\end{equation}
where $\alp_{0}$ and $\alp_{\calN}$ are from \ref{hyp:sol} and Definition~\ref{def:alp-N}, respectively.
Then there exists a function $A' = A'_{\med}(D,A)$ satisfying \eqref{eq:A'} such that the following holds.

Assume that
\begin{equation}\label{eq:med-ass.near.med}
		\phi = O_{\bfGmm}^{M}(A u^{-\alp}) \quad \hbox{ in } \calM_{\near} \cup \calM_{\med},\quad \phi = O_{\bfGmm}^M(A u^{-\alp +\nB} r^{-\nB})\quad \hbox{ in } \calM_{\wave},
\end{equation}
where $\max\{\f{d+7}2,d\} \leq M \leq \min\set{M_{0}, M_{c}}$ and $A > 0$.

 {\bf Case~1: Decay in the case of a one-term expansion.} Assume that
\begin{equation}\label{eq:med-ass.case.1.expansion}
\Phi = \rPhi_{0,0} + \rho_{1} \quad \hbox{ in } \calM_{\wave},
\end{equation}
where
\begin{equation}\label{eq:med-ass.case.1}
	\Phi_{0,0} = O_{\bfGmm}^{M}(A u^{-\alp+\nB}),\quad \rho_1 = O_{\bfGmm}^M(A u^{-\alp-\de_0+\nB}).
\end{equation}

Then
$$\phi = O^{M-\f{d+7}2}_{\bfGmm}(A' u^{-\alp-\de_0+\nB} r^{-\nB}) \hbox{ in } \calM_{\med}.$$

 {\bf Case~2: $\alp > \nB$ and there is improved decay in the expansion.} Assume that for an integer $1 \leq J \leq \min \set{J_{c}, J_{d}}$ such that
\begin{equation*}
	J - 1 + \nu_{\Box} < \alp < J + \nu_{\Box}
\end{equation*}
$\Phi$ admits the expansion
\begin{equation}\label{eq:med-ass.wave.0}
\Phi = \sum_{j=0}^{J -1} \sum_{k = 0}^{K_{j}} r^{-j} \log^{k} \left(\tfrac{r}{u}\right) \rPhi_{j,k} + \rho_{J} \quad \hbox{ in } \calM_{\wave},
\end{equation}
where
\begin{align}
	\rPhi_{j, k} &= O_{\bfGmm}^{M} (A u^{j- \alp - \dlt_{0} + \nu_{\Box}})& &\hbox{ for } 0 \leq j \leq J - 1,  \, 1 \leq k \leq K_{j} \label{eq:med-ass.wave.1}\\
	\rPhi_{(\leq j - \nu_{\Box}) j, 0} &= O_{\bfGmm}^{M} (A u^{j- \alp - \dlt_{0} + \nu_{\Box}})& &\hbox{ for } 0 \leq j \leq J - 1, \label{eq:med-ass.wave.2}\\
	\rPhi_{(\geq j - \nu_{\Box}+1) j, 0} &= O_{\bfGmm}^{M} (A u^{j- \alp + \nu_{\Box}})& &\hbox{ for } 0 \leq j \leq J - 1, \label{eq:med-ass.wave.4}
\end{align}
and the remainder obeys
\begin{equation}\label{eq:med-ass.wave.remainder}
\rho_{J} = O_{\bfGmm}^{M}(A r^{-\alp -\de_0 + \nu_{\Box}})
\quad \hbox{ in } \calM_{\wave}.
\end{equation}

Then
\begin{equation*}
	\begin{split}
	\phi = &\: O^{M-\max\{\f{d+7}2,d\}}_{\bfGmm}(A' u^{-\alp-\de_0+\nB} r^{-\nB}) \hbox{ in } \calM_{\med}.
	\end{split}
\end{equation*}

 {\bf Case~3: The decay is limited by expansion of the coefficient or data.} Assume that $J= \min \{J_{c},J_d\}$ and $\alp+\de_0 > \min \{J_{c}-1+\eta_c, J_d-1+\eta_d \} + \nB$. Suppose that $\Phi$ admits the expansion \eqref{eq:med-ass.wave.0} with $\Phi_{j,k}$ satisfying \eqref{eq:med-ass.wave.1}--\eqref{eq:med-ass.wave.4}. Let $\alp_{\mathfrak f} = \min\{J,J_{c}-1+\eta_{c},J_d-1+\eta_{d}\} +\nB$, and instead of \eqref{eq:med-ass.wave.remainder}, assume that the following holds for the remainder for some $0< \de_{\mathfrak f}' < \min\{\f 12, \eta_{c}, \eta_{d}\}-\de_0$:
 \begin{equation}\label{eq:med-ass.wave.remainder.case.3}
\rho_{J} = O_{\bfGmm}^{M}(A r^{-\alp_{\mathfrak f}+\de_{\mathfrak f}'+\nB} u^{-\de_{\mathfrak f}'} \max\{u^{\alp_{\mathfrak f}-J-\nB}\log^{K_J} u, 1 \})
\quad \hbox{ in } \calM_{\wave}.
\end{equation}

Then
\begin{equation}
	\begin{split}
	\phi = &\: O_{\bfGmm}^{M- \f{d+7}2}(A' u^{-\alp_{\mathfrak f}+\nB}r^{-\nB} \max\{u^{\alp_{\mathfrak f}-J-\nB} \log^{K_J}u, 1 \}) \hbox{ in } \calM_{\med}.
	\end{split}
\end{equation}

 {\bf Case~4: The decay is limited by the expansion of $\Phi$.} Suppose $J \leq \min\set{J_{c}, J_{d}}-1$ and $\alp < J+\nu_{\Box} < \alp + \dlt_{0}$. Assume that
 $\Phi$ admits the expansion
\begin{equation}\label{eq:med-ass.wave.0.case.4}
\Phi = \sum_{j=0}^{J } \sum_{k = 0}^{K_{J}} r^{-j} \log^{k} \left(\tfrac{r}{u}\right) \rPhi_{j,k} + \rho_{J+1} \quad \hbox{ in } \calM_{\wave},
\end{equation}
where for $0\leq j \leq J-1$, $\rPhi_{j, k}$ satisfies \eqref{eq:med-ass.wave.1}--\eqref{eq:med-ass.wave.4}, for $j = J$ (recall \eqref{eq:rcPhi-rPhi}),
\begin{align}
	\rcPhi_{J, k} &= O_{\bfGmm}^M(A)
	& & \hbox{ for } 1 \leq k \leq K_{J}, \label{eq:med-ass.wave-main-PhiJk} \\
	\rcPhi_{(\leq J - \nu_{\Box}) J, 0} &= O_{\bfGmm}^M(A) & & \label{eq:med-ass.wave-main-PhiJ0-low}\\
	\rcPhi_{(\geq J - \nu_{\Box}+1) J, 0}
	&= O_{\bfGmm}^{M} (A u^{J- \alp  + \nu_{\Box}}), & & \label{eq:med-ass.wave-main-PhiJ0-high}
\end{align}
and the remainder $\rho_{J+1}$ satisfies
\begin{equation}\label{eq:med-ass.wave-main-rho}
\rho_{J+1}
= O_{\bfGmm}^{M}(A r^{-J-\eta_{\mathfrak f}} u^{\eta_{\mathfrak f}} \log^{K_{J}} u )
+ O_{\bfGmm}^{M}(A r^{-\alp-\dlt_{0}+\nu_{\Box}}).
\end{equation}

Then
\begin{equation}
\begin{split}
	\phi(u,r,\theta) = &\: O_{\bfGmm}^{M-\f{d+7}2} (A' u^{-J} r^{-\nB} \log^{K_J} u ) \hbox{ in } \calM_{\med}.
\end{split}
\end{equation}

\end{proposition}

\subsection{Pointwise estimates for the Minkowskian inhomogeneous wave equation}\label{sec:Huygens}


The tools below are designed to imitate the use of the method of characteristics in the $3+1$-dimensional scalar (radial) problem, which is common to many existing approaches for proving Price's law.

In what follows, we write $\Box_{\bfm}^{-1} f$ to denote the forward solution (i.e., $(\psi, \rd_{t} \psi) = 0$ outside the light cone $\set{u \geq 0}$) to $\Box_{\bfm} \psi = f$.
\begin{lemma} [Strong Huygens principle]\label{lem:huygens}
Let $\supp f \subseteq \set{u \geq 0}$. Fix $U > 1$, and define
\begin{equation*}
	D_{U} = \set{(u, r, \tht) : u + 2r \geq U,\, u \geq 0}.
\end{equation*}
Then
\begin{equation*}
	\left. \Box_{\bfm}^{-1} f \right|_{\calC_{U}} = \left. \Box_{\bfm}^{-1}(\chf_{D_{U}} f) \right|_{\calC_{U}}.
\end{equation*}
Moreover, fix $1 \leq R \aleq U$. Let $C_U^R$ be as in \eqref{eq:CUR.def.small.R} and define
\begin{equation*}
	D_{U,R} = \Big\{ (u,r,\theta): 0\leq u \leq \f{3U}2,\, \f{U}2 \leq u+2r \leq \f 32(U+2R) \Big\}.
\end{equation*}
Then
\begin{equation*}
	\rst{\Box_\bfm^{-1} f}_{C_U^R} = \rst{\Box_\bfm^{-1} (\chf_{D_{U,R}} f)}_{C_U^R}.
\end{equation*}
\end{lemma}

Next, we state an integrated local energy decay estimates; see for instance \cite[Proposition~11]{MetTat}. We state the version that we use, where we noted that we can put in the localization of the error terms in $D_{U,R}$ using Lemma~\ref{lem:huygens}.
\begin{lemma} [Integrated local energy decay]\label{lem:iled-flat}
Let $f$ be compactly supported. Fix $1 \leq R \aleq U$. Then
\begin{equation*}
	\nrm{\Box_\bfm^{-1} f}_{LE(C_U^R)} \aleq \nrm{f}_{LE^{\ast}(D_{U,R})},
\end{equation*}
where, for any set $X$,
\begin{align}
	 \nrm{\varphi}_{LE(X)} = &\:  \nrm{\chf_{X} (|\rd\varphi| + |\varphi|)}_{L^{2} L^{2}(\bbR \times B_{1})} \notag \\
	&\:+ \sup_{\widetilde{R}\in 2^{\bbZ_{\geq 0}}} \Big(\widetilde{R}^{-\frac{1}{2}} \nrm{ \chf_{X}\rd \varphi}_{L^{2}L^{2}(\mathbb R \times B_{2\widetilde{R}} \setminus B_{\widetilde{R}})}
	+ \widetilde{R}^{-\frac{3}{2}} \nrm{\chf_{X} \varphi}_{L^{2}L^{2}(\mathbb R\times B_{2\widetilde{R}} \setminus B_{\widetilde{R}})} \Big), \label{eq:LE.def}\\
	\nrm{f}_{LE^{\ast}(X)} = &\: \nrm{\chf_{X} f}_{L^{2} L^{2}(\bbR \times B_{1})}
	+ \sum_{R' \in 2^{\bbZ_{\geq 0}}} (R')^{\frac{1}{2}}\nrm{\chf_{X} f}_{L^{2} L^{2}(\bbR \times B_{2R'} \setminus B_{R'})}.
\end{align}
\end{lemma}

\begin{lemma}[Pointwise estimates extracted from integrated local energy decay]\label{lem:ILED.pointwise}
Suppose $1\leq R\ls U$. Then
\begin{equation*}
\begin{split}
    &\: \sum_{I_u + I_r + |I| \leq s} |(u\rd_u)^{I_u} (r\rd_r)^{I_r} \bfOmg^{I}\Box_{\bfm}^{-1} F|(U,R,\Theta) \\
    \ls &\: R^{-\f{d-1}2} U^{-\f 12} \min\{U,R\} \sum_{I_u + I_r + |I| \leq s+\f{d+3}2} \nrm{(u\rd_u)^{I_u} (r\rd_r)^{I_r} \bfOmg^{I} F}_{LE^*(D_{U,R})}.
\end{split}
\end{equation*}
\end{lemma}
\begin{proof}
By Lemma~\ref{lem:kl-sob},
\begin{equation*}
\begin{split}
&\: U^{\f 12} R^{\f{d-3}2} \sum_{I_u + I_r + |I| \leq s}|(u\rd_u)^{I_u} (r\rd_r)^{I_r} \bfOmg^{I}\Box_{\bfm}^{-1} f|(U,R,\Theta) \\
\ls &\: R^{-\frac{3}{2}}  \nrm{\bfGmm^{(\leq s+\f{d+3}2)} (\Box_{\bfm}^{-1} f)}_{L^{2}(C_U^R)} + R^{-\f 12}\min\{U, R\} \sum_{I_u + I_r + |I| \leq s+\f{d+3}2} \nrm{(u\rd_u)^{I_u} (r\rd_r)^{I_r} \bfOmg^{I} f}_{L^{2}(C_U^R)},
\end{split}
\end{equation*}
where $\bfGmm$ on the right-hand side runs through $\bfGmm \in \{ \bfS, \bfOmg\}$.

Using that $[\Box_{\bfm}, \bfOmg] = 0$, $[\Box_{\bfm}, \bfS] = 2 \Box_{\bfm}$, we obtain $\bfOmg \Box_{\bfm}^{-1} F = \Box_{\bfm}^{-1} \bfOmg F$ and $\bfS\Box_{\bfm}^{-1} F = \Box_{\bfm}^{-1} \bfS F + 2F$. Therefore, iterating these commutation relations and using Lemma~\ref{lem:iled-flat}, we obtain
\begin{equation*}
\begin{split}
&\: U^{\f 12} R^{\f{d-3}2} \sum_{I_u + I_r + |I| \leq s}|(u\rd_u)^{I_u} (r\rd_r)^{I_r} \bfOmg^{I}\Box_{\bfm}^{-1} F|(U,R,\Theta) \\
\ls &\: R^{-\frac{3}{2}}  \nrm{ \Box_{\bfm}^{-1} (\bfGmm^{(\leq s+ \f{d+3}2)} F)}_{L^{2}(C_U^R)} + R^{-\f 12} \min\{U, R\} \sum_{I_u + I_r + |I| \leq s+\f{d+3}2} \nrm{(u\rd_u)^{I_u} (r\rd_r)^{I_r} \bfOmg^{I} F}_{L^{2}(C_U^R)} \\
\ls &\:  R^{-1} \min\{U, R\} \sum_{I_u + I_r + |I| \leq s+\f{d+3}2} \nrm{(u\rd_u)^{I_u} (r\rd_r)^{I_r} \bfOmg^{I} F}_{LE^{*}(D_{U,R})}.
\end{split}
\end{equation*}
Rearranging gives the desired result. \qedhere

\end{proof}

\begin{lemma}[Additional pointwise estimate when $d=3$]\label{lem:3d.pointwise}
Let $d=3$ and $\de \in (0,1)$. Suppose $F:\mathbb R^{3+1} \to \mathbb R$ is a sufficiently regular function on Minkowskian spacetime supported in $\{t\geq 0\}$ such that
\begin{equation}\label{eq:3d.pointwise.0}
F = O^{M}_{\bfGmm}(u^{-\alp} \brk{r}^{-2-\de}).
\end{equation}
Let $(U,R)$ and $U_{0}$ be such that $1\leq R$ and $1\leq U_0 \leq U$. Then
$$\Box_{\bfm}^{-1}(\chi_{> U_0} F)(U,R,\Theta) =_{\de^{-1}} O^{M-\f{d+3}2}_{\bfGmm}(\max\{U^{-\alp}, U_0^{-\alp}\} R^{-\de}).$$
\end{lemma}
\begin{proof}
We first prove the bound $\Box_{\bfm}^{-1} F$ itself (without derivatives bound). Since $d=3$, we can use the positivity of the fundamental solution so that we only need to estimate $\varphi$ which satisfies
\begin{equation}\label{eq:3d.pointwise.1}
\Box_{\bfm} \varphi = \chi_{> U_0} u^{-\alp} \brk{r}^{-2-\de},\quad (\varphi, \bfT \varphi)_{u+r =0} = (0,0).
\end{equation}
Since $\varphi$ is spherically symmetric, we have
\begin{equation}\label{eq:3d.pointwise.2}
\Big| -2 \rd_u \rd_r (r \varphi) + \rd_r^2 (r\varphi) \Big| \ls \chi_{> U_0} u^{-\alp} \brk{r}^{-1-\de} \ls \max\{U^{-\alp}, U_0^{-\alp}\} r^{-1-\de}.
\end{equation}
We integrate the inequality \eqref{eq:3d.pointwise.2} along the integral curves of $-2\rd_u + \rd_r$ and $\rd_r$ (and use that the data vanish and that $r\varphi(u,0,\theta)=0$) to obtain
\begin{equation}\label{eq:3d.pointwise.3}
\begin{split}
R|\varphi(U,R,\Theta)| \ls  &\: \max\{U^{-\alp}, U_0^{-\alp}\}  \int_{U_0}^U \int_{\f 12 (U-u)}^{\f 12 (U+2R-u)} r^{-1-\de} \, \ud r\, \ud u.
\end{split}
\end{equation}
We then compute
\begin{equation}\label{eq:3d.pointwise.4}
\begin{split}
&\:  \int_{U_0}^U \int_{\f 12 (U-u)}^{\f 12 (U+2R-u)} r^{-1-\de} \, \ud r\, \ud u \ls \de^{-1} \int_{U_0}^U \Big((U-u)^{-\de} - (U+2R-u)^{-\de} \Big)\, \ud u \\
\ls &\: \de^{-1} (R^{1-\de} + (U-U_0 +2R)^{1-\de} - (U-U_0)^{1-\de}).
\end{split}
\end{equation}
We claim that $(U-U_0 +2R)^{1-\de} - (U-U_0)^{1-\de} \ls R^{1-\de}$. Indeed, this is obvious if $U-U_0 \leq R$. If $U-U_0 \geq R$, then by the fundamental theorem of calculus,
$$(U-U_0 +2R)^{1-\de} - (U-U_0)^{1-\de} = R \int_{0}^{1} (1-\de) (U+2sR - U_0)^{-\de}\, \ud s\leq R (U-U_0)^{-\de} \leq R^{1-\de}.$$
Plugging this into \eqref{eq:3d.pointwise.4} and combining with \eqref{eq:3d.pointwise.3}, we obtain the pointwise bound
$$\Big|\Box_{\bfm}^{-1}(\chi_{> U_0} F)(U,R)\Big| \ls_{\de^{-1}} \max\{U^{-\alp}, U_0^{-\alp}\} R^{-\de}.$$

To obtain the higher derivatives bound, first commute with $\bfGmm \in \{ \bfS, \bfOmg\}$ to control $\bfGmm^I \Box_{\bfm}^{-1}(\chi_{> U_0} F)$ with the same bound as above. After integrating, this gives $L^2$ bounds
\begin{equation}\label{eq:3d.pointwise.5}
\|\bfGmm^I \Box_{\bfm}^{-1}(\chi_{> U_0} F)\|_{L^2(C^R_U)} \ls_{\de^{-1}}  \max\{U^{-\alp}, U_0^{-\alp}\} U^{\f 12} R^{\f d2 -\de} \quad |I| \leq M.
\end{equation}
On the other hand, using \eqref{eq:3d.pointwise.0}, we have
\begin{equation}\label{eq:3d.pointwise.6}
\|\bfGmm^I (\chi_{> U_0} F)\|_{L^2(C^R_U)} \ls  U^{-\alp} U^{\f 12} R^{\f d2-2 -\de} \quad |I| \leq M.
\end{equation}
Hence, the lemma follows from an application of Lemma~\ref{lem:kl-sob} using the bounds \eqref{eq:3d.pointwise.5} and \eqref{eq:3d.pointwise.6}. \qedhere
\end{proof}

\subsection{Contribution of incoming radiation} \label{subsec:huygens-corr}
The following basic computation will be useful for separating out the contribution of the incoming radiation.
\begin{lemma} \label{lem:med.incoming}
The following holds:
\begin{equation} \label{eq:med.incoming}
[\Box_{\bfm}, \chi_{>U_{0}}(u)] \psi = - 2 \chi_{>U_{0}}'(u) r^{-\nu_{\Box}} \rd_{r}(r^{\nu_{\Box}} \psi).
\end{equation}
\end{lemma}
\begin{proof}
This identity follows immediately from the standard formula (see \eqref{eq:m-1})
\begin{equation*}
\Box_{\bfm} = - 2\rd_u \rd_r + \rd_r^2 + \f{d-1}r (\rd_r - \rd_u ) + \f 1{r^2} \rslap.
\end{equation*}
\end{proof}

We also state and prove a lemma that will be used to relate the contribution of the incoming radiation with the Minkowski computations performed in Section~\ref{sec:Minkowski.wave}.

\begin{lemma}\label{lem:med.explicit.inho}
Suppose $\zeta:\mathbb R\to \mathbb R$ is compactly supported in $(-\infty, U_0)$. The following holds for $u - U_0 \geq 2$:
\begin{align}
& \Box_{\bfm} \Big( Y_{(\ell)}\int_{-\infty}^\infty \zeta(u') \varphi^{\bfm[1]}_{(\ell)j,k}(u-u',r) \, \ud u' \Big)
= - 2 \zeta(u) Y_{(\ell)} r^{-\nB} \partial_r (r^{-j} \log^k r), \label{eq:med.explicit.inho.1}\\
& \Box_{\bfm} \Big( Y_{(\ell)}\int_{-\infty}^\infty \zeta(u') \varphi^{\bfm[u']}_{(\ell)j,k}(u-u',r) \, \ud u' \Big)
= - 2 \zeta(u) Y_{(\ell)} r^{-\nB} \partial_r (r^{-j} \log^k (\tfrac ru)). \label{eq:med.explicit.inho.2}
\end{align}
\end{lemma}
\begin{proof}
Noticing that $Y_{(\ell)} \varphi^{\bfm[1]}_{(\ell)j,k}(u,r)$ satisfies the wave equation for $u\neq 0$ and that it has a jump discontinuity at $u=0$, we obtain
\begin{equation}
\begin{split}
 \Box_{\bfm} \Big( Y_{(\ell)}\int_{-\infty}^\infty \zeta(u') \varphi^{\bfm[a]}_{(\ell)j,k}(u-u',r) \, \ud u' \Big)
= &\: - \zeta(u) Y_{(\ell)} \Big(2\partial_r \varphi^{\bfm[a]}_{(\ell)j,k} + \f{d-1}r \varphi^{\bfm[a]}_{(\ell)j,k} \Big)(r,0) \\
= &\: - 2 \zeta(u) Y_{(\ell)} r^{-\nB} \partial_r (r^{-j} \log^k (\tfrac ra)),
\end{split}
\end{equation}
where in the last line we used \eqref{eq:varphi.m.data} with $a=1$. This proves \eqref{eq:med.explicit.inho.1}. The proof of \eqref{eq:med.explicit.inho.2} is similar. \qedhere
\end{proof}

\subsection{Controlling the error terms arising from the proof of the iteration lemma}\label{sec:med.error}

In this subsection, we control the error terms arising from the proof of the iteration lemma. In the proof of the iteration lemma, when estimating $\phi(U,R,\Theta)$, we will cut off $\phi$ with $\chi_{>U_0}$, where $U_0 \leq \eta_{1} U$. We choose $\eta_{1}>0$ to be sufficiently small (depending on $d$, $J_{\frkf}$, $K_{J_{\frkf}}$, $\min\{M_{0},M_{c}\}$) so that the Minkowski estimate \eqref{eq:phi.ell.j.k.upper} holds in $(\mathbb R^{d+1},\bfm)$ whenever $u \geq \eta_{1}a$ for $0\leq j\leq J_{\frkf}$, $0\leq k\leq K_{J_{\frkf}}$ and $I_r,I_u\leq \min\{M_{0},M_{c}\}$. Assume also that $\eta_{1} \leq \min\{ \f{\eta_{0}}4, \f 12\}$. \textbf{We now fix the choice of $\eta_{1}$.}

In the following proposition, we work under the same assumptions as Proposition~\ref{prop:med} and estimate most of the error terms. We will isolate one main term in $\phi$, which corresponds to the contribution from $\calM_{\wave}$.

\begin{proposition}\label{prop:med.error}
Let $0< \de_{0} \leq \min\{\f{\de_c}2,\f{\de_d}2,\f 12\}$ and $\alp \leq \alp_d$. In the case $\calN \neq 0$, assume \eqref{eq:alpN-med}.

Assume that
\begin{equation}\label{eq:med-ass.again}
	\phi = O_{\bfGmm}^{M}(A u^{-\alp}) \quad \hbox{ in } \calM_{\near} \cup \calM_{\med},\quad \phi = O_{\bfGmm}^M(A u^{-\alp +\nB} r^{-\nB})\quad \hbox{ in } \calM_{\wave},
\end{equation}
where $\f{d+7}2 \leq M \leq \min\set{M_{0}, M_{c}}$ and $A > 0$.

Then, for any $(U,R,\Theta)\in \calM_{\med}$ with $R\geq 2R_{\far}$ and for any $U_0 \leq \eta_1 U$ (where $\eta_1$ is as fixed above), the following holds with $A'$ satisfying \eqref{eq:A'}:
\begin{equation*}
	\begin{split}
	\phi(U,R,\Theta) = &\: - 2 \Box_\bfm^{-1} \Big(\chi_{>R_{1}}(r) \chi'_{>U_0}(u) r^{-\nB} \rd_r (r^{\nB} \phi)\Big)(U,R,\Theta) \\
	&\: + O^{M-\f{d+7}2}_{\bfGmm}(A' \max\{U^{-\alp},U_0^{-\alp}\} U^{\nB-\de_0} R^{-\nB}).
	\end{split}
\end{equation*}
\end{proposition}

\begin{proof}

We first localize $\phi$ as follows.

Fix a point $(U,R,\Theta) \in \calM_{\med}$ with $R \geq 2R_{\far}$. Let $1\leq U_0 \leq \eta_{1} U$. (For the upper bound, we will only later take $U_0 = \eta_{1} U$, but the more general estimates will be useful when we derive the sharp asymptotics.)

In order to bound $\phi$ at $(U,R,\Theta)$, we will derive and use the equation for $\Box_\bfm (\chi_{>2 R_{\far}}(r) \chi_{>U_0}(u) \phi)$. The derivation will be achieved in Step~1. For the terms appearing in this equation, we first isolate the main terms (Step~2), bound the error terms $e_i$ (Step~3) and then finally estimate the contribution $\Box_\bfm^{-1}e_i$ using Lemmas~\ref{lem:ILED.pointwise} and \ref{lem:3d.pointwise}.

The following facts are easy to derive, and will often be used without explicit comments.
\begin{itemize}
\item In the region $D_{U,R} \cap \{u \in [\frac{U_0}2,U_0]\}$, the $u$-value is comparable to $U_0$ and the $r$-value is comparable to $U$.
\item Everywhere in the region $D_{U,R}$, the $u$-value satisfies $U_0\ls u \ls U$ and the $r$-value satisfies $R_1 \ls r \ls U$.
\end{itemize}

\pfstep{Step~1: Computing $\Box_\bfm (\chi_{>2 R_{\far}}(r) \chi_{>U_0}(u) \phi)$} Since $\calP \phi = f + \calN(\phi)$, we have
\begin{equation}\label{eq:med.e1-e5}
\begin{split}
\Box_\bfm (\chi_{>2 R_{\far}}(r) \chi_{>U_0}(u) \phi) = &\: - \chi_{>2 R_{\far}}(r) [\chi_{>U_0}(u) , \Box_\bfm] \phi  \\
&\:  -[\chi_{>2 R_{\far}}(r), \Box_\bfm] (\chi_{>U_0}(u)\phi)\\
&\: + \chi_{>2 R_{\far}}(r) \chi_{>U_0}(u) (\Box_\bfm - \calP) \phi \\
&\: + \chi_{>2 R_{\far}}(r) \chi_{>U_0}(u) f \\
&\: + \chi_{>2 R_{\far}}(r) \chi_{>U_0}(u) \calN(\phi) \\
=: &\: e_1+ \ldots + e_5.
\end{split}
\end{equation}

\pfstep{Step~2: Computing the main term $e_1$} By Lemma~\ref{lem:med.incoming}, we have
\begin{equation}\label{eq:med-e1}
e_1 = - \chi_{>2 R_{\far}}(r) \chi_{>U_0}'(u) (2\rd_r \phi + \f{d-1}{r} \phi) = - 2 \chi_{>2 R_{\far}}(r) \chi_{>U_0}'(u) r^{-\nB} \rd_r (r^\nB\phi).
\end{equation}

%

\pfstep{Step~3: Estimating the error terms  $e_2$, $e_3$, $e_4$ and $e_5$}
Our goal in this step is to show that all the error terms obey the bound
\begin{equation}\label{eq:error.goal.med}
e_i(u,r,\theta) = O^{M-2}(A' u^{-\alp} r^{-2-\de_0}),\quad i = 2,3,4,5.
\end{equation}

\textbf{The term $e_{2}$.} Thanks to the commutator $[\chi_{>R_1}, \Box_{\bfm}]$, $e_2$ is supported in $\{r \leq R_1\}$. Hence, the bound \eqref{eq:error.goal.med} holds because of \eqref{eq:med-ass.again}.

\textbf{The term $e_{3}$.} For the term $(\calP - \Box_{\bfm})\phi$, we have (see \ref{hyp:med})
$$(\calP - \Box_{\bfm})\phi = O_{\bfGmm}^{M_c}(r^{-\de_{c}}) \rd^2 \phi + O_{\bfGmm}^{M_c}(r^{-\de_c-1}) \rd \phi + O_{\bfGmm}^{M_c}(r^{-\de_c-2}) \phi \hbox{ in } \calM_\med.$$
In $\calM_\wave$, we have slightly more precise estimates for the coefficients, namely that
$$(\calP - \Box_{\bfm})\phi = O_{\bfGmm}^{M_c}(u^{-\de_{c}}\log^{K_{c}'}\left(\tfrac{r}{u}\right)) \rd^2 \phi + O_{\bfGmm}^{M_c}(u^{-\de_c} r^{-1} \log^{K_{c}}\left(\tfrac{r}{u}\right)) \rd \phi + O_{\bfGmm}^{M_c}(u^{-\de_c} r^{-2}\log^{K_{c}'}\left(\tfrac{r}{u}\right)) \phi,$$
but moreover, after decomposing into the $\{ \rd_r, \rd_u, \f 1 r \mathring{\slashed\nabla}\}$ basis, either we have the good $\rd_r$, $\f 1 r \mathring{\slashed\nabla}$, or else for each $\rd_u$ there is an additional factor of $\f ur$; see Lemma~\ref{lem:conj-wave-wave}.

As a result, using also the assumed bound \eqref{eq:med-ass.again} for $\phi$, the following pointwise bound holds:
\begin{equation}\label{eq:med.e3.pointwise}
e_3 = \begin{cases}
O_{\bfGmm}^{M-2} (A u^{-\alp} r^{-2-\de_c}) & \hbox{ in } \calM_\med \\
O_{\bfGmm}^{M-2} (A u^{-\alp+\nB-\de_c} r^{-2-\nB}\log^{K_{c}'}\left(\tfrac{r}{u}\right)) = O_{\bfGmm}^{M-2} (A u^{-\alp}r^{-2-\f{\de_c}2}) & \hbox{ in } \calM_\wave
\end{cases},
\end{equation}
where in the last estimate in $\calM_\wave$, we used $u\ls r$ in $\calM_\wave$, and used $r^{-\de_c}$ to dominate the factor $\log^{K_{c}'}\left(\tfrac{r}{u}\right)$.

\textbf{The term $e_4$.} We turn to the inhomogeneous term $f$. Using \eqref{eq:near-f}--\eqref{eq:wave-f-rem}, we obtain
$$e_4 =
\begin{cases}
O_{\bfGmm}^{M_0}(D u^{-\alp_d} r^{-2-\de_d}) &\hbox{ in } \calM_{\med} \\
O_{\bfGmm}^{M_0}(D u^{-\alp_d} r^{-2-\f{\de_d}2}) & \hbox{ in } \calM_{\wave}
\end{cases} $$
where we have used that $u\ls r$ in $\calM_\wave$ and used a $r^{-\frac{\de_d}2}$ power to absorb the logarithms in \eqref{eq:wave-f-exp}.


\textbf{The term $e_5$.} We next turn to the nonlinear term. Using \eqref{eq:alpN-med}, \eqref{eq:sol-med}, \eqref{eq:sol-wave} and \eqref{eq:med-ass.near.med}, we apply Definition~\ref{def:alp-N} with $\alp_{\calN}' = \alp_{\calN} + \de_0$ to obtain
$$ e_5 =O_{\bfGmm}^{M-2} (A u^{-\alp} r^{-2-\de_0})  \hbox{ in } \calM_\med \cup \calM_\wave,$$
where we have used that $r\ls u$ in $\calM_\med$, $u\ls r$ in $\calM_\wave$.

\pfstep{Step~4: Obtaining pointwise bounds for $\Box_{\bfm}^{-1} e_i$} From \eqref{eq:error.goal.med}, we can straightforwardly obtain the desired pointwise bounds using Lemma~\ref{lem:3d.pointwise} (when $d=3$) and Lemma~\ref{lem:ILED.pointwise} (when $d \geq 5$).

In $d=3$, we use Lemma~\ref{lem:3d.pointwise} to show that \eqref{eq:error.goal.med} implies
\begin{equation}\label{eq:med.error.final.d=3}
\begin{split}
(\Box_{\bfm}^{-1} e_i)(U,R,\Theta) = &\: O_{\bfGmm}^{M-\f{d+7}2}(A' \max\{U^{-\alp}, U_0^{-\alp}\}R^{-\de_0}) \\
= &\: O_{\bfGmm}^{M-\f{d+7}2}(A' \max\{U^{-\alp}, U_0^{-\alp}\}U^{\f{d-1}2-\de_0}R^{-\f{d-1}2}),\quad i=2,3,4,5,\, d = 3.
\end{split}
\end{equation}

When $d \geq 5$, we first compute the $LE^*$ norm of $r^{-\omg}$ for $\omg \leq \f 52$ as follows:
\begin{equation}\label{eq:med.LE.error.romg.high.d.1}
\begin{split}
&\: \| r^{-\omg} \|_{LE^*(D_{U,R} \cap \{r\geq \f{R_1}2 \} \cap \{u\geq \f{U_0}2 \})} \\
\ls &\: \sum_{k: 2^k \leq \f{R_1}2} \Big( \int_0^{T} \| r^{\f 12} r^{-\omg} \|_{L^2(D_{U,R} \cap \{t=\tau\}\cap \{2^k \leq r \leq 2^{k+1}\}  \cap \{u\geq \f{U_0}2\})}^2 \, \ud \tau \Big)^{\f 12}\\
\ls &\:  \sum_{k: \f{R_1}2\leq 2^k \ls (T+2R)} (\int_0^T \, \ud \tau )^{\f 12} \Big( \int_{r=2^k}^{r=2^{k+1}} r^{-2\omg+1} r^{d-1} \, \ud r \Big)^{\f 12}  \ls U^{\f 12} \sum_{k: 2^k \ls U} 2^{\f{(d-2\omg+1)k}2} \ls U^{\f{d-2\omg+2}2},
\end{split}
\end{equation}
where in the final sum, we used that $d-2\omg+1 > 0$ because $d\geq 5$ and $\omg \leq \f 52$. Therefore,
\begin{equation}\label{eq:med.LE.error.romg.high.d.2}
\| u^{-\alp} r^{-2-\de_0} \|_{LE^*(D_{U,R} \cap \{r\geq \f{R_1}2 \} \cap \{u\geq \f{U_0}2 \})} \ls \max\{U^{-\alp}, U_0^{-\alp}\} U^{\f{d-2}2-\de_0}.
\end{equation}
Therefore, combining \eqref{eq:error.goal.med}, \eqref{eq:med.LE.error.romg.high.d.2} with Lemma~\ref{lem:ILED.pointwise}, we obtain
\begin{equation}\label{eq:med.error.final.dgeq5}
(\Box_{\bfm}^{-1} e_i)(U,R,\Theta) = O^{M-\f{d+7}2}_{\bfGmm}(A' \max\{U^{-\alp}, U_0^{-\alp}\} U^{\f{d-1}2-\de_0} R^{-\f{d-1}2}),\quad i=2,3,4,5,\, d \geq 5.
\end{equation}

Combining \eqref{eq:med.error.final.d=3} and \eqref{eq:med.error.final.dgeq5} yields the desired bound. \qedhere

%
%

\qedhere
\end{proof}


We now prove a general estimate for the wave equation on Minkowski spacetime. In the context of the proof of Proposition~\ref{prop:med} in Section~\ref{sec:iteration.intermediate.proof}, we will start Proposition~\ref{prop:med.error} so that we need to control the term $- 2 \Box_\bfm^{-1} \Big(\chi_{>R_{1}}(r) \chi'_{>U_0}(u) r^{-\nB} \rd_r (r^{\nB} \phi)\Big)(U,R,\Theta)$. We then expand $r^{\nB} \phi$ using \eqref{eq:med-ass.case.1.expansion}, \eqref{eq:med-ass.wave.0} or \eqref{eq:med-ass.wave.0.case.4}. Proposition~\ref{prop:rho.est} will be used to control the contribution from $\rho_1$, $\rho_J$ or $\rho_{J+1}$ in these expansions. Notice that in the proof of Proposition~\ref{prop:rho.est}, we rely on the strong Huygen's principle to ensure that the contribution comes from $\calM_{\far}$.

\begin{proposition}\label{prop:rho.est}
Let $\rho$ be a function in $\calM_\wave$ satisfying
\begin{equation}\label{eq:med.general.rho}
\rho = O_{\bfGmm}^{M}(A_{\rho} r^{-\alp_{\rho} + \nu_{\Box}} u^{\bt_{\rho}} \log^{K_{\rho}} u)
\quad \hbox{ in } \calM_{\wave}
\end{equation}
for some $A_{\rho} > 0$, $\alp_{\rho} \in \bbR$, $\bt_{\rho} \in \bbR$ and $K_{\rho} \in \bbZ_{\geq 0}$.

Suppose $(U,R,\Theta) \in \calM_\med$, $R \geq 2R_{\far}$ and $1\leq U_0 \leq \eta_{1} U$, where $\eta_{1} \leq \min\{ \f{\eta_{0}}4, \f 12\}$ as defined in the beginning of Section~\ref{sec:med.error}. Then
$$\Box_\bfm^{-1} \Big(\chi_{>2 R_{\far}}(r) \chi_{>U_0}'(u) r^{-\nB} \rd_r \rho \Big)(U,R,\Theta) = O_{\bfGmm}^{M-\f{d+7}2}(A_\rho U^{-\alp_\rho + \nB+ \f 12} U_0^{\bt_\rho -\f 12} R^{-\nB} \log^{K_\rho}U_0).$$
\end{proposition}
\begin{proof}
Using \eqref{eq:med.general.rho}, we have the pointwise bound
\begin{equation}\label{eq:rho.est.e.def}
 \chi_{>2 R_{\far}}(r) \chi_{>U_0}'(u) r^{-\nB} \rd_r \rho = O_{\bfGmm}^{M-1}(A_\rho r^{-\alp_\rho-1} u^{\bt_\rho-1} \log^{K_\rho} u).
\end{equation}
We compute
\begin{equation}\label{eq:rho.est.LE*}
\begin{split}
&\: \| r^{-\alp_\rho-1} u^{\bt_\rho-1} \log^{K_\rho} u\|_{LE^*(D_{U,R}\cap \{r \geq \f{R_1}2\} \cap \{\f{U_0}2 \leq u \leq U_{0}\})} \\
\ls &\: \underbrace{U^{-\alp_\rho-1} U_0^{\bt_\rho-1} \log^{K_\rho}{U_0}}_{\mathrm{pointwise\, bound}}\cdot \underbrace{U^{\f 12}}_{r^{\f 12} \mathrm{\,weight}} \cdot \underbrace{(U^{d} U_0)^{\f 12}}_{(\mathrm{Volume})^{\f 12}} = U^{-\alp_\rho+\nB} U_0^{\bt_\rho-\f 12} \log^{K_\rho} U_0.
\end{split}
\end{equation}
The desired conclusion then follows from applying Lemma~\ref{lem:ILED.pointwise} with \eqref{eq:rho.est.e.def} and \eqref{eq:rho.est.LE*}. \qedhere
\end{proof}

\subsection{Proof of iteration lemma (Proposition~\ref{prop:med})}\label{sec:iteration.intermediate.proof}
We now have all the tools to prove Proposition~\ref{prop:med}.
\begin{proof}[Proof of Proposition~\ref{prop:med}]
First note that on a finite-$r$ region, all the desired estimates simply follow as a consequence of \eqref{eq:med-ass.near.med}. Thus, throughout the proof, we take $(U,R,\Theta) \in \calM_{\med}$ with $R \geq 2R_{\far}$. We will \textbf{apply Proposition~\ref{prop:med.error} and Proposition~\ref{prop:rho.est} with $U_0 = \eta_1 U$}, where $\eta_{1}>0$ is as defined in the beginning of Section~\ref{sec:med.error}. In particular, we will use that $U_0$ and $U$ are comparable, often without explicit comments.

For the rest of the proof $(u,r,\theta)$ will denote the variables of the input while $(U,R,\Theta)$ will denote the variables of the output.

\pfstep{Step~1: Proof of Case 1} Using the assumption \eqref{eq:med-ass.near.med}, we apply Proposition~\ref{prop:med.error} to obtain
\begin{equation}\label{eq:pf.med.iteration.case.1}
	\begin{split}
	\phi(U,R,\Theta) = &\: - 2 \Box_\bfm^{-1} \Big(\chi_{>2 R_{\far}}(r) \chi'_{>U_0}(u) r^{-\nB} \rd_r (r^{\nB} \phi)\Big)(U,R,\Theta) + O_{\bfGmm}^{M-\f{d+7}2}(A' U^{-\alp+\nB-\de_0} R^{-\nB}),
	\end{split}
\end{equation}
where we have used that $U_0\sim U$.

It thus remains to control the first term in the right-hand side of \eqref{eq:pf.med.iteration.case.1}. We write $r^\nB \phi = \rPhi_{0,0} + \rho_1$ as in \eqref{eq:med-ass.case.1.expansion} in the assumptions. Whenever $U \geq 2(1-\eta_1)^{-1}$ (so that $U-U_0\geq 2$), we have by Lemma~\ref{lem:med.corrected} that
$$\Box_\bfm^{-1} \Big(\chi_{>2 R_{\far}}(r) \chi'_{>U_0}(u) r^{-\nB} \rd_r \Phi_{0,0}\Big)(U,R,\Theta) = 0$$
Using the assumption \eqref{eq:med-ass.case.1}, we apply Proposition~\ref{prop:rho.est} (with $\alp_\rho= \nB$, $\bt_\rho = -\alp-\de_0+\nB$, $K_\rho = 0$) we obtain
\begin{equation}\label{eq:med.basic.bound.for.drrho}
\Box_\bfm^{-1} \Big(\chi_{>2 R_{\far}}(r) \chi'_{>U_0}(u) r^{-\nB} \rd_r \rho \Big)(U,R,\Theta) = O_{\bfGmm}^{M-\f{d+7}2}(A  U^{-\alp-\de_0+\nB} R^{-\nB}).
\end{equation}

Combining the above estimates gives the desired conclusion.

\pfstep{Step~2: Proof of Case 2} This is essentially the same as in Case 1 except for controlling the main term using the expansion \eqref{eq:med-ass.wave.0}. As in Step~1, we first use Proposition~\ref{prop:med.error} to get \eqref{eq:pf.med.iteration.case.1}. It thus remains to control $- 2 \Box_\bfm^{-1} \Big(\chi_{>2 R_{\far}}(r) \chi'_{>U_0}(u) r^{-\nB} \rd_r (r^{\nB} \phi)\Big)$. We will separately bound the $\rPhi_{j,k}$ terms and the $\rho_{J}$ term in the expansion \eqref{eq:med-ass.wave.0}.

For the $\rPhi_{j,k}$ terms, we use the estimates in \eqref{eq:med-ass.wave.1}--\eqref{eq:med-ass.wave.4} together with the explicit form of the solution given in Lemma~\ref{lem:med.explicit.inho} and the bound \eqref{eq:phi.ell.j.k.upper} to obtain
\begin{equation}\label{eq:med.main.terms.with.expected.decay}
\begin{split}
&\: \sum_{j=0}^{J -1} \sum_{k = 0}^{K_{j}} \Box_\bfm^{-1} \Big(\chi_{>2 R_{\far}}(r) \chi'_{>U_0}(u) r^{-\nB} \rPhi_{j,k} \rd_r \Big( r^{-j} \log^{k} \left(\tfrac{r}{u}\right)\Big) \Big)(U,R,\Theta) \\
= &\: \sum_{j=0}^{J -1} \sum_{k = 0}^{K_{j}}  O_{\bfGmm}^{M-d}\Bigg(A U^{-j-\nB} U_0^{j-\alp-\de_0+\nB} \log^{K_\rho} \left( \tfrac{U-U_0}{U_0} \right) \Bigg) \\
= &\: \sum_{j=0}^{J -1} \sum_{k = 0}^{K_{j}}  O_{\bfGmm}^{M-d}(A U^{-\alp-\de_0}) = O_{\bfGmm}^{M-d}(A U^{-\alp-\de_0}),
\end{split}
\end{equation}
where $d$ derivatives are used so as to sum over spherical harmonics.

As for the contribution from $\rho_{J}$, we use the bound \eqref{eq:med-ass.wave.remainder} and apply Proposition~\ref{prop:rho.est} (with $\alp_\rho= \alp + \de_0$, $\bt_\rho = 0$, $K_\rho = 0$) to obtain the estimate \eqref{eq:med.basic.bound.for.drrho} when $\rho = \rho_{J}$. Putting everything together, we obtain the desired bound.

\pfstep{Step~3: Proof of Case 3} Compared to Case 2, the only difference is that we have the assumed bound \eqref{eq:med-ass.wave.remainder.case.3} (instead of \eqref{eq:med-ass.wave.remainder}) for $\rho_{J}$. We have two possibilities (since $\alp_{\mathfrak f} - J - \nB \leq 0$ by definition):
\begin{enumerate}
    \item If $\alp_{\mathfrak f} - J - \nB = 0$, then $\max\{u^{\alp_{\mathfrak f}-J-\nB}\log^{K_J} u, 1 \} = \log^{K_J} u$. We use Proposition~\ref{prop:rho.est} with $(\alp_\rho,\bt_\rho,K_\rho) = (\alp_{\mathfrak f}-\de_{\mathfrak f}',-\de_{\mathfrak f}',K_J)$ to obtain the bound
    \begin{equation}
    \begin{split}
        &\: \Box_\bfm^{-1} \Big(\chi_{>2 R_{\far}}(r) \chi'_{>U_0}(u) r^{-\nB} \rd_r \rho_J \Big)(U,R,\Theta) = O_{\bfGmm}^{M-\f{d+7}2}(A  U^{-\alp_{\mathfrak f}+\nB} R^{-\nB}\log^{K_J}U) \quad \hbox{in $\calM_{\med}$.}
    \end{split}
    \end{equation}
    \item If $\alp_{\mathfrak f} - J - \nB < 0$, then $\max\{u^{\alp_{\mathfrak f}-J-\nB}\log^{K_J} u, 1 \} = 1$. We use Proposition~\ref{prop:rho.est} with $(\alp_\rho,\bt_\rho,K_\rho) = (\alp_{\mathfrak f}-\de_{\mathfrak f}',-\de_{\mathfrak f}',0)$ to obtain
    \begin{equation}
    \begin{split}
        &\: \Box_\bfm^{-1} \Big(\chi_{>2 R_{\far}}(r) \chi'_{>U_0}(u) r^{-\nB} \rd_r \rho_J \Big)(U,R,\Theta) = O_{\bfGmm}^{M-\f{d+7}2}(A  U^{-\alp_{\mathfrak f}+\nB} R^{-\nB}) \quad \hbox{in $\calM_{\med}$.}
    \end{split}
    \end{equation}
\end{enumerate}

As in Step~2, all the other terms are $O^{M-\max\{\f{d+7}2, d\}}_{\bfGmm}(A' U^{-\alp-\de_0+\nB} R^{-\nB})$. Therefore, after noting that
$$O^{M-\max\{\f{d+7}2, d\}}_{\bfGmm}(A' U^{-\alp-\de_0+\nB} R^{-\nB})  = O_{\bfGmm}^{M-\max\{\f{d+7}2, d\}}(A'  U^{-\alp_{\mathfrak f}+\nB} R^{-\nB})$$
(which holds because $\alp+\de_0 > \min\{J_{c}-1+\eta_c,J_d-1+\eta_d\}+\nB> \alp_{\mathfrak f}$), we obtain the desired estimate.

\pfstep{Step~4: Proof of Case 4} First, we again use Proposition~\ref{prop:med.error} to get \eqref{eq:pf.med.iteration.case.1}. Then, notice that terms similar to those in Case 2 are bounded by $O_{\bfGmm}^{M-\max\{\f{d+7}2, d\}}(A' u^{-\alp-\de_0+\nB} r^{-\nB})$, which is better than what we need since $J < \alp + \de_0 -\nB$.

We will therefore only need to consider the new contributions to $- 2 \Box_\bfm^{-1} \Big(\chi_{>2 R_{\far}}(r) \chi'_{>U_0}(u) r^{-\nB} \rd_r (r^{\nB} \phi)\Big)$ coming from the four type of terms \eqref{eq:med-ass.wave-main-PhiJk}--\eqref{eq:med-ass.wave-main-rho}.

The term in \eqref{eq:med-ass.wave-main-PhiJ0-high} does not contribute to the intermediate zone by Lemmas~\ref{lem:med.explicit.inho} and \ref{lem:med.corrected}. For the terms in \eqref{eq:med-ass.wave-main-PhiJk} and \eqref{eq:med-ass.wave-main-PhiJ0-low}, we use Lemma~\ref{lem:med.explicit.inho} and \eqref{eq:phi.ell.j.k.upper} to obtain
\begin{equation}
\begin{split}
&\:  \sum_{k = 0}^{K_{J}} \Box_\bfm^{-1} \Big(\chi_{>2 R_{\far}}(r) \chi'_{>U_0}(u) r^{-\nB} \rPhi_{J,k} \rd_r \Big( r^{-J} \log^{k} \left(\tfrac{r}{u}\right)\Big) \Big)(U,R,\Theta) = O_{\bfGmm}^{M-d}(A U^{-J-\nB}\log^{K_J}U).
\end{split}
\end{equation}

Finally, we consider \eqref{eq:med-ass.wave-main-rho}. The only new term is that coming from $O_{\bfGmm}^{M}(A r^{-J-\eta_{\mathfrak f}} u^{\eta_{\mathfrak f}} \log^{K_J} u)$ term, for which we apply Proposition~\ref{prop:rho.est} with $A_\rho = A$, $\alp_\rho = J + \eta_{\mathfrak f} + \nB$, $\bt_\rho = \eta_{\mathfrak f}$ and $K_\rho = K_J$ to obtain
\begin{equation}
\begin{split}
 \Box_\bfm^{-1} \Big(\chi_{>2 R_{\far}}(r) \chi'_{>U_0}(u) r^{-\nB} \rd_r  \rho\Big)(U,R,\Theta)
= &\: O_{\bfGmm}^{M-\f{d+7}2}(A  U^{-J} R^{-\nB} \log^{K_J}U).
\end{split}
\end{equation}
Combining the above estimates and using $R\ls U$ yields the desired bound. \qedhere
\end{proof}

\section{Analysis in the wave zone: higher radiation fields} \label{sec:wave}
In this section, we perform the analysis in the wave zone. Our main iteration lemma (Proposition~\ref{prop:wave-main}) takes as the key input the improved $u$-decay for $\phi$ in the intermediate zone to improve the radiation field expansion for $\Phi = r^{\nu_{\Box}} \phi$ in the wave zone. We also state and prove an alternative iteration lemma, Proposition~\ref{prop:wave-0}, which is used at the initial stage of our proofs of the main theorems.

This section is organized as follows. The main results of this section, namely Propositions~\ref{prop:wave-main} and \ref{prop:wave-0} are stated in {\bf Section~\ref{sec:iteration.wave}}; the rest of the section is devoted to their proofs. In {\bf Section~\ref{subsec:recurrence}}, we analyze the formal recurrence equations satisfied by higher radiation fields, culminating in Proposition~\ref{prop:recurrence}. Along the way, we justify the properties of the recurrence equations assumed in Section~\ref{subsec:recurrence-formal}. In {\bf Section~\ref{subsec:expansion-error}}, we use the formally defined higher radiation fields to construct an approximate solution $\Phi_{<J}$, called the \emph{truncated expansion of $\Phi$ below order $J$}, and compute the corresponding error in the equation. In {\bf Section~\ref{subsec:remainder}}, we derive the key equation satisfied by the remainder $\rho_{J} = \Phi - \Phi_{<J}$. Then in \textbf{Section~\ref{subsec:wave-main-pf}} and \textbf{Section~\ref{subsec:wave-0-pf}}, we prove Propositions~\ref{prop:wave-main} and \ref{prop:wave-0}.

\subsection{Statement of the iteration lemmas} \label{sec:iteration.wave}
We begin by introducing some small parameters and conventions. Let $\eta_{d}$ and $\eta_{\mathfrak f}$ be positive numbers satisfying
\begin{gather}
\eta_{d} =\alp_{d} -\nu_{\Box}- J_{d}+1, \notag \\
0 < \eta_{\mathfrak f} < \min\set{\tfrac{1}{2}, J_{c}-1-J+\eta_{c}, J_{d}-1-J+\eta_{d}}. \label{eq:wave-main-etaf}
\end{gather}
Note that $0 < \eta_{d} \leq 1$. Let $\dlt_{0} > 0$ be a positive number that satisfy the following:
\begin{equation} \label{eq:dlt0-wave}
\dlt_{0} < \min\set{\tfrac{1}{2}, \eta_{c}, \eta_{d}}, \qquad \dlt_{0} \leq \min\set{\dlt_{c}, \dlt_{d}},
\end{equation}
Define also a string of nonnegative integers $0 = K_{0} \leq K_{1} \leq \ldots \leq K_{\min\set{J_{c}, J_{d}-1}}$ as in Lemma~\ref{lem:Kj-def} below; we remark that $K_{0} = \ldots = K_{\min\set{J_{c}, J_{d}-1}} = 0$ if there are no logarithms in the coefficients or in the data (i.e., $K_{c} = K_{d} = 0$). Also, in what follows {\bf we suppress the dependence of constants on $R_{\far}$.}

The main result of this section is as follows.

\begin{proposition}[Main iteration lemma in the wave zone] \label{prop:wave-main}
Let $\nu_{\Box} < \alp < \min\set{J_{c}, J_{d}} + \nu_{\Box}$, and if $\calN \neq 0$, assume also
\begin{equation} \label{eq:alpi-dlt0-nonlin}
	\alp \geq \alp_{\calN} + 2 \dlt_{0}.
\end{equation}
 Then there exist a positive integer $C'_{\wave}$ and a function $A'_{\wave}$ satisfying \eqref{eq:A'} such that the following holds. Assume that
\begin{equation} \label{eq:wave-main-hyp-med}
\phi = O_{\bfGmm}^{M}(A u^{-\alp}) \quad \hbox{ in } \calM_{\med},
\end{equation}
where $C'_{\wave} < M \leq \min \set{M_{0}, M_{c}}$ and $A > 0$.
For an integer $J$ such that $1 \leq J \leq \min\set{J_{c}, J_{d}}$ and
\begin{equation}\label{eq:alp.J.range}
J -1  + \nu_{\Box} < \alp < J + \nu_{\Box},
\end{equation}
assume that $\Phi$ admits the expansion
\begin{equation*}
\Phi = \sum_{j=0}^{J-1} r^{-j} \rPhi_{j,0} + \sum_{j=1}^{J-1} \sum_{k \in K_{j} \setminus \set{0}} r^{-j} \log^{k} \left(\tfrac{r}{u}\right) \rPhi_{j,k} + \rho_{J} \quad \hbox{ in } \calM_{\wave},
\end{equation*}
where $\rPhi_{j, k}$ with $1 \leq j \leq J-1$, $0 \leq k \leq K_{j}$ are determined by the recurrence equation \eqref{eq:recurrence-jk} with
\begin{align}
	\rPhi_{j, k} &= O_{\bfGmm}^{M} (A u^{j- \alp + \nu_{\Box}})& &\hbox{ for } 0 \leq j \leq J-1, \, 1 \leq k \leq K_{j}, \label{eq:wave-main-hyp-Phi-jk} \\
	\rPhi_{(\leq j - \nu_{\Box}) j, 0} &= O_{\bfGmm}^{M} (A u^{j- \alp + \nu_{\Box}})& &\hbox{ for } 0 \leq j \leq J - 1, \label{eq:wave-main-hyp-Phi-j0-low} \\
	\rPhi_{(\geq j - \nu_{\Box}+1) j, 0} &= O_{\bfGmm}^{M} (A u^{j- \alp +\dlt_{0} + \nu_{\Box}})& &\hbox{ for } 0 \leq j \leq J - 1, \label{eq:wave-main-hyp-Phi-j0-high}
\end{align}
and $\rPhi_{j, k} = 0$ for $1 \leq j \leq J-1$, $k > K_{j}$. Moreover, the remainder obeys
\begin{equation} \label{eq:wave-main-hyp-rhoJ}
\rho_{J} = O_{\bfGmm}^{M}(A r^{-\alp+\nu_{\Box}}) \quad \hbox{ in } \calM_{\wave}.
\end{equation}

Then with $M' = M - C'_{\wave}$ and $A' = A'_{\wave}(D, A)$, we have
\begin{align}
	\rPhi_{j, k} &= O_{\bfGmm}^{M'} (A' u^{j- \alp -\dlt_{0}+ \nu_{\Box}})& &\hbox{ for } 0 \leq j \leq J-1, \, 1 \leq k \leq K_{j}, \label{eq:wave-main-Phijk} \\
	\rPhi_{(\leq j - \nu_{\Box}) j, 0} &= O_{\bfGmm}^{M'} (A' u^{j- \alp -\dlt_{0} + \nu_{\Box}})& &\hbox{ for } 0 \leq j \leq J - 1, \label{eq:wave-main-Phij0-low} \\
	\rPhi_{(\geq j - \nu_{\Box}+1) j, 0} &= O_{\bfGmm}^{M'} (A' u^{j- \alp  + \nu_{\Box}})& &\hbox{ for } 0 \leq j \leq J - 1, \label{eq:wave-main-Phij0-high}
\end{align}
Moreover, for the remainder $\rho_{J}$, the following holds.
\begin{itemize}
\item {\bf Case~1: $J \leq \min\set{J_{c}, J_{d}}-1$ and $\alp + \dlt_{0} < J+\nu_{\Box}$.}
Then we have
\begin{equation} \label{eq:wave-main-rhoJ}
\rho_{J} = O_{\bfGmm}^{M'}(A' r^{-\alp-\dlt_{0}+\nu_{\Box}}) \quad \hbox{ in } \calM_{\wave}.
\end{equation}

\item {\bf Case~2: $J \leq \min\set{J_{c}, J_{d}}-1$ and $\alp + \dlt_{0}> J+\nu_{\Box}$.}
In this case, we may extend the expansion by one more order. Define $\rPhi_{J, k}$ (for $0 \leq k \leq K_{J}$) by solving the recurrence equation \eqref{eq:recurrence-jk} and taking $\rPhi_{J, k} = 0$ for $k > K_{J}$. Then the following limits (recall \eqref{eq:change.rPhi.rcPhi}) are well-defined:
\begin{align}
\rcPhi_{J, K_{J}}(\infty) &:= \lim_{u \to \infty} \rPhi_{J, K_{J}}(u, \cdot), \label{eq:rcPhiJK-infty} \\
\rcPhi_{J, k}(\infty) &:= \lim_{u \to \infty} \left( \rPhi_{J, k}(u, \cdot) - \sum_{K = k+1}^{K_{J}} \binom{K}{k} \rcPhi_{J, K}(\infty) \right), \hbox{ for } k = K_{J}-1, \ldots, 1, \label{eq:rcPhiJk-infty} \\
\rcPhi_{(\leq J - \nu_{\Box}) J, 0}(\infty) &:= \lim_{u \to \infty} \left( \rPhi_{(\leq J - \nu_{\Box}) J, 0}(u, \cdot) - \sum_{K = 1}^{K_{J}} K \rcPhi_{(\leq J - \nu_{\Box}) J, K}(\infty) \right). \label{eq:rcPhiJ0-infty}
\end{align}
Moreover, we have
\begin{align}
	\rcPhi_{J ,k}(\infty) &= O_{\bfGmm}^{M'}(A') & & \hbox{ for all } 1 \leq k \leq K_{J}, \label{eq:wave-main-cPhiJk} \\
	\rcPhi_{(\leq J - \nu_{\Box}) J ,0}(\infty) &= O_{\bfGmm}^{M'}(A'). \label{eq:wave-main-cPhiJ0-low}
\end{align}
For $\rPhi_{J, k}$, we have
\begin{align}
	\rPhi_{J, k} &=
	\sum_{K = k+1}^{K_{J}} \binom{K}{k} \log^{K-k} u \, \rcPhi_{J, K}(\infty)
	& & \hbox{ for } 0 \leq k \leq K_{J}, \label{eq:wave-main-PhiJk.limit} \\
	&\peq + \rcPhi_{J, k}(\infty) + O_{\bfGmm}^{M'} (A' u^{J- \alp -\dlt_{0}+ \nu_{\Box}})& & \notag \\
	\rPhi_{(\leq J - \nu_{\Box}) J, 0} &=
	\sum_{K = 1}^{K_{J}} \log^{K} u \, \rcPhi_{(\leq J - \nu_{\Box}) J, K}(\infty) & & \label{eq:wave-main-PhiJ0-low.limit}\\
	&\peq + \rcPhi_{(\leq J - \nu_{\Box}) J, 0}(\infty) + O_{\bfGmm}^{M'} (A' u^{J- \alp -\dlt_{0} + \nu_{\Box}}), & & \notag \\
	\rPhi_{(\geq J - \nu_{\Box}+1) J, 0}
	&= O_{\bfGmm}^{M'} (A' u^{J- \alp  + \nu_{\Box}}). & & \label{eq:wave-main-PhiJ0-high.limit}
\end{align}
To state the estimate for the remainder
\begin{equation*}
\rho_{J+1} = \rho_{J} -  \sum_{k=0}^{K_{J}} r^{-J} \log^{k} (\tfrac{r}{u}) \rPhi_{J, k},
\end{equation*}
we divide further into two sub-cases:

\begin{itemize}
\item {\bf Case~2(a): $((\rcPhi_{J, k}(\infty))_{k=1}^{K_{J}}, \rcPhi_{(\leq J - \nu_{\Box}) J, 0}(\infty)) = 0$.}
We have,
\begin{align}
\rho_{J+1}
&= O_{\bfGmm}^{M'}(A' r^{-\alp-\dlt_{0}+\nu_{\Box}}) \quad \hbox{ in } \calM_{\wave}. \label{eq:wave-main-rhoJ+1}
\end{align}

\item {\bf Case~2(b): $((\rcPhi_{J, k}(\infty))_{k=1}^{K_{J}}, \rcPhi_{(\leq J - \nu_{\Box}) J, 0}(\infty)) \neq 0$.}
Then for any $\eta_{\mathfrak f}$ satisfying \eqref{eq:wave-main-etaf}, we have
\begin{align}
\rho_{J+1}
&= O_{\bfGmm}^{M'}(A' r^{-J-\eta_{\mathfrak f}} u^{\eta_{\mathfrak f}} \log^{K_{J}} u )
+ O_{\bfGmm}^{M'}(A' r^{-\alp-\dlt_{0}+\nu_{\Box}}) \quad \hbox{ in } \calM_{\wave}, \label{eq:wave-main-rhoJ+1-final}
\end{align}
where the implicit constants may depend on $\eta_{\mathfrak f}$.
\end{itemize}

\item{\bf Case~3: $J = \min\set{J_{c}, J_{d}}$ and $\alp + \dlt_{0}< \min\set{J_{c}-1+\eta_{c}, J_{d}-1+\eta_{d}}+\nu_{\Box}$.}
Then we have
\begin{equation} \label{eq:wave-main-rhoJ-case3}
\rho_{J} = O_{\bfGmm}^{M'}(A' r^{-\alp-\dlt_{0}+\nu_{\Box}}) \quad \hbox{ in } \calM_{\wave}.
\end{equation}

\item{\bf Case~4: $J = \min\set{J_{c}, J_{d}}$ and $\alp + \dlt_{0} > \min\set{J_{c}-1+\eta_{c}, J_{d}-1+\eta_{d}} +\nu_{\Box}$.} 
Assume also that $\alp + \dlt_{0} \neq \max\set{J_{c}-1 + \eta_{c}, J_{d} - 1 + \eta_{d}} + \nu_{\Box}$. Set $\alp_{\mathfrak f} = \min \set{J, J_{c}-1 + \eta_{c}, J_{d} - 1 + \eta_{d}} + \nu_{\Box}$. Then for any $0 < \dlt_{\mathfrak f}' < \min \{\f 12, \eta_{c}, \eta_{d}\} - \de_0$, we have
\begin{equation} \label{eq:wave-main-rhoJ-final}
\begin{aligned}
	\rho_{J}
	&= O_{\bfGmm}^{M'}(A' r^{-\alp_{\mathfrak f} + \dlt_{\mathfrak f}' + \nu_{\Box}} u^{-\dlt_{\mathfrak f}'} \max\set{u^{\alp_{\mathfrak f} - J - \nu_{\Box} } \log^{K_{J}} u, 1})
\end{aligned} \quad \hbox{ in } \calM_{\wave}.
\end{equation}
Here, $K_{J} = K_{J} (d, K_{c}, K_{d}, K_{J-1}, \calN ,\min\set{J_{c}, J_{d}})$ and $K_{J} = 0$ if $K_{c} = K_{d} = 0$. Moreover, all implicit constants may depend on $\dlt_{\mathfrak f}'$.
\end{itemize}
\end{proposition}

In each case, the control on the $r$-decay of the remainder $\rho_{J}$ improves, and sometimes the radiation field expansion (i.e., the expansion of $\Phi$ in terms of $r^{-j} \log^{k}(\tfrac{r}{u}) \rPhi_{j, k}$) in $\calM_{\wave}$ improves as well. Case~1 is when only the remainder bound improves. Case~2 is when the order of the expansion is improved as well; in Case~2(a), the iteration can continue, while in Case~2(b), it must stop. Cases~3 and 4 concern the scenario when the order of the expansion is already the maximum allowed by the coefficients and/or the data. Case~3 is when only the remainder bound improves, and Case~4 marks the end of the iteration under this scenario.

Note that Proposition~\ref{prop:wave-main} applies only when $\alp > \nu_{\Box}$, which might not be necessarily the case at the initial stage of the proof of the main upper bound theorems. Furthermore, even if $\alp_{0} > \nu_{\Box}$, where $\alp_{0}$ is the initial decay rate from \ref{hyp:sol}, we do not have the compatible bounds for $\rPhi_{j, k}$ and $\rho_{J}$ at the beginning (i.e., \eqref{eq:wave-main-hyp-Phi-jk}--\eqref{eq:wave-main-hyp-rhoJ} with $\alp = \alp_{0}$). For these reasons, we introduce the following alternative version of the wave-zone iteration lemma, to be used at the initial stage of the proof of the main theorems.
\begin{proposition}[Initial iteration lemma in the wave zone] \label{prop:wave-0}
Let $0 \leq \nu_{0} < \frac{1}{2}$, $0 < \eta_{w} < \min \set{\frac{1}{2}, \alp_{d} - \nu_{\Box}}$ and $\alp < \nu_{\Box} + \eta_{w} - \dlt_{0}$; if $\calN \neq 0$, assume also
\begin{equation} \label{eq:alp0-dlt0-nonlin}
\alp_{0} \geq \alp_{\calN} + 2 \dlt_{0}.
\end{equation}
Then there exist a positive integer $C_{\wave, 0}$ and a function $A'_{\wave, 0}$ satisfying \eqref{eq:A'} such that the following holds. Assume that
\begin{equation} \label{eq:wave-0-hyp-med}
\phi = O_{\bfGmm}^{M}(A u^{-\alp}) \quad \hbox{ in } \calM_{\med},
\end{equation}
where $C_{\wave, 0} < M \leq \min\set{M_{0}, M_{c}}$ and $A > 0$. Assume one of the following in $\calM_{\wave}$:
\begin{itemize}
\item {\bf Assumption~1:} We have
\begin{equation*}
	\Phi = O_{\bfGmm}^{M}(A r^{\nu_{0}} u^{-\nu_{0} -\alp+\nu_{\Box}}); \quad \hbox{or}
\end{equation*}
\item {\bf Assumption~2:} We have
\begin{equation*}
	\Phi = \rPhi_{0, 0} + \rho_{1},
\end{equation*}
where the Friedlander radiation field $\rPhi_{0, 0}$ obeys
\begin{equation} \label{eq:wave-0-hyp-Phi-00}
	\rPhi_{0, 0} = O_{\bfGmm}^{M}(A u^{-\alp+\dlt_{0}+\nu_{\Box}})
\end{equation}
while the remainder $\rho_{1}$ obeys $\rho_{1}(u, r, \tht) \to 0$ as $r \to \infty$ for each $u \geq 1$, $\tht \in \bbS^{d-1}$, as well as
\begin{equation} \label{eq:wave-0-hyp-rho-1}
	\rho_{1} = O_{\bfGmm}^{M}(A \min\set{r^{-\alp + \nu_{\Box}}, u^{-\alp+\nu_{\Box}}}).
\end{equation}
\end{itemize}

Then, under either assumption, with $M' = M - C_{\wave, 0}$ and $A' = A'_{\wave, 0}(D, A)$, we have $\Phi = \rPhi_{0, 0} + \rho_{1}$ where, in $\calM_{\wave}$,
\begin{align}
	\rPhi_{0, 0} &= O_{\bfGmm}^{M'}(A' u^{-\alp+\nu_{\Box}}) , \label{eq:wave-0-Phi-00}\\
	\rho_{1} &= O_{\bfGmm}^{M'}(A' r^{-\eta_{w}}u^{-\alp-\dlt_{0}+\eta_{w}+\nu_{\Box}}). \label{eq:wave-0-rho-1}
\end{align}
\end{proposition}

Note that it is always possible to choose $\eta_{w} > \dlt_{0}$, so that Proposition~\ref{prop:wave-0} is always applicable when $\alp \leq \nu_{\Box}$ (in fact, even when $\alp$ exceeds $\nu_{\Box}$ by a bit). Indeed, when $\alp_{d} - \nu_{\Box} > \frac{1}{2}$, this is obvious in view of \eqref{eq:dlt0-wave}. When $\alp_{d} - \nu_{\Box} \leq \frac{1}{2}$, we have $J_{d} = 1$ and $\eta_{d} = \alp_{d} - \nu_{\Box}$, so the condition on $\eta_{w}$ becomes $0 < \eta_{w} < \eta_{d}$, which is again acceptable.

Proposition~\ref{prop:wave-0} has been formulated with two possible assumptions in order to easily initiate the iteration in Section~\ref{sec:upper} (more specifically, Assumption~1 is needed only for beginning the iteration). We also point out that \eqref{eq:alp0-dlt0-nonlin} requires the initial decay rate $\alp_{0}$ to exceed $\alp_{\calN}$, not $\alp$ (cf.~\eqref{eq:alpi-dlt0-nonlin}). In our proofs of the main theorems, we will stop using Proposition~\ref{prop:wave-0} once $\alp$ is sufficiently larger than $\nu_{\Box}$ and pass to Proposition~\ref{prop:wave-main}.

The remainder of this section is devoted to the proof of Propositions~\ref{prop:wave-main} and \ref{prop:wave-0}.

\subsection{Recurrence equations for the higher radiation fields}  \label{subsec:recurrence}
In Section~\ref{subsec:recurrence-formal}, we derived the recurrence equations
\begin{equation*} \tag{\ref{eq:recurrence-jk}}
\begin{aligned}
	\rd_{u} \rPhi_{j, k}&=
	(k+1) u^{-1} \rPhi_{j, k+1}
	+ \frac{k+1}{j} \left( \rd_{u} \rPhi_{j, k+1} - (k+2) u^{-1} \rPhi_{j, k+2} \right)
\\
&\peq
- \frac{1}{2j}\Big( (j-1)j - \frac{(d-1)(d-3)}{4} + \rslap \Big) \rPhi_{j-1, k}   \\
&\peq
+ \frac{1}{2j} (j-1)(k+1) \rPhi_{j-1, k+1} - \frac{1}{2j}(k+2)(k+1) \rPhi_{j-1, k+2}
+ \frac{1}{2j} \rF_{j, k},
\end{aligned}
\end{equation*}
where $\rF_{j, k}$ is the coefficient of $r^{-j-1} \log^{k} (\tfrac{r}{u})$ in the formal expansion in $\calM_{\wave}$ of the right-hand side of
\begin{equation*} \tag{\ref{eq:Phi-eq-0}}
	 Q_{0} \Phi = - \bfh^{\alp \bt} \nb_{\alp} \rd_{\bt} \Phi - \bfC^{\alp} \rd_{\alp} \Phi - W \Phi + r^{\frac{d-1}{2}} \calN(r^{-\frac{d-1}{2}} \Phi) + r^{\frac{d-1}{2}} f.
\end{equation*}

The first goal of this subsection is to prove the following properties of $\rF_{j, k}$, which were assumed in Section~\ref{subsec:recurrence-formal}:
\begin{lemma} \label{lem:recurrence-forcing-basic}
For $1 \leq j \leq \min\set{J_{c}, J_{d}} - 1$ and $k \geq 0$, the coefficients $\rF_{j, k}$ have the following properties:
\begin{enumerate}
\item $\rF_{j, k}$ is determined from $\rPhi_{j', k'}$ with only $j' \leq j-1$;
\item If there exists $K > 0$ such that $\rPhi_{j', k'} = 0$ for all $j' \leq j-1$ and $k' > K$, then $\rF_{j, k} = 0$ for all $k$ greater than some finite number $\td{K}_{j}$ that depends on $d$, $K_{c}$, $\calN$, $j$ and $K$.
\end{enumerate}
\end{lemma}

We also prove the following quantitative bounds on $\rF_{j, k}$.

\begin{lemma} \label{lem:recurrence-forcing}
Assume $\rPhi_{j, k} = O_{\bfGmm}^{M}(A u^{j-\alp + \nu_{\Box}})$ for $j' \leq j-1$. When $\calN \neq 0$, assume also
\begin{equation} \label{eq:alp-dlt0-nonlin}
	\alp \geq \alp_{\calN} + 2 \dlt_{0}.
\end{equation}
Then there exists $A' = A'(D, A)$ such that, for $1 \leq j \leq \min\set{J_{c}, J_{d}} - 1$ and $k \geq 0$,
\begin{equation*}
\rF_{j, k} = O_{\bfGmm}^{M-2}(A' u^{j-1-\alp-\dlt_{0} + \nu_{\Box}}) + \rf_{j+1+\nu_{\Box}, k},
\end{equation*}
where $A'$ and the implicit constant in $O_{\bfGmm}^{M-C'}$ may depend on the coefficients, $\calN$, $\dlt_{0}$, $j$ and $k$. Moreover, $A'$ satisfies \eqref{eq:A'}.
\end{lemma}

\begin{proof} [Proof of Lemmas~\ref{lem:recurrence-forcing-basic} and \ref{lem:recurrence-forcing}]
The coefficient $\rF_{j, k}$ consists of
\begin{equation} \label{eq:Fjk}
\rF_{j, k} = \rL_{j, k} + \rN_{j, k} + \rf_{j+1+\nu_{\Box}, k},
\end{equation}
where $\rL_{j, k}$ and $\rN_{j, k}$ are the contributions of the linear and nonlinear part, respectively, of the right-hand side of \eqref{eq:Phi-eq-0}. To prove the lemmas, we need to study the structure of $\rL_{j, k}$ and $\rN_{j, k}$.

\pfstep{Structure of $\rL_{j, k}$}
We first describe the linear contribution $\rL_{j, k}$. Recall from Lemma~\ref{lem:conj-wave} that the coefficients of the linear part of the right-hand side of \eqref{eq:Phi-eq-0} admit the expansions \eqref{eq:wave-h-uu-phg}--\eqref{eq:wave-W-phg}. We also note the following computation for each monomial:
\begin{align}
	\rd_{u} \left(r^{-(j+\nu)} \log^{k} (\tfrac{r}{u}) \rPhi_{j, k} \right)
	&= r^{-(j+\nu)} \log^{k} (\tfrac{r}{u}) \rd_{u} \rPhi_{j, k}
	- k r^{-(j+\nu)} \log^{k-1} (\tfrac{r}{u}) u^{-1} \rPhi_{j, k}, \label{eq:monomial-du} \\
	\rd_{r} \left(r^{-(j+\nu)} \log^{k} (\tfrac{r}{u}) \rPhi_{j, k} \right)
	&= -(j+\nu) r^{-(j+\nu+1)} \log^{k} (\tfrac{r}{u}) \rPhi_{j, k}
	+ k r^{-(j+\nu+1)} \log^{k-1} (\tfrac{r}{u}) \rPhi_{j, k},  \label{eq:monomial-dr}\\
	\rd_{u}^{2} \left(r^{-(j+\nu)} \log^{k} (\tfrac{r}{u}) \rPhi_{j, k}\right)
	&= r^{-(j+\nu)} \log^{k} (\tfrac{r}{u}) \rd_{u}^{2} \rPhi_{j, k}
	- 2k r^{-(j+\nu)} \log^{k-1} (\tfrac{r}{u}) u^{-1} \rd_{u} \rPhi_{j, k} \label{eq:monomial-duu} \\
	&\peq
	+ k r^{-(j+\nu)} \log^{k-1} (\tfrac{r}{u}) u^{-2} \rPhi_{j, k}
	+ k(k-1) r^{-(j+\nu)} \log^{k-2} (\tfrac{r}{u}) u^{-2} \rPhi_{j, k}, \notag \\
	\rd_{u} \rd_{r} \left(r^{-(j+\nu)} \log^{k} (\tfrac{r}{u}) \rPhi_{j, k}\right)
	&= - (j+\nu) r^{-(j+\nu+1)} \log^{k} (\tfrac{r}{u}) \rd_{u} \rPhi_{j, k} \label{eq:monomial-dur} \\
	&\peq
	+ k r^{-(j+\nu+1)} \log^{k-1} (\tfrac{r}{u}) \rd_{u} \rPhi_{j, k} \notag \\
	&\peq
	+ (j+\nu) k r^{-(j+\nu+1)} \log^{k-1} (\tfrac{r}{u}) u^{-1} \rPhi_{j, k} \notag \\
	&\peq
	- k(k-1) r^{-(j+\nu+1)} \log^{k-2} (\tfrac{r}{u}) u^{-1} \rPhi_{j, k}, \notag \\
	\rd_{r}^{2} \left(r^{-(j+\nu)} \log^{k} (\tfrac{r}{u}) \rPhi_{j, k}\right)
	&= (j+\nu)(j+\nu+1) r^{-(j+\nu+2)} \log^{k} (\tfrac{r}{u}) \rPhi_{j, k} \label{eq:monomial-drr} \\
	&\peq
	- (2j+2\nu+1) k r^{-(j+\nu+2)} \log^{k-1} (\tfrac{r}{u}) \rPhi_{j, k} \notag \\
	&\peq
	+ k (k-1) r^{-(j+\nu+2)} \log^{k-2} (\tfrac{r}{u}) \rPhi_{j, k}. \notag
 \end{align}
Recall also the non-zero Christoffel symbols associated with $\bfm$ in the $(u, r, \tht)$-coordinates from \eqref{eq:bondi-Gmm}. After a straightforward computation, we see that $\rL_{j, k}$ takes the following form:
\begin{align*}
	\rL_{j, k} &= \sum_{j', k'} \Big[
	- \rh^{uu}_{j - j' + 1, k - k'} \rd_{u}^{2} \rPhi_{j', k'}
	+ 2 k' \rh^{uu}_{j - j'+1, k - k' + 1} u^{-1} \rd_{u} \rPhi_{j', k'} \\
	&\peq \phantom{\sum_{j', k'} \Big(}
	- k' \rh^{uu}_{j - j' + 1, k - k' + 1} u^{-2} \rPhi_{j', k'}
	- k' (k'-1) \rh^{uu}_{j - j' + 1, k - k' + 2} u^{-2} \rPhi_{j', k'} \\
	&\peq \phantom{\sum_{j', k'} \Big(}
	+ j' \rh^{ur}_{j - j', k - k'} \rd_{u} \rPhi_{j', k'}
	- k' \rh^{ur}_{j - j', k - k' + 1} \rd_{u} \rPhi_{j', k'}  \\
	&\peq \phantom{\sum_{j', k'} \Big(}
	- j' k' \rh^{ur}_{j - j', k - k' + 1} u^{-1} \rPhi_{j', k'}
	+ k' (k'-1) \rh^{ur}_{j - j', k - k' + 2} u^{-1} \rPhi_{j', k'} \\
	&\peq \phantom{\sum_{j', k'} \Big(}
	- \rsh^{uA}_{j - j', k - k'} \rd_{\tht^{A}} \rd_{u} \rPhi_{j',  k'}
	+ k' \rsh^{uA}_{j - j', k - k' + 1} \rd_{\tht^{A}} u^{-1} \rPhi_{j',  k'} \\
	&\peq \phantom{\sum_{j', k'} \Big (}
	- j' (j'+1) \rh^{rr}_{j - j' - 1, k - k'} \rPhi_{j', k'} \\
	&\peq \phantom{\sum_{j', k'} \Big (}
	+ (2 j'+1) k' \rh^{rr}_{j - j' - 1, k - k' + 1} \rPhi_{j', k'}
	- k' (k'+1) \rh^{rr}_{j - j' - 1, k - k' + 2} \rPhi_{j', k'} \\
	&\peq \phantom{\sum_{j', k'} \Big (}
	+ (j'+1) \rsh^{r A}_{j - j' - 1, k - k'} \rd_{\tht^{A}} \rPhi_{j', k'}
	- k' \rsh^{r A}_{j - j' - 1, k - k' + 1} \rd_{\tht^{A}} \rPhi_{j', k'}
	+ \rsh^{A B}_{j - j' - 1, k - k'} \rsnb_{\rd_{\tht^{A}}} \rd_{\tht^{B}} \rPhi_{j', k'}\\
	&\peq \phantom{\sum_{j', k'} \Big (}
	+ \rsh^{A B}_{j - j', k - k'} \rsgmm_{AB} \rd_{u} \rPhi_{j', k'}
	- k' \rsh^{A B}_{j - j', k - k'+1} \rsgmm_{AB} u^{-1} \rPhi_{j', k'} \\
	&\peq \phantom{\sum_{j', k'} \Big (}
	+ j'  \rsh^{A B}_{j - j' - 1, k - k'}\rsgmm_{AB} \rPhi_{j', k'}
	- k'  \rsh^{A B}_{j - j' - 1, k - k'+1}\rsgmm_{AB} \rPhi_{j', k'} \\
	&\peq \phantom{\sum_{j', k'} \Big (}
	- \rbfC^{u}_{j - j' + 1, k - k'} \rd_{u} \rPhi_{j', k'}
	+ k' \rbfC^{u}_{j - j'+1, k - k' + 1} u^{-1} \rPhi_{j', k'}  \\
	&\peq \phantom{\sum_{j', k'} \Big (}
	+ j' \rbfC^{r}_{j - j', k - k'} \rPhi_{j', k'}
	- k' \rbfC^{r}_{j - j', k - k' +1} \rPhi_{j', k'}
	- \rsbfC^{A}_{j - j', k - k'} \rd_{\tht^{A}} \rPhi_{j', k'}
	- \rW_{j - j' +1, k-k'} \rPhi_{j', k'} \Big] .
\end{align*}
While the expression is long, there are only a few properties that we need to observe. First, recall from Section~\ref{subsec:conj-wave} that in general
$\rh^{\alp \bt}_{j'', k''} = 0$ for $j'' < 0$, $\rbfC^{\alp}_{j'', k''} = 0$ for $j'' < 1$ and $\rW_{j'', k''} = 0$ for $j'' < 2$
while for coefficients in front of $\rd_{u} \rPhi_{j', k'}$ or $\rd_{u}^{2} \rPhi_{j', k'}$, we have further vanishing properties
\begin{equation*}
\rh^{uu}_{j'', k''} = 0 \hbox{ for } j'' < 2, \quad
\rh^{ur}_{j'', k''}, \rsh^{uA}_{j'', k''}, \rh^{AB}_{j'', k''} \rsgmm_{AB} = 0 \hbox{ for } j'' < 1, \quad
\rbfC^{u}_{j'', k''} = 0 \hbox{ for } j'' < 2.
\end{equation*}
By inspecting the preceding expression, it follows that $\rL_{j, k}$ depends only on $\rPhi_{j', k'}$ with $j' \leq j-1$.

Second, since $\rh^{\alp \bt}_{j'', k''} = 0$ for $k'' > K_{c}$ and $\rbfC^{\alp}_{j'', k''}, \rW_{j'', k''} = 0$ for $k'' > K_{c}'$, it follows that if $\rPhi_{j', k'} = 0$ for all $j' \leq j-1$ and $k' > K$, then $\rL_{j, k} = 0$ for $k > K_{c}' + K$. Hence, the contribution of $\rL_{j, k}$ is acceptable for Lemma~\ref{lem:recurrence-forcing-basic}.

Finally, the size of each term can be estimated rather simply from our notation. Recall from \eqref{eq:wave-hjk-phg}, \eqref{eq:wave-Cjk-phg} and \eqref{eq:wave-Wjk-phg} that the $u$-decay of a coefficient depends only on its type (i.e., whether it is $\rh$, $\rbfC$ or $\rW$) and its $j$-index. By checking the sum of the $j$-index of the coefficient, the $j$-index of the coefficient $\rPhi_{j', k'}$ (i.e., $j'$), the number of $u^{-1}$'s applied and the number of $\rd_{u}$'s applied, it follows that
\begin{equation} \label{eq:recurrence-forcing-L}
	\rL_{j, k} = O_{\bfGmm}^{M-2}(A u^{j-1-\alp-\dlt_{c}+\nu_{\Box}}),
\end{equation}
where the loss $2$ comes from $\rd_{u}^{2}$ and $\rsnb_{\rd_{\tht^{A}}} \rd_{\tht^{B}}$. This estimate is acceptable for Lemma~\ref{lem:recurrence-forcing}.

\pfstep{Structure of $\rN_{j, k}$}
Recall the discussion of the structure of $\rN^{<J}_{j, k}$ in Section~\ref{subsec:nonlin-est}; observe that we simply have $\rN_{j, k} = \rN^{<J}_{j, k}$, where the RHS is computed with $\Phi_{<J}$ of the form \eqref{eq:Phi<J-formal} with $J > j$ and $K$ sufficiently large. By Lemma~\ref{lem:rNjk}.(1) and (2), it follows that $\rN_{j, k}$ depends only on $\rPhi_{j', k'}$ with $j' \leq j-1$, and that $\rN_{j, k} = 0$ if $\rPhi_{j', k'} = 0$ for all $j' \leq j-1$, $k' > K$ and $k > K_{c} + \frac{j+1+\nu_{\Box}}{\nu_{\Box}} K$. This concludes the proof of Lemma~\ref{lem:recurrence-forcing-basic}

Finally, by the assumption $\alp \geq \alp_{\calN} + 2 \dlt_{0}$ and Definition~\ref{def:alp-N}.(4) (with $\alp_{\calN}' = \alp_{\calN}+\dlt_{0}$), we have
\begin{equation} \label{eq:recurrence-forcing-N}
	\rN_{j, k} = O_{\bfGmm}^{M-2}(A_{\calN}(A) u^{j-1-\alp-\dlt_{0} + \nu_{\Box}}),
\end{equation}
which is acceptable for Lemma~\ref{lem:recurrence-forcing}. \qedhere
\end{proof}

In the following lemma, we use Lemma~\ref{lem:recurrence-forcing-basic} to define the maximum possible order $K_{j}$ of the logarithmic factors among terms of the form $r^{-j} \log^{k}(\tfrac{r}{u}) \rPhi_{j, k}$.
\begin{lemma} \label{lem:Kj-def}
There exists a sequence $K_{0} \leq K_{1} \leq K_{2} \leq \ldots \leq K_{\min\set{J_{c}, J_{d}}-1}$, depending only on $d$, $K_{c}$, $K_{d}$, $\calN$ and $\min\set{J_{c}, J_{d}}-1$, such that
\begin{enumerate}
\item $K_{0} = 0$.
\item $K_{j} \geq K_{d}$ for $j \geq 1$.
\item If $\rPhi_{j', k'} = 0$ for all $j' \leq j-1$ and $k' > K_{j-1}$, then $\rF_{j, k} = 0$ for all $k > K_{j}$.
\end{enumerate}
Moreover, no logarithms appear if there are none in the equation and the data, i.e.,
\begin{equation} \label{eq:no-log}
	K_{c} = K_{d} = 0 \imp K_{j} = 0 \quad \hbox{ for all } j.
\end{equation}
\end{lemma}
We omit the proof of this lemma, which is a simple application of Lemma~\ref{lem:recurrence-forcing-basic}. We note that, from the proof of  Lemmas~\ref{lem:recurrence-forcing-basic}--\ref{lem:recurrence-forcing}, we may arrange so that $K_{1} \leq \max\set{K_{c}', K_{d}}$.
\begin{remark} \label{rem:Kj-rem}
In fact, using the notation introduced in the proof of Lemmas~\ref{lem:recurrence-forcing-basic}--\ref{lem:recurrence-forcing}, the second property can be strengthened to: If $\rPhi_{j', k'} = 0$ for all $j' \leq j-1$ and $k' > K_{j-1}$, then every term
in $\rL_{j, k}$ and $\rN_{j, k}$ is zero for all $k > K_{j}$. This observation will be useful below.
\end{remark}

At this point, {\bf we define $0 = K_{0} \leq K_{1} \leq \ldots \leq K_{\min\set{J_{c}, J_{d}}-1}$ using Lemma~\ref{lem:Kj-def} and fix them for the remainder of the section.}

We are now ready to state and prove the main quantitative result on the higher radiation fields $\rPhi_{j, k}$. 
\begin{proposition} \label{prop:recurrence}
Let $0 \leq J - 1 \leq \min\set{J_{c}, J_{d}} - 1$. Assume that
\begin{align}
	\rPhi_{j, k} &= O_{\bfGmm}^{M} (A u^{j- \alp + \nu_{\Box}})& &\hbox{ for } 0 \leq j \leq J-1 \hbox{ and } 0 \leq k \leq K_{j}, \label{eq:recurrence-hyp-falloff} \\
	\rPhi_{j, k} &= 0& &\hbox{ for } 0 \leq j \leq J-1 \hbox{ and } k > K_{j}, \label{eq:recurrence-hyp-no-logs}
\end{align}
where
\begin{equation} \label{eq:recurrence-hyp-j}
	J-1+\nu_{\Box} < \alp < J+ \nu_{\Box}.
\end{equation}
Assume also that $\rPhi_{j, k}$ solves the recurrence equation \eqref{eq:recurrence-jk} for $0 \leq j \leq J-1$.
When $\calN \neq 0$, assume also \eqref{eq:alp-dlt0-nonlin}. Then there exists $C' > 0$, which depends on $J$, $(K_{j})_{j=0}^{\min\set{J_{c}, J_{d}}-1}$ and the $C'$ in Lemma~\ref{lem:recurrence-forcing}, and $A' = A'(D, A)$, which satisfies \eqref{eq:A'} and which depends on $J$, $(K_{j})_{j=0}^{\min\set{J_{c}, J_{d}}-1}$ and the $A'$ in Lemma~\ref{lem:recurrence-forcing}, such that following statements hold.
\begin{enumerate}
\item For $j \leq J-1$, we have
\begin{align*}
	\rPhi_{j, k} &= O_{\bfGmm}^{M - C'} (A' u^{j- \alp -\dlt_{0} + \nu_{\Box}})& &\hbox{ for } 1 \leq j \leq J-1, \, 1 \leq k \leq K_{j}, \\
	\rPhi_{(\leq j - \nu_{\Box}) j, 0} &= O_{\bfGmm}^{M - C'} (A' u^{j- \alp - \dlt_{0} + \nu_{\Box}}).& & \hbox{ for } 1 \leq j \leq J-1, \\
	\rPhi_{(\geq j - \nu_{\Box}+1) j, 0} &= O_{\bfGmm}^{M} (A u^{j- \alp + \nu_{\Box}}),& & \hbox{ for } 1 \leq j \leq J-1.
\end{align*}

\item Assume that $J \leq \min\set{J_{c}, J_{d}} - 1$ and $J + \nu_{\Box} > \alp + \dlt_{0}$. Define $\rPhi_{J, k}$ by requiring that it solves the recurrence equation \eqref{eq:recurrence-jk} and also that $\rPhi_{J, k} = 0$ for $k > K_{J}$. Then
\begin{align*}
	\rPhi_{J, k}
	&= O_{\bfGmm}^{M - C'} (A' u^{J- \alp -\dlt_{0} + \nu_{\Box}})& &\hbox{ for } 1 \leq k \leq K_{J}, \\
	\rPhi_{(\leq J - \nu_{\Box}) J, 0}
	&= O_{\bfGmm}^{M - C'} (A' u^{J- \alp - \dlt_{0} + \nu_{\Box}}),& & \\
	\rPhi_{(\geq J - \nu_{\Box}+1) J, 0}
	&= O_{\bfGmm}^{M - C'} (A' u^{J- \alp + \nu_{\Box}}).& &
\end{align*}
\item Assume that $J \leq \min\set{J_{c}, J_{d}} - 1$ and $J + \nu_{\Box} < \alp + \dlt_{0}$. Define $\rPhi_{J, k}$ by requiring that it solves the recurrence equation \eqref{eq:recurrence-jk} and also that $\rPhi_{J, k} = 0$ for $k > K_{J}$.

Then $\rcPhi_{J, k}(\infty)$ for $1 \leq k \leq K_{J}$ and $\rcPhi_{(\leq J - \nu_{\Box}) J, 0}(\infty)$ are well-defined according to \eqref{eq:rcPhiJK-infty}--\eqref{eq:rcPhiJ0-infty}, as uniform limits of functions of $\tht$ as $u \to \infty$. The limits $\rcPhi_{J, k}(\infty)$ and $\rcPhi_{(\leq J - \nu_{\Box}) J, 0}(\infty)$ satisfy (as functions of $\tht$)
\begin{align*}
	\rcPhi_{J ,k} (\infty) &= O_{\bfGmm}^{M-C'}(A') & & \hbox{ for all } 1 \leq k \leq K_{J}, \\
	\rcPhi_{(\leq J - \nu_{\Box}) J ,0}(\infty) &= O_{\bfGmm}^{M-C'}(A').
\end{align*}

Moreover, we have
\begin{align*}
	\rPhi_{J, k}
	&= \sum_{K = k}^{K_{J}} \binom{K}{k} \log^{K-k} u \rcPhi_{J, K}(\infty) + O_{\bfGmm}^{M - C'} (A' u^{J- \alp -\dlt_{0} + \nu_{\Box}})& &\hbox{ for } 1 \leq k \leq K_{J},
\end{align*}
and for $k = 0$, we have
\begin{align*}
	\rPhi_{(\leq J - \nu_{\Box}) J, 0}
	&= \sum_{K = 0}^{K_{J}} \log^{K} u \rcPhi_{(\leq J - \nu_{\Box}) J, K}(\infty)
	 + O_{\bfGmm}^{M - C'} (A' u^{J- \alp - \dlt_{0} + \nu_{\Box}}), \\
	\rPhi_{(\geq J - \nu_{\Box}+1) J, 0}
	&= O_{\bfGmm}^{M - C'} (A' u^{J- \alp + \nu_{\Box}}).
\end{align*}
\end{enumerate}
\end{proposition}
\begin{remark}
As written the two cases in (2) and (3) are non-exhaustive. But we shall use the freedom to choose the parameter $\dlt_{0}$ to avoid falling to the case $j - \alp - \dlt_{0} + \nu_{\Box} = 0$. Similarly, by perturbing the parameters a bit, we shall always make sure that $\alp - \nu_{\Box} \not \in \bbZ$ within our iteration procedure so that there exists $j$ such that \eqref{eq:recurrence-hyp-j} holds.
\end{remark}

\begin{remark}[Regular fall-off vs.~Newman--Penrose cancellation for the recurrence equations] \label{rem:np-exp}
Recall the recurrence equation \eqref{eq:recurrence-jk} for $\rPhi_{j, k}$. Lemma~\ref{lem:recurrence-forcing} says that $\rF_{j, k}$ -- which is the contribution of all non-Minkowskian parts of the equation $ \calP \phi = \calN(\phi) + f$ -- decays better than other terms occurring in $\rd_{u} \rPhi_{j, k}$, such as $\rPhi_{j-1, k}$. By induction, we expect that if $\rPhi_{j, k}$ decays as $u \to \infty$, then its decay rate should be $u^{j}$ times that of $\rPhi_{0}$. We shall call this the \emph{regular fall-off} of $\rPhi_{j, k}$. This expectation is the reason behind assuming \eqref{eq:recurrence-hyp-falloff} in Proposition~\ref{prop:recurrence}. Accordingly, we dub \eqref{eq:recurrence-hyp-falloff} the \emph{regular fall-off condition with decay exponent $\alp$}.

In this light, the most nontrivial assertion in Proposition~\ref{prop:recurrence}.(1) is that the $u$-decay of $\rPhi_{j, k}$ for $k \geq 1$, as well as $\rPhi_{(\ell) j, 0}$ for $0 \leq \ell \leq j - \nu_{\Box}$ always \emph{improves} compared to the regular fall-off. The former is simply because, under our assumptions, new logarithmic terms may dynamically be generated only through non-Minkowskian parts of $\calP \phi = \calN(\phi) + f$ (recall \eqref{eq:no-log}), which have better decay. The latter is a consequence of the remarkable fact that the only term in the equation for $\rd_{u} \rPhi_{(\ell) j, k}$ (i.e., $\bbS_{(\ell)}$ of the RHS of \eqref{eq:recurrence-jk}) is
\begin{equation} \label{eq:np-algebra}
- \frac{1}{2j}\Big( (j-1)j - \frac{(d-1)(d-3)}{4} + \ell(\ell+d-1) \Big) \rPhi_{(\ell) j-1, k}
=  -\frac{1}{2j} (j - \nu_{\Box} - \ell)(j+\nu_{\Box}+\ell-1) \rPhi_{(\ell) j-1, k}
\end{equation}
which \emph{vanishes} when $j = \nu_{\Box} + \ell$! By induction, $\rPhi_{(\ell) j, k}$ is improved for $j \geq \nu_{\Box} + \ell$. This cancellation is special to $d \geq 3$ and odd, and was also the ultimate reason behind the conservation of the Newman--Penrose constant in Section~\ref{sec:price}. For this reason, we call it the \emph{Newman--Penrose cancellation}. The same cancellation is responsible for the improved decay for $\rPhi_{J, k}$ and $\rPhi_{(\leq J - \nu_{\Box}) J, 0}$ in Statement~(2), as well as the convergence of the limits $\rcPhi_{J, k}(\infty)$ and $\rcPhi_{(\leq J - \nu_{\Box}) J, 0}(\infty)$ in Statement~(3). See also Remark~\ref{rem:np-exp-error} below.
\end{remark}

\begin{proof}
In the proof, we adopt the following conventions:
\begin{itemize}
\item $C'$ refers to a positive number that may differ from line to line, which may depend on $J$, $(K_{j})_{j=0}^{\min\set{J_{c}, J_{d}}-1}$ and the $C'$ in Lemma~\ref{lem:recurrence-forcing}. We furthermore assume that each $C'$ is always greater than or equal to all $C'$'s that occurred before in the proof.
 \item $A' = A'(D, A)$, where $A'(\cdot, \cdot)$ refers to a positive function that obeys \eqref{eq:A'} and may differ from line to line. We furthermore assume that each $A'(\cdot, \cdot)$ is always greater than or equal to (pointwisely) $A'(\cdot, \cdot)$'s that occurred before in the proof.
\end{itemize}
Moreover, we will freely use the following to remove $\bbS_{(\ell)}$ (up to losing a constant depending on $\ell$):
\begin{equation} \label{eq:S-ell-Linfty}
\nrm{\bbS_{(\ell)} g}_{L^{\infty}_{\tht}} \aleq_{\ell}\nrm{g}_{L^{\infty}_{\tht}},
\end{equation}
Indeed,
\begin{equation*}
	\nrm{\bbS_{(\ell)} g}_{L^{\infty}_{\tht}}
	\aleq \nrm{\bbS_{(\ell)} g}_{H^{\frac{d+1}{2}}_{\tht}}
	\aleq (1+\ell)^{\frac{d-1}{2}+1} \nrm{\bbS_{(\ell)} g}_{L^{2}_{\tht}}
	\aleq (1+\ell)^{\frac{d-1}{2}+1} \nrm{g}_{L^{2}_{\tht}}
	\aleq (1+\ell)^{\frac{d-1}{2}+1} \nrm{g}_{L^{\infty}_{\tht}}.
\end{equation*}

\pfstep{Step~1: Proof of (1)}
We set up an induction on $(j, k)$ ordered according to the size of $r^{-j} \log^{k} (\frac{r}{u})$ (with $u$ fixed) in the following way:
\begin{equation*}
	(1, K_{1}) < (1, K_{1}-1) < \ldots (1, 0) < (2, K_{2}) < (2, 0) < \ldots < (J-1, 0).
\end{equation*}
Let $(j, k)$ be one of the above indices. For the purpose of induction, we assume that the desired estimate $\rPhi_{j', k'} = O_{\bfGmm}^{M-C'}(A' u^{j'- \alp - \dlt_{0}+\nu_{\Box}})$ holds for all $(j', k') < (j, k)$, where we write $\rPhi_{(\leq j' - \nu_{\Box}) j', 0}$ instead when $k' = 0$. We divide further into two cases:

\pfstep{Step~1(a): $k \neq 0$}
Consider $\rd_{u} \rPhi_{j, k}$ given by the recurrence equation \eqref{eq:recurrence-jk}. By Lemma~\ref{lem:recurrence-forcing} and the induction hypothesis, the following estimates hold: For $\abs{I} \leq M - C'$, we have
\begin{equation*}
	\abs{\bfGmm^{I} \rd_{u} \rPhi_{j, k}} \aleq A' u^{j-1-\alp-\dlt_{0}+\nu_{\Box}}.
\end{equation*}
Here, the key point is that $\rd_{u} \rPhi_{j, k} - \frac{1}{2j} \rF_{j, k}$ only involves $\rPhi_{j', k'}$ with $(j', k') < (j, k)$ and $k' \neq 0$, which already exhibits an improved fall-off rate (by $-\dlt_{0}$). By a simple induction on $\abs{I}$ based on the fact that $[\rd_{u}, \bfGmm]$ is either $\rd_{u}$ or $0$, we also obtain
\begin{equation*}
	\abs{\rd_{u} \bfGmm^{I} \rPhi_{j, k}} \aleq A' u^{j-1-\alp-\dlt_{0}+\nu_{\Box}}.
\end{equation*}
Since $j \leq J-1$, by our hypothesis, $\bfGmm^{I} \rPhi_{j, k} \to 0$ as $u \to \infty$. Therefore, by the fundamental theorem of calculus and the preceding bound, the desired estimate for $\rPhi_{j, k}$ follows.

\pfstep{Step~1(b): $k = 0$}
Consider $\rd_{u} \rPhi_{(\leq j - \nu_{\Box}) j, 0}$ given by the recurrence equation \eqref{eq:recurrence-jk}. By Lemma~\ref{lem:recurrence-forcing} and the induction hypothesis, the following estimates hold: For $\abs{I} \leq M - C'$, we have
\begin{equation*}
	\abs*{\bfGmm^{I} \left(\rd_{u} \rPhi_{(\leq j-\nu_{\Box}) j, 0} + \frac{1}{2j} \sum_{\ell=0}^{j-\nu_{\Box}} (j - \nu_{\Box} - \ell)(j+\nu_{\Box}+\ell-1) \rPhi_{(\ell) j-1, 0} \right)} \aleq A' u^{j-1-\alp-\dlt_{0}+\nu_{\Box}}.
\end{equation*}
Here, we used \eqref{eq:np-algebra} to rewrite $((j-1)j - (\nu_{\Box}-1)\nu_{\Box}+\rslap) \rPhi_{(\ell) j-1, 0}$. The key observation here is that, thanks to factor $(j - \nu_{\Box} - \ell)$, the summand with $\ell = j - \nu_{\Box}$ on the left-hand side vanishes! Using the induction hypothesis on $\rPhi_{(\leq j-1-\nu_{\Box})j-1, 0}$, it follows that
\begin{equation*}
	\abs*{\bfGmm^{I} \rd_{u} \rPhi_{(\leq j-\nu_{\Box}) j, 0}} \aleq A' u^{j-1-\alp-\dlt_{0}+\nu_{\Box}}.
\end{equation*}
At this point, we may proceed as in Step~1(a) and conclude the desired estimate for $\rPhi_{(\leq j-\nu_{\Box}) j, 0}$.

Finally, we remark that the desired bound for $\rPhi_{(\geq j - \nu_{\Box}+1)j, 0} = (1 - \sum_{\ell =0}^{j-\nu_{\Box}} \bbS_{(\ell)})\rPhi_{j, 0}$ follows immediately.

\pfstep{Step~2: Proof of (2)}
In this case, by an induction on $k = K_{J}, \ldots, 1$ as in Step~1(a), one may show that, for $\abs{I} \leq M - C'$,
\begin{equation*}
\abs{\rd_{u} \bfGmm^{I} \rPhi_{J, k}} \aleq A' u^{J-1-\alp-\dlt_{0}+\nu_{\Box}},\quad 1 \leq k \leq K_j.
\end{equation*}
In this case, since $J - \alp - \dlt_{0} + \nu_{\Box} > 0$, we may integrate the above bound for $\rd_{u} \bfGmm^{I} \rPhi_{j, k}$ from $u = 1$ to obtain the desired bound. When $k = 0$, by arguing as in Step~1(b), one may show that, for $\abs{I} \leq M - C'$,
\begin{equation*}
\abs{\rd_{u} \bfGmm^{I} \rPhi_{(\leq J - \nu_{\Box}) J, 0}} \aleq A' u^{J-1-\alp-\dlt_{0}+\nu_{\Box}},
\end{equation*}
from which the desired bound also follows. Finally, for $\rPhi_{J, 0}$ itself, we have the worse bound (for $\abs{I} \leq M - C'$)
\begin{equation*}
\abs{\rd_{u} \bfGmm^{I} \rPhi_{J, 0}} \aleq A' u^{J-1-\alp+\nu_{\Box}}
\end{equation*}
due to the term $-\frac{1}{2j} ((j-1)j - (\nu_{\Box}-1)\nu_{\Box}+\rslap) \rPhi_{j-1, 0}$. Integrating from $u= 1$, we obtain
\begin{equation*}
\rPhi_{J, 0} = O_{\bfGmm}^{M - C'} (A' u^{J-\alp+\nu_{\Box}}),
\end{equation*}
from which the desired bound for $\rPhi_{(\geq J - \nu_{\Box}+1)J, 0}$ follows.

\pfstep{Step~3: Proof of (3)}
In this case, we need to be careful since $\lim_{u \to \infty} \rPhi_{J, k}(u, \tht)$ does \emph{not} exist in general for $k < K_{J}$; the culprits in the recurrence equation \eqref{eq:recurrence-jk} are the terms involving $u^{-1} \rPhi_{J, k+1}$ and $u^{-1} \rPhi_{J, k+2}$ on the right-hand side. To handle this issue, we begin by establishing the following bounds: for $k = K_{J}, \ldots, 1$ and $\abs{I} \leq M - C'$, we have
\begin{equation} \label{eq:recurrence-Jk-modified}
	\abs*{\bfGmm^{I}(\rd_{u} \rPhi_{J, k} - (k+1) u^{-1} \rPhi_{J, k+1})} \aleq A' u^{J-1-\alp-\dlt_{0}+\nu_{\Box}},
\end{equation}
and for $k= 0$ and $\abs{I} \leq M- C'$, we have
\begin{equation} \label{eq:recurrence-J0-modified}
	\abs*{\bfGmm^{I}(\rd_{u} \rPhi_{(\leq J-\nu_{\Box}) J, 0} - (k+1) u^{-1} \rPhi_{(\leq J-\nu_{\Box})  J, 1})} \aleq A' u^{J-1-\alp-\dlt_{0}+\nu_{\Box}}.
\end{equation}
The basic idea is to rewrite \eqref{eq:recurrence-jk} as
\begin{align*}
\rd_{u} \rPhi_{j, k}- (k+1) u^{-1} \rPhi_{j, k+1}
	&= \frac{k+1}{j} \left( \rd_{u} \rPhi_{j, k+1} - (k+2) u^{-1} \rPhi_{j, k+2} \right)  \\
&\peq
- \frac{1}{2j}\Big( (j-1)j - \frac{(d-1)(d-3)}{4} + \rslap \Big) \rPhi_{j-1, k}   \\
&\peq
+ \frac{1}{2j} (j-1)(k+1) \rPhi_{j-1, k+1} - \frac{1}{2j}(k+2)(k+1) \rPhi_{j-1, k+2}
+ \frac{1}{2j} \rF_{j, k}.
\end{align*}
Treating the first term on the right-hand side by an induction on $k = K_{J}, \ldots, 1$, and the rest as in Step~1, we arrive at \eqref{eq:recurrence-Jk-modified}. Then applying a similar argument to $\bbS_{(\leq J - \nu_{\Box})}$ of the $k=0$ case of the above equation (this projection is necessary to treat the second term on the right-hand side), we arrive at \eqref{eq:recurrence-J0-modified}.

Next, we perform an induction on $k = K_{J}, \ldots, 0$ to prove Statement~(3). Let $k \in \set{1, \ldots, K_{J}}$ and assume the following induction hypothesis: there exist $(\rcPhi_{J, K}(\infty))_{K=k+1}^{K_{J}}$ such that
\begin{equation*}
\rPhi_{J, k+1} - \sum_{K=k+1}^{K_{J}} \binom{K}{k+1} \log^{K-k-1} u \rcPhi_{J, K}(\infty) = O_{\bfGmm}^{M-C'}(A' u^{J-\alp-\dlt_{0}+\nu_{\Box}}).
\end{equation*}
We remark that this assumption is vacuously satisfied for $k = K_{J}$. Observe that
\begin{align*}
	&\rd_{u} \left( \rPhi_{J, k} - \sum_{K=k+1}^{K_{J}} \binom{K}{k} \log^{K-k} u \rcPhi_{J, K}(\infty) \right) \\
	&=   \rd_{u} \rPhi_{J, k} - \sum_{K=k+1}^{K_{J}} \binom{K}{k} (K-k) u^{-1} \log^{K-k-1} u \rcPhi_{J, K}(\infty) \\
	&=  (k+1) u^{-1} \rPhi_{J, k+1} - \sum_{K=k+1}^{K_{J}} \binom{K}{k} (K-k) u^{-1} \log^{K-k-1} u \rcPhi_{J, K}(\infty)
	+ O_{\bfGmm}^{M-C'}(A' u^{J-1-\alp-\dlt_{0}+\nu_{\Box}}) \\
	&=  \sum_{K=k+1}^{K_{J}} \binom{K}{k+1} (k+1) u^{-1} \log^{K-k-1} u \rcPhi_{J, K}(\infty)
	- \sum_{K=k+1}^{K_{J}} \binom{K}{k} (K-k) u^{-1} \log^{K-k-1} u \rcPhi_{J, K}(\infty) \\
	&\peq + (k+1) u^{-1} \left( \rPhi_{J, k+1} - \sum_{K=k+1}^{K_{J}} \binom{K}{k+1} \log^{K-k-1} u \rcPhi_{J, K}(\infty)\right)
	+ O_{\bfGmm}^{M-C'}(A' u^{J-1-\alp-\dlt_{0}+\nu_{\Box}}) \\
	&= O_{\bfGmm}^{M-C'}(A' u^{J-1-\alp-\dlt_{0}+\nu_{\Box}}),
\end{align*}
where we used \eqref{eq:recurrence-Jk-modified} for the second identity, and the induction hypothesis for the last identity.
Hence,  the following limit exists:
\begin{equation*}
	\rcPhi_{J, k}(\infty) = \lim_{u \to \infty}  \left( \rPhi_{J, k} - \sum_{K=k+1}^{K_{J}} \binom{K}{k} \log^{K-k} u \rcPhi_{J, K}(\infty) \right),
\end{equation*}
and the desired bound $\rcPhi_{J,k}(\infty) = O^{M-C'}_{\bfGmm}(A')$ holds.
Moreover, by the fundamental theorem of calculus (as well as another induction argument to commute $\rd_{u}$ and $\bfGmm^{I}$ as in Step~1), we obtain
\begin{align*}
	\rPhi_{J, k} - \sum_{K=k}^{K_{J}} \binom{K}{k} \log^{K-k} u \rcPhi_{J, K}(\infty) = O_{\bfGmm}^{M-C'}(A' u^{J-\alp-\dlt_{0}+\nu_{\Box}}).
\end{align*}
For $k = 0$, an analogous argument shows the existence of the limit
\begin{equation*}
	\rcPhi_{(\leq J-\nu_{\Box}) J, 0}(\infty) = \lim_{u \to \infty}  \left( \rPhi_{(\leq J - \nu_{\Box}) J, 0} - \sum_{K=1}^{K_{J}} \binom{K}{k} \log^{K-k} u \rcPhi_{(\leq J - \nu_{\Box}) J, K}(\infty) \right),
\end{equation*}
the bound $\rcPhi_{(\leq J - \nu_{\Box}) J ,0}(\infty) = O_{\bfGmm}^{M-C'}(A')$, and
\begin{align*}
	\rPhi_{(\leq J-\nu_{\Box}) J, 0} - \sum_{K=0}^{K_{J}} \log^{K} u \rcPhi_{(\leq J-\nu_{\Box}) J, K}(\infty) = O_{\bfGmm}^{M-C'}(A' u^{J-\alp-\dlt_{0}+\nu_{\Box}}).
\end{align*}
Finally, the bound for $\rPhi_{(\geq J - \nu_{\Box}+1) J, 0}$ follows as in Step~2. This completes the proof of Statement~(3). \qedhere
\end{proof}

\subsection{Truncated expansion and error estimate}  \label{subsec:expansion-error}
Our next aim is to define an approximation of $\Phi$ in $\calM_{\far} = \calM_{\med} \cup \calM_{\wave}$ using the expansion formed by the monomials $r^{-j} \log^{k} (\tfrac{r}{u}) \rPhi_{j, k}$ discussed in Section~\ref{subsec:recurrence}. While it is clear that such an expansion is advantageous in $\calM_{\wave}$ (where $\frac{r}{u}$ is large), this is not the case in $\calM_{\med}$ (where $\frac{r}{u}$ is small). Hence, we need to truncate the approximation in spacetime in an appropriate manner.

Let $\chi_{>\eta_{0}^{-1}}(\cdot) = \chi_{>1}(\eta_{0} (\cdot))$, where $\chi_{>1}(\cdot) : \bbR \to [0, 1]$ is a smooth nondecreasing function supported in $(\frac{1}{2}, \infty)$ and equals $1$ on $[1, \infty)$. Observe that $\chi_{>\eta_{0}^{-1}}\left( \tfrac{r}{u} \right) = 1$ on $\calM_{\wave}$. We define the \emph{truncated expansion of $\Phi$ below order $J$} to be
\begin{equation} \label{eq:Phi<J}
	\Phi_{<J} = \sum_{j=0}^{J-1} r^{-j} \rPhi_{j, 0} + \chi_{>\eta_{0}^{-1}}\left( \tfrac{r}{u} \right)  \sum_{j=0}^{J-1} \sum_{k =1}^{K_{j}} r^{-j} \log^{k} (\tfrac{r}{u}) \rPhi_{j, k},
\end{equation}
where, {\bf for $0 \leq j \leq J-1$, we assume that $\rPhi_{j, k} = 0$ for $k > K_{j}$ and $\rPhi_{j, k}$ solves the recurrence equation \eqref{eq:recurrence-jk} for $0 \leq k \leq K_{j}$}. We also define the corresponding \emph{truncated error} $E_{J}$ to be
\begin{equation} \label{eq:exp-error}
\begin{aligned}
	E_{J} &= Q_{0}\left(\Phi_{<J}\right)
	+  \chi_{>\eta_{0}^{-1}}\left( \tfrac{r}{u} \right) \left( \bfh^{\alp \bt} \nb_{\alp} \rd_{\bt} \Phi_{<J} + \bfC^{\alp} \rd_{\alp} \Phi_{<J} + W \Phi_{<J} \right) \\
	&\peq -  \chi_{>\eta_{0}^{-1}}\left( \tfrac{r}{u} \right) r^{\nu_{\Box}} \calN\left(r^{-\nu_{\Box}}\Phi_{<J}\right) -  \chi_{>\eta_{0}^{-1}}\left( \tfrac{r}{u} \right) r^{\nu_{\Box}} f.
\end{aligned}\end{equation}
Note that $E_{J} = \calQ(\Phi_{<J}) - r^{\nu_{\Box}} \calN(r^{-\nu_{\Box}}\Phi_{<J}) - r^{\nu_{\Box}} f$ in $\calM_{\wave}$, while $\Phi_{<J} = \sum_{j=0}^{J-1} r^{-j} \rPhi_{j, 0}$ and $E_{J} = Q_{0}(\sum_{j=0}^{J-1} r^{-j} \rPhi_{j, 0})$ outside the support of $\chi_{>\eta_{0}^{-1}}\left( \tfrac{r}{u} \right)$ (which is essentially $\calM_{\med}$). Such definitions turn out to be favorable in both $\calM_{\wave}$ and $\calM_{\med}$; see Remarks~\ref{rem:exp-error-med}, \ref{rem:rho-error-med}, as well as the useful fact that $\bfK^{J} \Phi_{<J} = 0$ outside the support of $\chi_{>\eta_{0}^{-1}}\left( \tfrac{r}{u} \right)$ (see Lemma~\ref{lem:rho-J-med}).

The goal of this subsection is to obtain the following lemmas concerning $E_{J}$. The first lemma concerns the region $\calM_{\wave}$, and it will be sufficient for all cases of Proposition~\ref{prop:wave-main} except Case~2.
\begin{lemma} \label{lem:exp-error-wave}
Let $0 \leq J - 1 \leq \min\set{J_{c}, J_{d}} - 1$ and
\begin{equation*}
	J - 1  + \nu_{\Box} < \alp < J + \nu_{\Box}.
\end{equation*}
Assume that $\rPhi_{j, k}$ satisfies \eqref{eq:recurrence-hyp-falloff}. When $\calN \neq 0$, assume \eqref{eq:alp-dlt0-nonlin} as well.  Then there exist $C' > 0$ and $A' = A'(D, A)$, which satisfies \eqref{eq:A'}, such that we have
\begin{equation} \label{eq:exp-error-wave}
\begin{aligned}
	E_{J} &= E_{J; \Box, k= 0} + E_{J; f} \\
	&\peq + O_{\bfGmm}^{M-C'}(A' r^{-J-1} \log^{K_{J}} (\tfrac{r}{u}) u^{J-1-\alp-\dlt_{0}+\nu_{\Box}}) \\
	&\peq + O_{\bfGmm}^{M-C'}(A' r^{-J_{c}-\eta_{c}} u^{J_{c}-2+\eta_{c}-\alp-\dlt_{0}+\nu_{\Box}})
\end{aligned}
\qquad \hbox{ in } \calM_{\wave},
\end{equation}
where
\begin{align}
E_{J; \Box, k = 0} &=
r^{-J-1}
 ((J-1) J + (\nu_{\Box}-1)\nu_{\Box} + \rslap)  \rPhi_{J-1, 0},  \label{eq:EJ-Box-k=0}\\
E_{J; f} &= - \sum_{j=J+1}^{J_{d}} \sum_{k=0}^{K_{d}} r^{-j} \log^{k} (\tfrac{r}{u}) \rf_{j+\frac{d-1}{2}, k}(u, \tht) - r^{\nu_{\Box}} \rem_{\alp_{d}+1}[f], \label{eq:EJ-f}\\
r^{\nu_{\Box}} \rem_{\alp_{d}+1}[f] &= O_{\bfGmm}^{M_{0}}(D r^{-\alp_{d}-1+\nu_{\Box}} u^{-1-\dlt_{d}}). \label{eq:EJ-f-rem}
\end{align}
Moreover, if $J \leq \min\set{J_{c}, J_{d}}-1$ then $K_{J}$ is as in Lemma~\ref{lem:Kj-def}, and if $J = \min\set{J_{c}, J_{d}}$ then $K_{J} = K_{J} (d, K_{c}, K_{d}, K_{J-1}, \calN ,\min\set{J_{c}, J_{d}})$ such that $K_{c} = K_{d} = 0 \imp K_{J} = 0$.
\end{lemma}
For the proof of Case~2 of Proposition~\ref{prop:wave-main}, we need the following variant of Lemma~\ref{lem:exp-error-wave}.
\begin{lemma} \label{lem:exp-error-f-wave}
Let $1 \leq J \leq \min\set{J_{c}, J_{d}}-1$ and
\begin{equation*}
	J - 1 + \nu_{\Box} < \alp < J  + \nu_{\Box} < \alp + \dlt_{0}.
\end{equation*}
Assume that $\rPhi_{j, k}$ satisfies \eqref{eq:recurrence-hyp-falloff}. When $\calN \neq 0$, assume \eqref{eq:alp-dlt0-nonlin} as well.
Then there exist $C' > 0$ and $A' = A'(D, A)$, which satisfies \eqref{eq:A'}, such that we have
\begin{equation} \label{eq:exp-error-f-wave}
\begin{aligned}
	E_{J+1} &= E_{J+1; \Box, k= 0}
	+ E_{J+1; f} \\
	&\peq + O_{\bfGmm}^{M-C'}(\td{\frkL} A' r^{-J-2} \log (\tfrac{r}{u}) \log^{K_{J}-1} r) \\
	&\peq + O_{\bfGmm}^{M-C'}(A' r^{-J-2} \log^{K_{J+1}} (\tfrac{r}{u}) u^{J-\alp-\dlt_{0}+\nu_{\Box}})\\
	&\peq + O_{\bfGmm}^{M-C'}(A' r^{-J_{c}-\eta_{c}} u^{J_{c}-2+\eta_{c}-\alp-\dlt_{0}+\nu_{\Box}})
\end{aligned}
\qquad \hbox{ in } \calM_{\wave},
\end{equation}
where we omit the third term on the right-hand side if $K_{J} = 0$.
Here, $E_{J+1; \Box, k = 0}$, $E_{J+1; f}$ and $K_{J+1}$ are as in Lemma~\ref{lem:exp-error-wave}, and
\begin{equation}\label{eq:tdfrkL.def}
	\td{\frkL} := \begin{cases}
	1 & \hbox{ if } \left( \rcPhi_{(\leq J - \nu_{\Box})J, 0}(\infty), (\rcPhi_{J, k}(\infty))_{k=1,\ldots, K_{J}} \right) \neq 0, \\
	0 & \hbox{ otherwise.}
	\end{cases}
\end{equation}
\end{lemma}

\begin{remark}[Cancellation for $\bfK^{J} E_{J; \Box, k=0}$] \label{rem:exp-error-wave}
Lemmas~\ref{lem:exp-error-wave} and \ref{lem:exp-error-f-wave} show that all terms in $E_{J}$ are clearly improved in $\calM_{\wave}$ except for the contribution of $E_{J; \Box, k =0}$.
Remarkably, this contribution is either improved or cancelled in our analysis of $\rho_{J}$. For $\rho_{(\ell) J}$ with $\ell \leq J - \nu_{\Box}$, we exploit the Newman--Penrose cancellation; see Remark~\ref{rem:np-exp-error}. For $\rho_{(\geq J - \nu_{\Box}+1)J}$, the good variable to work with turns out to be $\bfK^{J} \rho_{(\geq J - \nu_{\Box}+1) J}$ in view of the nice positivity property that its equation enjoys when multiplied by $r^{p} \rd_{r}$ (see Lemma~\ref{lem:Qj-rp} and Lemma~\ref{lem:rho-J-rp-eqn}). Hence, we only see $\bbS_{(\geq J - \nu_{\Box}+1)} (\bfK+2r)^{J} E_{J}$ in our analysis. But since $(\bfK+2r)^{J} r^{-J-1} = 0$, we have $(\bfK + 2r)^J E_{J; \Box, k= 0} = 0$.
\end{remark}

\begin{remark} [Newman--Penrose cancellation for the truncated expansion error] \label{rem:np-exp-error}
Another key source of improvement for $E_{J; \Box}$ in $\calM_{\wave}$ is the Newman--Penrose cancellation, which also arose in the context of recurrence equations; cf.~ Proposition~\ref{prop:recurrence} and Remark~\ref{rem:np-exp}. In view of estimating $\rho_{(\ell) J}$ with $\ell \leq J - \nu_{\Box}$ (for which Remark~\ref{rem:exp-error-wave} does not apply), consider $\bbS_{(\ell)} E_{J; \Box}$. Since $\rPhi_{(\ell) J-1, k}$ with $k \geq 1$ is always improved, the only possibly non-improved term in $\bbS_{(\ell)} E_{J; \Box}$ is
\begin{equation*}
( J - \nu_{\Box} - \ell ) ( J + \nu_{\Box} + \ell - 1 )  \rPhi_{(\ell) J-1, 0}.
\end{equation*}
Observe that when $\ell = J - \nu_{\Box}$, the non-improved term $( J - \nu_{\Box} - \ell ) ( J + \nu_{\Box} + \ell - 1 )  \rPhi_{(\ell) J-1, k}$ vanishes! This vanishing is the manifestation of the Newman--Penrose cancellation at the level of the expansion error. Moreover, when $\ell \leq J -1 - \nu_{\Box}$, then $\rPhi_{(\ell) J-1, k}$ is already improved thanks to the Newman--Penrose cancellation in the recurrence equations (see Proposition~\ref{prop:recurrence}); hence, $\bbS_{(\ell)} E_{J; \Box}$ is improved.
\end{remark}

To state the results in $\calM_{\med}$, we introduce the notation
\begin{align*}
\Phi_{<J, k=0} := \sum_{j=0}^{J-1} r^{-j} \rPhi_{j, 0}.
\end{align*}
In $\calM_{\med}$, we have the following results:
\begin{lemma} \label{lem:exp-error-med}
Under the same hypotheses as in Lemma~\ref{lem:exp-error-wave}, there exist $C' > 0$ and $A' = A'(D, A)$, which satisfies \eqref{eq:A'}, such that
\begin{equation} \label{eq:exp-error-med}
\begin{aligned}
	E_{J} &= Q_{0}(\Phi_{<J, k=0})  \\
	&\peq + O_{\bfGmm}^{M-C'}\left( A' \chi_{> (2 \eta_{0})^{-1}}(\tfrac{r}{u}) u^{-\alp-2-\dlt_{0}+\nu_{\Box}} \right)  \\
	&\peq + O_{\bfGmm}^{M_{0}}\left( D \chi_{> (2 \eta_{0})^{-1}}(\tfrac{r}{u}) u^{-\alp_{d}-2-\dlt_{d}+\nu_{\Box}} \right)
\end{aligned}
\qquad \hbox{ in } \calM_{\med}.
\end{equation}
\end{lemma}

\begin{lemma} \label{lem:exp-error-f-med}
Under the same hypotheses as in Lemma~\ref{lem:exp-error-f-wave}, there exist $C' > 0$ and $A' = A'(D, A)$, which satisfies \eqref{eq:A'}, such that
\begin{equation} \label{eq:exp-error-f-med}
\begin{aligned}
	E_{J+1} &= Q_{0}(\Phi_{<J+1, k=0})
	+ O_{\bfGmm}^{M-C'}\left( \td{\frkL} A' \chi_{> (2 \eta_{0})^{-1}}(\tfrac{r}{u}) u^{-J-2} \log^{K_{J}-1} u \right)  \\
	&\peq + O_{\bfGmm}^{M-C'}\left( A' \chi_{> (2 \eta_{0})^{-1}}(\tfrac{r}{u}) u^{-\alp-2-\dlt_{0}+\nu_{\Box}}  \right)  \\
	&\peq + O_{\bfGmm}^{M_{0}}\left( D \chi_{> (2 \eta_{0})^{-1}}(\tfrac{r}{u}) u^{-\alp_{d}-2-\dlt_{d}+\nu_{\Box}} \right)
\end{aligned}
\qquad \hbox{ in } \calM_{\med},
\end{equation}
where we omit the second term on the right-hand side if $K_{J} = 0$.
\end{lemma}

\begin{remark}[Cancellation in $E_{J}$ in $\calM_{\med}$] \label{rem:exp-error-med}
Lemmas~\ref{lem:exp-error-med} and \ref{lem:exp-error-f-med} show that all terms in $E_{J}$ are clearly improved, except for $Q_{0}(\Phi_{<J, k=0})$.
As we shall see, this non-improved contribution gets cancelled in our analysis of $\rho_{J}$. It turns out that we only need the expression for $E_{J}$ in $\calM_{\med}$ for higher spherical harmonics of $\rho_{J}$, since the lower spherical harmonics can be treated using the method of characteristics (Lemma~\ref{lem:Qj-char}) working entirely in $\calM_{\wave}$. More precisely, $E_{J}$ enters in our analysis only in the context of estimating $\bfK^{J} \rho_{(\geq J - \nu_{\Box}+1) J}$, which means that we only see $\bbS_{(\geq J - \nu_{\Box}+1)} (\bfK+2r)^{J} E_{J}$ in $\calM_{\med}$. Observe, finally, that $(\bfK+2r)^{J}$ eliminates any monomial of the form $r^{-j} \mathring{a}(u, \tht)$ with $2 \leq j \leq J+1$, and that $Q_{0}(\Phi_{<J, k=0})$ consists entirely of such terms!
\end{remark}

The remainder of this section is devoted to proofs of the above lemmas.
\begin{proof}[Proof of Lemmas~\ref{lem:exp-error-wave} and \ref{lem:exp-error-f-wave}]
In the entire proof, we work in $\calM_{\wave}$. Recall that in $\calM_{\wave}$, the expansion in \eqref{eq:Phi<J} takes the form
\begin{equation} \label{eq:Phi<J-wave}
	\Phi_{<J} = \sum_{j=0}^{J-1} \sum_{k =0}^{K_{j}} r^{-j} \log^{k} (\tfrac{r}{u}) \rPhi_{j, k}.
\end{equation}
Moreover, in $\calM_{\wave}$,
\begin{equation*}
	E_{J} = \calQ\left(\Phi_{<J}\right) -  r^{\nu_{\Box}} \calN\left(r^{-\nu_{\Box}}\Phi_{<J}\right) -  r^{\nu_{\Box}} f.
\end{equation*}
By construction (i.e., the recurrence equations for $\rPhi_{j, k}$ for $j \leq J-1$), \emph{all monomials of the form $r^{-j-1} \log^{k} (\tfrac{r}{u})$ with $j \leq J-1$ are cancelled}. We decompose
\begin{equation}\label{eq:EJ.decompose}
E_{J} = E_{J; \Box} + E_{J; \linear} + E_{J; \nonlinear} + E_{J; f},
\end{equation}
where the four terms on the right-hand side refer to contributions of $\Box_{\bfm}$, of $\calP - \Box_{\bfm}$, of $\calN$, and of $f$, respectively, after the above cancellation is taken into account; their precise definitions shall be given below.

\pfstep{Step~1: Minkowskian contribution}
The contribution of $\Box_{\bfm}$ in $E_{J}$ is given by
\begin{equation} \label{eq:EJ-Box}
\begin{aligned}
E_{J; \Box} &= \sum_{k=0}^{K_{J-1}} \left(\rd_{r}^{2} - \nu_{\Box} (\nu_{\Box}-1) r^{-2} + r^{-2} \rslap \right) \left(r^{-J+1} \log^{k} (\tfrac{r}{u}) \rPhi_{J-1, k} \right),
\end{aligned}
\end{equation}
which consists of all terms with $r^{-j-1} \log^{k} (\tfrac{r}{u})$ with $j \geq J$ in $Q_{0}(\Phi_{<J})$. Observe that the $k = 0$ summand coincides with \eqref{eq:EJ-Box-k=0}. Using Proposition~\ref{prop:recurrence}.(1) to estimate the $k \geq 1$ summation, we obtain
\begin{equation*}
	E_{J; \Box} = E_{J; \Box, k=0} + O_{\bfGmm}^{M-C'}(A' r^{-J-1} \log^{K_{J-1}} (\tfrac{r}{u})u^{J-1 - \alp -\dlt_{0}+\nu_{\Box}}),
\end{equation*}
which is sufficient for the proof of Lemma~\ref{lem:exp-error-wave}. For Lemma~\ref{lem:exp-error-f-wave}, we repeat the above proof with $J$ replaced by $J+1$. Here, using Proposition~\ref{prop:recurrence}.(3) to estimate $\rPhi_{J, k}$, we have
\begin{align*}
	E_{J+1; \Box} &= E_{J+1; \Box, k=0}
	+ \sum_{k=1}^{K_{J}} \sum_{K=k}^{K_{J}} \binom{K}{k} \left(\rd_{r}^{2} - \nu_{\Box} (\nu_{\Box}-1) r^{-2} + r^{-2} \rslap \right) \left(r^{-J+1} \log^{k} (\tfrac{r}{u}) \log^{K-k} u \rcPhi_{J, K} \right) \\
	&\peq + O_{\bfGmm}^{M-C'}(A' r^{-J-2} \log^{K_{J}} (\tfrac{r}{u})u^{J - \alp -\dlt_{0}+\nu_{\Box}}).
\end{align*}
Using $\log^{k-1}(\tfrac{r}{u}) \log^{K_{J}-k} u \leq (\log (\tfrac{r}{u}) + \log u)^{K_{J}-1} = \log^{K_{J}-1} r$ and $\rcPhi_{J, K}(\infty) = O_{\bfGmm}^{M-C'}(A')$, we see that the second term on the right-hand side is $O_{\bfGmm}^{M-C'}(\td{\frkL} A' r^{-J-2} \log \left(\tfrac{r}{u}\right)\log^{K_{J}-1} r)$ (and nonexistent when $K_{J} = 0$), which is acceptable.

\pfstep{Step~2: Non-Minkowskian, linear contribution}
We use
\begin{equation*}
	E_{J; \linear} = \sum_{j \geq J} \sum_{k \geq 0} r^{-j-1} \log^{k} (\tfrac{r}{u}) \rL^{<J}_{j, k} + E_{J; \linear, \rem}.
\end{equation*}
where
\begin{align*}
\rL^{<J}_{j, k} = \sum_{j', k' : j' < J,\, k' \leq K_{j'}} \left[\hbox{summand in the definition of $\rL_{j, k}$}\right].
\end{align*}

We first consider Lemma~\ref{lem:exp-error-wave}. As in \eqref{eq:recurrence-forcing-L}, we may show that \eqref{eq:recurrence-hyp-falloff} for $\rPhi_{j, k}$ with $0 \leq j \leq J-1$ implies
\begin{align*}
\rL^{<J}_{j, k} = O_{\bfGmm}^{M-2}(A u^{j-1-\alp-\dlt_{c}+\nu_{\Box}}).
\end{align*}
Hence,
\begin{align*}
 \sum_{j \geq J} \sum_{k \geq 0} r^{-j-1} \log^{k} (\tfrac{r}{u}) \rL^{<J}_{j, k}
 = O_{\bfGmm}^{M-C'}(A r^{-J-1} \log^{K_{J}} (\tfrac{r}{u}) u^{J-1-\alp-\dlt_{c}-\nu_{\Box}}),
\end{align*}
which is acceptable. We note that, if $J \leq \min\set{J_{c}, J_{d}}-1$, $K_{J}$ may be chosen to be the same as in Lemma~\ref{lem:Kj-def} thanks to Remark~\ref{rem:Kj-rem}.

For the remainder, we begin by writing
\begin{align*}
	E_{J; \linear, \rem}
	&= -\rem_{J_{c}+\eta_{c}}[\bfh^{uu}] \rd_{u}^{2} \Phi_{< J}
	-2 \rem_{J_{c}-1+\eta_{c}}[\bfh^{ur}] \rd_{u} \rd_{r} \Phi_{< J}
	-2 \rem_{J_{c}-1+\eta_{c}}[r \bfh^{uA}] r^{-1} \rd_{\tht^{A}} \rd_{u} \Phi_{< J} \\
	&\peq
	- \rem_{J_{c}-2+\eta_{c}}[\bfh^{rr}] \rd_{r}^{2} \Phi_{< J}
	- \rem_{J_{c}-2+\eta_{c}}[r \bfh^{rA}] r^{-1} \rd_{\tht^{A}} \rd_{r} \Phi_{< J}
	- \rem_{J_{c}-2+\eta_{c}}[r^{2} \bfh^{AB}] r^{-2} \rd_{\tht^{A}} \rd_{\tht^{B}} \Phi_{< J} \\
	&\peq
	-\rem_{J_{c}+\eta_{c}}[\bfC^{u}] \rd_{u} \Phi_{< J}
	-\rem_{J_{c}-1+\eta_{c}}[\bfC^{r}] \rd_{r} \Phi_{< J}
	-\rem_{J_{c}-1+\eta_{c}}[r \bfC^{A}] r^{-1} \rd_{\tht^{A}} \Phi_{< J} \\
	&\peq
	- \rem_{J_{c}+\eta_{c}}[W] \Phi_{< J}.
\end{align*}
By Lemma~\ref{lem:conj-wave-wave}, we have
\begin{align*}
E_{J; \linear, \rem}
= O_{\bfGmm}^{M-C}( A r^{-J_{c}-\eta_{c}} u^{J_{c}-2+\eta_{c}-\alp-\dlt_{c}+\nu_{\Box}}),
\end{align*}
which is acceptable.

For the proof of Lemma~\ref{lem:exp-error-f-wave}, we repeat the above argument with $J$ replaced by $J+1$. Observe that, since $\alp < J + \nu_{\Box}$, by Proposition~\ref{prop:recurrence}.(3) we still have
\begin{equation} \label{eq:rPhi-J-falloff}
	\rPhi_{J, k} = O_{\bfGmm}^{M-C'}(A' u^{J - \alp + \nu_{\Box}}) \hbox{ for } 0 \leq k \leq K_{J},
\end{equation}
and we obtain the analogous bounds
\begin{align*}
 \sum_{j \geq J+1} \sum_{k \geq 0} r^{-j-1} \log^{k} (\tfrac{r}{u}) \rL^{<J+1}_{j, k}
&= O_{\bfGmm}^{M-C'}(A r^{-J-2} \log^{K_{J+1}} (\tfrac{r}{u}) u^{J-\alp-\dlt_{c}-\nu_{\Box}}),
\\
E_{J+1; \linear, \rem}
&= O_{\bfGmm}^{M-C'}(A r^{-J_{c}-\eta_{c}} u^{J_{c}-2+\eta_{c}-\alp-\dlt_{c}+\nu_{\Box}}).
\end{align*}

\pfstep{Step~3: Nonlinear contribution}
The contribution of $\calN$ without the monomials $r^{-j-1} \log^{k}(\tfrac{r}{u})$ with $j \leq J-1$ takes the form
\begin{equation*}
	E_{J; \nonlinear} = \sum_{j \geq J} \sum_{k \geq 0} r^{-j-1} \log^{k} (\tfrac{r}{u}) \rN^{<J}_{j, k} + E_{J; \nonlinear, \rem},
\end{equation*}
where $\rN^{<J}_{j, k}$ and $E_{J; \nonlinear, \rem}$ are defined using the expression \eqref{eq:Phi<J-wave} for $\rPhi_{<J}$ (see Section~\ref{subsec:nonlin-est}). Under our hypothesis $\alp \geq \alp_{\calN} + 2 \dlt_{0}$, this contribution is at least as good as the linear contribution. Indeed, since $\alp \geq \alp_{\calN} + 2 \dlt_{0}$ and \eqref{eq:recurrence-hyp-falloff} hold, we may use the bounds in Definition~\ref{def:alp-N}.(4) with $\alp_{\calN}' = \alp_{\calN} + \dlt_{0}$. Hence,
\begin{align*}
	\sum_{j \geq J} \sum_{k \geq 0} r^{-j-1} \log^{k} (\tfrac{r}{u}) \rN^{<J}_{j, k} &= \sum_{j \geq J} \sum_{k \geq 0}  r^{-j-1} \log^{k} O_{\bfGmm}^{M-2}(C_{j, k} A_{\calN}(A) u^{j-\alp-\dlt_{0}+\nu_{\Box}}),
\end{align*}
and
\begin{align*}
	E_{J; \nonlinear, \rem}
	&= O_{\bfGmm}^{M-2}(A_{\calN}(A) r^{-J_{c}-\eta_{c}} u^{J_{c}-2+\eta_{c}-\alp-\dlt_{0}+\nu_{\Box}}),
\end{align*}
which are acceptable for Lemma~\ref{lem:exp-error-wave}. In case of \ref{lem:exp-error-f-wave}, we repeat the the above argument with $J$ replaced by $J+1$, using also \eqref{eq:rPhi-J-falloff} for $\rPhi_{J, k}$ with $0 \leq k \leq K_{J}$.

\pfstep{Step~4: Contribution of $f$}
Since all terms of order $(j, k)$ with $j < J$ are cancelled, by \ref{hyp:forcing}, it follows that
\begin{align*}
E_{J; f}
&= - \sum_{j=J+1}^{J_{d}} \sum_{k=0}^{K_{d}} r^{-j} \log^{k} (\tfrac{r}{u}) \rf_{j+\nu_{\Box}, k}(u, \tht)
	- r^{\nu_{\Box}} \rem_{\alp_{d}+1}[f] \\
&= - \sum_{j=J+1}^{J_{d}} \sum_{k=0}^{K_{d}} r^{-j} \log^{k} (\tfrac{r}{u}) \rf_{j+\nu_{\Box}, k}(u, \tht)
	 + O_{\bfGmm}^{M_{0}}(D r^{-\alp_{d}-1+\nu_{\Box}} u^{-1-\dlt_{d}}).
\end{align*}
This completes the proof of Lemma~\ref{lem:exp-error-wave}; for Lemma~\ref{lem:exp-error-f-wave}, we repeat this argument with $J$ replaced by $J+1$. \qedhere
\end{proof}

\begin{proof}[Proof of Lemma~\ref{lem:exp-error-med} and Lemma~\ref{lem:exp-error-f-med}]
We begin with Lemma~\ref{lem:exp-error-med}. We rewrite $E_{J}$ in $\calM_{\med}$ as
\begin{equation} \label{eq:exp-error-med:pf}
\begin{aligned}
	E_{J} &= Q_{0}\left(\sum_{j=0}^{J-1} r^{-j} \rPhi_{j, 0}\right) + Q_{0} \left( \chi_{>\eta_{0}^{-1}}\left( \tfrac{r}{u} \right) \sum_{j=0}^{J-1} \sum_{k=1}^{K_{j}} r^{-j} \log^{k} (\tfrac{r}{u})\rPhi_{j, k}\right)\\
	&\peq
	+  \chi_{>\eta_{0}^{-1}}\left( \tfrac{r}{u} \right) \left( \bfh^{\alp \bt} \nb_{\alp} \rd_{\bt} \Phi_{<J} + \bfC^{\alp} \rd_{\alp} \Phi_{<J} + W \Phi_{<J} \right) \\
	&\peq -  \chi_{>\eta_{0}^{-1}}\left( \tfrac{r}{u} \right) r^{\nu_{\Box}} \calN\left(r^{-\nu_{\Box}}\Phi_{<J}\right) -  \chi_{>\eta_{0}^{-1}}\left( \tfrac{r}{u} \right) r^{\nu_{\Box}} f.
\end{aligned}
\end{equation}
The first term on the right-hand side is already acceptable. All the remaining terms on the right-hand side of \eqref{eq:exp-error-med:pf} possess the cutoff $\chi_{>\eta_{0}^{-1}}\left( \tfrac{r}{u} \right)$; observe that $r \aeq u$ in $\supp [ \chi_{>\eta_{0}^{-1}}( \tfrac{r}{u} ) ] \cap \calM_{\med}$. Hence, the second term on the right-hand side of \eqref{eq:exp-error-med:pf} is acceptable thanks to Proposition~\ref{prop:recurrence}.(1) for $\rPhi_{j, k}$ with $j \leq J-1$. The last term is acceptable thanks to \ref{hyp:forcing}. To see that the remaining terms are acceptable, we first use \eqref{eq:recurrence-hyp-falloff} and observe that $\Phi_{<J} = O_{\bfGmm}^{M-C'}(u^{-\alp+\nu_{\Box}})$ in $\supp [ \chi_{>\eta_{0}^{-1}}( \tfrac{r}{u} ) ] \cap \calM_{\med}$. Using Lemma~\ref{lem:conj-wave-med} for $\bfh$, $\bfC$ and $W$, and $\alp \geq \alp_{\calN} + 2 \dlt_{0}$ combined with Definition~\ref{def:alp-N}.(2) (with $\alp_{\calN}'= \alp_{\calN} + \dlt_{0}$) for $\calN$, the desired bound follows.

To establish Lemma~\ref{lem:exp-error-f-med}, we repeat the above argument with $J$ replaced by $J+1$. Using Proposition~\ref{prop:recurrence}.(3), we have
\begin{equation} \label{eq:exp-error-f-med:pf}
\begin{aligned}
	E_{J+1} &= Q_{0}\left(\sum_{j=0}^{J} r^{-j} \rPhi_{j, 0}\right)
	+ Q_{0} \left( \chi_{>\eta_{0}^{-1}}\left( \tfrac{r}{u} \right) \sum_{k=1}^{K_{J}} \sum_{K=k}^{K_{J}} r^{-J} \log^{k} (\tfrac{r}{u})  \binom{K}{k} \log^{K-k} u \rcPhi_{J, K}(\infty) \right) \\
	&\peq
	+ Q_{0} \left( \chi_{>\eta_{0}^{-1}} \left( \tfrac{r}{u} \right) \sum_{k=1}^{K_{J}} r^{-J} \log^{k} (\tfrac{r}{u}) \left( \rPhi_{J, k} - \sum_{K=k}^{K_{J}} \binom{K}{k} \log^{K-k} u \rcPhi_{J, K}(\infty) \right) \right)\\
	&\peq
	+ Q_{0} \left( \chi_{>\eta_{0}^{-1}}\left( \tfrac{r}{u} \right) \sum_{j=0}^{J-1} \sum_{k=1}^{K_{j}} r^{-j} \log^{k} (\tfrac{r}{u})\rPhi_{j, k}  \right)\\
	&\peq
	+  \chi_{>\eta_{0}^{-1}}\left( \tfrac{r}{u} \right) \left( \bfh^{\alp \bt} \nb_{\alp} \rd_{\bt} \Phi_{<J+1} + \bfC^{\alp} \rd_{\alp} \Phi_{<J+1} + W \Phi_{<J+1} \right) \\
	&\peq -  \chi_{>\eta_{0}^{-1}}\left( \tfrac{r}{u} \right) r^{\nu_{\Box}} \calN\left(r^{-\nu_{\Box}}\Phi_{<J+1}\right) -  \chi_{>\eta_{0}^{-1}}\left( \tfrac{r}{u} \right) r^{\nu_{\Box}} f.
\end{aligned}
\end{equation}
Again, the first term on the right-hand side is already acceptable. The second term on the right-hand side of \eqref{eq:exp-error-f-med:pf} is $O_{\bfGmm}^{M-C'}(\td{\frkL} A' \chi_{>(2 \eta_{0})^{-1}} \left(\tfrac{r}{u}\right) u^{-J-2} \log^{K_{J}-1} u)$ by Proposition~\ref{prop:recurrence}.(3) and the fact that $k \geq 1$ in the summand. The third term is acceptable thanks to Proposition~\ref{prop:recurrence}.(1) for $\rPhi_{j, k}$ with $j \leq J-1$, and the last term is handled using \ref{hyp:forcing} as before. To show that the remaining terms are acceptable, we first use \eqref{eq:rPhi-J-falloff} (i.e., regular fall-off with decay exponent $\alp$) and \eqref{eq:recurrence-hyp-falloff} to conclude that $\Phi_{<J+1} = O_{\bfGmm}^{M - C'}(A' u^{-\alp+\nu_{\Box}})$ in $\supp [ \chi_{>\eta_{0}^{-1}}( \tfrac{r}{u} ) ] \cap \calM_{\med}$. From here on, the proof is identical to that of Lemma~\ref{lem:exp-error-med}.
\qedhere
\end{proof}

\subsection{Equation for the remainder} \label{subsec:remainder}
In $\calM_{\far}$\footnote{We remark that while Propositions~\ref{prop:wave-main} and \ref{prop:wave-0} only involve $\rho_{J}$ in $\calM_{\wave}$, we shall need its definition on the entire $\calM_{\far}$ for their proofs.}, we define the remainder $\rho_{J}$ as
\begin{equation} \label{eq:rhoJ-def}
	\rho_{J} = \Phi - \Phi_{<J},
\end{equation}
where $J \leq \min\set{J_{c}, J_{d}}$ so that $\Phi_{<J}$ is well-defined.
Our main goal here is to write down the equation for $\rho_{J}$ in a convenient form for the ensuing analysis. We will also introduce modifications $\crho_{J}$ and $\hrho{U}_{J}$ of $\rho_{J}$, which will be useful for the proofs of Cases~2 and~4 in Proposition~\ref{prop:wave-main}, respectively. 

\subsubsection{Preliminaries}\label{sec:remainder.prelim}
We begin with a useful property of $\rho_{J}$ outside the support of $\chi_{\eta_{0}^{-1}}(\tfrac{r}{u})$ (which is essentially $\calM_{\med}$), which is a consequence of the way we truncate the expansion $\Phi_{<J}$.
\begin{lemma} \label{lem:rho-J-med}
For any $J \geq 0$, we have
\begin{equation*}
\bfK^{J} \rho_{J} = \bfK^{J} \Phi \quad \hbox{ in } \calM_{\far} \setminus \supp \left[ \chi_{\eta_{0}^{-1}}(\tfrac{r}{u})\right].
\end{equation*}
\end{lemma}
\begin{proof}
This cancellation is not surprising. Recall that $\Phi - \rho_{J} = \Phi_{<J} = \sum_{j=0}^{J-1} r^{-j} \rPhi_{J, 0}$ in this region, where the right-hand side is a degree $J-1$ polynomial in the variable $z = r^{-1}$. Moreover, $\bfK = \rd_{z}$ in the $(u, z, \tht)$ coordinates. Hence $\bfK^{J} \Phi_{<J} = \rd_{z}^{J}(\hbox{polynomial of degree $J-1$ in $z$}) = 0$.
\end{proof}

Now we perform some preliminary computations concerning the equation that $\rho_{J}$ satisfies. By \eqref{eq:Phi-eq-0}, \eqref{eq:exp-error} and \eqref{eq:rhoJ-def}, we have
\begin{align*}
Q_{0} \rho_{J}
&= Q_{0} \Phi - Q_{0} \Phi_{<J} \\
&= - \bfh^{\alp \bt} \nb_{\alp} \rd_{\bt} \Phi - \bfC^{\alp} \rd_{\alp} \Phi - W \Phi
+ r^{-\nu_{\Box}} \calN(r^{-\nu_{\Box}} \Phi)
+ r^{\nu_{\Box}} f \\
&\peq
- E_{J}
+  \chi_{>\eta_{0}^{-1}}\left( \tfrac{r}{u} \right) \left( \bfh^{\alp \bt} \nb_{\alp} \rd_{\bt} \Phi_{<J} + \bfC^{\alp} \rd_{\alp} \Phi_{<J} + W \Phi_{<J} \right) \\
&\peq -  \chi_{>\eta_{0}^{-1}}\left( \tfrac{r}{u} \right) r^{\nu_{\Box}} \calN\left(r^{-\nu_{\Box}}\Phi_{<J}\right) -  \chi_{>\eta_{0}^{-1}}\left( \tfrac{r}{u} \right) r^{\nu_{\Box}} f \\
&= -E_{J}
-  \chi_{>\eta_{0}^{-1}}\left( \tfrac{r}{u} \right) \left[ \bfh^{\alp \bt} \nb_{\alp} \rd_{\bt} \rho_{J} + \bfC^{\alp} \rd_{\alp} \rho_{J} + W \rho_{J} \right] \\
&\peq +  \chi_{>\eta_{0}^{-1}}\left( \tfrac{r}{u} \right) \left[ r^{\nu_{\Box}} \calN\left(r^{-\nu_{\Box}}(\Phi_{<J} + \rho_{J}) \right) - r^{\nu_{\Box}} \calN\left(r^{-\nu_{\Box}}\Phi_{<J}\right)\right] \\
&\peq + \left( 1-\chi_{>\eta_{0}^{-1}}\left( \tfrac{r}{u} \right) \right) \left[ - \bfh^{\alp \bt} \nb_{\alp} \rd_{\bt} \Phi - \bfC^{\alp} \rd_{\alp} \Phi - W \Phi + r^{\nu_{\Box}} \calN(r^{-\nu_{\Box}} \Phi)  + r^{\nB} f \right].
\end{align*}
We define 
\begin{align}
e_{J; \wave} &= e_{J; \wave, \linear} + e_{J; \wave, \nonlinear} \label{eq:eJwave-def} \\
e_{J; \wave, \linear} &= -  \chi_{>\eta_{0}^{-1}}\left( \tfrac{r}{u} \right) \left[ \bfh^{\alp \bt} \nb_{\alp} \rd_{\bt} \rho_{J} + \bfC^{\alp} \rd_{\alp} \rho_{J} + W \rho_{J} \right] 			\label{eq:eJwave-linear-def} \\
e_{J; \wave, \nonlinear} &= +  \chi_{>\eta_{0}^{-1}}\left( \tfrac{r}{u} \right) \left[ r^{\nu_{\Box}} \calN\left(r^{-\nu_{\Box}}(\Phi_{<J} + \rho_{J}) \right) - r^{\nu_{\Box}} \calN\left(r^{-\nu_{\Box}}\Phi_{<J}\right)\right],		\label{eq:eJwave-nonlinear-def}
\end{align}
and
\begin{align}
e_{\med} &= e_{\med, \linear} + e_{\med, \nonlinear} + e_{\med, f}, 		\label{eq:eJmed-def} \\
e_{\med, \linear} &= \left( 1-\chi_{>\eta_{0}^{-1}}\left( \tfrac{r}{u} \right) \right) \left[ - \bfh^{\alp \bt} \nb_{\alp} \rd_{\bt} \Phi - \bfC^{\alp} \rd_{\alp} \Phi - W \Phi \right]  	\label{eq:eJmed-linear-def} \\
e_{\med, \nonlinear} &= \left( 1-\chi_{>\eta_{0}^{-1}}\left( \tfrac{r}{u} \right) \right) \left[ r^{\nu_{\Box}} \calN(r^{-\nu_{\Box}} \Phi) \right]	\label{eq:eJmed-nonlinear-def} \\
e_{\med, f} &= \left( 1-\chi_{>\eta_{0}^{-1}}\left( \tfrac{r}{u} \right) \right) r^{\nB} f, 	\label{eq:eJmed-f-def}
\end{align}
so that
\begin{equation} \label{eq:Q0-rhoJ}
\begin{aligned}
	Q_{0} \rho_{J} &= - E_{J} + e_{J; \wave} + e_{\med} \\
	 &= - E_{J} + e_{J; \wave, \linear} + e_{J; \wave, \nonlinear} \\
	 &\peq + e_{\med, \linear} + e_{\med, \nonlinear} + e_{\med, f}.
\end{aligned}
\end{equation}

\begin{remark} [Errors $e_{J; \wave}$ and $e_{\med}$] \label{rem:rho-error-med}
Observe that $e_{\med}$ is an error that vanishes in $\calM_{\wave}$, whereas $e_{J; \wave}$ is localized to the region $\eta_{0} r \ageq u$. Moreover, thanks to our definitions of $\Phi_{<J}$ and $e_{J}$, observe that $e_{\med}$ does not involve the expansion $\Phi_{< J}$, whose presence would have been not favorable in $\calM_{\med}$!
\end{remark}
Applying $(\bfK+2r)^{j}$, we obtain
\begin{align*}
Q_{j} \bfK^{j} \rho_{J}
&= -(\bfK+2r)^{j} E_{J} + (\bfK+2r)^{j} e_{J; \wave} + (\bfK+2r)^{j} e_{\med}.
\end{align*}
Applying $(\bfGmm + c_{\bfGmm})^{I}$, we have
\begin{equation} \label{eq:rho-J-eqn-0}
\begin{aligned}
Q_{j} \bfGmm^{I} \bfK^{j} \rho_{J}
&= - \left[ (\bfGmm+c_{\bfGmm})^{I} Q_{j} - Q_{j} \bfGmm^{I} \right] \bfK^{j} \rho_{J}
-(\bfGmm+c_{\bfGmm})^{I} (\bfK+2r)^{j} E_{J} \\
&\peq + (\bfGmm+c_{\bfGmm})^{I} (\bfK+2r)^{j} e_{J; \wave} + (\bfGmm+c_{\bfGmm})^{I} (\bfK+2r)^{j} e_{\med}.
\end{aligned}
\end{equation}

Finally, we may also apply $\bbS_{(\ell)}$ or $\bbS_{(\geq \ell)}$; note that
\begin{equation*}
[Q_{j}, \bbS_{(\ell)}] = 0, \quad
[\bfGmm, \bbS_{(\ell)}] = 0, \quad
[\bfGmm + c_{\bfGmm}, \bbS_{(\ell)}] = 0, \quad
[\bfK, \bbS_{(\ell)}] = 0, \quad
[\bfK+2r, \bbS_{(\ell)}] = 0.
\end{equation*}
Therefore,
\begin{align*}
Q_{j} \bfGmm^{I} \bfK^{j} \rho_{(\ell) J}
&= - \left[ (\bfGmm+c_{\bfGmm})^{I} Q_{j} - Q_{j} \bfGmm^{I} \right] \bfK^{j} \rho_{(\ell) J}
-\bbS_{(\ell)}(\bfGmm+c_{\bfGmm})^{I} (\bfK+2r)^{j} E_{J} \\
&\peq + \bbS_{(\ell)} (\bfGmm+c_{\bfGmm})^{I} (\bfK+2r)^{j} e_{J; \wave} + \bbS_{(\ell)} (\bfGmm+c_{\bfGmm})^{I} (\bfK+2r)^{j} e_{\med}, \\
Q_{j} \bfGmm^{I} \bfK^{j} \rho_{(\geq \ell) J}
&= - \left[ (\bfGmm+c_{\bfGmm})^{I} Q_{j} - Q_{j} \bfGmm^{I} \right] \bfK^{j} \rho_{(\geq \ell) J}
-\bbS_{(\geq \ell)}(\bfGmm+c_{\bfGmm})^{I} (\bfK+2r)^{j} E_{J} \\
&\peq + \bbS_{(\geq \ell)} (\bfGmm+c_{\bfGmm})^{I} (\bfK+2r)^{j} e_{J; \wave} + \bbS_{(\geq \ell)} (\bfGmm+c_{\bfGmm})^{I} (\bfK+2r)^{j} e_{\med}.
\end{align*}

The remainder $\rho_{J}$ that we discussed so far will be useful for the proof of Proposition~\ref{prop:wave-main} in Cases~1 and 3. However, in Case~2, when we wish to extend the expansion by one order, we need to use a slight modification of $\rho_{J}$. Assume $J \leq \min\set{J_{c}, J_{d}} -1$ so that $\rPhi_{J, k}$ ($0 \leq k \leq K_{J}$) is well-defined. We subtract off improved terms of the form $r^{-j} \log^{k} (\tfrac{r}{u}) \rPhi_{j, k}$ with $j = J$ from $\rho_{J}$ and form
\begin{equation} \label{eq:crho-def}
	\crho_{J} = \rho_{J} - \chi_{> \eta_{0}^{-1}}(\tfrac{r}{u}) \sum_{k=1}^{K_{J}} r^{-J} \log^{k} (\tfrac{r}{u}) \rPhi_{J, k} - r^{-J} \rPhi_{(\leq J - \nu_{\Box}) J, 0}.
\end{equation}

To compute the equation solved by $\crho_{J}$, the following alternative expression is useful:
\begin{equation*}
	\crho_{J} = \rho_{J+1} + r^{-J} \rPhi_{(\geq J - \nu_{\Box} +1) J, 0}.
\end{equation*}
For $\ell \leq J-\nu_{\Box}$, we have $\bbS_{(\ell)} \crho_{J} = \bbS_{(\ell)} \rho_{J+1}$. Hence, in lieu of \eqref{eq:Q0-rhoJ}, we have
\begin{equation} \label{eq:Q0-crhoJ-low}
Q_{0} \crho_{(\ell) J} = - \bbS_{(\ell)} E_{J+1} +\bbS_{(\ell)} e_{J+1; \wave} + \bbS_{(\ell)} e_{\med}.
\end{equation}
On the other hand, for higher spherical harmonics we have
\begin{equation} \label{eq:Q0-crhoJ-high}
\begin{aligned}
Q_{0} \crho_{(\geq J - \nu_{\Box}+1) J}
&= Q_{0} (r^{-J} \rPhi_{(\geq J-\nu_{\Box}+1) J, 0}) - \bbS_{(\geq J - \nu_{\Box}+1)} E_{J+1} \\
&\peq +\bbS_{(\geq J - \nu_{\Box}+1)} e_{J+1; \wave} + \bbS_{(\geq J - \nu_{\Box}+1)} e_{\med}.
\end{aligned}
\end{equation}
Concerning \eqref{eq:Q0-crhoJ-high}, we note the identity
\begin{equation} \label{eq:Q0-crhoJ-high-key}
Q_{0} (r^{-J} \rPhi_{(\geq J-\nu_{\Box}+1) J, 0}) - \bbS_{(\geq J - \nu_{\Box}+1)} E_{J+1; \Box, k=0}
= -2 J r^{-J-1} \rd_{u} \rPhi_{(\geq J-\nu_{\Box}+1) J, 0},
\end{equation}
which cancels after taking $(\bfK+2r)^{J}$ (see \eqref{eq:wave-main-2-rp-Ej+1-cancel} in the proof of Proposition~\ref{prop:wave-main} below).
We also note that $e_{J+1; \wave}$ may be rewritten in terms of $\crho_{J}$ as
\begin{equation} \label{eq:eJ+1wave-crho}
\begin{aligned}
e_{J+1; \wave}
&= -  \chi_{>\eta_{0}^{-1}}\left( \tfrac{r}{u} \right) \left[ \left(\bfh^{\alp \bt} \nb_{\alp} \rd_{\bt} + \bfC^{\alp} \rd_{\alp}  + W\right) \crho_{J} \right]  \\
&\peq +  \chi_{>\eta_{0}^{-1}}\left( \tfrac{r}{u} \right) \left[\left(\bfh^{\alp \bt} \nb_{\alp} \rd_{\bt} + \bfC^{\alp} \rd_{\alp} + W\right)  \left(r^{-J} \rPhi_{(\geq J - \nu_{\Box}+1) J, 0} \right)  \right]  \\
&\relphantom{=} +  \chi_{>\eta_{0}^{-1}}\left( \tfrac{r}{u} \right) \left[ r^{\nu_{\Box}} \calN\left(r^{-\nu_{\Box}} ( \check{\Phi}_{<J} + \crho_{J} ) \right) - r^{\nu_{\Box}} \calN(r^{-\nu_{\Box}} \check{\Phi}_{<J}) \right]  \\
&\relphantom{=} - \chi_{>\eta_{0}^{-1}}\left( \tfrac{r}{u} \right) \left[ r^{\nu_{\Box}} \calN\left(r^{-\nu_{\Box}} ( \check{\Phi}_{<J} + r^{-J} \rPhi_{(\geq J-\nu_{\Box}+1) J, 0}) \right) - r^{\nu_{\Box}} \calN(r^{-\nu_{\Box}} \check{\Phi}_{<J}) \right] \\
& =: \check{e}_{J+1; \wave, \mathrm{linear},1} + \check{e}_{J+1; \wave, \mathrm{linear},2} + \check{e}_{J+1; \wave, \mathrm{nonlinear},1} + \check{e}_{J+1; \wave, \mathrm{nonlinear},2},
\end{aligned}
\end{equation}
where
\begin{equation} \label{eq:cPhi<J}
\check{\Phi}_{<J} = \Phi_{<J} + \sum_{k=1}^{K_{J}} r^{-J} \log^{k}(\tfrac{r}{u}) \rPhi_{J, k} + r^{-J} \rPhi_{(\leq J-\nu_{\Box}) J, 0}.
\end{equation}

Arguing as before, we obtain, for the low spherical harmonics $\ell \leq J - \nB$:
\begin{align*}
Q_{j} \bfGmm^{I} \bfK^{j} \crho_{(\ell) J}
&= - \left[ (\bfGmm+c_{\bfGmm})^{I} Q_{j} - Q_{j} \bfGmm^{I} \right] \bfK^{j} \crho_{(\ell) J}
-\bbS_{(\ell)}(\bfGmm+c_{\bfGmm})^{I} (\bfK+2r)^{j} E_{J+1} \\
&\peq + \bbS_{(\ell)} (\bfGmm+c_{\bfGmm})^{I} (\bfK+2r)^{j} e_{J+1; \wave} + \bbS_{(\ell)} (\bfGmm+c_{\bfGmm})^{I} (\bfK+2r)^{j} e_{\med},
\end{align*}
and for the high spherical harmonics:
\begin{align*}
Q_{j} \bfGmm^{I} \bfK^{j} \crho_{(\geq J-\nu_{\Box}+1) J}
&= - \left[ (\bfGmm+c_{\bfGmm})^{I} Q_{j} - Q_{j} \bfGmm^{I} \right] \bfK^{j} \crho_{(\geq J-\nu_{\Box}+1) J}  \\
&\peq + \bbS_{(\geq J-\nu_{\Box}+1)} (\bfGmm+c_{\bfGmm})^{I} (\bfK+2r)^{j} \left( Q_{0} (r^{-J} \rPhi_{(\geq J-\nu_{\Box}+1) J, 0}) - E_{J+1} \right) \\
&\peq + \bbS_{(\geq J-\nu_{\Box}+1)} (\bfGmm+c_{\bfGmm})^{I} (\bfK+2r)^{j} e_{J+1; \wave} + \bbS_{(\geq J-\nu_{\Box}+1)} (\bfGmm+c_{\bfGmm})^{I} (\bfK+2r)^{j} e_{\med}.
\end{align*}

Finally, we introduce a modification of $\rho_{J}$ that takes into account the strong Huygens principle, which will be useful in the proof of Case~4 of Proposition~\ref{prop:wave-main}. Given $U > 2$, we define $\hrho{U}_{J}$ as
\begin{equation} \label{eq:hrho-J}
	\hrho{U}_{J} = r^{\nu_{\Box}} \Box_{\bfm}^{-1} \left[\chi_{> 1} (\tfrac{u + 2r}{\frac{1}{2} U}) \Box_{\bfm} (\chi_{>2R_{\far}}(r) \chi_{> 2}(u) r^{-\nu_{\Box}}  \rho_{J}) \right],
\end{equation}
using the identification of $\calM_{\far}$ and $\mathbb R\times (\mathbb R^{d} \setminus B_{R_\far})$ via the global coordinates, where $\Box_{\bfm}^{-1} f$ denotes the forward solution to $\Box_{\bfm} \psi = f$. We claim that
\begin{equation} \label{eq:hrho-J-huygens}
	\left. \chi_{>2R_{\far}}(r) \hrho{U}_{J} \right|_{\calC_{U}} = \left. \chi_{>2R_{\far}}(r) \rho_{J}  \right|_{\calC_{U}}.
\end{equation}
Indeed, by Lemma~\ref{lem:huygens} (strong Huygens principle),
\begin{align*}
	\left. \chi_{>2R_{\far}}(r) r^{-\nu_{\Box}}\hrho{U}_{J} \right|_{\calC_{U}}
	&= \left. \Box_{\bfm}^{-1} \left[\chf_{D_{U}} \cdot \chi_{> 1} (\tfrac{u + 2r}{\frac{1}{2} U}) \Box_{\bfm} (\chi_{>2R_{\far}}(r) \chi_{> 2}(u) r^{-\nu_{\Box}}  \rho_{J}) \right] \right|_{\calC_{U}} \\
	&= \left. \Box_{\bfm}^{-1} \left[\chf_{D_{U}} \Box_{\bfm} (\chi_{>2R_{\far}}(r) \chi_{> 2}(u) r^{-\nu_{\Box}}  \rho_{J}) \right] \right|_{\calC_{U}} \\
	&= \left. \Box_{\bfm}^{-1} \left[\Box_{\bfm} (\chi_{>2R_{\far}}(r) \chi_{> 2}(u) r^{-\nu_{\Box}}  \rho_{J}) \right] \right|_{\calC_{U}}
	= \left. \chi_{>2R_{\far}}(r) \chi_{> 2}(u) r^{-\nu_{\Box}}\rho_{J} \right|_{\calC_{U}},
\end{align*}
where we used the fact that $ \chi_{> 1} (\tfrac{u + 2r}{\frac{1}{2} U}) = 1$ on $D_{U}$, as well as the uniqueness of the forward solution.

Using Lemma~\ref{lem:med.incoming}, as well as $\Box_{\bfm} r^{-\nu_{\Box}} = r^{-\nu_{\Box}} Q_{0}$, we obtain the following equation for $\hrho{U}_{J}$:
\begin{equation*}
\begin{split}
	Q_{0} (\hrho{U}_{J})
	= &\: - 2 \chi_{> 1} (\tfrac{u + 2r}{\frac{1}{2} U}) \chi_{>2R_{\far}}(r) \chi_{> 2}'(u) \rd_{r} \rho_{J} \\
	&\: + \chi_{> 1} (\tfrac{u + 2r}{\frac{1}{2} U}) \chi_{> 2}(u) [\Box_{\bfm}, \chi_{>2R_{\far}}](r^{-\nB}\rho_J)
	+ \chi_{> 1} (\tfrac{u + 2r}{\frac{1}{2} U})  \chi_{>2R_{\far}}(r) \chi_{> 2}(u) Q_{0} \rho_{J}.
\end{split}
\end{equation*}
Arguing as before, we obtain
\begin{align*}
	Q_{j} \bfGmm^{I} \bfK^{j} (\hrho{U}_{(\geq \ell) J})
&= - \left[ (\bfGmm+c_{\bfGmm})^{I} Q_{j} - Q_{j} \bfGmm^{I} \right] \bfK^{j} \hrho{U}_{(\geq \ell) J}  \\
&\peq - 2 \bbS_{(\geq \ell)} (\bfGmm + c_{\bfGmm})^{I} (\bfK + 2r)^{j} \left[\chi_{> 1} (\tfrac{u + 2r}{\frac{1}{2} U}) \chi_{>2R_{\far}}(r) \chi_{> 2}'(u) \rd_{r} \rho_{J}\right] \\
&\peq +\bbS_{(\geq \ell)} (\bfGmm + c_{\bfGmm})^{I} (\bfK + 2r)^{j} \left[  \chi_{> 1} (\tfrac{u + 2r}{\frac{1}{2} U}) \chi_{> 2}(u) [\Box_{\bfm}, \chi_{>2R_{\far}}](r^{-\nB}\rho_J) \right]  \\
&\peq
+ \bbS_{(\geq \ell)} (\bfGmm + c_{\bfGmm})^{I} (\bfK + 2r)^{j} \left[ \chi_{> 1} (\tfrac{u + 2r}{\frac{1}{2} U}) \chi_{>2R_{\far}}(r) \chi_{> 2}(u) Q_{0} \rho_{J}\right].
\end{align*}
To estimate the right-hand side of this equation, it is useful to observe that
\begin{equation} \label{eq:hF-Gmm}
	\abs{\bfGmm^{I} (\chi_{> 1} (\tfrac{u + 2r}{\frac{1}{2} U}) F)} \aleq_{\abs{I}} \chf_{\set{u + 2r \geq \frac{1}{4} U}}\abs{\bfGmm^{\leq \abs{I}} F},
\end{equation}
which follows from the simple observation that $\max\set{\frac{r}{U}, \frac{u}{U}} \leq \frac{u + 2 r}{U} \leq \frac{1}{2} $ on $\supp \chi'_{> 1} (\tfrac{u + 2r}{\frac{1}{2} U})$. We remark that such a bound motivates the use of the smooth cutoff $\chi_{> 1}(\tfrac{u+2r}{\frac{1}{2} U})$ in \eqref{eq:hrho-J}.

\subsubsection{Main equations for $\rho_{J}$, $\crho_{J}$ and $\hrho{U}_{J}$}
We are now ready to state the main equations that we shall use to bound $\rho_{J}$ and $\crho_{J}$. For spherical harmonics up to $\ell \leq J - \nu_{\Box}$, we shall choose $j = \ell + \nu_{\Box}-1$ and work with the following:
\begin{lemma} \label{lem:rho-J-char-eqn}
For $\bfGmm^{I} = (r \rd_{r})^{I_{r \rd_{r}}} \bfS^{I_{\bfS}}$ and $\ell \leq J - \nB$, we have
\begin{equation} \label{eq:rho-J-char-eqn}
\begin{aligned}
\left( Q_{\ell + \nu_{\Box}-1}  - I_{r \rd_{r}} r^{-1} \rd_{r} \right)\bfGmm^{I} \bfK^{\ell + \nu_{\Box}-1} \rho_{(\ell) J}
&=  O_{\bfGmm}^{\infty} (r^{-1}) \rd_{r} \bfGmm^{\leq \abs{I}-1} \bfK^{\ell + \nu_{\Box}-1} \rho_{(\ell) J} \\
&\peq	+ O_{\bfGmm}^{\infty} (r^{-2}) \bfGmm^{\leq \abs{I}-1} \bfK^{\ell + \nu_{\Box}-1} \rho_{(\ell) J} \\
&\peq
- (\bfGmm+c_{\bfGmm})^{I} (\bfK+2r)^{\ell + \nu_{\Box}-1} \bbS_{(\ell)} E_{J} \\
&\peq + \bbS_{(\ell)} (\bfGmm+c_{\bfGmm})^{I} (\bfK+2r)^{\ell + \nu_{\Box}-1} e_{J; \wave} \\
&\peq + \bbS_{(\ell)} (\bfGmm+c_{\bfGmm})^{I} (\bfK+2r)^{\ell + \nu_{\Box}-1} e_{\med},
\end{aligned}
\end{equation}
and
\begin{equation} \label{eq:crho-J-char-eqn}
\begin{aligned}
\left( Q_{\ell + \nu_{\Box}-1}  - I_{r \rd_{r}} r^{-1} \rd_{r} \right)\bfGmm^{I} \bfK^{\ell + \nu_{\Box}-1} \crho_{(\ell) J}
&=  O_{\bfGmm}^{\infty} (r^{-1}) \rd_{r} \bfGmm^{\leq \abs{I}-1} \bfK^{\ell + \nu_{\Box}-1} \crho_{(\ell) J} \\
&\peq	+ O_{\bfGmm}^{\infty} (r^{-2}) \bfGmm^{\leq \abs{I}-1} \bfK^{\ell + \nu_{\Box}-1} \crho_{(\ell) J} \\
&\peq
- (\bfGmm+c_{\bfGmm})^{I} (\bfK+2r)^{\ell + \nu_{\Box}-1} \bbS_{(\ell)} E_{J+1} \\
&\peq + \bbS_{(\ell)} (\bfGmm+c_{\bfGmm})^{I} (\bfK+2r)^{\ell + \nu_{\Box}-1} e_{J+1; \wave} \\
&\peq + \bbS_{(\ell)} (\bfGmm+c_{\bfGmm})^{I} (\bfK+2r)^{\ell + \nu_{\Box}-1} e_{\med}.
\end{aligned}
\end{equation}
\end{lemma}
Here, the two important points are that (1) the left-hand sides are of the form to which Lemma~\ref{lem:Qj-char} is applicable and (2) the term $\bbS_{(\ell)} E_{J, \Box}$ is improved in the range $\ell \leq J - \nu_{\Box}$ thanks to the Newman--Penrose cancellation; see Remark~\ref{rem:np-exp-error}. We note the analogous equation for $\hrho{U}_{J}$ is unnecessary for $\ell \leq J - \nu_{\Box}$, since Lemma~\ref{lem:Qj-char} applied to \eqref{eq:rho-J-char-eqn} is already sufficient for the proof of Case~4 of Proposition~\ref{prop:wave-main}.

For spherical harmonics with $\ell \geq J - \nu_{\Box} + 1$, we choose $j = J$ and work with the following:

\begin{lemma} \label{lem:rho-J-rp-eqn}
For $\bfGmm^{I} = (r \rd_{r})^{I_{r \rd_{r}}} \bfS^{I_{\bfS}} \prod_{j, k : j < k} \bfOmg_{jk}^{I_{\bfOmg_{jk}}}$, we have
\begin{equation} \label{eq:rho-J-rp-eqn}
\begin{aligned}
Q_{J} \bfGmm^{I} \bfK^{J} \rho_{(\geq J - \nu_{\Box}+1) J}
&=
I_{r \rd_{r}} (r^{-1} \rd_{r} \bfGmm^{I} + r^{-2} \rslap \bfGmm^{I-(1, 0, \ldots)})\bfK^{J} \rho_{(\geq J - \nu_{\Box}+1) J} \\
&\peq + O_{\bfGmm}^{\infty} (r^{-1}) \rd_{r} \bfGmm^{\leq \abs{I}-1} \bfK^{J} \rho_{(\geq J - \nu_{\Box} + 1) J} \\
&\peq	+ O_{\bfGmm}^{\infty} (r^{-2}) \rslap \bfGmm^{\leq \abs{I}-2} \bfK^{J} \rho_{(\geq J - \nu_{\Box} + 1) J} \\
&\peq
	+ O_{\bfGmm}^{\infty} (r^{-2}) \bfGmm^{\leq \abs{I}-1} \bfK^{J} \rho_{(\geq J - \nu_{\Box} + 1) J} \\
&\peq -\bbS_{(\geq J - \nu_{\Box}+1)}(\bfGmm+c_{\bfGmm})^{I} (\bfK+2r)^{J} E_{J}   \\
&\peq
+ \bbS_{(\geq J - \nu_{\Box}+1)} (\bfGmm+c_{\bfGmm})^{I} (\bfK+2r)^{J} e_{J; \wave}  \\
&\peq + \bbS_{(\geq J - \nu_{\Box}+1)} (\bfGmm+c_{\bfGmm})^{I} (\bfK+2r)^{J} e_{\med},
\end{aligned}
\end{equation}
\begin{equation} \label{eq:crho-J-rp-eqn}
\begin{aligned}
Q_{J} \bfGmm^{I} \bfK^{J} \crho_{(\geq J - \nu_{\Box}+1) J}
&=
I_{r \rd_{r}} (r^{-1} \rd_{r}  \bfGmm^{I} + r^{-2} \rslap \bfGmm^{I-(1, 0, \ldots)})\bfK^{J} \crho_{(\geq J - \nu_{\Box}+1) J} \\
&\peq + O_{\bfGmm}^{\infty} (r^{-1}) \rd_{r} \bfGmm^{\leq \abs{I}-1} \bfK^{J} \crho_{(\geq J - \nu_{\Box} + 1) J} \\
&\peq	+ O_{\bfGmm}^{\infty} (r^{-2}) \rslap \bfGmm^{\leq \abs{I}-2} \bfK^{J} \crho_{(\geq J - \nu_{\Box} + 1) J} \\
&\peq
	+ O_{\bfGmm}^{\infty} (r^{-2}) \bfGmm^{\leq \abs{I}-1} \bfK^{J} \crho_{(\geq J - \nu_{\Box} + 1) J} \\
&\peq + \bbS_{(\geq J - \nu_{\Box}+1)}(\bfGmm+c_{\bfGmm})^{I} (\bfK+2r)^{J} \left( Q_{0} (r^{-J} \rPhi_{(\geq J-\nu_{\Box}+1) J, 0})  - E_{J+1}\right)   \\
&\peq
+ \bbS_{(\geq J - \nu_{\Box}+1)} (\bfGmm+c_{\bfGmm})^{I} (\bfK+2r)^{J} e_{J+1; \wave}  \\
&\peq + \bbS_{(\geq J - \nu_{\Box}+1)} (\bfGmm+c_{\bfGmm})^{I} (\bfK+2r)^{J} e_{\med},
\end{aligned}
\end{equation}
and, for any $U > 1$,
\begin{equation} \label{eq:hrho-J-rp-eqn}
\begin{aligned}
Q_{J} \bfGmm^{I} \bfK^{J} (\hrho{U}_{(\geq J - \nu_{\Box}+1) J})
&=
I_{r \rd_{r}} (r^{-1} \rd_{r} \bfGmm^{I} + r^{-2} \rslap \bfGmm^{I-(1, 0, \ldots)})\bfK^{J} (\hrho{U}_{(\geq J - \nu_{\Box}+1) J}) \\
&\peq + O_{\bfGmm}^{\infty} (r^{-1}) \rd_{r} \bfGmm^{\leq \abs{I}-1} \bfK^{J} (\hrho{U}_{(\geq J - \nu_{\Box}+1) J}) \\
&\peq	+ O_{\bfGmm}^{\infty} (r^{-2}) \rslap \bfGmm^{\leq \abs{I}-2} \bfK^{J} (\hrho{U}_{(\geq J - \nu_{\Box}+1) J}) \\
&\peq
	+ O_{\bfGmm}^{\infty} (r^{-2}) \bfGmm^{\leq \abs{I}-1} \bfK^{J} (\hrho{U}_{(\geq J - \nu_{\Box}+1) J}) \\
&\peq
- 2 (\bfGmm + c_{\bfGmm})^{I} (\bfK + 2r)^{J} \left[\chi_{> 1} (\tfrac{u + 2r}{\frac{1}{2} U}) \chi_{> 2}'(u) \rd_{r} \rho_{(\geq J - \nu_{\Box} +1) J} \right] \\
&\peq +\bbS_{(\geq \ell)} (\bfGmm + c_{\bfGmm})^{I} (\bfK + 2r)^{j} \left[  \chi_{> 1} (\tfrac{u + 2r}{\frac{1}{2} U}) \chi_{> 2}(u) [\Box_{\bfm}, \chi_{>2R_{\far}}](r^{-\nB}\rho_J) \right]  \\
&\peq -\bbS_{(\geq J - \nu_{\Box}+1)}(\bfGmm+c_{\bfGmm})^{I} (\bfK+2r)^{J} \left[ \chi_{> 1} (\tfrac{u + 2r}{\frac{1}{2} U}) \chi_{>2R_{\far}}(r) \chi_{> 2}(u)  E_{J}  \right] \\
&\peq
+ \bbS_{(\geq J - \nu_{\Box}+1)} (\bfGmm+c_{\bfGmm})^{I} (\bfK+2r)^{J} \left[ \chi_{> 1} (\tfrac{u + 2r}{\frac{1}{2} U})\chi_{>2R_{\far}}(r) \chi_{> 2}(u)  e_{J; \wave} \right] \\
&\peq + \bbS_{(\geq J - \nu_{\Box}+1)} (\bfGmm+c_{\bfGmm})^{I} (\bfK+2r)^{J} \left[ \chi_{> 1} (\tfrac{u + 2r}{\frac{1}{2} U}) \chi_{>2R_{\far}}(r) \chi_{> 2}(u)  e_{\med} \right],
\end{aligned}
\end{equation}
where we omit any term involving $\bfGmm^{I'}$ with $\abs{I'} < 0$.
\end{lemma}

Here, the two important points are that (1) the left-hand side is of the form to which Lemma~\ref{lem:Qj-rp} is applicable (Angelopoulos--Aretakis--Gajic positivity; see Remark~\ref{rem:aag-pos}) and (2) the only non-improved term in $E_{J, \Box}$, namely $r^{-J-1} ((j-1)j - (\nu_{\Box}-1)\nu_{\Box} + \rslap) \rPhi_{(\ell) J-1, 0}$, is cancelled by the application of $(\bfK+2r)^{J}$.

Observe that in Lemma~\ref{lem:rho-J-rp-eqn} and Lemma~\ref{lem:rho-J-char-eqn}, we rely on different structures of the equations for the low and high spherical harmonics. A remarkable feature of odd space dimensions, thanks to which our proof works, is the exact complementarity of the above two structures!

Lemmas~\ref{lem:rho-J-char-eqn} and \ref{lem:rho-J-rp-eqn} are quick consequences of the preceding computation and the commutation identity \eqref{eq:comm-Qj}; we omit the details.

\subsection{Proof of the main iteration lemma in the wave zone} \label{subsec:wave-main-pf}
We now put together the ingredients discussed so far and prove Proposition~\ref{prop:wave-main}.

\subsubsection{Conventions}
The following conventions are in effect in this subsection:
\begin{itemize}
\item All implicit constants, as well as $C'$ and $A'$ discussed below, may depend on $d$, $\eta_{0}$, $R_{\far}$, $\alp$ and $J$. Note also that $J$ is determined by $\alp$.

\item In what follows, $C'$ refers to a positive number that may differ from line to line. We furthermore assume that each $C'$ is always greater than or equal to all $C'$'s that occurred before in the proof. The final $C'$ is the $C'_{\wave}$ in Proposition~\ref{prop:wave-main}.

\item In what follows $A' = A'(D, A)$, where $A'(\cdot, \cdot)$ refers to a positive function that obeys \eqref{eq:A'} and may differ from line to line. We furthermore assume that each $A'(\cdot, \cdot)$ is always greater than or equal to (pointwisely) $A'(\cdot, \cdot)$'s that occurred before in the proof. The final $A'(\cdot, \cdot)$ is the $A'_{\wave}(\cdot, \cdot)$ in Proposition~\ref{prop:wave-main}.

\item Abusing the notation a bit, we will denote the projections of $\calC_{U}$, $\calD_{U_{0}}^{U}$, etc., (which are subsets of $\calM_{\far} \subseteq \bbR^{2}_{r, u} \times \bbS^{d-1}_{\tht}$) to $\bbR^{2}_{r, u}$ by the same symbols $\calC_{U}$, $\calD_{U_{0}}^{U}$, etc., respectively.

\item Unless otherwise specified, $U \geq 1$ and $(U, R, \Tht) \in \calM_{\wave}$.
\end{itemize}

\subsubsection{Higher radiation field bounds}
We begin by showing that the $u^{-\alp}$-decay of $\phi$ in $\calM_{\med}$ assumed in Proposition~\ref{prop:wave-main}, together with other hypotheses there, leads to an improved $u^{-j-\alp+\nu_{\Box}}$-decay for $\rPhi_{j, 0}$ with $0 \leq j \leq J - 1$ (at the cost of losing some vector field regularity). This $u$-decay rate shall be the \emph{regular fall-off} for the proof of Proposition~\ref{prop:wave-main} (see Remark~\ref{rem:np-exp}). The main challenge of the bulk of the proof is to obtain an improved fall-off for the remainder.

\begin{lemma} \label{lem:wave-reg-fall}
Assume the hypotheses of Proposition~\ref{prop:wave-main}. Then we have
\begin{equation} \label{eq:wave-rPhi-j0-reg}
	\rPhi_{j, 0} = O_{\bfGmm}^{M - C'} (A u^{j-\alp + \nu_{\Box}}) \quad \hbox{ for } 0 \leq j \leq J - 1.
\end{equation}
\end{lemma}

\begin{proof}
We perform an induction based on the identity
\begin{equation*}
\rPhi_{J'-1, 0}
= \frac{(-1)^{J'-1}}{(J'-1)!}\bfK^{J'-1} \left( \Phi - \chi_{> \eta_{0}^{-1}}(\tfrac{r}{u}) \sum_{j=0}^{J'-1} \sum_{k =1}^{K_{j}} r^{-j} \log^{k} (\tfrac{r}{u}) \rPhi_{j, k} - \rho_{J'} \right),
\end{equation*}
which holds \emph{everywhere} in $\calM_{\far}$.
The base case is $J' = J$. Note that, by our hypotheses, the right-hand side falls off at the rate $u^{-J+1-\alp+\nu_{\Box}}$ in $\calM_{\med} \cap \calM_{\wave}$ (where $r \aeq u$), and hence, so does $\rPhi_{J-1, 0}$. More precisely,
\begin{equation*}
	\rPhi_{J-1, 0} = O_{\bfGmm}^{M-J+1}(A u^{J-1-\alp+\nu_{\Box}}),
\end{equation*}
where the $(J - 1)$-loss of vector field regularity comes from converting vector fields into $r^{-J+1} \bfK^{J-1}$. Since
\begin{align*}
\rho_{J'-1} &= \rho_{J'} + r^{-J'+1} \rPhi_{J'-1, 0} + \chi_{> \eta_{0}^{-1}}(\tfrac{r}{u}) r^{-J_{i'}+1} \log^{k} (\tfrac{r}{u}) \sum_{k=1}^{K_{J'-1}} \rPhi_{J'-1, k},
\end{align*}
it follows that $\rho_{J-1}$ has the corresponding $u$-decay in $\calM_{\med} \cap \calM_{\wave}$, i.e.,
\begin{align*}
\rho_{J-1} = O_{\bfGmm}^{M-J+1}(A u^{-\alp+\nu_{\Box}}) \quad \hbox{ in } \calM_{\med} \cap \calM_{\wave}.
\end{align*}
Hence, we may descend one order and repeat the argument for $J' = J-1$, and so on. In conclusion, we have
\begin{equation*}
	\rPhi_{j,  0} = O_{\bfGmm}^{M- \tfrac{1}{2}(J-1+j)(J-j)} (A u^{j-\alp+\nu_{\Box}}) \quad \hbox{ for } 0 \leq j \leq J - 1. \qedhere
\end{equation*}
\end{proof}

At this point, the higher radiation field bounds in Proposition~\ref{prop:wave-main} follow immediately:
\begin{proof}[Proof of \eqref{eq:wave-main-Phijk}--\eqref{eq:wave-main-Phij0-high} in Proposition~\ref{prop:wave-main}]
Thanks to Lemma~\ref{lem:wave-reg-fall}, we may apply Proposition~\ref{prop:recurrence} with $j = J$ and $\alp = \alp$. By Statement~(1), we obtain the improved bounds for $\rPhi_{j, k}$ for $0 \leq j \leq J - 1$, $1 \leq k \leq K_{j}$ and $\rPhi_{(\leq j - \nu_{\Box}) j, 0}$ stated as \eqref{eq:wave-main-Phijk}--\eqref{eq:wave-main-Phij0-high} in Proposition~\ref{prop:wave-main} with $A' = D + A'(D, A)$ for some $A'$ satisfying \eqref{eq:A'}\footnote{We make a technical comment that we reserve the right to enlarge $A'$ and fix $A'_{\wave}$ until the end of this section.}. \qedhere
\end{proof}

\subsubsection{Remainder bounds}
We are left to establish the remainder bounds \eqref{eq:wave-main-rhoJ}, \eqref{eq:wave-main-rhoJ+1}, \eqref{eq:wave-main-rhoJ-case3} and \eqref{eq:wave-main-rhoJ-final}. To achieve this, we divide the proof according to the cases in Proposition~\ref{prop:wave-main}.

\begin{proof}[Proof of Case~1 ($J \leq \min\set{J_{c}, J_{d}}-1$ and $\alp + \dlt_{0} < J - \nu_{\Box}$)]
We proceed in several steps.

\pfstep{Step~1: Proof of the remainder bound assuming key error bounds}
Our basic approach is to split $\rho_{J} = \rho_{(\geq J - \nu_{\Box} +1) J} + \rho_{(\leq J - \nu_{\Box}) J}$, and use equations \eqref{eq:rho-J-rp-eqn} and \eqref{eq:rho-J-char-eqn}, respectively. To analyze \eqref{eq:rho-J-rp-eqn}, we will use the following error bounds, which will be proved in Step~2: For $\abs{I} \leq M - C'$, with $C'$ to be determined at the end of Step~2, and $- 4 J < p < 1$, which satisfies \eqref{eq:wave-main-1-p-lower1}, \eqref{eq:wave-main-1-p-lower2} and \eqref{eq:wave-main-1-p-lower3} below (the existence of a $p$ obeying these conditions will be shown in Step~2), we have
\begin{align}
 \int_{1}^{U} \left( \int_{\calC_{u} \cap \calM_{\wave}} r^{p} \abs*{\bbS_{(\geq J - \nu_{\Box} + 1)}(\bfGmm + c_{\bfGmm})^{I}(\bfK + 2r)^{J} E_{J} }^{2} \, \ud r \ud \rssgm(\tht) \right)^{\frac{1}{2}}\ud u
 & \aleq A' U^{\frac{p-1}{2} + J - \alp - \dlt_{0} + \nu_{\Box}},  \label{eq:wave-main-1-rp-error-EJ-wave} \\
\left( \iint_{\calD_{1}^{U} \cap \calM_{\med}} r^{p+1} \abs*{\bbS_{(\geq J - \nu_{\Box} + 1)}(\bfGmm + c_{\bfGmm})^{I}(\bfK + 2r)^{J} E_{J}}^{2} \, \ud u \ud r \ud \rssgm(\tht) \right)^{\frac{1}{2}}
&\aleq A' U^{\frac{p-1}{2} + J - \alp - \dlt_{0} + \nu_{\Box}}, \label{eq:wave-main-1-rp-error-EJ-med} \\
 \int_{1}^{U} \left( \int_{\calC_{u} \cap \calM_{\wave}} r^{p} \abs*{\bbS_{(\geq J - \nu_{\Box} + 1)}(\bfGmm + c_{\bfGmm})^{I}(\bfK + 2r)^{J} e_{J; \wave} }^{2} \, \ud r \ud \rssgm(\tht) \right)^{\frac{1}{2}}\ud u
 & \aleq A' U^{\frac{p-1}{2} + J - \alp - \dlt_{0} + \nu_{\Box}}, \label{eq:wave-main-1-rp-error-eJwave-wave} \\
\left( \iint_{\calD_{1}^{U} \cap \calM_{\med}} r^{p+1} \abs*{\bbS_{(\geq J - \nu_{\Box} + 1)}(\bfGmm + c_{\bfGmm})^{I}(\bfK + 2r)^{J} e_{J; \wave}}^{2} \, \ud u \ud r \ud \rssgm(\tht) \right)^{\frac{1}{2}}
&\aleq A' U^{\frac{p-1}{2} + J - \alp - \dlt_{0} + \nu_{\Box}},  \label{eq:wave-main-1-rp-error-eJwave-med} \\
\left( \iint_{\calD_{1}^{U} \cap \calM_{\med}} r^{p+1} \abs*{\bbS_{(\geq J - \nu_{\Box} + 1)}(\bfGmm + c_{\bfGmm})^{I}(\bfK + 2r)^{J} e_{\med}}^{2} \, \ud u \ud r \ud \rssgm(\tht) \right)^{\frac{1}{2}}
&\aleq A' U^{\frac{p-1}{2} + J - \alp - \dlt_{0} + \nu_{\Box}}.  \label{eq:wave-main-1-rp-error-emed}
\end{align}
To analyze \eqref{eq:rho-J-char-eqn}, we will use the following error bounds, which will be proved in Step~3: For $\abs{I} \leq M - C'$, with $C'$ to be determined at the end of Step~3, and $0 \leq \ell \leq J - \nu_{\Box}$, we have
\begin{align}
	& R^{\mu} \int_{1}^{U} \left. r^{-\mu} \nrm*{(\bfGmm + c_{\bfGmm})^{I}(\bfK + 2r)^{\ell+\nu_{\Box}-1} \bbS_{(\ell)} E_{J}}_{L^{\infty}_{\tht}} \right|_{(u, r) = (\sgm, R+\frac{U-\sgm}{2}) }\, \ud \sgm \notag \\
	&\aleq A' R^{\ell+\nu_{\Box}-1}R^{-J-1} \log^{K_{J}} ( \tfrac{R}{U} ) U^{J-\alp-\dlt_{0}+\nu_{\Box}},
 \label{eq:wave-main-1-char-error-EJ} \\
	& R^{\mu} \int_{1}^{U} \left. r^{-\mu} \nrm*{(\bfGmm + c_{\bfGmm})^{I}(\bfK + 2r)^{\ell+\nu_{\Box}-1} \bbS_{(\ell)} e_{J; \wave}}_{L^{\infty}_{\tht}} \right|_{(u, r) = (\sgm, R+\frac{U-\sgm}{2}) }\, \ud \sgm \notag \\
	& \aleq A' R^{\ell+\nu_{\Box}-1}R^{-J-1} \log^{K_{J}} ( \tfrac{R}{U} ) U^{J-\alp-\dlt_{0}+\nu_{\Box}}, \label{eq:wave-main-1-char-error-eJwave}
\end{align}
where $\mu = 2(\ell+\nu_{\Box}-1)+I_{r \rd_{r}}$.
In the remainder of this step, we will assume the above key error bounds and establish \eqref{eq:wave-main-rhoJ}.

\pfstep{Step~1(a): Estimates for high spherical harmonics}
Our aim is to show that, for $\abs{I} \leq M - C'$ and $-4 J < p < 1$ satisfying \eqref{eq:wave-main-1-p-lower1}, \eqref{eq:wave-main-1-p-lower2} and \eqref{eq:wave-main-1-p-lower3} below, we have
\begin{equation} \label{eq:wave-main-1-rp}
\begin{aligned}
& \sup_{u \in [1, U]} \left(\int_{\calC_{u}} \chi_{>2 R_{\far}} r^{p} (\rd_{r} \bfGmm^{I} \bfK^{J} \rho_{(\geq J - \nu_{\Box}+1) J})^{2}  \, \ud r \ud \rssgm(\tht) \right)^{\frac{1}{2}} \\
&+ \left( \iint_{\calD_{1}^{U}} \chi_{>2 R_{\far}} r^{p-1} (\rd_{r} \bfGmm^{I} \bfK^{J} \rho_{(\geq J - \nu_{\Box}+1) J})^{2} \, \ud u \ud r \ud \rssgm(\tht) \right)^{\frac{1}{2}} \\
&+ \left( \iint_{\calD_{U_{0}}^{U}} \chi_{>2 R_{\far}} r^{p-3} \left(\abs{\rsnb \bfGmm^{I} \bfK^{J} \rho_{(\geq J - \nu_{\Box}+1) J}}^{2}  + (\bfGmm^{I} \bfK^{J} \rho_{(\geq J - \nu_{\Box}+1) J})^{2} \right) \, \ud u \ud r \ud \rssgm(\tht) \right)^{\frac{1}{2}} \\
& \aleq A' U^{\frac{p-1}{2}+J-\alp-\dlt_{0}+\nu_{\Box}}.
\end{aligned}
\end{equation}

We will perform an induction on $\abs{I}$, beginning with the base case $\abs{I} = 0$.  In this case, \eqref{eq:rho-J-rp-eqn} becomes:
\begin{equation} \label{eq:rho-J-rp-eqn-I=0}
\begin{aligned}
Q_{J} \bfK^{J} \rho_{(\geq J - \nu_{\Box}+1) J}
&=
-\bbS_{(\geq J - \nu_{\Box}+1)} (\bfK+2r)^{J} E_{J}   \\
&\peq
+ \bbS_{(\geq J - \nu_{\Box}+1)} (\bfK+2r)^{J} e_{J; \wave}  \\
&\peq + \bbS_{(\geq J - \nu_{\Box}+1)} (\bfK+2r)^{J} e_{\med}.
\end{aligned}
\end{equation}
We apply Lemma~\ref{lem:Qj-rp} with $\Psi = \bfK^{J} \rho_{(\geq J - \nu_{\Box}+1) J}$, $I = 0$, $R_{1} = 2 R_{\far}$, $j = J$ and some $- 4 J < p < 1$ satisfying \eqref{eq:wave-main-1-p-lower1}, \eqref{eq:wave-main-1-p-lower2} and \eqref{eq:wave-main-1-p-lower3} below. The contributions of the first and second terms on the right-hand side of \eqref{eq:Qj-rp} are bounded using
\begin{align}
\left( \int_{\calC_{1}} \chi_{>2 R_{\far}} r^{p} (\rd_{r} \bfK^{J} \rho_{(\geq J - \nu_{\Box}+1) J})^{2}  \, \ud r \ud \rssgm(\tht) \right)^{\frac{1}{2}}
& \aleq D, \label{eq:wave-main-1-id-I=0} \\
R_{\far}^{\frac{p-1}{2}} \left( \iint_{\calD_{1}^{U} \cap \set{R_{\far} < r < 2 R_{\far}}} \abs{(\rd_{r}, r^{-1} \rsnb, r^{-1}) \bfK^{J} \rho_{(\geq J - \nu_{\Box}+1) J}}^{2} \, \ud u \ud r \ud \rssgm(\tht) \right)^{\frac{1}{2}}
&\aleq A R_{\far}^{\frac{p-2}{2}+J}. \label{eq:wave-main-1-bdry-I=0}
\end{align}
In the first inequality, we used \ref{hyp:id} and $J_{d} - 1\geq J$. In the second inequality, we used $\bfK^{J} \rho_{J} = \bfK^{J} \Phi$ in $\calM_{\med}$, as well as $\alp > \nu_{\Box} \geq 1$ to integrate in $u$. The contribution of the right-hand side of \eqref{eq:rho-J-rp-eqn-I=0} is handled by \eqref{eq:wave-main-1-rp-error-EJ-wave}--\eqref{eq:wave-main-1-rp-error-emed} (recall also that $e_{\med} = 0$ in $\calM_{\wave}$). Finally, observe that the last two terms on the right-hand side of \eqref{eq:Qj-rp} are zero since $\abs{I} = 0$. Using \eqref{eq:Qj-rp} and \eqref{eq:wave-main-1-p-lower1}, the desired bound \eqref{eq:wave-main-1-rp} when $\abs{I} = 0$ follows.

Next, we consider $0 < \abs{I} \leq M - C'$, where $C'$ is as in \eqref{eq:wave-main-1-rp-error-EJ-wave}--\eqref{eq:wave-main-1-rp-error-emed}. We assume, for the purpose of induction, that \eqref{eq:wave-main-1-rp} has been proved for all $I'$ with $\abs{I'} < \abs{I}$.
We apply Lemma~\ref{lem:Qj-rp} to \eqref{eq:rho-J-rp-eqn} with $\bfGmm^{I} \Psi = \bfGmm^{I}\bfK^{J} \rho_{(\geq J - \nu_{\Box}+1) J}$, $R_{1} = 2 R_{\far}$, $j = J$ and some $- 4 J < p < 1$ satisfying \eqref{eq:wave-main-1-p-lower1}, \eqref{eq:wave-main-1-p-lower2} and \eqref{eq:wave-main-1-p-lower3} below. As before, the contributions of the first and second terms on the right-hand side of \eqref{eq:Qj-rp} are bounded as follows (which are proved similarly as \eqref{eq:wave-main-1-id-I=0} and \eqref{eq:wave-main-1-bdry-I=0}):
\begin{align}
\left( \int_{\calC_{1}} \chi_{>2 R_{\far}} r^{p} (\rd_{r} \bfGmm^{I} \bfK^{J} \rho_{(\geq J - \nu_{\Box}+1) J})^{2}  \, \ud r \ud \rssgm(\tht) \right)^{\frac{1}{2}} \label{eq:wave-main-1-id}
& \aleq D, \\
R_{\far}^{\frac{p-1}{2}} \left( \iint_{\calD_{1}^{U} \cap \set{R_{\far} < r < 2 R_{\far}}} \abs{(\rd_{r}, r^{-1} \rsnb, r^{-1}) \bfGmm^{I} \bfK^{J} \rho_{(\geq J - \nu_{\Box}+1) J}}^{2} \, \ud u \ud r \ud \rssgm(\tht) \right)^{\frac{1}{2}}
&\aleq A R_{\far}^{\frac{p-2}{2}+J}. \label{eq:wave-main-1-bdry}
\end{align}
The contributions of $(\rd_{r}, r^{-1} \rsnb, r^{-1})\bfGmm^{\leq \abs{I}-1} \bfK^{J} \rho_{(\geq J - \nu_{\Box}+1) J}$ on the right-hand side of \eqref{eq:rho-J-rp-eqn} are handled by the induction hypothesis. The contributions of $E_{J}$, $e_{J; \wave}$ and $e_{\med}$ on the right-hand side of \eqref{eq:rho-J-rp-eqn} are handled by \eqref{eq:wave-main-1-rp-error-EJ-wave}--\eqref{eq:wave-main-1-rp-error-emed}. Next, observe that the second to last term on the right-hand side of \eqref{eq:Qj-rp} is bounded by the induction hypothesis. Finally, the last term on the right-hand side of \eqref{eq:Qj-rp} is zero by the induction hypothesis, plus an application of Cauchy--Schwarz to handle $\rd_{r}$. In conclusion, the desired bound \eqref{eq:wave-main-1-rp} for $\bfGmm^{I}  \bfK^{J} \rho_{(\geq J - \nu_{\Box}+1) J}$ follows.

Now that we have proved \eqref{eq:wave-main-1-rp}, we finally derive a bound on $\rho_{(\geq J - \nu_{\Box} +1) J}$.
Since $p < 1$, by \eqref{eq:wave-main-1-rp} (with a change of notation $(U, R, \Tht) \mapsto (u, r, \tht)$) and Lemma~\ref{lem:hardy-radial}, it follows that, for $r \geq 2 R_{\far}$,
\begin{align*}
\abs{\bfGmm^{I} \bfK^{J} \rho_{(\geq J - \nu_{\Box}+1) J}} (u, r, \tht)
&\aleq A' r^{-\frac{p-1}{2}} u^{\frac{p-1}{2} + J - \alp - \dlt_{0} + \nu_{\Box}} + (\tfrac{R_{\far}}{r})^{\frac{p-1}{2}}\abs{\bfGmm^{I}  \bfK^{J} \rho_{(\geq J - \nu_{\Box}+1) J}}(u, 2R_{\far}, \tht) \\
&\aleq A' r^{-\frac{p-1}{2}} u^{\frac{p-1}{2} + J - \alp - \dlt_{0} + \nu_{\Box}},
\end{align*}
where we used $\bfK^{J} \rho_{J} = \bfK^{J} \Phi$ in $\calM_{\med}$ and \eqref{eq:wave-main-hyp-med} in the second inequality. Using \eqref{eq:comm-K-Omg}--\eqref{eq:comm-K-rdr} (which says that $[\bfGmm, \bfK] = c'_{\bfGmm} \bfK$ for some constant $c'_{\bfGmm}$), we may use a simple induction on $\abs{I}$ to show that
\begin{align*}
\abs{\bfK^{J} \bfGmm^{I} \rho_{(\geq J - \nu_{\Box}+1) J}}
\aleq A' r^{-\frac{p-1}{2}} u^{\frac{p-1}{2} + J - \alp - \dlt_{0} + \nu_{\Box}}.
\end{align*}
Note also that, as a consequence of our hypothesis \eqref{eq:wave-main-hyp-rhoJ}, $\bfGmm^{I} \rho_{(\geq J - \nu_{\Box}+1) J} = o(r^{-J+1})$ as $r \to \infty$. By Taylor's theorem in the variable $z = \frac{1}{r}$, it follows that
\begin{align*}
\abs{\bfGmm^{I} \rho_{(\geq J - \nu_{\Box}+1) J}}
\aleq A' r^{-J-\frac{p-1}{2}} u^{\frac{p-1}{2} + J - \alp - \dlt_{0} + \nu_{\Box}}.
\end{align*}
Thanks to \eqref{eq:wave-main-1-p-lower1}, in $\calM_{\wave}$ we may bound $u^{\frac{p-1}{2} + J - \alp - \dlt_{0} + \nu_{\Box}} \aleq r^{\frac{p-1}{2} + J - \alp - \dlt_{0} + \nu_{\Box}}$. In conclusion,
\begin{align*}
\rho_{(\geq J - \nu_{\Box}+1) J}
= O_{\bfGmm}^{M - C'} (A' r^{- \alp - \dlt_{0} + \nu_{\Box}}),
\end{align*}
which is acceptable.

\pfstep{Step~1(b): Estimates for low spherical harmonics}
Let $\ell \in \set{0, \ldots, J - \nu_{\Box}}$. Our aim is to show that, for $0 \leq \abs{I} \leq M - C'$,
\begin{equation} \label{eq:wave-main-1-char}
	\abs{\rd_{r} \bfGmm^{I} \bfK^{\ell+\nu_{\Box}-1} \rho_{(\ell) J}}
	+ r^{-1} \abs{\bfGmm^{I} \bfK^{\ell+\nu_{\Box}-1} \rho_{(\ell) J}}
	\aleq A' r^{\ell+\nu_{\Box}-1} r^{-J-1} \log^{K_{J}} (\tfrac{r}{u})u^{J-\alp-\dlt_{0}+\nu_{\Box}}.
\end{equation}

As in Step~1(a), we perform an induction on $\abs{I}$, beginning with the case $\abs{I} = 0$. In this case, \eqref{eq:rho-J-char-eqn} becomes
\begin{equation} \label{eq:rho-J-char-eqn-I=0}
\begin{aligned}
Q_{\ell+\nu_{\Box} - 1} \bfK^{\ell+\nu_{\Box}-1} \rho_{(\ell) J}
&=
- (\bfK+2r)^{\ell + \nu_{\Box}-1} \bbS_{(\ell)} E_{J} \\
&\peq + \bbS_{(\ell)} (\bfK+2r)^{\ell + \nu_{\Box}-1} e_{J; \wave} \\
&\peq + \bbS_{(\ell)} (\bfK+2r)^{\ell + \nu_{\Box}-1} e_{\med}.
\end{aligned}
\end{equation}
We apply Lemma~\ref{lem:Qj-char} with $s = 0$. The contribution of the first term on the right-hand side of \eqref{eq:Qj-char} is bounded using \ref{hyp:id} (and $J-\alp_{d}-\dlt_{0}+\nu_{\Box})>0$ as follows:
\begin{align*}
	\abs*{ \left. (\tfrac{R}{r})^{\mu} \rd_r\bfK^{\ell+\nu_{\Box}-1} \rho_{(\ell) J}  \right|_{(u, r, \tht) = (1, R+\frac{U-1}{2}, \Tht)} }
	\aleq D R^{\ell+\nu_{\Box}-1} R^{-J-1} \log^{K_{d}} ( \tfrac{R}{U} ) U^{J-\alp_{d}-\dlt_{0}+\nu_{\Box}},
\end{align*}
where $\mu = 2(\ell+\nu_{\Box}-1)$. The contribution of the right-hand side of \eqref{eq:rho-J-char-eqn-I=0} is handled by \eqref{eq:wave-main-1-char-error-EJ} and \eqref{eq:wave-main-1-char-error-eJwave} (since $e_{\med} = 0$ in $\calM_{\wave}$). We conclude that
\begin{equation} \label{eq:wave-main-1-char-I=0-dr}
	\abs{\rd_{r} \bfK^{\ell+\nu_{\Box}-1} \rho_{(\ell) J}} \aleq A' r^{\ell+\nu_{\Box}-1} r^{-J-1} \log^{K_{J}} (\tfrac{r}{u})u^{J-\alp-\dlt_{0}+\nu_{\Box}}.
\end{equation}
By \eqref{eq:wave-main-hyp-rhoJ}, $\bfK^{\ell+\nu_{\Box}-1} \rho_{(\ell) J} = o(r^{-J+\ell+\nu_{\Box}})$ as $r \to \infty$. Using the fundamental theorem of calculus to write
\begin{equation*}
\bfK^{\ell+\nu_{\Box}-1} \rho_{(\ell) J}(u, r, \tht) = - \int_{r}^{\infty} \rd_{r} \bfK^{\ell+\nu_{\Box}-1} \rho_{(\ell) J}(u, r', \tht) \, \ud r',
\end{equation*}
and bounding the right-hand side using \eqref{eq:wave-main-1-char-I=0-dr}, we obtain
\begin{equation*}
	\abs*{\bfK^{\ell+\nu_{\Box}-1} \rho_{(\ell) J}} \aleq A' r^{\ell+\nu_{\Box}-1} r^{-J}\log^{K_{J}} (\tfrac{r}{u}) u^{J-\alp-\dlt_{0}+\nu_{\Box}},
\end{equation*}
which finishes the proof of \eqref{eq:wave-main-1-char} when $I = 0$.

Next, we consider $0 < \abs{I} \leq M - C'$, where $C'$ is as in \eqref{eq:wave-main-1-char-error-EJ}--\eqref{eq:wave-main-1-char-error-eJwave}. We assume, for the purpose of induction, that \eqref{eq:wave-main-1-char} has been proved for all $I'$ with $\abs{I'} < \abs{I}$.
We apply Lemma~\ref{lem:Qj-char} to \eqref{eq:rho-J-char-eqn} with $\Psi = \bfGmm^{I} \bfK^{\ell+\nu_{\Box}-1} \rho_{(\ell) J}$, $j = J$ and $s = I_{r \rd_{r}}$. As before, the contribution of the first term on the right-hand side of \eqref{eq:Qj-char} is bounded as follows:
\begin{align*}
	\abs*{ \left. (\tfrac{R}{r})^{\mu} \rd_{r} \bfGmm^{I} \bfK^{\ell+\nu_{\Box}-1} \rho_{(\ell) J}  \right|_{(u, r, \tht) = (1, R+\frac{U-1}{2}, \Tht)} }
	\aleq D R^{\ell+\nu_{\Box}-1} R^{-J-1} \log^{K_{d}} ( \tfrac{R}{U} ) U^{J-\alp_{d}-\dlt_{d}+\nu_{\Box}},
\end{align*}
where $\mu = 2(\ell+\nu_{\Box}-1)+I_{r \rd_{r}}$.
For the contributions of $(\rd_{r}, r^{-1})\bfGmm^{\leq \abs{I}-1} \bfK^{\ell+\nu_{\Box}-1} \rho_{(\ell) J}$ on the right-hand side of \eqref{eq:rho-J-char-eqn}, we have
\begin{align*}
& R^{\mu} \int_{1}^{U} \left. r^{-\mu-1} \nrm*{(\rd_{r}, r^{-1})\bfGmm^{\leq \abs{I}-1}\bfK^{\ell+\nu_{\Box}-1} \rho_{(\ell) J}}_{L^{\infty}_{\tht}} \right|_{(u, r) = (\sgm, R+\frac{U-\sgm}{2}) }\, \ud \sgm \notag \\
	& \aleq A' R^{\mu} \int_{1}^{U} (R + \tfrac{U-u}{2})^{-\mu+\ell-J+\nu_{\Box}-3} \log^{K_{J}} \left( \tfrac{R + \frac{U-u}{2}}{u} \right) u^{J-\alp-\dlt_{0}+\nu_{\Box}} \, \ud u \\
	& \aleq A' R^{\ell+\nu_{\Box}-1} R^{-J-1} \log^{K_{J}} ( \tfrac{R}{U} ) U^{J-\alp-\dlt_{0}+\nu_{\Box}}
\end{align*}
where we used the induction hypothesis. The contributions of $E_{J}$ and $e_{J; \wave}$ on the right-hand side of \eqref{eq:rho-J-rp-eqn} are handled by \eqref{eq:wave-main-1-char-error-EJ}--\eqref{eq:wave-main-1-char-error-eJwave}, whereas $e_{\med} = 0$ in our context (see Remark~\ref{rem:Qj-char-supp}). In conclusion, we obtain
\begin{equation*}
	\abs{\rd_{r} \bfGmm^{I} \bfK^{\ell+\nu_{\Box}-1} \rho_{(\ell) J}} \aleq A' r^{\ell+\nu_{\Box}-1} r^{-J-1} \log^{K_{J}} (\tfrac{r}{u})u^{J-\alp-\dlt_{0}+\nu_{\Box}}.
\end{equation*}
Arguing as in the case $\abs{I} = 0$ using the fundamental theorem of calculus and $\bfGmm^{I} \bfK^{\ell+\nu_{\Box}-1} \rho_{(\ell) J} = o(r^{-J+\ell+\nu_{\Box}})$ as $r \to \infty$, we obtain
\begin{equation*}
	\abs*{\bfGmm^{I} \bfK^{\ell+\nu_{\Box}-1} \rho_{(\ell) J}} \aleq A' r^{\ell+\nu_{\Box}-1} r^{-J} \log^{K_{J}} (\tfrac{r}{u}) u^{J-\alp-\dlt_{0}+\nu_{\Box}},
\end{equation*}
which finishes the proof of \eqref{eq:wave-main-1-char}.

Finally, let us derive a bound on $\rho_{(\leq J - \nu_{\Box}) J}$. Beginning with \eqref{eq:wave-main-1-char}, and using a simple induction argument on $\abs{I}$ based on the commutator identities \eqref{eq:comm-K-Omg}--\eqref{eq:comm-K-rdr}, we may obtain
\begin{equation*}
	\abs*{\bfK^{\ell+\nu_{\Box}-1} \bfGmm^{I} \rho_{(\ell) J}} \aleq A' r^{\ell+\nu_{\Box}-1} r^{-J} \log^{K_{J}} (\tfrac{r}{u}) u^{J-\alp-\dlt_{0}+\nu_{\Box}}.
\end{equation*}
Recall that, by \eqref{eq:wave-main-hyp-rhoJ}, $\bfGmm^{I} \rho_{(\ell) J} = o(r^{-J+1})$ as $r \to \infty$. By Taylor's theorem in the variable $z = \frac{1}{r}$, we have
\begin{align*}
\abs{\bfGmm^{I} \rho_{(\leq J - \nu_{\Box}) J}}
\aleq A' r^{-J} \log^{K_{J}} (\tfrac{r}{u}) u^{J-\alp-\dlt_{0}+\nu_{\Box}}.
\end{align*}
The desired remainder bound now follows since $\frac{r}{u} \ageq 1$ in $\calM_{\wave}$.

\pfstep{Step~2: Proof of error bounds for high spherical harmonics}
In this step, we establish \eqref{eq:wave-main-1-rp-error-EJ-wave}--\eqref{eq:wave-main-1-rp-error-emed}. We divide the integrated region into $\calM_{\med} \cup \calM_{\wave}$ and argue separately.

In what follows, we freely use
\begin{equation*}
	\int \abs{\bbS_{(\geq \ell)} g}^{2} \, \ud \rssgm(\tht) \aleq \int \abs{g}^{2} \, \ud \rssgm(\tht) \aleq \nrm{g}_{L^{\infty}_{\tht}}^{2},
\end{equation*}
in order to remove the projection $\bbS_{(\geq \ell)}$.

\pfstep{Step~2(a): Error estimate in $\calM_{\wave}$}
In this substep, we prove \eqref{eq:wave-main-1-rp-error-EJ-wave} and \eqref{eq:wave-main-1-rp-error-eJwave-wave}.
We begin with the contribution of $E_{J}$. In view of \eqref{eq:exp-error-wave}, we split $E_{J} = E_{J; \Box, k=0} + E_{J; f} + (E_{J} - E_{J; \Box, k=0} - E_{J; f})$. As discussed in Remark~\ref{rem:exp-error-wave}, the term $(\bfGmm + c_{\bfGmm})^{I}(\bfK + 2r)^{J} E_{J; \Box, k=0}$ (see \eqref{eq:EJ-Box-k=0}) vanishes thanks to the crucial cancellation $(\bfK + 2 r)^{J} r^{-J-1} = 0$!

To control the other two terms, we choose
\begin{equation} \label{eq:wave-main-1-p-lower1}
	p > 1 - 2 (J - \alp - \dlt_{0} + \nu_{\Box}).
\end{equation}

For $E_{J; f}$, we use \eqref{eq:EJ-f}, \eqref{eq:EJ-f-rem}, and \ref{hyp:forcing}. For $\abs{I} \leq M_{0} - J$,
\begin{align*}
& \int_{1}^{U} \left( \int_{\calC_{u} \cap \calM_{\wave}} r^{p} \abs*{\bbS_{(\geq J - \nu_{\Box} + 1)}(\bfGmm + c_{\bfGmm})^{I}(\bfK + 2r)^{J} E_{J; f} }^{2} \, \ud r \ud \rssgm(\tht) \right)^{\frac{1}{2}} \ud u \\
& \aleq \int_{1}^{U} \left( \int_{\calC_{u} \cap \calM_{\wave}} r^{p} \left(D r^{J} r^{-J-1} \log^{K_{d}}(\tfrac{r}{u}) u^{J-1-\alp_{d}-\dlt_{d}+\nu_{\Box}} \right)^{2} \, \ud r \right)^{\frac{1}{2}} \ud u \\
& \peq + \int_{1}^{U} \left( \int_{\calC_{u} \cap \calM_{\wave}} r^{p} \left( D r^{J-1-\alp_{d}+\nu_{\Box}} u^{-1-\dlt_{d}} \right)^{2} \, \ud r \right)^{\frac{1}{2}} \ud u \\
& \aleq \int_{1}^{U} D u^{\frac{p-1}{2} + J - 1 - \alp_{d} - \dlt_{d} + \nu_{\Box}} \, \ud u,
\end{align*}
where we used $p < 1$ and $J \leq J_{d} - 1 \leq \alp_{d} - \nu_{\Box}$ for the last inequality. Since $\alp \leq J + \nu_{\Box} \leq \alp_{d}$ and $\dlt_{0} \leq \dlt_{d}$, the last line may be bounded by $C D U^{\frac{p-1}{2} + J - \alp - \dlt_{0} + \nu_{\Box}}$ if $p$ obeys \eqref{eq:wave-main-1-p-lower1}.

For $E_{J} - E_{J; \Box,k=0} - E_{J; f}$, we use \eqref{eq:exp-error-wave}, so that when $\abs{I} \leq M - C' - J$,
\begin{align*}
& \int_{1}^{U} \left( \int_{\calC_{u} \cap \calM_{\wave}} r^{p} \abs*{\bbS_{(\geq J - \nu_{\Box} + 1)}(\bfGmm + c_{\bfGmm})^{I}(\bfK + 2r)^{J} \left[ E_{J} - E_{J; \Box,k=0} - E_{J; f} \right]}^{2} \, \ud r \ud \rssgm(\tht) \right)^{\frac{1}{2}} \ud u \\
& \aleq \int_{1}^{U} \left( \int_{\calC_{u} \cap \calM_{\wave}} r^{p} \left( A' r^{-1} \log^{K_{J}}(\tfrac{r}{u}) u^{J-1-\alp-\dlt_{0}+\nu_{\Box}} \right)^{2} \, \ud r \right)^{\frac{1}{2}} \ud u  \\
& \peq + \int_{1}^{U} \left( \int_{\calC_{u} \cap \calM_{\wave}} r^{p} \left( A' r^{J-J_{c}-\eta_{c}} u^{J_{c}-2+\eta_{c}-\alp-\dlt_{0}+\nu_{\Box}} \right)^{2} \, \ud r \right)^{\frac{1}{2}} \ud u \\
& \aleq A' \int_{1}^{U} u^{\frac{p-1}{2} + J-1-\alp-\dlt_{0}+\nu_{\Box}} \, \ud u,
\end{align*}
where we used $p < 1$ and $J \leq J_{c}-1$ for the last inequality. As before, the last line may be bounded by $C A' U^{\frac{p-1}{2} + J - \alp - \dlt_{0} + \nu_{\Box}}$ if $p$ obeys \eqref{eq:wave-main-1-p-lower1}.

Next, we treat the contribution of $e_{J; \wave}$, which is given by \eqref{eq:eJwave-def}. We begin with $e_{J; \wave, \linear}$ (see \eqref{eq:eJwave-linear-def}). We use Lemma~\ref{lem:conj-wave-wave} and the hypothesis \eqref{eq:wave-main-hyp-rhoJ} on $\rho_{J}$. Hence, for $\abs{I} \leq M - J - 2$,
\begin{align*}
& \int_{1}^{U} \left( \int_{\calC_{u} \cap \calM_{\wave}} r^{p} \abs*{\bbS_{(\geq J - \nu_{\Box} + 1)}(\bfGmm + c_{\bfGmm})^{I}(\bfK + 2r)^{J} e_{J; \wave, \linear}}^{2} \, \ud r \ud \rssgm(\tht)\right)^{\frac{1}{2}} \ud u \\
& = \int_{1}^{U} \left( \int_{\calC_{u} \cap \calM_{\wave}} r^{p} \abs*{\bbS_{(\geq J - \nu_{\Box} + 1)}(\bfGmm + c_{\bfGmm})^{I}(\bfK + 2r)^{J} \left[ \bfh^{\alp \bt} \nb_{\alp} \rd_{\bt} \rho_{J} + \bfC^{\alp} \rd_{\alp} \rho_{J} + W \rho_{J} \right]}^{2} \, \ud r \ud \rssgm(\tht)\right)^{\frac{1}{2}} \ud u \\
& \aleq \int_{1}^{U} \left( \int_{\calC_{u} \cap \calM_{\wave}} r^{p} \left( A r^{J} r^{-2} \log^{K_{c}'} (\tfrac{r}{u}) u^{-\dlt_{c}} r^{-\alp+\nu_{\Box}} \right)^{2} \, \ud r \right)^{\frac{1}{2}} \ud u  \\
& \aleq \int_{1}^{U} A u^{\frac{p-1}{2} + J - 1 - \alp - \dlt_{c} +\nu_{\Box}} \, \ud u,
\end{align*}
where we used $p < 1$ and $J - 1 - \alp + \nu_{\Box} < 0$ for the last inequality. Since $\dlt_{0} \leq \dlt_{c}$, this term may also be bounded by $C A U^{\frac{p-1}{2} + J - \alp - \dlt_{0} + \nu_{\Box}}$ if $p$ obeys \eqref{eq:wave-main-1-p-lower1}.

Next, we handle $e_{J; \wave, \nonlinear}$, which is defined in \eqref{eq:eJwave-nonlinear-def}.
Recall from Section~\ref{subsec:nonlin-est} that we may write
\begin{equation*}
\begin{aligned}
& e_{J; \wave, \nonlinear} \\
&=  \chi_{>\eta_{0}}(\tfrac{r}{u}) \left[ \bfh^{\alp \bt}_{\calN}(r^{-\nu_{\Box}} \Phi_{<J}; r^{-\nu_{\Box}} \rho_{J}) \nb_{\alp} \rd_{\bt} \rho_{J} + \bfC^{\alp}_{\calN}(r^{-\nu_{\Box}} \Phi_{<J}; r^{-\nu_{\Box}} \rho_{J}) \rd_{\alp} \rho_{J} + W_{\calN}(r^{-\nu_{\Box}} \Phi_{<J}; r^{-\nu_{\Box}} \rho_{J}) \rho_{J}\right].
\end{aligned}\end{equation*}
Since $\alp \geq \alp_{\calN} + 2 \dlt_{0}$, we may apply the estimates in Definition~\ref{def:alp-N}.(3) (with $\alp_{\calN}' = \alp_{\calN} + \dlt_{0}$). Using also the hypotheses for $\rPhi_{j, k}$ and $\rho_{J}$, for $\abs{I} \leq M - J - 2$, we have
\begin{align*}
& \int_{1}^{U} \left( \int_{\calC_{u} \cap \calM_{\wave}} r^{p} \abs*{\bbS_{(\geq J - \nu_{\Box} + 1)}(\bfGmm + c_{\bfGmm})^{I}(\bfK + 2r)^{J} e_{J; \wave, \nonlinear}}^{2} \, \ud r \ud \rssgm(\tht) \right)^{\frac{1}{2}} \ud u \\
& \aleq \int_{1}^{U} \left( \int_{\calC_{u} \cap \calM_{\wave}} r^{p} \left( A_{\calN}(A) r^{J} r^{-2} u^{-\dlt_{0}} r^{-\alp+\nu_{\Box}} \right)^{2} \, \ud r \right)^{\frac{1}{2}} \ud u  \\
& \aleq \int_{1}^{U} A_{\calN}(A) u^{\frac{p-1}{2} + J - 1 - \alp - \dlt_{0} +\nu_{\Box}} \, \ud u,
\end{align*}
where we used $p < 1$ and $J - 1 - \alp + \nu_{\Box} < 0$ for the last inequality. Provided that $p$ obeys \eqref{eq:wave-main-1-p-lower1}, this term may also be bounded by $C A U^{\frac{p-1}{2} + J - \alp - \dlt_{0} + \nu_{\Box}}$.

\pfstep{Step~2(b): Error estimate in $\calM_{\med}$}
In this substep, we prove \eqref{eq:wave-main-1-rp-error-EJ-med}, \eqref{eq:wave-main-1-rp-error-eJwave-med} and \eqref{eq:wave-main-1-rp-error-emed}.
As before, we begin with the contribution of $E_{J}$. In view of \eqref{eq:exp-error-med}, we split $E_{J} = Q_{0}(\Phi_{<J, k=0}) + (E_{J} - Q_{0}(\Phi_{<J, k=0}))$. As discussed in Remark~\ref{rem:exp-error-med}, the contribution of $Q_{0}(\Phi_{<J, k=0})$ vanishes, i.e.,
\begin{align*}
	(\bfK+2r)^{J} \left( Q_{0}(\Phi_{<J, k=0})  \right) = 0,
\end{align*}
since $(\bfK+2r)^{J} r^{-j} \mathring{a}(u, \tht)$ for $2 \leq j \leq J+1$, and $Q_{0}(\Phi_{<J, k=0})$ consists entirely of such terms. Hence, by \eqref{eq:exp-error-med},
\begin{align*}
& \left( \iint_{\calD_{1}^{U} \cap \calM_{\med}} r^{p+1} \abs*{\bbS_{(\geq J - \nu_{\Box} + 1)}(\bfGmm + c_{\bfGmm})^{I}(\bfK + 2r)^{J} E_{J}}^{2} \, \ud u \ud r \ud \rssgm(\tht)  \right) ^{\frac{1}{2}} \\
& \aleq \left( \iint_{\calD_{1}^{U} \cap \set{r \aeq u}} u^{p+1} \left( A' u^{J} u^{-\alp - 2 - \dlt_{0} + \nu_{\Box}}  \right)^{2} \, \ud u \ud r \right)^{\frac{1}{2}} \\
& \peq + \left( \iint_{\calD_{1}^{U} \cap \set{r \aeq u}} u^{p+1} \left( D u^{J} u^{-\alp_{d} - 2 - \dlt_{d} + \nu_{\Box}}  \right)^{2} \, \ud u \ud r \right)^{\frac{1}{2}} \\
& \aleq \left(\int_{1}^{U} \left(A' u^{\frac{p-1}{2}} u^{J - \alp - \frac{1}{2} - \dlt_{0} + \nu_{\Box}}\right)^{2} \, \ud u \right)^{\frac{1}{2}} + \left(\int_{1}^{U} \left(D u^{\frac{p-1}{2}} u^{J - \alp - \frac{1}{2} - \dlt_{d} + \nu_{\Box}}\right)^{2} \, \ud u \right)^{\frac{1}{2}}.
\end{align*}
Since $\dlt_{0} \leq \dlt_{d}$, the last line may be bounded by $C A' U^{\frac{p-1}{2} + J - \alp - \dlt_{0} + \nu_{\Box}}$ if $p$ obeys \eqref{eq:wave-main-1-p-lower1}.

We now handle the contribution of $e_{J; \wave}$ in \eqref{eq:eJwave-def}, which is supported in $\set{r \geq (2\eta_{0})^{-1} u}$. In $\calM_{\med} \cap \set{r \geq (2\eta_{0})^{-1} u}$, we have $r \aeq u$ and thus by Lemma~\ref{lem:wave-reg-fall} and \eqref{eq:wave-main-hyp-med},
\begin{equation*}
	\rho_{J} = \Phi - \Phi_{<J} = O_{\bfGmm}^{M - C'}(A u^{-\alp+\nu_{\Box}}).
\end{equation*}
From now on the argument is similar to that in $\calM_{\wave}$, suitably modified for $\calM_{\med}$: we use the above estimate with Lemma~\ref{lem:conj-wave-med} for $e_{J; \wave, \linear}$; whereas we use Definition~\ref{def:alp-N}.(2) with $\alp_{\calN}' = \alp_{\calN} + \dlt_{0}$ and Lemma~\ref{lem:wave-reg-fall} and \eqref{eq:wave-main-hyp-med} for $e_{J; \wave, \nonlinear}$. 
In conclusion, for $\abs{I} \leq M - J - C' - 2$, we have
\begin{align*}
& \left(\iint_{\calD_{1}^{U} \cap \calM_{\med}} r^{p+1} \abs*{\bbS_{(\geq J - \nu_{\Box} + 1)}(\bfGmm + c_{\bfGmm})^{I}(\bfK + 2r)^{J} (e_{J; \wave, \linear} + e_{J; \wave, \nonlinear})}^{2} \, \ud u \ud r \ud \rssgm(\tht) \right)^{\frac{1}{2}} \\
& \aleq \left(\iint_{\calD_{1}^{U} \cap \set{r \aeq u}} u^{p+1} \left( A u^{J} u^{-2} u^{-\dlt_{c}} u^{-\alp+\nu_{\Box}} \right)^{2} + u^{p+1} \left( A_{\calN}(A) u^{J} u^{-2} u^{-\dlt_{0}} u^{-\alp+\nu_{\Box}} \right)^{2} \, \ud u \ud r \right)^{\frac{1}{2}} \\
& \aleq \left(\int_{1}^{U} \left(A' u^{\frac{p-1}{2}} u^{J - \alp - \frac{1}{2} - \dlt_{0} + \nu_{\Box}}\right)^{2} \, \ud u \right)^{\frac{1}{2}}.
\end{align*}
where we used $\dlt_{0} \leq \dlt_{c}$ in the last inequality. The last line may be bounded by $C A' U^{\frac{p-1}{2} + J - \alp - \dlt_{0} + \nu_{\Box}}$ if $p$ obeys \eqref{eq:wave-main-1-p-lower1}.

It remains to estimate the contribution of $e_{\med}$, which is given by \eqref{eq:eJmed-def}. We first treat the contribution of the non-Minkowskian linear coefficients $e_{\med, \linear}$ given by \eqref{eq:eJmed-linear-def}. Using Lemma~\ref{lem:conj-wave-med} and the hypothesis for $\Phi$, for $\abs{I} \leq M - J - 2$, we have,
\begin{align*}
& \left( \iint_{\calD_{1}^{U} \cap \calM_{\med}} r^{p+1} \abs*{\bbS_{(\geq J - \nu_{\Box} + 1)}(\bfGmm + c_{\bfGmm})^{I}(\bfK + 2r)^{J} e_{\med, \linear}}^{2} \, \ud u \ud r \ud \rssgm(\tht) \right)^{\frac{1}{2}} \\
& \aleq \left( \iint_{\calD_{1}^{U} \cap \calM_{\med}} r^{p+1} \left( A r^{J} r^{-2-\dlt_{c}} r^{\nu_{\Box}} u^{-\alp} \right)^{2} \, \ud u \ud r \right)^{\frac{1}{2}}  \\
& \aleq \left(\int_{1}^{U} \left(A u^{\frac{p-1}{2}} u^{J - \alp - \frac{1}{2} - \dlt_{c} + \nu_{\Box}}\right)^{2} \, \ud u \right)^{\frac{1}{2}},
\end{align*}
where, for the last inequality, we need to impose
\begin{equation} \label{eq:wave-main-1-p-lower2}
	p > 2 - 2(J - \dlt_{c} +\nu_{\Box} ).
\end{equation}
Since $J \geq 1$, $\nu_{\Box} \geq 1$ and $\dlt_{c} \leq 1$, \eqref{eq:wave-main-1-p-lower2} holds if $p > 0$. Since $\dlt_{0} \leq \dlt_{c}$, the last line may be bounded by $C A U^{\frac{p-1}{2} + J - \alp - \dlt_{0} + \nu_{\Box}}$ if $p$ obeys \eqref{eq:wave-main-1-p-lower1} as well.

Next, we treat the contribution of the nonlinearity $e_{\med, \nonlinear}$ given by \eqref{eq:eJmed-nonlinear-def}. Recall from Section~\ref{subsec:nonlin-est} that we may write
\begin{equation*}
e_{\med, \nonlinear} = \left( 1-\chi_{>\eta_{0}^{-1}}\left( \tfrac{r}{u} \right) \right) \left[ \bfh^{\alp \bt}_{\calN}(0; r^{-\nu_{\Box}} \Phi) \nb_{\alp} \rd_{\bt} \Phi + \bfC^{\alp}_{\calN}(0; r^{-\nu_{\Box}} \Phi) \rd_{\alp} \Phi + W_{\calN}(r^{-\nu_{\Box}} 0; r^{-\nu_{\Box}} \Phi) \Phi \right].
\end{equation*}
We apply the estimates in Definition~\ref{def:alp-N}.(2) with $\alp_{\calN}' = \alp_{\calN} + \dlt_{0}$ (which is possible since $\alp_{\calN} \geq 2 \dlt_{0})$. Using also the hypothesis for $\Phi$, for $\abs{I} \leq M - J - 2$, we have
\begin{align*}
& \left( \iint_{\calD_{1}^{U} \cap \calM_{\med}} r^{p+1} \abs*{\bbS_{(\geq J - \nu_{\Box} + 1)}(\bfGmm + c_{\bfGmm})^{I}(\bfK + 2r)^{J}  e_{\med, \nonlinear}}^{2} \, \ud u \ud r \ud \rssgm(\tht) \right)^{\frac{1}{2}}\\
& \aleq \left( \iint_{\calD_{1}^{U} \cap \calM_{\med}} r^{p+1} \left( A' r^{J} r^{-2-\dlt_{0}} r^{\nu_{\Box}} u^{-\alp} \right)^{2} \, \ud u \ud r \right)^{\frac{1}{2}}  \\
& \aleq \left(\int_{1}^{U} \left(A' u^{\frac{p-1}{2}} u^{J - \alp - \frac{1}{2} - \dlt_{0} + \nu_{\Box}}\right)^{2} \, \ud u \right)^{\frac{1}{2}}.
\end{align*}
As in the case of $e_{\med, \linear}$, the last line may be bounded by $C A U^{\frac{p-1}{2} + J - \alp - \dlt_{0} + \nu_{\Box}}$ if $p$ obeys $p > 0$ and \eqref{eq:wave-main-1-p-lower1}.

Finally, for the contribution of $e_{\med, f}$ given by \eqref{eq:eJmed-f-def}, we use \ref{hyp:forcing}. For $\abs{I} \leq M_{0} - J$, we have
\begin{align*}
& \left( \iint_{\calD_{1}^{U} \cap \calM_{\med}} r^{p+1} \abs*{\bbS_{(\geq J - \nu_{\Box} + 1)}(\bfGmm + c_{\bfGmm})^{I}(\bfK + 2r)^{J} e_{\med, f}}^{2} \, \ud u \ud r \ud \rssgm(\tht) \right)^{\frac{1}{2}} \\
& \aleq \left( \iint_{\calD_{1}^{U} \cap \calM_{\med}} r^{p+1} \left( D r^{J} r^{-2-\dlt_{d}+\nu_{\Box}} u^{-\alp_{d}} \right)^{2} \, \ud u \ud r \right)^{\frac{1}{2}}  \\
& \aleq \left(\int_{1}^{U} \left(D u^{\frac{p-1}{2}} u^{J - \alp - \frac{1}{2} - \dlt_{d} + \nu_{\Box}}\right)^{2} \, \ud u \right)^{\frac{1}{2}},
\end{align*}
where, for the last inequality, we need to impose
\begin{equation} \label{eq:wave-main-1-p-lower3}
	p > 2 - 2(J - \dlt_{d} +\nu_{\Box} ).
\end{equation}
Since $J \geq 1$, $\nu_{\Box} \geq 1$ and $\dlt_{d} \leq 1$, \eqref{eq:wave-main-1-p-lower3} holds if $p > 0$. Since $\dlt_{0} \leq \dlt_{d}$, the last line may be bounded by $C D U^{\frac{p-1}{2} + J - \alp - \dlt_{0} + \nu_{\Box}}$ if $p$ obeys \eqref{eq:wave-main-1-p-lower1} as well.

\pfstep{Step~3: Proof of error bounds for low spherical harmonics}
Finally, we prove the error bounds \eqref{eq:wave-main-1-char-error-EJ}--\eqref{eq:wave-main-1-char-error-eJwave}. As in \eqref{eq:wave-main-1-char-error-EJ}--\eqref{eq:wave-main-1-char-error-eJwave}, we use the shorthand $\mu = 2(\ell+\nu_{\Box}-1)+I_{r \rd_{r}}$.
In what follows, we freely use \eqref{eq:S-ell-Linfty} to remove $\bbS_{(\ell)}$.  Moreover, we will suppress the dependence of implicit constants on $\ell$.

We begin with the contribution of $E_{J}$. In view of \eqref{eq:exp-error-wave}, we split $E_{J} = E_{J; \Box, k=0} + E_{J; f} + (E_{J} - E_{J; \Box, k=0} - E_{J; f})$. As discussed in Remark~\ref{rem:np-exp-error}, the term $E_{J; \Box, k=0}$ exhibits an improved fall-off thanks to Newman--Penrose cancellation. More precisely, note that $\bbS_{(\ell)}E_{J; \Box, k=0} = 0$ if $\ell = J - \nu_{\Box}$. Moreover, if $0 \leq \ell < J - \nu_{\Box}$, then by \eqref{eq:wave-main-Phijk}--\eqref{eq:wave-main-Phij0-low} (which has already been established), we have
\begin{align*}
\bbS_{(\ell)}E_{J; \Box, k=0} = O_{\bfGmm}^{M-C'}(A' r^{-J-1} u^{J-1-\alp-\dlt_{0}+\nu_{\Box}}).
\end{align*}
Therefore,
\begin{align*}
	& R^{\mu} \int_{1}^{U} (R + \tfrac{U-u}{2})^{-\mu} \nrm*{(\bfGmm + c_{\bfGmm})^{I}(\bfK + 2r)^{\ell+\nu_{\Box}-1} \bbS_{(\ell)} E_{J; \Box, k=0} (u, R + \tfrac{U-u}{2}, \tht)}_{L^{\infty}_{\tht}} \, \ud u \\
	& \aleq R^{\mu} \int_{1}^{U} A' (R + \tfrac{U-u}{2})^{-\mu+\ell+\nu_{\Box}-1-J-1} u^{J-1-\alp-\dlt_{0}+\nu_{\Box}} \, \ud u \\
	& \aleq A' R^{\ell+\nu_{\Box}-1-J-1} \int_{1}^{U} u^{J-1-\alp-\dlt_{0}+\nu_{\Box}} \, \ud u \\
	&\aleq A' R^{\ell+\nu_{\Box}-1} R^{-J-1} U^{J-\alp-\dlt_{0}+\nu_{\Box}},
\end{align*}
where we used $U\ls R$ on the third line and $J-\alp-\dlt_{0}+\nu_{\Box} > 0$ on the last line.

For the contribution of $E_{J; f}$, using \ref{hyp:forcing}, \eqref{eq:EJ-f} and \eqref{eq:EJ-f-rem}, we have
\begin{align*}
	& R^{\mu} \int_{1}^{U} (R + \tfrac{U-u}{2})^{-\mu} \nrm*{(\bfGmm + c_{\bfGmm})^{I}(\bfK + 2r)^{\ell+\nu_{\Box}-1} \bbS_{(\ell)} E_{J; f} (u, R + \tfrac{U-u}{2}, \tht)}_{L^{\infty}_{\tht}} \, \ud u \\
	& \aleq R^{\mu} \int_{1}^{U} D (R + \tfrac{U-u}{2})^{-\mu+\ell+\nu_{\Box}-1-J-1} \log^{K_{d}} \left( \frac{R+\tfrac{U-u}{2}}{u} \right) u^{J-1-\alp_{d}-\dlt_{d}+\nu_{\Box}} \, \ud u \\
	& \peq + R^{\mu} \int_{1}^{U} D (R + \tfrac{U-u}{2})^{-\mu+\ell+\nu_{\Box}-1-1-\alp_{d}+\nu_{\Box}} u^{-1-\dlt_{d}} \, \ud u \\
	& \aleq D R^{\ell+\nu_{\Box}-1-J-1} \int_{1}^{U} \log^{K_{d}} \left( \frac{R+\tfrac{U-u}{2}}{u} \right) u^{J-1-\alp_{d}-\dlt_{d}+\nu_{\Box}} \, \ud u \\
	& \peq + D R^{\ell+\nu_{\Box}-2-\alp_{d}+\nu_{\Box}} \int_{1}^{U} u^{-1-\dlt_{d}} \, \ud u \\
	&\aleq D R^{\ell+\nu_{\Box}-1} R^{-J-1} \log^{K_{J}} (\tfrac{R}{U}) U^{J-\alp-\dlt_{0}+\nu_{\Box}}.
\end{align*}
Here, we used $U\ls R$ for the second inequality and $\alp < J + \nu_{\Box} \leq J_{d} - 1 + \nu_{\Box} \leq \alp_{d}$, $\dlt_{0} \leq \dlt_{d}$, $K_{d} \leq K_{J}$ and $J -\alp_{0} - \dlt_{0} +\nu_{\Box} > 0$ for the last inequality.

Lastly, by Lemma~\ref{lem:exp-error-wave}, we have
\begin{align*}
	& R^{\mu} \int_{1}^{U} (R + \tfrac{U-u}{2})^{-\mu} \nrm*{(\bfGmm + c_{\bfGmm})^{I}(\bfK + 2r)^{\ell+\nu_{\Box}-1} \bbS_{(\ell)} (E_{J} - E_{J; \Box, k=0} - E_{J; f}) (u, R + \tfrac{U-u}{2}, \tht)}_{L^{\infty}_{\tht}} \, \ud u \\
	& \aleq R^{\mu} \int_{1}^{U} A' (R + \tfrac{U-u}{2})^{-\mu+\ell+\nu_{\Box}-1-J-1} \log^{K_{J}} \left( \frac{R+\tfrac{U-u}{2}}{u} \right) u^{J-1-\alp-\dlt_{0}+\nu_{\Box}} \, \ud u \\
	& \peq + R^{\mu} \int_{1}^{U} A' (R + \tfrac{U-u}{2})^{-\mu+\ell+\nu_{\Box}-1-J_{c}-\eta_{c}} u^{J_{c}-2+\eta_{c}-\alp-\dlt_{0}+\nu_{\Box}} \, \ud u \\
	& \aleq A' R^{\ell+\nu_{\Box}-1-J-1} \int_{1}^{U}  \log^{K_{J}} \left( \frac{R+\tfrac{U-u}{2}}{u} \right) u^{J-1-\alp-\dlt_{0}+\nu_{\Box}} \, \ud u \\
	& \peq + A' R^{\ell+\nu_{\Box}-1-J_{c}-\eta_{c}} \int_{1}^{U} u^{J_{c}-2+\eta_{c}-\alp-\dlt_{0}+\nu_{\Box}} \, \ud u \\
	& \aleq A' R^{\ell+\nu_{\Box}-1-J-1} \log^{K_{J}} \left( \tfrac{R}{U} \right) U^{J-\alp-\dlt_{0}+\nu_{\Box}}
	+ A' R^{\ell+\nu_{\Box}-1-J_{c}-\eta_{c}} U^{J_{c}-1+\eta_{c}-\alp-\dlt_{0}+\nu_{\Box}} \\
	& \aleq A' R^{\ell+\nu_{\Box}-1} R^{-J-1} \log^{K_{J}} \left( \tfrac{R}{U} \right) U^{J-\alp-\dlt_{0}+\nu_{\Box}}.
\end{align*}
Here, we used $U\ls R$ for the second inequality, $J_{c}-1+\eta_{c}-\alp-\dlt_{0}+\nu_{\Box} > J-\alp-\dlt_{0}+\nu_{\Box} > 0$ for the third inequality and $(\frac{U}{R})^{J_{c}-1+\eta_{c}-J} \aleq 1$ in the fourth inequality.

Next, we handle the contribution of $e_{J; \wave}$. We begin with the contribution of $e_{J; \wave, \linear}$. Using Lemma~\ref{lem:conj-wave-wave} and the hypothesis for $\rho_{J}$, for $\abs{I} \leq M - C'$, we have
\begin{align*}
	& R^{\mu} \int_{1}^{U} (R + \tfrac{U-u}{2})^{-\mu} \nrm*{(\bfGmm + c_{\bfGmm})^{I}(\bfK + 2r)^{\ell+\nu_{\Box}-1} \bbS_{(\ell)} e_{J; \wave, \linear} (u, R + \tfrac{U-u}{2}, \tht)}_{L^{\infty}_{\tht}} \, \ud u \\
	& \aleq R^{\mu} \int_{1}^{U} A (R + \tfrac{U-u}{2})^{-\mu + \ell+\nu_{\Box}-1-2-\alp+\nu_{\Box}} \log^{K_{c}'} \left( \frac{R+\tfrac{U-u}{2}}{u} \right) u^{-\dlt_{c}} \, \ud u \\
	& \aleq A R^{\ell+\nu_{\Box}-3-\alp+\nu_{\Box}} \int_{1}^{U}  \log^{K_{c}'} \left( \frac{R+\tfrac{U-u}{2}}{u} \right) u^{-\dlt_{c}} \, \ud u \\
	& \aleq A R^{\ell+\nu_{\Box}-3-\alp+\nu_{\Box}} \log^{K_{c}'} (\tfrac{R}{U}) U^{1-\dlt_{c}}
	 \aleq A R^{\ell+\nu_{\Box}-1} R^{-J-1} \log^{K_{J}} \left( \tfrac{R}{U} \right) U^{J-\alp_{0}-\dlt_{0}+\nu_{\Box}}.
\end{align*}
where we used $U\ls R$ in the second inequality, and $\dlt_{c} < 1$ in the third inequality and $K_{c}' \leq K_{J}$, $\dlt_{0} \leq \dlt_{c}$ and $(\frac{U}{R})^{\alp - \nu_{\Box} - J+1} \aleq 1$ in the fourth inequality.

For the contribution of $e_{J; \wave, \nonlinear}$, we use Definition~\ref{def:alp-N}.(3) with $\alp_{\calN}' = \alp_{\calN}+\dlt_{0}$ and the hypotheses on $\rPhi_{j, k}$ and $\rho_{J}$. Using either $\nu_{\Box} \geq 2$ for $d \geq 5$, or $\nu_{\Box} = 1$ and \eqref{eq:alp-N-wave-3d} for $d = 3$, we have, for $\abs{I} \leq M - C'$,
\begin{align*}
	& R^{\mu} \int_{1}^{U} (R + \tfrac{U-u}{2})^{-\mu} \nrm*{(\bfGmm + c_{\bfGmm})^{I}(\bfK + 2r)^{\ell+\nu_{\Box}-1} \bbS_{(\ell)} e_{J; \wave, \nonlinear} (u, R + \tfrac{U-u}{2}, \tht)}_{L^{\infty}_{\tht}} \, \ud u \\
	& \aleq R^{\mu} \int_{1}^{U} A' (R + \tfrac{U-u}{2})^{-\mu + \ell+\nu_{\Box}-1-2-\alp+\nu_{\Box}}  u^{-\dlt_{0}} \, \ud u \\
	 & \aleq A' R^{\ell+\nu_{\Box}-1} R^{-J-1} U^{J-\alp_{0}-\dlt_{0}+\nu_{\Box}},
\end{align*}
which is acceptable. \qedhere

\end{proof}

\begin{proof}[Proof of Case~2 ($J \leq \min\set{J_{c}, J_{d}}-1$ and $\alp + \dlt_{0} > J + \nu_{\Box}$)]
In this case, thanks to \eqref{eq:wave-main-Phijk}--\eqref{eq:wave-main-Phij0-high} (which we have established), Statement~(3) of Proposition~\ref{prop:recurrence} is applicable, from which the existence of the limits $\rcPhi_{J, k}(\infty)$ ($1 \leq k \leq K_{J}$) and $\rcPhi_{(\leq J - \nu_{\Box}) J, 0}(\infty)$, as well as \eqref{eq:wave-main-cPhiJk}--\eqref{eq:wave-main-cPhiJ0-low}, follows. To establish the remainder bound, we work with the modified remainder $\crho_{J}$ defined by \eqref{eq:crho-def}. By \eqref{eq:wave-main-hyp-rhoJ} and Proposition~\ref{prop:recurrence}, it follows that $\crho_{J}$ satisfies the same hypothesis as $\rho_{J}$ except for the replacement of $A$ by $A'(A)$ and some loss of vector field regularity:
\begin{equation} \label{eq:wave-main-crhoJ}
	\crho_{J} = O_{\bfGmm}^{M - C'}(A' r^{-\alp+\nu_{\Box}}).
\end{equation}

We now proceed as in the proof of Case~1.

\pfstep{Step~1: Proof of the remainder bound assuming key error bounds}
As before, our basic approach is to split $\crho_{J} = \crho_{(\geq J - \nu_{\Box} +1) J} + \crho_{(\leq J - \nu_{\Box}) J}$, and use equations \eqref{eq:crho-J-rp-eqn} and \eqref{eq:crho-J-char-eqn}, respectively. Thanks to the modification from $\rho_{J}$ to $\crho_{J}$, the error terms will have better $r$-decay.

We will now state key error bounds, which will be proved in Steps~2 and 3. To unify the proofs of Cases~2(a) and 2(b), we use the notation $\td{\frkL}$ as defined in Lemma~\ref{lem:exp-error-f-wave}.

To analyze \eqref{eq:crho-J-rp-eqn}, we will use the following error bounds, to be proved in Step~2: For $\abs{I} \leq M - C'$, with $C'$ to be determined at the end of Step~2, and $1 < p < 2$, which satisfies \eqref{eq:wave-main-2-p-upper1}, \eqref{eq:wave-main-2-p-upper2}, \eqref{eq:wave-main-2-p-lower1} and \eqref{eq:wave-main-2-p-upper3} below (the existence of a $p$ obeying these conditions will be shown in Step~2), we have
\begin{align}
& \int_{1}^{U} \left( \int_{\calC_{u} \cap \calM_{\wave}} r^{p} \abs*{\bbS_{(\geq J - \nu_{\Box} + 1)}(\bfGmm + c_{\bfGmm})^{I}(\bfK + 2r)^{J} \left( Q_{0} (r^{-J} \rPhi_{(\geq J-\nu_{\Box}+1) J, 0})  - E_{J+1} \right) }^{2} \, \ud r \ud \rssgm(\tht) \right)^{\frac{1}{2}}\ud u \notag \\
 & \aleq \td{\frkL} A' U^{\frac{p-1}{2}} \log^{K_{J}-1} U + A' U^{\frac{p-1}{2} + J - \alp - \dlt_{0} + \nu_{\Box}},  \label{eq:wave-main-2-rp-error-EJ+1-wave} \\
&\left( \iint_{\calD_{1}^{U} \cap \calM_{\med}} r^{p+1} \abs*{\bbS_{(\geq J - \nu_{\Box} + 1)}(\bfGmm + c_{\bfGmm})^{I}(\bfK + 2r)^{J} \left( Q_{0} (r^{-J} \rPhi_{(\geq J-\nu_{\Box}+1) J, 0})  - E_{J+1} \right)}^{2} \, \ud u \ud r \ud \rssgm(\tht) \right)^{\frac{1}{2}} \notag \\
&\aleq \td{\frkL} A' U^{\frac{p-1}{2}} \log^{K_{J}-1} U+ A' U^{\frac{p-1}{2} + J - \alp - \dlt_{0} + \nu_{\Box}},
 \label{eq:wave-main-2-rp-error-EJ+1-med} \\
& \int_{1}^{U} \left( \int_{\calC_{u} \cap \calM_{\wave}} r^{p} \abs*{\bbS_{(\geq J - \nu_{\Box} + 1)}(\bfGmm + c_{\bfGmm})^{I}(\bfK + 2r)^{J} e_{J+1; \wave} }^{2} \, \ud r \ud \rssgm(\tht) \right)^{\frac{1}{2}}\ud u \notag \\
& \aleq A' U^{\frac{p-1}{2} + J - \alp - \dlt_{0} + \nu_{\Box}}, \label{eq:wave-main-2-rp-error-eJ+1wave-wave} \\
&\left( \iint_{\calD_{1}^{U} \cap \calM_{\med}} r^{p+1} \abs*{\bbS_{(\geq J - \nu_{\Box} + 1)}(\bfGmm + c_{\bfGmm})^{I}(\bfK + 2r)^{J} e_{J+1; \wave}}^{2} \, \ud u \ud r \ud \rssgm(\tht) \right)^{\frac{1}{2}} \notag \\
&\aleq A' U^{\frac{p-1}{2} + J - \alp - \dlt_{0} + \nu_{\Box}}, \label{eq:wave-main-2-rp-error-eJ+1wave-med} \\
&\left( \iint_{\calD_{1}^{U} \cap \calM_{\med}} r^{p+1} \abs*{\bbS_{(\geq J - \nu_{\Box} + 1)}(\bfGmm + c_{\bfGmm})^{I}(\bfK + 2r)^{J} e_{\med}}^{2} \, \ud u \ud r \ud \rssgm(\tht) \right)^{\frac{1}{2}} \notag \\
& \aleq A' U^{\frac{p-1}{2} + J - \alp - \dlt_{0} + \nu_{\Box}}. \label{eq:wave-main-2-rp-error-emed}
\end{align}
To analyze \eqref{eq:crho-J-char-eqn}, we will use the following error bounds, which will be proved in Step~3: For $\abs{I} \leq M - C'$, with $C'$ to be determined at the end of Step~3, and $0 \leq \ell \leq J - \nu_{\Box}$, we have 
\begin{align}
	& R^{\mu} \int_{1}^{U} \left. r^{-\mu} \nrm*{(\bfGmm + c_{\bfGmm})^{I}(\bfK + 2r)^{\ell+\nu_{\Box}-1} \bbS_{(\ell)} E_{J+1}}_{L^{\infty}_{\tht}} \right|_{(u, r) = (\sgm, R+\frac{U-\sgm}{2}) }\, \ud \sgm \notag \\
	&\aleq  \td{\frkL} A' R^{\ell+\nu_{\Box}-1} R^{-J-2} (\log^{K_{J}} R) U + A' R^{\ell+\nu_{\Box}-1} R^{-J-1-\de_{0}} \log^{K_{J+1}} (\tfrac{R}{U}) U^{J-\alp+\nu_{\Box}},
 \label{eq:wave-main-2-char-error-EJ+1} \\
	& R^{\mu} \int_{1}^{U} \left. r^{-\mu} \nrm*{(\bfGmm + c_{\bfGmm})^{I}(\bfK + 2r)^{\ell+\nu_{\Box}-1} \bbS_{(\ell)} e_{J+1; \wave}}_{L^{\infty}_{\tht}} \right|_{(u, r) = (\sgm, R+\frac{U-\sgm}{2}) }\, \ud \sgm \notag \\
	& \aleq  A' R^{\ell+\nu_{\Box}-1} R^{-J-1-\de_{0}} \log^{K_{J+1}} (\tfrac{R}{U}) U^{J-\alp+\nu_{\Box}}, \label{eq:wave-main-2-char-error-eJ+1wave}
\end{align}
where $\mu = 2(\ell+\nu_{\Box}-1)+I_{r \rd_{r}}$.
In the remainder of this step, we will assume the above key error bounds and establish \eqref{eq:wave-main-rhoJ+1}.

\pfstep{Step~1(a): Estimates for high spherical harmonics}
Here we will show that, for $\abs{I} \leq M - C'$ and $1 < p < 2$ satisfying \eqref{eq:wave-main-2-p-upper1}, \eqref{eq:wave-main-2-p-upper2}, \eqref{eq:wave-main-2-p-lower1} and \eqref{eq:wave-main-2-p-upper3} below, we have
\begin{equation} \label{eq:wave-main-2-rp}
\begin{aligned}
& \sup_{u \in [1, U]} \left(\int_{\calC_{u}} \chi_{>2 R_{\far}} r^{p} (\rd_{r} \bfGmm^{I} \bfK^{J} \crho_{(\geq J - \nu_{\Box}+1) J})^{2}  \, \ud r \ud \rssgm(\tht) \right)^{\frac{1}{2}} \\
&+ \left( \iint_{\calD_{1}^{U}} \chi_{>2 R_{\far}} r^{p-1} (\rd_{r} \bfGmm^{I} \bfK^{J} \crho_{(\geq J - \nu_{\Box}+1) J})^{2} \, \ud u \ud r \ud \rssgm(\tht) \right)^{\frac{1}{2}} \\
&+ \left( \iint_{\calD_{U_{0}}^{U}} \chi_{>2 R_{\far}} r^{p-3} \left(\abs{\rsnb \bfGmm^{I} \bfK^{J} \crho_{(\geq J - \nu_{\Box}+1) J}}^{2}  + (\bfGmm^{I} \bfK^{J} \crho_{(\geq J - \nu_{\Box}+1) J})^{2} \right) \, \ud u \ud r \ud \rssgm(\tht) \right)^{\frac{1}{2}} \\
& \aleq \td{\frkL} A' U^{\frac{p-1}{2}} \log^{K_{J}} U + A' U^{\frac{p-1}{2} + J - \alp - \dlt_{0} + \nu_{\Box}}.
\end{aligned}
\end{equation}

As in Case~1, Step~1(a), the proof proceeds by an induction on $\abs{I}$. We will only give the details of the base case, since the induction step is similar as before. In the base case $\abs{I} = 0$, the main equation \eqref{eq:crho-J-rp-eqn} becomes
\begin{equation} \label{eq:crho-J-rp-eqn-I=0}
\begin{aligned}
Q_{J} \bfK^{J} \crho_{(\geq J - \nu_{\Box}+1) J}
&=
\bbS_{(\geq J - \nu_{\Box}+1)} (\bfK+2r)^{J} \left( Q_{0} (r^{-J} \rPhi_{(\geq J-\nu_{\Box}+1) J, 0})  - E_{J+1}\right)   \\
&\peq
+ \bbS_{(\geq J - \nu_{\Box}+1)} (\bfK+2r)^{J} e_{J+1; \wave}  \\
&\peq + \bbS_{(\geq J - \nu_{\Box}+1)} (\bfK+2r)^{J} e_{\med}.
\end{aligned}
\end{equation}
As before, we apply Lemma~\ref{lem:Qj-rp} with $I=0$, $R_{1} = 2 R_{\far}$, $j = J$ and $1 < p < 2$ also satisfying \eqref{eq:wave-main-2-p-upper1}, \eqref{eq:wave-main-2-p-upper2}, \eqref{eq:wave-main-2-p-lower1} and \eqref{eq:wave-main-2-p-upper3} below. The contributions of the first and second terms on the right-hand side of \eqref{eq:Qj-rp} are bounded using \eqref{eq:wave-main-1-id} and \eqref{eq:wave-main-1-bdry}, which are still applicable (even with the new choice of $p$ with and $\rho$ replaced by $\crho$). The contribution of the right-hand side of \eqref{eq:crho-J-rp-eqn-I=0} is handled by \eqref{eq:wave-main-2-rp-error-EJ+1-wave}, \eqref{eq:wave-main-2-rp-error-eJ+1wave-wave} and \eqref{eq:wave-main-2-rp-error-emed}. Finally, the last two terms on the right-hand side of \eqref{eq:Qj-rp} are zero since $\abs{I} = 0$. Using \eqref{eq:Qj-rp}, the desired bound \eqref{eq:wave-main-2-rp} in the base case $\abs{I} = 0$ follows. As remarked before, by an induction argument as in Case~1, Step~1(a) but using the same error bounds \eqref{eq:wave-main-2-rp-error-EJ+1-wave}, \eqref{eq:wave-main-2-rp-error-eJ+1wave-wave} and \eqref{eq:wave-main-2-rp-error-emed}, the general case $0 \leq \abs{I} \leq M - C'$ follows.

Next, we extract the limit of $\crho_{(\geq J - \nu_{\Box} + 1) J}$ as $r \to \infty$ and prove \eqref{eq:wave-main-rhoJ+1} for $\rho_{(\geq J-\nu_{\Box}+1) J+1}$. Since $p > 1$, by the fundamental theorem of calculus, Cauchy--Schwarz and \eqref{eq:wave-main-2-rp} with $\abs{I} = 0$ (with a change of notation $(U, R, \Tht) \mapsto (u, r, \tht)$), it follows that the following limit exists (and defines the left-hand side):
\begin{equation*}
	\rPhi'_{(\geq J - \nu_{\Box} + 1) J, 0} (u, \tht) := \frac{(-1)^{J}}{J!} \lim_{r \to \infty} \bfK^{J} \crho_{(\geq J - \nu_{\Box} + 1) J}(u, r, \tht).
\end{equation*}
Moreover, by \eqref{eq:wave-main-2-rp} with $\abs{I} > 0$, it is straightforward to justify
\begin{equation*}
	\bfGmm^{I} \rPhi'_{(\geq J - \nu_{\Box} + 1) J, 0} (u, \tht)= \frac{(-1)^{J}}{J!} \lim_{r \to \infty} \bfGmm^{I} \bfK^{J} \crho_{(\geq J - \nu_{\Box} + 1) J}(u, r, \tht).
\end{equation*}
In addition, by the fundamental theorem of calculus and \eqref{eq:wave-main-2-rp}, we have
\begin{align*}
&\abs{\bfGmm^{I} \bfK^{J} \crho_{(\geq J - \nu_{\Box}+1) J}(u, r, \tht) - (-1)^{J} (J!)\bfGmm^{I} \rPhi'_{(\geq J - \nu_{\Box} + 1) J, 0}(u, r, \theta)} \\
&\aleq \td{\frkL} A' r^{-\frac{p-1}{2}} u^{\frac{p-1}{2}} \log^{K_{J}} u + A' r^{-\frac{p-1}{2}} u^{\frac{p-1}{2} + J - \alp - \dlt_{0} + \nu_{\Box}}.
\end{align*}
Introducing
\begin{equation*}
	\rho'_{J+1} = \crho_{J} - r^{-J} \rPhi'_{(\geq J - \nu_{\Box} + 1) J, 0},
\end{equation*}
and noticing that $\bfK^{J} r^{-J} = (-1)^{J} (J)!$, the preceding estimate may be rewritten as
\begin{align*}
\abs{\bfGmm^{I} \bfK^{J} \rho'_{(\geq J - \nu_{\Box}+1) J+1}}
\aleq \td{\frkL} A' r^{-\frac{p-1}{2}} u^{\frac{p-1}{2}} \log^{K_{J}} u + A' r^{-\frac{p-1}{2}} u^{\frac{p-1}{2} + J - \alp - \dlt_{0} + \nu_{\Box}}.
\end{align*}
At this point, arguing as in the proof of Case~1, Step~1(a) using \eqref{eq:comm-K-Omg}--\eqref{eq:comm-K-rdr}, an induction on $\abs{I}$ and Taylor's theorem in $z = \frac{1}{r}$, it follows that
\begin{align*}
\abs{\bfGmm^{I} \rho'_{(\geq J - \nu_{\Box}+1) J+1}}
\aleq \td{\frkL} A' r^{-\frac{p-1}{2}-J} u^{\frac{p-1}{2}} \log^{K_{J}} u + A' r^{-\frac{p-1}{2}-J} u^{\frac{p-1}{2} + J - \alp - \dlt_{0} + \nu_{\Box}}.
\end{align*}
Thanks to \eqref{eq:wave-main-2-p-lower1}, in $\calM_{\wave}$ we may bound $u^{\frac{p-1}{2} + J - \alp - \dlt_{0} + \nu_{\Box}} \aleq r^{\frac{p-1}{2} + J - \alp - \dlt_{0} + \nu_{\Box}}$. In conclusion,
\begin{align*}
\rho'_{(\geq J - \nu_{\Box}+1) J+1}
=  O_{\bfGmm}^{M-C'}(\td{\frkL}  A' r^{-\frac{p-1}{2}-J} u^{\frac{p-1}{2}} \log^{K_{J}} u)
+ O_{\bfGmm}^{M-C'}(A' r^{- \alp - \dlt_{0} + \nu_{\Box}}).
\end{align*}

In Step~1(c) below, we will show that, in fact, $\rPhi'_{(\geq J - \nu_{\Box} + 1) J, 0} = \rPhi_{(\geq J - \nu_{\Box} + 1) J, 0}$, so that $\rho'_{(\geq J - \nu_{\Box}+1) J+1} = \rho_{(\geq J - \nu_{\Box}+1) J+1}$.

\pfstep{Step~1(b): Estimates for low spherical harmonics}
Let $\ell \in \set{0, \ldots, J - \nu_{\Box}}$. For $\crho_{(\ell) J}$, the following bound holds: for $0 \leq \abs{I} \leq M - C'$,
\begin{equation} \label{eq:wave-main-2-char}
\begin{aligned}
	&\abs{\rd_{r} \bfGmm^{I} \bfK^{\ell+\nu_{\Box}-1} \crho_{(\ell) J}}
	+ r^{-1} \abs{\bfGmm^{I} \bfK^{\ell+\nu_{\Box}-1} \crho_{(\ell) J}}  \\
	&\aleq \td{\frkL} A' r^{\ell+\nu_{\Box}-1} r^{-J-2} (\log^{K_{J}} r)  u + A' r^{\ell+\nu_{\Box}-1} r^{-J-1-\de_{0}} \log^{K_{J+1}} (\tfrac{r}{u}) u^{J-\alp+\nu_{\Box}}.
\end{aligned}
\end{equation}
Indeed, \eqref{eq:wave-main-2-char} may be proved exactly like \eqref{eq:wave-main-1-char} in Case~1, Step~1(b), where we use the error bounds \eqref{eq:wave-main-2-char-error-EJ+1} and \eqref{eq:wave-main-2-char-error-eJ+1wave} instead of \eqref{eq:wave-main-1-char-error-EJ} and \eqref{eq:wave-main-1-char-error-eJwave}, respectively. We omit the straightforward details.

Next, we derive a bound on $\crho_{(\leq J - \nu_{\Box}) J}$. Beginning with \eqref{eq:wave-main-2-char}, and using a simple induction argument on $\abs{I}$ based on the commutator identities \eqref{eq:comm-K-Omg}--\eqref{eq:comm-K-rdr}, we may prove
\begin{equation*}
	\abs*{\bfK^{\ell+\nu_{\Box}-1} \bfGmm^{I} \crho_{(\ell) J}}
	\aleq \td{\frkL} A' r^{\ell+\nu_{\Box}-1} r^{-J-1} (\log^{K_{J}} r) u + A' r^{\ell+\nu_{\Box}-1} r^{-J-\de_{0}} \log^{K_{J+1}} (\tfrac{r}{u}) u^{J-\alp+\nu_{\Box}}.
\end{equation*}
Recall that, by \eqref{eq:wave-main-crhoJ}, $\bfGmm^{I} \crho_{(\ell) J} = o(r^{-J+1})$ as $r \to \infty$. By Taylor's theorem in the variable $z = \frac{1}{r}$, we have
\begin{equation*}
\begin{split}
\abs{\bfGmm^{I} \crho_{(\leq J - \nu_{\Box}) J}}
\aleq &\: \td{\frkL} A' r^{-J-1} (\log^{K_{J}} r) u + A'r^{-J-\dlt_{0}} \log^{K_{J+1}} (\tfrac{r}{u}) u^{J-\alp+\nu_{\Box}} \\
\aleq &\: \td{\frkL} A' r^{-J-1} (\log^{K_{J}} r) u + A' r^{-\alp-\de_{0}+\nB},
\end{split}
\end{equation*}
where we used $J-\alp+\nB >0$ (by \eqref{eq:alp.J.range}) in the last line.

\pfstep{Step~1(c): Conclusion of the proof}
By Steps~1(a) and 1(b) and the definition of $\crho_{J}$, it follows that
\begin{align*}
	\Phi = \sum_{j=0}^{J-1} \sum_{k=0}^{K_{j}} r^{-j} \log^{k}(\tfrac{r}{u}) \rPhi_{j, k} + \sum_{k=1}^{K_{J}} r^{-J} \log^{k}(\tfrac{r}{u}) \rPhi_{J, k} +r^{-J} (\rPhi_{(\leq J - \nu_{\Box}) J, 0} + \rPhi'_{(\geq J - \nu_{\Box}+1) J, 0}) + \rho'_{J+1}
\end{align*}
where $\rho'_{J+1} = \crho_{(\leq J - \nu_{\Box}) J} + \rho'_{(\geq J - \nu_{\Box}+1) J+1}$ satisfies
\begin{equation*}
    \begin{split}
	   \rho'_{J+1}
	   = &\: O_{\bfGmm}^{M-C'}(\td{\frkL} A' r^{-J-\frac{p-1}{2}} u^{\frac{p-1}{2}} \log^{K_{J}} u )
	   + O_{\bfGmm}^{M-C'}( A' r^{-\alp-\de_{0}+\nB}),
    \end{split}
\end{equation*}
with $1 < p < 2$ satisfying \eqref{eq:wave-main-2-p-upper1}, \eqref{eq:wave-main-2-p-upper2}, \eqref{eq:wave-main-2-p-lower1} and \eqref{eq:wave-main-2-p-upper3} (note that the bounds from Step~1(b) are more favorable).

Importantly, note that $\bfGmm^{I} \rho'_{J+1} = o(r^{-J})$ as $r \to \infty$ for $\abs{I} \leq M- C'$.
We plug $\Phi$ into the equation $\calQ(\Phi) = r^{\nu_{\Box}} \calN(r^{-\nu_{\Box}} \Phi) + r^{\nu_{\Box}} f$ and expand in terms of the form $r^{-j-1} \log^{k}(\frac{r}{u})$. Note that all terms with either $j < J$ or $j = J$ and $k > 0$ are cancelled thanks to the recurrence equation satisfied by $\rPhi_{j, k}$ for $j < J$ and $j = J$ and $k > 0$ (cf.~the proofs of Lemmas~\ref{lem:recurrence-forcing-basic} and \ref{lem:recurrence-forcing}). Hence, the terms with $j = J$, $k = 0$ must now cancel each other, which implies that $\rPhi'_{(\geq J - \nu_{\Box}+1) J, 0} + \rPhi_{(\leq J - \nu_{\Box}) J, 0}$ and $\rPhi_{J, 0}$ have the same $u$-partial derivative (which is determined only by $f$ and $\rPhi_{j, k}$ with $j \leq J - 1$) with the same initial data. Hence $\rPhi_{(\geq J - \nu_{\Box}+1) J, 0} = \rPhi'_{(\geq J - \nu_{\Box}+1) J, 0}$ and $\rho_{J+1} = \rho'_{J+1}$.

In conclusion, the desired remainder bound \eqref{eq:wave-main-rhoJ+1} in Case~2(a) follows. In Case~2(b), the above analysis establishes \eqref{eq:wave-main-rhoJ+1-final} with $\eta_{\mathfrak f} = \frac{p-1}{2}$. In view of \eqref{eq:wave-main-2-p-upper1}, \eqref{eq:wave-main-2-p-upper2}, \eqref{eq:wave-main-2-p-upper3} (and \eqref{eq:wave-main-2-p-lower1}), $\eta_{\frkf}$ can be chosen to be arbitrarily close to (but smaller than)  the right-hand side of \eqref{eq:wave-main-etaf}, where the higher the $\eta_{\mathfrak f}$ the stronger the bound \eqref{eq:wave-main-rhoJ+1-final}. The proof of Case~2 is thus complete.

\pfstep{Step~2: Proof of error bounds for high spherical harmonics}
We work with the same conventions as in Case~1, Step~2.

\pfstep{Step~2(a): Error estimate in $\calM_{\wave}$}  We begin with the contribution of $Q_{0} (r^{-J} \rPhi_{(\geq J-\nu_{\Box}+1) J, 0}) - \bbS_{(\geq J - \nu_{\Box}+1)} E_{J+1}$. Here, we use the splitting
\begin{align*}
	&Q_{0} (r^{-J} \rPhi_{(\geq J-\nu_{\Box}+1) J, 0}) - \bbS_{(\geq J - \nu_{\Box}+1)}E_{J+1} \\
	&= Q_{0} (r^{-J} \rPhi_{(\geq J-\nu_{\Box}+1) J, 0}) - \bbS_{(\geq J - \nu_{\Box}+1)} E_{J+1; \Box, k=0} \\
	&\peq - \bbS_{(\geq J - \nu_{\Box}+1)} E_{J+1; f} \\
	&\peq - \bbS_{(\geq J - \nu_{\Box}+1)}(E_{J+1} - E_{J+1; \Box, k=0} - E_{J+1; f}).
\end{align*}
To treat the first line of the right-hand side, recall the identity \eqref{eq:Q0-crhoJ-high-key}. Since $(\bfK+2r)^{J} r^{-J-1} = 0$, we have
\begin{equation} \label{eq:wave-main-2-rp-Ej+1-cancel}
(\bfK+2r)^{J} \left( Q_{0} (r^{-J} \rPhi_{(\geq J-\nu_{\Box}+1) J, 0}) - \bbS_{(\geq J - \nu_{\Box}+1)} E_{J+1; \Box, k=0} \right) = 0.
\end{equation}

For $E_{J; f}$, we use \eqref{eq:EJ-f}, \eqref{eq:EJ-f-rem} and \ref{hyp:forcing}. For $\abs{I} \leq M_{0} - J$,
\begin{align*}
& \int_{1}^{U} \left( \int_{\calC_{u} \cap \calM_{\wave}} r^{p} \abs*{\bbS_{(\geq J - \nu_{\Box} + 1)}(\bfGmm + c_{\bfGmm})^{I}(\bfK + 2r)^{J} E_{J+1; f} }^{2} \, \ud r \ud \rssgm(\tht) \right)^{\frac{1}{2}} \ud u \\
& \aleq \int_{1}^{U} \left( \int_{\calC_{u} \cap \calM_{\wave}} r^{p} \left(D r^{J} r^{-J-2} \log^{K_{d}}(\tfrac{r}{u}) u^{J-\alp_{d}-\dlt_{d}+\nu_{\Box}} \right)^{2} \, \ud r \right)^{\frac{1}{2}} \ud u \\
& \peq + \int_{1}^{U} \left( \int_{\calC_{u} \cap \calM_{\wave}} r^{p} \left( D r^{J-1-\alp_{d}+\nu_{\Box}} u^{-1-\dlt_{d}} \right)^{2} \, \ud r \right)^{\frac{1}{2}} \ud u \\
& \aleq \int_{1}^{U} D u^{\frac{p-1}{2} + J - 1 - \alp_{d} - \dlt_{d} + \nu_{\Box}} \, \ud u,
\end{align*}
where, for the last inequality, we impose
\begin{equation} \label{eq:wave-main-2-p-upper1}
	p < 2
\end{equation}
and
\begin{equation} \label{eq:wave-main-2-p-upper2}
	p < 1 + 2 (\alp_{d} - J - \nu_{\Box}) = 1 + 2 (J_{d} - 1 - J + \eta_{d}).
\end{equation}
Using $\alp < J + \nu_{\Box} \leq J_{d} - 1 + \nu_{\Box} < \alp_{d}$ and $\dlt_{0} \leq \dlt_{d}$, we may bound
\begin{align*}
\int_{1}^{U} D u^{\frac{p-1}{2} + J - 1 - \alp_{d} - \dlt_{d} + \nu_{\Box}} \, \ud u
\aleq C D U^{\frac{p-1}{2} + J - \alp - \dlt_{0} + \nu_{\Box}},
\end{align*}
provided that $p$ obeys
\begin{equation} \label{eq:wave-main-2-p-lower1}
	p > 1 + 2 (\alp + \dlt_{0} - J - \nu_{\Box} ).
\end{equation}
We note that \eqref{eq:wave-main-2-p-upper1} and \eqref{eq:wave-main-2-p-lower1} may be imposed simultaneously, since $\alp + \dlt_{0} - J - \nu_{\Box} < \dlt_{0} < \frac{1}{2}$; \eqref{eq:wave-main-2-p-upper2} and \eqref{eq:wave-main-2-p-lower1} may also be simultaneously imposed since
\begin{equation*}
	\dlt_{0} \leq \eta_{d} = \alp_{d} - \nu_{\Box} - J_{d} + 1 \leq \alp_{d} - \nu_{\Box} - J < \alp_{d} - \alp
\end{equation*}


For the remainder, we apply the estimate in Lemma~\ref{lem:exp-error-f-wave} so that the following estimate holds:
For $\abs{I} \leq M-J-C'$, we have 
\begin{align*}
& \int_{1}^{U} \left( \int_{\calC_{u} \cap \calM_{\wave}} r^{p} \abs*{\bbS_{(\geq J - \nu_{\Box} + 1)}(\bfGmm + c_{\bfGmm})^{I}(\bfK + 2r)^{J} \left[ E_{J+1} - E_{J+1; \Box, k=0} - E_{J+1; f} \right]}^{2} \, \ud r \ud \rssgm(\tht) \right)^{\frac{1}{2}} \ud u \\
&\aleq  \td{\frkL} \int_{1}^{U} \left( \int_{\calC_{u} \cap \calM_{\wave}} r^{p} \left( A' r^{J} r^{-J-2} \log(\tfrac{r}{u}) \log^{K_{J}-1} r  \right)^{2} \, \ud r \right)^{\frac{1}{2}} \ud u  \\
& \peq + \int_{1}^{U} \left( \int_{\calC_{u} \cap \calM_{\wave}} r^{p} \left( A' r^{J} r^{-J-2} \log^{K_{J+1}} (\tfrac r u)  u^{J - \alp -\de_0 + \nB} \right)^{2} \, \ud r \right)^{\frac{1}{2}} \ud u \\
& \peq +  \int_{1}^{U} \left( \int_{\calC_{u} \cap \calM_{\wave}} r^{p} \left( A' r^{J} r^{-J_{c}-\eta_{c}} u^{J_{c}-2+\eta_{c}-\alp-\de_0+\nB}  \right)^{2} \, \ud r \right)^{\frac{1}{2}} \ud u \\
& \aleq \td{\frkL} A' \int_{1}^{U} u^{\frac{p-1}{2} - 1} \log^{K_{J}-1} u \, \ud u + A' \int_{1}^{U} u^{\frac{p-1}{2} +J-1-\alp-\dlt_{0}+\nu_{\Box}} \, \ud u,
\end{align*}
where we used \eqref{eq:wave-main-2-p-upper1} and
\begin{equation} \label{eq:wave-main-2-p-upper3}
	p < 1 + 2 (J_{c}-1 + \eta_{c} - J).
\end{equation}
The last line may be bounded by
\begin{equation*}
C \td{\frkL} A' U^{\frac{p-1}{2}} \log^{K_{J}-1} U + C A' U^{\frac{p-1}{2}+J-\alp-\dlt_{0}+\nu_{\Box}},
\end{equation*}
provided that $p$ obeys \eqref{eq:wave-main-2-p-lower1} and $p>1$ (which is in turn a consequence of \eqref{eq:wave-main-2-p-lower1} since $\alp+\de_0-J-\nB >0$).
To check that the new condition \eqref{eq:wave-main-2-p-upper3} can be imposed together with \eqref{eq:wave-main-2-p-upper1}--\eqref{eq:wave-main-2-p-lower1}, it suffices to check that \eqref{eq:wave-main-2-p-upper3} and \eqref{eq:wave-main-2-p-lower1} can be imposed simultaneously, which is the casesince $\dlt_{0} < \eta_{c}$, $J\leq J_{c}-1$, and $\alp < J + \nu_{\Box}$.

Next, we treat the contribution of $e_{J+1; \wave}$. Here, we use the decomposition in \eqref{eq:eJ+1wave-crho}. For the linear terms, we use Lemma~\ref{lem:conj-wave-wave}, the bound \eqref{eq:wave-main-crhoJ} for $\crho_{J}$, and Proposition~\ref{prop:recurrence}.(3) for $\rPhi_{(\geq J-\nB+1)J,0}$. Thus, $\abs{I} \leq M - J - C'$,
\begin{align*}
& \int_{1}^{U} \left( \int_{\calC_{u} \cap \calM_{\wave}} r^{p} \abs*{\bbS_{(\geq J - \nu_{\Box} + 1)}(\bfGmm + c_{\bfGmm})^{I}(\bfK + 2r)^{J} \check{e}_{J+1; \wave, \mathrm{linear},1} }^{2} \, \ud r \ud \rssgm(\tht)\right)^{\frac{1}{2}} \ud u \\
& + \int_{1}^{U} \left( \int_{\calC_{u} \cap \calM_{\wave}} r^{p} \abs*{\bbS_{(\geq J - \nu_{\Box} + 1)}(\bfGmm + c_{\bfGmm})^{I}(\bfK + 2r)^{J} \check{e}_{J+1; \wave, \mathrm{linear},2}}^{2} \, \ud r \ud \rssgm(\tht)\right)^{\frac{1}{2}} \ud u \\
& \aleq \int_{1}^{U} \left( \int_{\calC_{u} \cap \calM_{\wave}} r^{p} \left( A' r^{J} r^{-2} \log^{K_{c}'} (\tfrac{r}{u}) u^{-\dlt_{c}} r^{-\alp+\nu_{\Box}} \right)^{2} \, \ud r \right)^{\frac{1}{2}} \ud u  \\
&\peq + \int_{1}^{U} \left( \int_{\calC_{u} \cap \calM_{\wave}} r^{p} \left( A' r^{J} r^{-2} \log^{K_{c}'} (\tfrac{r}{u}) u^{-\dlt_{c}} r^{-J} u^{J-\alp+\nu_{\Box}} \right)^{2} \, \ud r \right)^{\frac{1}{2}} \ud u \\
& \aleq \int_{1}^{U} A' u^{\frac{p-1}{2} + J - 1 - \alp - \dlt_{c} +\nu_{\Box}} \, \ud u,
\end{align*}
where we used \eqref{eq:wave-main-2-p-upper1} and $J-\alp+\nB < \dlt_{0} < \frac{1}{2}$ for the last inequality. Since $\dlt_{0} \leq \dlt_{c}$, this term may also be bounded by $C A' U^{\frac{p-1}{2} + J - \alp - \dlt_{0} + \nu_{\Box}}$ if $p$ obeys \eqref{eq:wave-main-2-p-lower1}.

Finally, we turn to the nonlinear terms of $e_{J+1; \wave}$ in \eqref{eq:eJ+1wave-crho}. Using Definition~\ref{def:alp-N}.(3) with $\alp'_{\calN} = \alp_{\calN}+\de_0$, and then use the bound \eqref{eq:wave-main-crhoJ} for $\crho_{J}$, and Proposition~\ref{prop:recurrence}.(3) for $\rPhi_{(\geq J-\nB+1)J,0}$, we argue as above to obtain, for $\abs{I} \leq M - J - C'$,
\begin{align*}
& \int_{1}^{U} \left( \int_{\calC_{u} \cap \calM_{\wave}} r^{p} \abs*{\bbS_{(\geq J - \nu_{\Box} + 1)}(\bfGmm + c_{\bfGmm})^{I}(\bfK + 2r)^{J} \check{e}_{J+1; \wave, \mathrm{nonlinear},1} }^{2} \, \ud r \ud \rssgm(\tht) \right)^{\frac{1}{2}} \ud u \\
& + \int_{1}^{U} \left( \int_{\calC_{u} \cap \calM_{\wave}} r^{p} \abs*{\bbS_{(\geq J - \nu_{\Box} + 1)}(\bfGmm + c_{\bfGmm})^{I}(\bfK + 2r)^{J} \check{e}_{J+1; \wave, \mathrm{nonlinear},2} }^{2} \, \ud r \ud \rssgm(\tht) \right)^{\frac{1}{2}} \ud u \\
& \aleq \int_{1}^{U} A' u^{\frac{p-1}{2} + J - 1 - \alp - \dlt_{0}, +\nu_{\Box}} \, \ud u \\
& \aleq A' U^{\frac{p-1}{2} + J - \alp - \dlt_{0} + \nu_{\Box}},
\end{align*}
where we have again used \eqref{eq:wave-main-2-p-upper1}, $J-\alp+\nB<\f 12$ and \eqref{eq:wave-main-2-p-lower1}.

We have thus concluded the proof of \eqref{eq:wave-main-2-rp-error-EJ+1-wave} and \eqref{eq:wave-main-2-rp-error-eJ+1wave-wave}.

\pfstep{Step~2(b): Error estimate in $\calM_{\med}$}
As in Step~2(a), we begin with the contribution of $Q_{0}(r^{-J} \rPhi_{(\geq J-\nu_{\Box}+1) J, 0}) - \bbS_{(\geq J-\nB+1)}E_{J+1}$. In view of \eqref{eq:exp-error-f-med}, we split
\begin{align*}
Q_{0}(r^{-J} \rPhi_{(\geq J-\nu_{\Box}+1) J, 0}) - \bbS_{(\geq J-\nu_{\Box}+1)} E_{J+1}
&= Q_{0}(r^{-J} \rPhi_{(\geq J-\nu_{\Box}+1) J, 0}) - \bbS_{(\geq J-\nu_{\Box}+1)} Q_{0}(\Phi_{<J +1, k=0}) \\
&\peq - \bbS_{(\geq J-\nu_{\Box}+1)} (E_{J+1} - Q_{0}(\Phi_{<J+1, k=0})).
\end{align*}
Note that in the first two terms, all the $r^{-J-2}$ are cancelled. Thus their contribution vanishes after taking $(\bfK+2r)^{J}$. For the remainder, by \eqref{eq:exp-error-f-med}, we have, for $|I|\leq M - J - C'$,
\begin{align*}
& \left( \iint_{\calD_{1}^{U} \cap \calM_{\med}} r^{p+1} \abs*{\bbS_{(\geq J - \nu_{\Box} + 1)}(\bfGmm + c_{\bfGmm})^{I}(\bfK + 2r)^{J} (E_{J+1}- Q_{0}(\Phi_{<J+1, k=0}))}^{2} \, \ud u \ud r \ud \rssgm(\tht)  \right) ^{\frac{1}{2}} \\
& \aleq \td{\frkL} \left( \iint_{\calD_{1}^{U} \cap \set{r \aeq u}} u^{p+1} \left( A' u^{J} u^{-J-2} \log^{K_{J}-1} u  \right)^{2} \, \ud u \ud r \right)^{\frac{1}{2}} \\
&\peq + \left( \iint_{\calD_{1}^{U} \cap \set{r \aeq u}} u^{p+1} \left( A' u^{J} u^{-\alp - 2 - \dlt_{0} + \nu_{\Box}}  \right)^{2} \, \ud u \ud r \right)^{\frac{1}{2}} \\
& \peq + \left( \iint_{\calD_{1}^{U} \cap \set{r \aeq u}} u^{p+1} \left( D u^{J} u^{-\alp_{d} - 2 - \dlt_{d} + \nu_{\Box}}  \right)^{2} \, \ud u \ud r \right)^{\frac{1}{2}} \\
& \aleq \td{\frkL} \left(\int_{1}^{U} \left(A' u^{\frac{p-1}{2}} u^{-\frac{1}{2}} \log^{K_{J}-1} u \right)^{2} \, \ud u \right)^{\frac{1}{2}}\\
&\peq + \left(\int_{1}^{U} \left(A' u^{\frac{p-1}{2}} u^{J - \alp - \frac{1}{2} - \dlt_{0} + \nu_{\Box}}\right)^{2} \, \ud u \right)^{\frac{1}{2}} + \left(\int_{1}^{U} \left(D u^{\frac{p-1}{2}} u^{J - \alp - \frac{1}{2} - \dlt_{d} + \nu_{\Box}}\right)^{2} \, \ud u \right)^{\frac{1}{2}}.
\end{align*}
Since $\dlt_{0} \leq \dlt_{d}$, the last line may be bounded by
\begin{equation*}
C \td{\frkL} A' U^{\frac{p-1}{2}} \log^{K_{J}-1} U + C (D + A') U^{\frac{p-1}{2} + J - \alp - \dlt_{0} + \nu_{\Box}}
\end{equation*}
provided that $p$ obeys $p > 1$ and \eqref{eq:wave-main-2-p-lower1}. Combining this with the discussion in the beginning of this step, we have thus proven \eqref{eq:wave-main-2-rp-error-EJ+1-med}.

We now handle the contribution of $e_{J+1; \wave}$, which is supported in $\set{r \geq (2\eta_{0})^{-1} u}$. As in Step~2(a), for the linear part, we use Lemma~\ref{lem:conj-wave-wave}, the bound \eqref{eq:wave-main-crhoJ} for $\crho_{J}$, and Proposition~\ref{prop:recurrence}.(3) for $\rPhi_{(\geq J-\nB+1)J,0}$; while for the nonlinear part, we use Definition~\ref{def:alp-N}.(3) with $\alp'_{\calN} = \alp_{\calN}+\de_0$, and then use the bound \eqref{eq:wave-main-crhoJ} for $\crho_{J}$, and Proposition~\ref{prop:recurrence}.(3) for $\rPhi_{(\geq J-\nB+1)J,0}$. Hence, using also $\de_{0} \leq \de_{c}$, for $|I| \leq M - J - C'$,
\begin{align*}
    & \left( \iint_{\calD_{1}^{U} \cap \calM_{\med}} r^{p+1} \abs*{\bbS_{(\geq J - \nu_{\Box} + 1)}(\bfGmm + c_{\bfGmm})^{I}(\bfK + 2r)^{J} e_{J+1; \wave}}^{2} \, \ud u \ud r \ud \rssgm(\tht)  \right) ^{\frac{1}{2}} \\
    & \aleq \left( \iint_{\calD_{1}^{U} \cap \set{r \aeq u}} u^{p+1} \left( A' u^{J} u^{-2} u^{-\dlt_{0}} u^{-\alp+\nu_{\Box}}  \right)^{2} \, \ud u \ud r \right)^{\frac{1}{2}} \\
    & \aleq \left(\int_{1}^{U} \left(A' u^{\f{p-1}2} u^{J-\alp-\f 12 -\de_{0}+\nB} \right)^2\, \ud u\right)^{\f 12} \ls A' U^{\frac{p-1}{2} + J - \alp - \dlt_{0} + \nu_{\Box}},
\end{align*}
where we have used \eqref{eq:wave-main-2-p-lower1}. 

Finally, it remains to treat the contribution of $e_{\med}$ given by \eqref{eq:eJmed-def}. Here, observe that we may proceed exactly as in Case~1, Step~2(b) and establish
\begin{align*}
& \left( \iint_{\calD_{1}^{U} \cap \calM_{\med}} r^{p+1} \abs*{\bbS_{(\geq J - \nu_{\Box} + 1)}(\bfGmm + c_{\bfGmm})^{I}(\bfK + 2r)^{J} \left[ e_{\med} \right]}^{2} \, \ud u \ud r \ud \rssgm(\tht) \right)^{\frac{1}{2}}
 \aleq A' U^{\frac{p-1}{2}+ J - \alp - \dlt_{0} + \nu_{\Box}},
\end{align*}
provided that $p$ obeys \eqref{eq:wave-main-1-p-lower2} and \eqref{eq:wave-main-1-p-lower3}, as well as \eqref{eq:wave-main-2-p-lower1} (which is the same as \eqref{eq:wave-main-1-p-lower1}). Observe as before that \eqref{eq:wave-main-1-p-lower2} and \eqref{eq:wave-main-1-p-lower3} automatically hold if $p > 0$, and hence that they do not place any further restrictions on $p$ in this case.

\pfstep{Step~3: Proof of error bounds for low spherical harmonics}
In this step, we establish the error bounds \eqref{eq:wave-main-2-char-error-EJ+1}--\eqref{eq:wave-main-2-char-error-eJ+1wave}.

We begin with the contribution of $E_{J+1}$. In view of \eqref{eq:exp-error-f-wave}, we split $E_{J+1} = E_{J+1; \Box, k=0} + E_{J+1; f} + (E_{J+1} - E_{J+1; \Box, k=0} - E_{J+1; f})$.

For $E_{J+1; \Box, k=0}$, we take advantage of $\ell \leq J - \nu_{\Box}$ and use \eqref{eq:wave-main-cPhiJk}, \eqref{eq:wave-main-cPhiJ0-low}, \eqref{eq:wave-main-PhiJ0-low.limit} to obtain
\begin{align*}
\bbS_{(\ell)}E_{J+1; \Box, k=0}
&= r^{-J-2} (J+1 - \nu_{\Box}-\ell)(J+1 + \nu_{\Box}+ \ell - 1) \rPhi_{(\ell) J, 0} \\
& = O_{\bfGmm}^{M-C'}(\td{\frkL} A' r^{-J-2} \log^{K_{J}} u)
+ O_{\bfGmm}^{M-C'}(A' r^{-J-2} u^{J-\alp-\dlt_{0}+\nu_{\Box}}).
\end{align*}
Hence,
\begin{align*}
	& R^{\mu} \int_{1}^{U} (R + \tfrac{U-u}{2})^{-\mu} \nrm*{(\bfGmm + c_{\bfGmm})^{I}(\bfK + 2r)^{\ell+\nu_{\Box}-1} \bbS_{(\ell)} E_{J+1; \Box, k=0} (u, R + \tfrac{U-u}{2}, \tht)}_{L^{\infty}_{\tht}} \, \ud u \\
	& \aleq \td{\frkL} A'  R^{\ell+\nB-1-J-2} \int_{1}^{U} \log^{K_J} u  \, \ud u + A' R^{\ell+\nB-1-J-2} \int_{1}^{U} u^{J-\alp-\de_0+\nB}  \, \ud u \\
	& \aleq \td{\frkL} A' R^{\ell+\nu_{\Box}-1} R^{-J-2} (\log^{K_{J}} R) U
	 + A' R^{\ell+\nu_{\Box}-1} R^{-J-2} U^{J + 1- \alp - \dlt_{0} + \nu_{\Box}},
\end{align*}
where we used $J + 1- \alp - \dlt_{0} + \nu_{\Box} > 0$ (by \eqref{eq:dlt0-wave}, \eqref{eq:alp.J.range}).

Next, the contribution of $E_{J+1; f}$ is estimated as in Case~1, Step~3 using \ref{hyp:forcing}, \eqref{eq:EJ-f} and \eqref{eq:EJ-f-rem}, with $J$ replaced by $J+1$:
\begin{align*}
	& R^{\mu} \int_{1}^{U} (R + \tfrac{U-u}{2})^{-\mu} \nrm*{(\bfGmm + c_{\bfGmm})^{I}(\bfK + 2r)^{\ell+\nu_{\Box}-1} \bbS_{(\ell)}E_{J+1; f} (u, R + \tfrac{U-u}{2}, \tht)}_{L^{\infty}_{\tht}} \, \ud u \\
	& \aleq D R^{\ell+\nu_{\Box}-1-J-2} \int_{1}^{U} \log^{K_{d}} \left( \frac{R+\tfrac{U-u}{2}}{u} \right) u^{J-\alp_{d}-\dlt_{d}+\nu_{\Box}} \, \ud u \\
	& \peq + D R^{\ell+\nu_{\Box}-2-\alp_{d}+\nu_{\Box}} \int_{1}^{U} u^{-1-\dlt_{d}} \, \ud u \\
	&\aleq D R^{\ell +\nu_{\Box} - 1} R^{-J-2} \log^{K_{d}} (\tfrac{R}{U}) U^{J+1-\alp-\dlt_{0}+\nu_{\Box}}
	+ D R^{\ell+\nu_{\Box}-1} R^{-1-\alp_{d}+\nB} \\
    &\aleq D R^{\ell+\nu_{\Box}-1} R^{-J-1-\de_0} \log^{K_{d}} (\tfrac{R}{U}) U^{J-\alp+\nB},
\end{align*}
where we used $\alp< J+\nB \leq J_{d} -1 + \nB < \alp_{d}$, $\de_0\leq \de_{d}$, $J+1-\alp-\de_0+\nB>0$ in the second inequality and used $
\alp_{d} -\nB - J \geq \alp_{d} - \nB - J_{d} +1 = \eta_{d} \geq \de_0$, $J-\alp+\nB >0$, in the last inequality.

Lastly, we bound the contribution of $E_{J+1} - E_{J+1; \Box, k=0} - E_{J+1; f}$ by Lemma~\ref{lem:exp-error-f-wave} as follows: for $\abs{I} \leq M - C'$,
\begin{align*}
	& R^{\mu} \int_{1}^{U} (R + \tfrac{U-u}{2})^{-\mu} \nrm*{(\bfGmm + c_{\bfGmm})^{I}(\bfK + 2r)^{\ell+\nu_{\Box}-1} \bbS_{(\ell)} (E_{J+1} - E_{J+1; \Box} - E_{J+1; f}) (u, R + \tfrac{U-u}{2}, \tht)}_{L^{\infty}_{\tht}} \, \ud u \\
	& \aleq R^{\mu} \int_{1}^{U} \td{\frkL} A' (R + \tfrac{U-u}{2})^{-\mu+\ell+\nB-1-J-2} \log \Big(\f{R+\tfrac{U-u}2}{u}\Big) \log^{K_J-1} (R+\tfrac{U-u}2) \, \ud u\\
    & \peq + R^{\mu} \int_{1}^{U} A' (R + \tfrac{U-u}{2})^{-\mu+\ell+\nB-1-J-2} \log^{K_{J+1}} \Big(\f{R+\tfrac{U-u}2}{u}\Big) u^{J-\alp-\de_0+\nB} \, \ud u \\
    & \peq + R^{\mu} \int_{1}^{U} A' (R + \tfrac{U-u}{2})^{-\mu+\ell+\nB-1-J_{c}-\eta_c} u^{J_{c}-2+\eta_{c}-\alp-\de_0+\nB} \, \ud u \\
    & \aleq \td{\frkL} A' R^{\ell+\nB-1-J-2} (\log^{K_J} R)U + A' R^{\ell+\nB-1-J-1-\de_{0}} \log^{K_{J+1}} (\tfrac RU) U^{J-\alp+\nB}, 
\end{align*}
where we used $J+1-\alp-\de_0+\nB > 0$, $J_{c}-1-\alp-\de_0+\nB+\eta_c \geq J - \alp + \nB >0$, $1-\de_0>0$ and $J_{c}-1-J+\eta_{c}-\de_0 \geq 0$ in the last line.

Next, we handle the contribution of $e_{J+1; \wave}$ using \eqref{eq:eJ+1wave-crho}. For the $\check{e}_{J+1;\wave,\mathrm{linear},1}$ term, we use \eqref{eq:wave-main-crhoJ} and proceed as the corresponding term in Case~1, Step~3 to obtain, for $\abs{I} \leq M - C'$,
\begin{align*}
	& R^{\mu} \int_{1}^{U} (R + \tfrac{U-u}{2})^{-\mu} \nrm*{(\bfGmm + c_{\bfGmm})^{I}(\bfK + 2r)^{\ell+\nu_{\Box}-1} \bbS_{(\ell)} \check{e}_{J+1;\wave,\mathrm{linear},1} (u, R + \tfrac{U-u}{2}, \tht)}_{L^{\infty}_{\tht}} \, \ud u \\
	& \aleq A' R^{\ell+\nu_{\Box}-1} R^{-\alp-2+\nu_{\Box}} \log^{K_{c}'} (\tfrac{R}{U}) U^{1-\dlt_{c}} \\
	& \aleq A' R^{\ell+\nu_{\Box}-1} R^{-J-1-\dlt_{0}} \log^{K_{c}'} (\tfrac{R}{U}) U^{J-\alp+\nu_{\Box}},
\end{align*}
where in the last line we used $\de_{c} \leq \de_{0}$ and $1-\de_0+J-\alp+\nB >0$.

To treat the $\check{e}_{J+1;\wave,\mathrm{linear},2}$ term in \eqref{eq:eJ+1wave-crho}, we use $r^{-J} \rPhi_{(\geq J - \nu_{\Box}+1) J, 0} = O_{\bfGmm}^{M-C'}(r^{-J} u^{J -\alp+\nu_{\Box}})$ (by Proposition~\ref{prop:recurrence}.(3)). We may proceed as the preceding bound and obtain, for $\abs{I} \leq M - C'$,
\begin{align*}
	& R^{\mu} \int_{1}^{U} (R + \tfrac{U-u}{2})^{-\mu} \nrm*{(\bfGmm + c_{\bfGmm})^{I}(\bfK + 2r)^{\ell+\nu_{\Box}-1} \bbS_{(\ell)} \check{e}_{J+1;\wave,\mathrm{linear},2} (u, R + \tfrac{U-u}{2}, \tht)}_{L^{\infty}_{\tht}} \, \ud u \\
	& \aleq A' R^{\ell+\nu_{\Box}-1} R^{-J-2} \log^{K_{c}'} (\tfrac{R}{U}) U^{J+1-\alp-\dlt_{c}+\nu_{\Box}} \\
    & \aleq A' R^{\ell+\nu_{\Box}-1} R^{-J-1-\de_0} \log^{K_{c}'} (\tfrac{R}{U}) U^{J-\alp+\nu_{\Box}},
\end{align*}
where we used $J+1-\alp-\dlt_{c}+\nu_{\Box} \geq J+1-\alp-\dlt_{0}+\nu_{\Box} >0$ in the first inequality and $\de_{c} \leq \de_{0}$ in the second inequality.

Turning to $\check{e}_{J+1;\wave,\mathrm{nonlinear},1}$ and $\check{e}_{J+1;\wave,\mathrm{nonlinear},2}$, we note that, as in Case 1, Step 3,  after using Definition~\ref{def:alp-N}.(3) with $\alp_{\calN}' = \alp_{\calN}+\de_0$, these terms can be estimated in exactly the same manner as $\check{e}_{J+1;\wave,\mathrm{linear},1}$ and $\check{e}_{J+1;\wave,\mathrm{linear},2}$. (In fact, they are slightly better and do not involve logarithms.) Hence,
\begin{align*}
    & R^{\mu} \int_{1}^{U} (R + \tfrac{U-u}{2})^{-\mu} \nrm*{(\bfGmm + c_{\bfGmm})^{I}(\bfK + 2r)^{\ell+\nu_{\Box}-1} \bbS_{(\ell)} \check{e}_{J+1;\wave,\mathrm{nonlinear},1} (u, R + \tfrac{U-u}{2}, \tht)}_{L^{\infty}_{\tht}} \, \ud u \\
    & + R^{\mu} \int_{1}^{U} (R + \tfrac{U-u}{2})^{-\mu} \nrm*{(\bfGmm + c_{\bfGmm})^{I}(\bfK + 2r)^{\ell+\nu_{\Box}-1} \bbS_{(\ell)} \check{e}_{J+1;\wave,\mathrm{nonlinear},2} (u, R + \tfrac{U-u}{2}, \tht)}_{L^{\infty}_{\tht}} \, \ud u \\
    & \aleq A' R^{\ell+\nu_{\Box}-1} R^{-J-1-\de_0} U^{J-\alp+\nu_{\Box}},
\end{align*}
which is acceptable. \qedhere
\end{proof}

\begin{proof}[Proof of Case~3 ($J = \min \{J_{c}, J_{d}\}$ and $\alp + \dlt_{0} < \min \set{J_{c}-1+\eta_{c}, J_{d}-1+\eta_{d}} + \nB$)]
The proof in this case is very similar to Case~1; we leave the details to the reader.\end{proof}

\begin{proof}[Proof of Case~4 ($J = \min \{J_{c}, J_{d}\}$ and $\alp + \dlt_{0} > \min \set{J_{c}-1+\eta_{c}, J_{d}-1+\eta_{d}} + \nB$)]
The proof in this case is similar to Case~1, but in order to obtain sharp bounds, we need to utilize the sharp Huygens principle to limit the integration domain in the application of Lemma~\ref{lem:Qj-rp}. The idea is to use $\hrho{U}$ defined in \eqref{eq:hrho-J}.

\pfstep{Step~1: Proof of the remainder bound assuming key error bounds}
It is sufficient to establish establish \eqref{eq:wave-main-rhoJ-final} on $\calC_{U} \cap \calM_{\wave}$ for each fixed $U > 3200 \eta_{0}^{-1}$. In view of \eqref{eq:hrho-J-huygens}, we have the identity
\begin{equation} \label{eq:wave-main-4-huygens}
\rho_{J} = \hrho{U}_{(\geq J - \nu_{\Box} + 1) J} + \rho_{(\leq J - \nu_{\Box}) J} \quad \hbox{ on } \calC_{U} \cap \calM_{\wave}.
\end{equation}
For the terms on the right-hand side, we use equations \eqref{eq:hrho-J-rp-eqn} and \eqref{eq:rho-J-char-eqn}, respectively.
To analyze \eqref{eq:hrho-J-rp-eqn}, we will use the following error bounds, which will be proved in Step~2: For $\abs{I} \leq M - C'$, with $C'$ to be determined at the end of Step~2, $p$ satisfying
\begin{align}
	p &< 1 - 2 (J - (J_{c}-1 + \eta_{c})), \label{eq:wave-main-4-p-upper1} \\
	p &< 1 - 2 (J - (J_{d} - 1 + \eta_{d})), \label{eq:wave-main-4-p-upper2} \\
	p &> -1 + 2\de_{0}, \label{eq:wave-main-4-p-lower0}
\end{align}
and $\chi_{**}(u,r)$, $\chi_{*}(u,r)$ denoting
\begin{equation}
    \chi_{**}(u,r) =\chi_{>2R_{\far}(r)} \chi_{>1}(\tfrac{u + 2r}{\frac{1}{2} U}),\quad  \chi_*(u,r) = \chi_{**}(u,r) \chi_{> 2}(u),
\end{equation}
we have 
\begin{align}
& \int_{1}^{U} \left( \int_{\calC_{u} \cap \calM_{\wave}} r^{p} \abs*{\bbS_{(\geq J - \nu_{\Box} + 1)}(\bfGmm + c_{\bfGmm})^{I}(\bfK + 2r)^{J} \left[ \chi_*(u,r) E_{J})\right] }^{2} \, \ud r \ud \rssgm(\tht) \right)^{\frac{1}{2}}\ud u \notag  \\
 & \aleq A' U^{\frac{p-1}{2}} \max\set{\log^{K_{J}} U, U^{J - (J_{c}-1 + \eta_{c})}, U^{J - (J_{d} - 1 + \eta_{d})}, U^{J - \alp - \dlt_{0} + \nu_{\Box}}},  \label{eq:wave-main-4-rp-error-EJ-wave} \\
&\left( \iint_{\calD_{1}^{U} \cap \calM_{\med}} r^{p+1} \abs*{\bbS_{(\geq J - \nu_{\Box} + 1)}(\bfGmm + c_{\bfGmm})^{I}(\bfK + 2r)^{J}\left[ \chi_*(u,r) E_{J}\right] }^{2} \, \ud u \ud r \ud \rssgm(\tht) \right)^{\frac{1}{2}} \notag\\
&\aleq A' U^{\frac{p-1}{2}} \max\set{U^{J - (J_{d} - 1 + \eta_{d})-\dlt_{0}}, U^{J - \alp - \dlt_{0} + \nu_{\Box}}}, \label{eq:wave-main-4-rp-error-EJ-med} \\
 &\int_{1}^{U} \left( \int_{\calC_{u} \cap \calM_{\wave}} r^{p} \abs*{\bbS_{(\geq J - \nu_{\Box} + 1)}(\bfGmm + c_{\bfGmm})^{I}(\bfK + 2r)^{J} \left[ \chi_*(u,r) e_{J; \wave} \right]}^{2} \, \ud r \ud \rssgm(\tht) \right)^{\frac{1}{2}}\ud u \notag \\
 & \aleq A' U^{\frac{p-1}{2} + J - \alp - \dlt_{0} + \nu_{\Box}}, \label{eq:wave-main-4-rp-error-eJwave-wave} \\
&\left( \iint_{\calD_{1}^{U} \cap \calM_{\med}} r^{p+1} \abs*{\bbS_{(\geq J - \nu_{\Box} + 1)}(\bfGmm + c_{\bfGmm})^{I}(\bfK + 2r)^{J} \left[ \chi_*(u,r) e_{J; \wave} \right]}^{2} \, \ud u \ud r \ud \rssgm(\tht) \right)^{\frac{1}{2}} \notag \\
&\aleq A' U^{\frac{p-1}{2} + J - \alp - \dlt_{0} + \nu_{\Box}},  \label{eq:wave-main-4-rp-error-eJwave-med} \\
&\left( \iint_{\calD_{1}^{U} \cap \calM_{\med}} r^{p+1} \abs*{\bbS_{(\geq J - \nu_{\Box} + 1)}(\bfGmm + c_{\bfGmm})^{I}(\bfK + 2r)^{J} \left[ \chi_*(u,r) e_{\med}\right]}^{2} \, \ud u \ud r \ud \rssgm(\tht) \right)^{\frac{1}{2}} \notag \\
&\aleq A' U^{\frac{p-1}{2}} \max\set{U^{J - (J_{d} - 1 + \eta_{d})-\dlt_{0}}, U^{J - \alp - \dlt_{0} + \nu_{\Box}}},  \label{eq:wave-main-4-rp-error-emed} \\
 &\int_{1}^{U} \left( \int_{\calC_{u} \cap \calM_{\wave}} r^{p} \abs*{\bbS_{(\geq J - \nu_{\Box} + 1)}(\bfGmm + c_{\bfGmm})^{I}(\bfK + 2r)^{J} \left[ \chi_{**}(u,r) \chi'_{> 2}(u) \rd_{r} \rho_{(\geq J - \nu_{\Box} + 1) J} \right]}^{2} \, \ud r \ud \rssgm(\tht) \right)^{\frac{1}{2}}\ud u \notag \\
 &\aleq A' U^{\frac{p-1}{2}}\max\set{\log^{K_{J}} U, U^{J - (J_{c}-1 + \eta_{c})}, U^{J - (J_{d} - 1 + \eta_{d})}},  \label{eq:wave-main-4-rp-error-id} \\
 &\left( \iint_{\calD_{1}^{U} \cap \set{2 R_{\far} < r < 4 R_{\far}}} R_{\far}^{p-1} \left(\abs{\rd_{r} \bfGmm^{I} {}^{[U]}\rho_J}^{2} + r^{-2} \abs{\rsnb \bfGmm^{I} {}^{[U]}\rho_J}^{2} + r^{-2} (\bfGmm^{I} {}^{[U]}\rho_J)^{2}\right)\, \ud u \ud r \ud \rssgm(\tht) \right)^{\frac{1}{2}}  \notag \\
 &\aleq A' U^{\frac{p-1}{2} + J - \alp - \dlt_{0} + \nu_{\Box}}. \label{eq:wave-main-4-rp-error-R}
\end{align}
Observe that, thanks to $U > 3200 \eta_{0}^{-1}$, we have $\supp \chi_{> 1}(\tfrac{u + 2r}{\frac{1}{2} U}) \chi'_{> 2}(u) \subseteq \calM_{\wave}$.

To analyze \eqref{eq:rho-J-char-eqn}, we will use the following error bounds, which will be proved in Step~3: For $\abs{I} \leq M - C'$, with $C'$ to be determined at the end of Step~3, and $0 \leq \ell \leq J - \nu_{\Box}$, we have
\begin{align}
	& R^{\mu} \int_{1}^{U} \left. r^{-\mu} \nrm*{(\bfGmm + c_{\bfGmm})^{I}(\bfK + 2r)^{\ell+\nu_{\Box}-1} \bbS_{(\ell)} E_{J}}_{L^{\infty}_{\tht}} \right|_{(u, r) = (\sgm, R+\frac{U-\sgm}{2}) }\, \ud \sgm \notag \\
	&\aleq A' R^{\ell+\nu_{\Box}-1} \max\{R^{- J - 1} \log^{K_{J}} R,
	R^{-(J_{c}-1 + \eta_{c})-1}, R^{-(J_{d} -1 + \eta_{d}) -1}, \notag \\
	&\phantom{\aleq A' R^{\ell+\nu_{\Box}-1} \max\{} \,
	R^{-(J_{c}-1 + \eta_{c})-1} U^{J_{c}-1 +\eta_{c} - \alp - \dlt_{0} + \nu_{\Box}}, R^{-J-1} \log^{K_{d}} (\tfrac{R}{U}) U^{J-(J_{d}-1-\eta_{d})-\dlt_{d}}, \notag \\
	&\phantom{\aleq A' R^{\ell+\nu_{\Box}-1} \max\{} R^{- J - 1} \log^{K_{J}} (\tfrac{R}{U}) U^{J - \alp - \dlt_{0} + \nu_{\Box}} \}, \label{eq:wave-main-4-char-error-EJ} \\
	& R^{\mu} \int_{1}^{U} \left. r^{-\mu} \nrm*{(\bfGmm + c_{\bfGmm})^{I}(\bfK + 2r)^{\ell+\nu_{\Box}-1} \bbS_{(\ell)} e_{J; \wave}}_{L^{\infty}_{\tht}} \right|_{(u, r) = (\sgm, R+\frac{U-\sgm}{2}) }\, \ud \sgm \notag \\
	& \aleq A' R^{\ell+\nu_{\Box}-1} R^{-\alp+\nu_{\Box}-2} \log^{K_{c}'} (\tfrac{R}{U}) U^{1-\dlt_{0}}, \label{eq:wave-main-4-char-error-eJwave}
\end{align}
where $\mu = 2(\ell+\nu_{\Box}-1)+I_{r \rd_{r}}$.
In the remainder of this step, we will assume the above key error bounds and establish \eqref{eq:wave-main-rhoJ-final}. Since the argument is very similar to Case~1,~Step~1(a), we shall simply sketch the argument and focus only on the numerology.

\pfstep{Step~1(a): Estimates for high spherical harmonics}
We proceed as in the proof of \eqref{eq:wave-main-1-rp} in Case~1,~Step~1(a), which involves an application of Lemma~\ref{lem:Qj-rp} with $R_1 = 4 R_{\far}$ to the equation~\eqref{eq:hrho-J-rp-eqn} for $\hrho{U}$. Here, we do not have the initial data term (cf.~\eqref{eq:wave-main-1-id}) on $\calC_{1}$ thanks to the $\chi_{>2}(u)$ cutoff. Note also that we need not consider the term with the commutator $[\Box_{\bfm}, \chi_{>2R_{\far}}]$ on the right-hand side of \eqref{eq:hrho-J-rp-eqn}, since that term is supported in $r \leq 2R_{\far}$. Thus, using \eqref{eq:wave-main-4-rp-error-R} for the second term on the right-hand side of \eqref{eq:Qj-rp} and using the bounds \eqref{eq:wave-main-4-rp-error-EJ-wave}--\eqref{eq:wave-main-4-rp-error-id} claimed above for the third and fourth terms on the right-hand side of \eqref{eq:Qj-rp}, we obtain the following: for $0 \leq \abs{I} \leq M - C'$, we have
\begin{align*}
& \sup_{u \in [1, U]} \left(\int_{\calC_{u}} \chi_{>4 R_{\far}} r^{p} (\rd_{r} \bfGmm^{I} \bfK^{J} (\hrho{U}_{(\geq J - \nu_{\Box}+1) J}))^{2}  \, \ud r \ud \rssgm(\tht) \right)^{\frac{1}{2}} \\
&+ \left( \iint_{\calD_{1}^{U}} \chi_{>4 R_{\far}} r^{p-1} (\rd_{r} \bfGmm^{I} \bfK^{J} \hrho{U}_{(\geq J - \nu_{\Box}+1) J}))^{2} \, \ud u \ud r \ud \rssgm(\tht) \right)^{\frac{1}{2}} \\
&+ \left( \iint_{\calD_{1}^{U}} \chi_{>4 R_{\far}} r^{p-3} \left(\abs{\rsnb \bfGmm^{I} \bfK^{J} \hrho{U}_{(\geq J - \nu_{\Box}+1) J})}^{2}  + (\bfGmm^{I} \bfK^{J} \hrho{U}_{(\geq J - \nu_{\Box}+1) J}))^{2} \right) \, \ud u \ud r \ud \rssgm(\tht) \right)^{\frac{1}{2}} \\
& \aleq A' U^{\frac{p-1}{2}} \max\set{\log^{K_{J}} U, U^{J - (J_{c}-1 + \eta_{c})}, U^{J - (J_{d} - 1 + \eta_{d})}, U^{J - \alp - \dlt_{0} + \nu_{\Box}}}.
\end{align*}
Then by the argument involving Lemma~\ref{lem:hardy-radial} and Taylor's theorem as in the end of Case~1,~Step~1(a), we obtain the following estimate in $\calM_{\wave}$: for $\abs{I} \leq M - C'$,
\begin{equation*}
	\abs{\bfGmm^{I} (\hrho{U}_{(\geq J - \nu_{\Box} + 1) J})(U, R, \Tht)} \aleq A' R^{-J - \frac{p-1}{2}} U^{\frac{p-1}{2}} \max\set{\log^{K_{J}} U, U^{J - (J_{c}-1 + \eta_{c})}, U^{J - (J_{d} - 1 + \eta_{d})}},
\end{equation*}
where we used $U^{J - \alp - \dlt_{0} + \nu_{\Box}} \leq \max\set{U^{J - (J_{c}-1 + \eta_{c})}, U^{J - (J_{d} - 1 + \eta_{d})}}$ by hypothesis.

We now choose the parameter $p$. Keeping in mind $J = \min\set{J_{c}, J_{d}}$,  we note that \eqref{eq:wave-main-4-p-upper1}--\eqref{eq:wave-main-4-p-lower0} is guaranteed by the condition
\begin{equation}\label{eq:last.step.p.choice}
	J - 1 +\de_0  < J + \tfrac{p-1}{2} < \alp_{\mathfrak f} -\nB,
\end{equation}
where $\alp_{\mathfrak f} = \min\set{J, J_{c}-1 + \eta_{c}, J_{d} - 1 + \eta_{d}} + \nu_{\Box}$ as in the statement of Proposition~\ref{prop:wave-main}. Note that there exists $p$ satisfying \eqref{eq:last.step.p.choice} since $\de_0 < \min \{\f 12, \eta_{c}, \eta_{d}\}$. For every $0 < \dlt_{\mathfrak f}' < \min \{\f 12, \eta_{c}, \eta_{d}\} - \de_0$ as in the statement of the theorem, we choose $J + \tfrac{p-1}{2} = \alp_{\mathfrak f} - \dlt_{\mathfrak f}' - \nu_{\Box}$. Also using \eqref{eq:wave-main-4-huygens}, we obtain
\begin{align*}
	\abs{\bfGmm^{I} \rho_{(\geq J - \nu_{\Box} + 1) J}(U, R, \Tht)}
	&= \abs{\bfGmm^{I} (\hrho{U}_{(\geq J - \nu_{\Box} + 1) J})(U, R, \Tht)} \\
	&\aleq A' R^{-\alp_{\mathfrak f} + \dlt_{\mathfrak f}' + \nu_{\Box}} U^{\alp_{\mathfrak f} -\dlt_{\mathfrak f}' - \nu_{\Box} - J} \max\set{\log^{K_{J}} U, U^{J - (J_{c}-1 + \eta_{c})}, U^{J - (J_{d} - 1 + \eta_{d})}} \\
	&\aleq A' R^{-\alp_{\mathfrak f} + \dlt_{\mathfrak f}' + \nu_{\Box}} U^{-\dlt_{\mathfrak f}'} \max\set{U^{\alp_{\mathfrak f} - J - \nu_{\Box}} \log^{K_{J}} U, 1},
\end{align*}
which is acceptable for the proof of \eqref{eq:wave-main-rhoJ-final}.

\pfstep{Step~1(b): Estimates for low spherical harmonics}
Proceeding as in the proof of in Case~1, Step~1(b), which involves application of Lemma~\ref{lem:Qj-char} to the equation \eqref{eq:rho-J-char-eqn} for $\rho_{(\ell)}$, but using the bounds \eqref{eq:wave-main-4-char-error-EJ} and \eqref{eq:wave-main-4-char-error-eJwave} claimed above, we obtain the following: for $0 \leq \ell \leq J - \nu_{\Box}$ and $\abs{I} \leq M - C'$, we have
\begin{align*}
	&\abs{\rd_{r} \bfGmm^{I} \bfK^{\ell+\nu_{\Box}-1} \rho_{(\ell) J}(U, R, \Tht)}
	+ r^{-1} \abs{\bfGmm^{I} \bfK^{\ell+\nu_{\Box}-1} \rho_{(\ell) J}(U, R, \Tht)} \\
	&\aleq A' R^{\ell+\nu_{\Box}-1} \max\{R^{- J - 1} \log^{K_{J}} R,
	R^{-(J_{c}-1 + \eta_{c})-1}, R^{-(J_{d} -1 + \eta_{d}) -1}, \notag \\
	&\phantom{\aleq A' R^{\ell+\nu_{\Box}-1} \max\{} \,
	R^{-(J_{c}-1 + \eta_{c})-1} U^{J_{c}-1 +\eta_{c} - \alp - \dlt_{0} + \nu_{\Box}},
	R^{-J-1} \log^{K_{d}} ( \tfrac{R}{U} ) U^{J-(J_{d}-1+\eta_{d})-\dlt_{d}}, \\
	&\phantom{\aleq A' R^{\ell+\nu_{\Box}-1} \max\{} \,
	R^{- J - 1} \log^{K_{J}} (\tfrac{R}{U}) U^{J - \alp - \dlt_{0} + \nu_{\Box}}, R^{-\alp+\nu_{\Box}-2} \log^{K_{c}'} (\tfrac{R}{U})U^{1-\dlt_{0}} \},
\end{align*}
where the third-to-last term inside $\max$ arises from the first term on the right-hand side of \eqref{eq:Qj-char} (i.e., contribution of $\rho_{J}$ on $\calC_{1}$), which is bounded exactly as in Case~1, Step~1(b) (note also that the right-hand side of \eqref{eq:wave-main-4-char-error-eJwave} is bounded from above by the contribution of the last term inside $\max$).

By \eqref{eq:wave-main-hyp-rhoJ}, we have $\bfGmm^{I} \rho_{J} = o(r^{-J +1})$ as $r \to \infty$ for $\abs{I} \leq M - C'$. Hence, using Taylor's theorem in the variable $z = \frac{1}{r}$ as in Case~1, Step~1(b), and noticing that $\ell+\nu_{\Box}-1- \min\set{J_{c}-1+\eta_{c}, J_{d}-1+\eta_{d}, J}  < 0$ (since $\ell+\nu_{\Box} -1 \leq J - 1 = \min\set{J_{c}-1, J_{d}-1}$), we obtain the following: for $\abs{I} \leq M - C'$, we have
\begin{equation} \label{eq:wave-main-4-char}
\begin{aligned}
	\abs{\bfGmm^{I}\rho_{(\leq J - \nu_{\Box}) J}(U, R, \Tht)}
	&\aleq A' \max\{R^{- J} \log^{K_{J}} R,
	R^{-(J_{c}-1 + \eta_{c})}, R^{-(J_{d} -1 + \eta_{d})}, \notag \\
	&\phantom{\aleq A' \max\{} \,
	R^{-(J_{c}-1 + \eta_{c})} U^{J_{c}-1 +\eta_{c} - \alp - \dlt_{0} + \nu_{\Box}},
	R^{-J} \log^{K_{d}} ( \tfrac{R}{U} ) U^{J-(J_{d}-1+\eta_{d})-\dlt_{d}}, \\
	&\phantom{\aleq A' \max\{} \,
	R^{- J} \log^{K_{J}} (\tfrac{R}{U}) U^{J - \alp - \dlt_{0} + \nu_{\Box}}, R^{-\alp+\nu_{\Box}-2} \log^{K_{c}'} (\tfrac{R}{U})U^{1-\dlt_{0}} \}.
\end{aligned}
\end{equation}
Since $\alp_{\mathfrak f}  - \nu_{\Box} = \min\set{J, J_{c}-1 + \eta_{c}, J_{d} - 1 + \eta_{d}}$ (and also remembering $\dlt_{d} > 0$ and $\alp + \dlt_{0} > \min\set{J_{c}-1 + \eta_{c}, J_{d} - 1 + \eta_{d}} +\nB > \alp_{\mathfrak f}$), we may derive from the above bound that
\begin{align*}
\abs{\bfGmm^{I}\rho_{(\leq J - \nu_{\Box}) J}(U, R, \Tht)}
\aleq A' R^{-\alp_{\mathfrak f} + \dlt_{\mathfrak f}' + \nu_{\Box}} U^{-\dlt_{\mathfrak f}'} \max\set{U^{\alp_{\mathfrak f} - J - \nu_{\Box}} \log^{K_{J}} U, 1},
\end{align*}
which is acceptable for the proof of \eqref{eq:wave-main-rhoJ-final}.

\pfstep{Step~2: Proof of error bounds for high spherical harmonics}
The aim of this step is to establish \eqref{eq:wave-main-4-rp-error-EJ-wave}--\eqref{eq:wave-main-4-rp-error-id}. Since the argument is significantly different from Case~1, Step~2, we shall give a detailed proof.

Throughout this step, we work with the same conventions as in Case~1, Step~2. We shall often use the inclusions
\begin{equation*}
\supp \chi_{> 1}(\tfrac{u + 2r}{\frac{1}{2} U})  \subseteq \set{u + 2r \geq \tfrac{1}{4} U} \subseteq \set{u \geq \tfrac{1}{8} U } \cup \set{r \geq \tfrac{1}{16} U}
\end{equation*}
to split the integration domain. Moreover, we will freely use \eqref{eq:hF-Gmm} to dispose the cutoff $\chi_{> 1}(\tfrac{u + 2r}{\frac{1}{2} U})$ in our estimates. We also remind the reader that by hypothesis, $\alp + \dlt_{0}$ is not equal to $J + \nu_{\Box}$, $J_{c}-1 + \eta_{c}$, nor $J_{d} - 1 + \eta_{d}$. This fact will ensure that no logarithms appear in various upper bounds for integrals below.

\pfstep{Step~2(a): Error estimate in $\calM_{\wave}$}
In this substep, we prove \eqref{eq:wave-main-4-rp-error-EJ-wave} and \eqref{eq:wave-main-4-rp-error-eJwave-wave}. To simplify the exposition, given any function $F$ defined on $\calM_{\wave}$, we introduce the shorthand
\begin{equation*}
\frkJ_{p, \wave}^{I, J}[F] = \int_{1}^{U} \left( \int_{\calC_{u} \cap \calM_{\wave}} r^{p} \abs*{\bbS_{(\geq J - \nu_{\Box} + 1)}(\bfGmm + c_{\bfGmm})^{I}(\bfK + 2r)^{J} \left[ \chi_{> 1}(\chi_*(u,r) F)\right] }^{2} \, \ud r \ud \rssgm(\tht) \right)^{\frac{1}{2}}\ud u.
\end{equation*}
Using \eqref{eq:hF-Gmm}, we may show that, for $m \geq \abs{I} + J + C'$,
\begin{align*}
\frkJ_{p, \wave}^{I, J}[F] \aleq J_{p, \wave, \early}^{m, J}[F] + J_{p, \wave, \late}^{m, J}[F]
\end{align*}
where the terms on the right-hand side denote the simpler integrals
\begin{align*}
J_{p, \wave, \early}^{m, J}[F] &= \int_{1}^{\frac{1}{4} U} \left( \int_{\max\set{\frac{1}{16} U, \eta_{0}^{-1} u}}^{\infty} r^{p} \nrm*{r^{J} \bfGmm^{(\leq m)} F}_{L^{\infty}_{\tht}}^{2} \, \ud r \right)^{\frac{1}{2}}\ud u, \\
J_{p, \wave, \late}^{m, J}[F] &= \int_{\frac{1}{8} U}^{U} \left( \int_{\frac{1}{8} \eta_{0}^{-1} U}^{\infty} r^{p} \nrm*{r^{J} \bfGmm^{(\leq m)} F }_{L^{\infty}_{\tht}}^{2} \, \ud r \right)^{\frac{1}{2}}\ud u.
\end{align*}
Since $(\bfK + 2r)^{J} E_{J; \Box, k=0} = 0$, \eqref{eq:wave-main-4-rp-error-EJ-wave} can be view as estimates for $\frkJ_{p, \wave}^{I, J}[F]$ with $F = E_{J} - E_{J;\Box,k=0} - E_{J;f}$ and $F = E_{J;f}$. Similarly, \eqref{eq:wave-main-4-rp-error-eJwave-wave} is an estimate for $\frkJ_{p, \wave}^{I, J}[F]$ with $F = e_{J ; \wave}$. So to prove these estimates, it suffices to estimate $J_{p, \wave, \early}^{m, J}[F]$ and $J_{p, \wave, \late}^{m, J}[F]$ for such $F$'s.

We begin with the contribution of $E_{J}$. As before, we split $E_{J} = E_{J; \Box, k=0} + E_{J; f} + (E_{J} - E_{J; \Box, k=0} - E_{J; f})$ and observe that $(\bfK + 2r)^{J} E_{J; \Box, k=0} = 0$. Using \eqref{eq:exp-error-wave}, we have, for $m \leq M - C' - J$,
\begin{align*}
& J_{p, \wave, \early}^{m, J}[E_{J} - E_{J; \Box, k=0} - E_{J; f}] \\
& \aleq \int_{1}^{\frac{1}{4} U} \left( \int_{\frac{1}{16} U}^{\infty} r^{p} \left( A' r^{J} r^{-J-1} \log^{K_{J}} (\tfrac{r}{u}) u^{J - 1 - \alp - \dlt_{0} + \nu_{\Box}}\right)^{2} \, \ud r \right)^{\frac{1}{2}} \ud u\\
&\peq + \int_{1}^{\frac{1}{4} U} \left( \int_{\frac{1}{16} U}^{\infty} r^{p} \left( A' r^{J} r^{-J_{c}-\eta_{c}} u^{J_{c}-2 + \eta_{c} -\alp - \dlt_{0} + \nu_{\Box}} \right)^{2} \, \ud r \right)^{\frac{1}{2}} \ud u \\
& \aleq A' U^{\frac{p-1}{2}} \int_{1}^{\frac{1}{4} U}  \log^{K_{J}} (\tfrac{U}{u}) u^{J - 1 - \alp - \dlt_{0} + \nu_{\Box}} \, \ud u
+ A' U^{\frac{p-1}{2}+ J - (J_{c}-1 + \eta_{c})} \int_{1}^{\frac{1}{4} U} u^{J_{c}-2 + \eta_{c} - \alp - \dlt_{0} + \nu_{\Box}} \, \ud u \\
& \aleq A' U^{\frac{p-1}{2}} \max\set{\log^{K_{J}} U, U^{J - (J_{c}-1 + \eta_{c})}, U^{J - \alp - \dlt_{0} + \nu_{\Box}}},
\end{align*}
where we used $p < 1$ and \eqref{eq:wave-main-4-p-upper1} for the second inequality. Similarly, for $m \leq M - C' - J$,
\begin{align*}
& J_{p, \wave, \late}^{m, J}[E_{J} - E_{J; \Box, k=0} - E_{J; f}] \\
& \aleq \int_{\frac{1}{8} U}^{U} \left( \int_{\frac{1}{8} \eta_{0}^{-1} U}^{\infty} r^{p} \left( A' r^{J} r^{-J-1} \log^{K_{J}} (\tfrac{r}{u}) u^{J - 1 - \alp - \dlt_{0} + \nu_{\Box}}\right)^{2} \, \ud r \right)^{\frac{1}{2}} \ud u\\
&\peq + \int_{\frac{1}{8} U}^{U} \left( \int_{\frac{1}{8} \eta_{0}^{-1} U}^{\infty} r^{p} \left( A' r^{J} r^{-J_{c}-\eta_{c}} u^{J_{c}-2 + \eta_{c} -\alp - \dlt_{0} + \nu_{\Box}} \right)^{2} \, \ud r \right)^{\frac{1}{2}} \ud u \\
& \aleq A' U^{\frac{p-1}{2}} U^{J - \alp - \dlt_{0} + \nu_{\Box}},
\end{align*}
where we used $p < 1$ and \eqref{eq:wave-main-4-p-upper1}. For $E_{J; f}$, we use \eqref{eq:EJ-f}, \eqref{eq:EJ-f-rem}, and \ref{hyp:forcing}. We have, for $m \leq M_{0} - J$,
\begin{align*}
& J_{p, \wave, \early}^{m, J}[E_{J; f}] \\
& \aleq \int_{1}^{\frac{1}{4} U} \left( \int_{\frac{1}{16} U}^{\infty} r^{p} \left( D r^{J} r^{-J-1} \log^{K_{J}} (\tfrac{r}{u}) u^{J - 1 - (J_{d} - 1 + \eta_{d})} \right)^{2} \, \ud r \right)^{\frac{1}{2}} \ud u\\
&\peq + \int_{1}^{\frac{1}{4} U} \left( \int_{\frac{1}{16} U}^{\infty} r^{p} \left( D r^{J} r^{-(J_{d} - 1 + \eta_{d})-1} u^{-1-\dlt_{d}} \right)^{2} \, \ud r \right)^{\frac{1}{2}} \ud u \\
& \aleq D U^{\frac{p-1}{2}} \int_{1}^{\frac{1}{4} U}  \log^{K_{J}} (\tfrac{U}{u}) u^{J - 1 - (J_{d} - 1 + \eta_{d})} \, \ud u
+ D U^{\frac{p-1}{2}+ J - (J_{d} - 1 + \eta_{c})} \int_{1}^{\frac{1}{4} U} u^{-1-\dlt_{d}} \, \ud u \\
& \aleq D U^{\frac{p-1}{2}} \max\set{\log^{K_{J}} U, U^{J - (J_{d} - 1 + \eta_{d})}},
\end{align*}
where we used $p < 1$ and \eqref{eq:wave-main-4-p-upper2} for the second inequality. Similarly, for $m \leq M_0 - J$,
\begin{align*}
J_{p, \wave, \late}^{m, J}[E_{J; f}]
& \aleq \int_{\frac{1}{8} U}^{U} \left( \int_{\frac{1}{8} \eta_{0}^{-1} U}^{\infty} r^{p} \left( D r^{J} r^{-J-1} \log^{K_{J}} (\tfrac{r}{u}) u^{J - 1 - (J_{d} - 1 + \eta_{d}) }\right)^{2} \, \ud r \right)^{\frac{1}{2}} \ud u\\
&\peq + \int_{\frac{1}{8} U}^{U} \left( \int_{\frac{1}{8} \eta_{0}^{-1} U}^{\infty} r^{p} \left( D r^{J} r^{-(J_{d} -1 + \eta_{d}) -1} u^{-1-\dlt_{d}}\right)^{2} \, \ud r \right)^{\frac{1}{2}} \ud u \\
& \aleq D U^{\frac{p-1}{2} + J - (J_{d} - 1 + \eta_{d})} + D U^{\frac{p-1}{2} + J - (J_{d} - 1 + \eta_{d}) - \dlt_{d}}
\end{align*}
where we used $p < 1$ and \eqref{eq:wave-main-4-p-upper2}. This proves \eqref{eq:wave-main-4-rp-error-EJ-wave}.

Next, we handle the contribution of $e_{J; \wave}$ (see \eqref{eq:eJwave-def}), which we split into $e_{J; \wave, \linear}$ and $e_{J; \wave, \nonlinear}$ as before. For the former, we use Lemma~\ref{lem:conj-wave-wave} and the hypothesis \eqref{eq:wave-main-hyp-rhoJ} on $\rho_{J}$. We have, for $m \leq M - J - 2$,
\begin{align*}
J_{p, \wave, \early}^{m, J}[e_{J; \wave, \linear}]
& \aleq \int_{1}^{\frac{1}{4} U} \left( \int_{\frac{1}{16} U}^{\infty} r^{p}  \left( A r^{J} r^{-2} \log^{K_{c}'} (\tfrac{r}{u}) u^{-\dlt_{c}} r^{-\alp + \nu_{\Box}} \right)^{2} \, \ud r \right)^{\frac{1}{2}} \ud u \\
& \aleq A U^{\frac{p-1}{2} + J - 1 - \alp + \nu_{\Box}} \int_{1}^{\frac{1}{4} U} \log^{K_{c}'}(\tfrac{U}{u}) u^{-\dlt_{c}} \, \ud u \\
& \aleq A U^{\frac{p-1}{2} + J - \alp - \dlt_{c} + \nu_{\Box}},
\end{align*}
where we used $p < 1$ and $\alp > J + \nu_{\Box} - 1$ in the second inequality. Similarly, for $m \leq M - J - 2$,
\begin{align*}
J_{p, \wave, \late}^{m, J}[e_{J; \wave, \linear}]
& \aleq \int_{\frac{1}{8} U}^{U} \left( \int_{\frac{1}{8} \eta_{0}^{-1} U}^{\infty} r^{p}  \left( A r^{J} r^{-2} \log^{K_{c}'} (\tfrac{r}{u}) u^{-\dlt_{c}} r^{-\alp + \nu_{\Box}} \right)^{2} \, \ud r \right)^{\frac{1}{2}} \ud u \\
& \aleq A U^{\frac{p-1}{2} + J - \alp - \dlt_{c} + \nu_{\Box}},
\end{align*}
where we used $p < 1$ and $\alp > J + \nu_{\Box} - 1$. For $e_{J; \wave, \nonlinear}$, we apply the estimates in Definition~\ref{def:alp-N}.(3) (with $\alp_{\calN}' = \alp_{\calN} + \dlt_{0}$). We have, for $m \leq M - J - 2$,
\begin{align*}
J_{p, \wave, \early}^{m, J}[e_{J; \wave, \nonlinear}]
& \aleq \int_{1}^{\frac{1}{4} U} \left( \int_{\frac{1}{16} U}^{\infty} r^{p}  \left( A' r^{J} r^{-2} u^{-\dlt_{0}} r^{-\alp + \nu_{\Box}} \right)^{2}  \, \ud r \right)^{\frac{1}{2}} \ud u\\
& \aleq A' U^{\frac{p-1}{2} + J - 1 - \alp + \nu_{\Box}} \int_{1}^{\frac{1}{4} U} u^{-\dlt_{0}} \, \ud u \\
& \aleq A' U^{\frac{p-1}{2} + J - \alp - \dlt_{0} + \nu_{\Box}},
\end{align*}
where we used $p < 1$ and $\alp > J + \nu_{\Box} - 1$ in the second inequality. Similarly, for $m \leq M - J - 2$,
\begin{align*}
J_{p, \wave, \late}^{m, J}[e_{J; \wave, \nonlinear}]
& \aleq \int_{\frac{1}{8} U}^{U} \left( \int_{\frac{1}{8} \eta_{0}^{-1} U}^{\infty} r^{p}  \left( A' r^{J} r^{-2} u^{-\dlt_{0}} r^{-\alp + \nu_{\Box}} \right)^{2}  \, \ud r \right)^{\frac{1}{2}} \ud u \\
& \aleq A' U^{\frac{p-1}{2} + J - \alp - \dlt_{0} + \nu_{\Box}},
\end{align*}
where we used $p < 1$ and $\alp > J + \nu_{\Box} - 1$. Recalling finally that $\de_{0} \leq \de_{c}$, we have thus proven \eqref{eq:wave-main-4-rp-error-eJwave-wave}.

\pfstep{Step~2(b): Error estimate in $\calM_{\med}$}
In this substep, we prove \eqref{eq:wave-main-4-rp-error-EJ-med}, \eqref{eq:wave-main-4-rp-error-eJwave-med} and \eqref{eq:wave-main-4-rp-error-emed}. As before, given any function $F$ defined on $\calM_{\med}$, we introduce the shorthand
\begin{equation*}
\frkJ_{p, \med}^{I, J}[F] = \left( \iint_{\calD_{1}^{U} \cap \calM_{\med}} r^{p+1} \abs*{\bbS_{(\geq J - \nu_{\Box} + 1)}(\bfGmm + c_{\bfGmm})^{I}(\bfK + 2r)^{J}\left[ \chi_*(u,r) F \right] }^{2} \, \ud u \ud r \ud \rssgm(\tht) \right)^{\frac{1}{2}}.
\end{equation*}
Note that, since $U > 3200 \eta_{0}^{-1}$, we have
\begin{equation}\label{eq:set.Huygens.Mmed.cutoff}
	\supp \chi_{> 1}(\tfrac{u + 2r}{\frac{1}{2} U}) \cap \calM_{\med} \subseteq \set{u \geq \tfrac{1}{1600} \eta_{0} U} \cap \calM_{\med}
\end{equation}
Using \eqref{eq:hF-Gmm}, we may show that, for $m \geq \abs{I} + J + C_{d}$,
\begin{align*}
\frkJ_{p, \med}^{I, J}[F] \aleq J_{p, \med}^{m, J}[F],
\end{align*}
where
\begin{align*}
J_{p, \med}^{m, J}[F] &=
\left( \int_{\frac{1}{1600} \eta_{0} U}^{U} \int_{R_{\far}}^{100 \eta_{0}^{-1} u} r^{p+1} \nrm*{r^{J} (\bfGmm + c_{\bfGmm})^{(\leq m)} F}_{L^{\infty}_{\tht}}^{2} \, \ud r \ud u  \right)^{\frac{1}{2}}.
\end{align*}

We begin with \eqref{eq:wave-main-4-rp-error-EJ-med}. We decompose $E_{J} = Q_{0}(\Phi_{<J, k=0}) + (E_{J} - Q_{0}(\Phi_{<J, k=0}))$ and note that $(\bfK + 2 r)^{J} Q_{0}(\Phi_{<J, k=0})$. By \eqref{eq:exp-error-med}, we estimate, for $m \leq M - C'$,
\begin{align*}
&J_{p, \med}^{m, J}[E_{J} - Q_{0}(\Phi_{<J, k=0})] \\
&\aleq   \left( \int_{\frac{1}{1600} \eta_{0} U}^{U} \int_{R_{\far}}^{100 \eta_{0}^{-1} u} r^{p+1} \left( A' \chi_{> (2 \eta_{0})^{-1}}(\tfrac{r}{u}) r^{J} u^{-\alp -2-\dlt_{0}+\nu_{\Box}}  \right)^{2} \, \ud r \ud u  \right)^{\frac{1}{2}} \\
&\peq + \left( \int_{\frac{1}{1600} \eta_{0} U}^{U} \int_{R_{\far}}^{100 \eta_{0}^{-1} u} r^{p+1} \left( D \chi_{> (2 \eta_{0})^{-1}}(\tfrac{r}{u}) r^{J} u^{-(J_{d} - 1 + \eta_{d}) -2-\dlt_{d}}  \right)^{2} \, \ud r \ud u  \right)^{\frac{1}{2}} \\
&\aleq A' U^{\frac{p-1}{2} + J - \alp - \dlt_{0} + \nu_{\Box}}
+ D U^{\frac{p-1}{2} + J - (J_{d} - 1 + \eta_{d}) - \dlt_{d}},
\end{align*}
where we used
\begin{equation} \label{eq:wave-main-4-p-lower1}
	p > - 2 - 2J,
\end{equation}
which is implied by \eqref{eq:wave-main-4-p-lower0} (i.e., $p > -1 + 2\de_{0}$) since $J \geq 1$.

Next, we turn to \eqref{eq:wave-main-4-rp-error-eJwave-med}. We decompose $e_{J; \wave}$ into $e_{J; \wave, \linear} + e_{J; \wave, \nonlinear}$ as in \eqref{eq:eJwave-def}, then use Lemma~\ref{lem:conj-wave-med} and Definition~\ref{def:alp-N}.(2) (with $\alp_{\calN}' = \alp_{\calN} + \dlt_{0}$), respectively, for each term. Also taking advantage of the cutoff $\chi_{> (2 \eta_{0})^{-1}}(\tfrac{r}{u})$, we estimate, for $m \leq M -J - 2$,
\begin{align*}
&J_{p, \med}^{m, J}[e_{J; \wave}] \\
&\aleq \left( \int_{\frac{1}{1600} \eta_{0} U}^{U} \int_{R_{\far}}^{100 \eta_{0}^{-1} u} r^{p+1} \left( A' \chi_{> (2 \eta_{0})^{-1}}(\tfrac{r}{u}) r^{J} u^{-\alp -2-\dlt_{c}+\nu_{\Box}}  \right)^{2} \, \ud r \ud u  \right)^{\frac{1}{2}} \\
&\peq + \left( \int_{\frac{1}{1600} \eta_{0} U}^{U} \int_{R_{\far}}^{100 \eta_{0}^{-1} u} r^{p+1} \left( D \chi_{> (2 \eta_{0})^{-1}}(\tfrac{r}{u}) r^{J} u^{-\alp-2-\dlt_{0}+\nu_{\Box}}  \right)^{2} \, \ud r \ud u  \right)^{\frac{1}{2}} \\
&\aleq A' U^{\frac{p+1}{2} + J - \alp - \dlt_{0} + \nu_{\Box}} + D U^{\frac{p+1}{2} + J - \alp - \dlt_{0} + \nu_{\Box}},
\end{align*}
where we used $\dlt_{0} \leq \dlt_{c}$. 

Finally, we prove \eqref{eq:wave-main-4-rp-error-emed}. We split $e_{\med}$ into $e_{\med, \linear}$, $e_{\med, \nonlinear}$ and $e_{\med, f}$ as in \eqref{eq:eJmed-def}. For $e_{\med, \linear}$, we use Lemma~\ref{lem:conj-wave-med} and the hypothesis for $\Phi$. We have, for $m \leq M -J- 2$,
\begin{align*}
J_{p, \med}^{m, J}[e_{\med, \linear}]
& \aleq \left( \int_{\frac{1}{1600} \eta_{0} U}^{U} \int_{R_{\far}}^{100 \eta_{0}^{-1} u} r^{p+1} \left(A r^{J} r^{-2-\dlt_{c}} r^{\nu_{\Box}} u^{-\alp} \right)^{2} \, \ud r \ud u \right)^{\frac{1}{2}} \\
& \aleq \left( \int_{\frac{1}{1600} \eta_{0} U}^{U} u^{p+2} \left(A u^{J} u^{-2-\dlt_{0}} u^{\nu_{\Box}} u^{-\alp} \right)^{2} \, \ud u \right)^{\frac{1}{2}} \\
&\aleq A U^{\frac{p-1}{2}+J - \alp - \dlt_{0} +\nu_{\Box}},
\end{align*}
where, in the second inequality, we used $\dlt_{0} \leq \dlt_{c}$ and the lower bound
\begin{equation} \label{eq:wave-main-4-p-lower2}
	p > 2 - 2 (J - \dlt_{0} + \nu_{\Box}),
\end{equation}
which is satisfied thanks to \eqref{eq:wave-main-4-p-lower0} (i.e., $p > -1 +2\de_{0}$) since $J, \,\nu_{\Box} \geq 1$. 
For $e_{\med, \nonlinear}$, we apply the estimates in Definition~\ref{def:alp-N}.(2) with $\alp_{\calN}' = \alp_{\calN} + \dlt_{0}$, as well as the hypothesis for $\Phi$. Then for $m \leq M -J - 2$, we have
\begin{align*}
J_{p, \med}^{m, J}[e_{\med, \nonlinear}]
& \aleq \left( \int_{\frac{1}{1600} \eta_{0} U}^{U} \int_{R_{\far}}^{100 \eta_{0}^{-1} u} r^{p+1} \left(A' r^{J} r^{-2-\dlt_{0}} r^{\nu_{\Box}} u^{-\alp} \right)^{2} \, \ud r \ud u \right)^{\frac{1}{2}} \\
& \aleq A' U^{\frac{p-1}{2}+J - \alp - \dlt_{0} +\nu_{\Box}},
\end{align*}
where we used the lower bound \eqref{eq:wave-main-4-p-lower2} for $p$ in the second inequality. Finally, for $e_{\med, f}$, we use \ref{hyp:forcing} to show that, for $m \leq M_{0}$,
\begin{align*}
J_{p, \med}^{m, J}[e_{\med, f}]
& \aleq \left( \int_{\frac{1}{1600} \eta_{0} U}^{U} \int_{R_{\far}}^{100 \eta_{0}^{-1} u} r^{p+1} \left(D r^{J} r^{-2-\dlt_{d}+\nu_{\Box}} u^{-(J_{d} - 1 + \eta_{d}) - \nu_{\Box}}
 \right)^{2} \, \ud r \ud u \right)^{\frac{1}{2}} \\
 &\aleq \left( \int_{\frac{1}{1600} \eta_{0} U}^{U} u^{p+2} \left(D u^{J} u^{-2-\dlt_{0}+\nu_{\Box}} u^{-(J_{d} - 1 + \eta_{d}) - \nu_{\Box}}  \right)^{2} \ud u \right)^{\frac{1}{2}} \\
 &\aleq D U^{\frac{p-1}{2} + J  - (J_{d} - 1 + \eta_{d}) - \dlt_{0} + \nu_{\Box}},
\end{align*}
where we used $\dlt_{0} \leq \dlt_{d}$ and the lower bound \eqref{eq:wave-main-4-p-lower2} for $p$ in the second inequality.

\pfstep{Step~2(c): Error estimate in $\set{1 < u < 2}$}
Finally, we prove \eqref{eq:wave-main-4-rp-error-id}. Using $U > 3200 \eta_{0}^{-1}$, for $m \geq \abs{I} + J + C'$, we may estimate using the fundamental theorem of calculus that
\begin{align*}
(\hbox{LHS of  \eqref{eq:wave-main-4-rp-error-id}})
&\aleq
\int_{1}^{2} \left( \int_{\frac{1}{16} U}^{\infty} r^{p} \nrm*{r^{J} \bfGmm^{(\leq m)} \rd_{r} \rho_{(\geq J - \nu_{\Box} + 1) J}}_{L^{\infty}_{\tht}}^{2} \, \ud r \ud \rssgm(\tht) \right)^{\frac{1}{2}}\ud u \\
&\aleq \left( \int_{\frac{1}{16} U}^{\infty} r^{p} \nrm*{r^{J} \rd_{r} \bfGmm^{(\leq m)} \rho_{(\geq J - \nu_{\Box} + 1) J}(1, r, \tht)}_{L^{\infty}_{\tht}}^{2} \, \ud r \ud \rssgm(\tht) \right)^{\frac{1}{2}} \\
&\peq + \sup_{u \in [1, 2]} \left( \int_{\frac{1}{16} U}^{\infty} r^{p} \nrm*{r^{J} \rd_{u} \rd_{r}  \bfGmm^{(\leq m)} \rho_{(\geq J - \nu_{\Box} + 1) J}(u, r, \tht)}_{L^{\infty}_{\tht}}^{2} \, \ud r \ud \rssgm(\tht) \right)^{\frac{1}{2}}.
\end{align*}
By \ref{hyp:id}, we have, for $m \leq M_{0} - 1$,
\begin{equation} \label{eq:wave-main-4-rp-error-id-1}
	\abs{\rd_{r} \bfGmm^{(\leq m)} \rho_{(\geq J - \nu_{\Box} + 1) J}} (1,r,\theta)
	\aleq D r^{-J-1} \log^{K_{d}} r
		+ D r^{-(J_{d} - 1 + \eta_{d}) -1}.
\end{equation}
On the other hand, using \eqref{eq:rho-J-eqn-0}, the commutation identity \eqref{eq:comm-Qj} and the definition of $Q_{0}$ in \eqref{eq:Qj}, we may derive the following equation for $- 2 \rd_{u} \rd_{r} \bfGmm^{I} \rho_{J}$:
\begin{equation} \label{eq:rho-J-eqn}
\begin{aligned}
- 2 \rd_{u} \rd_{r} \bfGmm^{I} \rho_{(\geq J - \nu_{\Box} + 1) J}
&=
O_{\bfGmm}^{\infty} (1) \rd_{r}^{2} \bfGmm^{\leq \abs{I}} \rho_{(\geq J - \nu_{\Box} + 1) J}
+ O_{\bfGmm}^{\infty} (r^{-2}) \rslap \bfGmm^{\leq \abs{I}} \rho_{(\geq J - \nu_{\Box} + 1) J}  \\
&\peq
+ O_{\bfGmm}^{\infty} (r^{-1}) \rd_{r} \bfGmm^{\leq \abs{I}} \rho_{(\geq J - \nu_{\Box} + 1) J}
+ O_{\bfGmm}^{\infty} (r^{-2}) \bfGmm^{\leq \abs{I}} \rho_{(\geq J - \nu_{\Box} + 1) J} \\
&\peq + \bbS_{(\geq J - \nu_{\Box} + 1)}(\bfGmm+c_{\bfGmm})^{I} (- E_{J}  + e_{J; \wave} + e_{\med}).
\end{aligned}
\end{equation}
Since the domain of integration lies inside $\calM_{\wave}$, we only estimate the right-hand side of \eqref{eq:rho-J-eqn} in $\calM_{\wave}$. For this, we use the following: the hypothesis \eqref{eq:wave-main-hyp-rhoJ} for the first four terms; \eqref{eq:exp-error-wave}--\eqref{eq:EJ-f-rem}, as well as \eqref{eq:wave-main-Phij0-high} and \ref{hyp:forcing} for $E_{J}$; \eqref{eq:eJwave-def}--\eqref{eq:eJwave-nonlinear-def}, as well as Lemma~\ref{lem:conj-wave-wave}, \eqref{eq:wave-main-hyp-rhoJ} and Definition~\ref{def:alp-N}.(3) (with $\alp_{\calN}' = \alp_{\calN} + \dlt_{0}$) for $e_{J, \wave}$; and finally that $e_{\med} = 0$ in $\calM_{\wave}$. We arrive at (for $m \leq M - C'$)
\begin{equation} \label{eq:wave-main-4-rp-error-id-2}
\abs{\rd_{u} \rd_{r} \bfGmm^{(\leq m)} \rho_{(\geq J - \nu_{\Box} + 1) J}}
\aleq A' r^{-\alp - 2} + A' r^{-J - 1} \log^{K_{J}} r + A'  r^{-(J_{c}-1 + \eta_{c})-1} + D r^{-(J_{d} - 1 + \eta_{d})-1}.
\end{equation}
Note that the first term on the right-hand side may be absorbed into the second term since $\alp > J - 1$. Using \eqref{eq:wave-main-4-rp-error-id-1} and \eqref{eq:wave-main-4-rp-error-id-2} to continue to estimate for $(\hbox{LHS of  \eqref{eq:wave-main-4-rp-error-id}})$, we obtain, for $m \leq M - C'$,
\begin{align*}
(\hbox{LHS of  \eqref{eq:wave-main-4-rp-error-id}})
\aleq A' U^{\frac{p-1}{2}} \max\set{\log^{K_{J}} U, U^{J - (J_{c}-1 + \eta_{c})}, U^{J - (J_{d} - 1 + \eta_{d})}},
\end{align*}
where we used the upper bounds \eqref{eq:wave-main-4-p-upper1} and \eqref{eq:wave-main-4-p-upper2} for $p$. This proves \eqref{eq:wave-main-4-rp-error-id}.

\pfstep{Step~2(d): Error estimate in $\{2R_{\far} < r < 4 R_{\far}\}$} Finally, we prove \eqref{eq:wave-main-4-rp-error-R}. We first need a preliminary estimate for $^{[U]}\rho_J$: we claim that
\begin{equation}\label{eq:claim.Urho}
    {}^{[U]}\rho_J(u,r,\theta) = O_{\bfGmm}^{M-C'}(A \max\{r u^{-\f 12} U^{-\alp+\nB}, U^{-\alp+\de_0} r^{-\de_0+\nB}\}),\quad \hbox{when $R_{\far} \leq r \ls u$}.
\end{equation}

To derive \eqref{eq:claim.Urho}, we decompose
\begin{equation}\label{eq:Urho.decomposition}
    \begin{split}
    ^{[U]}\rho_J = &\: r^{\nB}\Box_{\bfm}^{-1}\Big(\chi_{>\eta_0^{-1}}(\tfrac ru)\Box_{\bfm} (r^{-\nB}{}^{[U]}\rho_J)\Big) + r^{\nB}\Box_{\bfm}^{-1}\Big((1 - \chi_{>\eta_0^{-1}}(\tfrac ru)) \Box_{\bfm}(r^{-\nB}{}^{[U]}\rho_J)\Big) \\
    =: &\: ^{[U]}\rho_{J,\wave} + ^{[U]}\rho_{J,\med}.
    \end{split}
\end{equation}
By definition of $^{[U]}\rho_J$ (see \eqref{eq:hrho-J}), the bounds in Proposition~\ref{prop:recurrence}, \eqref{eq:wave-main-hyp-med} and \eqref{eq:wave-main-hyp-rhoJ}, and \eqref{eq:hF-Gmm}, we have
\begin{equation}
    \Box_{\bfm} (r^{-\nB} {}^{[U]}\rho_{J,\wave}) = O_{\bfGmm}^{M-C'}(A r^{-\alp-1} u^{-1}) ,\quad
     \Box_{\bfm} (r^{-\nB} {}^{[U]}\rho_{J,\med}) =   O_{\bfGmm}^{M-C'}(A u^{-\alp} r^{-2}).
\end{equation}

Note that $\Box_{\bfm} (r^{-\nB} {}^{[U]}\rho_{J,\wave})$ is supported in $\calM_{\wave}$. Also note that $\mathrm{supp}(\chi_{>1}(\tfrac{u+2r}{\frac 12 U})) \cap \calM_{\wave}\subset \{ r \geq \f{1}{4(2+\eta_0)}U \}$. Hence, for $|I| \leq M - C'$,
\begin{equation}
    \begin{split}
        &\: \| \bfGmm^{I} \Box_{\bfm} (r^{-\nB} {}^{[U]}\rho_{J,\wave}) \|_{LE^*(\{1\leq u \leq U\})} \\
        \ls &\: \sum_{k: 2^k \geq \f{1}{4(2+\eta_0)}U} \Big(\int_{1}^{U} \int_{2^k}^{2^{k+1}} (Ar^{-\alp-1} u^{-1})^2 r^{2\nB+1}\, \ud r \, \ud u \Big)^{\f 12}  \\
        \ls &\: \sum_{k: 2^k \geq \f{1}{4(2+\eta_0)}U} \Big( \int_{2^k}^{2^{k+1}} (Ar^{-\alp-1} )^2 r^{2\nB+1}\, \ud r \Big)^{\f 12}
        \ls \sum_{k: 2^k \geq \f{1}{4(2+\eta_0)}U} A 2^{-k(\alp-\nB)} \ls A U^{-\alp+\nB},
    \end{split}
\end{equation}
where we used $\alp > \nB$. By Lemma~\ref{lem:ILED.pointwise}, it follows that
\begin{equation}\label{eq:Urho.bound.1}
    ^{[U]}\rho_{J,\wave}(u,r,\theta) = O_{\bfGmm}^{M-C'}(A r u^{-\f 12} U^{-\alp+\nB}),\quad \hbox{when $R_{\far} \leq r \ls u$}.
\end{equation}

For the contribution from ${}^{[U]}\rho_{J,\med}$, we handle $d \geq 5$ and $d = 3$ separately. For $d \geq 5$, we use \eqref{eq:set.Huygens.Mmed.cutoff} and that $r\leq 100 \eta_0^{-1} u \leq 100 \eta_0^{-1} U$ on $\mathrm{supp}(\Box_{\bfm} (r^{-\nB} {}^{[U]}\rho_{J,\med})) \cap \{u\leq U\} \subset \calM_{\med} \cap \{u\leq U\}$. Hence, for $|I| \leq M - C'$,
\begin{equation}
    \begin{split}
        &\: \| \bfGmm^{I} \Box_{\bfm} (r^{-\nB} {}^{[U]}\rho_{J,\med}) \|_{LE^*(\{1\leq u \leq U\})} \\
        \ls &\: \sum_{k: \frac 12 R_{\far} \leq 2^k \leq 100\eta_0^{-1} U} \Big(\int_{\f 1{1600} \eta_{0} U}^{U} \int_{2^k}^{2^{k+1}} (A u^{-\alp} r^{-2})^2 r^{2\nB+1}\, \ud r \, \ud u \Big)^{\f 12}  \\
        \ls &\: \Big( \sum_{k: 2^k \leq 100\eta_0^{-1} U} 2^{\f k2} \Big)^{\f 12} \Big(\int_{\f 1{1600} \eta_{0} U}^{U}  \int_{\frac 12 R_{\far}}^{200\eta_0^{-1} U} (A u^{-\alp} r^{-2})^2 r^{2\nB+\f 12}\, \ud r \, \ud u \Big)^{\f 12} \\
        \ls &\: U^{\f 14} A U^{-\alp+\f 12} \Big( \int_{\f 12 R_{\far}}^{200\eta_0^{-1} U} r^{2\nB-\f 72} \, \ud r \Big)^{\f 12} \ls A U^{\f 14} U^{-\alp+\f 12} U^{\nB - \f 54} = A U^{-\alp+\nB-\f 12},
    \end{split}
\end{equation}
where in the fourth inequality we used that $d\geq 5 \implies 2\nB-\f 52 >0$. By Lemma~\ref{lem:ILED.pointwise}, it follows that
\begin{equation}\label{eq:Urho.bound.2}
    ^{[U]}\rho_{J,\med}(u,r,\theta) = O_{\bfGmm}^{M-C'}(A r u^{-\f 12} U^{-\alp+\nB-\f 12})\quad \hbox{when $R_{\far} \leq r \ls u$ and $d \geq 5$}.
\end{equation}
We now turn to $^{[U]}\rho_{J,\med}$ in the case $d = 3$. By \eqref{eq:set.Huygens.Mmed.cutoff} it follows that $u\geq \f{1}{1600}\eta_0 U$ on $\mathrm{supp}({}^{[U]}\rho_{J,\med})$. Hence, we can apply Lemma~\ref{lem:3d.pointwise} with $U_0 = \f{1}{1600}\eta_0 U$ and $\de = \de_0$ so that
\begin{equation}\label{eq:Urho.bound.3}
    ^{[U]}\rho_{J,\med}(u,r,\theta) = O_{\bfGmm}^{M-C'}(A U^{-\alp+\de_0} r^{-\de_0+\nB}) \quad \hbox{when $d = 3$},
\end{equation}
where we used that by the finite speed of propagation, $u\geq U$ on the support of $^{[U]}\rho_{J,\med}$. Combining \eqref{eq:Urho.decomposition}, \eqref{eq:Urho.bound.1}, \eqref{eq:Urho.bound.2} and \eqref{eq:Urho.bound.3}, we conclude the proof of the claim \eqref{eq:claim.Urho}.

We now use the bounds \eqref{eq:claim.Urho} for $^{[U]}\rho_{J}$ to control the term \eqref{eq:wave-main-4-rp-error-R}: for $|I| \leq M - C'$,
\begin{equation}
    \begin{split}
       \hbox{(LHS of \eqref{eq:wave-main-4-rp-error-R})}
       \ls &\: A \max\{  U^{-\alp+\nB}, U^{-\alp+\de_0+\f 12}\} \ls A' U^{\f{p-1}2} U^{J-\alp-\de_0+\nB},
    \end{split}
\end{equation}
where in the final inequality, we used $J\geq 1$, $\de_0 \leq \f 12$, $\nB \geq 1$ and \eqref{eq:wave-main-4-p-lower0} so that $\f{p-1}2+J-\de_0 >0$ and $\f{p-1}2 + J + \nB - \f 12 -2\de_0 >0$. (Recall also that the implicit constants are allowed to depend on $R_{\far}$.)

\pfstep{Step~3: Proof of error bounds for low spherical harmonics}
In this step, we establish \eqref{eq:wave-main-4-char-error-EJ} and \eqref{eq:wave-main-4-char-error-eJwave}. We only sketch the proofs, since they closely follow Case~1, Step~3; only the last estimates below differ due to the different hypotheses. From now on, take $\ell \leq J - \nB$.

We begin with the contribution of $E_{J}$. As before, in view of \eqref{eq:exp-error-wave}, we split $E_{J} = E_{J; \Box, k=0} + E_{J; f} + (E_{J} - E_{J; \Box, k=0} - E_{J; f})$. By \eqref{eq:wave-main-Phij0-low}, we have, for $\abs{I} \leq M - C'$,
\begin{align*}
	& R^{\mu} \int_{1}^{U} (R + \tfrac{U-u}{2})^{-\mu} \nrm*{(\bfGmm + c_{\bfGmm})^{I}(\bfK + 2r)^{\ell+\nu_{\Box}-1} \bbS_{(\ell)} E_{J; \Box, k=0} (u, R + \tfrac{U-u}{2}, \tht)}_{L^{\infty}_{\tht}} \, \ud u \\
	& \aleq A' R^{\ell+\nu_{\Box}-1-J-1} \int_{1}^{U} u^{J-1-\alp-\dlt_{0}+\nu_{\Box}} \, \ud u \\
	&\aleq A' R^{\ell+\nu_{\Box}-1} \max\set{R^{-J-1}, R^{-J-1} U^{J-\alp-\dlt_{0}+\nu_{\Box}}}.
\end{align*}
For the contribution of $E_{J; f}$, using \ref{hyp:forcing}, \eqref{eq:EJ-f} and \eqref{eq:EJ-f-rem}, we have, for $\abs{I} \leq M - C'$,
\begin{align*}
	& R^{\mu} \int_{1}^{U} (R + \tfrac{U-u}{2})^{-\mu} \nrm*{(\bfGmm + c_{\bfGmm})^{I}(\bfK + 2r)^{\ell+\nu_{\Box}-1} \bbS_{(\ell)} E_{J; f} (u, R + \tfrac{U-u}{2}, \tht)}_{L^{\infty}_{\tht}} \, \ud u \\
	& \aleq D R^{\ell+\nu_{\Box}-1-J-1} \int_{1}^{U} \log^{K_{d}} \left( \frac{R+\tfrac{U-u}{2}}{u} \right) u^{J-1-\alp_{d}-\dlt_{d}+\nu_{\Box}} \, \ud u \\
	& \peq + D R^{\ell+\nu_{\Box}-2-\alp_{d}+\nu_{\Box}} \int_{1}^{U} u^{-1-\dlt_{d}} \, \ud u \\
	&\aleq D R^{\ell+\nu_{\Box}-1} \max\set{R^{-J-1} \log^{K_{d}}R , R^{-J-1} \log^{K_{d}} (\tfrac{R}{U}) U^{J-(J_{d}-1-\eta_{d})-\dlt_{d}}, R^{- (J_{d} -1 + \eta_{d}) - 1}}.
\end{align*}
Lastly, by Lemma~\ref{lem:exp-error-wave}, we have, for $\abs{I} \leq M - C'$,
\begin{align*}
	& R^{\mu} \int_{1}^{U} (R + \tfrac{U-u}{2})^{-\mu} \nrm*{(\bfGmm + c_{\bfGmm})^{I}(\bfK + 2r)^{\ell+\nu_{\Box}-1} \bbS_{(\ell)} (E_{J} - E_{J; \Box, k=0} - E_{J; f}) (u, R + \tfrac{U-u}{2}, \tht)}_{L^{\infty}_{\tht}} \, \ud u \\
	& \aleq A' R^{\ell+\nu_{\Box}-1-J-1} \int_{1}^{U}  \log^{K_{J}} \left( \frac{R+\tfrac{U-u}{2}}{u} \right) u^{J-1-\alp-\dlt_{0}+\nu_{\Box}} \, \ud u \\
	& \peq + A' R^{\ell+\nu_{\Box}-1-J_{c}-\eta_{c}} \int_{1}^{U} u^{J_{c}-2+\eta_{c}-\alp-\dlt_{0}+\nu_{\Box}} \, \ud u \\
	& \aleq A' R^{\ell+\nu_{\Box}-1} \max\set{R^{-J-1} \log^{K_{J}} R, R^{-J-1}\log^{K_{J}} \left( \tfrac{R}{U} \right) U^{J-\alp-\dlt_{0}+\nu_{\Box}}} \\
	&\peq + A' R^{\ell+\nu_{\Box}-1} \max\set{R^{-(J_{c}-1+\eta_{c})-1}, R^{-(J_{c}-1+\eta_{c})-1}U^{J_{c}-1+\eta_{c}-\alp-\dlt_{0}+\nu_{\Box}}},
\end{align*}
which completes the proof of \eqref{eq:wave-main-4-char-error-EJ}.

Next, we handle the contribution of $e_{J; \wave}$, which we split into $e_{J; \wave, \linear}$ and $e_{J; \wave, \nonlinear}$ as before (see \eqref{eq:eJwave-def}). Using Lemma~\ref{lem:conj-wave-wave} and the hypothesis for $\rho_{J}$, for $\abs{I} \leq M - C'$, we have
\begin{align*}
	& R^{\mu} \int_{1}^{U} (R + \tfrac{U-u}{2})^{-\mu} \nrm*{(\bfGmm + c_{\bfGmm})^{I}(\bfK + 2r)^{\ell+\nu_{\Box}-1} \bbS_{(\ell)} e_{J; \wave, \linear} (u, R + \tfrac{U-u}{2}, \tht)}_{L^{\infty}_{\tht}} \, \ud u \\
	& \aleq R^{\mu} \int_{1}^{U} A (R + \tfrac{U-u}{2})^{-\mu + \ell+\nu_{\Box}-1-2-\alp+\nu_{\Box}} \log^{K_{c}'} \left( \frac{R+\tfrac{U-u}{2}}{u} \right) u^{-\dlt_{c}} \, \ud u \\
	& \aleq A R^{\ell+\nu_{\Box}-3-\alp+\nu_{\Box}} \int_{1}^{U}  \log^{K_{c}'} \left( \frac{R+\tfrac{U-u}{2}}{u} \right) u^{-\dlt_{c}} \, \ud u \\
	& \aleq A R^{\ell+\nu_{\Box}-1} R^{-\alp+\nu_{\Box}-2} \log^{K_{c}'} (\tfrac{R}{U}) U^{1-\dlt_{c}}.
\end{align*}
For the contribution of $e_{J; \wave, \nonlinear}$, we use Definition~\ref{def:alp-N}.(3) with $\alp_{\calN}' = \alp_{\calN}+\dlt_{0}$ and the hypotheses on $\rPhi_{j, k}$ and $\rho_{J}$. We have, for $\abs{I} \leq M - C'$,
\begin{align*}
	& R^{\mu} \int_{1}^{U} (R + \tfrac{U-u}{2})^{-\mu} \nrm*{(\bfGmm + c_{\bfGmm})^{I}(\bfK + 2r)^{\ell+\nu_{\Box}-1} \bbS_{(\ell)} e_{J; \wave, \nonlinear} (u, R + \tfrac{U-u}{2}, \tht)}_{L^{\infty}_{\tht}} \, \ud u \\
	& \aleq R^{\mu} \int_{1}^{U} A' (R + \tfrac{U-u}{2})^{-\mu + \ell+\nu_{\Box}-1-2-\alp+\nu_{\Box}}  u^{-\dlt_{0}} \, \ud u \\
	 & \aleq A' R^{\ell+\nu_{\Box}-1} R^{-\alp+\nu_{\Box}-2} U^{1-\dlt_{0}},
\end{align*}
which completes the proof of \eqref{eq:wave-main-4-char-error-eJwave}. \qedhere
\end{proof}

\subsection{Proof of the initial iteration lemma in the wave zone} \label{subsec:wave-0-pf}
Finally, we prove Proposition~\ref{prop:wave-0}. The proposition is proved by applying Lemma~\ref{lem:Qj-rp} ($r^{p}$ multiplier) to the equation for $\Phi$. This is similar to, but slightly simpler than, the high spherical harmonics case of Proposition~\ref{prop:wave-main}.

\subsubsection{Preliminaries}
As mentioned above, for Proposition~\ref{prop:wave-0}, the main variable we work with is $\Phi$. The main equation for $\Phi$ is given in the following lemma:
\begin{lemma} \label{lem:Q0-Phi}
For $\bfGmm^{I} = (r \rd_{r})^{I_{r \rd_{r}}} \bfS^{I_{\bfS}} \prod_{j, k : j < k} \bfOmg_{jk}^{I_{\bfOmg_{jk}}}$, we have,
\begin{equation} \label{eq:Q0-Phi}
\begin{aligned}
	\left( Q_{0} \bfGmm^{I} - I_{r \rd_{r}} ( r^{-1} \rd_{r}\bfGmm^{I} + r^{-2} \rslap \bfGmm^{I - (1, 0, \ldots)} ) \right) \Phi
	&= O_{\bfGmm}^{\infty}(r^{-1}) \rd_{r} \bfGmm^{\leq \abs{I} - 1} \Phi \\
	&\peq + O_{\bfGmm}^{\infty}(r^{-2}) \rslap \bfGmm^{\leq \abs{I} - 2} \Phi \\
	&\peq + O_{\bfGmm}^{\infty}(r^{-2}) \bfGmm^{\leq \abs{I} - 1} \Phi \\
	&\peq - (\bfGmm+c_{\bfGmm})^{I} \left(\bfh^{\alp \bt} \nb_{\alp} \rd_{\bt} \Phi + \bfC^{\alp} \rd_{\alp} \Phi + W \Phi \right) \\
	&\peq + (\bfGmm+c_{\bfGmm})^{I} \left(r^{\nu_{\Box}} \calN(r^{-\nu_{\Box}} \Phi) + r^{\nu_{\Box}} f\right).
\end{aligned}
\end{equation}
\end{lemma}
\begin{proof}
Recall from Section~\ref{subsec:conj-wave} that $\Phi$ obeys the following equation:
\begin{equation*}
	Q_{0} \Phi = - \left(\bfh^{\alp \bt} \nb_{\alp} \rd_{\bt} \Phi - \bfC^{\alp} \rd_{\alp} \Phi + W \Phi \right) + r^{\nu_{\Box}} \calN(r^{-\nu_{\Box}} \Phi) + r^{\nu_{\Box}} f.
\end{equation*}
Applying $(\bfGmm+c_{\bfGmm})^{I}$ and using the commutator identity \eqref{eq:comm-Qj}, we obtain \eqref{eq:Q0-Phi}. \qedhere
\end{proof}

The following error bounds will be used in the proof of Proposition~\ref{prop:wave-0}.
\begin{lemma} \label{lem:wave-0f-error}
There exists $C' > 0$ and a function $A'$ satisfying \eqref{eq:A'} such that the following holds. Assume that, for $M \in \bbZ_{\geq 0}$ with $M \geq C'$, $\alp \in \bbR$ and $0 \leq \nu_{0} < \frac{1}{2}$,
\begin{align*}
	\phi &= O_{\bfGmm}^{M}(A u^{-\alp})  & & \hbox{ in } \calM_{\med}, \\
	\Phi &= O_{\bfGmm}^{M}(A r^{\nu_{0}} u^{-\nu_{0} -\alp+\nu_{\Box}})  & & \hbox{ in } \calM_{\wave}.
\end{align*}
Then for $\abs{I} \leq M - C'$ and $1 < p < 2$, we have
\begin{align}
 \int_{\frac{U'}{2}}^{U'} \left( \int_{\calC_{u} \cap \calM_{\wave}} r^{p} \abs*{(\bfGmm + c_{\bfGmm})^{I}\left(\bfh^{\alp \bt} \nb_{\alp} \rd_{\bt} \Phi + \bfC^{\alp} \rd_{\alp} \Phi + W \Phi \right) }^{2} \, \ud r \ud \rssgm(\tht) \right)^{\frac{1}{2}}\ud u
 & \aleq A' (U')^{\frac{p-1}{2} - \alp - \dlt_{0} + \nu_{\Box}} , \label{eq:wave-0f-error-lin-wave} \\
\left( \iint_{\calD_{\frac{U'}{2}}^{U'} \cap \calM_{\med}} r^{p+1} \abs*{(\bfGmm + c_{\bfGmm})^{I}\left(\bfh^{\alp \bt} \nb_{\alp} \rd_{\bt} \Phi + \bfC^{\alp} \rd_{\alp} \Phi + W \Phi \right)}^{2} \, \ud u \ud r \ud \rssgm(\tht) \right)^{\frac{1}{2}}
&\aleq A' (U')^{\frac{p-1}{2} - \alp - \dlt_{0} + \nu_{\Box}} , \label{eq:wave-0f-error-lin-med} \\
 \int_{\frac{U'}{2}}^{U'} \left( \int_{\calC_{u} \cap \calM_{\wave}} r^{p} \abs*{(\bfGmm + c_{\bfGmm})^{I} \left( r^{\nu_{\Box}} \calN(r^{-\nu_{\Box}} \Phi) \right) }^{2} \, \ud r \ud \rssgm(\tht) \right)^{\frac{1}{2}}\ud u
 & \aleq A' (U')^{\frac{p-1}{2} - \alp - \dlt_{0} + \nu_{\Box}} , \label{eq:wave-0f-error-nonlin-wave} \\
\left( \iint_{\calD_{\frac{U'}{2}}^{U'} \cap \calM_{\med}} r^{p+1} \abs*{(\bfGmm + c_{\bfGmm})^{I} \left( r^{\nu_{\Box}} \calN(r^{-\nu_{\Box}} \Phi) \right)}^{2} \, \ud u \ud r \ud \rssgm(\tht) \right)^{\frac{1}{2}}
&\aleq A' (U')^{\frac{p-1}{2} - \alp - \dlt_{0} + \nu_{\Box}} . \label{eq:wave-0f-error-nonlin-med}
\end{align}
Moreover, for $\abs{I} \leq M - C'$, $1 < p < \min \set{1 + 2 (\alp_{d} - \nu_{\Box}), 2}$, we have
\begin{align}
 \int_{\frac{U'}{2}}^{U'} \left( \int_{\calC_{u} \cap \calM_{\wave}} r^{p} \abs*{(\bfGmm + c_{\bfGmm})^{I} (r^{\nB} f) }^{2} \, \ud r \ud \rssgm(\tht) \right)^{\frac{1}{2}}\ud u
 & \aleq D (U')^{\frac{p-1}{2} - \alp_{d} - \dlt_{0} + \nu_{\Box}}, \label{eq:wave-0f-error-f-wave} \\
\left( \iint_{\calD_{\frac{U'}{2}}^{U'} \cap \calM_{\med}} r^{p+1} \abs*{(\bfGmm + c_{\bfGmm})^{I} (r^{\nB} f)}^{2} \, \ud u \ud r \ud \rssgm(\tht) \right)^{\frac{1}{2}}
&\aleq D (U')^{\frac{p-1}{2} - \alp_{d} - \dlt_{0} + \nu_{\Box}}.  \label{eq:wave-0f-error-f-med}
\end{align}
\end{lemma}

\begin{proof}
In what follows, we shall omit the prime in $U'$ and simply write $U$.

\pfstep{Step~1: Error estimate in $\calM_{\wave}$}
We begin with \eqref{eq:wave-0f-error-lin-wave}. For $\abs{I} \leq M - 2$, we have
\begin{align*}
& \int_{\frac{U}{2}}^{U} \left( \int_{\calC_{u} \cap \calM_{\wave}} r^{p} \abs*{(\bfGmm + c_{\bfGmm})^{I} \left[ \bfh^{\alp \bt} \nb_{\alp} \rd_{\bt} \Phi + \bfC^{\alp} \rd_{\alp} \Phi + W \Phi \right]}^{2} \, \ud r \ud \rssgm(\tht)\right)^{\frac{1}{2}} \ud u \\
& \aleq \int_{\frac{U}{2}}^{U} \left( \int_{\calC_{u} \cap \calM_{\wave}} r^{p} \left( A r^{-2 + \nu_{0}} \log^{K_{c}'} (\tfrac{r}{u}) u^{-\dlt_{c}} u^{-\nu_{0} -\alp + \nu_{\Box}} \right)^{2} \, \ud r \right)^{\frac{1}{2}} \ud u  \\
& \aleq \int_{\frac{U}{2}}^{U} A u^{\frac{p-1}{2} - 1 - \alp - \dlt_{c} + \nu_{\Box}} \, \ud u
 \aleq A U^{\frac{p-1}{2} - \alp - \dlt_{0} + \nu_{\Box}},
\end{align*}
where we used $1 < p < 2$, $\nu_{0} < \frac{1}{2}$ and $\dlt_{0} \leq \dlt_{c}$.

Next, we prove \eqref{eq:wave-0f-error-nonlin-wave}. We write
\begin{align*}
r^{\nu_{\Box}} \calN(\phi) = \bfh_{\calN}^{\mu \nu}(0; \phi) \nb_{\mu} \rd_{\nu} \Phi + \bfC_{\calN}^{\mu}(0; \phi) \rd_{\mu} \Phi + W_{\calN}(0; \phi) \Phi,
\end{align*}
where $\bfh_{\calN}$, $\bfC_{\calN}$ and $W_{\calN}$ are as in Definition~\ref{def:alp-N}. Using \ref{hyp:sol} and the condition $\alp_{0} \geq \alp_{\calN} + 2 \dlt_{0}$, we may apply \eqref{eq:alp-N-wave} (as well as \eqref{eq:alp-N-wave-3d} if $d =3$). Then the numerology is rather similar to the previous case, and we may argue as follows:
\begin{align*}
& \int_{\frac{U}{2}}^{U} \left( \int_{\calC_{u} \cap \calM_{\wave}} r^{p} \abs*{(\bfGmm + c_{\bfGmm})^{I} \left[ r^{\nu_{\Box}} \calN(r^{-\nu_{\Box}} \Phi) \right]}^{2} \, \ud r \ud \rssgm(\tht) \right)^{\frac{1}{2}} \ud u \\
& \aleq \int_{\frac{U}{2}}^{U} \left( \int_{\calC_{u} \cap \calM_{\wave}} r^{p} \left( A_{\calN}(A) r^{-2+\nu_{0}} u^{-\dlt_{0}} u^{-\nu_{0} -\alp+\nu_{\Box}} \right)^{2} \, \ud r \right)^{\frac{1}{2}} \ud u  \\
& \aleq \int_{\frac{U}{2}}^{U} A_{\calN}(A) u^{\frac{p-1}{2} - 1 - \alp - \dlt_{0} + \nu_{\Box}} \, \ud u
 \aleq A_{\calN}(A) U^{\frac{p-1}{2} - \alp - \dlt_{0} + \nu_{\Box}},
\end{align*}
where we used $1 < p < 2$ and $\nu_{0} < \frac{1}{2}$.

For the contribution of the forcing term $f$ (i.e., \eqref{eq:wave-0f-error-f-wave}), we use \ref{hyp:forcing}. For $\abs{I} \leq M_{0}$,
\begin{align*}
& \int_{\frac{U}{2}}^{U} \left( \int_{\calC_{u} \cap \calM_{\wave}} r^{p} \abs*{(\bfGmm + c_{\bfGmm})^{I} ( r^{\nu_{\Box}} f )}^{2} \, \ud r \ud \rssgm(\tht) \right)^{\frac{1}{2}} \ud u \\
& \aleq \int_{\frac{U}{2}}^{U} \left( \int_{\calC_{u} \cap \calM_{\wave}} r^{p} \left( D r^{-2} \log^{K_{d}} (\tfrac{r}{u}) u^{-\alp_{d}-\dlt_{d}+\nu_{\Box}} \right)^{2} \, \ud r \right)^{\frac{1}{2}} \ud u \\
& \peq + \int_{\frac{U}{2}}^{U} \left( \int_{\calC_{u} \cap \calM_{\wave}} r^{p} \left( D r^{-\alp_{d}-1+\nu_{\Box}} u^{-1-\dlt_{d}} \right)^{2} \, \ud r \right)^{\frac{1}{2}} \ud u \\
& \aleq \int_{\frac{U}{2}}^{U} D u^{\frac{p-1}{2} - 1 - \alp_{d} - \dlt_{0} + \nu_{\Box}} \, \ud u
\aleq D U^{\frac{p-1}{2} - \alp_{d} - \dlt_{0} + \nu_{\Box}},
\end{align*}
where we used $1 < p < \min\set{2, 1 + 2(\alp_{d} - \nu_{\Box})}$ and $\dlt_{0} \leq \dlt_{d}$.

\pfstep{Step~2: Error estimate in $\calM_{\med}$}
We first prove \eqref{eq:wave-0f-error-lin-med}. For $\abs{I} \leq M - 2$, we have
\begin{align*}
& \left( \iint_{\calD_{\frac{U}{2}}^{U} \cap \calM_{\med}} r^{p+1} \abs*{(\bfGmm + c_{\bfGmm})^{I}  \left[ \bfh^{\alp \bt} \nb_{\alp} \rd_{\bt} \Phi + \bfC^{\alp} \rd_{\alp} \Phi + W \Phi \right]}^{2} \, \ud u \ud r \ud \rssgm(\tht) \right)^{\frac{1}{2}} \\
& \aleq \left( \iint_{\calD_{\frac{U}{2}}^{U} \cap \calM_{\med}} r^{p+1} \left( A r^{-2-\dlt_{0}} r^{\nu_{\Box}} u^{-\alp} \right)^{2} \, \ud u \ud r  \right)^{\frac{1}{2}} \\
& \aleq \left( \int_{\frac{U}{2}}^{U} (A u^{\frac{p-1}{2} - \frac{1}{2} - \alp - \dlt_{0} + \nu_{\Box}})^{2} \, \ud u \right)^{\frac{1}{2}}
 \aleq A U^{\frac{p-1}{2} - \alp - \dlt_{0} + \nu_{\Box}},
\end{align*}
where we used $\dlt_{0} \leq \dlt_{c}$ in the first inequality, then $p > 1$ and $\dlt_{0} < \frac{1}{2}$ for the second inequality.

Next, we prove \eqref{eq:wave-0f-error-nonlin-med}. As before, thanks to \ref{hyp:sol} and $\alp_{0} \geq \alp_{\calN} + 2 \dlt_{0}$, \eqref{eq:alp-N-med} with $(\phi; \psi) = (0; \phi)$ is applicable. Then the numerology is similar to the previous case, and we may argue as follows:
\begin{align*}
& \left(\iint_{\calD_{\f{U}2}^{U} \cap \calM_{\med}} r^{p+1} \abs*{(\bfGmm + c_{\bfGmm})^{I} \left[  r^{\nu_{\Box}} \calN(r^{-\nu_{\Box}} \Phi) \right]}^{2} \, \ud u \ud r \ud \rssgm(\tht) \right)^{\frac{1}{2}}\\
& \aleq \left( \iint_{\calD_{\f{U}2}^{U} \cap \calM_{\med}} r^{p+1} \left( A' r^{-2-\dlt_{0}} r^{\nu_{\Box}} u^{-\alp} \right)^{2} \, \ud u \ud r  \right)^{\frac{1}{2}} \\
& \aleq \left( \int_{\f{U}2}^{U}  (A' u^{\frac{p-1}{2} - \frac{1}{2} - \alp - \dlt_{0} + \nu_{\Box}})^{2} \, \ud u \right)^{\frac{1}{2}}
 \aleq A' U^{\frac{p-1}{2} - \alp - \dlt_{0} + \nu_{\Box}},
\end{align*}
where we again used $p > 1$ and $\dlt_{0} < \frac{1}{2}$.

Finally, for the contribution of $f$ (i.e., \eqref{eq:wave-0f-error-f-med}), we use \ref{hyp:forcing}.
\begin{align*}
& \left(\iint_{\calD_{\frac{U}{2}}^{U} \cap \calM_{\med}} r^{p+1} \abs*{(\bfGmm + c_{\bfGmm})^{I} \left[  r^{\nu_{\Box}} f  \right]}^{2} \, \ud u \ud r \ud \rssgm(\tht) \right)^{\frac{1}{2}} \\
& \aleq \left( \iint_{\calD_{\frac{U}{2}}^{U} \cap \calM_{\med}} r^{p+1} \left( D r^{-2-\dlt_{0}+\nu_{\Box}} u^{-\alp_{d}} \right)^{2} \, \ud u \ud r \right)^{\frac{1}{2}} \\
& \aleq \left( \int_{\frac{U}{2}}^{U} (A u^{\frac{p-1}{2} - \frac{1}{2} - \alp_{d} - \dlt_{d} + \nu_{\Box}})^{2} \, \ud u \right)^{\frac{1}{2}}
 \aleq A U^{\frac{p-1}{2} - \alp_{d} - \dlt_{0} + \nu_{\Box}}.
\end{align*}
where we used $\dlt_{0} \leq \dlt_{d}$ in the first inequality, then $p > 1$ and $\dlt_{0} < \frac{1}{2}$ for the second inequality.

Here, for the second inequality to be valid, we require
\begin{equation} \label{eq:wave-0-p-lower2}
	p > 2 - 2(\nu_{\Box} - \dlt_{d}),
\end{equation}
and for the last inequality, we used $p > 1$, $\alp < \nu_{\Box} < \alp_{d}$, $\dlt_{0} \leq \dlt_{d}$ and $\dlt_{0} < \frac{1}{2}$. \qedhere

\end{proof}

\subsubsection{Proof of Proposition~\ref{prop:wave-0}}
The proof consists of several steps.

\pfstep{Step~1: Initial reduction}
First, we reduce Assumption~2 to Assumption~1. Observe that
\begin{equation*}
	\rPhi_{0, 0} = \Phi - \rho_{1}
\end{equation*}
holds \emph{everywhere} in $\calM_{\far}$. Hence, by \eqref{eq:wave-0-hyp-med} and \eqref{eq:wave-0-hyp-rho-1}, it follows that
\begin{equation} \label{eq:Phi-00-reg}
	\rPhi_{0, 0} = O_{\bfGmm}^{M}(A u^{-\alp+\nu_{\Box}}),
\end{equation}
which in particular establishes \eqref{eq:wave-0-Phi-00} with $\nu_{0} = 0$.

\pfstep{Step~2: Proof of the proposition from Lemmas~\ref{lem:Q0-Phi} and ~\ref{lem:wave-0f-error}}
Recall from Lemma~\ref{lem:Q0-Phi} the equation solved by $\bfGmm^{I} \Phi$ (simply set $\psi = 0$ and $\calU = \calM_{\far}$). Our plan is to apply Lemma~\ref{lem:Qj-rp}. To estimate the contribution of the last two terms on the right-hand side, we apply Lemma~\ref{lem:wave-0f-error} with $p = 1 + 2 \eta_{w}$ (so that $\frac{p-1}{2} = \eta_{w}$). By summing up the dyadic bounds in Lemma~\ref{lem:wave-0f-error} from $U' = 2$ to $U' = U$, and using $\alp < \nu_{\Box} + \eta_{w} - \dlt_{0}$, the following bounds hold:
\begin{align}
 \int_{1}^{U} \left( \int_{\calC_{u} \cap \calM_{\wave}} r^{p} \abs*{(\bfGmm + c_{\bfGmm})^{I}\left(\bfh^{\alp \bt} \nb_{\alp} \rd_{\bt} \Phi + \bfC^{\alp} \rd_{\alp} \Phi + W \Phi \right) }^{2} \, \ud r \ud \rssgm(\tht) \right)^{\frac{1}{2}}\ud u
 & \aleq A' U^{\eta_{w} - \alp - \dlt_{0} + \nu_{\Box}} , \label{eq:wave-0-error-lin-wave} \\
\left( \iint_{\calD_{1}^{U} \cap \calM_{\med}} r^{p+1} \abs*{(\bfGmm + c_{\bfGmm})^{I}\left(\bfh^{\alp \bt} \nb_{\alp} \rd_{\bt} \Phi + \bfC^{\alp} \rd_{\alp} \Phi + W \Phi \right)}^{2} \, \ud u \ud r \ud \rssgm(\tht) \right)^{\frac{1}{2}}
&\aleq A' U^{\eta_{w} - \alp - \dlt_{0} + \nu_{\Box}} , \label{eq:wave-0-error-lin-med} \\
 \int_{1}^{U} \left( \int_{\calC_{u} \cap \calM_{\wave}} r^{p} \abs*{(\bfGmm + c_{\bfGmm})^{I} \left( r^{\nu_{\Box}} \calN(r^{-\nu_{\Box}} \Phi) \right) }^{2} \, \ud r \ud \rssgm(\tht) \right)^{\frac{1}{2}}\ud u
 & \aleq A' U^{\eta_{w} - \alp - \dlt_{0} + \nu_{\Box}} , \label{eq:wave-0-error-nonlin-wave} \\
\left( \iint_{\calD_{1}^{U} \cap \calM_{\med}} r^{p+1} \abs*{(\bfGmm + c_{\bfGmm})^{I} \left( r^{\nu_{\Box}} \calN(r^{-\nu_{\Box}} \Phi) \right)}^{2} \, \ud u \ud r \ud \rssgm(\tht) \right)^{\frac{1}{2}}
&\aleq A' U^{\eta_{w} - \alp - \dlt_{0} + \nu_{\Box}} , \label{eq:wave-0-error-nonlin-med} \\
 \int_{1}^{U} \left( \int_{\calC_{u} \cap \calM_{\wave}} r^{p} \abs*{(\bfGmm + c_{\bfGmm})^{I} (r^{\nB} f) }^{2} \, \ud r \ud \rssgm(\tht) \right)^{\frac{1}{2}}\ud u
 & \aleq A' U^{\eta_{w} - \alp - \dlt_{0} + \nu_{\Box}}, \label{eq:wave-0-error-f-wave} \\
\left( \iint_{\calD_{1}^{U} \cap \calM_{\med}} r^{p+1} \abs*{(\bfGmm + c_{\bfGmm})^{I} (r^{\nB} f)}^{2} \, \ud u \ud r \ud \rssgm(\tht) \right)^{\frac{1}{2}}
&\aleq A' U^{\eta_{w} - \alp - \dlt_{0} + \nu_{\Box}}.  \label{eq:wave-0-error-f-med}
\end{align}

Now we have all the tools to prove \eqref{eq:wave-0-rho-1}. Proceeding as in the proof of Proposition~\ref{prop:wave-main}, Case~1, Step~1(a) in Section~\ref{subsec:wave-main-pf} using Lemma~\ref{lem:Qj-rp}, \eqref{eq:wave-0-error-lin-wave}--\eqref{eq:wave-0-error-f-med}, the obvious bounds (for $\abs{I} \leq M - C'$)
\begin{align}
\left( \int_{\calC_{1}} \chi_{>2 R_{\far}} r^{p} (\rd_{r} \bfGmm^{I} \Phi)^{2}  \, \ud r \ud \rssgm(\tht) \right)^{\frac{1}{2}} \label{eq:wave-0-id}
& \aleq D, \\
R_{\far}^{\frac{p-1}{2}} \left( \iint_{\calD_{1}^{U} \cap \set{R_{\far} < r < 2 R_{\far}}} \abs{(\rd_{r}, r^{-1} \rsnb, r^{-1}) \bfGmm^{I} \Phi}^{2} \, \ud u \ud r \ud \rssgm(\tht) \right)^{\frac{1}{2}}
&\aleq A R_{\far}^{\frac{p-2}{2}}. \label{eq:wave-0-bdry}
\end{align}
as well as an induction on $\abs{I}$, we obtain the following bound: for $\abs{I} \leq M - C'$ and $p$ chosen as above, we have
\begin{equation} \label{eq:wave-0-rp}
\begin{aligned}
& \sup_{u \in [1, U]} \left(\int_{\calC_{u}} \chi_{>2 R_{\far}} r^{p} (\rd_{r} \bfGmm^{I} \Phi)^{2}  \, \ud r \ud \rssgm(\tht) \right)^{\frac{1}{2}}
+ \left( \iint_{\calD_{1}^{U}} \chi_{>2 R_{\far}} r^{p-1} (\rd_{r} \bfGmm^{I} \Phi)^{2} \, \ud u \ud r \ud \rssgm(\tht) \right)^{\frac{1}{2}} \\
&+ \left( \iint_{\calD_{U_{0}}^{U}} \chi_{>2 R_{\far}} r^{p-3} \left(\abs{\rsnb \bfGmm^{I} \Phi}^{2}  + (\bfGmm^{I} \Phi)^{2} \right) \, \ud u \ud r \ud \rssgm(\tht) \right)^{\frac{1}{2}} \\
& \aleq A' U^{\eta_{w}-\alp-\dlt_{0}+\nu_{\Box}}.
\end{aligned}
\end{equation}

Now, since $p > 1$, by the fundamental theorem of calculus, Cauchy--Schwarz and \eqref{eq:wave-0-rp} with $\abs{I} = 0$ (with a change of notation $(U, R, \Tht) \mapsto (u, r, \tht)$), it follows that the following limit exists:
\begin{equation*}
		\rPhi_{0, 0}(u, \tht) = \lim_{r \to \infty} \Phi(u, r, \tht)
\end{equation*}
where we remind the reader that the Friedlander radiation field $\rPhi_{0, 0}(u, \tht)$ is, by definition, the above limit. By \eqref{eq:wave-0-rp} with $\abs{I} > 0$, it also follows that, for $\abs{I} \leq M - C'$,
\begin{equation*}
		\bfGmm^{I} \rPhi_{0, 0}(u, \tht) = \lim_{r \to \infty} \bfGmm^{I} \Phi(u, r, \tht).
\end{equation*}
By Lemma~\ref{lem:hardy-radial}, it follows that $\rho_{1} = \Phi - \Phi_{0, 0}$ obeys \eqref{eq:wave-0-rho-1} with $\eta_{w} = \frac{p-1}{2}$. \hfill \qedsymbol

\section{Iteration and proof of the main upper bound theorem} \label{sec:upper}
In this section, we put together the iteration lemmas in the different regions of spacetime proven in the last three sections to establish Main Theorems~\ref{thm:upper} (see \textbf{Section~\ref{sec:pf.of.thm.1}}) and Main Theorem~\ref{thm:upper-sphsymm} (see \textbf{Section~\ref{sec:pf.of.thm.3}}). In the process, we also establish Lemma~\ref{lem:radiation.field} (see \textbf{Section~\ref{sec:lem.radiation.field}}).

\subsection{Proof of Main Theorem~\ref{thm:upper}}\label{sec:pf.of.thm.1}

\begin{proof}[Proof of Main Theorem~\ref{thm:upper}]

\pfstep{Step~1: Basic setup for an induction argument}
First, we can take $\alp_0$, $\de_0$ smaller if necessary so that $\alp_0\geq \alp_{\calN}+2\de_0$ continue to hold, but that $\alp_0 +\de_0 < \nB + \eta_w$ and $\alp_0 \notin \mathbb Z$ for some $0< \eta_w < \min \{\f 12, \alp_d-\nB\}$. (Note that this is possible since $\alp_{\calN} \leq \nB$ by part (2) of Proposition~\ref{prop:alp-N}; see discussions after Proposition~\ref{prop:wave-0}.)

We set up an induction as follows. For every $0< \de_0< \min\{\f 12, \eta_c, \eta_d, \de_c,\de_d \}$, we define $\{ \alp_i \}_{i\geq 1}$ recursively by
then let
$$\alp_{i+1} = \alp_i + \de_0,\quad \hbox{for } i\geq 0.$$
For a choice of $\de_0$ as above, define also
$$I_{\nB} = \min \{i : \alp_i > \nB\}, \quad I_{\mathrm{total}} = \min\{ i : \alp_i > \alp_{\mathfrak f} \}.$$
Note that since $\alp_{\mathfrak f} > \nB$, we have $I_{\nB} \leq I_{\mathrm{total}}$.

By changing $\de_0$ slightly if necessary, we can assume $\alp_1,\ldots, \alp_{I_{\mathrm{total}}} \notin \mathbb Z \cup \{J_{c}-1+\eta_c\} \cup \{ J_d-1+\eta_d\}$. From now on fix such a $\de_0$.

We will iteratively improve our decay estimates starting from the assumed decay in Main Theorem~\ref{thm:upper}. In each step, the decay rate improves from $\alp_i$ to $\alp_{i+1}$. We also introduce $\{A_i\}_{i=0}^{I_{\mathrm{total}}}$ and $\{M'_i\}_{i=0}^{I_{\mathrm{total}}}$ to measure the amplitude and count the number of derivatives respectively. More precisely, we define $A_0$ as in assumption \ref{hyp:sol} and
\begin{equation}
A_i = A'(A_{i-1}),\quad i = 1,\ldots, I_{\mathrm{total}},
\end{equation}
where $A'$ can be chosen to be a function satisfying \eqref{eq:A'} so that Propositions~\ref{prop:near-med}, \ref{prop:med}, \ref{prop:wave-main} and \ref{prop:wave-0} all apply. Also, let $M'_0 = M_0 - C_{\wave, 0}$ (where $M_0$ is as in the assumption of Main Theorem~\ref{thm:upper} and $C_{\wave, 0}$ is as in Proposition~\ref{prop:wave-0}) and define
\begin{equation}
M'_{i} = \lfloor b M'_{i-1} \rfloor,\quad i=1,\ldots, I_{\mathrm{total}},
\end{equation}
where $b>0$ is a sufficiently small constant chosen so that if $M_0$ is sufficiently large, then in the application of Propositions~\ref{prop:near-med}, \ref{prop:med}, \ref{prop:wave-main} and \ref{prop:wave-0}, we can handle the necessary loss of derivatives in these propositions and so that $M'_{I_{\mathrm{total}}} \geq 0$.

\pfstep{Step~2: Initial iteration} By induction on $i$, we prove the following statement for $i = 0,1,\ldots, I_{\nB} -1$:
\begin{itemize}
\item $\Phi = r\phi$ admits the following expansion in $\calM_\wave$
\begin{equation}\label{eq:main.induct.1.wave.expansion}
	\Phi = \rPhi_{0, 0} + \rho_{1},
\end{equation}
with $\rho_{1}(u, r, \tht) \to 0$ as $r \to \infty$ for each $u \geq 1$, $\tht \in \bbS^{d-1}$, and satisfies the following estimates:
\begin{equation} \label{eq:main.induct.1.wave}
	\rPhi_{0, 0} = O_{\bfGmm}^{M'_{i}}(A_{i} u^{-\alp_{i}+\nu_{\Box}}),\quad\rho_{1} = O_{\bfGmm}^{M'_{i}}(A_{i} u^{-\alp_{i+1}+\nu_{\Box}}).
\end{equation}
\item In $\calM_\near\cup \calM_\med$, $\phi$ satisfies the following estimates:
\begin{equation}\label{eq:main.induct.1.near-med}
\phi = O_{\bfGmm}^{M'_i}(A_i \tau^{-\alp_i}) \quad \hbox{in $\calM_\near\cup \calM_\med$}.
\end{equation}
\end{itemize}

To obtain the base case ($i=0$), first notice that \eqref{eq:main.induct.1.near-med} already holds for $i=0$ thanks to \eqref{eq:sol-near}--\eqref{eq:sol-med} in assumption \ref{hyp:sol} of the theorem. We then apply Proposition~\ref{prop:wave-0} with Assumption~1 (and with $\alp = \alp_0$, which holds because of assumption \ref{hyp:sol}) so that we obtain \eqref{eq:main.induct.1.wave} with $i=0$.

We now carry out the induction. Assume that \eqref{eq:main.induct.1.wave.expansion}, \eqref{eq:main.induct.1.wave} and \eqref{eq:main.induct.1.near-med} all hold when $i=I$ for some $I = 0,1,\ldots, I_{\nB}-2$.

Using the induction hypotheses, we apply Case~1 of Proposition~\ref{prop:med} and Proposition~\ref{prop:near-med} (in that order) with $\alp = \alp_I$ to obtain that \eqref{eq:main.induct.1.near-med} holds for $i=I+1$. We then apply Proposition~\ref{prop:wave-0} with Assumption~2 and $\alp = \alp_{I+1}$; \eqref{eq:wave-0-hyp-med} holds because it has just been established and Assumption~2 holds by the induction hypothesis. Hence, we obtain that \eqref{eq:main.induct.1.wave.expansion}--\eqref{eq:main.induct.1.wave} hold for $i = I+1$ as a consequence of \eqref{eq:wave-0-hyp-med}.  This completes the initial induction.

\pfstep{Step~3: Handling the exceptional $I_{\mathrm{total}} = I_{\nB}$ case} We treat the cases $I_{\mathrm{total}} = I_{\nB}$ separately. If $I_{\mathrm{total}} = I_{\nB}$, then $\alp_{\mathfrak f} = \alp_{d} \in (\nB, \nB+1)$. In this case, we apply Case~4 of Proposition~\ref{prop:wave-main}, Case~3 of Proposition~\ref{prop:med}, and then Proposition~\ref{prop:near-med} to obtain
\begin{equation}\label{eq:upper.main.together.1}
\phi = O_{\bfGmm}^{M'_{I_{\mathrm{total}}}}(A_{I_{\mathrm{total}}} \min\{ \tau^{-\alp_d}, \tau^{-\alp_d+\nB} r^{-\nB}\}).
\end{equation}
In this case, we then end the argument here.

\pfstep{Step~4: The main induction} Assume now that $I_{\mathrm{total}} > I_{\nB}$. For $i = I_{\nB}, \ldots, I_{\mathrm{total}} - 1$, we define $1 \leq J_i \leq \min \{ J_{c}, J_d\}$ to be the unique integer such that
$$J_i - 1 + \nB < \alp_i < J_i + \nB.$$
(This is well-defined since $\alp_i \neq \mathbb Z$.)

If $I_{\nB} = I_{\mathrm{total}}$, we skip this step and go directly to Step~4. Otherwise, we will induct on the following statements for $i = I_{\nB}, I_{\nB}+ 1,\ldots, I_{\mathrm{total}}-1$:
\begin{itemize}
\item For $J_i$ as above, $\Phi = r^\nB \phi$ admits the expansion
\begin{equation}\label{eq:wave-main-hyp-Phi-expansion.Thm1}
\Phi = \sum_{j=0}^{J_{i}-1} r^{-j} \rPhi_{j,0} + \sum_{j=1}^{J_{i}-1} \sum_{k \in K_{j} \setminus \set{0}} r^{-j} \log^{k} \left(\tfrac{r}{u}\right) \rPhi_{j,k} + \rho_{J_{i}} \quad \hbox{ in } \calM_{\wave},
\end{equation}
where $\rPhi_{j, k}$ with $1 \leq j \leq J_{i}-1$, $0 \leq k \leq K_{j}$ are determined by the recurrence equation \eqref{eq:recurrence-jk} with
\begin{align}
	\rPhi_{j, k} &= O_{\bfGmm}^{M'_{i}} (A_{i} u^{j- \alp_{i} + \nu_{\Box}})& &\hbox{ for } 0 \leq j \leq J_{i}-1, \, 1 \leq k \leq K_{j}, \label{eq:wave-main-hyp-Phi-jk.Thm1} \\
	\rPhi_{(\leq j - \nu_{\Box}) j, 0} &= O_{\bfGmm}^{M'_{i}} (A_{i} u^{j- \alp_{i} + \nu_{\Box}})& &\hbox{ for } 0 \leq j \leq J_{i} - 1, \label{eq:wave-main-hyp-Phi-j0-low.Thm1} \\
	\rPhi_{(\geq j - \nu_{\Box}+1) j, 0} &= O_{\bfGmm}^{M'_{i}} (A_{i} u^{j- \alp_{i} +\dlt_{0} + \nu_{\Box}})& &\hbox{ for } 0 \leq j \leq J_{i} - 1, \label{eq:wave-main-hyp-Phi-j0-high.Thm1}
\end{align}
and $\rPhi_{j, k} = 0$ for $1 \leq j \leq J_{i}-1$, $k > K_{j}$. Moreover, the remainder obeys
\begin{equation} \label{eq:wave-main-hyp-rhoJ.Thm1}
\rho_{J_{i}} = O_{\bfGmm}^{M'_{i}}(A_{i} r^{-\alp_{i}+\nu_{\Box}}) \quad \hbox{ in } \calM_{\wave}.
\end{equation}
\item In $\calM_\near\cup \calM_\med$, $\phi$ satisfies the following estimates:
\begin{equation}\label{eq:Thm1-main-assumption-near-med}
\phi = O_{\bfGmm}^{M'_i}(A_i \tau^{-\alp_i}) \quad \hbox{in $\calM_\near\cup \calM_\med$}.
\end{equation}
\end{itemize}

To start the induction, we check the base case when $i=I_{\nB}$. By definition of $I_{\nB}$, we have $J_{I_{\nB}}=1$. Therefore, in $\calM_\wave$ we only need to check \eqref{eq:wave-main-hyp-Phi-j0-high.Thm1} for $(i,j) = (I_{\nB},0)$ and \eqref{eq:wave-main-hyp-rhoJ.Thm1} for $(i,J_i) = (I_{\nB},1)$. These follow from the $i = I_{\nB}-1$ case of \eqref{eq:main.induct.1.wave} in the previous step. Then for $\calM_{\near}\cup \calM_{\med}$, we use Case~1 of Proposition~\ref{prop:med} and Proposition~\ref{prop:near-med} to obtain the $i = I_{\nB}$ case of \eqref{eq:Thm1-main-assumption-near-med}.

If $I_{\mathrm{total}} = I_{\nB}+1$, then we are done. Otherwise, we now carry out the induction: assume that \eqref{eq:wave-main-hyp-Phi-expansion.Thm1}--\eqref{eq:Thm1-main-assumption-near-med} hold for some $I = I_{\nB}, I_{\nB}+ 1,\ldots, I_{\mathrm{total}}-2$.

Using \eqref{eq:wave-main-hyp-Phi-expansion.Thm1}--\eqref{eq:wave-main-hyp-rhoJ.Thm1} and \eqref{eq:Thm1-main-assumption-near-med} for $i = I$, we can apply Case~1, 2(a) or 3 (depending on whether $J_I \leq \min\{J_{c}-1, J_d-1\}$ and whether $\alp_{I+1} > J_I+\nB$) of Proposition~\ref{prop:wave-main} to show that \eqref{eq:wave-main-hyp-Phi-expansion.Thm1}--\eqref{eq:wave-main-hyp-rhoJ.Thm1} hold for $i= I+1$. We then apply Case~2 of Proposition~\ref{prop:med} and then Proposition~\ref{prop:near-med} to prove \eqref{eq:Thm1-main-assumption-near-med} for $i = I+1$. This completes the main induction.

\pfstep{Step~5: End of the argument} There are two ways that the argument ends: either one reaches the decay rate determined by the expansion of $\phi$ (Case~1), or one reaches the decay rate which is determined by the coefficient/data/inhomogeneity (Case~2).

\textbf{Case~1: $J_{\mathfrak f} \leq \min \{J_{c}-1,J_d-1\}$.} We apply Case~2(b) of Proposition~\ref{prop:wave-main} with $\alp = \alp_{I_{\mathrm{total}}-1}$. The assumptions hold because of what we established in Step~4 above. We then apply Case~4 of Proposition~\ref{prop:med} and then Proposition~\ref{prop:near-med} to obtain
\begin{equation}\label{eq:upper.main.together.2}
\phi = O_{\bfGmm}^{M'_{I_{\mathrm{total}}}}(A_{I_{\mathrm{total}}} \min\{ \tau^{-J_{\mathfrak f}-\nB}, \tau^{-J_{\mathfrak f}} r^{-\nB}\} \log^{K_{J_{\mathfrak f}}}\tau).
\end{equation}

\textbf{Case~2: $J_{\mathfrak f} = \min \{ J_{c}, J_d \}$.} 
Case~4 of Proposition~\ref{prop:wave-main} implies that $\Phi$ admits the expansion
\begin{equation*}
\Phi = \sum_{j=0}^{J-1} r^{-j} \rPhi_{j,0} + \sum_{j=1}^{J-1} \sum_{k \in K_{j} \setminus \set{0}} r^{-j} \log^{k} \left(\tfrac{r}{u}\right) \rPhi_{j,k} + \rho_{J} \quad \hbox{ in } \calM_{\wave},
\end{equation*}
where \eqref{eq:wave-main-Phijk}--\eqref{eq:wave-main-Phij0-high} hold for the higher radiation fields and $\rho_{J}$ satisfies the estimate
$$\rho_{J} = O_{\bfGmm}^{M_{I_{\mathrm{total}}}'}(A_{I_{\mathrm{total}}} r^{-\alp_{\mathfrak f}+\de_{\mathfrak f}'+\nB} u^{-\de_{\mathfrak f}'} \log^{K_{\mathfrak f}}u)\quad \hbox{in $\calM_{\wave}$},$$
where, after recalling \eqref{eq:alp.def}--\eqref{eq:Kf.def}, we have used $\max\set{u^{\alp_{\mathfrak f} - J - \nu_{\Box} } \log^{K_{J}} u, 1} = \log^{K_{\mathfrak f}}u$. In particular, since $\alp_{\mathfrak f} - \de_{\mathfrak f}' > \nB$ (since $J\geq 1$ and $\de_{\mathfrak f}' < \min \{\f 12, \eta_{c},\eta_{d}\}$), the above estimates imply
\begin{equation}\label{eq:upper.final.strange.case.1}
\phi = O_{\bfGmm}^{M_{I_{\mathrm{total}}}'}(A_{I_{\mathrm{total}}} r^{-\nB} u^{-\alp_{\mathfrak f}+\nB} \log^{K_{\mathfrak f}}u),\quad \hbox{in $\calM_{\wave}$},
\end{equation}
where we used $u\ls r$ in $\calM_{\wave}$.

On the other hand, with the estimates in $\calM_{\wave}$ as input, and using Case~3 of Proposition~\ref{prop:med}, we obtain that
\begin{align}\label{eq:upper.final.strange.case.2}
\phi = &\: O_{\bfGmm}^{M_{I_{\mathrm{total}}}'}(A_{I_{\mathrm{total}}} u^{-\alp_{\mathfrak f}+\nB}r^{-\nB} \log^{K_{\mathfrak f}} u) \hbox{ in } \calM_{\med}.
\end{align}

Finally, applying Proposition~\ref{prop:near-med} with $\alp = \alp_{I_{\mathrm{total}}-1}$, $\alp+\de_0 = \alp_{\mathfrak f}$ (in particular, we have taken a slightly smaller $\de_0$ here, but only for this step) and $K = K_{\mathfrak f}$, we also obtain
\begin{equation}\label{eq:upper.final.strange.case.3}
\phi = O_{\bfGmm}^{M_{I_{\mathrm{total}}}'}(A_{I_{\mathrm{total}}} \tau^{-\alp_{\mathfrak f}} \log^{K_{\mathfrak f}} \tau), \quad \hbox{in $\calM_{\med}\cup \calM_{\near}$}.
\end{equation}
 Combining \eqref{eq:upper.final.strange.case.1} and \eqref{eq:upper.final.strange.case.3} yields the desired estimate in this case.

\pfstep{Step~6: Putting everything together} Now fix $A_{\mathfrak f} = A_{I_{\mathrm{total}}}$. Notice that for $m_0$ sufficiently large and $a_0$ sufficiently small, if $M_0 \geq m_0$, then we have $M'_{I_{\mathrm{total}}} \geq \lfloor a_0 M_0\rfloor \geq 0$.  We can then fix $m_0$ and $a_0$. Recalling also the definition of $\alp_{\mathfrak f}$, the desired upper bound then follows from the estimates \eqref{eq:upper.main.together.1}, \eqref{eq:upper.main.together.2}, \eqref{eq:upper.final.strange.case.1} and \eqref{eq:upper.final.strange.case.3}. \qedhere

\end{proof}

\subsection{Proof of Lemma~\ref{lem:radiation.field}}\label{sec:lem.radiation.field}

We have essentially proven Lemma~\ref{lem:radiation.field} in the course of the proof of Main Theorem~\ref{thm:upper} in the previous subsection. We collect the relevant arguments below. 
    \begin{proof}[Proof of Lemma~\ref{lem:radiation.field}]
    We first prove part (1) of the lemma, i.e., the decay of the radiation field. By Main Theorem~\ref{thm:upper}, we in particular have $r^{\nB}\phi = O(A_{\mathfrak f}' u^{-\alp_{\mathfrak f}+\nB} \log^{K_{\mathfrak f}} u)$ in $\calM_{\wave}$. Taking the $r\to \infty$ limit gives $\Phi_{0,0} = O(A_{\mathfrak f}' u^{-\alp_{\mathfrak f}+\nB} \log^{K_{\mathfrak f}} u)$. Since $\alp_{\mathfrak f} > \nB$, this implies $\lim_{u\to \infty} \Phi_{0,0}(u)=0$.

    Part (2) of the lemma is a restatement of Lemma~\ref{lem:Kj-def}.

    Finally, part (3) of the lemma is a consequence of Case~2 in Proposition~\ref{prop:wave-main}.
    \end{proof}

\subsection{Proof of Main Theorem~\ref{thm:upper-sphsymm}}\label{sec:pf.of.thm.3}

We now turn to the setting of linear equations with spherically symmetric coefficients, where we have an improved upper bound when $r \ls \tau$. The key to the proof is Proposition~\ref{prop:near-med-sphsymm}.
\begin{proof}[Proof of Main Theorem~\ref{thm:upper-sphsymm}]
First note that the upper bound of $\tau^{-\alp_{\mathfrak f}+\nB} r^{-\nB} \log^{K_{J_{\mathfrak f}}} \tau$ is already given in Main Theorem~\ref{thm:upper}. It thus suffices to prove an improved upper bound when $r \ls \tau$.

We set up an induction as follows. Fix $\de_0 \in (0, \min\{\de_c,\de_d,\f 12\}]$ such that $\f{\ell}{\de_0} \in \mathbb N$. We prove inductively in $i$ that
\begin{equation}
\phi = O_{\bfGmm}^{M_i}(D \tau^{-\alp-i\de_0} \brk{r}^{i\de_0}) \hbox{ in } \calM_{\near} \cup \calM_{\med},\quad i = 0,1,\ldots, \f{\ell}{\de_0}.
\end{equation}
The base case ($i=0$) is an immediate consequence of Main Theorem~\ref{thm:upper} and the inductive step follows from an application of Proposition~\ref{prop:near-med-sphsymm}. \qedhere
\end{proof}

\section{Proof of the main sharp asymptotics theorem} \label{sec:lower}

Our goal in this section is to prove Main Theorem~\ref{thm:lower} and Main Theorem~\ref{thm:lower-sphsymm}.

The section will be organized as follows. We begin with some improved estimates for the Minkowskian wave equation in \textbf{Section~\ref{sec:improved.Mink}}. After introducing some conventions for the proof in \textbf{Section~\ref{sec:notation.in.lower}}, we then turn to the proof of Main Theorem~\ref{thm:lower}. First, we use Proposition~\ref{prop:wave-main} as input and prove the sharp asymptotics in $\calM_\med$ and $\calM_\wave$ in \textbf{Section~\ref{sec:med.wave.sharp}}. After that, we propagate the estimates inwards to obtain the sharp asymptotics in $\calM_{\near}$ in \textbf{Section~\ref{sec:near.sharp}}. Finally, we put together the estimates in \textbf{Section~\ref{sec:lower.everything}} to complete the proof of Main Theorem~\ref{thm:lower} and then prove Main Theorem~\ref{thm:lower-sphsymm} in \textbf{Section~\ref{sec:lower.SS}}.

\subsection{Improved error estimates for the Minkowskian wave equation}\label{sec:improved.Mink}

In this subsection, we derive some improved estimates for the Minkowskian wave equation. These estimates will then be used in Section~\ref{sec:med.wave.sharp} to control the error terms arising in the argument for the sharp asymptotics.

We first obtain an improved version of Proposition~\ref{prop:rho.est}. This makes use of our main upper bound theorem (Main Theorem~\ref{thm:upper}) in the special case of the linear wave equation on Minkowski spacetime.
\begin{proposition}\label{prop:rho.est.improved}
Let $\rho$ be a function in $\calM_\wave$ satisfying
\begin{equation}\label{eq:med.general.rho.improved}
\rho = O_{\bfGmm}^{M}(A_{\rho} r^{-\alp_{\rho} + \nu_{\Box}} u^{\bt_{\rho}} \log^{K_{\rho}} u)
\quad \hbox{ in } \calM_{\wave}
\end{equation}
for some $A_{\rho} > 0$, $\alp_{\rho} > \f{d-1}2$, $\bt_{\rho} \in \bbR$ and $K_{\rho} \in \bbZ_{\geq 0}$.

Suppose $(U,R,\Theta) \in \calM_\far$, $R_1 = 2R_{\far}$ and $1\leq U_0 \leq \eta_{1} U$, where $\eta_{1} \leq \min\{ \f{\eta_{0}}4, \f 12\}$ is the small constant chosen in the beginning of Section~\ref{sec:med.error}. Then, for every $\de'''>0$, there exist
$$m_{\bfm} = m_{\bfm}(d, \de''', \alp_\rho, \bt_\rho, K_\rho, \eta_{0}, \eta_{1}, R_{\far}) \in \mathbb N,\quad a_{\bfm}=a_{\bfm}(d, \de''', \alp_\rho, \bt_\rho, K_\rho, \eta_{0}, \eta_{1}, R_{\far}) \in (0,1)$$ such that the following holds if $M \geq m_{\bfm}$:
$$\Box_\bfm^{-1} \Big(\chi_{>R_{1}}(r) \chi_{>U_0}'(u) r^{-\nB} \rd_r \rho \Big)(U,R,\Theta) = O_{\bfGmm}^{\lfloor a_{\bfm} M \rfloor}\Big( A_\rho U^{-\alp_\rho} U_0^{\bt_\rho+\de'''} \log^{K_\rho}U_0 \min\{1, U^{\nB}R^{-\nB}\} \Big),$$
where the implicit constant is allowed to depend on $d, \alp_\rho, \de''', \bt_\rho, K_\rho, \eta_{0}, \eta_{1}, R_{\far}$.
\end{proposition}
\begin{proof}
Fix $\de'''>0$ and fix $(U,R,\Theta) \in \calM_{\far}$. We apply our main upper bound theorem (Main Theorem~\ref{thm:upper}) for the equation $\Box_{\bfm} \varphi = f$ with $0$ initial data and $f = \chi_{>R_{1}}(r) \chi_{>U_0}'(u) r^{-\nB} \rd_r \rho$. Since we are only in the output at $(U,R,\Theta)$, by the strong Huygens principle (Lemma~\ref{lem:huygens}), we could equivalently consider $\varphi = \Box_{\bfm}^{-1} (\chi_{>1}(\tfrac{u+2r}{\f 12U})f)$. Since $U_0 \leq \eta_{1} U$, for $U$ sufficiently large, $\mathrm{supp}(\chi_{>1}(\tfrac{u+2r}{\f 12U}) f) \subset \calM_{\wave}$. Moreover, $f$ satisfies the assumptions \eqref{eq:wave-f-exp}, \eqref{eq:wave-f-rem} with $D=A_\rho U_0^{\bt_\rho+\de'''} \log^{K_\rho} U_0$, $M_0 = M-1$, $\alp_d = \alp_\rho$ and $\de_d = \de'''$. (Notice that since this is just the linear wave equation on Minkowski, the assumptions \eqref{eq:sol-near}--\eqref{eq:sol-wave} are satisfied here using standard methods.)

Therefore, the desired estimate follows from Main Theorem~\ref{thm:upper} (which we have already proven). \qedhere
\end{proof}

Our next estimate does not follow directly from the statement of Main Theorem~\ref{thm:upper}, but can be obtained using similar tools as in Section~\ref{sec:med} and Section~\ref{sec:wave}. 

\begin{proposition}\label{prop:Mink.est.for.lower}
Suppose $F:\mathbb R^{d+1} \to \mathbb R$ is supported in $u\geq \f{U_0}2$ and $r\geq R_{\far}$ and satisfies the following estimate:
\begin{equation}\label{eq:Mink.est.for.lower.assumptions}
F(u,r,\theta) = \begin{cases}
 O_{\bfGmm}^{M}(A r^{-2-\de} u^{-\alp}) & \hbox{in $\calM_{\med}$} \\
  O_{\bfGmm}^{M}(A r^{-\nB-2} u^{-\alp-\de+\nB} ) & \hbox{in $\calM_{\wave}$}
  \end{cases}
\end{equation}
for some $K \in \mathbb Z_{\geq 0}$, $\de \in (0,\f 12]$ and $\alp\in \mathbb R$ satisfying $\alp \geq \nB+1$.

Then there exists $C\in \mathbb N$ such that if $M\geq C$, the following estimate holds in $\calM_{\far}$:
\begin{equation}\label{eq:Mink.est.for.lower.final}
\Box_{\bfm}^{-1} F(u,r,\th) = O_{\bfGmm}^{M-C}(A U_0^{-\alp}u^{\nB-\de} r^{-\nB})\quad \hbox{in $\calM_{\far}$}.
\end{equation}
\end{proposition}
\begin{proof}
As before, the number of derivatives $C$ is allowed to increase from line to line.

\pfstep{Step~1: Estimates in $\calM_{\med}$} The estimates needed in $\calM_{\med}$ are the same as those already established in the proof of Proposition~\ref{prop:med.error} so we will be brief. When $d =3$, the estimate \eqref{eq:Mink.est.for.lower.assumptions} together with Lemma~\ref{lem:3d.pointwise} immediately imply $\Box_{\bfm}^{-1} F(u,r,\th) = O_{\bfGmm}^{M-C}(A U_0^{-\alp} r^{-\de})= O_{\bfGmm}^{M-C}(A U_0^{-\alp}u^{-\de+\nB-\f 12} r^{-\nB+\f 12})$ since $r \ls u$.  When $d \geq 5$, arguing as in \eqref{eq:med.LE.error.romg.high.d.2} in Proposition~\ref{prop:med.error}, we see that \eqref{eq:Mink.est.for.lower.assumptions} implies $\| \bfGmm^{\leq M} F\|_{LE^*(D_{U,R})} \ls A U_0^{-\alp} U^{\f{d-2}2-\de}$. Using Lemma~\ref{lem:ILED.pointwise}, this in turn implies $(\Box_{\bfm}^{-1} F)(u,r,\th) = O_{\bfGmm}^{M-C}(A U_0^{-\alp} u^{\f{d-3}2-\de} r^{-\f{d-3}2})$. Thus, combining the above estimates, and using that $r \ls u$ in $\calM_{\med}$, we have obtained
\begin{equation}\label{eq:Mink.est.for.lower.med}
\Box_{\bfm}^{-1} F = O_{\bfGmm}^{M-C}(A U_0^{-\alp}u^{-\de+\nB-\f 12} r^{-\nB+\f 12})\quad \hbox{in $\calM_{\med}$}.
\end{equation}
Notice that this is better than the desired estimate \eqref{eq:Mink.est.for.lower.final} when restricted to $\calM_{\med}$.

\pfstep{Step~2: $r^p$ estimates in $\calM_{\wave}$} We prove an auxiliary bound (see \eqref{eq:Mink.est.for.lower.main.rp} below) using the $r^p$ estimate. Pick $p = 1+2\de$ so that
\begin{equation}\label{eq:p.for.lower}
p-3 <0, \quad \f{p-1}2 -\alp - \de + \nB  <0,\quad p+1 +2(\nB -2-\de)+1 >0.
\end{equation}
We apply the $r^p$ estimate in Lemma~\ref{lem:Qj-rp} with the $p$ above and with $\Psi = \Box_{\bfm}^{-1} F$, $j = 0$, $R_{1} = 2R_{\far}$, and $U_0$ replaced by $\f{U_0}2$. We first consider the $|I| = 0$ case and estimate each term on the right-hand side of \eqref{eq:Qj-rp}. The first term (with $U_0$ replaced by $\f{U_0}2$) vanishes since $F$ is supported in $u\geq U_0$. The second term is supported in a finite $r$ region can be bounded by
$$ U_0^{-\alp}U^{-\de+\nB}$$ after integrating the pointwise bound \eqref{eq:Mink.est.for.lower.med}. For the error terms in the third and fourth terms, we split the error term up in $\calM_{\wave}$ and $\calM_{\med}$ and control them as follows:
\begin{equation}
\begin{split}
&\:  \int_{\f{U_{0}}2}^{U} \left( \int_{\calC_{u} \cap \calM_{\wave}} r^{p} \abs*{ (r^{\nB} F) }^{2} \, \ud r \ud \rssgm(\tht) \right)^{\frac{1}{2}}\ud u \\
\ls &\: \int_{\f{U_{0}}2}^{U} \left( \int_{\calC_{u} \cap \calM_{\wave}} r^{p} \left( A  r^{-2}  u^{-\alp-\de+\nB} \right)^2 \, \ud r \right)^{\f 12}  \, \ud u\\
\ls &\: \int_{\f{U_{0}}2}^{U}  u^{\f{p-1}2-1-\alp-\de+\nB}\, \ud u \ls U_{0}^{\f{p-1}2-\alp-\de+\nB},
\end{split}
\end{equation}
where we used the first two inequalities in \eqref{eq:p.for.lower}, and
\begin{equation}
\begin{split}
&\: \left( \iint_{\calD_{\f{U_{0}}2}^{U} \cap \calM_{\med}} r^{p+1} \abs*{(r^{\nB} F)}^{2} \, \ud u \ud r \ud \rssgm(\tht) \right)^{\frac{1}{2}} \\
\ls &\:  \left(  \iint_{\calD_{\f{U_{0}}2}^{U} \cap \calM_{\med}} r^{p+1}\left( A r^{-2-\de} r^{\nB} u^{-\alp} \right)^2 \, \ud u \ud r \right)^{\frac{1}{2}} \\
\ls &\: \left(\int_{\f{U_{0}}2}^{U} (Au^{\f{p-1}2-\f 12 - \alp - \de+\nB})^2 \, \ud u \right)^{\f 12} \ls A U_{0}^{\f{p-1}2-\alp-\de+\nB},
\end{split}
\end{equation}
where we used the last two inequalities in \eqref{eq:p.for.lower}. By Lemma~\ref{lem:Qj-rp}, it follows that
\begin{equation}
\begin{split}
& \sup_{u \in [\f{U_{0}}2, U]} \left(\int_{\calC_{u}} \chi_{>2 R_{\far}} r^{p} (\rd_{r} (\Box_{\bfm}^{-1} F) )^{2} \, \ud r \ud \rssgm(\tht) \right)^{\frac{1}{2}}   \ls A U_{0}^{\f{p-1}2-\alp}U^{-\de+\nB}.
\end{split}
\end{equation}

For $|I|>0$, we carry out an induction in $|I|$. The first four terms on the right-hand side of \eqref{eq:Qj-rp} can be estimated exactly as above. The fifth term can be bounded above by $U_{0}^{\f{p-1}2-\alp}U^{-\de+\nB}$ using the induction hypothesis. The limit in the sixth term is easily controlled using both the induction hypothesis and the bound on the left-hand side of \eqref{eq:Qj-rp}. We have thus proven for $\bfGmm \in \set{r \rd_{r}, \bfS, \bfOmg}$ that
\begin{equation}\label{eq:Mink.est.for.lower.main.rp}
\begin{split}
& \sup_{u \in [\f{U_{0}}2, U]} \left(\int_{\calC_{u}} \chi_{>2 R_{\far}} r^{p} (\rd_{r} \bfGmm^{I} (\Box_{\bfm}^{-1} F) )^{2} \, \ud r \ud \rssgm(\tht) \right)^{\frac{1}{2}}   \ls A U_{0}^{\f{p-1}2-\alp}U^{-\de+\nB}.
\end{split}
\end{equation}

\pfstep{Step~3: Pointwise estimates in $\calM_{\wave}$} To obtain pointwise estimates in $\calM_{\wave}$, we combine \eqref{eq:Mink.est.for.lower.main.rp} with the bound \eqref{eq:Mink.est.for.lower.med} in $\calM_{\med}$.

Using \eqref{eq:Mink.est.for.lower.main.rp} and Lemma~\ref{lem:hardy-radial}, we obtain that
\begin{equation}\label{eq:Mink.est.for.lower.Hardy}
\Big| \bfGmm^{I} (r^{\nB} \Box_{\bfm}^{-1} F)(u,r,\theta) - \limsup_{r'\to \infty} \bfGmm^{I} (r^{\nB} \Box_{\bfm}^{-1} F)(u,r',\theta)\Big| \ls A U_{0}^{\f{p-1}2-\alp}u^{-\de+\nB} r^{-\f{p-1}2} ,\quad r \geq 2R_{\far}.
\end{equation}
Denoting the radiation field of $\Box_{\bfm}^{-1} F$ by $\rPhi_{0}$, we know that $\lim_{r\to \infty} r^{\nB} \bfGmm^{I} \Big[ (\Box_{\bfm}^{-1} F)(r,u,\theta) - r^{-\nB} \rPhi_{0}(u,\theta) \Big] = 0$. In particular, this implies $\lim_{r\to \infty} r \rd_{r} \bfGmm^{I} (r^{\nB} \Box_{\bfm}^{-1} F)(u,r',\theta) = 0$. Thus, \eqref{eq:Mink.est.for.lower.Hardy} implies that
\begin{equation}\label{eq:Mink.est.for.lower.dr.bound}
\Big| \rd_r \bfGmm^{I} (r^{\nB} \Box_{\bfm}^{-1} F)(u,r,\theta) \Big| \ls A U_{0}^{\f{p-1}2-\alp} u^{-\de+\nB} r^{-\f{p-1}2-1} ,\quad r \geq 2R_{\far}.
\end{equation}

We now derive the desired estimate in $\calM_{\wave}$. Fix $(u,r,\th) \in \calM_{\wave}$. Integrating along a constant-$(u,\th)$ curve from $(u,r= 50 \eta_0^{-1} u,\th) \in \calM_{\med}$ to $(u,r,\th)$, applying the fundamental theorem of calculus, and using the estimates \eqref{eq:Mink.est.for.lower.med} and \eqref{eq:Mink.est.for.lower.dr.bound}, we obtain
\begin{equation*}
\begin{split}
\Big| \bfGmm^{I} (r^{\nB} \Box_{\bfm}^{-1} F)(u,r,\theta) \Big| = &\:  \Big| \bfGmm^{I} (r^{\nB} \Box_{\bfm}^{-1} F)(u,50\eta_0^{-1}u,\theta) + \int_{50\eta_0^{-1}u}^r \rd_r \bfGmm^{I} (r^{\nB} \Box_{\bfm}^{-1} F)(u,r',\theta) \, \ud r' \Big| \\
\ls &\: A u^{\nB-\de} U_0^{-\alp} + A U_0^{\f{p-1}2-\alp}u^{-\de+\nB} \int_{50\eta_0^{-1}u}^r (r')^{-\f{p-1}2-1} \, \ud r' \\
\ls &\: A u^{\nB-\de} U_0^{-\alp} + A U_0^{\f{p-1}2-\alp}u^{-\de+\nB} u^{-\f{p-1}2} \ls A u^{\nB-\de} U_0^{-\alp},
\end{split}
\end{equation*}
where we have used $-\f{p-1}2<0$ and $U_0\leq 2u$ on $\supp(\Box_{\bfm}^{-1} F)$ (by finite speed of propagation). Commuting $[\bfGmm^{I},r^{\nB}]$ and dividing by $r^{\nB}$, we obtain that for $|I| \leq M-C$,
$$\Big|\bfGmm^{I} (\Box_{\bfm}^{-1} F)(u,r,\theta)\Big| \ls A U_0^{-\alp}u^{\nB-\de} r^{-\nB}.$$
Since we can take $\bfGmm \in \set{r \rd_{r}, \bfS, \bfOmg}$, this implies
$$ (\Box_{\bfm}^{-1} F)(u,r,\theta) = O_{\bfGmm}^{M-C}(A U_0^{-\alp}u^{\nB-\de} r^{-\nB}).$$
Combining this with \eqref{eq:Mink.est.for.lower.med}, we obtain \eqref{eq:Mink.est.for.lower.final}. \qedhere
\end{proof}

\subsection{Remarks on notations and application of Main Theorem~\ref{thm:upper}}\label{sec:notation.in.lower}

We now turn to the proof of Main Theorem~\ref{thm:lower}. \textbf{For the remainder of the section, we work under the assumptions of Main Theorem~\ref{thm:lower}.} In particular,

In the argument below, we will freely use our main upper bound theorem (Main Theorem~\ref{thm:upper}), which we have established above, as a starting point. Moreover, we will apply estimates derived in the proof of Main Theorem~\ref{thm:upper}, e.g., we will directly apply Case~2b of Proposition~\ref{prop:wave-main} below.

When applying Main Theorem~\ref{thm:upper}, we will use $\lfloor a_0 M_0\rfloor$ to denote the number of derivatives and $A_{\mathfrak f}$ to denote the size of $\phi$ (which is consistent with the notation of Main Theorem~\ref{thm:upper}). In the argument below, applications of the iteration lemmas will lead to loss of derivatives. Since the loss of derivatives is always a fraction of the total number of derivatives, \textbf{we will slightly abuse notation to write the number of derivatives in the error terms as $\lfloor a'_0 M_0 \rfloor$ where in fact $a'_0$ may decrease from line to line.}

Similarly, we slightly abuse notation
\begin{itemize}
\item \textbf{to allow $A'_{\mathfrak f}$, which is used to measure the size of the error terms,  is allowed to increase from line to line;}
\item \textbf{to allow $m'_0$, which is a necessary lower bound for $M_0$, to increase from line to line;}
\item \textbf{to allow $\de_{\mathfrak f}$, which is used to measure how much the error term is better than the main term, to decrease from line to line} until it is fixed in the beginning of Section~\ref{sec:near.sharp}.
\end{itemize}

In particular, we will assume without loss of generality that
\begin{equation}\label{eq:de_f.least}
\de_{\mathfrak f} \leq \min\{ \f 1{16}, \f{\de_c}4, \f{\de_0}4, \alp_d - \alp_{\mathfrak f}\}.
\end{equation}
(Recall that $\alp_d > \alp_{\mathfrak f}$ by the assumption \eqref{eq:lower-hyp-alpf}.)

\subsection{Sharp asymptotics in the wave and intermediate zone}\label{sec:med.wave.sharp}

Under the assumptions of Main Theorem~\ref{thm:lower}, we derive the sharp asymptotics in $\calM_{\far}$ in this subsection. See already Proposition~\ref{prop:wave.med.precise}. In particular, we take $\alp_{\mathfrak f} = J_{\frkf} + \nB$ with $J_{\frkf}$ as in the statement of Main Theorem~\ref{thm:lower}.

We need to subtract from $\Phi$ a good approximate solution, and take into account the strong Huygens principle. Let $U > 1$. Fix $\de_0 >0$ satisfying \eqref{eq:dlt0-wave} and define
\begin{equation}\label{eq:U0.def}
U_0 = U^{1-\bt},\quad \bt = \f{\de_0}{4(\alp_{\mathfrak f}+\de_0)}.
\end{equation} 
For $0 \leq j \leq J_{\mathfrak f}$, $0 \leq k \leq K_{j}$, take $\rPhi_{j, k}$ as in Main Theorem~\ref{thm:upper}. Define
\begin{equation} \label{eq:varphi-f}
	\varphi_{\mathfrak f} = \Box_{\bfm}^{-1} \left(-2 \chi_{>2 R_{\far}}(r) \chi_{>U_{0}}'(u) r^{-\nu_{\Box}} \rd_{r} \left(\sum_{j=0}^{J_{\mathfrak f}} \sum_{k=0}^{K_{j}} \rPhi_{j, k}(u) r^{-j} \log^{k}(\tfrac{r}{u})\right)\right).
\end{equation}

The proof of the sharp asymptotics has two steps. First, in Lemma~\ref{lem:varphif.is.this}, we show using the computations in Section~\ref{sec:Minkowski.wave} that
$$\varphi_{\mathfrak f} \approx \sum_{k=1}^{K_{J_{\mathfrak f}}} \varphi^{\bfm[1]}_{J_{\mathfrak f},k}[\rcPhi_{J_{\mathfrak f},k}(\infty)](u,r,\th)
 + \varphi^{\bfm[1]}_{J_{\mathfrak f},0}[\rcPhi_{(\leq J_{\mathfrak f} - \nB)J_{\mathfrak f},0}(\infty)](u,r,\theta).$$ Then, in Lemma~\ref{lem:varphif.is.good}, we use the Minkowskian wave equation estimates in Section~\ref{sec:improved.Mink} to show $\varphi_{\mathfrak f}$ is in turn a good approximation of $\phi$. Together, these steps give the sharp asymptotics.

\begin{lemma}\label{lem:varphif.is.this}
For $U_0$ as in \eqref{eq:U0.def} and $\varphi_{\frkf}$ as in \eqref{eq:varphi-f}, the following estimate holds:
\begin{equation*}
\begin{split}
 \varphi_{\frkf}(u,r,\theta)
 = &\: \sum_{k=1}^{K_{J_{\mathfrak f}}} \varphi^{\bfm[1]}_{J_{\mathfrak f},k}[\rcPhi_{J_{\mathfrak f},k}(\infty)](u,r,\th)
 + \varphi^{\bfm[1]}_{J_{\mathfrak f},0}[\rcPhi_{(\leq J_{\mathfrak f} - \nB)J_{\mathfrak f},0}(\infty)](u,r,\theta)  \\
 &\: \quad\qquad + O_{\bfGmm}^{\lfloor a'_0 M_0 \rfloor}(A'_{\mathfrak f} \min\{ u^{-\alp_{\mathfrak f}-\de_{\mathfrak f}+\nB} r^{-\nB}, u^{-\alp_{\mathfrak f}-\de_{\mathfrak f}}\}) \quad \hbox{in $\calM_{\far}$}.
\end{split}
\end{equation*}
\end{lemma}
\begin{proof}
We consider the $j\leq J_{\frkf}-1$ and $j = J_{\frkf}$ cases in the sum in \eqref{eq:varphi-f} separately.

We first consider the sum over $j\leq J_{\frkf}-1$. By the assumption \eqref{eq:lower-hyp-alpf}, $J_{\frkf} \leq \min\{J_{c}, J_{d}\} - 1$ and $\alp_{\frkf} = J_{\frkf} + \nB$. Using the bounds obtained in the proof of Case~1 in Step~5 of the proof of Main Theorem~\ref{thm:upper} Using the upper bound result in Main Theorem~\ref{thm:upper}, the assumptions of Case~2 of Proposition~\ref{prop:wave-main} hold with $\de_0$ fixed above and with $J = J_{\frkf}$, $\alp = \alp_{\frkf}-\f{\de_0}8$ (where $\f{\de_0}8$ is used to remove the logarithm). Hence, after applying Case~2 of Proposition~\ref{prop:wave-main}, it follows that the bounds \eqref{eq:wave-main-Phijk} and \eqref{eq:wave-main-Phij0-low} gives
\begin{align}
	\rPhi_{j, k} &= O_{\bfGmm}^{M'} (A' u^{j- \alp_{\frkf} -\f{7\dlt_{0}}8+ \nu_{\Box}})& &\hbox{ for } 0 \leq j \leq J_{\frkf}-1, \, 1 \leq k \leq K_{j}, \label{eq:wave-main-Phijk.relabelled} \\
	\rPhi_{(\leq j - \nu_{\Box}) j, 0} &= O_{\bfGmm}^{M'} (A' u^{j- \alp_{\frkf} -\f{7\dlt_{0}}8 + \nu_{\Box}})& &\hbox{ for } 0 \leq j \leq J_{\frkf} - 1, \label{eq:wave-main-Phij0-low.relabelled} 
\end{align}
Using these bounds for $\rPhi_{j,k}$, we now argue as in \eqref{eq:med.main.terms.with.expected.decay}, but considering the whole $\calM_{\far}$ and taking into account the new $U_0$, i.e., we use the decay estimates in \eqref{eq:wave-main-Phijk.relabelled}--\eqref{eq:wave-main-Phij0-low.relabelled}, Lemma~\ref{lem:med.explicit.inho}, \eqref{eq:phi.ell.j.k.upper} and \eqref{eq:explicit.wave.upper} to obtain
\begin{equation}\label{eq:med.main.terms.with.expected.decay.U0}
\begin{split}
&\: \sum_{j=0}^{J_{\mathfrak f} -1} \sum_{k = 0}^{K_{j}} \Box_\bfm^{-1} \Big(\chi_{>2 R_{\far}}(r) \chi'_{>U_0}(u) r^{-\nB} \rPhi_{j,k} \rd_r \Big( r^{-j} \log^{k} \left(\tfrac{r}{u}\right)\Big) \Big)(U,R,\Theta) \\
= &\: \sum_{j=0}^{J_{\mathfrak f} -1} \sum_{k = 0}^{K_{j}}  O_{\bfGmm}^{\lfloor a'_0 M_0 \rfloor}\Bigg(A'_{\mathfrak f} \min\{U^{-j}R^{-\nB}, U^{-j-\nB}\} U_0^{j-\alp_{\mathfrak f}-\f{7\de_0}8 +\nB} \log^{K_\rho} \left( \tfrac{U-U_0}{U_0} \right) \Bigg) \\
= &\: \sum_{j=0}^{J_{\mathfrak f} -1} \sum_{k = 0}^{K_{j}}  O_{\bfGmm}^{\lfloor a'_0 M_0 \rfloor}(A'_{\mathfrak f} \min\{R^{-\nB}, U^{-\nB}\} U^{-j} U^{(j-\alp_{\mathfrak f}-\f{7\de_0}8+\nB)(1-\bt)}\log^{K_\rho} U) \\
= &\:  O_{\bfGmm}^{\lfloor a'_0 M_0 \rfloor}(A'_{\mathfrak f} \min\{U^{-\alp_{\mathfrak f}-\f{\de_0}2+\nB} R^{-\nB}, U^{-\alp_{\mathfrak f}-\f{\de_0}2}\}).
\end{split}
\end{equation}
Here, we used that $-(\alp_{\mathfrak f}+\f{7\de_0}8)(1-\bt) = -\alp_{\mathfrak f}-\f{7\de_0}8+\f{7\de_0(\alp_{\mathfrak f}+\f{7\de_0}8)}{32(\alp_{\mathfrak f}+\de_0)} < -\alp_{\mathfrak f}-\f{21\de_0}{32}$, and then used $U^{-\f{5\de_0}{32}}$ to dominate the $\log^{K_\rho} U$ terms.

For the $j=J_{\mathfrak f}$ case, it is convenient to use \eqref{eq:rcPhi-rPhi} to write in terms of the renormalized radiation fields:
\begin{equation*}
	\sum_{k=0}^{K_{J_{\mathfrak f}}} \rPhi_{J_{\mathfrak f}, k}(u) \log^{k}(\tfrac{r}{u}) = \sum_{k=0}^{K_{J_{\mathfrak f}}} \rcPhi_{J_{\mathfrak f}, k}(u) \log^{k} r.
\end{equation*}
Then a very similar argument as above, except for using \eqref{eq:wave-main-PhiJk.limit}, \eqref{eq:wave-main-PhiJ0-low.limit} instead of \eqref{eq:wave-main-Phijk}, \eqref{eq:wave-main-Phij0-low}, gives
\begin{equation}
\begin{split}
&\: \sum_{k = 0}^{K_{J_{\frkf}}} \Box_\bfm^{-1} \Big(\chi_{>2 R_{\far}}(r) \chi'_{>U_0}(u) r^{-\nB} (\rcPhi_{J_{\frkf},k}(u) - \rcPhi_{J_{\frkf},k}(\infty))  \rd_r \Big( r^{-J_{\frkf}} \log^{k} \left(\tfrac{r}{u}\right)\Big) \Big)(U,R,\Theta) \\
&\: + \Box_\bfm^{-1} \Big(\chi_{>2 R_{\far}}(r) \chi'_{>U_0}(u) r^{-\nB} (\rcPhi_{J_{\frkf},0}(u) - \rcPhi_{(\leq J_{\frkf}-\nB)J_{\frkf},0}(\infty))  \rd_r \Big( r^{-J_{\frkf}} \log^{k} \left(\tfrac{r}{u}\right)\Big) \Big)(U,R,\Theta) \\
= &\:  O_{\bfGmm}^{\lfloor a'_0 M_0 \rfloor}(A'_{\mathfrak f} \min\{U^{-\alp_{\mathfrak f}-\f{\de_0}2+\nB} R^{-\nB}, U^{-\alp_{\mathfrak f}-\f{\de_0}2}\}).
\end{split}
\end{equation}
On the other hand, using Lemma~\ref{lem:med.explicit.inho} and \eqref{eq:profile.move.around}, we may write
\begin{align*}
	& \left[\Box_\bfm^{-1} \Big(\chi_{>2 R_{\far}}(r) \chi'_{>U_0}(u) r^{-\nB} \rcPhi_{J_{\frkf},k}(\infty)  \rd_r \Big( r^{-J_{\frkf}} \log^{k} \left(\tfrac{r}{u}\right)\Big) \Big) - \varphi^{\bfm[1]}_{J_{\mathfrak f},k}[\rcPhi_{J_{\mathfrak f},k}(\infty)]\right] (U,R,\Theta) \\
	&= O_{\bfGmm}^{\lfloor a'_0 M_0 \rfloor}(A_{\frkf}'  U_{0} \min\{ R^{-\nB} U^{-J_{\mathfrak f} - 1}, U^{-J_{\mathfrak f} - 1 - \nB} \}\log^{k} U), \\
	&\left[ \Box_\bfm^{-1} \Big(\chi_{>2 R_{\far}}(r) \chi'_{>U_0}(u) r^{-\nB} \rcPhi_{(\leq J_{\frkf}-\nB)J_{\frkf},0}(\infty)  \Big( \rd_r  r^{-J_{\frkf}} \Big) \Big) - \varphi^{\bfm[1]}_{J_{\mathfrak f}, 0}[\rcPhi_{(\leq J_{\mathfrak f} - \nu_{\Box}) J_{\mathfrak f}, 0}(\infty)]\right] (U,R,\Theta) \\
	&= O_{\bfGmm}^{\lfloor a'_0 M_0 \rfloor}(A_{\frkf}' U_{0} \min\{ R^{-\nB} U^{-J_{\mathfrak f} - 1}, U^{-J_{\mathfrak f} - 1 - \nB} \}).
\end{align*}
Recalling \eqref{eq:U0.def}, these terms are thus $O_{\bfGmm}^{\lfloor a'_0 M_0 \rfloor}(A_{\frkf}' \min\{ R^{-\nB} U^{-J_{\mathfrak f} - \f{\de_0}{4(\alp_{\mathfrak f}+\de_0)} }, U^{-J_{\mathfrak f} -\f{\de_0}{4(\alp_{\mathfrak f}+\de_0)} - \nB} )$. 

Finally, combining all the above estimates, choosing $\de_{\frkf} \leq \min \{ \f{\de_0}2, \f{\de_0}{4(\alp_{\mathfrak f}+\de_0)}\}$, and relabelling $(U,R.\Theta)$ as $(u,r,\theta)$, we obtain the desired bound. \qedhere
\end{proof}

The next lemma shows that $\varphi_{\frkf}$ is a good approximation of $\phi$.
\begin{lemma}\label{lem:varphif.is.good}
For $U_0$ as in \eqref{eq:U0.def} and $\varphi_{\frkf}$ as in \eqref{eq:varphi-f}, the following estimate holds:
$$(\phi - \varphi_{\frkf})(u,r,\theta) = O_{\bfGmm}^{\lfloor a'_0 M_0 \rfloor}(A'_{\mathfrak f} u^{-\alp_{\mathfrak f}-\de_{\mathfrak f}+\nB} r^{-\nB}) \quad \hbox{in $\calM_{\far}$}.$$
\end{lemma}
\begin{proof}
Let $(U,R,\Theta) \in \calM_{\far}$ be given with $U>1$. It suffices to assume that $R\geq 2 R_{\far}$ for otherwise the estimate already follows from Main Theorem~\ref{thm:upper}, Lemma~\ref{lem:varphif.is.this} and the bounds in Section~\ref{sec:Minkowski.wave}. 

We compute using the definition of $\varphi_{\frkf}$ in \eqref{eq:varphi-f} that
\begin{equation}\label{eq:for.diff.in.precise.rate}
    \begin{split}
        &\: \Box_{\bfm} (\chi_{>2R_{\mathrm{far}}} \chi_{>U_0} \phi - \varphi_{\frkf}) \\
        = &\: \chi_{>2R_{\mathrm{far}}} \chi_{>U_0} \Box_{\bfm}\phi  + [\Box_{\bfm}, \chi_{>R_{\far}}](\chi_{>U_{0}} \phi) + \chi_{>2R_{\mathrm{far}}} [\Box_{\bfm}, \chi_{>U_{0}}] \phi - \Box_{\bfm} \varphi_{\frkf} \\
        = &\: \chi_{>2R_{\mathrm{far}}} \chi_{>U_0} \Box_{\bfm}\phi +  [\Box_{\bfm}, \chi_{>2R_{\far}}](\chi_{>U_{0}} \phi) - 2 r^{-\nB} \chi_{>2R_{\far}}  \chi_{> U_{0}}'(u) \rd_{r} \rho_{J_{\mathfrak f}+1},
    \end{split}
\end{equation}
where we used Lemma~\ref{lem:med.incoming} as well as $\rho_{J_{\mathfrak f}+1} = \Phi - \sum_{j=0}^{J_{\mathfrak f}} \sum_{k=0}^{K_{j}} \rPhi_{j, k}(u) r^{-j} \log^{k}(\tfrac{r}{u})$ in the final line.

Arguing as in the terms $e_2,\ldots, e_5$ in \eqref{eq:med.e1-e5}, but using the bounds in the main upper bound theorem (Main Theorem~\ref{thm:upper}), we obtain
\begin{equation}
    \begin{split}
        &\: \chi_{>2R_{\mathrm{far}}} \chi_{>U_0} \Box_{\bfm}\phi + [\Box_{\bfm}, \chi_{>2R_{\far}}](\chi_{>U_{0}} \phi) \\
        = &\: \begin{cases}
             O_{\bfGmm}^{\lfloor a'_0 M_0 \rfloor} ( A_{\frkf}' r^{-2-\de_{0}} u^{-J_{\frkf}-\nB}\log^{K_{\frkf}} u) & \hbox{in $\calM_{\med}$} \\
            O_{\bfGmm}^{\lfloor a'_0 M_0 \rfloor} (A_{\frkf}' r^{-\nB-2} u^{-J_{\frkf}-\de_0}\log^{K_{\frkf}} u) & \hbox{in $\calM_{\wave}$}
        \end{cases} \\
        = &\: \begin{cases}
             O_{\bfGmm}^{\lfloor a'_0 M_0 \rfloor} ( A_{\frkf}' r^{-2-\de_{0}} u^{-J_{\frkf}-\nB+\f{\de_0}2}) & \hbox{in $\calM_{\med}$} \\
            O_{\bfGmm}^{\lfloor a'_0 M_0 \rfloor} (A_{\frkf}' r^{-\nB-2} u^{-J_{\frkf}-\f{\de_0}2}) & \hbox{in $\calM_{\wave}$}
        \end{cases}
    \end{split}
\end{equation}
We now apply Proposition~\ref{prop:Mink.est.for.lower} with $\alp = J_{\frkf}+\nB-\f{\de_0}2$ and $\de = \de_0$ to obtain that
\begin{equation}\label{eq:for.varphif.error}
\begin{split}
&\: \Bigg(\Box_{\bfm}^{-1} \Big( \chi_{>2R_{\mathrm{far}}} \chi_{>U_0} \Box_{\bfm}\phi + [\Box_{\bfm}, \chi_{>2R_{\far}}](\chi_{>U_{0}} \phi) \Big) \Bigg)(U,R,\Theta) \\
= &\: O_{\bfGmm}^{\lfloor a'_0 M_0 \rfloor}(A_{\frkf}' U_0^{-J_{\frkf}-\nB+\f{\de_0}2} U^{\nB-\de_0} R^{-\nB}).
\end{split}
\end{equation}
Recalling \eqref{eq:U0.def} and computing as follows
\begin{equation}
\begin{split}
 \nB-\de_0 - (1-\f{\de_0}{4(J_{\frkf}+\nB+\de_{0})})(J_{\frkf}+\nB - \f{\de_{0}}2)
< &\: -J_{\frkf}-\de_{0} +\f{\de_0}2 + \f{\de_0(J_{\frkf}+\nB)}{4(J_{\frkf}+\nB+\de_{0})} < -J_{\frkf}-\f{\de_0}4,
\end{split}
\end{equation}
it follows from \eqref{eq:for.varphif.error} that
\begin{equation}
\begin{split}
 \Bigg(\Box_{\bfm}^{-1} \Big( \chi_{>2R_{\mathrm{far}}} \chi_{>U_0} \Box_{\bfm}\phi + [\Box_{\bfm}, \chi_{>2R_{\far}}](\chi_{>U_{0}} \phi) \Big) \Bigg)(U,R,\Theta)
= O_{\bfGmm}^{\lfloor a'_0 M_0 \rfloor}(A_{\frkf}'  U^{-J_{\frkf}-\f{\de_0}4} R^{-\nB}).
\end{split}
\end{equation}

For the final term in \eqref{eq:for.diff.in.precise.rate}, we use \eqref{eq:wave-main-rhoJ+1-final} to write $\rho_{J_{\mathfrak f}+1} = \rho_{J_{\mathfrak f}+1}^{(1)} + \rho_{J_{\mathfrak f}+1}^{(2)}$ with
$$\rho_{J_{\mathfrak f}+1}^{(1)} = O_{\bfGmm}^{\lfloor a_0' M_0 \rfloor}(A_{\frkf}' r^{-\alp_{\mathfrak f}-\eta_{\mathfrak f}+\nB} u^{\eta_{\mathfrak f}} \log^{K_{J_{\mathfrak f}}} u),\quad \rho_{J_{\mathfrak f}+1}^{(2)} = O_{\bfGmm}^{\lfloor a_0' M_0 \rfloor}(A_{\frkf}' r^{-\alp_{\mathfrak f}-\de_0+\nB}).$$

For $\rho_{J_{\mathfrak f}+1}^{(1)}$, we apply Proposition~\ref{prop:rho.est.improved} with $\alp_\rho = \alp_{\mathfrak f}+\eta_{\mathfrak f}$, $\bt_\rho = \eta_{\mathfrak f}$ and $K_\rho = K_{J_{\mathfrak f}}$ to obtain
\begin{equation}\label{eq:med.main.terms.rho.1}
\begin{split}
&\: \Box_\bfm^{-1} \Big(\chi_{>2 R_{\far}}(r) \chi_{>U_0}'(u) r^{-\nB} \rd_r \rho_{J_{\mathfrak f}+1}^{(1)} \Big)(U,R,\Theta) \\
= &\: O_{\bfGmm}^{\lfloor a'_0 M_0 \rfloor}(A'_{\mathfrak f} U^{-\alp_{\mathfrak f} - \eta_{\mathfrak f}} U^{(\eta_{\mathfrak f} + \de''')(1-\bt)} \log^{K_J}U \min\{1, U^{\nB} R^{-\nB} \}) \\
= &\: O_{\bfGmm}^{\lfloor a'_0 M_0 \rfloor}(A'_{\mathfrak f} U^{-\alp_{\mathfrak f} -\de_{\mathfrak f}} \min\{1, U^{\nB} R^{-\nB} \}),
\end{split}
\end{equation}
where given any $\eta_{\mathfrak f}>0$, we have chosen $\de'''>0$ to be sufficiently small such that $\eta_{\mathfrak f} - (\eta_{\mathfrak f} + \de''')(1-\bt) =: 2\de_w >0$. After using $U^{-\de_w} \log^{K_J}U \ls 1$, we can then take $\de_{\mathfrak f} \leq \de_w$.

For $\rho_{J+1}^{(2)}$, we apply Proposition~\ref{prop:rho.est.improved} with $\alp_\rho = \alp_{\mathfrak f} + \de_0$, $\bt_\rho =0$ and $K_\rho = K_{J_{\mathfrak f}}$
\begin{equation}\label{eq:med.main.terms.rho.2}
\begin{split}
&\: \Box_\bfm^{-1} \Big(\chi_{>2 R_{\far}}(r) \chi_{>U_0}'(u) r^{-\nB} \rd_r \rho_{J_{\mathfrak f}+1}^{(2)} \Big)(U,R,\Theta) \\
= &\: O_{\bfGmm}^{\lfloor a'_0 M_0 \rfloor}(A'_{\mathfrak f} U^{-\alp_{\mathfrak f}-\de_0} \log^{K_{J_{\mathfrak f}}} U \min\{1, U^{\nB} R^{-\nB} \}) = O_{\bfGmm}^{\lfloor a'_0 M \rfloor}(A'_{\mathfrak f} U^{-\alp_{\mathfrak f}-\f{\de_0}2} \min\{1, U^{\nB} R^{-\nB} \}),
\end{split}
\end{equation}
where we have used $U^{-\f{\de_0}2}$ to absorb the logarithms.

Combining all the estimates above gives the desired conclusion after relabelling $(U,R,\Theta)$ as $(u,r,\th)$. \qedhere
\end{proof}

\begin{proposition}\label{prop:wave.med.precise}
Under the assumptions of Main Theorem~\ref{thm:lower} and the conventions for $a_0'$, $m_0'$, $\de_{\mathfrak f}$ and $A'_{\mathfrak f}$ in Section~\ref{sec:notation.in.lower}, the following estimate holds:
\begin{equation}\label{eq:med.precise}
\begin{split}
 \phi(u,r,\theta)
 = &\: \sum_{k=1}^{K_{J_{\mathfrak f}}} \varphi^{\bfm[1]}_{J_{\mathfrak f},k}[\rcPhi_{J_{\mathfrak f},k}(\infty)](u,r,\th)
 + \varphi^{\bfm[1]}_{J_{\mathfrak f},0}[\rcPhi_{(\leq J_{\mathfrak f} - \nB)J_{\mathfrak f},0}(\infty)](u,r,\theta)  \\
 &\: \quad\qquad + O_{\bfGmm}^{\lfloor a'_0 M_0 \rfloor}(A'_{\mathfrak f} u^{-\alp_{\mathfrak f}-\de_{\mathfrak f}+\nB} r^{-\nB}) \quad \hbox{in $\calM_{\far}$}.
\end{split}
\end{equation}
\end{proposition}
\begin{proof}
This is a direct consequence of Lemma~\ref{lem:varphif.is.this} and Lemma~\ref{lem:varphif.is.good}. \qedhere
\end{proof}

\subsection{Sharp asymptotics in the near-intermediate zone}\label{sec:near.sharp}

Finally, we turn to the sharp asymptotics estimates in the near-intermediate zone. \textbf{At this point, we fix the constant $\de_{\mathfrak f}>0$ so that the estimates in Proposition~\ref{prop:wave.med.precise} hold.}

\begin{proposition}[Sharp asymptotics in the near-intermediate zone]\label{prop:sharp.near}
Suppose the assumptions of Main Theorem~\ref{thm:lower} hold. Fix $\de_{\mathfrak f} >0$ as above and take $\de_a \in (0, \f{2\de_{\mathfrak f}}{d-1})$.

Then, under the conventions for $a_0'$, $m_0'$ and $A'_{\mathfrak f}$ in Section~\ref{sec:notation.in.lower}, the following estimate holds:
\begin{equation}
 \phi(\tau,r,\theta) = \sum_{k,k'=0}^{K_{J_{\mathfrak f}}} c_{k,k'} \mathfrak L_k \tau^{-J_{\mathfrak f} -\nB}(\log^{k'} \tfrac \tau 2)\psi( x) + O_{\bfGmm}^{\lfloor a'_0 M_0 \rfloor}(\tau^{-\alp_{\mathfrak f}-\de_{\mathfrak f}}) \quad \hbox{in $\{r \leq \tau^{1-\de_a}\}$},
\end{equation}
where the constants $c_{k,k'} = \mathfrak b^{(J_{\mathfrak f},k,0)}_{k-k'} \f{\Big( \f{d-3}2 \Big)!}{(d-2)!} 2^{J_{\mathfrak f}+\nB}$, with $\mathfrak b^{(J_{\mathfrak f},k,0)}_{k'}$ given in Lemma~\ref{lem:med.explicit.1}.
\end{proposition}
\begin{proof}
Let us only write down the case where $\rd_{(\mathrm{t})} \calM = \0$. The $\rd_{(\mathrm{t})} \calM \neq \0$ case is essentially the same after taking into account the different spatial profiles in \eqref{eq:psi.def}.

Take $0\leq i \leq \lfloor a_0' M_0 \rfloor$. We argue as in \eqref{eq:P0.cutoff}, except that (1) we carefully take away the main terms and (2) we choose a slightly smaller $\tau_0$. Writing $\calP_{0}$, $\calR_{0}$ for $^{(\infty)}\calP_{0}$, $^{(\infty)}\calR_{0}$, and using that $\calP_{0}(1 - \calR_{0} \calP_{0} 1) = 0$, we obtain a slightly different version of \eqref{eq:P0.cutoff}:
\begin{equation}\label{eq:P0.cutoff.lower}
\begin{split}
&\peq \calP_{0} \Big(\chi_{<\tau_0} (r) \bfT^{i} \phi - \sum_{k,k'=0}^{K_{J_{\mathfrak f}}} c_{k,k'} \chi_{<\tau_0}(r) \mathfrak L_k \bfT^{i} (\tau^{-\alp_{\mathfrak f}}\log^{k'} \tfrac \tau 2) (1 - \calR_0 \calP_0 1) \Big) \\
&= - [\chi_{<\tau_0}(r), \calP_{0}] \Big( \bfT^{i} \phi - \sum_{k,k'=0}^{K_{J_{\mathfrak f}}} c_{k,k'} \mathfrak L_k \bfT^{i} (\tau^{-\alp_{\mathfrak f}}\log^{k'} \tfrac \tau 2) (1 - \calR_0 \calP_0 1) \Big)\\
&\peq  + \chi_{<\tau_0}(r) (\calP_{0} - \calP) (\bfT^{i} \phi) \\
&\peq + \chi_{<\tau_0}(r) \bfT^{i} f \\
&\peq + \chi_{<\tau_0}(r) \bfT^{i} \calN(\phi) \\
&\peq + \chi_{<\tau_0}(r) [\calP_{0}, \bfT^i] \phi \\
&=: \widetilde{e}_{1} + e_{2}+ \ldots + e_{5},
\end{split}
\end{equation}
where $e_{2},\ldots,e_{5}$ are exactly as in \eqref{eq:P0.cutoff}, and $\widetilde{e}_1$ is a new term.

\textbf{From now on, we choose}
\begin{equation}\label{eq:tau-0.def}
\tau_0 = \f 12 \tau^{1-\de_a},\quad \de_a \leq \f{2\de_{\mathfrak f}}{d-1}.
\end{equation}
We now estimate the terms $\widetilde{e}_{1}, e_{2}, \ldots, e_{5}$.

We first consider $\widetilde{e}_1$ (for which we will derive a better estimate than \eqref{eq:med.induction.e1}.) We split the term as
\begin{equation}
\widetilde{e}_{1,1} := - [\chi_{<\tau_0}(r), \calP_{0}] \Big( \bfT^{i} \phi - \sum_{k,k'=0}^{K_{J_{\mathfrak f}}} c_{k,k'} \mathfrak L_k \bfT^{i} (\tau^{-\alp}\log^{k'} \tfrac \tau 2) \Big), \quad \widetilde{e}_{1,2} := \widetilde{e}_{1} - \widetilde{e}_{1,1}.
\end{equation}
Starting with the estimates in Proposition~\ref{prop:wave.med.precise} and computing the first two terms in \eqref{eq:med.precise} using Proposition~\ref{prop:med.explicit.precise} with $j=J_{\mathfrak f}$, $\ell=0$, $\de_m = \de_a$, we obtain
\begin{equation}\label{eq:asymp.proof.phi.expand.in.int}
\begin{split}
&\:  \phi(u,r,\theta) \\
 = &\: \sum_{k=1}^{K_{J_{\mathfrak f}}} \varphi^{\bfm[1]}_{J_{\mathfrak f},k}[\rcPhi_{J_{\mathfrak f},k}(\infty)](u,r,\th)
 + \varphi^{\bfm[1]}_{J_{\mathfrak f},0}[\rcPhi_{(\leq J_{\mathfrak f} - \nB)J_{\mathfrak f},0}(\infty)](u,r,\theta) + O_{\bfGmm}^{\lfloor a'_0 M_0 \rfloor}(A'_{\mathfrak f} u^{-\alp_{\mathfrak f}-\de_{\mathfrak f}+\nB} r^{-\nB}) \\
= &\: \sum_{k=0}^{K_{J_{\mathfrak f}}} c_{k,k'} \mathfrak L_k u^{-\alp_{\mathfrak f}} \log^{k'}(\tfrac u2) +O_{\bfGmm}^{\lfloor a'_0 M_0 \rfloor}(A'_{\mathfrak f} u^{-\alp_{\mathfrak f}-\f{\de_{a}}2} )  + O_{\bfGmm}^{\lfloor a'_0 M_0 \rfloor}(A'_{\mathfrak f} u^{-\alp_{\mathfrak f}- \de_{\mathfrak f} +\nB} r^{-\nB})\\
&\: \qquad\qquad\qquad\qquad\qquad \qquad\qquad\qquad\qquad\qquad \qquad\qquad\qquad\qquad\qquad \hbox{in $\calM_\med \cap \{r \leq u^{1-\de_a} \}$}.
\end{split}
\end{equation}
(For this estimate, recall that $\alp_{\mathfrak f} = J_{\mathfrak f} + \nB$ and use that $u^{-\f{\de_a}2}\log^{K_{J_{\mathfrak f}}} (\tfrac u2) \ls 1$ to remove the logarithm in the error term in Proposition~\ref{prop:med.explicit.precise}.)

Since $\tau_0 = \f 12 \tau^{1-\de_a}$ (by \eqref{eq:tau-0.def}), the derivative of $\chi_{<\tau_0}(r)$ is supported in the region $\{ \f{\tau^{1-\de_a}}2 \leq r \leq \tau^{1-\de_a}\} \subset \calM_\med\cap \{r \leq \tau^{1-\de_a} \}$. Hence, estimating the commutator as in \eqref{eq:P0.chi.commute}, using the estimate \eqref{eq:asymp.proof.phi.expand.in.int}, and noting that $\tau = u$ in this region, we obtain
\begin{equation}\label{eq:near.sharp.te11}
\begin{split}
\sum_{s\leq \lfloor a_0' M_0 \rfloor - i} |(\brk{r} \urd)^s \widetilde{e}_{1,1}| \ls &\: A'_{\mathfrak f} \tau_0^{-2} \max\{\tau^{-\alp_{\mathfrak f}-i-\de_{\mathfrak f}+\nB} \brk{r}^{-\nB}, \tau^{-\alp_{\mathfrak f} -i-\f{\de_a}2} \} \\
\ls &\: A'_{\mathfrak f} \brk{r}^{-2} \max\{ \tau^{-\alp_{\mathfrak f}-i-\de_{\mathfrak f}} (\tfrac {\tau}{\tau_0})^{\nB}, \tau^{-\alp_{\mathfrak f} -i-\f{\de_a}2} \} \\
\ls &\: A'_{\mathfrak f} \brk{r}^{-2} \max\{ \tau^{-\alp_{\mathfrak f}-i-\de_{\mathfrak f}+\nB\de_a}, \tau^{-\alp_{\mathfrak f} -i-\f{\de_a}2} \} \\
\ls &\: A'_{\mathfrak f} \brk{r}^{-2} \max\{ \tau^{-\alp_{\mathfrak f}-i-\f{\de_{\mathfrak f}}2}, \tau^{-\alp_{\mathfrak f} -i-\f{\de_a}2} \},
\end{split}
\end{equation}
where we have used that $\brk{r}$ and $\tau_0$ are comparable on the support of $\chi'_{<\tau_0}(r)$ and that \eqref{eq:tau-0.def} holds.

On the other hand, since $\calP_0 1 = O^{M_c}(\brk{r}^{-2-\de_c})$, it follows that $\|\calP_0 1 \|_{\ell^1_r H^{M_c,-\f d2+\f{\de_c}2+2}} \ls 1$. Thus, by Corollary~\ref{cor:near-pointwise}, we have $\brk{r}^{\f{\de_c}2}|(\brk{r}\urd)^{\leq M_c -s_c-\lfloor \f d2 \rfloor +1}\calR_0 \calP_0 1|  \ls 1$. Using the $\brk{r}$ and $\tau_0$ are comparable on the support of $\chi'_{<\tau_0}(r)$, and using \eqref{eq:de_f.least} and \eqref{eq:tau-0.def}, we thus obtain
\begin{equation}\label{eq:near.sharp.te12}
\sum_{s\leq \lfloor a_0' M_0 \rfloor - i} |(\brk{r} \urd)^s \widetilde{e}_{1,2}| \ls A'_{\mathfrak f} \tau^{-\alp_{\mathfrak f} - i} \tau_0^{-\f{\de_c}2} \brk{r}^{-2} \log^{K_{J_{\mathfrak f}}} \tfrac \tau 2 \ls A'_{\mathfrak f} \tau^{-\alp_{\mathfrak f} - i - \de_{\mathfrak f}} \brk{r}^{-2}.
 \end{equation}

Therefore, combining \eqref{eq:near.sharp.te11} and \eqref{eq:near.sharp.te12} (and using $\de_a \leq \f{2\de_{\mathfrak f}}{d-1}$), $\widetilde{e}_1$ satisfies
\begin{equation}\label{eq:near.sharp.te1}
\sum_{s\leq \lfloor a_0' M_0 \rfloor - i} |(\brk{r} \urd)^s \widetilde{e}_{1}| \ls A'_{\mathfrak f} \tau^{-\alp_{\mathfrak f}-i-\f{\de_a}2} \brk{r}^{-2}.
\end{equation}

For the remaining terms $e_2,\ldots, e_5$, we use the estimates \eqref{eq:med.induction.e2}, \eqref{eq:med.induction.e3}, \eqref{eq:med.induction.e4}, \eqref{eq:med.induction.e51} and \eqref{eq:med.induction.e523} derived in the proof of Proposition~\ref{prop:near-med}. Let us note that the derivation of \eqref{eq:med.induction.e2}--\eqref{eq:med.induction.e4} and \eqref{eq:med.induction.e51}--\eqref{eq:med.induction.e523} relies only on \eqref{eq:near-med-ass-near} (but not \eqref{eq:near-med-ass-med}). In our setting, using the upper bound in Main Theorem~\ref{thm:upper}, we know that \eqref{eq:near-med-ass-near} and \eqref{eq:near-med-ass-med} hold for $\alp +\de_0 = \alp_{\mathfrak f}$. We can thus apply these estimates with $\alp = \alp_{\mathfrak f} - \de_0$ and $j = 2\nB -1$. Recalling that \eqref{eq:de_f.least} holds for $\de_{\mathfrak f}$, we thus obtain
\begin{equation}\label{eq:near.sharp.e2...e5}
\begin{split}
&\: \sum_{s\leq \lfloor a_0' M_0 \rfloor - i} \Big(  |(\brk{r}\urd)^s e_2| + \ldots + |(\brk{r}\urd)^s e_5| \Big)\\
\ls &\: \max\{A'_{\mathfrak f} \tau^{-\alp_{\mathfrak f}-i-\f 12} \brk{r}^{-\f 32}, A'_{\mathfrak f} \tau^{-\alp_{\mathfrak f}-i-\de_{\mathfrak f}} \brk{r}^{-2-\de_0},  A'_{\mathfrak f} \tau^{-\alp_{\mathfrak f}-i-\f{3\de_{\mathfrak f}}2} \brk{r}^{-2} \}.
\end{split}
\end{equation}
Using \eqref{eq:near.how.to.integrate}, \eqref{eq:near.how.to.integrate.2} with $\gamma = -\f d2$, we have
\begin{equation}
\| \chi_{<\tau_0} \brk{r}^{-\f 32} \|_{\ell^{1}_{r} \calH^{0,-\f d2 +2}} \ls \tau_0^{\f 12},\quad \| \chi_{<\tau_0} \brk{r}^{-2} \|_{\ell^{1}_{r} \calH^{0,-\f d2 +2}} \ls \log \tau_0,\quad \| \chi_{<\tau_0} \brk{r}^{-2-\de_0} \|_{\ell^{1}_{r} \calH^{0,-\f d2 +2}} \ls 1.
\end{equation}
Therefore, using \eqref{eq:near.sharp.te1} and \eqref{eq:near.sharp.e2...e5}, we obtain
\begin{equation}\label{eq:P0.cutoff.lower.est}
\begin{split}
&\: \sum_{s\leq \lfloor a_0' M_0 \rfloor - i} \Big( \| (\brk{r}\urd)^s \widetilde{e}_1\|_{\ell^{1}_{r} \calH^{0,-\f d2 +2}} + \sum_{i=2}^5 \| (\brk{r}\urd)^s e_i\|_{\ell^{1}_{r} \calH^{0,-\f d2 +2}} \Big) \\
\ls &\:  \max\{ A'_{\mathfrak f} \tau^{-\alp_{\mathfrak f}-i-\f 12} \tau_0^{\f 12}, A'_{\mathfrak f} \tau^{-\alp_{\mathfrak f}-i-\de_{\mathfrak f}}, A'_{\mathfrak f} \tau^{-\alp_{\mathfrak f}-i-\f{\de_a}2} \log \tau_0 \} \ls A'_{\mathfrak f} \tau^{-\alp_{\mathfrak f}-i-\f{\de_a}4},
\end{split}
\end{equation}
where in the last inequality, we used $\tau^{-\f 12} \tau_0^{\f 12} \ls \tau^{-\de_a}$ (by \eqref{eq:tau-0.def}), $\de_a \leq \de_{\mathfrak f}$ (by \eqref{eq:tau-0.def}) and $\tau^{-\f{\de_a}2} \log \tau_0 \ls \tau^{-\f{\de_a}4}$. Finally, we apply Corollary~\ref{cor:near-pointwise} using the equation \eqref{eq:P0.cutoff.lower} and the estimate \eqref{eq:P0.cutoff.lower.est} to obtain
$$\sum_{s+i \leq \lfloor a'_0 M_0 \rfloor} \Bigg| (\brk{r}\urd)^s (\tau\bfT)^i \Big( \phi(\tau,r,\theta) - \sum_{k,k'=0}^{K_{J_{\mathfrak f}}} c_{k,k'} \mathfrak L_k \tau^{-J_{\mathfrak f} -\nB}(\log^{k'} \tfrac \tau 2)\psi( x) \Big) \Bigg| \ls A'_{\mathfrak f} \tau^{-J_{\mathfrak f}-\nB-\f{\de_a}4},$$
where as discussed in Section~\ref{sec:notation.in.lower}, we take $a'_0$ smaller if neceessary. \qedhere

\end{proof}

\subsection{Putting everything together}\label{sec:lower.everything}

\begin{proof}[Proof of Main Theorem~\ref{thm:lower}]
The sharp asymptotics \eqref{eq:asymptotics.wave.zone} in the $r \geq R_{\far}$ region is given by Proposition~\ref{prop:wave.med.precise}. The sharp asymptotics \eqref{eq:asymptotics} in the $r\leq \tau^{1-\de_a}$ region is given by Proposition~\ref{prop:sharp.near}. \qedhere
%
\end{proof}

\subsection{Sharp asymptotics on spherically symmetric spacetimes}\label{sec:lower.SS}

\begin{proof}[Proof of Main Theorem~\ref{thm:lower-sphsymm}]
We now indicate the necessary modifications of the proof of Main Theorem~\ref{thm:lower} in order to prove Main Theorem~\ref{thm:lower-sphsymm}. First, we note that since \eqref{eq:asymptotics.wave.zone} is already proven in Main Theorem~\ref{thm:lower}, it suffices to establish \eqref{eq:asymptotics-sphsymm}.

In order to prove \eqref{eq:asymptotics-sphsymm}, we need two modifications of the proof of Proposition~\ref{prop:sharp.near}. (Again, we only consider only the case where $\rd_{(\mathrm{t})} \calM = \0$ since the other case is completely analogous.)
\begin{enumerate}
\item Instead of \eqref{eq:P0.cutoff.lower}, we consider the following equation:
\begin{equation}
\begin{split}
&\: \calP_{0} \Big(\chi_{<\tau_0} (r) \bfT^{i} \phi - \sum_{k,k'=0}^{K_{J_{\mathfrak f}}} c^{(\ell)}_{k,k'} \chi_{<\tau_0}(r) \mathfrak L_k \bfT^{i} (\tau^{-\alp_{\mathfrak f}}\log^{k'} \tfrac \tau 2) (r^\ell Y_{(\ell)} -  \mathcal R_0 \mathcal P_0 (r^\ell Y_{(\ell)})) \Big) \\
= &\: - [\chi_{<\tau_0}(r), \calP_{0}] \Big( \bfT^{i} \phi - \sum_{k,k'=0}^{K_{J_{\mathfrak f}}} c_{k,k'} \mathfrak L_k \bfT^{i} (\tau^{-\alp_{\mathfrak f}}\log^{k'} \tfrac \tau 2) (r^\ell Y_{(\ell)} -  \mathcal R_0 \mathcal P_0 (r^\ell Y_{(\ell)})) \Big) \\
&\: + \chi_{<\tau_0}(r) (\calP_{0} - \calP) (\bfT^{i} \phi) + \chi_{<\tau_0}(r) \bfT^{i} f + \chi_{<\tau_0}(r) \bfT^{i} \calN(\phi) + \chi_{<\tau_0}(r) [\calP_{0}, \bfT^i] \phi.
\end{split}
\end{equation}
Notice that this holds because $\calP_{0} (r^\ell Y_{(\ell)} -  \mathcal R_0 \mathcal P_0 (r^\ell Y_{(\ell)}))$. One key point here is that
$$\mathcal P_0 (r^\ell Y_{(\ell)}) = O^{M_c}(\brk{r}^{\ell-2-\de_c})$$
because $\Delta_{\bfe} (r^\ell Y_{(\ell)}) = 0$ on Euclidean space.
\item The second modification is that when applying Corollary~\ref{cor:near-pointwise}, we need to use the improvement for higher $\ell$ modes (i.e., the second estimate in the corollary) so that we can invert $\calP_0$ for terms with less $\brk{r}$ decay or with suitable $\brk{r}$ growth. \qedhere
\end{enumerate}
\end{proof}

\section{Analysis in the exterior region: weighted energy estimates}\label{sex:ext}

In this final section, we prove the exterior stability theorem (Theorem~\ref{thm:ext-stab}). By the discussion in Section~\ref{sec:thm.Cauchy}, this then implies Main Theorem~\ref{thm:Cauchy}, where the initial data are posed on an asymptotically flat (as opposed to asymptotically null) hypersurface.

As a preliminary, we give the precise definition of admissible decay exponent in \textbf{Section~\ref{sec:precise.def.ade.ext}}.

We then turn to the proof of Theorem~\ref{thm:ext-stab}. The main ingredient for the exterior stability theorem is an energy estimate in the exterior region. This will be proven in \textbf{Section~\ref{sec:EE}}; see Theorem~\ref{prop:en-ext} in . We then use the energy estimate to carry out a bootstrap argument in \textbf{Section~\ref{sec:bootstrap}}. Finally, we conclude the proof of Theorem~\ref{thm:ext-stab} in \textbf{Section~\ref{sec:theorem.ext.stab}}.

\subsection{Precise definition of admissible decay exponent}\label{sec:precise.def.ade.ext}

In $\calM_{\ext}$, we define a notion of admissible decay exponent for the nonlinear term. This is similar to Definition~\ref{def:alp-N}, except that we slightly modify the definition so that it is easier to be applied to derive $L^2$ (as opposed to $L^\infty$) estimates. In Definition~\ref{def:ext-alp-N} below, we use the notation $O_{\bfGmm}$ for the nonlinearity as introduced in Section~\ref{sec:nonlinearity.assumptions}, suitably modified to that we use \eqref{eq:O-Gmm-ext-scalar}--\eqref{eq:O-Gmm-ext} instead of \eqref{eq:O-Gmm-far-scalar}--\eqref{eq:O-Gmm-far}.

\begin{definition} [Admissible decay exponent] \label{def:ext-alp-N}
In $\calM_{\ext}$, we say that $\alp_{\calN}'$ is an \emph{admissible decay exponent} for $\calN$ if there exists a nonnegative nondecreasing function $A_{\calN} = A_{\calN}(A)$ with $A_{\calN} = O(A^{2})$ as $A \to 0$ such that, for every $\dlt_{0} > 0$, nonnegative integer $M \leq M_{c}$, and $A \geq 0$, the following holds:
\begin{enumerate}
\item (Estimates in $\calM_{\ext}$)
For $\bfh_{\calN}(\phi; \psi)$, $\bfC_{\calN}(\phi; \psi)$ and $W_{\calN}(\phi; \psi)$ defined by \eqref{eq:calN.expansion} and for $\phi$ and $\psi$ satisfying
\begin{equation*}
	\phi, \psi = O_{\bfGmm}^{M}(A \mbrk{u}^{-\alp+\nu_{\Box}} r^{-\nu_{\Box}})  \quad \hbox{ in } \calM_{\ext}
\end{equation*}
with $\alp \geq \alp_{\calN}' + \dlt_{0}$, we have, in $\calM_{\ext}$,
\begin{equation} \label{eq:alp-N-ext}
\begin{aligned}
	\bfh_{\calN}(\phi; \psi) &= O_{\bfGmm}^{M-1}(\tfrac{1}{A} A_{\calN}(A) \mbrk{u}^{-\dlt_{0}+\nu_{\Box}} r^{-\nu_{\Box}}), \\
	\bfC_{\calN}(\phi; \psi) &= O_{\bfGmm}^{M-1}(\tfrac{1}{A} A_{\calN}(A) \mbrk{u}^{-1-\dlt_{0}+\nu_{\Box}} r^{-\nu_{\Box}}), \\
	W_{\calN}(\phi; \psi) &= O_{\bfGmm}^{M-1}(\tfrac{1}{A} A_{\calN}(A) \mbrk{u}^{-2-\dlt_{0}+\nu_{\Box}} r^{-\nu_{\Box}}).
\end{aligned}
\end{equation}
Furthermore, if $d = 3$, then for the same $\phi$ and $\psi$ we also have, in $\calM_{\ext}$,
\begin{equation} \label{eq:alp-N-ext-3d}
\begin{aligned}
	\bfh_{\calN}^{uu}(\phi; \psi) &= O_{\bfGmm}^{M-1}(\tfrac{1}{A} A_{\calN}(A) \mbrk{u}^{-\dlt_{0}+2} r^{-2}), \\
	\bfC_{\calN}^{u}(\phi; \psi) &= O_{\bfGmm}^{M-1}(\tfrac{1}{A} A_{\calN}(A) \mbrk{u}^{1-\dlt_{0}} r^{-2}), \\
	W_{\calN}(\phi; \psi) &= O_{\bfGmm}^{M-1}(\tfrac{1}{A} A_{\calN}(A) \mbrk{u}^{-\dlt_{0}} r^{-2}).
\end{aligned}
\end{equation}

\item (Additional estimates in $\calM_{\ext}$ related to higher radiation fields) For every $J \in \bbZ_{\geq 0}$, $\vec{K} \in (\bbZ_{\geq 0})^{J}$ and $\rPhi_{j, k}$ for $0 \leq j \leq J$, $0 \leq k \leq K_{j}$ satisfying
\begin{equation*}
	\rPhi_{j, k} =  O_{\bfGmm}^{M}(A \mbrk{u}^{j-\alp+\nu_{\Box}})
\end{equation*}
with $\alp \geq \alp_{\calN}' + \dlt_{0}$, we have
\begin{equation} 
	\calN(r^{-\nu_{\Box}} \Phi_{<J}) = \sum_{j=0}^{\infty} \sum_{k=0}^{\infty} \rN^{<J}_{j, k}(u, \tht) r^{-j-1} \log^{k}(\tfrac{r}{\mbrk{u}}) + E_{J; \nonlinear, \rem}
\end{equation}
satisfying the estimates
\begin{align}
	\rN^{<J}_{j, k} &= O_{\bfGmm}^{M-2}(C_{j, k} A_{\calN}(A) \mbrk{u}^{j-1-\alp-\dlt_{0}+\nu_{\Box}}), \label{eq:ext-alp-N-rNjk<J} \\
	E_{J; \nonlinear, \rem}
	&= O_{\bfGmm}^{M-2}(A_{\calN}(A) r^{-J_{c}-\eta_{c}} \mbrk{u}^{J_{c}-2+\eta_{c}-\alp-\dlt_{0}+\nu_{\Box}}). \label{eq:ext-alp-N-rem}
\end{align}

\item (Estimates in $\calM_{\ext}$ used for $L^2$ estimates)
For $\phi$ satisfying
\begin{equation}\label{eq:ext-nonlinear.phi.assumption}
	\phi = O_{\bfGmm}^{\lfloor \f M2\rfloor + 2}(A \mbrk{u}^{-\alp+\nu_{\Box}} r^{-\nu_{\Box}})  \quad \hbox{ in } \calM_{\ext}
\end{equation}
with $\alp \geq \alp_{\calN}' + \dlt_{0}$, we have, in $\calM_{\ext}$, for any $|I|\leq M$,
\begin{equation}\label{eq:ext-alpN-def-est}
\begin{split}
&\:\Big| \bfGmm^{I} \calN - \quasi^{\mu \nu} \rd_\mu \rd_\nu \bfGmm^{I} \phi \Big| \\
\ls &\:\tfrac{1}{A} A_{\calN}(A) \Big(\mbrk{u}^{-\de_0} r^{-2} |\bfGmm^{\leq |I|} \phi| + \mbrk{u}^{1-\de_0} r^{-2} |\rd_u \bfGmm^{\leq |I|} \phi| +\mbrk{u}^{-\de_0} r^{-1} (|\rd_r \bfGmm^{\leq |I|} \phi| + r^{-1} |\rsnb \bfGmm^{\leq |I|} \phi|)   \Big).
\end{split}
\end{equation}
\end{enumerate}

\end{definition}

\begin{remark}\label{rmk:ext-alpN}
Similar to Proposition~\ref{prop:alp-N}, when $\calN$ is a monomial, $\alp_{\calN}' = \max\set{0, \frac{2-o-\bt}{n-1}}$ is admissible, where $o$, $n$, $\bt$ are the total number of derivatives, total number of factors and the decay rate of the coefficients, respectively. See Proposition~\ref{prop:alp-N} for details.

In particular, according to assumption \ref{hyp:ext-N}, $\alp_{\calN}' = 2$ is always admissible. This fact can also be checked directly. Indeed, assuming \eqref{eq:ext-nonlinear.phi.assumption} holds, we use \ref{hyp:ext-N} to obtain
\begin{equation}\label{eq:ext-check-alpN.1}
\begin{split}
\Big| \bfGmm^{I} \calN - \quasi^{\mu \nu} \rd_\mu \rd_\nu \bfGmm^{I} \phi \Big|
\ls A \mbrk{u}^{-\alp + \nB} r^{-\nB} |\bfGmm^{\leq |I|} \phi| + A \mbrk{u}^{-\alp + \nB} r^{-\nB} |\rd \bfGmm^{\leq |I|} \phi|.
\end{split}
\end{equation}
When $d=3$, using also the null condition \ref{hyp:nonlin-null}, we obtain
\begin{equation}\label{eq:ext-check-alpN.2}
\begin{split}
&\: \Big| \bfGmm^I \Big(  \calN(p, \phi,\ud \phi \Big) - \quasi^{\mu \nu} (p, \phi, \ud \phi) \rd_{\mu} \rd_{\nu} \bfGmm^{I} \phi \Big| \\
\ls &\:  D \mbrk{u}^{-\alp + 1} r^{-2} |\bfGmm^{\leq |I|} \phi| + D \mbrk{u}^{-\alp + 2} r^{-2} |\rd \bfGmm^{\leq |I|} \phi| + D \mbrk{u}^{-\alp + \nB} r^{-\nB} (|\rd_r \bfGmm^{\leq |I|} \phi| + r^{-1} |\rsnb \bfGmm^{\leq |I|} \phi|).
\end{split}
\end{equation}
Using \eqref{eq:ext-check-alpN.1} and \eqref{eq:ext-check-alpN.2}, it is easy to check that the bound \eqref{eq:ext-alpN-def-est} holds when $\alp \geq 2+ \de_0$.
\end{remark}

\subsection{Weighted energy estimate in the exterior zone}\label{sec:EE}

For the remainder of the section, {\bf we work under the assumptions of Theorem~\ref{thm:ext-stab}}. For notational convenience, however, we will restrict to the case $U_{0} := U_{h}(t_{0}, R_{0}) \leq -1$ (instead of $U_{h}(t_{0}, R_{0}) \leq \frac{3}{2}$ allowed in Theorem~\ref{thm:ext-stab}). The general case clearly follows after a simple translation, i.e., consider $u_h - \frac{5}{2}$ instead of $u_h$.

The goal of this subsection is to prove the following weighted energy estimate.
\begin{proposition}[Main weighted energy estimate in the exterior zone] \label{prop:en-ext}
Given $1 < p < 2$, $\alp > \nu_{\Box} + \frac{p-1}{2}$, $\dlt_{e} > 0$, $\dlt_{h} > 0$ and $A_{h}$, there exists $\eps_{e} = \eps_{e}(p, \alp, \dlt_{e}, \dlt_{h}) > 0$ and $R_{e} = R_{e}(p, \alp, \dlt_{e}, \dlt_{h}, A_{h}) \geq 1$ such that the following holds. Define $u_{h}$ as in \eqref{eq:ext-uh} with $(\dlt_{h}, A_{h})$ satisfying
\begin{equation*}
	A_{h} \geq \eps_{e}, \quad \dlt_{h} \leq \dlt_{e},
\end{equation*}
Given $R_{0} > 0$, let $U_{0}$ be the value of $u_{h}$ when $(x^{0}, r) = (t_{0}, R_{0})$ (i.e., $U_{0} = U_{h}(t_{0}, R_{0})$). Assume that
\begin{equation*}
	R_{0} \geq R_{e}, \quad U_{0} \leq -1.
\end{equation*}
Moreover, given $0 < t_{0} < T$, assume that the coefficients of $\calP$ satisfy the following bounds in $\set{t_{0} \leq x^{0} \leq T, \, u_{h} \leq U_{0}}$ with all implicit constants equal to $1$:
\begin{equation} \label{eq:en-ext-coeff}
\begin{gathered}
	(\bfg^{-1})^{uu} = O_{\bfGmm}^{1}(\eps_{e} r^{-\dlt_{e}} (\tfrac{r}{\mbrk{u}})^{-1 - (p-1)} ), \\
	(\bfg^{-1})^{ur} = -1 + O_{\bfGmm}^{1}(\eps_{e} r^{-\dlt_{e}} (\tfrac{r}{\mbrk{u}})^{-\frac{p-1}{2}}), \quad
	r (\bfg^{-1})^{uA} = O_{\bfGmm}^{1}(\eps_{e} r^{-\dlt_{e}} (\tfrac{r}{\mbrk{u}})^{-\frac{1}{2}-\frac{p-1}{2}}), \\
	(\bfg^{-1})^{rr} = 1 + O_{\bfGmm}^{1}(\eps_{e} \mbrk{u}^{-\dlt_{e}}), \quad
	r (\bfg^{-1})^{rA} = O_{\bfGmm}^{1}(\eps_{e} \mbrk{u}^{-\dlt_{e}}), \\
	r^{2} (\bfg^{-1})^{AB} = \rsgmm^{AB} + O_{\bfGmm}^{1}(\eps_{e} \mbrk{u}^{-\dlt_{e}}), \quad
	\rd_{r} \log \sqrt{- \det \bfg^{-1}} = - \frac{d-1}{r} + O_{\bfGmm}^{0}(\eps_{e} r^{-\dlt_{e}-1} (\tfrac{r}{\mbrk{u}})^{-\frac{p-1}{2}} ), \\
	\bfB^{u} = O_{\bfGmm}^{0}(\eps_{e} r^{-\dlt_{e}-1} (\tfrac{r}{\mbrk{u}})^{-\frac{p-1}{2}} ), \quad
	\bfB^{r} = O_{\bfGmm}^{0}(\eps_{e} r^{-\dlt_{e}} \mbrk{u}^{-1}), \quad
	r \bfB^{A} = O_{\bfGmm}^{0}(\eps_{e} \mbrk{u}^{-\dlt_{e}-1} (\tfrac{r}{\mbrk{u}})^{-\frac{1}{2}}), \\
	V = O_{\bfGmm}^{0}(\eps_{e} r^{-1-\dlt_{e}} \mbrk{u}^{-1} (\tfrac{r}{\mbrk{u}})^{-\frac{p-1}{2}}).
\end{gathered}
\end{equation}
Then for any $\psi$ and $g$ such that $\calP \psi = g$, we have
\begin{align}
	&\sup_{t \in [t_{0}, T]} \int_{\set{x^{0} = t, \, u_{h} \leq U_{0}}} \mbrk{u}^{2 (\alp - \nu_{\Box})-1} \bigg[ (\mbrk{u} \rd_{u} \psi)^{2} + (\mbrk{u} \rd_{r} \psi)^{2} + \left( (\tfrac{r}{\mbrk{u}})^{-1} \psi \right)^{2} \notag \\
	&\phantom{\sup_{t \in [t_{0}, T]} \int_{\set{x^{0} = t, \, u_{h} \leq U_{0}}} \mbrk{u}^{2 (\alp - \nu_{\Box})-1} \bigg[}
	+ \left( (\tfrac{r}{\mbrk{u}})^{-\frac{1}{2}+\frac{p-1}{2}} r^{1-\nu_{\Box}} \rd_{r} (r^{\nu_{\Box}} \psi) \right)^{2} + \left( (\tfrac{r}{\mbrk{u}})^{-\frac{1}{2}+\frac{p-1}{2}}\abs{\rsnb \psi}\right)^{2} \bigg]  \ud \sgm \notag \\
	&+\sup_{R \in 2^{\bbZ_{\geq 0}} R_{0}} \iint_{\set{t_{0} \leq x^{0} \leq T, \, u_{h} \leq U_{0}, \, R \leq r \leq 2 R}} \frac{\mbrk{u}^{2 (\alp - \nu_{\Box})-1}}{r} \left[(\mbrk{u} \rd_{u} \psi)^{2} + (r \rd_{r} \psi)^{2} +  \psi^{2} \right]  \ud \mathrm{V} \notag \\
	&+\iint_{\set{t_{0} \leq x^{0} \leq T, \, u_{h} \leq U_{0}}} \frac{\mbrk{u}^{2 (\alp - \nu_{\Box})-1}}{r} \left[ \left( (\tfrac{r}{\mbrk{u}})^{\frac{p-1}{2}} r^{1-\nu_{\Box}} \rd_{r} (r^{\nu_{\Box}} \psi) \right)^{2} + \left( (\tfrac{r}{\mbrk{u}})^{-\frac{1}{2}+\frac{p-1}{2}}\abs{\rsnb \psi}\right)^{2} \right]\, \ud \mathrm{V} \notag \\
	&\aleq \int_{\set{x^{0} = t_{0}, \, r \geq R_{0}}} \mbrk{u}^{2 (\alp - \nu_{\Box})-1} \bigg[ (\mbrk{u} \rd_{u} \psi)^{2} + (\mbrk{u} \rd_{r} \psi)^{2} \notag \\
	&\phantom{\aleq \int_{\set{x^{0} = t_{0}, \, r \geq R_{0}}} \mbrk{u}^{2 (\alp - \nu_{\Box})-1} \bigg[}
	+ \left( (\tfrac{r}{\mbrk{u}})^{-\frac{1}{2}+\frac{p-1}{2}} r^{1-\nu_{\Box}} \rd_{r} (r^{\nu_{\Box}} \psi) \right)^{2} + \left( (\tfrac{r}{\mbrk{u}})^{-\frac{1}{2}+\frac{p-1}{2}}\abs{\rsnb \psi}\right)^{2}  \bigg] \ud \sgm \label{eq:en-ext}\\
	&\peq + \left( \sum_{R \in 2^{\bbZ_{\geq 0}} R_{0}} \left(  \iint_{\set{t_{0} \leq x^{0} \leq T, \, u_{h} \leq U_{0}, \, R \leq r \leq 2 R}} r \mbrk{u}^{2 (\alp - \nu_{\Box})+1} g^{2} \, \ud \mathrm{V} \right)^{\frac{1}{2}} \right)^{2}, \notag
\end{align}
where the implicit constant depends on $p$, $\alp$, $\dlt_{e}$, $\dlt_{h}$ and $A_{h}$, but is independent of $t_{0}$, $T$ and $R_{0}$.
\end{proposition}
\begin{remark}\label{rmk:ext.improved.coeff}
The assumptions \eqref{eq:en-ext-coeff} can be further weakened in a couple of ways. First, $r^{-\dlt_{e}}$ may be removed by assuming appropriate summability on $r$-dyadic regions. Second, if we drop the first term on the left-hand side of \eqref{eq:en-ext} (i.e., the flux term on level-$x^{0}$ hypersurfaces), then the assumption on $(\bfg^{-1})^{rr}$ and $r (\bfg^{-1})^{rA}$ may be weakened to
\begin{equation*}
	(\bfg^{-1})^{rr} = 1 + O_{\bfGmm}^{1}(\eps_{e} r^{-\dlt_{e}} \tfrac{r}{\mbrk{u}}), \quad
	r (\bfg^{-1})^{rA} = O_{\bfGmm}^{1}(\eps_{e} \mbrk{u}^{-\dlt_{e}} (\tfrac{r}{\mbrk{u}})^{\frac{1}{2}});
\end{equation*}
see Step~2 in the proof of Proposition~\ref{prop:en-ext} for the key computation.
Note, however, that this modification allows for level-$x^{0}$ hypersurfaces to be non-spacelike, which is the reason why we lose control of the flux term on level-$x^{0}$ hypersurfaces in \eqref{eq:en-ext-coeff}. We have chosen to avoid this modification to keep other parts of this paper simpler (for instance, the proof of Theorem~\ref{thm:ext-stab}).
\end{remark}

Here and in the remainder of this section, $\ud \mathrm{V}$ and $\ud \sgm$ are the standard volume and induced volume densities on the Minkowski spacetime, i.e., $\ud \mathrm{V} = \ud x^{0} \ldots \ud x^{d}$ and $\ud \sgm = \ud x^{1} \ldots \ud x^{d}$ on $\set{x^{0} = t}$ in rectangular coordinates. We also use the following notation concerning the coefficients of $\calP$ expressed in the $(u, r, \tht)$ coordinates. Namely, we write
\begin{equation*}
	\calP \psi = \Box_{\bfm} \psi + \bfh^{\alp \bt} \rd_{\alp} \rd_{\bt} \psi + b^{\alp} \rd_{\alp} \psi + V \psi,
\end{equation*}
where $\bfh = \bfg^{-1} - \bfm^{-1}$ as in \eqref{eq:bfh-def} and
\begin{equation*}
b^{\gmm} = \bfh^{\alp \bt} {}^{(\bfm)} \Gmm_{\alp \bt}^{\gmm} + \left( {}^{(\bfg)} \Gmm_{\alp \bt}^{\gmm} - {}^{(\bfm)} \Gmm^{\gmm}_{\alp \bt} \right) (\bfg^{-1})^{\alp \bt} + \bfB^{\gmm},
\end{equation*}
which may be easily verified. We also introduce $\Psi := r^{\nu_{\Box}} \psi$ and $G := r^{\nu_{\Box}} g$. In the $(u, r, \tht)$ coordinates, we write
\begin{equation*}
	(Q_{0} + \bfh^{\alp \bt} \rd_{\alp} \rd_{\bt} + c^{\alp} \rd_{\alp} + W) \Psi = F,
\end{equation*}
where
\begin{equation}\label{eq:ext-Q_0}
	Q_{0} = - 2 \rd_{u} \rd_{r} + \rd_{r}^{2} + \frac{1}{r^{2}} \rslap - \left(\frac{d-1}{2}\right)\left(\frac{d-3}{2}\right) \frac{1}{r^{2}},
\end{equation}
and $\bfC$, $W$ are defined as in \eqref{eq:bfC-def}--\eqref{eq:W-def} and $c^{\gmm} = \bfh^{\alp \bt} {}^{(\bfm)} \Gmm_{\alp \bt}^{\gmm} + \bfC^{\gmm}$.

Hypothesis \eqref{eq:en-ext-coeff} ensures that the following bounds hold for the coefficients:
\begin{lemma} \label{lem:en-ext-coeff-bondi}
Under the hypothesis of Proposition~\ref{prop:en-ext}, the following bounds hold in $\set{t_{0} \leq x^{0} \leq T, \, u_{h} \leq U_{0}}$:
\begin{equation} \label{eq:en-ext-coeff-bondi}
\begin{gathered}
	\bfh^{uu} = O_{\bfGmm}^{1}(\eps_{e} r^{-\dlt_{e}} (\tfrac{r}{\mbrk{u}})^{-1 - (p-1)}), \quad
	\bfh^{ur} = O_{\bfGmm}^{1}(\eps_{e} r^{-\dlt_{e}} (\tfrac{r}{\mbrk{u}})^{-\frac{p-1}{2}}), \quad
	r \bfh^{uA} = O_{\bfGmm}^{1}(\eps_{e} r^{-\dlt_{e}} (\tfrac{r}{\mbrk{u}})^{-\frac{1}{2}-\frac{p-1}{2}}), \\
	\bfh^{rr} = O_{\bfGmm}^{1}(\eps_{e} \mbrk{u}^{-\dlt_{e}}), \quad
	r \bfh^{rA} = O_{\bfGmm}^{1}(\eps_{e} \mbrk{u}^{-\dlt_{e}}), \quad
	r^{2} \bfh^{AB} = O_{\bfGmm}^{1}(\eps_{e} \mbrk{u}^{-\dlt_{e}}), \\
	b^{u} = O_{\bfGmm}^{0}(\eps_{e} r^{-\dlt_{e}-1} (\tfrac{r}{\mbrk{u}})^{-\frac{p-1}{2}} ), \quad
	b^{r} = O_{\bfGmm}^{0}(\eps_{e} r^{-\dlt_{e}} \mbrk{u}^{-1}), \quad
	r b^{A} = O_{\bfGmm}^{0}(\eps_{e} \mbrk{u}^{-\dlt_{e}-1} (\tfrac{r}{\mbrk{u}})^{-\frac{1}{2}}), \\
	c^{u} = O_{\bfGmm}^{0}(\eps_{e} r^{-\dlt_{e}-1} (\tfrac{r}{\mbrk{u}})^{-\frac{p-1}{2}} ), \quad
	c^{r} = O_{\bfGmm}^{0}(\eps_{e} r^{-\dlt_{e}} \mbrk{u}^{-1}), \quad
	r c^{A} = O_{\bfGmm}^{0}(\eps_{e} \mbrk{u}^{-\dlt_{e}-1} (\tfrac{r}{\mbrk{u}})^{-\frac{1}{2}}), \\
	V = O_{\bfGmm}^{0}(\eps_{e} r^{-1-\dlt_{e}} \mbrk{u}^{-1} (\tfrac{r}{\mbrk{u}})^{-\frac{p-1}{2}}), \quad
	W = O_{\bfGmm}^{0}(\eps_{e} r^{-1-\dlt_{e}} \mbrk{u}^{-1} (\tfrac{r}{\mbrk{u}})^{-\frac{p-1}{2}}).
\end{gathered}
\end{equation}
\end{lemma}
This lemma is proved with an argument similar to those in Section~\ref{subsec:conj-wave}. We note, in particular, that the assumption for $\rd_{r} \log \sqrt{- \det \bfg^{-1}}$ in \eqref{eq:en-ext-coeff} ensures that $b^{u} - \bfB^{u} = O_{\bfGmm}^{0}(\eps_{e} r^{-\dlt_{e}-1} (\tfrac{r}{\mbrk{u}})^{-\frac{p-1}{2}} )$. We omit the details.

For a vector field $\bfY$ and a scalar function $\upsilon$, we have the following associated multiplier identity, which we will use to treat the non-Minkowskian part of $\calP$:
\begin{lemma} \label{lem:ext-mult}
Consider coefficients $\td{\bfh}^{\alp \bt}$ defined in an open subset $O$ of $\bbR^{d+1}$ in the $(u, r, \tht)$ coordinates. For any $\psi : O \to \bbR$ and $\Psi : O \to \bbR$, we have
\begin{align*}
(\td{\bfh}^{\alp \bt} \rd_{\alp} \rd_{\bt} \psi)(\bfY^{\gmm} \rd_{\gmm} \psi + \upsilon \psi) r^{d-1}
& = \rd_{\alp}\left[ \td{\bfh}^{\alp \bt} \rd_{\bt} \psi(\bfY^{\gmm} \rd_{\gmm} \psi + \upsilon \psi) r^{d-1}
- \tfrac{1}{2} \td{\bfh}^{\bt \gmm} \bfY^{\alp} \rd_{\bt} \psi \rd_{\gmm} \psi r^{d-1}\right] \\
&\peq - r^{-(d-1)} \left[ \rd_{\alp} (r^{d-1} \td{\bfh}^{\alp \bt} \bfY^{\gmm})
- \tfrac{1}{2} \rd_{\alp} (r^{d-1} \td{\bfh}^{\bt \gmm} \bfY^{\alp}) \right] \rd_{\bt} \psi \rd_{\gmm} \psi r^{d-1} \\
&\peq - \td{\bfh}^{\alp \bt} \upsilon \rd_{\bt} \psi \rd_{\alp} \psi r^{d-1}
- r^{-(d-1)}\rd_{\alp} (r^{d-1} \td{\bfh}^{\alp \bt} \upsilon)  \rd_{\bt} \psi \psi r^{d-1}.
\end{align*}
\end{lemma}
The relevance of the weight $r^{d-1}$ in the identity for $\psi$ is, of course, that the Minkowskian volume form $\ud V$ takes the form $r^{d-1} \ud u \ud r \ud \tht$ in the $(u, r, \tht)$ coordinates. Moreover, $\Psi$ stands for the conjugated variable $\Psi = r^{\nu_{\Box}} \psi$, in which (the square root of) this weight is already included.

\begin{proof}
The proof is a standard algebraic computation. Indeed, for $\psi$, we compute
\begin{align*}
(\td{\bfh}^{\alp \bt} \rd_{\alp} \rd_{\bt} \psi)(\bfY^{\gmm} \rd_{\gmm} \psi + \upsilon \psi) r^{d-1}
& = \rd_{\alp}\left[ \td{\bfh}^{\alp \bt} \rd_{\bt} \psi(\bfY^{\gmm} \rd_{\gmm} \psi + \upsilon \psi) r^{d-1}\right] \\
&\peq - \frac{1}{2} \td{\bfh}^{\alp \bt} \bfY^{\gmm} \rd_{\gmm} (\rd_{\alp} \psi \rd_{\bt} \psi) r^{d-1}
- \rd_{\alp} (r^{d-1} \td{\bfh}^{\alp \bt} \bfY^{\gmm}) \rd_{\bt} \psi \rd_{\gmm} \psi \\
&\peq - \td{\bfh}^{\alp \bt} \upsilon \rd_{\bt} \psi \rd_{\alp} \psi r^{d-1}
- \rd_{\alp} (r^{d-1} \td{\bfh}^{\alp \bt} \upsilon)  \rd_{\bt} \psi \psi \\
& = \rd_{\alp}\left[ \td{\bfh}^{\alp \bt} \rd_{\bt} \psi(\bfY^{\gmm} \rd_{\gmm} \psi + \upsilon \psi) r^{d-1}\right]
- \rd_{\gmm} \left[\frac{1}{2} \td{\bfh}^{\alp \bt} \bfY^{\gmm} \rd_{\alp} \psi \rd_{\bt} \psi r^{d-1}\right] \\
&\peq + \frac{1}{2} \rd_{\gmm} (r^{d-1} \td{\bfh}^{\alp \bt} \bfY^{\gmm})  \rd_{\alp} \psi \rd_{\bt} \psi
- \rd_{\alp} (r^{d-1} \td{\bfh}^{\alp \bt} \bfY^{\gmm}) \rd_{\bt} \psi \rd_{\gmm} \psi \\
&\peq - \td{\bfh}^{\alp \bt} \upsilon \rd_{\bt} \psi \rd_{\alp} \psi r^{d-1}
- \rd_{\alp} (r^{d-1} \td{\bfh}^{\alp \bt} \upsilon)  \rd_{\bt} \psi \psi. \qedhere
\end{align*}
\end{proof}

We now state vector field multiplier estimates for $\Box_{\bfm}$ that are needed in our proof. We start with the basic energy identity.
\begin{lemma} [Energy identity] \label{lem:ext-en}
Let $\bfT^{\gmm} \rd_{\gmm} := \rd_{u}$ in the $(u, r, \tht)$ coordinates. Then the following identity holds:
\begin{align*}
 \Box_{\bfm} \psi \bfT^{\gmm} \rd_{\gmm} \psi r^{d-1}
= \rd_{\alp} \left[(\bfm^{-1})^{\alp \bt} \rd_{\bt} \psi \rd_{u} \psi r^{d-1}
- \frac{1}{2} (\bfm^{-1})^{\bt \gmm} \bfT^{\alp} \rd_{\bt} \psi \rd_{\gmm} \psi r^{d-1} \right].
\end{align*}
Moreover, define $u_{h}$ as in \eqref{eq:ext-uh} and assume that $U_{\ast} \leq U_{h}(t_{0}, R_{\ast}) < 0$ with $t_{0} \geq 0$, $\gmm_{h}(R_{\ast}) \leq \frac{1}{2}$ and $\gmm_{h}'(R_{\ast}) \leq \frac{1}{4}$. For $t_{1}  > t_{0}$, we have
\begin{align*}
&\int_{\set{x^{0} = t_{1}, \, u_{h} \leq U_{\ast}}} \left[(\rd_{u} \psi)^{2} + (\rd_{r} \psi)^{2} + r^{-2} \abs{\rsnb \psi}^{2} \right] \, \ud \sgm \\
&+ \int_{\set{t_{0} \leq x^{0} \leq t_{1}, \, u_{h} = U_{\ast}}} \left[A_{h} r^{-1-\dlt_{h}} (\rd_{u} \psi)^{2} + (\rd_{r} \psi)^{2} + r^{-2} \abs{\rsnb \psi}^{2} \right] r^{d-1} \, \ud r \ud \rssgm \\
&\aleq \int_{\set{x^{0} = t_{0}, \, u_{h} \leq U_{\ast}}} \left[(\rd_{u} \psi)^{2} + (\rd_{r} \psi)^{2} + r^{-2} \abs{\rsnb \psi}^{2} \right] \, \ud \sgm
	+ \abs*{\iint_{\set{t_{0} \leq x^{0} \leq t_{1}, \, u_{h} \leq U_{\ast}}} \Box_{\bfm} \psi \bfT^{\mu} \rd_{\mu} \psi \, \ud \mathrm{V}}
\end{align*}
\end{lemma}

\begin{proof}
The first identity is the standard energy identity for $\Box_{\bfm}$ and $\bfT$. The second statement follows by applying the divergence theorem (or Stokes's theorem) to the first identity, using the facts that $\gmm_{h}(r) \leq \gmm_{h}(R_{\ast}) \leq \frac{1}{2}$ and $A_{h} r^{-1-\dlt_{h}} = \gmm_{h}'(r) \leq \gmm_{h}'(R_{\ast}) \leq \frac{1}{4}$ in $\set{t_{0} \leq x^{0} \leq t_{1}, \, u_{h} \leq U_{\ast}}$.
\end{proof}
Next, we recall vector field multipliers for the integrated local energy decay (or Morawetz estimate). 
\begin{lemma} [Multiplier for integrated local energy decay]\label{lem:ext-iled}
There exist a family $\set{(\bfX_{R}, \xi_{R})}_{R \in 2^{\bbZ_{\geq 0}}}$ of vector field-scalar function pairs such that the following holds. Define $u_{h}$ as in \eqref{eq:ext-uh} and assume that $U_{\ast} \leq U_{h}(t_{0}, R_{\ast}) < 0$ with $t_{0} \geq 0$, $\gmm_{h}(R_{\ast}) \leq \frac{1}{2}$ and $\gmm_{h}'(R_{\ast}) \leq \frac{1}{4}$. For any $t_{1} > t_{0}$, we have
\begin{align*}
	&\iint_{\set{t_{0} \leq x^{0} \leq t_{1}, u_{h} \leq U_{\ast}, \, r \leq 2}} \left[(\rd_{u} \psi)^{2} + (\rd_{r} \psi)^{2} + r^{-2} \abs{\rsnb \psi}^{2} + r^{-2} \psi^{2} \right] \, \ud \mathrm{V} \\
	&\peq + \sup_{R \in 2^{\bbZ_{\geq 0}}} \iint_{\set{t_{0} \leq x^{0} \leq t_{1}, \, u_{h} \leq U_{\ast}, \, R \leq r \leq 2R}} r^{-1} \left[(\rd_{u} \psi)^{2} + (\rd_{r} \psi)^{2} + r^{-2} \abs{\rsnb \psi}^{2} + r^{-2} \psi^{2} \right] \, \ud \mathrm{V} \\
	&\aleq \int_{\set{x^{0} = t_{0}, \, u_{h} \leq U_{\ast}}} \left[(\rd_{u} \psi)^{2} + (\rd_{r} \psi)^{2} + r^{-2} \abs{\rsnb \psi}^{2} \right] \, \ud \sgm \\
	&\peq + \abs*{\iint_{\set{t_{0} \leq x^{0} \leq t_{1}, \, u_{h} \leq U_{\ast}}} \Box_{\bfm} \psi \bfT^{\mu} \rd_{\mu} \psi \, \ud \mathrm{V}}
	+ \sup_{R \in 2^{\bbZ_{\geq 0}}} \abs*{\iint_{\set{t_{0} \leq x^{0} \leq t_{1}, \, u_{h} \leq U_{\ast}}} \Box_{\bfm} \psi (\bfX_{R}^{\mu} \rd_{\mu} \psi + \xi_{R} \psi) \, \ud \mathrm{V}}
\end{align*}
where, in the $(x^{0}, x^{1}, \ldots x^{d})$ coordinates,
\begin{align*}
	\bfX_{R}^{\gmm} \rd_{\gmm} \psi + \xi_{R} \psi
	&= \sum_{j=1}^{d} \left( \rd_{x^{j}} (x^{j} \bt_{R}(\abs{x}) \psi) + \bt_{R}(\abs{x}) x^{j} \rd_{x^{j}} \psi \right) + b_{R}(\abs{x}) \\
	&= 2 \sum_{j=1}^{d} \bt_{R}(\abs{x}) x^{j} \rd_{x^{j}} \psi  + \left(\sum_{j=1}^{d} \rd_{x^{j}} (x^{j} \bt_{R}(\abs{x})) + b_{R}(\abs{x})\right) \psi
\end{align*}
with
\begin{gather*}
	\abs*{\frac{\ud^{k}}{\ud s^{k}}\bt_{R}(s)} \aeq (1+s)^{-1-k} \hbox{ for } s > 0, \, k = 0, 1, \\
	\abs*{\frac{\ud^{k}}{\ud s^{k}} b_{R}(s)} \aleq (1+s)^{-1-k} \hbox{ for } s > 0, \, k = 0, 1.
\end{gather*}
\end{lemma}
\begin{proof}
The multiplier $\bfX_{R}^{\gmm} \rd_{\gmm} + \xi_{R}$ is exactly $i (Q + \psi)$ in \cite{MetTat}. More precisely, using the notation from Section~5 in that paper, $Q$ is as in \cite[Lemma~4]{MetTat} with a positive slowly varying envelope $\alp_{m}$ satisfying $\sum_{m \geq 0 } \alp_{m} = 1$ and $\alp_{\log_{2} R} \aeq 1$. Note that $\bt_{R}(s) = \dlt_{0} \phi(\dlt_{0} s)$ in the notation of \cite[Lemma~4]{MetTat}. Moreover, $b_{R}(s) = i \psi(s)$ with the same choice of $\alp_{m}$ (while the improved derivative bound is not stated, we may easily ensure this property). Finally, the boundary terms that arise from integration by parts on level-$u_{h}$ and $x^{0}$ hypersurfaces are bounded using the energy identity (Lemma~\ref{lem:ext-en}). \qedhere
\end{proof}

We need the following multiplier identity, which is a slight variant of the $r^{p}$-energy of Dafermos--Rodnianski \cite{DRNM}. In particular, we include a $u$-dependent weight $\varpi(u)$ to generate the crucial term $- \varpi'(u) \wp(r) (\rd_{r} \Psi)^{2}$ in the first identity below, which is a variant of the ghost weight method of Alinhac \cite{sA2001}.
\begin{lemma}[Weighted $r^{p}$-energy identity] \label{lem:ext-rp}
In the $(u, r, \tht)$ coordinates, the following identities hold:
\begin{align*}
&r^{\nu_{\Box}} \Box_{\bfm} \psi (- \varpi(u) \wp(r) \rd_{r} \Psi) \\
&= \rd_{u} \left[ \varpi(u) \wp(r) (\rd_{r} \Psi)^{2}   \right] + \rd_{r} \left[ \frac{1}{2} \varpi(u) \wp(r) \left(- (\rd_{r} \Psi)^{2} + r^{-2} \abs{\rsnb \Psi}^{2} + \nu_{\Box}(\nu_{\Box} - 1) r^{-2} \Psi^{2} \right)\right] + \rsdiv(\ldots) \\
&\peq - \varpi'(u) \wp(r) (\rd_{r} \Psi)^{2}
+ \frac{1}{2} \varpi(u) \wp'(r) (\rd_{r} \Psi)^{2}
+ \frac{1}{2} \varpi(u) (r^{-2} \wp(r))' \left( \abs{\rsnb \Psi}^{2} + \nu_{\Box} (\nu_{\Box} - 1) \Psi^{2} \right),
\end{align*}
and
\begin{align*}
&r^{\nu_{\Box}} (\calP - \Box_{\bfm}) \psi (- \varpi(u) \wp(r) \rd_{r} \Psi) \\
&= \rd_{u} \left[ - \varpi(u) \wp(r) \bfh^{u \bt} \rd_{\bt} \Psi \rd_{r} \Psi \right]
+ \rd_{r} \left[ \frac{1}{2} \varpi(u) \wp(r) \left(- 2 \bfh^{r \bt} \rd_{\bt} \Psi \rd_{r} \Psi + \bfh^{\alp \bt} \rd_{\alp} \Psi \rd_{\bt} \Psi \right)\right] + \rsdiv(\ldots) \\
&\peq
+ \varpi'(u) \wp(r) \bfh^{u \bt} \rd_{\bt} \Psi  \rd_{r} \Psi
+ \varpi(u) \wp(r)  \left( (\rd_{\alp} \bfh^{\alp \gmm} - c^{\gmm}) \rd_{\gmm} \Psi - W \Psi \right) \rd_{r} \Psi \\
&\peq + \varpi(u) \wp'(r) \bfh^{r \bt} \rd_{\bt} \Psi  \rd_{r} \Psi
- \frac{1}{2} \varpi(u) \wp'(r) \bfh^{\alp \bt} \rd_{\alp} \Psi \rd_{\bt} \Psi
- \frac{1}{2} \varpi(u) \wp(r) \rd_{r} \bfh^{\alp \bt} \rd_{\alp} \Psi \rd_{\bt} \Psi .
\end{align*}
\end{lemma}
\begin{proof}
In the $(u, r, \tht)$ coordinates,
\begin{equation*}
	(Q_{0} + \bfh^{\alp \bt} \rd_{\alp} \rd_{\bt} + c^{\alp} \rd_{\alp} + W) \Psi = r^{\nu_{\Box}} \calP \psi.
\end{equation*}
We compute
\begin{align*}
Q_{0} \Psi \varpi(u) \wp(r) \rd_{r} \Psi
&= \left(- 2 \rd_{u} \rd_{r} \Psi + \rd_{r}^{2} \Psi + \frac{1}{r^{2}} \rslap \Psi - \nu_{\Box} (\nu_{\Box} - 1) \frac{1}{r^{2}} \Psi \right) \varpi(u) \wp(r) \rd_{r} \Psi \\
&= - \rd_{u} \left(\varpi(u) \wp(r) (\rd_{r} \Psi)^{2} \right) + \rsdiv(\ldots) \\
&\peq  - \rd_{r} \left(- \frac{1}{2} \varpi(u) \wp(r) (\rd_{r} \Psi)^{2} + \frac{1}{2} \varpi(u) r^{-2} \wp(r) \abs{\rsnb \Psi}^{2} + \frac{1}{2} \nu_{\Box}(\nu_{\Box} - 1) \varpi(u) r^{-2} \wp(r) \Psi^{2} \right) \\
&\peq + \varpi'(u) \wp(r) (\rd_{r} \Psi)^{2} - \frac{1}{2} \varpi(u) \wp'(r) (\rd_{r} \Psi)^{2}  +   \frac{1}{2} \varpi(u) (r^{-2} \wp(r))' \left( \abs{\rsnb \Psi}^{2} + \nu_{\Box} (\nu_{\Box} - 1) \Psi^{2} \right),
\end{align*}
which proves the first identity. To prove the second identity, we begin with the contribution of the term $\bfh^{\alp \bt} \rd_{\alp} \rd_{\bt} \Psi$:
\begin{align*}
\bfh^{\alp \bt} \rd_{\alp} \rd_{\bt} \Psi \varpi(u) \wp(r) \rd_{r} \Psi
&= \rd_{\alp} \left(\varpi(u) \wp(r)  \bfh^{\alp \bt} \rd_{\bt} \Psi \rd_{r} \Psi \right)
- \varpi(u) \wp(r)  \rd_{\alp} \bfh^{\alp \bt} \rd_{\bt} \Psi \rd_{r} \Psi \\
&\peq
- \varpi'(u) \wp(r) \bfh^{u \bt} \rd_{\bt} \Psi  \rd_{r} \Psi
- \varpi(u) \wp'(r) \bfh^{r \bt} \rd_{\bt} \Psi  \rd_{r} \Psi
- \frac{1}{2} \varpi(u) \wp(r) \bfh^{\alp \bt} \rd_{r} (\rd_{\alp} \Psi \rd_{\bt} \Psi) \\
&= \rd_{\alp} \left(\varpi(u) \wp(r) \bfh^{\alp \bt} \rd_{\bt} \Psi \rd_{r} \Psi \right)
- \rd_{r} \left(\frac{1}{2} \varpi(u) \wp(r) \bfh^{\alp \bt} \rd_{\alp} \Psi \rd_{\bt} \Psi \right) \\
&\peq - \varpi(u) \wp(r)  \rd_{\alp} \bfh^{\alp \bt} \rd_{\bt} \Psi \rd_{r} \Psi
- \varpi'(u) \wp(r) \bfh^{u \bt} \rd_{\bt} \Psi  \rd_{r} \Psi
- \varpi(u) \wp'(r) \bfh^{r \bt} \rd_{\bt} \Psi  \rd_{r} \Psi \\
&\peq + \frac{1}{2} \varpi(u) \wp(r) \bfh^{\alp \bt} \rd_{\alp} \Psi \rd_{\bt} \Psi
+ \frac{1}{2} \varpi(u) \wp(r) \rd_{r} \bfh^{\alp \bt} \rd_{\alp} \Psi \rd_{\bt} \Psi.
\end{align*}
The incorporation of first and zeroth order coefficients $c^{\alp} \rd_{\alp}$ and $W$ (respectively) is straightforward since it does not require any additional manipulations. \qedhere
\end{proof}

We are ready to prove Proposition~\ref{prop:en-ext}.
\begin{proof}[Proof of Proposition~\ref{prop:en-ext}]
\noindent {\it Step~0.}
We begin by making some reductions and preparations. By taking $R_{e}$ to be sufficiently large, we may ensure that $\gmm_{h}(R_{e}) \leq \frac{1}{2}$ and $\gmm_{h}'(R_{e}) \leq \frac{1}{4}$ (which ensures that Lemmas~\ref{lem:ext-en} and \ref{lem:ext-iled} are applicable). Choosing $R_{e}$ larger still, we may also ensure that
\begin{equation} \label{eq:en-ext-uh}
\bfg^{-1}(\ud u_{h}, \ud u_{h}) < 0, \quad \frac{3}{4} < \frac{\mbrk{u_{h}}}{\mbrk{u}} <\frac{5}{4}
\end{equation}
in $\set{t_{0} \leq x^{0} \leq T, \, u_{h} \leq U_{0}}$. We extend the coefficients of $\calP - \Box_{\bfm}$ to the spacetime slab $\set{t_{0} \leq x^{0} \leq T}$ so that $\calP = \Box_{\bfm}$ for $u_{h} \geq \frac{1}{2} U_{0}$ and the extended coefficients satisfy the bounds in Proposition~\ref{prop:en-ext} in $\set{t_{0} \leq x^{0} \leq T}$ (with a possibly larger implicit constant); indeed, this is achieved by applying the usual Sobolev extension and using the smooth cutoff $\chi_{> -U_{0}}(-u_{h})$. We may also modify the definition of $u_{h}$ in the region $\set{u_{h} \geq U_{0}}$ so that $u_{h}$ is smooth and \eqref{eq:en-ext-uh} holds everywhere in $\set{t_{0} \leq x^{0} \leq T}$. Indeed, this is achieved by smoothly transitioning to $u_{h} = u - c_{h} \mbrk{u} r^{-\dlt_{h}}$ in $\set{u \geq \frac{1}{8} U_{0}, \, r \geq 1}$ for $c_{h} > 0$ sufficiently small, then smoothly transitioning to $u_{h} = x^{0}$ near $r = 0$.
Throughout the proof, we will suppress the dependence of constants on $p$, $\alp$, $\dlt_{e}$, $\dlt_{h}$ and $A_{h}$. To keep the dyadic decompositions simpler, we also assume that $R_{0}, U_{0} \in 2^{\bbZ}$ (the general case follows by a minor modification of the argument below).

Let $\psi$ and $g$ be functions on $\set{t_{0} \leq x^{0} \leq T}$. For $U < 0$, define
\begin{align*}
\frkE_{U}^{2}[(\psi, \rd_{x^{0}} \psi)|_{\set{x^{0} = t}}] &:= \int_{\set{x^{0} = t, \, 2 U \leq u_{h} \leq U}} \bigg[ (\mbrk{U} \rd_{u} \psi)^{2} + (\mbrk{U} \rd_{r} \psi)^{2} + \left( (\tfrac{r}{\mbrk{U}})^{-1} \psi \right)^{2} \notag \\
	&\phantom{:= \int_{\set{x^{0} = t, \, 2U \leq u_{h} \leq U}} \bigg[}
	+ \left( (\tfrac{r}{\mbrk{U}})^{-\frac{1}{2}+\frac{p-1}{2}} r^{1-\nu_{\Box}} \rd_{r} (r^{\nu_{\Box}} \psi) \right)^{2} + \left( (\tfrac{r}{\mbrk{U}})^{-\frac{1}{2}+\frac{p-1}{2}}\abs{\rsnb \psi}\right)^{2} \bigg]  \ud \sgm, \\
	\frkI_{U}[g]^{2} &:= \left( \sum_{R \in 2^{\bbZ} : R \geq R_{0}} \left(  \iint_{\set{t_{0} \leq x^{0} \leq T, \, 2 U \leq  u_{h} \leq U, \, R \leq r \leq 2 R}} \mbrk{u}^{2} r g^{2} \, \ud \mathrm{V} \right)^{\frac{1}{2}} \right)^{2}, \\
\frkM_{U}^{2}[\psi] &:= \sup_{R \in 2^{\bbZ} : R \geq R_{0}} \iint_{\set{t_{0} \leq x^{0} \leq T, \, 2 U \leq u_{h} \leq U, \, R \leq r \leq 2 R}} \frac{1}{r} \frkm^{2}[\psi]  \ud \mathrm{V}
	+\iint_{\set{t_{0} \leq x^{0} \leq T, \, u_{h} \leq U}} \frac{1}{r} \frkm_{p}^{2}[\psi] \, \ud \mathrm{V}, \\
\end{align*}
where
\begin{align*}
\frkm^{2}[\psi] &:= (\mbrk{u} \rd_{u} \psi)^{2} + (r \rd_{r} \psi)^{2} +  \psi^{2}, \\
\frkm_{p}^{2}[\psi] &:= \left( (\tfrac{r}{\mbrk{u}})^{\frac{p-1}{2}} r^{1-\nu_{\Box}} \rd_{r} (r^{\nu_{\Box}} \psi) \right)^{2} + \left( (\tfrac{r}{\mbrk{u}})^{-\frac{1}{2}+\frac{p-1}{2}}\abs{\rsnb \psi}\right)^{2}.
\end{align*}

Next, given a family of vector field-scalar function pairs  $\set{(\bfY_{\bfalp}, \upsilon_{\bfalp})}_{\bfalp \in \calA}$ to be specified below in Step~1, we define
\begin{align*}
\frkH_{U}[\psi]^{2} &:= \sup_{\bfalp \in \calA} \left( \sum_{R \in 2^{\bbZ} : R \geq R_{0}} \left(  \iint_{\set{t_{0} \leq x^{0} \leq T, \, 2 U \leq  u_{h} \leq U, \, R \leq r \leq 2 R}} \frac{1}{r} \frkh_{\bfalp}^{2}[\psi] \, \ud \mathrm{V} \right)^{\frac{1}{2}} \right)^{2} \\
&\relphantom{:=}
+ \left( \sum_{R \in 2^{\bbZ} : R \geq R_{0}} \left(  \iint_{\set{t_{0} \leq x^{0} \leq T, \, 2 U \leq  u_{h} \leq U, \, R \leq r \leq 2 R}} \frac{1}{r} \frkh_{p}^{2}[\psi] \, \ud \mathrm{V} \right)^{\frac{1}{2}} \right)^{2},
\end{align*}
where
\begin{align*}
	\mbrk{u}^{-2} r^{-1} \frkh_{\bfalp}^{2}[\psi]
	&:= \sum_{\alp, \bt, \gmm} \bigg[ \abs*{r^{-(d-1)} \rd_{\alp} (r^{d-1} \bfh^{\alp \bt} \bfY^{\gmm}_{\bfalp}) \rd_{\bt} \psi \rd_{\gmm} \psi}
	+ \abs*{r^{-(d-1)} \rd_{\alp} (r^{d-1} \bfh^{\bt \gmm} \bfY^{\alp}_{\bfalp}) \rd_{\bt} \psi \rd_{\gmm} \psi} \bigg] \\
	&\relphantom{:=}
	+ \sum_{\alp, \gmm} \abs*{b^{\alp} \bfY^{\gmm}_{\bfalp} \rd_{\alp} \psi \rd_{\gmm} \psi}
	+ \sum_{\gmm} \abs*{V \bfY^{\gmm}_{\bfalp} \psi \rd_{\gmm} \psi} \\
	&\relphantom{:=}
	+ \sum_{\alp, \bt} \bigg[ \abs*{\bfh^{\alp \bt} \upsilon \rd_{\bt} \psi \rd_{\alp} \psi} + \abs*{r^{-(d-1)}\rd_{\alp} (r^{d-1} \bfh^{\alp \bt} \upsilon_{\bfalp})  \rd_{\bt} \psi \psi} \bigg]
	+ \sum_{\alp} \abs*{b^{\alp} \upsilon_{\bfalp} \rd_{\alp} \psi \psi}
	+ \abs*{V \upsilon_{\bfalp} \psi^{2}}, \\
	\mbrk{u}^{-2} r^{-1} \frkh_{p}^{2}[\psi]
	&:= \sum_{\bt} \mbrk{u}^{-1} (\tfrac{r}{\mbrk{u}})^{p} \abs*{\bfh^{u \bt} \rd_{\bt} \Psi \rd_{r} \Psi} r^{-(d-1)}
	+ \sum_{\alp, \gmm} (\tfrac{r}{\mbrk{u}})^{p} \abs*{\rd_{\alp} \bfh^{\alp \gmm} \rd_{\gmm} \Psi \rd_{r} \Psi} \\
	&\relphantom{:=}
	+ \sum_{\alp, \bt} \mbrk{u}^{-1} (\tfrac{r}{\mbrk{u}})^{p-1} \left( \abs*{\bfh^{\alp \bt}} + \abs*{r \rd_{r} \bfh^{\alp \bt}}\right)\abs*{\rd_{\alp} \Psi \rd_{\bt} \Psi} r^{-(d-1)} \\
	&\relphantom{:=}
	+ \sum_{\gmm} (\tfrac{r}{\mbrk{u}})^{p} \abs*{c^{\gmm} \rd_{\gmm} \Psi \rd_{r} \Psi}
	+ (\tfrac{r}{\mbrk{u}})^{p} \abs*{W \Psi \rd_{r} \Psi}.
\end{align*}

In what follows, we will make sure that, for some constant $A_{m} > 0$ (independent of $\bfalp$), all members of the family $\set{(\bfY_{\bfalp}, \upsilon_{\bfalp})}_{\bfalp \in \calA}$ satisfy (with respect to the $(x^{0}, x^{1}, \ldots, x^{d})$ coordinates)
\begin{equation} \label{eq:en-ext-mult-intr}
	\abs{\bfY_{\bfalp}^{\mu}} + \abs{\rd_{\alp} \bfY_{\bfalp}^{\mu}} \leq A_{m}, \quad \abs{\upsilon_{\bfalp}} \leq A_{m} \hbox{ in } \set{r \leq \tfrac{1}{2}},
\end{equation}
as well as the following properties in $\set{r \geq \frac{1}{2}}$ (with respect to the $(u, r, \tht)$ coordinates):
\begin{equation}  \label{eq:en-ext-mult}
\bfY_{\bfalp}^{u} = O_{\bfGmm}^{1}(A_{m}), \quad
\bfY_{\bfalp}^{r} = O_{\bfGmm}^{1}(A_{m}), \quad
\bfY_{\bfalp}^{\tht^{A}} = 0, \quad
\upsilon_{\bfalp} = O_{\bfGmm}(A_{m} \tfrac{1}{r}),
\end{equation}
where all the implicit constants are $1$.

\smallskip \noindent {\it Step~1.}  Given $\alp \in \bbR$, define
\begin{align*}
\frkE_{\alp; U}^{2}[(\psi, \rd_{x^{0}} \psi)|_{\set{x^{0} = t}}] &= \mbrk{U}^{2 (\alp - \nu_{\Box}) -1} \frkE_{U}^{2}[(\psi, \rd_{x^{0}} \psi)|_{\set{x^{0} = t}}], \\
\frkE_{\alp; \ell^{\infty}_{\leq U_{0}}}^{2}[(\psi, \rd_{x^{0}} \psi)|_{\set{x^{0} = t}}] &= \sup \set*{\frkE_{\alp; U}^{2}[(\psi, \rd_{x^{0}} \psi)|_{\set{x^{0} = t}}] : \tfrac{- U}{-U_{0}} \in 2^{\bbZ_{\geq 0}}},
\end{align*}
and also $\frkM_{\alp; U}^{2}[\psi]$, $\frkM_{\alp; \ell^{\infty}_{\leq U_{0}}}^{2}[\psi]$, $\frkI_{\alp; U}^{2}[g]$, $\frkI_{\alp; \ell^{\infty}_{\leq U_{0}}}^{2}[g]$, $\frkH_{\alp; U}^{2}[\psi]$ and $\frkH_{\alp; \ell^{\infty}_{\leq U_{0}}}^{2}[\psi]$ analogously. When the objects $\psi$ and $g$ are clearly specified, we will often use the  shorthands
\begin{gather*}
\frkE_{\alp; U}^{2}(t) = \frkE_{\alp; U}^{2}[(\psi, \rd_{x^{0}} \psi)|_{\set{x^{0} = t}}],  \quad
\frkM_{\alp; U}^{2} = \frkM_{\alp; U}^{2}[\psi], \quad
\frkI_{\alp; U}^{2} = \frkI_{\alp; U}^{2}[g], \quad
\frkH_{\alp; U}^{2} = \frkH_{\alp; U}^{2}[\psi],
\end{gather*}
and we similarly define $\frkE_{\alp; \ell^{\infty}_{\leq U_{0}}}^{2}(t)$, $\frkM_{\alp; \ell^{\infty}_{\leq U_{0}}}^{2}$, $\frkI_{\alp; \ell^{\infty}_{\leq U_{0}}}^{2}$ and $\frkH_{\alp; \ell^{\infty}_{\leq U_{0}}}^{2}$.

Let $\psi, g$ be functions on the extended spacetime region $\set{t_{0} \leq x^{0} \leq T}$ solving $\calP \psi = g$. For $\alp > \nu_{\Box} + \frac{p-1}{2}$, we claim that
\begin{align}
\sup_{t_{0} \leq t \leq T} \frkE_{\alp; \ell^{\infty}_{\leq \frac{1}{2} U_{0}}}^{2}(t)
+ \frkM_{\alp; \ell^{\infty}_{\leq \frac{1}{2} U_{0}}}^{2} \aleq \frkE_{\alp; \ell^{\infty}_{\leq \frac{1}{2} U_{0}}}^{2}(t_{0}) + \frkI_{\alp; \ell^{\infty}_{\leq \frac{1}{2} U_{0}}}^{2} + \frkH_{\alp; \ell^{\infty}_{\leq \frac{1}{2} U_{0}}}^{2}
+ \eps_{e} \sup_{t_{0} \leq t \leq T} \frkE_{\alp; \ell^{\infty}_{\leq \frac{1}{2} U_{0}}}^{2}(t). \label{eq:en-ext:pf:1}
\end{align}
This claim will be proved by putting together Lemmas~\ref{lem:ext-en}--\ref{lem:ext-rp} and using finite speed of propagation. We remark that the reason why we localize to $u_{h} \leq \frac{1}{2} U_{0}$ is that $\bfg^{-1} = \bfm^{-1}$ for $u_{h} \geq \frac{1}{2} U_{0}$ by our extension procedure.

We begin with Lemmas~\ref{lem:ext-en} and \ref{lem:ext-iled}. We let $\set{\bfY_{\bfalp}}_{\bfalp \in \calA} = \set{\bfT} \cup \set{\bfX_{R}}_{R \in 2^{\bbZ_{\geq 0}}}$ and $\set{\upsilon_{\bfalp}}_{\bfalp \in \calA} = \set{0} \cup \set{\xi_{R}}_{R \in 2^{\bbZ_{\geq 0}}}$; note that \eqref{eq:en-ext-mult-intr}--\eqref{eq:en-ext-mult} hold with a constant $A_{m}$ independent of $\bfalp$. Let $U \leq U_{0}$. We apply Lemmas~\ref{lem:ext-en} and \ref{lem:ext-iled} with $U_{\ast} = \frac{1}{2} U$ to the solution $\psi_{\leq U} := \chi_{> -U}(-u_{h}) \psi$ solving $\calP \psi_{\leq U} = \chi_{> -U}(- u_{h}) g + [\calP, \chi_{> -U}(- u_{h})] \psi$ and $(\psi_{\leq U}, \rd_{x^{0}} \psi_{\leq U})|_{\set{x^{0} = t_{0}}} = \chi_{> -U}(-u_{h}) (\psi, \rd_{x^{0}} \psi)|_{\set{x^{0} = t_{0}}}$. Multiplying the resulting expression by $\mbrk{U}^{2(\alp - \nu_{\Box}) + 1}$ and using finite speed of propagation (recall that level-$u_{h}$ hypersurfaces are spacelike for $u_{h} \leq U_{0}$), we obtain 
\begin{align}
&\sup_{t_{0} \leq t \leq T} \int_{\set{x^{0} = t, \, u_{h} \leq U}} \mbrk{U}^{2 (\alp - \nu_{\Box}) - 1} \bigg[ (\mbrk{U} \rd_{u} \psi)^{2} + (\mbrk{U} \rd_{r} \psi)^{2} + \left( \tfrac{\mbrk{U}}{r} \psi \right)^{2} + \left( \tfrac{\mbrk{U}}{r} \abs{\rsnb \psi} \right)^{2} \bigg] \, \ud \sgm \notag \\
&+ \sup_{R \in 2^{\bbZ} : R \geq R_{0}} \iint_{\set{t_{0} \leq x^{0} \leq T, \, u_{h} \leq U, \, R \leq r \leq 2 R}} \frac{\mbrk{U}^{2(\alp - \nu_{\Box}) - 1}}{r} \bigg[ (\mbrk{U} \rd_{u} \psi)^{2} + (\mbrk{U} \rd_{r} \psi)^{2} + \left( \tfrac{\mbrk{U}}{r} \psi \right)^{2} + \left( \tfrac{\mbrk{U}}{r} \abs{\rsnb \psi} \right)^{2} \bigg]  \ud \mathrm{V} \notag \\
&\aleq
\int_{\set{x^{0} = t_{0}, \, u_{h} \leq \frac{1}{2} U}} \mbrk{U}^{2 (\alp - \nu_{\Box}) - 1} \bigg[ (\mbrk{U} \rd_{u} \psi)^{2} + (\mbrk{U} \rd_{r} \psi)^{2} + \left( \tfrac{\mbrk{U}}{r} \psi \right)^{2} + \left( \tfrac{\mbrk{U}}{r} \abs{\rsnb \psi} \right)^{2} \bigg] \, \ud \sgm \notag\\
&\peq + \sup_{t_{0} \leq t_{1} \leq T, \, \bfalp \in \calA} \abs*{\iint_{\set{t_{0} \leq x^{0} \leq t_{1}, \,  u_{h} \leq \frac{1}{2} U}} \mbrk{U}^{2(\alp - \nu_{\Box}) + 1} (\chi_{> -U}(-u_{h}) g + [\calP, \chi_{> -U}(- u_{h})] \psi) (\bfY^{\gmm}_{\bfalp} \rd_{\gmm} \psi_{\leq U} + \upsilon_{\bfalp} \psi_{\leq U}) \, \ud \mathrm{V} }\notag \\
&\peq + \sup_{t_{0} \leq t_{1} \leq T, \, \bfalp \in \calA} \abs*{\iint_{\set{t_{0} \leq x^{0} \leq t_{1}, \,  u_{h} \leq \frac{1}{2} U}}  \mbrk{U}^{2(\alp - \nu_{\Box}) + 1}   (\calP - \Box_{\bfm}) \psi_{\leq U} (\bfY^{\gmm}_{\bfalp} \rd_{\gmm} \psi_{\leq U} + \upsilon_{\bfalp} \psi_{\leq U}) \, \ud \mathrm{V}} \notag \\
&\aleq \frkE_{\alp; \ell^{\infty}_{\leq \frac{1}{2} U_{0}}}^{2}(t_{0}) + \frkI_{\alp; \ell^{\infty}_{\leq \frac{1}{2} U_{0}}} \frkM_{\alp; \ell^{\infty}_{\leq \frac{1}{2} U_{0}}} + \frkH_{\alp; \ell^{\infty}_{\leq \frac{1}{2} U_{0}}}^{2}  + \eps_{e} \sup_{t_{0} \leq t \leq T} \frkE_{\alp; \ell^{\infty}_{\leq \frac{1}{2} U_{0}}}^{2}(t) \label{eq:en-ext:pf:1:1} 
\end{align}
where the last inequality follows from the definitions of $\frkE_{\alp; \ell^{\infty}_{\leq \frac{1}{2} U_{0}}}^{2}(t_{0})$, $\frkI_{\alp; \ell^{\infty}_{\leq \frac{1}{2} U_{0}}}^{2}$, $\frkM_{\alp; \ell^{\infty}_{\leq \frac{1}{2} U_{0}}}^{2}$ and $\frkH_{\alp; \ell^{\infty}_{\leq \frac{1}{2} U_{0}}}^{2}$ and Lemma~\ref{lem:ext-mult} for the contribution of $(\calP-\Box_{\bfm}) \psi_{\leq U}$. We remark that last term on the right-hand side of \eqref{eq:en-ext:pf:1:1} bounds the boundary terms on level-$x^{0}$ hypersurfaces arising from Lemma~\ref{lem:ext-mult}, while no boundary terms on $u_{h} = \frac{1}{2} U$ arise in this procedure because of the support property of $\psi_{\leq U}$. 

For $U = \frac{1}{2} U_{0}$, we apply Lemmas~\ref{lem:ext-en} and \ref{lem:ext-iled} with $U_{\ast} = \frac{1}{2} U_{0}$ directly to the solution $\psi$ and multiplying the resulting expression by $\mbrk{U_{0}}^{2(\alp - \nu_{\Box}) + 1}$. Proceeding as before, we obtain
\begin{align} 
&\sup_{t_{0} \leq t \leq T} \int_{\set{x^{0} = t, \, u_{h} \leq \frac{1}{2} U_{0}}} \mbrk{U_{0}}^{2 (\alp - \nu_{\Box}) - 1} \bigg[ (\mbrk{U_{0}} \rd_{u} \psi)^{2} + (\mbrk{U_{0}} \rd_{r} \psi)^{2} + \left( \tfrac{\mbrk{U_{0}}}{r} \psi \right)^{2} + \left( \tfrac{\mbrk{U_{0}}}{r} \abs{\rsnb \psi} \right)^{2} \bigg] \, \ud \sgm \notag \\
&+ \sup_{R \in 2^{\bbZ} : R \geq R_{0}} \iint_{\set{t_{0} \leq x^{0} \leq T, \, u_{h} \leq \frac{1}{2} U_{0}, \, R \leq r \leq 2 R}} \frac{\mbrk{U_{0}}^{2(\alp - \nu_{\Box}) - 1}}{r} \bigg[ (\mbrk{U_{0}} \rd_{u} \psi)^{2} + (\mbrk{U_{0}} \rd_{r} \psi)^{2} + \left( \tfrac{\mbrk{U_{0}}}{r} \psi \right)^{2} + \left( \tfrac{\mbrk{U_{0}}}{r} \abs{\rsnb \psi} \right)^{2} \bigg]  \ud \mathrm{V} \notag \\
&\aleq
\int_{\set{x^{0} = t_{0}, \, u_{h} \leq \frac{1}{2} U_{0}}} \mbrk{U_{0}}^{2 (\alp - \nu_{\Box}) - 1} \bigg[ (\mbrk{U_{0}} \rd_{u} \psi)^{2} + (\mbrk{U_{0}} \rd_{r} \psi)^{2} + \left( \tfrac{\mbrk{U_{0}}}{r} \psi \right)^{2} + \left( \tfrac{\mbrk{U_{0}}}{r} \abs{\rsnb \psi} \right)^{2} \bigg] \, \ud \sgm \notag\\
&\peq + \sup_{t_{0} \leq t_{1} \leq T, \, \bfalp \in \calA} \abs*{\iint_{\set{t_{0} \leq x^{0} \leq T, \,  u_{h} \leq \frac{1}{2} U_{0}}} \mbrk{U_{0}}^{2(\alp - \nu_{\Box}) + 1} g (\bfY^{\gmm}_{\bfalp} \rd_{\gmm} \psi_{\leq U} + \upsilon_{\bfalp} \psi_{\leq U}) \, \ud \mathrm{V} }\notag \\
&\peq + \sup_{t_{0} \leq t_{1} \leq T, \, \bfalp \in \calA} \abs*{\iint_{\set{t_{0} \leq x^{0} \leq T, \,  u_{h} \leq \frac{1}{2} U_{0}}}  \mbrk{U_{0}}^{2(\alp - \nu_{\Box}) + 1}   (\calP - \Box_{\bfm}) \psi (\bfY^{\gmm}_{\bfalp} \rd_{\gmm} \psi + \upsilon_{\bfalp} \psi) \, \ud \mathrm{V}} \notag \\
&\aleq \frkE_{\alp; \ell^{\infty}_{\leq \frac{1}{2} U_{0}}}^{2}(t_{0}) + \frkI_{\alp; \ell^{\infty}_{\leq \frac{1}{2} U_{0}}} \frkM_{\alp; \ell^{\infty}_{\leq \frac{1}{2} U_{0}}} + \frkH_{\alp; \ell^{\infty}_{\leq \frac{1}{2} U_{0}}}^{2}
+ \eps_{e} \sup_{t_{0} \leq t \leq T} \frkE_{\alp; \ell^{\infty}_{\leq \frac{1}{2} U_{0}}}^{2}(t).\label{eq:en-ext:pf:1:1'}
\end{align}
We remark that, when bounding the integral involving $(\calP - \Box_{\bfm}) \psi$ by the last two terms on the right-hand side of \eqref{eq:en-ext:pf:1:1'} using Lemma~\ref{lem:ext-mult}, no boundary term on $u_{h} = \frac{1}{2} U_{0}$ arises since $\calP - \Box_{\bfm} = 0$ there according to our extension procedure in Step~0.

Next, we apply Lemma~\ref{lem:ext-rp}. For $U \leq U_{0}$, we apply Lemma~\ref{lem:ext-rp} with $\wp(r) = r^{p}$ and
\begin{equation*}
\varpi(u) = \mbrk{U}^{2 (\alp - \nu_{\Box}) - (p-1)} \chi_{> -U}(-u_{h}) \left(1 + \frac{1}{100 \veps_{\varpi}^{-1} \mbrk{U}} \int_{-\infty}^{-u} (\chi_{> - \veps_{\varpi} U} - \chi_{> -4 \veps_{\varpi}^{-1} U})(s) \, \ud s  \right).
\end{equation*}
Observe that $\chi_{>\veps_{\varpi} (-U)} - \chi_{>\veps_{\varpi}^{-1} (-4U)} \geq 0$ is supported in $s \in [-\tfrac{1}{2} \veps_{\varpi} U, -4\veps_{\varpi}^{-1}U]$ and equals one on $[-U, -2U]$. Hence, $\varpi(u) \aeq \mbrk{U}^{2 (\alp - \nu_{\Box}) - (p-1)}$ everywhere, but $\varpi'(u) \geq \frac{1}{100} \mbrk{U}^{2 (\alp - \nu_{\Box}) - (p-1)-1}$ for $u \in [2 \veps_{\varpi}^{-1} U, \veps_{\varpi} U]$. We fix our choice of $\veps_{\varpi} > 0$ so that $\set{2U \leq u_{h} \leq U} \subseteq \set{2 \veps_{\varpi}^{-1} U \leq u \leq \veps_{\varpi} U}$. As a result, we obtain 
\begin{align}
&\sup_{t_{0} \leq t \leq T} \int_{\set{x^{0} = t, \, u_{h} \leq U}} \mbrk{U}^{2 (\alp - \nu_{\Box}) - 1} \bigg[ \left( (\tfrac{r}{\mbrk{U}})^{-\frac{1}{2}+\frac{p-1}{2}} r^{1-\nu_{\Box}} \rd_{r} (r^{\nu_{\Box}} \psi) \right)^{2} + \left( (\tfrac{r}{\mbrk{U}})^{-\frac{1}{2}+\frac{p-1}{2}}\abs{\rsnb \psi}\right)^{2} \bigg]  \ud \sgm \notag \\
&+ \iint_{\set{t_{0} \leq x^{0} \leq T, \, u_{h} \leq U}} \frac{\mbrk{U}^{2 (\alp - \nu_{\Box})-1}}{r} \left[ \left( (\tfrac{r}{\mbrk{U}})^{-\frac{1}{2}+\frac{p-1}{2}} r^{1-\nu_{\Box}} \rd_{r} (r^{\nu_{\Box}} \psi) \right)^{2} + \left( (\tfrac{r}{\mbrk{U}})^{-\frac{1}{2}+\frac{p-1}{2}}\abs{\rsnb \psi}\right)^{2} \right]\, \ud \mathrm{V} \notag \\
&+ \iint_{\set{t_{0} \leq x^{0} \leq T, \, 2U \leq u_{h} \leq U}} \frac{\mbrk{U}^{2 (\alp - \nu_{\Box})-1}}{r} \left( (\tfrac{r}{\mbrk{U}})^{\frac{p-1}{2}} r^{1-\nu_{\Box}} \rd_{r} (r^{\nu_{\Box}} \psi) \right)^{2} \, \ud \mathrm{V} \notag \\
&\aleq
\int_{\set{x^{0} = t_{0}, \, u_{h} \leq \frac{1}{2} U}} \mbrk{U}^{2 (\alp - \nu_{\Box}) - 1} \bigg[ (\mbrk{U} \rd_{u} \psi)^{2} + (\mbrk{U} \rd_{r} \psi)^{2} + \left( \tfrac{\mbrk{U}}{r} \psi \right)^{2} + \left( \tfrac{\mbrk{U}}{r} \abs{\rsnb \psi} \right)^{2} \bigg] \, \ud \sgm \notag\\
&\peq + \abs*{\iint_{\set{t_{0} \leq x^{0} \leq T, \,  u_{h} \leq \frac{1}{2} U}} r^{\nu_{\Box}} g (- \varpi(u) r^{p} \rd_{r} \Psi) \, \ud \mathrm{V}}
+ \abs*{\iint_{\set{t_{0} \leq x^{0} \leq T, \,  u_{h} \leq \frac{1}{2} U}} r^{\nu_{\Box}} (\calP - \Box_{\bfm}) \psi (- \varpi(u) r^{p} \rd_{r} \Psi) \, \ud \mathrm{V}} \notag \\
&\aleq \frkE_{\alp; \ell^{\infty}_{\leq \frac{1}{2} U_{0}}}^{2}(t_{0}) + \frkI_{\alp; \ell^{\infty}_{\leq \frac{1}{2} U_{0}}} \frkM_{\alp; \ell^{\infty}_{\leq \frac{1}{2} U_{0}}} + \frkH_{\alp; \ell^{\infty}_{\leq \frac{1}{2} U_{0}}}^{2} + \eps_{e} \sup_{t_{0} \leq t \leq T} \frkE_{\alp; \ell^{\infty}_{\leq \frac{1}{2} U_{0}}}^{2}(t), \label{eq:en-ext:pf:1:2}
\end{align}
where the last inequality is proved as before, but using the second identity in Lemma~\ref{lem:ext-rp} for the contribution of $(\calP - \Box_{\bfm}) \psi$. We note that the third term on the left-hand side arises from the term $-\varpi'(u) \wp(r) (\rd_{r} \Psi)^{2}$ in Lemma~\ref{lem:ext-rp}. 

For $U = \frac{1}{2} U_{0}$, we apply Lemma~\ref{lem:ext-rp} with $\wp(r) = r^{p}$ and
\begin{equation*}
\varpi(u) = \mbrk{U_{0}}^{2 (\alp - \nu_{\Box}) - (p-1)} \chf_{>\frac{1}{2} U_{0}}(-u_{h}) \left(1 + \frac{1}{100 \veps_{\varpi}^{-1} \mbrk{U_{0}}} \int_{-\infty}^{-u} (\chi_{> -\frac{1}{2} \veps_{\varpi} U_{0}} - \chi_{> -2 \veps_{\varpi}^{-1} U_{0}})(s) \, \ud s  \right).
\end{equation*}
Then we obtain \eqref{eq:en-ext:pf:1:2} with $U = \frac{1}{2} U_{0}$. Here, we remark that the boundary term on $u_{h} = \frac{1}{2} U_{0}$ arising from the integration of the first identity in Lemma~\ref{lem:ext-rp} has a favorable sign, whereas that from the second identity in Lemma~\ref{lem:ext-rp} vanishes since $\calP = \Box_{\bfm}$ there according to our extension procedure in Step~0.

Finally, we also observe that
\begin{align}
&\sup_{U \leq \frac{1}{2} U_{0}: -U \in 2^{\bbZ_{\geq 0}}} \sup_{R \in 2^{\bbZ} : R \geq R_{0}} \iint_{\set{t_{0} \leq x^{0} \leq T, \, 2U \leq u_{h} \leq U, \, R \leq r \leq 2R}} \frac{\mbrk{U}^{2 (\alp - \nu_{\Box})-1}}{r} \bigg[ \psi^{2} + (r \rd_{r} \psi)^{2} \bigg] \, \ud \mathrm{V} \notag \\
&\aleq \sup_{U' \in 2^{\bbZ}} \mbrk{U'}^{2 (\alp - \nu_{\Box})-1} \int_{U'}^{2U'} \int_{\bbS^{d-1}} \sup_{r} \abs*{\chf_{\set{t_{0} \leq x^{0} \leq T, u_{h} \leq \frac{1}{2} U_{0}}} \bigg[ \psi^{2} + (r \rd_{r} \psi)^{2} \bigg](u, r, \tht) r^{d-1}} \, \ud \sgm(\tht) \ud u \label{eq:en-ext:pf:1:3} \\
&\aleq \sup_{U' \in 2^{\bbZ}} \mbrk{U'}^{2 (\alp - \nu_{\Box})-1} \int_{U'}^{2U'} \int_{\bbS^{d-1}} \chf_{\set{u_{h} \leq \frac{1}{2} U_{0}}} ( \psi^{2} r^{d-1} )|_{\set{x^{0} = t_{0}}}\, \ud \sgm(\tht) \ud u \notag \\
&\peq + \sup_{U' \in 2^{\bbZ}} \mbrk{U'}^{2 (\alp - \nu_{\Box})-1} \int_{U'}^{2U'} \int_{\bbS^{d-1}} \int_{0}^{\infty} \chf_{\set{t_{0} \leq x^{0} \leq T, u_{h} \leq \frac{1}{2} U_{0}}} \left( (\tfrac{r}{\mbrk{U'}})^{\frac{p-1}{2}} r^{1-\nu_{\Box}} \rd_{r} (r^{\nu_{\Box}} \psi) \right)^{2} \, r^{d-1} \frac{\ud r}{r} \ud u \ud \sgm(\tht)   \notag \\
&\aleq \frkE_{\alp; \ell^{\infty}_{\leq \frac{1}{2} U_{0}}}^{2}(t_{0})
+ \sup_{U \leq \frac{1}{2} U_{0}: -U \in 2^{\bbZ_{\geq 0}}} \iint_{\set{t_{0} \leq x^{0} \leq T, \, 2U \leq u_{h} \leq U}} \frac{\mbrk{U}^{2 (\alp - \nu_{\Box})-1}}{r} \left( (\tfrac{r}{\mbrk{U}})^{\frac{p-1}{2}} r^{1-\nu_{\Box}} \rd_{r} (r^{\nu_{\Box}} \psi) \right)^{2} \, \ud \mathrm{V}. \notag
\end{align}
Here, the first and third inequalities are straightforward since $u \aeq u_{h}$. To prove the second inequality, we begin by bounding $\psi(u, \cdot, \tht)$ using the fundamental theorem of calculus in the $r$-variable for $r^{\nu_{\Box}} \psi(u, \cdot, \tht)$, then bounding $r \rd_{r} \psi$ by $r^{1-\nu_{\Box}} \rd_{r} (r^{\nu_{\Box}} \psi)$ and $\psi$. Adding up \eqref{eq:en-ext:pf:1:1}, \eqref{eq:en-ext:pf:1:2} and \eqref{eq:en-ext:pf:1:3}, then taking the supremum over $U \leq \frac{1}{2} U_{0}$ ($- U \in 2^{\bbZ_{\geq 0}}$), the desired claim \eqref{eq:en-ext:pf:1} follows.

\smallskip \noindent {\it Step~2.}
In this step we will prove that
\begin{equation} \label{eq:en-ext:pf:2}
\frkH_{\alp; \ell^{\infty}_{\leq \frac{1}{2} U_{0}}}^{2} \aleq \eps_{e} \frkM_{\alp; \ell^{\infty}_{\leq \frac{1}{2} U_{0}}}^{2}.
\end{equation}
Then, for $\eps_{e}$ sufficiently small, the two last terms on the right-hand side of \eqref{eq:en-ext:pf:2} may be absorbed into the left-hand side, thereby leading to
\begin{equation} \label{eq:en-ext:pf:1'}
\sup_{t_{0} \leq t \leq T} \frkE_{\alp; \ell^{\infty}_{\leq \frac{1}{2} U_{0}}}^{2}(t)
+ \frkM_{\alp; \ell^{\infty}_{\leq \frac{1}{2} U_{0}}}^{2} \aleq \frkE_{\alp; \ell^{\infty}_{\leq \frac{1}{2} U_{0}}}^{2}(t_{0}) + \frkI_{\alp; \ell^{\infty}_{\leq \frac{1}{2} U_{0}}}^{2}.
\end{equation}

Inspecting the definitions of $\frkM_{\alp}$, $\frkH_{\alp}$, $\frkh_{\bfalp}$ and $\frkh_{p}$, we see that \eqref{eq:en-ext:pf:3} would hold if the following is established for $\alp, \bt, \gmm \in \set{u, r, 1, \ldots, d-1}$:
\begin{align*}
	\mbrk{u} r (\tfrac{r}{\mbrk{u}})^{p} \abs*{\bfh^{u \bt} \rd_{\bt} \Psi \rd_{r} \Psi} r^{-(d-1)} &\aleq \eps_{e} \left( r^{-\min\set{\dlt_{e}, \frac{p-1}{2}}} \frkm^{2} [\psi] + \frkm_{p}^{2} [\psi]\right), \\
	\mbrk{u}^{2} r(\tfrac{r}{\mbrk{u}})^{p} \abs*{\rd_{\alp} \bfh^{\alp \gmm} \rd_{\gmm} \Psi \rd_{r} \Psi} r^{-(d-1)} & \aleq \eps_{e} \left( r^{-\min\set{\dlt_{e}, \frac{p-1}{2}}} \frkm^{2} [\psi] + \frkm_{p}^{2} [\psi]\right), \\
	\mbrk{u} r (\tfrac{r}{\mbrk{u}})^{p-1} \left( \abs*{\bfh^{\alp \bt}} + \abs*{r \rd_{r} \bfh^{\alp \bt}}\right)\abs*{\rd_{\alp} \Psi \rd_{\bt} \Psi} r^{-(d-1)}& \aleq \eps_{e} \left( r^{-\min\set{\dlt_{e}, \frac{p-1}{2}}} \frkm^{2} [\psi] + \frkm_{p}^{2} [\psi]\right), \\
	\mbrk{u}^{2} r (\tfrac{r}{\mbrk{u}})^{p} \abs*{c^{\gmm} \rd_{\gmm} \Psi \rd_{r} \Psi} r^{-(d-1)} &\aleq \eps_{e} \left( r^{-\min\set{\dlt_{e}, \frac{p-1}{2}}} \frkm^{2} [\psi] + \frkm_{p}^{2} [\psi]\right), \\
	\mbrk{u}^{2} r(\tfrac{r}{\mbrk{u}})^{p} \abs*{W \Psi \rd_{r} \Psi} r^{-(d-1)} & \aleq \eps_{e} \left( r^{-\min\set{\dlt_{e}, \frac{p-1}{2}}} \frkm^{2} [\psi] + \frkm_{p}^{2} [\psi]\right), \\
	\mbrk{u}^{2} r \abs*{r^{-(d-1)} \rd_{\alp} (r^{d-1} \bfh^{\alp \bt} \bfY^{\gmm}_{\bfalp}) \rd_{\bt} \psi \rd_{\gmm} \psi} &\aleq \eps_{e} \left( r^{-\min\set{\dlt_{e}, \frac{p-1}{2}}} \frkm^{2} [\psi] + \frkm_{p}^{2} [\psi]\right), \\
	\mbrk{u}^{2} r \abs*{r^{-(d-1)} \rd_{\alp} (r^{d-1} \bfh^{\bt \gmm} \bfY^{\alp}_{\bfalp}) \rd_{\bt} \psi \rd_{\gmm} \psi} &\aleq \eps_{e} \left( r^{-\min\set{\dlt_{e}, \frac{p-1}{2}}} \frkm^{2} [\psi] + \frkm_{p}^{2} [\psi]\right), \\
	\mbrk{u}^{2} r \abs*{b^{\alp} \bfY^{\gmm}_{\bfalp} \rd_{\alp} \psi \rd_{\gmm} \psi}
	&\aleq \eps_{e} \left( r^{-\min\set{\dlt_{e}, \frac{p-1}{2}}} \frkm^{2} [\psi] + \frkm_{p}^{2} [\psi]\right), \\
	\mbrk{u}^{2} r \abs*{V \bfY^{\gmm}_{\bfalp} \psi \rd_{\gmm} \psi} & \aleq \eps_{e} \left( r^{-\min\set{\dlt_{e}, \frac{p-1}{2}}} \frkm^{2} [\psi] + \frkm_{p}^{2} [\psi]\right), \\
	\mbrk{u}^{2} r \abs*{\bfh^{\alp \bt} \upsilon \rd_{\bt} \psi \rd_{\alp} \psi} &\aleq \eps_{e} \left( r^{-\min\set{\dlt_{e}, \frac{p-1}{2}}} \frkm^{2} [\psi] + \frkm_{p}^{2} [\psi]\right), \\
	\mbrk{u}^{2} r \abs*{r^{-(d-1)}\rd_{\alp} (r^{d-1} \bfh^{\alp \bt} \upsilon_{\bfalp})  \rd_{\bt} \psi \psi} &\aleq \eps_{e} \left( r^{-\min\set{\dlt_{e}, \frac{p-1}{2}}} \frkm^{2} [\psi] + \frkm_{p}^{2} [\psi]\right), \\
	\mbrk{u}^{2} r \abs*{b^{\alp} \upsilon_{\bfalp} \rd_{\alp} \psi \psi} &\aleq \eps_{e} \left( r^{-\min\set{\dlt_{e}, \frac{p-1}{2}}} \frkm^{2} [\psi] + \frkm_{p}^{2} [\psi]\right), \\
	\mbrk{u}^{2} r \abs*{V \upsilon_{\bfalp} \psi^{2}} & \aleq \eps_{e} \left( r^{-\min\set{\dlt_{e}, \frac{p-1}{2}}} \frkm^{2} [\psi] + \frkm_{p}^{2} [\psi]\right).
\end{align*}
These bounds follow from \eqref{eq:en-ext-coeff-bondi} and \eqref{eq:en-ext-mult-intr}--\eqref{eq:en-ext-mult} (for $Y_{\bfalp}$ and $\upsilon_{\bfalp}$) by straightforward (albeit numerous) case-by-case computations. Here, we will only record the cases that saturate each bound in \eqref{eq:en-ext-coeff-bondi} and omit the details for the other (more favorable) cases:
\begin{itemize}
\item $\bfh^{uu} = O_{\bfGmm}^{1}(\eps_{e} r^{-\dlt_{e}} (\tfrac{r}{\mbrk{u}})^{-1 - (p-1)})$ $\imp$
\begin{align*}
	\mbrk{u} r (\tfrac{r}{\mbrk{u}})^{p-1} \left( \abs*{\bfh^{uu}} + \abs*{r \rd_{r} \bfh^{uu}}\right)\abs*{\rd_{u} \Psi \rd_{u} \Psi} r^{-(d-1)} \aleq \eps_{e} r^{-\dlt_{e}} \mbrk{u}^{2} (\rd_{u} \psi)^{2} \aleq \eps_{e} r^{-\dlt_{e}} \frkm^{2}[\psi].
\end{align*}
\item $\bfh^{ur} = O_{\bfGmm}^{1}(\eps_{e} r^{-\dlt_{e}} (\tfrac{r}{\mbrk{u}})^{-\frac{p-1}{2}})$ $\imp$
\begin{align*}
	\mbrk{u} r (\tfrac{r}{\mbrk{u}})^{p-1} \left( \abs*{\bfh^{ur}} + \abs*{r \rd_{r} \bfh^{ur}}\right)\abs*{\rd_{u} \Psi \rd_{r} \Psi} r^{-(d-1)}& \aleq \eps_{e} r^{-\dlt_{e}} \mbrk{u} r (\tfrac{r}{\mbrk{u}})^{\frac{p-1}{2}} \abs{\rd_{u} \psi \rd_{r} \Psi} r^{-\frac{d-1}{2}} \aleq \eps_{e} r^{-\dlt_{e}} \frkm[\psi] \frkm_{p}[\psi].
\end{align*}
\item $r \bfh^{uA} = O_{\bfGmm}^{1}(\eps_{e} r^{-\dlt_{e}} (\tfrac{r}{\mbrk{u}})^{-\frac{1}{2}-\frac{p-1}{2}})$ $\imp$
\begin{align*}
	\mbrk{u} r (\tfrac{r}{\mbrk{u}})^{p-1} \left( \abs*{\bfh^{uA}} + \abs*{r \rd_{r} \bfh^{uA}}\right)\abs*{\rd_{u} \Psi \rd_{A} \Psi} r^{-(d-1)}& \aleq \eps_{e} r^{-\dlt_{e}} \mbrk{u} (\tfrac{r}{\mbrk{u}})^{-\frac{1}{2}+\frac{p-1}{2}} \abs{\rd_{u} \psi \rd_{A} \psi} \aleq \eps_{e} r^{-\dlt_{e}} \frkm[\psi] \frkm_{p}[\psi].
\end{align*}
\item $\bfh^{rr}$: For this step, we may even allow for the weaker assumption $\bfh^{rr} = O_{\bfGmm}^{1}(\eps_{e} r^{-\dlt_{e}} \tfrac{r}{\mbrk{u}})$, under which
\begin{align*}
	\mbrk{u}^{2} r \abs*{r^{-(d-1)} \rd_{r} (r^{d-1} \bfh^{r r} \bfY^{u}_{\bfalp}) \rd_{r} \psi \rd_{u} \psi} &\aleq \eps_{e} r^{-\dlt_{e}} \tfrac{r}{\mbrk{u}} \mbrk{u}^{2} \abs{\rd_{r} \psi \rd_{u} \psi}
	\aleq \eps_{e} r^{-\dlt_{e}} \frkm^{2}[\psi], \\
	\mbrk{u}^{2} r \abs*{r^{-(d-1)} \rd_{u} (r^{d-1} \bfh^{r r} \bfY^{u}_{\bfalp}) \rd_{r} \psi \rd_{r} \psi} &\aleq \eps_{e} r^{-\dlt_{e}} \tfrac{r}{\mbrk{u}} \mbrk{u} r (\rd_{r} \psi)^{2}
	\aleq \eps_{e} r^{-\dlt_{e}} \frkm^{2}[\psi].
\end{align*}
\item $r \bfh^{rA}$: For this step, we may allow for the weaker assumption $r \bfh^{r A} = O_{\bfGmm}^{1}(\eps_{e} \mbrk{u}^{-\dlt_{e}} (\tfrac{r}{\mbrk{u}})^{\frac{1}{2}})$, under which
\begin{align*}
	\mbrk{u} r (\tfrac{r}{\mbrk{u}})^{p-1} \left( \abs*{\bfh^{r A}} + \abs*{r \rd_{r} \bfh^{r A}}\right)\abs*{\rd_{r} \Psi \rd_{A} \Psi} r^{-(d-1)}& \aleq \eps_{e} \mbrk{u}^{-\dlt_{e}} (\tfrac{r}{\mbrk{u}})^{\frac{1}{2}} \mbrk{u} (\tfrac{r}{\mbrk{u}})^{p-1} \abs{\rd_{r} \Psi \rd_{A} \psi} r^{-\frac{d-1}{2}} \aleq \eps_{e} \mbrk{u}^{-\dlt_{e}} \frkm_{p}^{2}[\psi], \\
	\mbrk{u}^{2} r \abs*{r^{-(d-1)} \rd_{r} (r^{d-1} \bfh^{r A} \bfY^{u}_{\bfalp}) \rd_{A} \psi \rd_{u} \psi}
	&\aleq \eps_{e} \mbrk{u}^{-\dlt_{e}} (\tfrac{r}{\mbrk{u}})^{\frac{1}{2}} \tfrac{\mbrk{u}^{2}}{r} \abs{\rd_{A} \psi} \abs{\rd_{u} \psi}
	\aleq \eps_{e} r^{-\min\set{\dlt_{e}, \frac{p-1}{2}}} \frkm_{p}[\psi] \frkm[\psi] \\
	\mbrk{u}^{2} r \abs*{r^{-(d-1)} \rd_{u} (r^{d-1} \bfh^{r A} \bfY^{u}_{\bfalp}) \rd_{r} \psi \rd_{A} \psi}
	&\aleq \eps_{e} \mbrk{u}^{-\dlt_{e}} (\tfrac{r}{\mbrk{u}})^{\frac{1}{2}} \mbrk{u} \abs{\rd_{r} \psi \rd_{A} \psi}
	\aleq \eps_{e} r^{-\min\set{\dlt_{e}, \frac{p-1}{2}}} \frkm_{p}[\psi] \frkm[\psi].
\end{align*}
\item $r^{2} \bfh^{AB} = O_{\bfGmm}^{1}(\eps_{e} \mbrk{u}^{-\dlt_{e}})$ $\imp$
\begin{align*}
	\mbrk{u} r (\tfrac{r}{\mbrk{u}})^{p-1} \left( \abs*{\bfh^{A B}} + \abs*{r \rd_{r} \bfh^{A B}}\right)\abs*{\rd_{A} \Psi \rd_{B} \Psi} r^{-(d-1)}& \aleq \eps_{e} \mbrk{u}^{-\dlt_{e}}\frac{\mbrk{u}}{r} (\tfrac{r}{\mbrk{u}})^{p-1} \abs{\rd_{A} \psi \rd_{B} \psi} \aleq \eps_{e} \mbrk{u}^{-\dlt_{e}} \frkm_{p}^{2}[\psi], \\
	\mbrk{u}^{2} r \abs*{r^{-(d-1)}\rd_{A} (r^{d-1} \bfh^{A B} \upsilon_{\bfalp})  \rd_{B} \psi \psi} &\aleq \eps_{e} \mbrk{u}^{-\dlt_{e}} \tfrac{\mbrk{u}^{2}}{r^{2}} \abs{\psi \rd_{B} \psi}
	\aleq \eps_{e} r^{-\min\set{\dlt_{e}, \frac{3}{2}+\frac{p-1}{2}}} \frkm[\psi] \frkm_{p}[\psi].
\end{align*}
\item $b^{u}, c^{u} = O_{\bfGmm}^{0}(\eps_{e} r^{-\dlt_{e}-1} (\tfrac{r}{\mbrk{u}})^{-\frac{p-1}{2}} )$ $\imp$
\begin{align*}
	\mbrk{u}^{2} r (\tfrac{r}{\mbrk{u}})^{p} \abs*{c^{u} \rd_{u} \Psi \rd_{r} \Psi} r^{-(d-1)} &\aleq \eps_{e} r^{-\dlt_{e}-1} (\tfrac{r}{\mbrk{u}})^{-\frac{p-1}{2}} \mbrk{u}^{2} r (\tfrac{r}{\mbrk{u}})^{p} \abs{\rd_{u} \psi \rd_{r} \Psi} r^{-\frac{d-1}{2}} \aleq \eps_{e} r^{-\dlt_{e}} \frkm[\psi] \frkm_{p}[\psi].
\end{align*}
\item $b^{r}, c^{r} = O_{\bfGmm}^{0}(\eps_{e} r^{-\dlt_{e}} \mbrk{u}^{-1})$ $\imp$
\begin{align*}
	\mbrk{u}^{2} r (\tfrac{r}{\mbrk{u}})^{p} \abs*{c^{r} \rd_{r} \Psi \rd_{r} \Psi} r^{-(d-1)} &\aleq \eps_{e} r^{-\dlt_{e}} \mbrk{u}^{-1} \mbrk{u}^{2} r (\tfrac{r}{\mbrk{u}})^{p} (\rd_{r} \Psi)^{2} r^{-(d-1)} \aleq \eps_{e} r^{-\dlt_{e}} \frkm_{p}^{2}[\psi], \\
	\mbrk{u}^{2} r \abs*{b^{r} \bfY^{\gmm}_{\bfalp} \rd_{r} \psi \rd_{\gmm} \psi}
	&\aleq \eps_{e} r^{-\dlt_{e}} \mbrk{u}^{-1} \mbrk{u}^{2} r \abs{\rd_{r} \psi} (\abs{\rd_{u} \psi} + \abs{\rd_{r} \psi}) \aleq \eps_{e} r^{-\dlt_{e}} \frkm^{2}[\psi].
\end{align*}
\item $r b^{A}, r c^{A} = O_{\bfGmm}^{0}(\eps_{e} \mbrk{u}^{-\dlt_{e}-1} (\tfrac{r}{\mbrk{u}})^{-\frac{1}{2}})$ $\imp$
\begin{align*}
	\mbrk{u}^{2} r (\tfrac{r}{\mbrk{u}})^{p} \abs*{c^{A} \rd_{A} \Psi \rd_{r} \Psi} r^{-(d-1)} &\aleq \eps_{e} \mbrk{u}^{-\dlt_{e}-1} (\tfrac{r}{\mbrk{u}})^{-\frac{1}{2}} \mbrk{u}^{2} (\tfrac{r}{\mbrk{u}})^{p} \abs{\rd_{A} \psi \rd_{r} \Psi} r^{-\frac{d-1}{2}} \aleq \eps_{e} \mbrk{u}^{-\dlt_{e}} \frkm_{p}^{2}[\psi], \\
	\mbrk{u}^{2} r \abs*{b^{A} \bfY^{\gmm}_{\bfalp} \rd_{A} \psi \rd_{\gmm} \psi}
	&\aleq \eps_{e} \mbrk{u}^{-\dlt_{e}-1} (\tfrac{r}{\mbrk{u}})^{-\frac{1}{2}} \mbrk{u}^{2} \abs{\rd_{A} \psi}(\abs{\rd_{u} \psi} + \abs{\rd_{r} \psi})
	\aleq \eps_{e} r^{-\min\set{\dlt_{e}, \frac{p-1}{2}}} \frkm_{p}[\psi] \frkm[\psi].
\end{align*}
\item $V, W = O_{\bfGmm}^{0}(\eps_{e} r^{-1-\dlt_{e}} \mbrk{u}^{-1} (\tfrac{r}{\mbrk{u}})^{-\frac{p-1}{2}})$ $\imp$
\begin{align*}
	\mbrk{u}^{2} r(\tfrac{r}{\mbrk{u}})^{p} \abs*{W \Psi \rd_{r} \Psi} r^{-(d-1)} & \aleq \eps_{e} r^{-1-\dlt_{e}} \mbrk{u}^{-1} (\tfrac{r}{\mbrk{u}})^{-\frac{p-1}{2}} \mbrk{u}^{2} r (\tfrac{r}{\mbrk{u}})^{p} \abs{\psi \rd_{r} \Psi} r^{-\frac{d-1}{2}} \aleq \eps_{e} r^{-\dlt_{e}} \frkm[\psi] \frkm_{p}[\psi].
\end{align*}
\end{itemize}

\smallskip \noindent {\it Step~3.} Consider now $\psi$ and $g$ defined in the original spacetime region $\set{t_{0} \leq x^{0} \leq T, \, u_{h} \leq U_{0}}$. We begin by extending the conclusion of Step~1 and 2 to this case, i.e., we claim that
\begin{equation} \label{eq:en-ext:pf:1''}
\sup_{t_{0} \leq t \leq T} \frkE_{\alp; \ell^{\infty}_{\leq U_{0}}}^{2}(t)
+ \frkM_{\alp; \ell^{\infty}_{\leq U_{0}}}^{2} \aleq \frkE_{\alp; \ell^{\infty}_{\leq U_{0}}}^{2}(t_{0}) + \frkI_{\alp; \ell^{\infty}_{\leq U_{0}}}^{2}.
\end{equation}
To prove this, we consider an extension of $(\psi, \rd_{x^{0}} \psi)|_{\set{x^{0} = t_{0}}}$ on $\set{x^{0}=t_{0}, \, u_{h} \leq U_{0}}$ (resp.~$g$ on $\set{t_{0} \leq x^{0} \leq T, \, u_{h} \leq U_{0}}$) to $\set{x^{0} = t_{0}}$ (resp.~$\set{t_{0} \leq x^{0} \leq T}$) such that
\begin{align*}
	(\psi, \rd_{x^{0}} \psi)|_{\set{x^{0} = t_{0}}} &= 0 \quad \hbox{ for } u_{h} \geq \frac{1}{2} U_{0}, & \frkE_{\frac{1}{2}U_{0}}^{2}[(\psi, \rd_{x^{0}} \psi)|_{\set{x^{0} = t_{0}}}] &\aleq \frkE_{U_{0}}^{2}[(\psi, \rd_{x^{0}} \psi)|_{\set{x^{0} = t_{0}}}],  \\
	g &= 0 \quad \hbox{ for } u_{h} \geq \frac{1}{2} U_{0}, & \frkI_{\frac{1}{2}U_{0}}[g]^{2} &\aleq \frkI_{U_{0}}[g]^{2}.
\end{align*}
Indeed, such an extension can be constructed by the usual Sobolev extension operator and applying smooth cutoff $\chi_{> -U_{0}}(-u_{h})$ in case of $(\psi, \rd_{x^{0}} \psi) |_{\set{x^{0} = t_{0}}}$. In case of $g$, it suffices to simply extend it by zero. Denote by $\td{\psi}$ the solution to $\calP \td{\psi} = g$ in $\set{t_{0} \leq x^{0} \leq T}$, where both $g$ and the initial data are extended as above. By finite speed of propagation, $\td{\psi}$ agrees with $\psi$ in $\set{t_{0} \leq x^{0} \leq T, \, u_{h} \leq U_{0}}$. Hence, if we apply \eqref{eq:en-ext:pf:1'} to $\td{\psi}$, then its left-hand side bounds the left-hand side of \eqref{eq:en-ext:pf:1''} from above. Moreover, by the above extension procedire, the right-hand side of \eqref{eq:en-ext:pf:1'} is bounded by that of \eqref{eq:en-ext:pf:1''} (up to a positive constant). This completes the proof of \eqref{eq:en-ext:pf:1''}.

\smallskip \noindent {\it Step~4.} Define
\begin{align*}
\frkE_{\alp}^{2}(t) &:= \int_{\set{x^{0} = t, \, u_{h} \leq U_{0}}} \mbrk{u}^{2 (\alp - \nu_{\Box})-1} \bigg[ (\mbrk{u} \rd_{u} \psi)^{2} + (\mbrk{u} \rd_{r} \psi)^{2} + \left( (\tfrac{r}{\mbrk{u}})^{-1} \psi \right)^{2} \notag \\
	&\relphantom{:= \int_{\set{x^{0} = t, \, u_{h} \leq U_{0}}} \mbrk{u}^{2 (\alp - \nu_{\Box})-1} \bigg[}
	+ \left( (\tfrac{r}{\mbrk{u}})^{-\frac{1}{2}+\frac{p-1}{2}} r^{1-\nu_{\Box}} \rd_{r} (r^{\nu_{\Box}} \psi) \right)^{2} + \left( (\tfrac{r}{\mbrk{u}})^{-\frac{1}{2}+\frac{p-1}{2}}\abs{\rsnb \psi}\right)^{2} \bigg]  \ud \sgm, \\
\frkM_{\alp}^{2} &:= \sup_{R \in 2^{\bbZ} : R \geq R_{0}} \iint_{\set{t_{0} \leq x^{0} \leq T, \, u_{h} \leq U_{0}, \, R \leq r \leq 2 R}} \frac{\mbrk{u}^{2 (\alp - \nu_{\Box})-1}}{r} \left[(\mbrk{u} \rd_{u} \psi)^{2} + (r \rd_{r} \psi)^{2} +  \psi^{2} \right]  \ud \mathrm{V} \notag \\
	&\relphantom{:=}
	+\iint_{\set{t_{0} \leq x^{0} \leq T, \, u_{h} \leq U_{0}}} \frac{\mbrk{u}^{2 (\alp - \nu_{\Box})-1}}{r} \left[ \left( (\tfrac{r}{\mbrk{u}})^{\frac{p-1}{2}} r^{1-\nu_{\Box}} \rd_{r} (r^{\nu_{\Box}} \psi) \right)^{2} + \left( (\tfrac{r}{\mbrk{u}})^{-\frac{1}{2}+\frac{p-1}{2}}\abs{\rsnb \psi}\right)^{2} \right]\, \ud \mathrm{V}, \\
	\frkI_{\alp}^{2} &:= \left( \sum_{R \in 2^{\bbZ} : R \geq R_{0}} \left(  \iint_{\set{t_{0} \leq x^{0} \leq T, \, u_{h} \leq U_{0}, \, R \leq r \leq 2 R}} \mbrk{u}^{2(\alp - \nu_{\Box}) + 1} r g^{2} \, \ud \mathrm{V} \right)^{\frac{1}{2}} \right)^{2}.
\end{align*}
We claim that
\begin{align}
\sup_{t_{0} \leq t \leq T} \frkE_{\alp}^{2}(t)
+ \frkM_{\alp}^{2} \aleq \frkE_{\alp}^{2}(t_{0}) + \frkI_{\alp}^{2}. \label{eq:en-ext:pf:3}
\end{align}
Note that this estimate is equivalent to \eqref{eq:en-ext}, so verifying this claim would complete the proof.

The idea is to perform a simple interpolation argument based on Step~3. We introduce a smooth partition of unity $1 = \sum_{j \in \bbZ} \td{\chi}(2^{-j}(\cdot))$ on $(0, \infty)$ such that $\supp \td{\chi} \subseteq (\frac{1}{2}, 2)$ and $\sum_{j \leq j_{0}} \td{\chi}(2^{-j}(s)) = 1$ for $0 < s < 2^{j_{0}}$. Then we let $\psi_{U'}$ be the solution to $\calP \psi_{U'} = \td{\chi}((-U')^{-1}(-u_{h})) g$ with $(\psi_{U'}, \rd_{x^{0}} \psi_{U'})(t_{0}) = \td{\chi}((-U')^{-1} (-u_{h})) (\psi, \rd_{x^{0}} \psi)(t_{0})$. By linearity and finite speed of propagation, we have
\begin{equation*}
	\psi = \sum_{U' : (-U') = 2^{\bbZ_{\geq 0}} (- U_{0})} \psi_{U'}.
\end{equation*}
Given $\alp > \nu_{\Box} + \frac{p-1}{2}$, fix $\veps > 0$ such that $\alp - 2\veps > \nu_{\Box} + \frac{p-1}{2}$ as well. Applying \eqref{eq:en-ext:pf:1''} to $\psi_{U'}$, we obtain
\begin{align*}
&\sup_{t_{0} \leq t \leq T} \frkE_{\alp \pm \veps; \ell^{\infty}_{\leq U_{0}}}[(\psi_{U'}, \rd_{x^{0}} \psi_{U'})|_{\set{x^{0} = t}}]
+ \frkM_{\alp \pm \veps; \ell^{\infty}_{\leq U_{0}}}[\psi_{U'}] \\
&\aleq \frkE_{\alp \pm \veps; \ell^{\infty}_{\leq U_{0}}}[(\psi_{U'}, \rd_{x^{0}} \psi_{U'})|_{\set{x^{0} = t}}]
+ \frkI_{\alp \pm \veps; \ell^{\infty}_{\leq U_{0}}}[g_{U'}] \\
&\aleq \mbrk{U'}^{\pm \veps} \left(\sum_{U'' \aeq U'} \frkE_{\alp; U''}(t_{0})
+ \frkI_{\alp; U''}\right),
\end{align*}
where $\sum_{U'' \aeq U'}$ is the summation over $\set{U'' : (-U')^{-1}(-U'') \in \set{2^{-1}, 1}}$. It follows that
\begin{align*}
\sup_{t_{0} \leq t \leq T} \frkE_{\alp; U}[(\psi_{U'}, \rd_{x^{0}} \psi_{U'})|_{\set{x^{0} = t}}]
+ \frkM_{\alp; U}[\psi_{U'}]
\aleq \min\set*{\left(\tfrac{\mbrk{U}}{\mbrk{U'}}\right)^{\veps}, \left(\tfrac{\mbrk{U}}{\mbrk{U'}}\right)^{-\veps}} \left(\sum_{U'' \aeq U'} \frkE_{\alp; U''}(t_{0})
+ \frkI_{\alp; U''}\right).
\end{align*}
Summing up in $U'$ (to bound $\psi$ using sublinearity) and performing the $U''$ summation first, we obtain
\begin{align*}
	\sup_{t_{0} \leq t \leq T} \frkE_{\alp; U}(t) + \frkM_{\alp; U}
	\aleq \sum_{U' : (-U_{0})^{-1} U' \in 2^{\bbZ_{\geq 0}}} \min\set*{\left(\tfrac{\mbrk{U}}{\mbrk{U'}}\right)^{\veps}, \left(\tfrac{\mbrk{U}}{\mbrk{U'}}\right)^{-\veps}} \left(\frkE_{\alp; U'}(t_{0})
+ \frkI_{\alp; U'}\right).
\end{align*}
By Young's inequality, it now follows that
\begin{align*}
	\sum_{U : (-U_{0})^{-1} U \in 2^{\bbZ_{\geq 0}}} \sup_{t_{0} \leq t \leq T} \frkE_{\alp; U}^{2}(t) + \frkM_{\alp; U}^{2}
	\aleq  \sum_{U' : (-U_{0})^{-1} U' \in 2^{\bbZ_{\geq 0}}} \left(\frkE_{\alp; U'}^{2}(t_{0})
+ \frkI_{\alp; U'}^{2}\right),
\end{align*}
and this clearly implies \eqref{eq:en-ext:pf:3}, as desired. \qedhere
\end{proof}

\subsection{Bootstrap argument and the main energy estimate}\label{sec:bootstrap}

We introduce a bootstrap argument for our main energy estimate. (The bootstrap is only necessary for the nonlinear term. If we consider only a linear equation, a bootstrap argument is not needed.) For $T > t_0$, and $u_{h}$ as in \eqref{eq:ext-uh}, define the bootstrap region
\begin{equation}\label{eq:def.BT}
\calB_T:= \{(u,r,\th)\in \calM_{\ext}: t_0 \leq x^{0} \leq T,\, u_h \leq U_{0} \},
\end{equation}
where, according to the convention discussed at the beginning of Section~\ref{sec:EE}, $U_{0} \leq -1$.
Assume that for some $T > x^{0}$, the solution exists and remains regular in the region $\calB_T$ and the following estimate holds:
\begin{equation}\label{eq:BA}
	\phi = O_{\bfGmm}^{M_{0} - \f{d+7}2}(\sqrt{D} \mbrk{u}^{-(\alp-\nu_{\Box})} r^{-\nu_{\Box}}) \quad \hbox{ in } \calB_T,
\end{equation}
where, importantly, the implicit constant in $O_{\bfGmm}^{M_{0} - \f{d+7}2}$ is independent of $T$.

We will use the energy estimates in Proposition~\ref{prop:en-ext} together with the version of Klainerman's Sobolev inequality in Proposition~\ref{prop:ext-sob} below to improve the constant in \eqref{eq:BA} from $\sqrt{D}$ to $D$.

For this purpose, we commute the equation for $\phi$ with $\bfGmm^{I}$ and write
\begin{equation}\label{eq:ext-commuted.eqn}
\begin{split}
&\: \Big(\Box_m + (\bfh^{\alp\bt} - \calH^{\alp\bt}) \rd_\alp \rd_\bt + b^{\alp} \rd_\alp + V \Big) \bfGmm^{I} \phi \\
= &\: [\Box_m, \bfGmm^{I}] \phi +  \Big[\bfh^{\alp\bt} \rd_\alp \rd_\bt + b^\alp \rd_\alp + V, \bfGmm^{I} \Big] \phi
 + \Big(\bfGmm^I \Big( \calN(p,\phi,\ud \phi) \Big) - \quasi^{\mu \nu} (p, \phi, \ud \phi) \rd_{\mu} \rd_{\nu} \bfGmm^{I} \phi \Big)+ \bfGmm^I f,
\end{split}
\end{equation}
where here, and for the remainder of the subsection, we take $\bfGmm \in \{ \bfT, \bfS, \bfOmg, \bfL\}$.

In preparation of the main energy estimates in Proposition~\ref{prop:EE.ext} below, we control the terms arising on the right-hand side of \eqref{eq:ext-commuted.eqn} in the three lemmas below. In these lemmas and in the remainder of the section, we will use $\dlt_{0} := \min\set{\dlt_{c}, \dlt_{d}, \frac{1}{2} (\alp - \alp_{\calN})}$ (we remind the reader that $\alp_{\calN} = - \infty$ when $\calN = 0$).
\begin{lemma}\label{lem:bootstrap.linear}
Under the assumptions \ref{hyp:ext-g}, \ref{hyp:ext-B}, \ref{hyp:ext-V}, the following estimate holds in $\calM_{\ext}$ for every $|I| \leq M_{0}-2$:
\begin{equation}
\begin{split}
&\: \Big| \Big[ \bfh^{\alp\bt} \rd_\alp \rd_\bt + b^\alp \rd_\alp + V, \bfGmm^{I} \Big] \phi\Big| \\
\ls &\: A_c r^{-2}\mbrk{u}^{1-\de_c} |\rd_{u} \bfGmm^{\leq |I|} \phi| + A_c r^{-1} \mbrk{u}^{-\de_c} (|\rd_{r} \bfGmm^{\leq |I|} \phi| + r^{-1} |\rsnb \bfGmm^{\leq |I|} \phi|) + A_c r^{-2} \mbrk{u}^{-\de_c} |\bfGmm^{\leq |I|} \phi|.
\end{split}
\end{equation}
\end{lemma}
\begin{proof}
After recalling that $M_{0} \leq M_{c}$, this is an immediate consequence of \ref{hyp:ext-g}, \ref{hyp:ext-B}, \ref{hyp:ext-V}. \qedhere
\end{proof}

\begin{lemma}\label{lem:bootstrap.nonlinear}
Suppose that $M_{0} \geq \f{d+9}{2}$. Under the bootstrap assumption \eqref{eq:BA}, for every $|I| \leq M_{0}-2$, the nonlinear term $\calN$ satisfies the following estimate in $\calB_{T}$:
\begin{equation}
\begin{split}
&\: \Big| \bfGmm^I \Big( \calN(p,\phi,\ud \phi) \Big) - \quasi^{\mu \nu} (p, \phi, \ud \phi) \rd_{\mu} \rd_{\nu} \bfGmm^{I} \phi \Big| \\
\ls &\:  \sqrt{D} \Big( r^{-2} \mbrk{u}^{1-\de_0}  |\rd_u \bfGmm^{\leq |I|} \phi| +  r^{-1} \mbrk{u}^{-\de_{0}} (|\rd_r \bfGmm^{\leq |I|} \phi| + r^{-1} |\rsnb \bfGmm^{\leq |I|} \phi|) + r^{-2} \mbrk{u}^{-\de_0} |\bfGmm^{\leq |I|} \phi| \Big).
\end{split}
\end{equation}
\end{lemma}
\begin{proof}
Recall that $\alp_d > \alp_{\calN}$. Moreover, since $J_{c} \in \mathbb Z_{\geq 2}$, $\eta_{c}>0$ and $\alp_{\calN} \leq 2$ (see Remark~\ref{rmk:ext-alpN}), we must have $J_{c} - 1+\eta_{c}+\nB \geq 2+\eta_{c} > \alp_{\calN}$. Thus, $\alp > \alp_{\calN}$. Since $\de_0 \leq \f 12(\alp -\alp_{\calN})$, the desired estimate then follows from the definition of minimal decay exponent in Definition~\ref{def:ext-alp-N} and Definition~\ref{def:ext-alp-N-min}. (Note that the condition $M_{0} \geq \f{d+9}{2}$ ensures that $M_0 - \f{d+7}{2}\geq \lfloor \frac{M_0-2}2 \rfloor + 2$ so that \eqref{eq:BA} implies \eqref{eq:ext-nonlinear.phi.assumption} in Definition~\ref{def:ext-alp-N} holds.) \qedhere
\end{proof}

\begin{lemma}\label{lem:bootstrap.f}
For $|I|\leq M_0$, the following estimate holds in $\calM_{\ext}$:
$$\Big| \bfGmm^I f \Big| \ls D r^{-\f 32 - \f{\de_0}2-\nB} \mbrk{u}^{-\alp_{d} - \f{\de_{0}}2 +\nB- \f 12}.$$
\end{lemma}
\begin{proof}
It is easy to read off from \ref{hyp:ext-forcing} that $|\bfGmm^I f |\ls D r^{-2-\nB} \mbrk{u}^{-\alp_{d} - \dlt_{d} +\nB}$. The desired bound then follows from $r^{-1} \ls \mbrk{u}^{-1}$  and $\dlt_{0} \leq \dlt_{d}$. \qedhere
\end{proof}

\begin{lemma}\label{lem:ext-effective.commuted}
Let $\calP = \Box_{\bfm} + (\bfh^{\alp\bt} - \calH^{\alp\bt} )\rd_\alp\rd_\bt + b^\alp \rd_\alp + V$ and $\bfGmm \in \{ \bfT, \bfS, \bfOmg, \bfL\}$. Then the following holds for $|I|\leq \min\{M_0,M_c\}-2$:
\begin{equation}\label{eq:ext-effective.commuted.wave.equation}
\begin{split}
	|\calP \bfGmm^I \phi | \ls g_1 + D r^{-\f 32 - \f{\de_0}2-\nB} \mbrk{u}^{-\alp_{d} - \f{\de_{0}}2 +\nB- \f 12},
\end{split}
\end{equation}
where
$$g_1 = \max\{ A_c, \sqrt{D}\} \Big(r^{-2}\mbrk{u}^{1-\de_0} |\rd_{u} \bfGmm^{\leq |I|} \phi| + r^{-1} \mbrk{u}^{-\de_0} (|\rd_{r} \bfGmm^{\leq |I|} \phi| + r^{-1} |\rsnb \bfGmm^{\leq |I|} \phi|) +  r^{-2} \mbrk{u}^{-\de_0} |\bfGmm^{\leq |I|} \phi|\Big).$$
\end{lemma}
\begin{proof}
By the commutation properties of $[\Box_{\bfm}, \bfGmm]$ in Lemma~\ref{lem:ext-vf-comm}, it follows that $[\Box_{\bfm}, \bfGmm^I] \phi \ls |\Box_{\bfm} \bfGmm^{\leq |I|-1} \phi|$. Thus, starting with the equation \eqref{eq:ext-effective.commuted.wave.equation} and using the estimates in Lemma~\ref{lem:bootstrap.linear}--Lemma~\ref{lem:bootstrap.f}, we can argue inductively in $|I|$ to obtain the desired conclusion (recall that $0 < \dlt_{0} \leq \dlt_{c}$). \qedhere
\end{proof}

\begin{proposition}\label{prop:EE.ext}
Under the bootstrap assumption \eqref{eq:BA}, the following estimate holds in $\calB_{T}$ for every $|I|\leq M_{0}-2$:
\begin{align*}
	&\sup_{t \in [t_{0}, T]} \int_{\set{x^{0} = t}\cap \calB_T} \mbrk{u}^{2 (\alp - \nu_{\Box})-1} \bigg[ (\mbrk{u} \rd_{u} \bfGmm^I \phi)^{2} + (\mbrk{u} \rd_{r} \bfGmm^I \phi)^{2} + \left( (\tfrac{r}{\mbrk{u}})^{-1} \bfGmm^I \phi \right)^{2} \notag \\
	&\phantom{\sup_{t \in [t_{0}, T]} \int_{\set{x^{0} = t}\cap \calB_T} \mbrk{u}^{2 (\alp - \nu_{\Box})-1} \bigg[}
	+ \left( (\tfrac{r}{\mbrk{u}})^{\frac{p-1}{2}} r^{1-\nu_{\Box}} \rd_{r} (r^{\nu_{\Box}} \bfGmm^I \phi) \right)^{2} + \left( (\tfrac{r}{\mbrk{u}})^{-\frac{1}{2}+\frac{p-1}{2}}\abs{\rsnb \bfGmm^I \phi}\right)^{2} \bigg]  \ud \sgm \notag \\
	&+\sup_{R \in 2^{\bbZ_{\geq 0}} R_{0}} \iint_{\set{R \leq r \leq 2 R}\cap \calB_T} \frac{\mbrk{u}^{2 (\alp - \nu_{\Box})-1}}{r} \left[(\mbrk{u} \rd_{u} \bfGmm^I \phi)^{2} + (r \rd_{r} \bfGmm^I \phi)^{2} +  \bfGmm^I \phi^{2} \right]  \ud \mathrm{V} \notag \\
	&+\iint_{\calB_T} \frac{\mbrk{u}^{2 (\alp - \nu_{\Box})-1}}{r} \left[ \left( (\tfrac{r}{\mbrk{u}})^{\frac{p-1}{2}} r^{1-\nu_{\Box}} \rd_{r} (r^{\nu_{\Box}} \bfGmm^I \phi) \right)^{2} + \left( (\tfrac{r}{\mbrk{u}})^{-\frac{1}{2}+\frac{p-1}{2}}\abs{\rsnb \bfGmm^I \phi}\right)^{2} \right]\, \ud \mathrm{V} \aleq D^2,
\end{align*}
where the implicit constant is independent of $T$.
\end{proposition}
\begin{proof}
We now apply the energy estimate in Proposition~\ref{prop:en-ext} for equation \eqref{eq:ext-effective.commuted.wave.equation}. For the initial data term in Proposition~\ref{prop:en-ext}, we use \ref{hyp:ext-id} to obtain
\begin{equation}
\begin{split}
&\int_{\set{x^{0} = t_{0}}\cap \calB_T} \mbrk{u}^{2 (\alp - \nu_{\Box})-1} \bigg[ (\mbrk{u} \rd_{u} \bfGmm^I \phi)^{2} + (\mbrk{u} \rd_{r} \bfGmm^I \phi)^{2} \\
	&\phantom{\aleq \int_{\set{x^{0} = t_{0}, \, r \geq R_{0}}} \mbrk{u}^{2 (\alp - \nu_{\Box})-1} \bigg[}
	+ \left( (\tfrac{r}{\mbrk{u}})^{\frac{p-1}{2}} r^{1-\nu_{\Box}} \rd_{r} (r^{\nu_{\Box}} \bfGmm^I \phi) \right)^{2} + \left( (\tfrac{r}{\mbrk{u}})^{-\frac{1}{2}+\frac{p-1}{2}}\abs{\rsnb \bfGmm^I \phi}\right)^{2}  \bigg] \ud \sgm \ls D^2.
\end{split}
\end{equation}

Next, we turn to the terms on the right-hand side of \eqref{eq:ext-effective.commuted.wave.equation}. For $g_1$ as in Lemma~\ref{lem:ext-effective.commuted}, we have
\begin{equation*}
\begin{split}
&\: \left( \sum_{R \in 2^{\bbZ_{\geq 0}} R_{0}} \left(  \iint_{\set{t_{0} \leq x^{0} \leq T, \, u_{h} \leq U_{0}, \, R \leq r \leq 2 R}} r \mbrk{u}^{2 (\alp - \nu_{\Box})+1} g_1^{2} \, \ud \mathrm{V} \right)^{\frac{1}{2}} \right)^{2} \\
\ls &\: B \sup_{R \in 2^{\bbZ_{\geq 0}} R_{0}} \iint_{\set{R \leq r \leq 2 R}\cap \calB_T} \frac{\mbrk{u}^{2 (\alp - \nu_{\Box})-1}}{r} \left[(\mbrk{u} \rd_{u} \bfGmm^{\leq |I|} \phi)^{2} + (r \rd_{r} \bfGmm^{\leq |I|} \phi)^{2} \right]  \ud \mathrm{V} \\
&\: + B  \sup_{R \in 2^{\bbZ_{\geq 0}} R_{0}} \iint_{\set{R \leq r \leq 2 R}\cap \calB_T} \frac{\mbrk{u}^{2 (\alp - \nu_{\Box})-1}}{r}  \left( (\bfGmm^{\leq |I|}  \phi)^{2}  + (\tfrac{r}{\mbrk{u}})^{-\frac{1}{2}+\frac{p-1}{2}}\abs{\rsnb \bfGmm^{\leq |I|} \phi}\right)^{2} \ud \mathrm{V}.
\end{split}
\end{equation*}
where $B = \max\{A_c^2 R_0^{-\de_0}, D R_0^{-\de_0}\}$. After choosing $R_0$ large enough and $D$ small enough, this term can be controlled by the energy itself and can be absorbed to the left-hand side of the energy estimates.

The second term on the right-hand side of \eqref{eq:ext-effective.commuted.wave.equation} can then be estimated as follows:
\begin{equation*}
\begin{split}
&\: \left( \sum_{R \in 2^{\bbZ_{\geq 0}} R_{0}} \left(  \iint_{\set{R \leq r \leq 2 R}\cap \calB_T} r \mbrk{u}^{2 (\alp - \nu_{\Box})+1} (D r^{-1- \f{\de_0}2-\nB} \mbrk{u}^{-\alp_{d} - \f{\de_{0}}2 +\nB- 1})^{2} \, \ud \mathrm{V} \right)^{\frac{1}{2}} \right)^{2} \\
\ls &\: D^2 \left( \sum_{R \in 2^{\bbZ_{\geq 0}} R_{0}} \Big(\int_{R}^{2R} \int_{-\infty}^{2} \mbrk{u}^{-1-\de_0} r^{-1-\de_0} \, \ud u \, \ud r\Big)^{\f 12} \right)^{2} \ls D^2.
\end{split}
\end{equation*}

Combining the above estimates yields the desired conclusion. \qedhere
\end{proof}

In order to carry out the bootstrap argument, we will need to obtain pointwise estimates using the $L^2$ bounds above. For this purpose, we formulate a (spacetime) version of Klainerman's Sobolev inequality in the exterior region. This is similar to that in Section~\ref{subsec:kl-sob}, but is slightly modified to be adapted to our setting. 
\begin{proposition}\label{prop:ext-sob}
Let $t_{0} \geq 0$, $T \geq 1$ and $U \leq 0$. For all $(u, r, \tht) \in \set{t_{0} \leq x^{0} \leq T, \, u_{h} \leq U}$, we have
\begin{equation}\label{eq:KS.ext.goal}
\begin{split}
& \sum_{I_u + I_r + |I| \leq s} \abs{(\mbrk{u}\rd_u)^{I_u} (r\rd_r)^{I_r} \bfOmg^{I} h(u, r, \tht)}  \\
&\aleq \mbrk{u}^{-\alp+\nu_{\Box}} r^{-\nu_{\Box}} \Bigg[ \sup_{R \in 2^{\bbZ_{\geq 0}} R_{0}} \left( \iint_{\set{t_{0} \leq x^{0} \leq T, \, u_{h} \leq U, \, R \leq r \leq 2 R}} \frac{\mbrk{u'}^{2 (\alp - \nu_{\Box})-1}}{r'} \left( \bfGmm^{(\leq s+\frac{d+3}{2})} h \right) ^{2}   \ud \mathrm{V} \right)^{\frac{1}{2}} \\
&\relphantom{\aleq \mbrk{u}^{-\alp+\nu_{\Box}} r^{-\nu_{\Box}} \Bigg[}
+ \sup_{t_{0} \leq t \leq T} \left( \int_{\set{x^{0} = t, \, u_{h} \leq U}} \mbrk{u'}^{2 (\alp - \nu_{\Box}) - 1} \left((\tfrac{r'}{\mbrk{u'}})^{-1} \bfGmm^{(\leq s+\frac{d+1}{2})} h \right)^{2} \ud \sgm \right)^{\frac{1}{2}} \Bigg].
\end{split}
\end{equation}
\end{proposition}
\begin{proof}
This follows from applications of Lemma~\ref{lem:rescaled-sob}. Fix a point $p = (u,r,\th) \in \set{t_{0} \leq x^{0} \leq T, \, u_{h} \leq U}$.
\begin{enumerate}
\item When $r \leq 2 T$, we apply a \emph{spacetime} embedding as follows. Using $r\leq T$, we can find a ball (or half-ball) $B$ in $\mathbb R^d$ with radius $O(r)$ that contains $(r,\th)$, and a $u$-interval $I$ of length $O(\mbrk{u})$ that contains $u$ so that $I\times B\subset \set{t_{0} \leq x^{0} \leq T, \, u_{h} \leq U}$. By Lemma~\ref{lem:rescaled-sob} (applied both to $B$ and $I$),
$$\sum_{I_u + I_r + |I| \leq s} \abs{(\mbrk{u}\rd_u)^{I_u} (r\rd_r)^{I_r} \bfOmg^{I} h(u, r, \tht)}  \ls r^{-\f d2} \mbrk{u}^{-\f 12} \Big(\iint_{I\times B} (\bfGmm^{(\leq s+ \f{d+1}2)} h)^2\, \ud V\Big)^{\f 12},$$
which is controlled by the first term on the right-hand side of \eqref{eq:KS.ext.goal}.
\item In the $r > 2 T$ case, we use \emph{spatial} embedding. Since $(u,r,\th) \in \set{t_{0} \leq x^{0} \leq T, \, u_{h} \leq U}$, $r > 2 T$ implies that $\mbrk{u} \aeq r$. Consider the constant $x^{0}$ hypersurface $\{x^0 = t(u,r)\}$, and find a $d$-dimensional ball (or half-ball) $B \subset \{x^0 = t(u,r), u \leq U\}$ containing $p$ with radius $O(r) = O(\mbrk{u})$. Then Lemma~\ref{lem:rescaled-sob} implies
$$\sum_{I_u + I_r + |I| \leq s} \abs{(\mbrk{u}\rd_u)^{I_u} (r\rd_r)^{I_r} \bfOmg^{I} h(u, r, \tht)}  \ls \mbrk{u}^{-\f d2} \Big(\int_{B} (\bfGmm^{(\leq s+ \f{d+1}2)} h)^2\, \ud \sigma \Big)^{\f 12},$$
which is controlled by the second term on the right-hand side of \eqref{eq:KS.ext.goal} (after using again that $\mbrk{u} \aeq r$). \qedhere
\end{enumerate}
\end{proof}

We now turn to the main bootstrap argument.

\begin{proposition}\label{prop:bootstrap}
The following hold:
\begin{enumerate}
\item The unique smooth solution to \eqref{eq:main-eq} with initial data given in Theorem~\ref{thm:ext-stab} remains regular in the region $\{x^0 \geq t_0,\, u_h \leq U_{0}\}$.
\item The following energy estimates hold for $\psi = \bfGmm^I \phi$, $\bfGmm\in\{\bfT, \bfS, \bfOmg, \bfL\}$ and $|I| \leq M_{0}-2$:
\begin{align}
	&\sup_{t \in [t_{0}, \infty]} \int_{\set{x^{0} = t, \, u_{h} \leq U_{0}}} \mbrk{u}^{2 (\alp - \nu_{\Box})-1} \bigg[ (\mbrk{u} \rd_{u} \psi)^{2} + (\mbrk{u} \rd_{r} \psi)^{2} + \left( (\tfrac{r}{\mbrk{u}})^{-1} \psi \right)^{2} \notag \\
	&\phantom{\sup_{t \in [t_{0},\infty} \int_{\set{x^{0} = t, \, u_{h} \leq U_{0}}} \mbrk{u}^{2 (\alp - \nu_{\Box})-1} \bigg[}
	+ \left( (\tfrac{r}{\mbrk{u}})^{\frac{p-1}{2}} r^{1-\nu_{\Box}} \rd_{r} (r^{\nu_{\Box}} \psi) \right)^{2} + \left( (\tfrac{r}{\mbrk{u}})^{-\frac{1}{2}+\frac{p-1}{2}}\abs{\rsnb \psi}\right)^{2} \bigg]  \ud \sgm \notag \\
	&+\sup_{R \in 2^{\bbZ_{\geq 0}} R_{0}} \iint_{\set{t_{0} \leq x^{0} \leq \infty, \, u_{h} \leq U_{0}, \, R \leq r \leq 2 R}} \frac{\mbrk{u}^{2 (\alp - \nu_{\Box})-1}}{r} \left[(\mbrk{u} \rd_{u} \psi)^{2} + (r \rd_{r} \psi)^{2} +  \psi^{2} \right]  \ud \mathrm{V} \notag \\
	&+\iint_{\set{t_{0} \leq x^{0} \leq \infty, \, u_{h} \leq U_{0}}} \frac{\mbrk{u}^{2 (\alp - \nu_{\Box})-1}}{r} \left[ \left( (\tfrac{r}{\mbrk{u}})^{\frac{p-1}{2}} r^{1-\nu_{\Box}} \rd_{r} (r^{\nu_{\Box}} \psi) \right)^{2} + \left( (\tfrac{r}{\mbrk{u}})^{-\frac{1}{2}+\frac{p-1}{2}}\abs{\rsnb \psi}\right)^{2} \right]\, \ud \mathrm{V} \ls D^2. \notag
\end{align}
\item The following pointwise estimate holds in the region $\{x^0 \geq t_0,\, u_h \leq U_{0}\}$:
\begin{equation}\label{eq:ext-pointwise}
\phi = O_{\bfGmm}^{M_{0}-\f{d+7}{2}}(D \mbrk{u}^{-(\alp-\nu_{\Box})} r^{-\nB}).
\end{equation}
\end{enumerate}

\end{proposition}
\begin{proof}
We now complete the bootstrap argument. Starting with the bootstrap assumption, we apply Proposition~\ref{prop:EE.ext} to obtain
\begin{equation}
\sup_{R \in 2^{\bbZ_{\geq 0}} R_{0}} \iint_{\set{R \leq r \leq 2 R}\cap \calB_T} \frac{\mbrk{u}^{2 (\alp - \nu_{\Box})-1}}{r} ( \bfGmm^{\leq M_{0}-2} \phi)^{2}  \ud \mathrm{V} \ls D^2.
\end{equation}
Using this estimate with Proposition~\ref{prop:ext-sob}, we then obtain
\begin{equation}\label{eq:BA.done!}
	\phi = O_{\bfGmm}^{M_{0} - \f{d+7}2}(D \mbrk{u}^{-(\alp-\nu_{\Box})} r^{-\nu_{\Box}}) \quad \hbox{ in } \calB_T,
\end{equation}
which improves the bootstrap assumption \eqref{eq:BA}. A standard continuity argument shows that the solution remains regular in the full region $\calB_T$ for any $T> t_0$, i.e., the solution remains regular in $\calM_{\ext}$. This proves part (1).

Now we take the estimate in Proposition~\ref{prop:EE.ext} and take $T\to \infty$ to obtain the estimate in part (2).  Finally, after noting that $\{x^0\geq t_0,\,u_h\leq U_{0} \} = \cup_{T\geq t_0} \calB_T$, the pointwise estimate \eqref{eq:ext-pointwise} in part (3) has already been established in \eqref{eq:BA.done!} above. \qedhere
\end{proof}

\subsection{Proof of Theorem~\ref{thm:ext-stab}}\label{sec:theorem.ext.stab}

From now on, we prove our estimates allowing for loss of derivatives. We will use the notation $O_{\bfGmm}^{M_{0}-C}$ to denote estimates allowing for losing finitely many derivatives. The constant $C$ in $O_{\bfGmm}^{M_{0}-C}$ could increase from line to line, as long as it depends only on $d$, $\de_{c}$, $\de_{d}$, $J_{c}$ and $J_{d}$.

\begin{proposition}\label{prop:ext.part.3}
\begin{enumerate}
\item The Friedlander radiation field $\rPhi_0(u,\th) = \lim_{r\to \infty} r^{\nB} \phi(u,r,\th)$ is well-defined in $\calM_{\ext} \cap \set{u \leq U_{0}}$ and satisfies the estimate
\begin{equation}\label{eq:ext-radiation.field}
\rPhi_0 = O_{\bfGmm}^{M_{0}-C}(D \mbrk{u}^{-(\alp-\nu_{\Box})} ).
\end{equation}
\item From $\rPhi_{0}$, define $\rPhi_{j, k}(u, \tht)$ for $0 \leq k \leq K_{j}$, $0 \leq j \leq J - 1$, where $J= \min\set{J_{c}, J_{d}}$, using the formal recurrence equations \eqref{eq:ext-recurrence-jk} with the initial condition $\lim_{u\to -\infty} \rPhi_{j,k}(u,\th) = 0$. Then $\rPhi_{j,k}$ is well-defined for $u\leq U_{0}$ and obeys the following estimates:
\begin{equation}\label{eq:ext-higher.radiation}
\rPhi_j = O_{\bfGmm}^{M_{0}-C}(D \mbrk{u}^{-\alp+\nB+j}), \quad \forall j = 1,\ldots, J-1,\, \forall k = 0,\ldots, K_j.
\end{equation}
\item For $J = \min\{J_c,J_d\}$, define $\rho_J$ and $\Phi_{<J}$ in $\calM_{\ext} \cap \set{u \leq U_{0}}$ by
\begin{equation}\label{eq:ext-def.rhoJ}
r^{\nB} \phi(u,r,\th) = \Phi_{<J} + \rho_J,\quad \Phi_{<J} = \sum_{j=0}^{J-1} \sum_{k=0}^{K_j} r^{-j} \log^k (\tfrac r{\mbrk{u}}) \rPhi_{j,k},
\end{equation}
where $\rPhi_{j,k}$ are as in part (2) above.
Then $\rho_{J}$ satisfies the estimate
\begin{equation}\label{eq:ext.rho.final.goal}
\rho_J = \begin{cases}
O^{M_{0}-C}_{\bfGmm}(D r^{-\min\set{J_{c}-1+\eta_{c}, J_{d}-1+\eta_{d}}}) & \hbox{if $\eta_{d} <1$} \\
O^{M_{0}-C}_{\bfGmm}(D r^{-\min\set{J_{c}-1+\eta_{c}, J_{d}-1+\eta_{d}}}\log^{K_J} (\tfrac r{\mbrk{u}})\log r) & \hbox{if $\eta_{d} =1$}
\end{cases}.
\end{equation}
\end{enumerate}
\end{proposition}
\begin{proof}
\pfstep{Step~1: Estimates for the Friendlander radiation field}
Using the estimate in \eqref{eq:ext-pointwise}, together with \ref{hyp:ext-g}, \ref{hyp:ext-B}, \ref{hyp:ext-V}, \ref{hyp:ext-N} and \ref{hyp:ext-forcing}, we obtain
$$Q_0(r^{\nB}\phi) = O^{M_0-C}_{\bfGmm}(D\mbrk{u}^{-\alp+\nB-\de_{0}}r^{-2}).$$
Recalling the form of $Q_0$ in coordinates in \eqref{eq:ext-Q_0}, we deduce that
\begin{equation}\label{eq:ext-Qeqn.estimated}
\rd_u \rd_r (r^{\nB}\phi) =  O^{M_0-C}_{\bfGmm}(D\mbrk{u}^{-\alp+\nB}r^{-2}).
\end{equation}
Since $\alp >\nB$, there exists $\eta_{e} \in (0,1)$ such that $-\alp+\nB - 1+\eta_{e} <-1$. We then use $\mbrk{u} \ls r$ to bound \eqref{eq:ext-Qeqn.estimated} above by
\begin{equation}\label{eq:ext-Qeqn.estimated.2}
\rd_u \rd_r (r^{\nB}\phi) =  O^{M_0-C}_{\bfGmm}(D\mbrk{u}^{-\alp+\nB-1+\eta_{e}}r^{-1+\eta_{e}}).
\end{equation}
Using Proposition~\ref{prop:ext-sob} (with $T\to t_0$), the initial bound $\rd_r (r^{\nB}\phi) |_{x^0 = t_0} = O_{\bfGmm}^{M_0-C}(D \mbrk{u}^{-\alp+\nB} r^{-\nB})$ holds. It follows from integrating \eqref{eq:ext-Qeqn.estimated.2} that
$$\rd_r (r^{\nB} \phi) = O^{M_0-C}_{\bfGmm}(D\mbrk{u}^{-\alp+\nB+\eta_{e}}r^{-1-\eta_{e}}).$$
Since $r^{-1-\eta_{e}}$ is integrable for large $r$, this implies that
$\rPhi_0(u,r) = \lim_{r\to \infty} r^{\nB} \phi$ is well-defined. Moreover,
\begin{equation}\label{eq:ext-rPhi0.rho1}
\rPhi_0(u,r) = O^{M_{0}-C}_{\bfGmm}(D\mbrk{u}^{-(\alp-\nB)}),\quad \rho_1 := r^{\nB} \phi - \rPhi_0 = O^{M_0-C}_{\bfGmm}(D\mbrk{u}^{-\alp+\nB+\eta_{e}}r^{-\eta_{e}}).
\end{equation}
In particular, this proves point (1) of the proposition.

\pfstep{Step~2: Estimates for higher radiation fields} To control the higher radiation fields, we note that analogues of Lemma~\ref{lem:conj-wave-med} and Lemma~\ref{lem:recurrence-forcing-basic} continue to hold $\calM_{\ext}$. Thus, for $j \leq J = \min\{J_c,J_d\}$, the recurrence equations \eqref{eq:ext-recurrence-jk} imply that for $j \geq 1$,
\begin{equation}\label{eq:ext-recurrence.simple}
\begin{split}
\rd_{u} \rPhi_{j,k} = &\: \sum_{\substack{j' \leq j-1 \\ k' \leq K_{j'}}} O_{\bfGmm}^{\min\{M_0-C,M_c-2\}}(\mbrk{u}^{j-j'-1})  \sum_{I_u + I_{\bfOmg} \leq 2} (\mbrk{u}\rd_{u})^{I_u} \bfOmg^{I_{\bfOmg}}\rPhi_{j',k'} \\
&\: + \sum_{k\leq k' \leq K_{j'}} O_{\bfGmm}^{\infty}(\mbrk{u}^{-1}) \sum_{I_u + I_{\bfOmg} \leq 2} (\mbrk{u}\rd_{u})^{I_u} \bfOmg^{I_{\bfOmg}} \rPhi_{j,k'}.
\end{split}
\end{equation}

Using \eqref{eq:ext-recurrence.simple}, we can set up an induction in the following order:
\begin{equation*}
	(0,0) < (1, K_{1}) < (1, K_{1}-1) < \ldots (1, 0) < (2, K_{2}) < (2, 0) < \ldots < (J, 0).
\end{equation*}
The bound for $\rPhi_{j,k}$ has already been established in \eqref{eq:ext-radiation.field}. Then, by induction, and integrating \eqref{eq:ext-recurrence.simple} from $u=-\infty$, we obtain the estimate \eqref{eq:ext-higher.radiation}. This proves part (2) of the proposition.

\pfstep{Step~3(a): Error terms in the equation for the remainder} Part (3) of the proposition will be proven in two substeps. Denoting $\Phi = r^{\nB} \phi$ and computing as in Section~\ref{sec:remainder.prelim}, we have
\begin{equation}\label{eq:Q0rhoJ}
\begin{split}
-2\rd_u \rd_r \rho_{J} =&\:  -(Q_{0}  + 2\rd_u \rd_r) \rho_{J}+ Q_{0} \rho_{J} = -(Q_{0}  + 2\rd_u \rd_r) \rho_{J} + Q_{0} \rho_{J} + Q_{0} \Phi - Q_{0} \Phi_{<J} \\
=&\: -(Q_{0}  + 2\rd_u \rd_r) \rho_{J} -E_{J}
-   \left[ \bfh^{\alp \bt} \nb_{\alp} \rd_{\bt} \rho_{J} + \bfC^{\alp} \rd_{\alp} \rho_{J} + W \rho_{J} \right] \\
&\: +  r^{\nu_{\Box}} \calN\left(r^{-\nu_{\Box}}(\Phi_{<J} + \rho_{J}) \right) - r^{\nu_{\Box}} \calN\left(r^{-\nu_{\Box}}\Phi_{<J}\right),
\end{split}
\end{equation}
where, as in Section~\ref{sec:remainder.prelim}, $E_{J}$ is defined as
\begin{equation*}
	E_{J} = \calP \Phi_{<J} - r^{\nu_{\Box}} \calN(r^{-\nu_{\Box}} \Phi_{<J}) - r^{\nu_{\Box}} f.
\end{equation*}
To bound $E_{J}$, we argue in a similar manner as in Lemma~\ref{lem:exp-error-wave}. In particular, we decompose
\begin{equation}
E_{J} = E_{J; \Box} + E_{J; \linear} + E_{J; \nonlinear} + E_{J; f},
\end{equation}
as defined by \eqref{eq:EJ.decompose} and bound each term as in Lemma~\ref{lem:exp-error-wave}.

Similarly to \eqref{eq:EJ-Box}, the Minkowskian contribution is given by
$$E_{J; \Box} = \sum_{k=0}^{K_{J-1}} \left(\rd_{r}^{2} - \nu_{\Box} (\nu_{\Box}-1) r^{-2} + r^{-2} \rslap \right) \left(r^{-J+1} \log^{k} (\tfrac{r}{\mbrk{u}}) \rPhi_{J-1, k} \right).$$
Using the bound \eqref{eq:ext-higher.radiation}, we bound $E_{J; \Box} = O_{\bfGmm}^{M_{0}-C}(D r^{-J-1} \log^{K_{J}} (\tfrac{r}{\mbrk{u}})  \mbrk{u}^{-\alp+\nB+J-1} )$.

For the non-Minkowskian linear contribution $E_{J; \linear}$, arguing as in Step~2 of the proof of Lemma~\ref{lem:exp-error-wave}, but using the assumptions \ref{hyp:ext-g}, \ref{hyp:ext-B} and \ref{hyp:ext-V} and the bounds \eqref{eq:ext-radiation.field}, \eqref{eq:ext-higher.radiation}, we obtain
\begin{align*}
E_{J; \linear}
=  O_{\bfGmm}^{M-C}(D r^{-J-1} \log^{K_J}(\tfrac r{\mbrk{u}}) \mbrk{u}^{J-1-\alp-\de_c-\nB})+ O_{\bfGmm}^{M-C}(D r^{-J_{c}-\eta_{c}} \mbrk{u}^{J_{c}-2+\eta_{c}-\alp-\dlt_{c}+\nu_{\Box}}).
\end{align*}

For the nonlinear contribution $E_{J; \nonlinear}$, we argue as in Step~3 of the proof of Lemma~\ref{lem:exp-error-wave}, but use \eqref{eq:ext-alp-N-rNjk<J}, \eqref{eq:ext-alp-N-rem} and that $D^2 \leq D$ to obtain
\begin{align*}
E_{J; \nonlinear}
=  O_{\bfGmm}^{M-C}(D r^{-J-1} \log^{K_J}(\tfrac r{\mbrk{u}}) \mbrk{u}^{J-1-\alp-\de_0-\nB})+ O_{\bfGmm}^{M-C}(D r^{-J_{c}-\eta_{c}} \mbrk{u}^{J_{c}-2+\eta_{c}-\alp-\dlt_{0}+\nu_{\Box}}).
\end{align*}

For the contribution from $f$, we note that $E_{J;f} = f - \sum_{j=2}^{J} \sum_{k=0}^{K_{J}} r^{-j-\nu_{\Box}} \log^{k} (\tfrac{r}{\mbrk{u}}) \rf_{j, k}(u, \tht)$ (with $\rf_{j,k}$ as in \eqref{eq:ext-quasijk}) so that \ref{hyp:ext-forcing} implies
$$E_{J;f} = O_{\bfGmm}^{M_0}(D r^{-J-1} \log^{K_{d}} (\tfrac r{\mbrk{u}}) \mbrk{u}^{J-1-\alp_d-\de_0+\nB}) + O_{\bfGmm}^{M_0}(D r^{-1-\alp_{d}} \mbrk{u}^{-1-\de_0}).$$

Combining the above, we thus obtain
\begin{equation}\label{eq:ext-EJ.bound}
\begin{split}
	E_{J} &= O_{\bfGmm}^{M_{0}-C} (D r^{-J-1} \log^{K_{J}} (\tfrac{r}{\mbrk{u}}) \mbrk{u}^{J-1-\alp+\nu_{\Box}}) \\
	&\peq + O_{\bfGmm}^{M_{0}-C} (D r^{-J_{c} - \eta_{c}}  \mbrk{u}^{J_{c}+\eta_{c}-2-\de_c-\alp+\nu_{\Box}}) \\
	&\peq + O_{\bfGmm}^{M_{0}-C} (D r^{-J_{d} - \eta_{d}} \mbrk{u}^{-1-\dlt_{d}}).
\end{split}
\end{equation}

\pfstep{Step~3(b): Estimates for the remainder} Since $r^{-1} \ls \mbrk{u}^{-1}$, it follows from \eqref{eq:ext-rPhi0.rho1} and \eqref{eq:ext-higher.radiation} that
\begin{equation}\label{eq:ext.rho.initial}
\rho_J = O^{M_{0}-C}_{\bfGmm}(D\mbrk{u}^{-\alp+\nB+\eta_{e}}r^{-\eta_{e}}).
\end{equation}
We iteratively improve the bound \eqref{eq:ext.rho.initial} to get to our desired estimate. Our main claim is that for any $\nu >0$ such that $\nu-\alp+\nu_{\Box} \notin \mathbb Z$,
\begin{equation}\label{eq:ext.rho.improvement}
\begin{split}
&\: \rho_J = O^{M_{0}-C}_{\bfGmm}(D\mbrk{u}^{\nu-\alp+\nB}r^{-\nu}) \\
\implies &\: \rho_J = \begin{cases}
O^{M_{0}-C}_{\bfGmm}(D\max\{\mbrk{u}^{\nu+\f 12-\alp+\nB} r^{-\nu-\f 12} , r^{-\min\set{J_{c}-1+\eta_{c}, J_{d}-1+\eta_{d}}}\}) & \hbox{if $\eta_{d} <1$} \\
O^{M_{0}-C}_{\bfGmm}(D\max\{\mbrk{u}^{\nu+\f 12-\alp+\nB} r^{-\nu-\f 12} , r^{-\min\set{J_{c}-1+\eta_{c}, J_{d}-1+\eta_{d}}} \log^{K_J} (\tfrac r{\mbrk{u}})\log r\}) & \hbox{if $\eta_{d} =1$}
\end{cases}.
\end{split}
\end{equation}
Starting from \eqref{eq:ext.rho.initial}, we use \eqref{eq:ext.rho.improvement} repeatedly so as to achieve the desired estimate \eqref{eq:ext.rho.final.goal}.

It thus remains to prove \eqref{eq:ext.rho.improvement}. For this purpose, we assume that for some $\nu > 0$,
\begin{equation}\label{eq:ext.rho.improvement.assumption}
\begin{split}
\rho_J = O^{M_{0}-C}_{\bfGmm}(D\mbrk{u}^{\nu-\alp+\nB}r^{-\nu}).
\end{split}
\end{equation}
By \eqref{eq:ext.rho.improvement.assumption}, we use \eqref{eq:ext-Q_0} and \ref{hyp:ext-g}, \ref{hyp:ext-B}, \ref{hyp:ext-V} to obtain
\begin{align}
(Q_0 + 2\rd_u \rd_r)\rho_J = O^{M_{0}-C}_{\bfGmm}(D\mbrk{u}^{\nu-\alp+\nB}r^{-\nu-2}), \label{eq:ext-rhoJ.error.2} \\
\bfh^{\alp \bt} \nb_{\alp} \rd_{\bt} \rho_{J} - \bfC^{\alp} \rd_{\alp} \rho_{J} - W \rho_{J}  = O^{M_{0}-C}_{\bfGmm}(D\mbrk{u}^{\nu-\alp+\nB}r^{-\nu-2}\log^{K_{c}'} (\tfrac{r}{\mbrk{u}})).\label{eq:ext-rhoJ.error.3}
\end{align}
For the nonlinear term, we apply Definition~\ref{def:ext-alp-N}.(1) and the estimates on $\rPhi_{j,k}$ and $\rho_J$ to obtain
\begin{equation}\label{eq:ext-rhoJ.error.4}
r^{\nu_{\Box}}\calN(r^{-\nu_{\Box}} (\Phi_{<J} + \rho_{J})) - r^{\nu_{\Box}}\calN(r^{-\nu_{\Box}} \Phi_{<J}) = O^{M_{0}-C}_{\bfGmm}(D\mbrk{u}^{\nu-\alp+\nB}r^{-\nu-2}).
\end{equation}

Thus, plugging the estimates \eqref{eq:ext-EJ.bound}, \eqref{eq:ext-rhoJ.error.2}, \eqref{eq:ext-rhoJ.error.3} and \eqref{eq:ext-rhoJ.error.4} into \eqref{eq:Q0rhoJ}, we obtain
\begin{equation*}
\begin{split}
	- 2 \rd_{u} \rd_{r} \rho_{J} =&\: O_{\bfGmm}^{M_{0}-C}(D r^{-\nu-2} \log^{K_{c}'} (\tfrac{r}{\mbrk{u}})\mbrk{u}^{\nu-\alp+\nu_{\Box}}) \\
	&\: + O_{\bfGmm}^{M_{0}-C} (D r^{-J-1} \log^{K_{J}} (\tfrac{r}{\mbrk{u}}) \mbrk{u}^{J-1-\alp+\nu_{\Box}}) \\
	&\: + O_{\bfGmm}^{M_{0}-C} (D r^{-J_{c} - \eta_{c}}  \mbrk{u}^{J_{c}+\eta_{c}-2-\de_{c}-\alp+\nu_{\Box}}) \\
	&\: + O_{\bfGmm}^{M_{0}-C} (D r^{-J_{d} - \eta_{d}} \mbrk{u}^{-1-\dlt_{d}}).
\end{split}
\end{equation*}
Let us first consider the case $\eta_{d} <1$. Integrating from the initial hypersurface, we obtain
\begin{align*}
\rd_{r} \rho_{J} &= O_{\bfGmm}^{M_{0}-C} (D r^{-\nu-2}  \max\set{\log^{K_{c}'} (\tfrac{r}{\mbrk{u}}) \mbrk{u}^{\nu-\alp+\nu_{\Box}+1}, r^{\nu-\alp+\nu_{\Box}+1}}) \\
&\peq + O_{\bfGmm}^{M_{0}-C} (D r^{-J-1}  \max\{ \log^{K_{J}} (\tfrac{r}{\mbrk{u}}) \mbrk{u}^{J-\alp+\nu_{\Box}}, r^{J-\alp+\nu_{\Box}}\}) \\
&\peq + O_{\bfGmm}^{M_{0}-C} (D r^{-J-\eta_{c}}  \max\{  \mbrk{u}^{J_{c}-1+\eta_{c}-\alp-\f{\de_{c}}2+\nu_{\Box}}, r^{J_{c}-1+\eta_{c}-\alp-\f{\de_c}2+\nu_{\Box}}\}) \\
	&\peq + O_{\bfGmm}^{M_{0}-C} (D r^{-J_{d} - \eta_{d}} \mbrk{u}^{-\dlt_{d}}),
\end{align*}
where we used that  $\nu-\alp+\nu_{\Box} \notin \mathbb Z$, $J-\alp+\nu_{\Box} \neq 0$ and noted that $| r - t_0| \ls r$. Then since 
$$(\tfrac{\mbrk{u}}{r})^{\f 12} \log^{K_{c}'} (\tfrac{r}{\mbrk{u}}),\,(\tfrac{\mbrk{u}}{r})^{1-\eta_c} \log^{K_{c}'} (\tfrac{r}{\mbrk{u}})\ls 1,$$ we obtain
\begin{equation}\label{eq:ext-rho.to.be.integrated}
\begin{split}
\rd_{r} \rho_{J} = &\: O_{\bfGmm}^{M_{0}-C} (D r^{-\nu-\f 32}  \max\set{\mbrk{u}^{\nu-\alp+\nu_{\Box}+\f 12}, r^{\nu-\alp+\nu_{\Box}+\f 12}}) \\
&\: + O_{\bfGmm}^{M_{0}-C} (D r^{-J- \eta_c}  \max\{ \mbrk{u}^{J-\alp-1+ \eta_c +\nu_{\Box}}, r^{J-\alp-1+ \eta_c +\nu_{\Box}}\})  + O_{\bfGmm}^{M_{0}-C} (D r^{-J_d-\eta_d} ) \\
= &\: O_{\bfGmm}^{M_{0}-C}(D r^{-\nu-\f 32}\mbrk{u}^{\nu-\alp+\nu_{\Box}+\f 12}) + O_{\bfGmm}^{M_{0}-C} (D r^{-\min\{J_c+\eta_c, J_{d} + \eta_{d} \}}),
\end{split}
\end{equation}
where in the last line we used \eqref{eq:ext-alp.def}.

Since $\rho_{J} \to 0$ as $r \to \infty$ (by \eqref{eq:ext.rho.improvement}), we may integrate \eqref{eq:ext-rho.to.be.integrated} from infinity to obtain
\begin{equation}
\begin{split}
\rho_{J} = &\: O_{\bfGmm}^{M_{0}-C}(D r^{-\nu-\f 12}\mbrk{u}^{\nu-\alp+\nu_{\Box}+\f 12}) + O_{\bfGmm}^{M_{0}-C} (D r^{-\min\{J_c - 1 + \eta_c, J_{d} - 1 + \eta_{d}\}}),
\end{split}
\end{equation}
which is what we wanted to prove when $\eta_{c},\,\eta_{d} <1$ (see \eqref{eq:ext.rho.improvement}).

In the case $\eta_{d} = 1$, instead of \eqref{eq:ext-rho.to.be.integrated}, we have an addition term of $$O_{\bfGmm}^{M_{0}-C} (D r^{-J-1} \log^{K_{J}}(\tfrac r{\mbrk{u}}) \log r).$$ Arguing as before but carrying the extra factor of $\log^{K_{J}}(\tfrac r{\mbrk{u}}) \log r$, we obtain the desired estimate \eqref{eq:ext.rho.improvement} for this case. \qedhere
\end{proof}

\subsection{Putting everything together}

We are now ready to put everything together to conclude the proof of Theorem~\ref{thm:ext-stab}.
\begin{proof}[Proof of Theorem~\ref{thm:ext-stab}]
For Part (1) of Theorem~\ref{thm:ext-stab}, that $\ud u_{h}$ is timelike is an immediate consequence of the assumptions of the theorem and the direct computation
\begin{equation*}
	\ud u_{h} = \left( 1 - \gmm_{h}(r) \tfrac{\ud}{\ud s}\mbrk{s} \big|_{s=u} \right) \ud u + 2 A_{h} \mbrk{u} r^{-1-\dlt_{h}} \ud r.
\end{equation*}
The statement $\f 34 < \f{\mbrk{u_h}}{\mbrk{u}} < \f 54$ follows from the fact that $u$ and $u_h$ can be chosen to be arbitrarily close.

Part (2) of Theorem~\ref{thm:ext-stab} follows from Proposition~\ref{prop:bootstrap} after choosing $C_{e\ast} = \f{d+7}2$ and $m_{e\ast} = \f{d+9}2$.

Finally, part (3) of Theorem~\ref{thm:ext-stab} follows from Proposition~\ref{prop:ext.part.3} after we choose $C_{e\ast}'$ sufficiently large so that \eqref{eq:ext-radiation.field}, \eqref{eq:ext-higher.radiation} and \eqref{eq:ext.rho.final.goal} hold with $C = C_{e\ast}'$ and take $m_{e\ast}' = \max\{C_{e\ast}', \f{d+9}{2}\}$. \qedhere
\end{proof}

\appendix

\section{Notation and parameters}\label{sec:app}
\begin{itemize}
\item $\bbZ_{\geq 0}$, $\bbN$: $\bbZ_{\geq 0} = \set{0, 1, 2, \ldots}$ and $\bbN = \set{1, 2, \ldots}$.
\item $d$: spatial dimension of manifold $\calM$. $d\in \bbN$ odd with $d \geq 3$.
\item $\calM_{\near}$, $\calM_{\far}$, $\calM_{\med}$, $\calM_{\wave}$ and $\calM_{\ext}$: $\calM = \calM_{\near} \cup \calM_{\far} \cup \calM_{\ext}$ and $\calM_{\far} = \calM_{\med} \cup \calM_{\wave}$; see Section~\ref{sec:decomposition}.
\item $R_{\far}$: a parameter used in the definition of $\calM_{\far}$ and $\calM_{\near}$.
\item $\eta_{0}$: $\eta_{0} = \f 14$ used in the definition of $\calM_{\med}$ and $\calM_{\wave}$.
\item $\eta_{1}$: small parameter for cutoff function in $\calM_{\med}$. See beginning of Section~\ref{sec:med.error}.
\item $\phi$: solution to the main equation \eqref{eq:main-eq}.
\item $\Phi$, $\rPhi_{j,k}$, $\rcPhi_{j,k}$: $\Phi = r^{\nB} \phi$, while $\rPhi_{j,k}$ and $\rcPhi_{j,k}$ are the higher radiation field (see \eqref{eq:formal.expansion}, \eqref{eq:ext-formal.expansion}) and the renormalized higher radiation field (see \eqref{eq:rcPhi-rPhi}) respectively.
\item $f$: denotes the inhomogeneous term in the main equation \eqref{eq:main-eq}.
\item $\calN$, $\calH$, $\calF$: $\calN$ is the nonlinearity in the main equation \eqref{eq:main-eq}, while $\quasi$ and $\semi$ are the quasilinear and semilinear terms; see \ref{hyp:nonlin}.
\item $\csemi$: $\csemi = \semi + \quasi^{\mu \nu}  \left( {}^{(\bfg)}\Gmm^{\alp}_{\mu \nu} - {}^{(\bfm)} \Gmm^{\alp}_{\mu \nu} \right) \rd_{\alp} \phi$; it is the semilinear nonlinearity that arises by using the Minkowskian covariant derivative.
\item $\alp_{0}$: initial decay rate assumed on the solution $\phi$.
\item $\alp_{d}$: decay rate of the initial data and $f$.
\item $\alp_{\calN}$: minimal decay exponent for $\calN$; this is the minimal decay of $\phi$ needed to regard the linearization of $\calN$ as obeying the same decay assumptions as the coefficients in $\calP$. See Definition~\ref{def:alp-N-min}, Proposition~\ref{prop:alp-N}, Definition~\ref{def:alp-N} and Definition~\ref{def:ext-alp-N-min}.
\item $\alp_{\calN}'$: admissible decay exponent for $\calN$; see Definition~\ref{def:alp-N} and Definition~\ref{def:ext-alp-N}.
\item $\alp_{\mathfrak f}$: polynomial part of the final decay rate of the solution $\phi$ obtained by the main theorems.
\item $K_{\mathfrak f}$: logarithmic part of the final decay rate of the solution $\phi$ obtained by the main theorems.
\item $K_j$: the maximal growth rate of the logarithmic part for monomials of the form $\rPhi_{j,k} r^{-j} \log(\tfrac ru)$ or $\rPhi_{j,k} r^{-j} \log(\tfrac r{\mbrk{u}})$; see Lemma~\ref{lem:Kj-def}.
\item $K_{c}$: an integer counting the maximum power of log terms in the coefficients and the nonlinearity; see \ref{hyp:wave-g}--\ref{hyp:wave-V}\ref{hyp:nonlin-wave}, \ref{hyp:ext-g}, \ref{hyp:ext-B}, \ref{hyp:ext-V} and \ref{hyp:ext-N}.
\item $K_{c}'$: an integer counting the maximum power of log terms in the coefficients after writing the equations in terms of $Q_0$; see Lemma~\ref{lem:conj-wave-wave}.
\item $K_{d}$: an integer counting the maximum power of log terms in the initial data and $f$.
\item $\dlt_{c}$: a small parameter quantifying the decay of the coefficients towards the Minkowskian values.
\item $\dlt_{d}$: a small parameter used in the assumptions for $f$.
\item $\dlt_{a}$: $\dlt_{a} \in (0,\f{2\de_{\mathfrak f}}{d-1})$, a small parameter used to define spacetime regions in Theorems~\ref{thm:lower} and \ref{thm:lower-sphsymm}.
\item $\dlt_{0}$: a small parameter used in the inductive improvement of the decay rate satisfying \eqref{eq:dlt0-wave}.
\item $\dlt_{\mathfrak f}$: a small parameter used to quantify the improved $\tau$-decay of the remainder in Theorems~\ref{thm:lower} and \ref{thm:lower-sphsymm}.
\item $\dlt_{\mathfrak f}'$: a small parameter used to quantify the improved $u$-decay of the remainder in Proposition~\ref{prop:wave-main}, Case~4.
\item $\dlt_{m}$: a small parameter used to define spacetime regions in Section~\ref{sec:Minkowski.wave}
\item $\dlt_{h}$: a small parameter used in the definition of $u_{h}$; see \eqref{eq:ext-uh}.
\item $\dlt_{e}$: a small parameter quantifying the decay of the coefficients (towards the Minkowskian values) in the exterior energy estimate, Proposition~\ref{prop:en-ext}.
\item $\eps_{e}$: a small parameter quantifying the size of the coefficients (relative to the Minkowskian values) in the exterior energy estimate, Proposition~\ref{prop:en-ext}.
\item $\bt$: a small parameter for defining $U_0 = U^{1-\bt}$ in the proof of the sharp asymptotics theorem; see \eqref{eq:U0.def}.
\item $\eta_{c} \in (0,1]$: a small parameter quantifying the improved $r$-decay of the remainder of the coefficients in $\calM_{\wave}$
\item $\eta_{d} =\alp_{d} -\nu_{\Box}- J_{d}+1 \in (0,1]$.
\item $\eta_{\mathfrak f}$: a small parameter quantifying the improved $r$-decay of the remainder in Proposition~\ref{prop:wave-main}, Case~2.
\item $\eta_{\mathfrak f}'$: a small parameter satisfying \eqref{eq:wave-main-etaf} quantifying the improved $r$-decay in Proposition~\ref{prop:wave-main}, Case~2(b).
\item $\eta_{w}$: a small parameter quantifying the improved $r$-decay of the remainder in Proposition~\ref{prop:wave-0}.
\item $\eta_{e}$: a small parameter quantifying the $r$-decay in Proposition~\ref{prop:ext.part.3}.
\item $\nu_{\Box}$: sharp $r$-decay rate of free waves on $\bbR^{d+1}$, i.e., $\nu_{\Box} = \frac{d-1}{2}$.
\item $\nu_{0}$: $r^{-\frac{d-1}{2} + \nu_{0}}$ is the initial $r$-decay in $\calM_{\wave}$ assumed on the solution.
\item $s_{c}$: regularity loss in the stationary bound.
\item $\bfm$: Minkowski metric on $\bbR^{d+1}$.
\item $\nb$: covariant derivative on $(\bbR^{d+1}, \bfm)$.
\item $\rsgmm$: metric on the unit round sphere $\bbS^{d-1}$.
\item $\rssgm$: volume form on the unit round sphere $\bbS^{d-1}$.
\item $\rslap$: angular Laplacian on the unit round sphere $\bbS^{d-1}$.
\item $\rsnb$: covariant derivative on the unit round sphere $\bbS^{d-1}$.
\item $(x^{0}, x^{1}, \ldots, x^{d})$: globally defined rectangular coordinates on $\calM$. If we write $x$, it refers only to the spatial coordinates $(x^{1}, \ldots, x^{d})$. 
\item $\tau_{0}$, $U_{0}$: cutoff parameters used in near-intermediate zone (Section~\ref{sec:near}) and intermediate zone (Section~\ref{sec:med}), respectively.
\item $(u, r, \tht)$: Bondi-type coordinates in $\calM_{\far} \cup \calM_{\ext}$, defined as defined as $r = (\sum_{j=1,\ldots d} (x^{j})^{2})^{\frac{1}{2}}$, $u = x^{0} - r + 3 R_{\far}$, and $\tht = \frac{x}{r} \in \bbS^{d-1}$.
\item $\tau$: a smooth function that agrees with $x^{0}$ when $r \leq 2 R_{\far}$ and $u$ when $r \geq 4 R_{\far}$; see \eqref{eq:tau.def}.
\item $\mbrk{u}$: a smooth function that is even, nondecreasing in $(0, \infty)$, equals $u$ for $\abs{u} > 1$ and $\frac{1}{2}$ for $\abs{u} < \frac{1}{2}$; see Section~\ref{sec:E-assumptions}.
\item $u_{h}$: a modification of $u$ such that $\ud u_{h}$ is timelike; see \eqref{eq:ext-uh}.
\item $U_{h}(t, R)$: the value of $u_{h}$ when $(x^{0}, r) = (t, R)$.
\item $A_{0}$: the initially assumed size of solution in \ref{hyp:sol}.
\item $A_{\mathfrak f}$, $A'_{\mathfrak f}$: the upper bound on the size of the solution in Theorems~\ref{thm:upper} and \ref{thm:lower}, respectively.
\item $A_{\calN}$: a positive function used to quantify the size of the nonlinearity; see Definitions~\ref{def:ext-alp-N} and~\ref{def:alp-N}.
\item $A_{h}$: a positive number used in the definition of $u_{h}$; see \eqref{eq:ext-uh}.
\item $m_{0}$, $m'_{0}$, $m_{0, \ell}$ and $m_{0, \ell}'$: the lower bound on the total number of derivatives of the initial data, inhomogeneous terms and the coefficients needed to apply Theorems~\ref{thm:upper}, \ref{thm:lower}, \ref{thm:upper-sphsymm} and \ref{thm:lower-sphsymm}, respectively.
\item $M_0$, $M_{c}$: parameters for the total number of derivatives of the initial data, inhomogeneous terms (for $M_0$) and the coefficients (for $M_c$).
\item $D$: a parameter quantifying the size of the initial data and the inhomogeneous term.
\item $\calC_{U}$, $\ul{\calC}_{V}$, $\calD_{U_{0}}^{U}$: regions in $\calM_{\far}$ defined with respect to the approximate double null coordinates
\item $D_{U, R}$: region in $\calM_{\far}$ arising from the strong Huygens principle.
\item $C_{T}^{U}$, $C_{T}^{R}$: regions in $\calM_{\far}$.
\item $\calP$: main linear operator.
\item ${}^{(\tau)} \calP$: main linear operator with coefficients frozen at $\tau = const$.
\item ${}^{(\tau)} \calP_{0}$: zero time frequency part of ${}^{(\tau)} \calP$, i.e., obtained by dropping $\rd_{\tau}$ from ${}^{(\tau)} \calP$; see \ref{hyp:ult-stat} and \ref{hyp:ult-stat'}.
\item ${}^{(\tau)} \calR_{0}$: right-inverse of ${}^{(\tau)} \calP_{0}$; see \ref{hyp:ult-stat} and \ref{hyp:ult-stat'}.
\item $\Box_{\bfg}$: the d'Alembertian associated with the metric $\bfg$.
\item $\Box_{\bfm}$: the Minkowskian d'Alembertian.
\item $\Delta_{\bfe}$: the Euclidean Laplacian.
\item $\calQ$: conjugated linear operator $r^{\nu_{\Box}} \calP = \calQ r^{\nu_{\Box}}$.
\item $Q_{0}$: conjugated Minkowskian d'Alembertian $r^{\nu_{\Box}} \Box_{\bfm} = Q_{0} r^{\nu_{\Box}}$.
\item $\mathbb{S}_{(\ell)}$: projection to the $\ell$-th spherical harmonics.
\item $n_{\calN}$: minimal degree of the nonlinearity.
\item $\td{\frkL}$: notation to keep track of $\left( \rcPhi_{(\leq J - \nu_{\Box})J, 0}(\infty), (\rcPhi_{J, k}(\infty))_{k=1,\ldots, K_{J}} \right) \neq 0$; see \eqref{eq:tdfrkL.def}.
\item $O_{\bfGmm}^{N}$: notation for vector field regularity bounds; see \eqref{eq:O-Gmm-far-scalar}, \eqref{eq:O-Gmm-far}, \eqref{eq:O-Gmm-ext-scalar}, \eqref{eq:O-Gmm-ext}.
\item $\chi_{>1} : \bbR \to [0, 1]$ is a smooth nondecreasing function supported in $(\frac{1}{2}, \infty)$ and equals $1$ on $[1, \infty)$.
\item $\chi_{> \lmb}(\cdot) = \chi_{>1}(\lmb^{-1} (\cdot))$.
 \item $\bfT$: almost stationary vector field on $\calM$, equals $\rd_{x^{0}}$ in the global coordinates on $\calM$.
 \item $\bfS$: scaling vector field in $\calM_{\far}\cup \calM_{\ext}$; $u \rd_{u} + r \rd_{r}$ in Bondi coordinates.
 \item $\bfOmg_{ab}$: rotation vector field in $\calM_{\far} \cup \calM_{\ext}$, equals $x^a \rd_{x^b} - x^b \rd_{x^a}$ in the global coordinates.
 \item $\bfK$: $r^{2} \rd_{r}$ in Bondi coordinates on $\calM_{\far}$.
 \item $\bfL_{a}$: Lorentz boost vector fields in $\calM_{\ext}$, equals $(x^{0} + 3 R_{\far}) \rd_{a} + x^{a} \rd_{0}$ in the global coordinates.
 \item ${}^{(\tau)} \Sgm$: level-$\tau$ hypersurface when $\tau$ is finite, and the moduli space of integral curves of $\bfT$ when $\tau = \infty$ (which is diffeomorphic to ${}^{(\tau_{0})}\Sgm$ with $\tau_{0}$ any finite number); see \ref{hyp:ult-stat} and \ref{hyp:ult-stat'}.
\item $\ell^{p} \calH^{s, \gmm}$: weighted Sobolev spaces on ${}^{(\tau)} \Sgm$; see Definition~\ref{def:w-soblev}.

\end{itemize}

\subsubsection*{Schematic notation} The following notation may mean different specific expressions in different places.
\begin{itemize}
\item $\rd^{(m)}$: an $m$-th order partial derivative with respect to the global coordinate system $(x^{0}, x^{1}, \ldots, x^{d})$, unless otherwise stated
\item $\rd^{(\leq m)}$: an $m'$-th order partial derivative with $m' \leq m$ with respect to the global coordinate system $(x^{0}, x^{1}, \ldots, x^{d})$, unless otherwise stated
\item $\tdGmm^{(\leq m)}$: an $m'$-fold application of vector fields $\tdGmm \in \set{\bfS, \bfOmg}$ with $m' \leq m$
\item $\bfGmm^{(\leq m)}$: an $m'$-fold application of vector fields $\bfGmm \in \set{\bfS, \bfOmg, r \rd_{r}}$ with $m' \leq m$
\item $\urd^{(\leq m)}$: an $m'$-th order partial derivative with $m' \leq m$ with respect to the coordinate system $(x^{1}, \ldots, x^{d})$ on ${}^{(\tau)} \Sgm$
\end{itemize}

\bibliographystyle{DLplain}
\bibliography{tails}

\end{document}